\documentclass[3p,shortbib,onecolumn]{elsarticle}
\pdfoutput=1 
    
\usepackage{
    amsmath,
    amssymb,
    graphicx}
\usepackage[utf8]{inputenc}
\usepackage[backref=page]{hyperref} 
\hypersetup{colorlinks=true,citecolor=blue} 
\usepackage{slashed}
\usepackage{bbold}
\usepackage{enumitem} 

\usepackage{tabularx}
\usepackage{bm}
\usepackage[export]{adjustbox}

\newcommand{\be}{\begin{equation}}
\newcommand{\ee}{\end{equation}}
\newcommand{\ba}{\begin{align}}
\newcommand{\ea}{\end{align}}
\newcommand{\bi}{\begin{itemize}}
\newcommand{\ei}{\end{itemize}}
\newcommand{\tsf}[1]{\textsf{#1}}
\newcommand{\tsfs}[1]{{\tiny \textsf{#1}}}

\newcommand{\mbf}[1]{\mathbf{#1}}
\newcommand{\trm}[1]{\textrm{#1}}
\newcommand{\bracket}[2]{\bra{#1}\,#2\rangle} 
\newcommand{\bra}[1]{\langle\,#1\,|}          
\newcommand{\ket}[1]{|\,#1\,\rangle}          
\newcommand{\ud}{{\mathrm{d}}} 
\newcommand{\bs}[1]{\boldsymbol{#1}}
\newcommand{\eps}{\varepsilon}
\newcommand{\prob}{\textsf{P}}
\newcommand{\rate}{\textsf{R}}
\newcommand{\vphi}{\varphi}
\newcommand{\Ai}{\textrm{Ai}}
\newcommand{\eqnref}[1]{Eq. (\ref{#1})}
\newcommand{\figref}[1]{Fig. \ref{#1}}
\newcommand{\secref}[1]{Sec. \ref{#1}}
\newcommand{\Eschwinger}{E_{S}}
\renewcommand{\Im}[1]{\mathrm{Im}}
\newcommand{\LCm}{{\scriptscriptstyle -}} 
\newcommand{\LCp}{{\scriptscriptstyle +}}
\newcommand{\LCpm}{{\scriptscriptstyle \pm}}
\newcommand{\LCmp}{{\scriptscriptstyle \mp}}

\newcommand{\LCperp}{{\scriptscriptstyle \perp}}
\usepackage[usenames,dvipsnames]{xcolor}
\usepackage{tikz} 

\newcommand{\nphi}{\bar{\phi}}

\newcommand{\ilfrac}[2]{\frac{#1}{#2}}

\newcommand{\stokes}{\mathbf{n}}
\newcommand{\phaselengthparameter}{\Phi}
\newcommand{\dynamicphase}{\Psi}
\let\vec\mathbf

\hfuzz=\maxdimen \tolerance=10000
\vfuzz=\maxdimen \tolerance=10000
\begin{document}

\begin{frontmatter}

\title{Advances in QED with intense background fields}

\author[AFadd]{A.~Fedotov}
\ead{amfedotov@mephi.ru}
\author[AIadd]{A.~Ilderton}
\ead{anton.ilderton@ed.ac.uk}
\author[FKadd1,FKadd2,FKadd3]{F.~Karbstein}
\ead{felix.karbstein@uni-jena.de}
\author[BKadd]{B.~King}
\ead{b.king@plymouth.ac.uk}
\author[FKadd1,FKadd2]{D.~Seipt}
\ead{d.seipt@hi-jena.gsi.de}
\author[HTadd]{H.~Taya}
\ead{hidetoshi.taya@riken.jp}
\author[GTadd1,GTadd2]{G.~Torgrimsson}
\ead{greger.torgrimsson@umu.se}

\address[AFadd]{National Research Nuclear University MEPhI, Kashirskoe sh. 31, Moscow, 115409, Russia}
\address[AIadd]{Higgs Centre, School of Physics and Astronomy,
University of Edinburgh, EH9 3JZ, Scotland, UK}
\address[FKadd1]{Helmholtz-Institut Jena, Fr\"obelstieg 3, 07743 Jena, Germany}
\address[FKadd2]{GSI Helmholtzzentrum f\"ur Schwerionenforschung, Planckstra\ss e 1, 64291 Darmstadt, Germany}
\address[FKadd3]{Theoretisch-Physikalisches Institut, Friedrich-Schiller-Universit\"at Jena, Max-Wien-Platz 1, 07743 Jena, Germany}
\address[BKadd]{Centre for Mathematical Sciences, University of Plymouth, PL4 8AA, UK}
\address[HTadd]{RIKEN iTHEMS, RIKEN, Wako 351-0198, Japan}
\address[GTadd1]{Helmholtz-Zentrum Dresden-Rossendorf, Bautzner Landstra\ss e 400, 01328 Dresden, Germany}
\address[GTadd2]{Department of Physics, Ume{\aa} University, SE-901 87 Ume\aa, Sweden}

\begin{abstract}
Upcoming and planned experiments combining increasingly intense lasers and energetic particle beams will access new regimes of nonlinear, relativistic, quantum effects. This improved experimental capability has driven substantial progress in QED in intense background fields. We review here the advances made during the last decade, with a focus on theory and phenomenology. As ever higher intensities are reached, it becomes necessary to consider processes at higher orders in both the number of scattered particles and the number of loops, and to account for non-perturbative physics (e.g.~the Schwinger effect), with extreme intensities requiring resummation of the loop expansion. {In addition to} increased intensity, experiments will reach higher accuracy, {and these improvements are being matched by developments in theory such as in} approximation frameworks, the description of finite-size effects, and the range of physical phenomena analysed. Topics on which there has been substantial progress include: radiation reaction, spin and polarisation, nonlinear quantum vacuum effects and connections to other fields including physics beyond the Standard Model.

\end{abstract}

\end{frontmatter}
\setcounter{tocdepth}{2}
\tableofcontents{}

\newpage

\section{Introduction}\label{sec:intro}
In high-intensity laser pulses, electrons can be accelerated to relativistic velocities over a single laser wavelength. Such lasers, made possible by chirped pulse amplification~\cite{Strickland85} for which the 2018 Nobel prize was awarded~\cite{nobellink}, have great potential not only for applications in the sciences, industry and medicine, but also as a tool to probe fundamental quantum physics. Pulses of light in which the photon density surpasses one photon per Compton wavelength cubed are now routinely produced at modern laser facilities. {They provide a means of experimental investigation complementary to} accelerator searches for probing the `intensity frontier' \cite{Proceedings:2012ulb} of the Standard Model.

Laser light is well described by a coherent state, in which the interaction of laser photons with charged matter adds coherently. In high intensity laser pulses, the charge-field coupling becomes large enough that the perturbative hierarchy is disrupted and the interaction between the charge and the laser must be accounted for to all orders in perturbation theory, or non-perturbatively. This is an example of `non-perturbativity at weak coupling'; while the fine structure constant $\alpha$ remains small, the photon density $\rho$ in a laser scales as $\rho\sim\xi^2/\alpha$ in which the dimensionless intensity parameter, $\xi$, defined below, nowadays easily exceeds unity, and the effective charge-field coupling is $\sqrt{\alpha\rho} \sim \xi \gg1$.

This situation is in contrast to existing high-precision tests of quantum electrodynamics (QED), where electromagnetic fields are low intensity and calculations can be performed perturbatively. For example, the electron anomalous magnetic moment and fine-structure constant have been measured to agree with theory up to order $O(\alpha^{5})$, or better than one part in a billion\footnote{{A precise measurement of the electron magnetic moment in highly charged ions allows precise tests of bound state QED \cite{Sturm:2011uow,Sailer:2022azt}, for instance on the reliability of perturbative expansions in $Z\alpha$ \cite{2015JPCRD..44c1205S,Yerokhin:2017sfg}. {Also note the the electric fields probed by the ground state of high-$Z$ atoms can be of the order of the critical, or `Schwinger', field of QED, see below Eq.~(\ref{eq:chi:def}).}}}~\cite{Hanneke:2010au, Aoyama:2017uqe}.
Other recent high-profile tests of QED in ultra-peripheral heavy ion collisions, where light-by-light scattering (ATLAS \cite{ATLAS:2017fur,ATLAS:2019azn} and CMS \cite{CMS:2018erd}) and linear Breit-Wheeler pair-creation (STAR \cite{STAR:2019wlg}) have been measured for the first time, are found to be consistent with perturbative QED calculations \cite{dEnterria:2013zqi,Li:2019yzy}.

An intense, or strong, electromagnetic field (characterised more precisely below) can be regarded as a coherent state of high occupation number or, by the correspondence principle, an essentially classical field. For high occupation number, one can also neglect back-reaction on the field and consider it as fixed. (Again, this approximation eventually breaks down, as will be discussed.) The appropriate theory for studying e.g.~high-intensity laser-matter interactions is therefore quantum field theory (QFT) in an external, or background, field (sometimes the term `strong-field QED' (SFQED) is used).

This is a subject as old as QFT itself~\cite{Heisenberg:1936nmg,Schwinger:1951nm,Toll:1952rq}, and within it lies, to illustrate, the topic of Schwinger pair-production from an external field (also called the Sauter-Schwinger effect)~\cite{Sauter:1931zz, Schwinger:1951nm}. This is perhaps the most familiar, and `prototype', example of a non-perturbative QFT effect which can admit an analytic treatment. Significant progress has been made in understanding the theory and phenomenology of the Schwinger effect in more realistic backgrounds, with a particular focus on models of colliding laser pulses, which offer one route toward eventual experimental measurement~\cite{Bulanov:2010ei}.  Despite this progress, there remain many unanswered questions about the time-resolved Schwinger effect; when are the pairs produced (or `become real'), what can one say about the behaviour of the system at non-asymptotic times, and what can be measured~\cite{Tanji:2008ku, Dabrowski:2014ica,Dabrowski:2016tsx}? These are questions which go to the heart of quantum mechanics and which arise also in tunneling ionisation~\cite{Landsman:2014auv} and in cosmological scenarios~\cite{Parker:2012at,Birrell:1982ix}. We return to the Schwinger effect below.

Much of the work on which this review builds, began in the 1960s~\cite{1962JMP.....3...59R,Nikishov:1964zza,Nikishov:1964zzab,Brown:1964zzb,Kibble:1965zza,Frantz:1965}, shortly after, and inspired by, the invention of the laser itself. In most of these early papers~\cite{Nikishov:1964zza,Nikishov:1964zzab}, laser fields were modelled as monochromatic plane waves, or their low frequency limit, constant crossed fields.  This setup allowed for analytic progress in the calculation of scattering observables while, crucially, treating the strong background exactly. However, the model neglects the finite duration of a real laser pulse as well as the structure transverse to its propagation direction, focusing effects, which go hand in hand with shorter pulses and higher intensities. (Notable exceptions are the prescient early papers~\cite{Brown:1964zzb,Kibble:1965zza}.) As such, much of the work in subsequent decades has concentrated on the incorporation of more realistic structure in the modelling of laser pulses (``finite size effects") in tandem with experimental developments.

Considering more realistic descriptions of laser fields (even by introducing only e.g.~a finite pulse duration) has allowed for progress in many areas. For example, at the time of the E144 experiment on pair production in the collision of a laser and a high-energy electron beam~\cite{E144:1996enr}, there was no complete theory of the process being investigated, namely `nonlinear trident pair production', as even the tree level amplitude had not been calculated exactly in strong backgrounds. This illustrates the extreme complexity of amplitudes in strong fields, even for low numbers of scattered particles (four in the case of trident). This situation has now changed; by going beyond the simplest models, and doing so from a QFT perspective, we now have a fairly complete understanding of the nonlinear trident process~\cite{Dinu:2017uoj,King:2018ibi,Mackenroth:2018smh}.

Indeed, it has been found in many cases that the simple constant and monochromatic backgrounds originally studied can often \textit{obscure} the physics rather than exemplify it. Going beyond this has allowed the resolution of long-standing problems, such as the nature of the much discussed `effective mass' in a laser pulse~\cite{Harvey:2012ie}, and the refinement of models and approximation schemes needed to plan and analyse experiments in the high intensity regime, the advent of which (see Sec.~\ref{sec:intro:experiments} below) has been a significant driving factor for the field. Predictions for signals of vacuum birefringence have, for example, been made more precise and refined in preparation for upcoming experiments which aim to measure this subtle consequence of light-by-light scattering~\cite{Battesti:2012hf,King:2015tba,Karbstein:2019oej}, see Sec.~\ref{sec:LBL}.

Experimental developments have also renewed interest in the behaviour of QED in extremely strong fields, far beyond what we can realise today. This has lead to a great deal of activity surrounding the Ritus-Narozhny conjecture, which suggests that at sufficiently high intensities, not only is the charge-field coupling large, but the fine structure constant itself becomes enhanced by intensity effects~\cite{Ritus:1970radiative,Fedotov:2017conjecture}. The implication is that QED becomes `fully non-perturbative' in such a regime, requiring all loop orders to be resummed in order to yield reliable results. This conjecture, as well as the application of resurgence in QFT~\cite{Dorigoni:2014hea,Dunne:2016nmc}, has inspired new interest in the behaviour of higher loop processes in strong fields, and their resummation.

\begin{figure}[t!!]
    \centering
    \includegraphics[width=0.7\columnwidth]{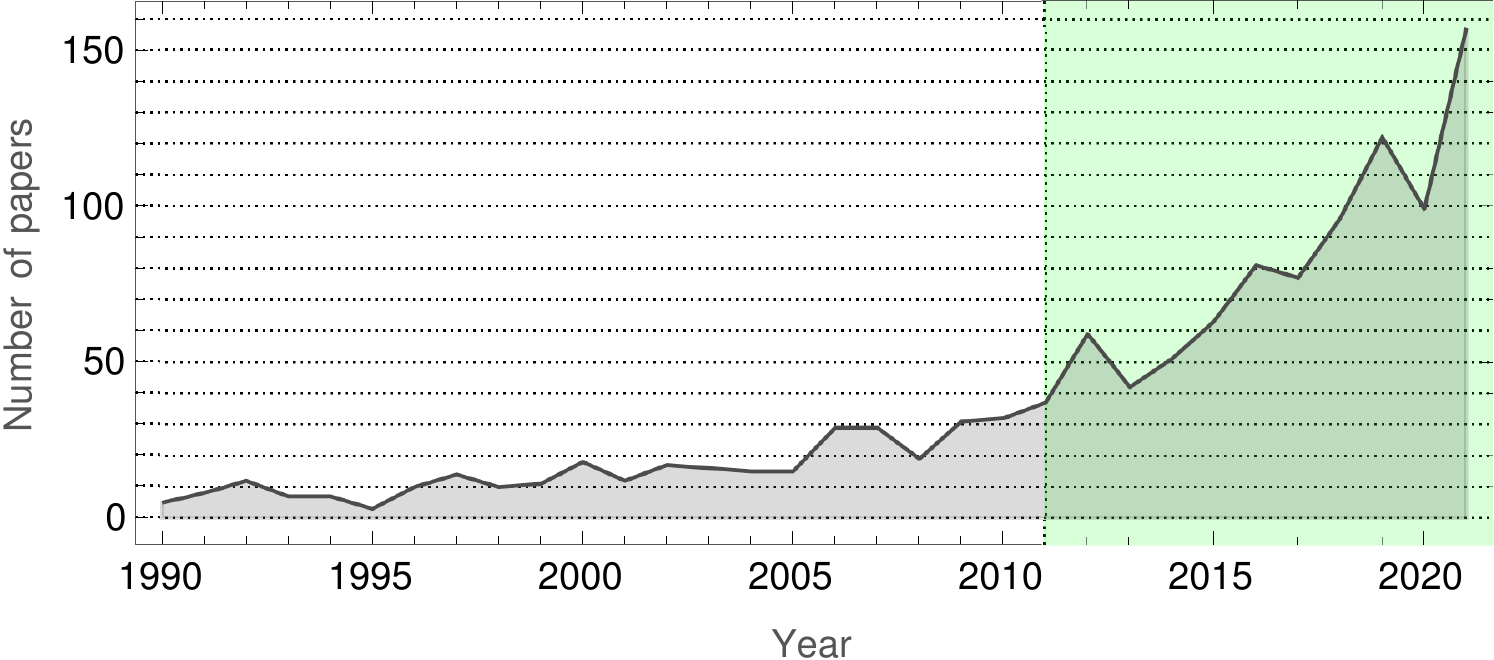}
    \caption{
    \label{FIG:ABSPLOT}
    Indicative bibliometric search using NASA-ADS, for at least one of the following terms occurring in the abstract: ``strong field QED'', ``nonlinear QED'', ``nonlinear Compton'', ``nonlinear Breit-Wheeler'', ``locally constant field'', ``Schwinger effect'', ``Schwinger pair''.
    The shaded region is the last decade, on which the current review is focussed.
    }
\end{figure}

This review has been written now because of both technological progress and increasing research activity; see Fig.~\ref{FIG:ABSPLOT} for some estimate of the increase in papers published in recent years. QED in intense fields has also begun to attract attention from the high energy physics community, and experiments colliding conventionally-accelerated electron beams with intense laser pulses will be performed in E320 \cite{Salgado:2021fgt} at SLAC and LUXE \cite{Altarelli:2019zea,Abramowicz:2019gvx,Heinemann2020,Abramowicz:2021zja} at DESY in the near future. At the same time, a new generation of laser facilities have recently come online (CoRELS \cite{2021Optic...8..630Y}), are being commissioned (ELI \cite{Turcu:2016dxm}), are being built (SEL \cite{sel18}) or are in the process of consultation (MP3 \cite{MP3ref}). Therefore, a review of recent theory developments is timely for experts and newcomers alike.

\subsection{Experimental landscape}\label{sec:intro:experiments}
QED in intense background fields can be tested in a number of ways.   We give here an overview of the region of parameter space that has been, and will be, probed in experiment. This section also introduces some standard parameters for quantifying total particle yields that will be used throughout the review.

The intensity of a background field is often quantified using the dimensionless and gauge-invariant `classical nonlinearity parameter', (also `intensity parameter' despite being proportional to the square-root of intensity), $\xi$, (in the literature also `$a_{0}$' or occasionally `$\eta$').
This parameter occurs naturally in scattering calculations as the dimensionless charge-field coupling, and can be written in a more physical way as $\xi = e E \lambda_{c}/ \hbar \omega$, where $-e<0$ is the electron charge, $\lambda_{c}=\hbar/mc$ is the Compton wavelength for the electron, with mass $m$ and $E\geq0$ is electric field strength. In this form, $\xi$ is the work done by the background, over the Compton wavelength of the electron in units of the background photon energy. The probability of leading-order perturbative calculations of tree-level processes is proportional to $\xi^{2}$, and so $\xi^{2}$ can be understood as an approximate measure of the number of photons interacting with an electron. It can be defined for a plane wave in a manifestly gauge-invariant way \cite{Heinzl:2008rh} as $\xi=\left(e^{2}\langle (-p\cdot {\mathcal F})^2\rangle/[m k\cdot p]^2\right)^{1/2}$, where ${\mathcal F}$ is the classical field strength tensor, $p$ is the momentum of a probe particle, $k$ is the wavevector of the background and  $\langle \cdot \rangle$ indicates a phase cycle average over the phase $\varphi=k\cdot x$.

The energy of the collision between probe and {plane wave} background, can be quantified using a dimensionless linear quantum parameter, $\eta$ (in the literature also `$b_{0}$'), which for a plane-wave background takes the form $\eta= \hbar k\cdot p/m^{2}c^{4}$. {The parameter $\eta$ is therefore equal to the laser frequency in the rest frame of an accelerated charge.}
{If the particle is a photon, $\eta$ is half the centre of mass energy when in a collision with a single laser photon; the threshold for linear Breit-Wheeler is $\eta\geq 2$.}

The `quantum nonlinear parameter', $\chi$ (occasionally $\Upsilon$ or $\eta$ in the literature) can be written as
\be
    \label{eq:chi:def}
    \chi=\frac{e\hbar [-(p\cdot {\mathcal F})^2]^{1/2}}{m^{3}c^{4}}
\ee
and applied to a general background field. It can be interpreted in many ways: {for an electron or positron, it can be phrased} as the work done by the background, over the Compton wavelength {in the particle's rest frame}, in units of the particle's rest energy; as the ratio of the electric field to the Schwinger limit $E_{\rm S} = m^2c^3/(e\hbar)$ in the rest frame of an accelerated charge; as the proper acceleration; as the world line curvature of a particle moving under the Lorentz forces times the Compton wavelength \cite{Seipt:2019dnn}. Therefore when $\chi \sim \mathcal{O}(1)$, quantum nonlinear processes such as pair creation, should become probable. These three parameters, $\xi$, $\eta$ and $\chi$, are related in a plane wave background via $\chi = \xi \eta$.

Several important experimental tests of QED in intense background fields have been performed by colliding proton beams with amorphous media and oriented crystals, where the high energy of the proton beams and the strong static inter-planar crystalline fields combine to give a quantum nonlinearity parameter of the order of $\chi =0.1 \ldots 7$ \cite{Nielsen:2021ppf}. The NA63 experiment collides electrons and protons with energies $O(100)\,\trm{GeV}$ provided by Cern's SPS (Super Proton Synchrotron) with fixed targets of different proton numbers, $Z$. It has had widespread success in measuring strong-field QED effects in the crystal's background field which can vary, depending on the transverse momentum variation \cite{Wistisen:2019eza}, from undulator-like to synchrotron-like. In the last decade, NA63 has measured the quantum suppression of synchrotron radiation in the range $\chi=0.05\ldots 0.7$ \cite{CERNNA63:2012zsc}, measured the Landau-Pomeranchuk-Migdal (LPM) effect of radiation suppression due to multiple Compton scattering within the photon formation length \cite{CERNNA63:2012hqm,CERNNA63:2013ahd}, measured radiation reaction \cite{DiPiazza:2015oia} in the classical limit (verifying the Landau-Lifshitz equation \cite{Nielsen:2021ppf}) and has also observed quantum effects \cite{Wistisen:2017pgr,Wistisen:2019eza}. 

\begin{figure}[t!!]
    \centering
    \includegraphics[width=0.8\columnwidth]{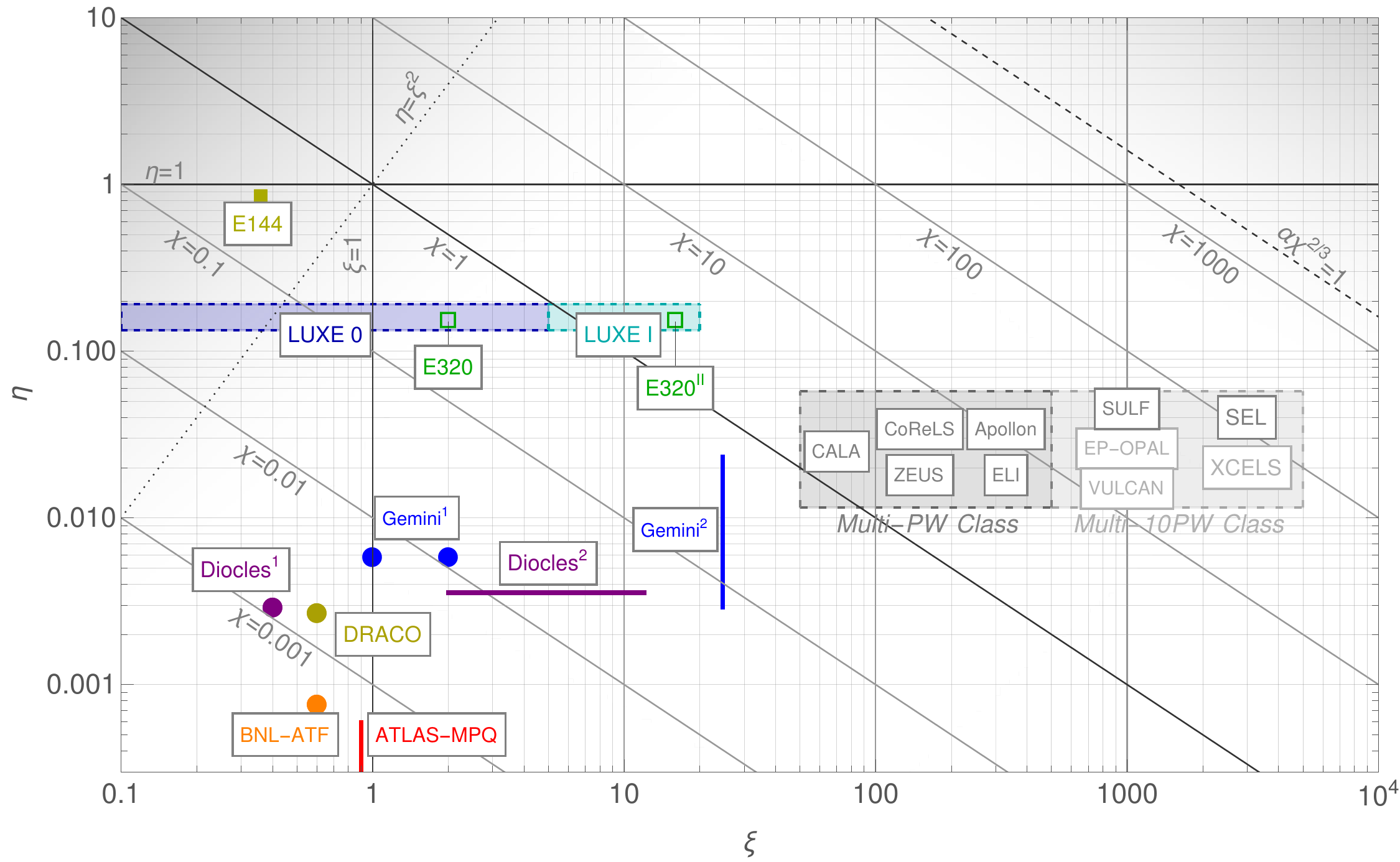}
    \caption{
    \label{FIG:PARAMPLOT}
    Laser-particle experiments. Solid lines and markers indicate reported experimental results; dashed lines and empty markers indicate planned experiments. The specific experimental values plotted, in $(\xi,\eta)$ co-ordinates, are: Apollon \cite{papadopoulos2016}; ATLAS-MPQ $(0.9,5\ldots 6\times 10^{-4})$ \cite{2015PhRvL.114s5003K}; BNL-ATF $(0.6, 7.5\times10^{-4})$ \cite{Sakai:2015mra}; CALA \cite{2021NJPh...23j5002S}; CoReLS \cite{2021Optic...8..630Y}; DIOCLES$^1$ $(0.4,0.0029)$ \cite{Chen:2013mba}; DIOCLES$^2$ $(2\ldots12,0.0036)$ \cite{Yan2017}; DRACO $(0.6,0.0027~[230\,\trm{MeV}])$  \cite{Hannasch:2021kyh}; E144 $(0.36,0.83)$ \cite{Bamber:1999zt}; E320 $(2,0.15)$ \cite{Salgado:2021fgt}; E320$^{\tsf{II}}$ (proposed upgrade) $(16,0.15)$ \cite{meuren2019probing}; ELI \cite{Turcu:2016dxm}; {EP-OPAL \cite{Zuegel:14}}; Gemini$^{1}$ $(1\ldots2,0.006)$ \cite{Sarri:2014gea}; Gemini$^{2}$ $(24.7,0.003\ldots0.02)$ \cite{Cole:2017zca,Poder:2017dpw}; LUXE 0 $(0.1\ldots5,0.13\ldots0.19)$ \cite{Abramowicz:2021zja}; LUXE 1 $(5\ldots20,0.10\ldots0.19)$ \cite{Abramowicz:2021zja}; {VULCAN \cite{Hernandez_Gomez_2010}}; SEL \cite{sel18}; {SULF \cite{Li:18}}; XCELS \cite{Mukhin_2021}; ZEUS \cite{Nees21}. The grey `Multi-PW Class' and `Multi-10PW Class' regions correspond to typical values of $\xi$ that can be produced at these laser facilities, and values of $\eta$ correspond to $1\ldots5\,\trm{GeV}$ particle beams colliding at 20 degrees with the laser pulse, na\"ively assuming the electron reaches the peak intensity at the focus of the laser pulse.
    }
\end{figure}

Although they have seen much success {reaching intensity parameters as large as $\xi \sim O(10^{2})$ \cite{DiPiazza:2019vwb}}, {a possible future} limitation of using oriented crystals is the maximum intensity parameter that can be produced. {In contrast, intensities reachable at the next generation of multi-PW lasers can in principle soon exceed $\xi\sim O(10^{2})$, and at multi-PW lasers, could exceed $\xi\sim O(10^{3})$. For example, if a laser of wavelength $800\,\trm{nm}$ is focussed with linear polarisation to an intensity of $10^{23}\,\trm{Wcm}^{-2}$, the intensity parameter corresponds to $\xi\approx 150$. Therefore, in the near future, lasers will push} further into the $\xi \gg 1$ parameter region allowing the `quantum nonlinear' region of $\chi \sim O(1)$ to be probed with lasers for the first time. This region can also be probed by using currently available lasers and instead making the probe energy higher. A summary of laser-particle experiments\footnote{Fixed-target experiments colliding particle beams with crystals have achieved $\chi\sim O(1)$, but due to the difficulty of defining an `intensity parameter', have not been included on the parameter plot, which is for laser-particle experiments only.} and facilities with their target parameters are given in \figref{FIG:PARAMPLOT}. (A comprehensive review of high power laser systems can be found in \cite{danson19}, which also contains laser landscape plots, as does the recent review \cite{Gonoskov:2021hwf}.)

The experiments in \figref{FIG:PARAMPLOT} can be divided into two groups: those with an electron beam produced by a conventional accelerator (E144, E320, LUXE), and the rest using an electron beam produced by laser wakefield acceleration. Whilst the former allow for higher precision measurements due to the lower emittance of the electron beam, the latter will be employed at high intensity laser facilities that will probe higher values of $\xi$. One set of QED laser experiments that do not appear on \figref{FIG:PARAMPLOT} are the tests of light-by-light scattering using real photons. Since these experiments typically employ optical or x-ray photons, the energy parameter is rather low. Furthermore, since they are probing the perturbative, weak-field limit (four-photon scattering: $2\to2$ scattering at one loop), the physics being tested is somewhat different to that in intense backgrounds. Examples here include HIBEF (Helmholtz International Beamline for Extreme Fields) which will collide $12.9\,\trm{keV}$ x-ray photons from the European XFEL with $1.55\,\trm{eV}$ photons from a $\trm{PW}$-class intense optical laser \cite{Schlenvoigt:2016jrd}; an experiment at the SACLA \cite{yabashi2015overview} x-ray free electron laser which tested light-by-light scattering cross-section at a centre-of-mass energy of $6.5\,\trm{keV}$ \cite{Inada:2014srv} and laser-cavity experiments using quasi-constant magnetic fields such as PVLAS~\cite{Ejlli:2020yhk} and BMV \cite{Cadene:2013bva}.

Typical signatures of QED effects in e.g.~nonlinear Compton scattering (photon emission from an electron in a background field) include the influence of the electron's effective mass on the harmonics in radiation spectra in energy \cite{2015PhRvL.114s5003K} and transverse momentum \cite{Sakai:2015mra} (depending if $\chi \ll 1$, these effects can also be described using nonlinear classical electrodynamics); the ellipticity of the transverse momentum distribution of photons \cite{Yan2017} and the correlation of the scattered electron and photon energies \cite{Cole:2017zca}.

Upcoming lasers will access much higher values of $\xi$, and so the phenomenological aspects of radiation reaction in laser fields have been a topic of much research in recent years. Progress in this area has recently been reviewed in~\cite{Gonoskov:2021hwf}, which covers radiation reaction approaches in PIC codes, kinetic equations~\cite{Neitz:2013qba,Yoffe:2015mba},
    and novel effects in particle dynamics due to radiation reaction such as trapping~\cite{Gonoskov:2013aoa},
    straggling~\cite{Blackburn:2014cig} and quenching~\cite{Harvey:2016uiy}. Another important development has been planned experiments that combine lasers with conventional accelerators, such as LUXE~\cite{Abramowicz:2021zja} and E320~\cite{E320ref}, which can probe how strong-field QED processes behave as $\xi$ is increased from below to above $\xi=1$. This has led to a development of approximation frameworks \cite{Heinzl:2020ynb,Nielsen:2021nuo} that are valid at intermediate values of $\xi$ and hence beyond the locally constant field approximation (LCFA).

\subsection{Outline and scope of the review} 

This review will cover the fundamental theory of QED in intense background fields. The focus is on progress made during, roughly, the past 10 years. We will not review work in the decades before this, which is covered by previous reviews~\cite{Ritus1985,Dunne:2005sx,Marklund:2006my,2009RPPh...72d6401E,DiPiazza:2011tq,Narozhny:2015vsb}, although we will naturally refer to original papers when introducing basic concepts. {The review covers material published/arXived before $28^\text{th}$ February 2022.}

Sec.~\ref{sec:prelim} provides an introduction to relevant theory concepts {(other recent pedagogical overviews are given in \cite{Seipt:2017ckc,Karbstein:2019oej}
)}. We review the by-now well established methods of strong-field QED, in particular the Furry picture expansion of scattering amplitudes, and the primary solvable cases of the Schwinger effect in constant fields, and scattering in plane wave backgrounds.

In Sec.~\ref{sec:first} we review progress in studying the simplest `first order' (three-point tree level) processes of strong-field QED in plane wave backgrounds. Despite this being a long-studied topic, several important advances have been made in the last decade: the consequences of including a physical finite pulse rather than assuming an unphysical infinitely long wave, how single or multiple finite pulses affect particle spectra through interference effects, the control of particle polarisation, and higher-precision calculations. These advances have an impact on the material in later sections.

In Sec.~\ref{sec:second} we review second-order processes including e.g.~four point scattering amplitudes. Typically challenging to calculate, progress has centred around: computation in finite pulses; the separation of on-shell/off-shell contributions, or two-step/one-step  contributions, where a finite pulse duration can play an important role; polarisation sums and calculation of various limiting cases. In Sec.~\ref{sec:approx} approximation schemes employed in numerical simulation are reviewed. Advances in the last decade include a better understanding of the validity range of common approximations; development of new approximations that are more accurate for specific applications and further development of the phase space distribution formalism centred on the equal-time Wigner function. Some of these approximations are used to calculate higher-point, and higher-loop order, amplitudes, which are reviewed in Sec.~\ref{sec:higher}. This section includes recent results from all-orders and resummed perturbation theory, in particular their application to the topics of radiation reaction and polarisation in higher-order processes.

Sec.~\ref{sec:LBL} covers light-by-light scattering and its observable effects. Recent progress includes the calculation of higher-order and all-orders results in strong magnetic fields, the discovery that one-particle-reducible processes are in fact non-zero, and an improvement in the modelling of collisions involving focussed backgrounds. Strong electric fields are the focus of Sec.~\ref{sec:Schwinger}, in which we review the non-perturbative creation of electron-positron pairs, or Schwinger effect. Advances in this area include fields with spacetime inhomogeneities leading to dynamical-assistance and quantum-interference effects; radiative corrections beyond one-loop; and pair creation at non-asymptotic times, relating to the Stokes phenomenon in mathematics. We also mention connection to condensed-matter phenomena and highlight some experimental results made there over the decade. In Sec.~\ref{sec:RN} we review the Ritus-Narozhny conjecture on the breakdown of the Furry expansion. Although dating from the 1980s, this conjecture has received renewed attention in the last decade, and recent developments in understanding higher-order effects and finite pulses are being used to better understand the nature and possible resolution of the conjecture. In Sec.~\ref{sec:beyondPW} we review progress made in going beyond the case of constant and plane wave backgrounds. This includes both exact solutions and approximate methods such as reduction of order and WKB (Wentzel–Kramers–Brillouin), as well as improvements in the understanding of back-reaction and depletion. 

Sec.~\ref{sec:beyondQED} covers progress made in applying the methods of strong-field QED to electroweak physics, Yang-Mills theory, quantum chromodynamics (QCD), gravity, and physics beyond the Standard Model, including the possibilities of new particle searches using intense laser fields. We conclude in Sec.~\ref{sec:conclusions} with some open questions and possible directions for future research. For the convenience of the reader we summarise our conventions and key notation  in Sec.~\ref{sec:conventions}.

There are several topics which naturally connect to studies of QED in intense background fields. Here, our focus is on advances in fundamental theory and its calculation; we direct the reader to the following reviews and articles of adjacent fields.

Some advances detailed in the current review are routinely employed in numerical simulation of high-power laser-\emph{plasma} physics, recent reviews of which can be found in~\cite{Gonoskov:2021hwf,Zhang:2020lxl}. The link to plasma physics can be established via e.g.~kinetic theory with more information to be found in the recent reviews~\cite{Zhang:2020lxl,BrodinZamanian2021}, also~\cite{Fauth:2021nwe} and references therein. Intense background fields occur in many contexts. Reviews of strong-field atomic physics can be found in \cite{2014JPhB...47t4001P,2019RPPh...82k6001A}. For strong fields in astrophysics see e.g.~\cite{Lai:2014nma}, for lab-based astrophysics see~\cite{2016RPPh...79d6901M,Kim:2019joy}, and for magnetar environments see~\cite{Turolla:2015mwa,Kaspi:2017fwg,Kim:2021kif}.     The physics of ultra-relativistic particles in strong magnetic fields is very similar to that in constant crossed fields, which will recur throughout this review, see~\cite{Kuznetsov:2004tb} for a discussion of both fields in astroparticle physics.  The properties of QCD in strong magnetic fields is a broad topic which lies beyond the scope of this review, but for which see~\cite{Kharzeev:2013jha, Hattori:2016emy}.

\section{Theoretical preliminaries}\label{sec:prelim}
We set $\hbar=c=\epsilon_0=1$ throughout this review, although $\hbar$ may occasionally be reinstated when needed.

\subsection{The Furry expansion}\label{sec:intro:Furry}
\begin{figure}[t!]
    \centering
    \includegraphics[width=0.5\columnwidth]{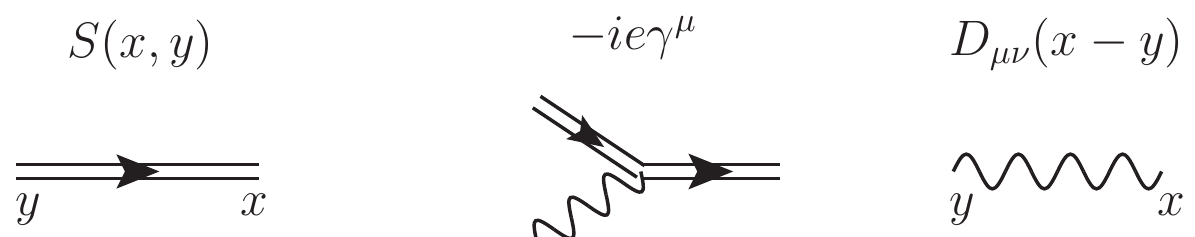}
    \caption{
    \label{FIG:FURRY}
    {In a background field, the Feynman rules are formulated in position space, due to the general loss of overall momentum conservation. The fermion propagator $S(x,y)$ (left) is the inverse of the Dirac operator in the background, i.e.~obeys (\ref{eq:DiracBG}), and is represented by a double line. The lack of translation invariance means that the propagator is a function of both the spacetime coordinates which it connects. The vertex (middle) and photon propagator (right) are as in QED without background.}}
\end{figure}
{Our starting point is the QED Lagrangian for electrons and positrons (Dirac fermion $\psi$) interacting with photons, (gauge field $A_\mu$), all in the presence of a background, or external, electromagnetic field with gauge potential~$\mathcal{A}_\mu$,}
\be\label{action-with-background}
	\begin{split}
		\mathcal{L} = 
		-\frac{1}{4}F^{\mu\nu}F_{\mu\nu} &+ \bar\psi (i\slashed{\partial}  -m\big) \psi - e\bar\psi\big( \slashed{A} +  \slashed{\mathcal{A}}\big)\psi\;,
\end{split}\ee
{in which $F_{\mu\nu}$ is the field strength of $A_\mu$,} and we suppress both gauge fixing terms and counterterms. The absence of a kinetic term for $\mathcal{A}_\mu$ reflects the assumption that either the background obeys Maxwell's equations in vacuum\footnote{{Scattering in an \emph{on-shell} background is equivalent to scattering in vacuum but with the same coherent state of photons in the initial and final asymptotic states, which is a more complete way of introducing the background. The fact that these coherent pieces are the same is the statement that we neglect depletion of the field, see Sec.~\ref{sec:beyondPW:backreaction}.}}, {or that its source (which is easily added, see e.g.~\cite{Raicher:2013cja} in the context of plasma physics) is not relevant to the processes being considered; this will be case for us, as we will mostly be interested in changes to quantum processes due to the presence of strong backgrounds, rather than quantum-induced changes on the dynamics of the background. Exceptions to this are the focus of Sec.~\ref{sec:LBL}, see also~\cite{Gies:2016yaa}, and related issues appear in Sec.~\ref{sec:RN}.}

\begin{figure}[t!]
 \centering
 \includegraphics[width=0.45\columnwidth]{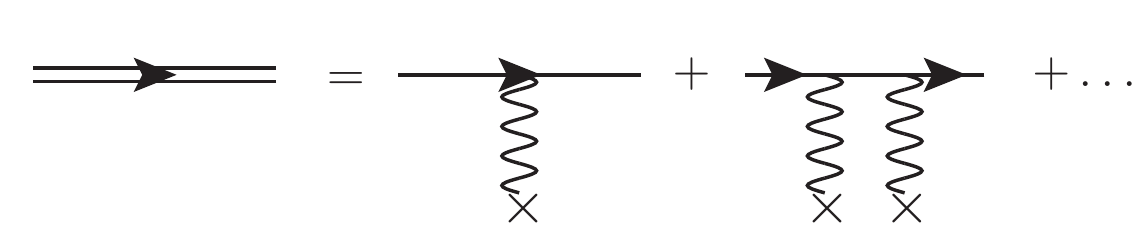}
\caption{\label{FIG:Volkovexpansion} The diagrammatic expansion of the Furry picture propagator in terms of ordinary (position space) Feynman diagrams in vaccum; each external photon line represents an interaction with the external field via the usual vertex $-ie\mathcal{A}_\mu \gamma^\mu$.}\end{figure}

The question is how to deal with the new term in (\ref{action-with-background}) and so calculate scattering amplitudes (and from them spectra, etc). Note that because $\mathcal{A}$ is a prescribed field, it defines (with explicit examples given below) a dimensionless effective coupling $\xi \sim e \mathcal{A}/m$. A field with coupling $\xi>1$ is \textit{strong}, and its interaction with matter cannot be treated by perturbation theory in $\xi$. Let us then rewrite~(\ref{action-with-background}) in terms of the background covariant derivative $\mathcal{D}_\mu := \partial_\mu +ie \mathcal{A}_\mu$, as
\be\label{Furry-action}
	\begin{split}
		\mathcal{L} =
		-\frac{1}{4}F^{\mu\nu}F_{\mu\nu} &+ \bar\psi (i\slashed{\mathcal{D}}  -m\big) \psi - \bar\psi e \slashed{A}\psi\;.
	\end{split}
	\ee
This trivial reorganisation of terms implicitly defines the `Furry picture' expansion of scattering amplitudes~\cite{Furry:1951zz}. The first two terms in~(\ref{Furry-action}) are still quadratic in the \textit{dynamical} fields and in this sense represent a ``free theory''. They contain all dependence on $\xi$, and so must be treated exactly. The third term in~(\ref{Furry-action}) is the original cubic interaction, controlled {by the usual small coupling $e$ or, at the level of cross-sections and probabilities\footnote{{We write $\alpha$ as shorthand for $\alpha(m^2)$, renormalised at the electron mass scale.}}, $\alpha$;}  it is to be treated in perturbation theory as normal. Treating the terms as such, we immediately obtain the \textit{position space} Feynman rules shown in Fig.~\ref{FIG:FURRY}; propagators are given by inverting the quadratic terms, interactions by the cubic term. The only difference to the usual position space rules of QED is therefore that the ``dressed'' fermion propagator $S$ is  the inverse of the Dirac operator in the background $\mathcal{A}$,
\be\label{eq:DiracBG}
    \big(i\slashed{\mathcal{D}} -m\big)S(x,y) = i \delta^4(x-y) \;,
\ee
and is in general a complicated function of two spacetime arguments (momentum not being conserved in the presence of a background). The propagator $S$ has an expansion in powers of the background, or strictly in powers of $e\mathcal{A}$, which corresponds directly to Feynman diagram expansion shown in Fig.~\ref{FIG:Volkovexpansion}. We will return to properties of this expansion in Sec.~\ref{sec:intro:PW}. 
Correlation functions generated by the Feynman rules are converted into S-matrix elements by applying the Lehmann-Symanzik-Zimmermann (LSZ) reduction formula to the external propagators, turning them into asymptotic particle wavefunctions (which are solutions of the Dirac equation in the background $\mathcal{A}$).

{To investigate the Furry expansion and the implied physics in more detail, we next need to consider the background field invariants}
\be\label{eq:invariants}
   \mathcal{S} := -\frac{1}{4} {\mathcal F}_{\mu\nu} {\mathcal F}^{\mu\nu}  = \frac12({\bf E}^2 -{\bf B}^2) \,,
   \qquad
   \mathcal{P}:=-\frac14 {\mathcal F}_{\mu\nu} \tilde{ {\mathcal F}}^{\mu\nu}  = {\bf E}\cdot{\bf B} \;,
\ee
in which\footnote{Note that a common alternative notation is to write these invariants as $\mathcal{F}$ and $\mathcal{G}$ respectively.} $\mathcal{F}_{\mu\nu} =\partial_\mu \mathcal{A}_\nu-\partial_\nu \mathcal{A}_\mu$. If $\mathcal{P}=0$ and $\mathcal{S}>0$ we can go to a frame where there is only an electric field. If $\mathcal{P}\not=0$ we can go to a frame in which there are both electric and magnetic fields, and they are parallel. In both cases there is a nonzero probability for the field to spontaneously produce (with or without the presence of any other probe particles) electron-positron pairs; in the case of a constant, or slowly varying, electric field for which it was originally studied, this is the Schwinger effect~\cite{Schwinger:1951nm,Dunne:2004nc}, {to which we turn now.}

\subsection{The Schwinger effect}\label{sec:ht1} 
The Feynman rules receive additional contributions when the field is capable of Schwinger pair production. {To illustrate (for details see the book~\cite{Nikishov:1985,Fradkin:1991zq}),} consider the three-point correlation function of the fields $A_\mu$, $\psi$ and $\bar\psi$, starting from the path integral and the Lagrangian (\ref{Furry-action}):
\be\label{pleasekillme}
    \int\!\!D A D\psi D{\bar\psi}\,  \exp\bigg(i\int {\rm d}^4x\,\mathcal{L}\bigg) \bar\psi \psi A_\mu \;.
\ee
This correlation function contributes to the amplitudes for e.g.~$\gamma \to e^\LCp e^\LCm$ in the presence of a background field (`nonlinear Breit Wheeler') and $e\to e\gamma$ (`nonlinear Compton scattering'), both to be discussed in  Sec.~\ref{sec:first}. {Consider the leading order Furry picture expansion of this correlator; we expand (\ref{pleasekillme}) to first order in $e$, which means expanding in powers of the final, interaction, term in (\ref{Furry-action}), containing the dynamical photon field $A$ (while} the strong field coupling $e{\mathcal A}\sim\xi$ is treated exactly). This gives
\be
(\ref{pleasekillme}) =
\int \!\!D A D\psi D{\bar\psi}\,\exp\bigg(i\int{\rm d}^4x\,\bigg\{ -\frac14 F^2 + {\bar\psi}(i\slashed{\mathcal{D}} -m)\psi \bigg\} \bigg)\bigg[-ie\int{\rm d}^4x\, {\bar\psi}\slashed{A}\psi \bigg]\bar\psi \psi A_\mu +\, \mathcal{O}(e^2)\; \nonumber
\ee
{As for QED in vacuum, we perform the Gaussian integral, with replaces all possible pairs of fields with the propagators in Fig.~\ref{FIG:FURRY} (Wick contraction), interacting via the three-point vertex. The same integral also generates determinants of the (gauge fixed) wave operator for $A_\mu$, and of the Dirac operator in the background; the former gives the same irrelevant prefactor as in vacuum, the latter requires further discussion so will be explicitly retained. Schematically, we have:}
\be\label{disconnfig}
(\ref{pleasekillme}) = \raisebox{-16pt}{\includegraphics[width=0.53\textwidth]{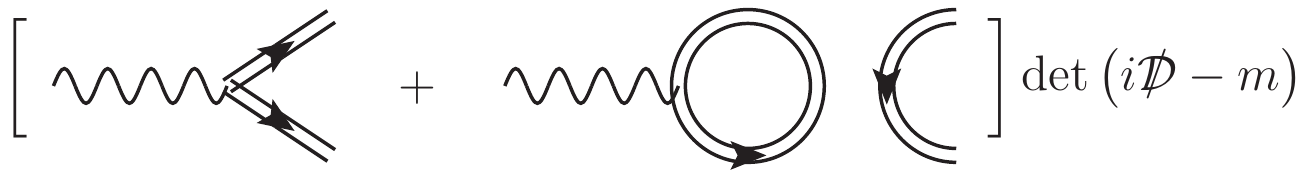} \raisebox{16pt}{\, $+\, \mathcal{O}(e^2)\;.$}}
\ee
Along with the (expected) connected three-point diagram, we have a nonzero disconnected diagram; the right hand part describes an interaction between the external field and the fermions (which may contribute to scattering without emission or, see below, Schwinger pair creation), while the left hand `tadpole' part cannot {(despite old claims to the contrary) in general} be normal-ordered away as in vacuum, since it describes the physical process of photon absorption or vacuum emission by the background~\cite{Karbstein:2014fva,Karbstein:2015qwa,Gies:2017ygp,Blinne:2018uyn}, see Sec.~\ref{sec:LBL}.

{A second difference compared to vacuum calculations is the contribution of} the determinant of the Dirac operator in the background, which multiplies all diagrams. This is the bubble diagram
\be\label{eq:bubble}
    \ln \text{det}(i\slashed{\mathcal{D}} -m) = \raisebox{-10pt}{\includegraphics[width=1cm]{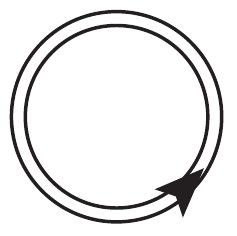}} 
\ee
which defines the famous one-loop Heisenberg-Euler effective action ${\Gamma}^{\rm 1\mathchar`-loop}_{\rm HE}$ via $\text{det}(i\slashed{\mathcal{D}} -m) = \exp \big\{ i {\Gamma}^{\rm 1\mathchar`-loop}_{\rm HE}\big\}$, see below and Sec.~\ref{sec:LBL}. This diagram, along with other bubbles generated at higher orders in $e$, no longer contributes simply a phase which can be divided out, as happens in the absence of background fields. Rather the bubble diagrams contain nontrivial physics, and the determinant is the leading order probability amplitude for the process in which one begins in vacuum in the asymptotic past, turns on the field (without any probe photon or other particle), turns it off, and asks whether the system is still in vacuum in the asymptotic future. This is the so-called \textit{vacuum persistence amplitude}
\be
    \bracket{0;\text{out}}{0;\text{in}} = \int\!\!D A\!\!\int\!\!D\psi\!\!\int\!\!D{\bar\psi}\,\,  e^{i\int{\rm d}^4x\,\mathcal{L}} = \text{det}(i\slashed{\mathcal{D}} -m) + \mathcal{O}(e^2) \;.
\ee
Mod-squared, it gives a vacuum persistence probability\footnote{The terms `vacuum persistence' and `vacuum decay' are very commonly used, but perhaps not the most accurate, as there is a field present, so we are not in vacuum.} which is in general \emph{less than one} precisely because pairs may be produced. The possibility of pair creation was first pointed out by Sauter in 1931~\cite{Sauter:1931zz} as a resolution to the Klein paradox in relativistic quantum mechanics~\cite{Klein:1929zz}.  Sauter's idea was extended to QED by Schwinger in 1951~\cite{Schwinger:1951nm}.  Using the proper time method, Schwinger explicitly computed the vacuum persistence probability in a constant electric field in terms of the one-loop Heisenberg-Euler effective action as $|\langle {\rm 0;out} | {\rm 0;in} \rangle |^2 = \exp\{ - 2\, {\rm Im}\,{\Gamma}^{\rm 1\mathchar`-loop}_{\rm HE} \}$, with the imaginary part given by
\begin{align}
	2\,{\rm Im}\,{\Gamma}^{\rm 1\mathchar`-loop}_{\rm HE} 
		= V \sum_{r} \int {\rm d}^3{\mbf p} \frac{1}{(2\pi)^3} \sum_{n=1}^\infty \frac{1}{n} \exp \left[ -n\pi \frac{m^2+{\mbf p}_\perp^2}{eE} \right] 
		= 2VT \frac{(eE)^2}{(2\pi)^3} \sum_{n=1}^\infty \frac{1}{n^2} \exp \left[ -n\pi \frac{m^2}{eE}  \right] ,  \label{eq:ht3}
\end{align}
where $\sum_r$ indicates a spin sum and $VT$ is the spacetime volume.  The vacuum persistence probability is precisely equal to one minus the total probability of creating pairs.  {The corresponding `vacuum decay rate' is $2\,{\rm Im}\,{\Gamma}^{\rm 1\mathchar`-loop}_{\rm HE}/T$. The Schwinger effect is of particular interest because it exhibits non-perturbative physics -- this is made explicit by (\ref{eq:ht3}), where expanding the exponential in powers of $e$ yields zero to every and all orders of perturbation theory.}

The number of created pairs was first computed by Nikishov \cite{Nikishov:1969tt}. By solving the Dirac equation in a constant electric field and using the Green function technique, the number of pairs $N$ in the final (asymptotic) state was found to be
\begin{align}
	N 
		= V \sum_{r} \int {\rm d}^3 \mbf{p} \frac{1}{(2\pi)^3} \exp \left[ -\pi \frac{m^2+{\mbf p}_\perp^2}{eE} \right] 
		= 2VT \frac{(eE)^2}{(2\pi)^3} \exp \left[ -\pi \frac{m^2}{eE} \right] . \label{eq:ht5}
\end{align}   
The number $N$ is {exponentially suppressed for weak fields $E$ below the Schwinger field $E \lesssim E_{\rm S} = m^2/e$}, while it grows quadratically in the formal limit of $E/E_{\rm S} \to \infty$. The number of created pairs (\ref{eq:ht5}) admits a simple physical interpretation that the Schwinger effect is driven by quantum tunneling.  This mechanism is analogous to the Landau-Zener effect in materials (see Sec.~\ref{sec:Schwinger:newCondMatt} for more about the condensed-matter analogue of the Schwinger effect).  Namely, in the presence of strong electric fields, there occurs a level crossing between the Dirac sea and the positive energy continuum.  An electron in the Dirac sea can then tunnel into the positive energy continuum, leaving behind a `hole' in the Dirac sea, which is a positron, and thereby a pair creation occurs.  The tunneling probability can be estimated via WKB as an exponential of an integral of the potential barrier over the classically forbidden region.  This roughly equals to a product between height ($\sim 2m$) and length ($\sim 2m/eE$) of the barrier, which reproduces the exponent (\ref{eq:ht5}).   The formulas (\ref{eq:ht3}) and (\ref{eq:ht5}) can be extended to charged particles with general spin statistics, other than electrons; see~\cite{Weisskopf:406571, Vanyashin:1965ple,Marinov:1972nx}, {and see~\cite{Frob:2014zka} for the Schwinger effect in de Sitter space as a model for false vacuum decay.}

In the strong-field limit, a sizeable amount of pairs are created, and backreaction on the external field, as well as interactions among created pairs, becomes important; such effects have not been taken into account here or in the vacuum decay rate (\ref{eq:ht3}), but will be discussed in Sec.~\ref{sec:Schwinger} and Sec.~\ref{sec:beyondPW}. Note that the rate of pair creation in a constant field is given by $N/T$, which is not the same as the rate of vacuum decay $2\,{\rm Im}\,{\Gamma}^{\rm 1\mathchar`-loop}_{\rm HE}/T$ -- these are different quantities \cite{Cohen:2008wz}.  One origin of this difference is the contribution of quantum correlations between created pairs~\cite{Fukushima:2009er, Gelis:2015kya}.  These correlations can be neglected in the weak field limit, where only the $n=1$ term dominates the vacuum decay probability; in this approximation the vacuum decay rate and the pair creation rate coincide. 

{It can be challenging to derive exact results analogous to (\ref{eq:ht3}) and (\ref{eq:ht5}) in more realistic background fields. Here the worldline instanton method, within the `first quantised' approach to field theory, has proven particularly useful in the calculation of effective actions and related quantities, see Sec.~\ref{sec:Schwinger} and Sec.~\ref{sec:beyondPW}, as well as {the reviews~\cite{Dunne:2005sx,Edwards:2019eby} and references therein.} However, using the `locally constant field approximation' (Sec.~\ref{sec:Schwinger} and Sec.~\ref{sec:approx}) one can generate a simple formula which extends the constant-field result (\ref{eq:ht5}) to arbitrary (but sufficiently slowly varying) background fields. It has been argued that this is sufficient for estimating the pair yield in the Schwinger effect for all-optical setups involving multiple, colliding laser pulses~\cite{Bulanov:2004de,Bulanov:2010ei,Gonoskov:2013ada}. In such scenarios pre-exponential factors (essentially volumetric effects) in the Schwinger formula can compensate for the extreme smallness of the exponential factor, suggesting the possibility of observing the Schwinger effect even with fields of strength below the Schwinger field~\cite{Bulanov:2004de, Bulanov:2010ei}.} 

Returning now to (\ref{disconnfig}), a general scattering process in a background may be accompanied by the spontaneous or stimulated production of any number of pairs. We observe that in such processes, the presence of nontrivial disconnected diagrams invalidates the equivalence between counting numbers of loops and counting powers of $\alpha$, see~\cite{Gies:2016yaa} and Sec.~\ref{sec:LBL}. In~(\ref{disconnfig}), for example, there are both one-loop and tree-level contributions at leading order. Due to such the contributions the number of diagrams contributing to any process will grow rapidly as one goes to higher orders in the Furry expansion, making calculations of exclusive processes daunting\footnote{{We mention, to avoid confusion, that it is common to define `relative' correlation functions as those in which the vacuum persistence amplitude/determinant factor is divided out, see e.g.~\cite{Fradkin:1991zq}. This may help to simplify expressions, but it is important to remember that this factor is part of physical probabilities, and must ultimately be retained.}}; as we will review in Sec.~\ref{sec:Schwinger} it is  simpler, and physically motivated, to consider inclusive observables. However, for the backgrounds to be discussed in the initial part of this review, we can neglect the Schwinger effect, either because no such process is possible, or because it is extremely unlikely. In this case, the only contribution to (\ref{disconnfig}) is the expected connected diagram, and the vacuum persistence factor is well-approximated by unity.

\subsection{Plane wave backgrounds}
\label{sec:intro:PW}
The practical use of the Furry expansion rests on being able to invert the Dirac operator either exactly, or approximately but \emph{without} treating $\xi$ perturbatively. We review here an important case for which this is possible, namely plane waves. Other cases, both exact and approximate, are discussed in Sec.~\ref{sec:beyondPW}.

Let $n_\mu$ be a lightlike vector, $n^2=0$, and $\epsilon^j_\mu$ ($j=1,2$) the two independent spacelike vectors 
transverse to $n_\mu$, so $\epsilon^i\cdot\epsilon^j = -\delta^{ij}$ and $n\cdot \epsilon^j=0$. A general plane wave, obeying Maxwell's equations in vacuum, is then 
\be\label{plane-wave-F}
    \mathcal{F}_{\mu\nu} = E_j(n\cdot x) \big( n_\mu  \epsilon^j_\nu - \epsilon^j_\mu n_\nu\big) \;,
\ee
in which the two arbitrary functions $E_j$ are the electric field components of the wave. The electric and magnetic fields are of equal magnitude and orthogonal, hence the invariants (\ref{eq:invariants}) vanish, which defines a `null' field. We say that $\mathcal{F}_{\mu\nu}$ is a function of `lightfront time' $n\cdot x$~\cite{Bakker:2013cea}.  Unless otherwise stated, our focus is on the physical situation that the $E_j$ are `sandwich' fields that vanish asymptotically, either through adiabatic switching or compact support; {such `sandwich' fields then admit good scattering boundary conditions.}

Starting from (\ref{plane-wave-F}), a given plane wave is more conveniently represented as follows. Let $\omega$ be a characteristic frequency scale of the wave\footnote{{The reader is encouraged to check that all observables are formally independent of this arbitrary \textit{choice}, as follows from the scale invariance of plane waves. Of course, any \emph{particular} plane wave will come with some characteristic frequency/length scale which can appear in observables. It is then natural to use this scale to define $\omega$ in order to e.g.~estimate the size of relevant parameters and effects.}}, and define $k_\mu = \omega n_\mu$ and $\varphi=k\cdot x$. Writing a prime for a $\varphi$ derivative, we package the fields and the electron charge into the four-vector $a_\mu$ defined by
\be\label{eqn:this-is-the-potential}
a'_\mu(\varphi) = eE_j(\varphi)\epsilon^j_\mu \;, \qquad \text{and} \qquad a_\mu(-\infty)=0 \;.
\ee
As such $a_\mu(\varphi)$ has the interpretation of the work done by the field on a particle of charge $e$, up to lightfront time $\phi$~\cite{Dinu:2012tj}. Then we have
\be\label{plane-wave-a}
    e\mathcal{F}_{\mu\nu}(x) = k_\mu a'_\nu(\varphi) - a'_\mu(\varphi) k_\nu \;.
\ee
\subsubsection{{A primer on lightfront coordinates}}\label{A primer on lightfront coordinates}
\begin{figure}[t!]
    \centering
    \includegraphics[width=0.4\textwidth]{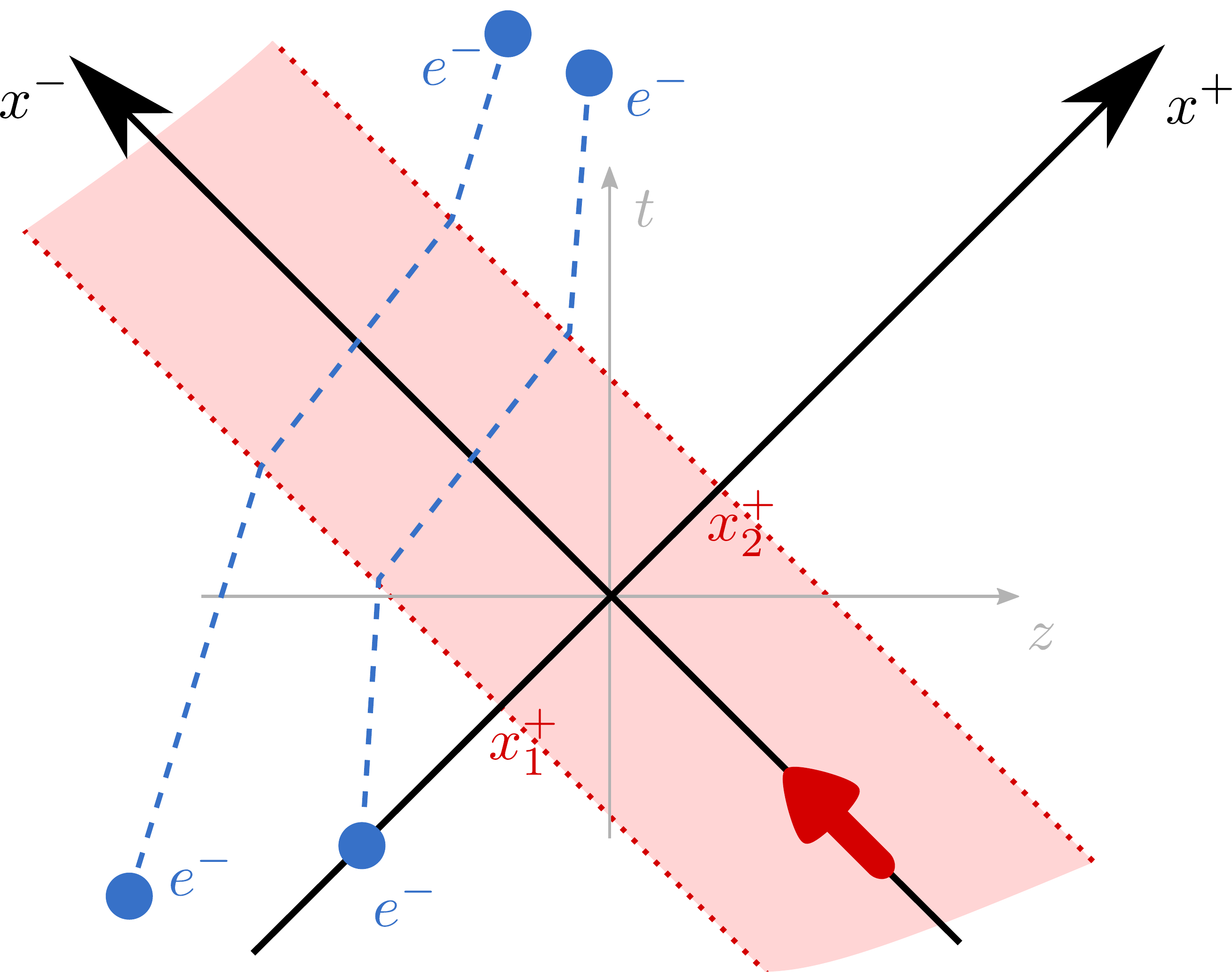}
    \caption{\label{fig:lightfront-pic} {A pulse of light moving in the negative $z$-direction, modelled as a plane wave of finite duration, is shown as the red region. The plane wave is homogeneous and infinitely extended in the transverse directions (not shown). All massive particles, see the blue tracks, enter and leave the wave at the same lightfront times $x^\LCp_1$ and $x^\LCp_2$ respectively, though these can correspond to different $t$ and $z$. Figure taken from~\cite{Seipt:2017ckc}.}}
\end{figure}

{It may help to visualise the physical setup by choosing explicit coordinates, which we can always do such that $n_\mu = (1,0,0,1)$ and so $n_\mu x^\mu=t+z$, while $(\epsilon^1_\mu x^\mu, \epsilon^2_\mu x^\mu) = (x,y)$. The field (\ref{plane-wave-F}) is then a pulse of light of finite extent moving \emph{down} the $z$-axis. The electric and magnetic fields point into the directions $x$ and $y$. It is quite natural to parameterise physical processes in plane wave backgrounds by lightfront time because, as illustrated in Fig~\ref{fig:lightfront-pic}, `sandwich' fields then admit good scattering boundary conditions and all massive particles enter and leave the background at (the same) finite lightfront times.}
{From the above choice, we define the ``lightfront coordinates" which are used throughout this review and for all four-vectors $v^\mu$ (including position and momentum) by:
\be\label{eq:lightfront-conventions}
    v^\LCpm = v^0 \pm v^3 \;,
    \qquad
    v_\LCpm = \frac12 v^\LCmp = \frac12 (v_0 \pm v_3) \;,\qquad
    {\bf v}^\LCperp =(v^1,v^2) \;,
    \qquad
    v_\LCperp = -v^\LCperp \;,
\ee
in which all the signs are dictated by our `mostly-minus' metric.  $x^\LCperp$ are referred to as the transverse coordinates while $x^\LCm$ is the longitudinal position.  The momentum component $p^\LCp=2p_\LCm$ is referred to as the longitudinal lightfront momentum, with either `longitudinal' or `lightfront' frequently dropped when the context is clear. As will be discussed below, total $p^\LCp$ is conserved in any scattering process on a plane wave background; $p^3$ is not conserved, nor is it a natural variable in plane wave or lightfront contexts, hence it never appears (and should not be confused with the longitudinal momentum, even though the plane wave, seen as a pulse of light, travels in the negative $z$-direction).}

{It is also common to use the `$\perp$' label in summation convention, so that a repeated $\perp$ is an instruction to sum over the transverse indices, i.e.~$p^\LCperp q^\LCperp = p^1 q^1+p^2 q^2$. Using this, the scalar product of two vectors is $x\cdot y = \tfrac12 x^\LCp y^\LCm +\tfrac12 x^\LCm y^\LCp - x^\LCperp y^\LCperp$ and the mass-shell condition is thus $p^2= p^\LCp p^\LCm - p^\LCperp p^\LCperp = m^2$. The transverse coordinates are sometimes set in bold as 2-vectors, so ${\bf p}\cdot {\bf q} = p^\LCperp q^\LCperp = p^1 q^1 +p^2 q^2$. For more on lightfront coordinates and lightfront field theory see~\cite{Heinzl:2000ht,Brodsky:1997de,Bakker:2013cea}.}

\subsubsection{Classical dynamics}\label{sec:prelim:plane:classical}
It is useful to consider the classical dynamics of a charged particle incident on (\ref{plane-wave-a}); the orbit $x^\mu$ of the particle is determined by the Lorentz force equation,
\be\label{eq:Lorentz}
    \dot{\pi}_\mu(\tau) = \frac{e}{m}\mathcal{F}_{\mu\nu}(x(\tau)) \pi^\nu(\tau) \;,
\ee
in which $\tau$ is proper time and $\pi_\mu = m\dot{x}_\mu$ is the kinetic (i.e. on-shell, physical) momentum. The solution may be parameterised in terms of lightfront time as
\be\label{pi-def}
	\pi_{p}^\mu(\varphi) = p^\mu - a^\mu(\varphi) + \frac{2p\cdot a(\varphi)-a^2(\varphi)}{2k\cdot p} k^\mu \;,
\ee
in which~$p_\mu$ is the initial momentum before entering the background. Integrating with respect to~$\ud\varphi/ (k\cdot p)$ yields the orbit itself, for an analysis of which using Frenet-Serret theory see~\cite{Seipt:2019dnn}. {To illustrate the benefit of lightfront coordinates the reader is invited to attempt to explicitly parameterise the momentum and orbit in terms of $t$ rather than $x^\LCp$.}

{We note from (\ref{pi-def}) that only \textit{one} component of kinematic momentum, $n\cdot \pi_p = n\cdot p =p^\LCp$ is conserved. However, both this longitudinal component and the transverse components of \textit{canonical} momentum $P^\mu = \pi_{p}^\mu(\varphi) + a^\mu$, are conserved; these three conserved quantities make particle motion in plane waves integrable (see also Sec.~\ref{sec:beyondPW}).} The momentum (\ref{pi-def}) obeys a `velocity memory effect'~\cite{Dinu:2012tj,Bieri:2013hqa,Zhang:2017geq}; that is, the final momentum $\pi_{p}^\mu(\infty)$ can differ from the initial momentum $\pi_{p}^\mu(-\infty) \equiv p^\mu$, and the particle retains a memory of its interaction with the wave: this is due simply to the fact that even when the electromagnetic fields turn off, $a_\mu$ becomes an, in general \textit{nonzero}, {constant $a^\infty_\mu:=a_\mu(\infty)$. It is therefore {limiting to impose} that $a_\mu(\infty)=0$ from the outset, as is sometimes done. {Whether or not `real' laser fields have $a_\mu(\infty)\not=0$ (sometimes called a DC component) is long-debated but irrelevant here; real lasers are not plane waves. Including $a_\mu(\infty)$ allows one to access simple models of vacuum acceleration and velocity memory effects which are of interest beyond strong field QED:} the analogous gravitational memory effect has, for example}, received attention as a possible detection mechanism for gravitational waves~\cite{Zhang:2017rno}, there is a related colour-memory effect in Yang-Mills~\cite{Pate:2017vwa}, and the connection between the memory effect, infra-red behaviour and gauge dependence in QED is explored in~\cite{Ilderton:2020rgk}.
Memory effects are one corner of the `infra-red triangle', which in recent years has revealed deep {and fundamental} structures in gauge theory and gravity~\cite{Strominger:2017zoo}. 

\subsubsection{Quantum dynamics}
\label{sec:prelim:plane:quantum}
With this we turn to the quantum theory. Observe that the classical `work done' $a_\mu(\varphi)$ is a candidate {\it gauge} potential for the plane wave. We adopt this choice of potential from here on. It has the benefit of making the physics manifest, as (i) it relates the potential to a known classical quantity, the integral of the electric field, and (ii) it leads to the explicit presence of the classical momentum $\pi_\mu$ in scattering amplitudes, which aids interpretation\footnote{For the special case of $1$ to $1$ scattering of a particle of momentum $p_\mu$, the gauge $\mathcal{A}\cdot p$ also presents simplifications~\cite{Lavelle:2020ijh}.}.

{We now write down the fermion propagator in a background plane wave:
\be\label{eq:VolkovProp}
    S(x,y) = i\int\!\frac{\ud^4p}{(2\pi)^4}\, \bigg(1+\frac{\slashed{k}\slashed{a}(k\cdot x)}{2k\cdot p}\bigg) \frac{\slashed{p}+m}{p^2-m^2+i0}\bigg(1+\frac{\slashed{a}(k\cdot y)\slashed{k}}{2k\cdot p}\bigg) \displaystyle e^{-ip\cdot (x-y) - i\int\limits^{k\cdot x}_{k\cdot y}\!\ud\phi\, \frac{2a(\phi)\cdot p-a^2(\phi)}{2k\cdot p}}\;.
\ee
From this ``Volkov''~\cite{Wolkow:1935zz} wavefunctions representing external legs in scattering amplitudes are handed out unambiguously by LSZ reduction of (\ref{eq:VolkovProp}).} The wavefunction for an incoming electron of initial momentum $p_\mu$ is, with $u_p$ the usual free electron spinor,
\be\label{volkov-e-in}
{\psi}_{p}(x) = \bigg(1+\frac{\slashed{k}\slashed{a}(\varphi)}{2k\cdot p}\bigg) u_{p} \displaystyle e^{-ip\cdot x - i\int\limits^{\varphi}_{-\infty}\ud\phi\, \frac{2a(\phi)\cdot p-a^2(\phi)}{2k\cdot p}}\;,
\ee
and that for a positron is
\be\label{volkov-pos-in}
{\bar{\psi}}^{(\LCp)}_{p}(x) ={\bar v}_p \bigg(1+\frac{\slashed{k}\slashed{a}(\varphi)}{2k\cdot p}\bigg) \displaystyle e^{-ip\cdot x - i\int\limits^{\varphi}_{-\infty}\ud\phi\, \frac{-2a(\phi)\cdot p-a^2(\phi)}{2k\cdot p}}\;,
\ee
with $v_p$ the free positron spinor. For outgoing states we should account for the memory effect. Define $\delta a_\mu = a_\mu - a_\mu(\infty)$, then an outgoing electron with momentum $p$ is described by\footnote{It is possible to rewrite the outgoing solutions, expressing them in terms of the kinematic momentum $\pi_\mu$ in the asymptotic future, for a particle of momentum $p_\mu$ in the asymptotic past. In this representation, which is used for some of the formulae in Sec.~\ref{sec:second}, the \emph{explicit} dependence on $a_\infty$ is absorbed into $\pi$.} (note the integral limits)~\cite{Kibble:1965zza,Dinu:2012tj}
\be\label{volkov-e-out}
{\bar\psi}_p(x) = \bar{u}_{p}\bigg(1+\frac{\delta \slashed{a}(\varphi)\slashed{k}}{2k\cdot p}\bigg) e^{ i(p+a(\infty))\cdot x + i\int\limits_{\infty}^{\varphi}\ud\phi\,
\frac{2\delta a(\phi)\cdot p-\delta a^2(\phi)}{2k\cdot p} 
} \;.
\ee
For fields such that $a_\mu(\infty)=0$ there is no memory effect, $\delta a_\mu = a_\mu$, and so incoming and outgoing wavefunctions are related simply by conjugation (up to an irrelevant phase).

Note that LSZ automatically generates the correct scattering boundary conditions; the in/out wavefunctions reduce to \emph{free} wavefunctions in the limit $\varphi\to -\infty$ / $\varphi\to +\infty$ respectively. The Volkov wavefunctions are, as is easily checked, solutions of the Dirac equation in the background plane wave, which is how they are usually introduced. {They are trivially normalised at equal \textit{lightfront} time~\cite{Neville:1971uc}. Despite claims in the literature, there are no non-Volkov solutions to the Dirac equation in a plane wave background; they are a complete set, which is again trivial to show at any lightfront time.  (The distinction between $x^\LCp$ and $t$ becomes more acute in the quantum theory; as is clear from Fig.~\ref{fig:lightfront-pic} there is no sense in which states at fixed $t$ can be set up as free states, as they always interact with the background.)} Wavepackets of Volkov states are easily constructed and analysed in scattering processes, see~\cite{Neville:1971uc,Ilderton:2012qe,Angioi:2016vir,Seipt:2017ckc}.

Being null, a plane wave background is not capable of Schwinger pair production, and the Volkov wavefunctions are \textit{single particle} wavefunctions. The exponents appearing therein are the classical Hamilton-Jacobi action, and reproduce the time-dependent classical momentum $\pi_\mu$ via $e^{i \ldots} \mathcal{D}_\mu e^{-i\ldots} = \pi_\mu$, in which the exponentials are those in (\ref{volkov-e-in})--(\ref{volkov-e-out}). While the background can clearly change the momentum of a particle, the situation with spin is different. The spin structure in the Volkov solutions (and observables calculated from them) is often approached from a semi-classical perspective, which leads to discussions of e.g.~the appropriate spin operator, the behaviour of the {classical spin vector, and so on.} A quantum approach, on the other hand, shows directly that the physics {can be made} very simple {in a plane wave}.

{Given the central dependence of quantities on lightfront time, a natural basis of fermion states is that of definite lightfront helicity. These states have the property that the helicity is equal to the expectation value of the spin in the $z$-direction (so $\pm1$), in \emph{all} Lorentz frames, so that we may talk of spin and helicity interchangeably~\cite{Chiu:2017ycx}. With this choice of free spinors, the spinor structure in (\ref{volkov-e-in}) can be shown to obey}
\be\label{spinspin}
        \bigg(1+\frac{\slashed{k}\slashed{a}(\varphi)}{2k\cdot p}\bigg) u_{p} = u_{\pi_p(\varphi)} \;,
    \ee
which is \textit{not} notation -- {the Volkov spinor (\ref{spinspin}) is identically equal to a \textit{free} spinor carrying the on-shell, time-dependent momentum $\pi_\mu$, so $\slashed{\pi} u_\pi = m u_\pi$.} {In particular, the (suppressed) spin/helicity label is unaffected and a basis of Volkov spinors (\ref{spinspin})} is obtained simply by replacing $p_\mu$ with $\pi_\mu$ component-wise {in the free spinor basis
(for the construction of which see e.g.~\cite{Ilderton:2020gno}).
What this means physically is that} {propagation through a plane wave cannot change the quantum spin state of an electron~\cite{Ilderton:2020gno}.} {(See also~\cite{Adamo:2021hno} for a demonstration of this in SUSY theories, using the spinor-helicity formalism.)} The emission or absorption of photons, and loop effects, can all however flip the spin, {as will be discussed in} Sec.~\ref{sec:first} and Sec.~\ref{sec:second}.
   
The field-dependent structure in (\ref{spinspin}) therefore accounts simply for the fact that the covariant spin vector $\Sigma_\mu$ must rotate to maintain orthogonality with the time-dependent particle momentum\footnote{Compare with a semi-classical approach using the Bargmann-Michel-Telegdi (BMT) equation~\cite{Bargmann:1959gz} \emph{without} contribution from the anomalous magnetic moment; the solution of this equation in a plane wave describes exactly the precession of $\Sigma_\mu$ needed to maintain $\pi\cdot\Sigma=0$, though the spin state is in fact not changing.} $\pi_\mu$, so $\pi\cdot \Sigma = 0$. This, together with the fact that the Volkov exponent is the classical Hamilton-Jacobi action, underlines the result that the Volkov wavefunctions are semiclassical-exact. {For representations of the Volkov solutions designed to make this explicit, including investigation of other spin bases}, see~\cite{DiPiazza:2021rum,DiPiazza:2022lij}.

{In the next section of this review we will consider S-matrix elements of Volkov solutions, photons, and their associated propagators in detail, analysing both their formal properties and the phenomenology they lead to.} We will see that, just like in the classical theory, three components of canonical momentum are conserved at each interaction vertex. The loss of just one momentum conservation law is enough, however, to lead to significant increases in complexity even in, say, three and four point tree level amplitudes.
 
{Before moving on, though, we comment briefly on large-distance, infra-red, behaviour of scattering in plane wave backgrounds, as it is related to ideas we have already met.} Consider a situation in which the background electromagnetic fields $\mathcal{F}_{\mu\nu}$ do \emph{not} switch off asymptotically. In some such cases, observables can be defined as limits of those in sandwich waves, but in other cases not. Amplitudes in a monochromatic (hence exactly periodic) background, for example, can be recovered from those in `wavetrain' backgrounds containing a finite number of cycles, in the limit where the number of cycles is taken to infinity~\cite{Heinzl:2010vg,King:2020hsk,Tang:2021qht}. (Working directly in the monochromatic limit allows for alternative physical interpretations, see e.g.~\cite{Lavelle:2014mka,Lavelle:2015jxa,Lavelle:2017dzx,Lavelle:2019lys,Lavelle:2019vcz} for continuing investigations.) On the other hand, amplitudes in `constant crossed fields' (CCF), meaning $E_j$ constant in (\ref{plane-wave-F}), are not recovered as the long-pulse limit of fields which are nonzero and constant for a finite time~\cite{Dinu:2012tj}; the latter exhibit the well-known logarithmic infra-red behaviour of QED, whereas the CCF misses all such effects and incorrectly implies that there is only an integrable singularity in the IR.


\section{First-order processes}\label{sec:first}

    We now consider the first specific class of quantum electrodynamical interactions of particles in a strong plane-wave background field: those that are of first order in the Furry expansion, Eq.~\eqref{Furry-action}, in a plane-wave background. By `first order', in this section, we specifically refer to tree-level processes with a single (dressed) interaction vertex, where all asymptotic in and out states are on their respective mass shells, i.e. they describe real incoming and outgoing particles. Their S-matrix elements can be given as $ S^{(1)} =  \langle \mathrm{out} | \hat S^{(1)}[A] | \mathrm{in} \rangle $, where $\hat S^{(1)}[A] = -ie\int \ud^4x : \bar\psi \slashed A \psi:$ and all operators are in the Furry picture, as described in Sec.~\ref{sec:intro:Furry}, and normal ordered (as indicated by $::$). (For applications of worldline methods to scattering in plane wave backgrounds see~\cite{Ilderton:2016qpj,Edwards:2021uif,Esposti:2021wsh}.) {For a discussion of first order processes in other `simple' backgrounds (such as constant magnetic fields relevant for astrophysics) we refer to~\cite{Kuznetsov:2004tb} and references therein.}

\begin{figure}
        \centering
        \includegraphics[width=0.8\columnwidth]{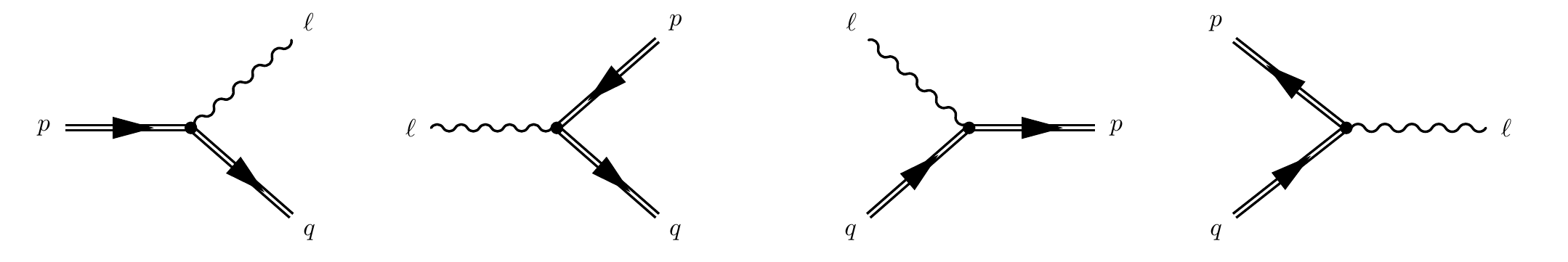}
        \caption{Feynman diagrams for the four first-order processes (left to right): Nonlinear Compton, nonlinear Breit-Wheeler, photon absorption, and one-photon pair annihilation.}
        \label{fig:first:graphs}
    \end{figure}

    Staying within proper QED in a plane-wave background, there is a total for four distinct first-order processes, consisting of the interaction of two `dressed' fermion lines, represented by the Volkov states, Eqs.~\eqref{volkov-e-in}--\eqref{volkov-e-out} and one ordinary photon line. Nonlinear Compton scattering (NLC, $e\to e\gamma$, Sec.~\ref{sec:first:nlc}) and nonlinear Breit-Wheeler pair production (NBW, $\gamma\to e^\LCp e^\LCm $, Sec.~\ref{sec:first:nbw}) are both $1\to2$ first-order strong-field QED processes; pair annihilation ($e^\LCp e^\LCm \to \gamma $) and photon absorption ($e\gamma\to e$, see Sec.~\ref{sec:first:absorption}) are their time-inverse $2\to1$ processes, see Figure \ref{fig:first:graphs}. The S-matrix elements for those processes are related by crossing symmetry. All those processes have in common that they are \emph{field induced} in the sense that their probabilities vanish in the absence of the background field ${\mathcal A}\to 0$, i.e.~as $\xi\to0$. This is because in QED without a background field the single vertex Feynman diagrams are kinematically forbidden for all particles on-shell.
    
    The (formally calculated) probabilities $\prob$ for the first order processes are not bounded. For instance, for $\xi\gg1$ the nonlinear Compton scattering rates, $\mathsf{R}$, have the well-known scaling behaviour \cite{Nikishov:1964zza,Ritus1985}
    \begin{align} \label{eq:first:NLCrate}
        \mathsf{R} \sim
        \begin{cases}
        \: \alpha \chi       \,,           & \chi \ll 1  \,,\\
        \: \alpha \chi^{2/3} \,,           & \chi \gg 1  \,.
        \end{cases}
    \end{align}
    Hence, the probability, which can be written as $\prob \sim \mathsf{R} \times  \phaselengthparameter$ for pulse phase length scale $\phaselengthparameter$, exceeds unity for sufficiently long pulses $\phaselengthparameter\gg1$ and/or high intensity $\xi\gg1$ since $\chi\propto\xi$. {This does not imply some inconsistency in strong-field QED, but rather is one of its central features:} $\prob >1$ indicates that higher-order $\mathcal O(\alpha^{n>1})$ ($n$-vertex) processes become important in these cases. This means one has to calculate processes with higher multiplicity final states, as well as loop corrections. However, the complete and exact evaluation of all higher-order terms of the Furry expansion within QED is not feasible for most cases. Therefore, one has to find suitable approximations to calculate those terms. This will be the subject of later sections in this review, see e.g.~Secs.~\ref{sec:second}, \ref{sec:approx} and \ref{sec:higher}.

    {In this section we aim at discussing experimentally observable signals for the most fundamental first-order processes. In order to circumvent the issue of higher orders for now, we introduce the notion of a \emph{`thin target'}, where the conditions of the scattering are such that multi-vertex processes are strongly suppressed.}
    This places a restriction on the background field pulses being not too strong and/or not too long. In this case, the scattering probabilities and spectra one calculates by squaring the first-order S-matrix are much smaller than unity and hence can---to a good approximation---be in principle be directly observed in an experiment. If the target is not thin, the probability for higher-order $n$-vertex processes are non-negligible. Nonetheless, the first order processes can often serve as building blocks to construct approximations for higher order processes, as is discussed in detail for instance in Sections \ref{sec:second} and \ref{sec:higher} of this review, or used as ingredients for simulations (see Sec.~\ref{sec:approx} or the recent Review~\cite{Gonoskov:2021hwf}). In this section the focus will be on observables, such as the differential spectra for the final particles and their polarisation properties in the thin target regime.

    \subsection{General characterisation of $1\to2$ first order processes}
    \label{sec:first_order_general_props}

    As mentioned above, the one-vertex processes for all particles on-shell are forbidden kinematically due to four-momentum conservation in vacuum. In the presence of a background plane wave these processes are allowed since the particles can exchange momentum with the background. For nonlinear Compton scattering and nonlinear Breit-Wheeler pair-creation the particles absorb momentum from the background, while for the $2\to 1$ processes momentum is deposited into the plane wave mode. Here we briefly discuss the general kinematics and properties of the $1\to 2$ first order processes while maximally exploiting the lightfront symmetry of the problem. The discussion of the inverse $2\to1$ processes is deferred to Sec.~\ref{sec:first:absorption}. We write for the four-wavevector of the plane wave $k_\mu=\omega(1,0,0,1)$, i.e.~$k^\LCm =2k_\LCp=2\omega$ is its only non-vanishing lightfront component, and the plane wave phase reads $\varphi=k\cdot x = \omega x^\LCp$ (cf.~Eq.~\eqref{eq:lightfront-conventions} for the adopted conventions for light-front variables). Lightfront momentum conservation at the vertex thus provides three constraints
    \begin{align} \label{eq:emc-lf}
        p_\mathrm{in}^\LCp = p_1^\LCp + p_2^\LCp \,, \qquad
        \vec p_\mathrm{in}^\LCperp = \vec p_1^\LCperp + \vec p_2^\LCperp \,,
    \end{align}
    with $\vec p^\LCperp = (p^x,p^y)$.
    
    It is convenient to introduce dimensionless and normalised momentum variables. By singling out `particle 1' as the one to be measured, we define the lightfront momentum transfer fraction $s$ and normalised transverse momentum $\vec r^\LCperp$ as follows:
    \begin{align} \label{eq:first:srperp}
        s :=  \frac{p_1^\LCp}{p_\mathrm{in}^\LCp} =  \frac{k \cdot p_1}{ k \cdot p_\mathrm{in} }  \,, \qquad \qquad
        \vec r^\LCperp := \frac{\vec p_1^\LCperp}{ms} \,.
    \end{align}
    The other particle `2' will be integrated out as will become apparent below. The physical ranges of those dimensionless variables are $s\in(0,1)$ and $\vec r^\LCperp \in \mathbb R^2$. The exchange of $P^\LCm$-momentum between the particles and the background field during the interaction does not yield a fourth conservation law.

\begin{figure}[!ht]
        \centering
         \includegraphics[width=7cm]{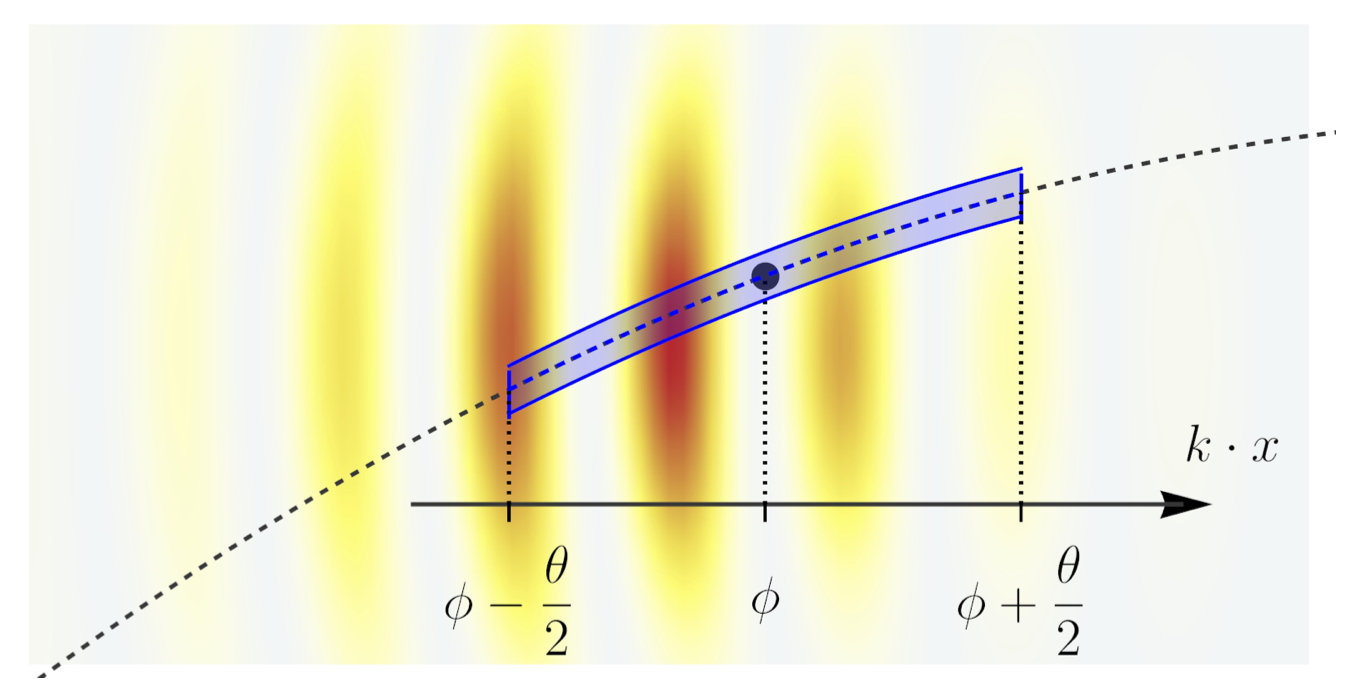} 
         \caption{Sketch of integration window defined by the interference phase $\theta$ around the average phase position $\phi$ for a probe charge propagating through a laser pulse.}
        \label{fig:lcfafig1}
    \end{figure}

    In a plane-wave background, one can formally write the S-matrix element for a first-order process using the four-dimensional lightfront delta function $\delta^{(4)}( t k - P ) = 2 \delta( t k^\LCm - P^\LCm ) \delta(P^\LCp)\delta^{(2)}(\vec P^\LCperp)=2 \delta( t k^\LCm - P^\LCm ) \delta^{(3)}_{l.f.}(P)$ as follows:
	\begin{align}
    S&=  
    	-ie (2\pi)^4 \int \! \frac{\ud t}{2\pi} 
    	\: \delta^{(4)}( t k - P  ) \: \mathcal M_t \,, 
	\end{align}
	where $P^\mu =   p_{1}^\mu + p_{2}^\mu - p_\mathrm{in}^\mu$, and the integral over $t$ takes into account the non-conservation of $P^\LCm$-momentum. After performing the integration over $\ud t$, the amount of lightfront momentum exchange between the background field and the particles is fixed by the kinematics, $t \to \nu = \nu(s,\vec r^\LCperp) = P^\LCm/k^\LCm $. For fixed $\nu$, one can \emph{formally} write down four-momentum conservation as $p_\mathrm{in} +\nu k = p_1 + p_2$. The amplitude $\mathcal M=\mathcal M_{t=\nu}$ is an integral over the laser phase $\varphi$ and contains all dependence on the properties of the laser pulse such as its pulse duration, shape and polarisation etc., as well as the spin and polarisation properties of the involved particles in the in and out states. The specific form of $\mathcal M$ for each of the first order processes will be discussed below.
	 The phase-space integrated probability for the process under consideration is then given by
    \begin{align}
        \prob 
                & = 
                \frac{1}{2p_\mathrm{in}^\LCp} \int \!  \ud \tilde{p}_1 \,  \ud \tilde{p}_2 \: |S|^2 \,,
    \end{align}
    where the Lorentz-invariant on-shell phase space element in lightfront coordinates is
    \be\label{eq:onshell:measure}
    \ud \tilde{p} = \frac{\ud p^\LCp \ud^2 \vec p^\LCperp }{(2\pi)^3 2 p^\LCp}.
    \ee
    The conservation of three lightfront momentum components in Eq.~\eqref{eq:emc-lf} allows one to completely integrate out the momentum $p_2$. \footnote{{Volume factors can be formally treated by the normalisation $\delta^{(3)}_{l.f.}(0)=1/(2\pi)^3$. Alternatively this can be done introducing wavepackets for the initial states~\cite{Ilderton:2012qe}.}} Employing \eqref{eq:first:srperp}, the triple differential probability for the $1\to 2$ processes can be expressed as 
    \begin{align} \label{eq:first-probability-general}
        \frac{\ud^3\prob}{\ud s\, \ud^2\vec r_\LCperp}  & = 
        \frac{ \alpha  }{16\pi^2  m^2 \eta^2 } 
        \frac{s }{1-s }  \: |\mathcal M |^2 \,,
    \end{align}
    with fine structure constant $\alpha = e^2/4\pi$, and quantum energy parameter $\eta =p_\mathrm{in} \cdot k /m^2$. The squared amplitude is a double integral over the phases of $\mathcal M$ and the complex conjugate, $|\mathcal M|^2 = \int \ud\varphi\, \ud \varphi' M(\varphi) M^*(\varphi')$, which is often conveniently expressed in terms of:
    \begin{align} \label{eqn:phidefs1}
    \phi = \frac{\varphi+\varphi'}{2}\,, \qquad \theta = \varphi-\varphi' \,,
    \end{align}
    where $\phi$ is the average phase and the phase difference $\theta$ is also called the interference phase. The relations between these variables are sketched in \figref{fig:lcfafig1}.

    In order to make contact between strong-field QED and perturbative QED calculations a cross section can be defined by dividing the probability by the flux of laser photons per transverse area, $\sigma = \prob / N_\perp$, where 
    $N_\perp = \omega^{-2} \int_{-\infty}^\infty \! \ud\varphi \: T^{00}(\varphi) $ where $T^{\mu\nu}$ is the energy momentum tensor of the plane-wave background field. For an infinitely long plane wave $N_\perp$ contains an infinite phase length (`volume factor') that cancels the corresponding factor in $\prob$, rendering $\sigma$ finite. For instance, one can show that one obtains the Klein-Nishina cross section in the limit $\xi\to0$ of the nonlinear Compton scattering cross section. However, due to the nonlinearity of the strong-field QED interaction, this `cross section' still depends explicitly on the value of $\xi$ in general. Hence, this concept is of limited utility in strong fields.

    \subsection{Nonlinear Compton scattering}
    \label{sec:first:nlc}

    Nonlinear Compton scattering (NLC) is given by the reaction $e(p)\to e(q) \: \gamma(\ell)$, where the particles' four-momenta are given in brackets.
    The corresponding scattering amplitude reads
    \begin{align} \label{eq:first:MNLC}
    \mathcal M = \int\! \ud \varphi \: 
    \bar u_{\pi_q} \slashed \epsilon^*  u_{\pi_p} \: e^{i \dynamicphase (\varphi)}
    \end{align}
    where the dynamic phase $\dynamicphase =  \int_{\varphi_0}^\varphi \ud \varphi' \: \dynamicphase'(\varphi')$
    with
    \begin{align}
        \dynamicphase'(\varphi) = \frac{\ell \cdot \pi_p(\varphi)}{k\cdot q} = \frac{s}{2\eta(1-s)} \left[ 
        1 + \left(\vec r^\LCperp-\frac{ \boldsymbol 
    \pi^\LCperp_p(\varphi)}{m}\right)^2 \right] \,.
    \end{align}
    The the emitted photon frequency $\omega'=\ell^0$ can be expressed as
    \begin{align} \label{eq:omega-l}
    \omega'(\nu) = \frac{\nu k\cdot p}{(p+\nu k)\cdot \upsilon}\,,
    \end{align}
    with $\upsilon^\mu=\ell^\mu/\ell^0$, $\nu = \ell\cdot p/k\cdot q = \frac{s}{2\eta(1-s)} (1+\rho_\LCperp^2)$ and $\boldsymbol{\rho}_\LCperp = \vec r_\LCperp-\vec p_\LCperp/m$, thus $s=\frac{2\nu\eta}{1+2\nu\eta + \rho_\LCperp^2}$. The normalised transverse momentum is a function of the emission angles ($\vartheta,\varpi$): $(r^x,r^y)=\frac{p^\LCp}{2m} (\cos\varpi,\sin\varpi)\tan\frac{\vartheta}{2}$, which is approximated for $\vartheta \ll 1$ and $p^\LCp\gg m$ as $r^j \approx \gamma \vartheta_j$ \cite{Blackburn:2019lgk}, where $\gamma$ is the initial electron Lorentz factor.

    The differential photon emission probability follows by evaluating Eq.~\eqref{eq:first-probability-general}, which is the completely polarisation resolved probability {(see also \secref{sec:first:nlc:spin})}. In order to acquire the unpolarised probability one has to sum over all polarisation states of the final particles and average over the initial electron polarisation as usual in scattering calculations in QFT. (However, the laser polarisation cannot be treated in a similar manner since the interaction with the background is nonlinear.)

    {The total NLC probability can be calculated analytically by performing the momentum integrals before the phase integrals \cite{baier98}. This is based on the observation that the transverse momentum integrals $\ud^2\vec r_\LCperp$ are Gaussian and hence may be performed immediately. Specifying e.g. linear polarisation of the background, $\vec a^\LCperp = m\xi(f(\varphi),0)$, using the form presented in Ref.~\cite{Dinu:2013hsd}, one obtains}
    \begin{align}\label{eq:first:rpGauss1} 
        \prob = \frac{\alpha}{2\pi \eta} \int \! \frac{\ud\varphi\ud\varphi'}{i(\theta + i0^+)}
        \int_0^1 \ud s \: 
        \left( 
            1 - \frac{\xi^2 g(s)  }{2} \,  \theta^2 \langle f'\rangle^2 
        \right) 
        \, \exp \left\{ \frac{i u \mu\theta}{2\eta} \right\} \,,
    \end{align}
    where $f'=\ud f/\ud\varphi$, $u=s/(1-s)$, and $g(s)=1+su/2$. The normalised Kibble mass $\mu$ is defined as
    \begin{align} \label{eq:first:kibble}
      \mu  = \frac{\langle \pi_p \rangle^2}{m^2} = 1 - \frac{\langle a^2 \rangle}{m^2}  + \frac{\langle a \rangle^2}{m^2} = 1 +\xi^2 (\langle f^2 \rangle  - \langle f \rangle^2 )  \,, 
      \qquad \langle f\rangle=\frac{1}{\theta} \intop_{\phi-\theta/2}^{\phi+\theta/2} \!  f(t)\: \ud t \,,
    \end{align}
    with the floating phase average $\langle f\rangle$ \cite{Brown:1964zzb,Kibble:1975vz,Harvey:2012ie}. Note that the term with the `1' in \eqref{eq:first:rpGauss1} needs to be regularised before the phase integrals can be evaluated numerically \cite{Dinu:2013hsd}. The integral over lightfront momentum transfer fraction $s$ can also be performed analytically, such that the total probability is expressed just as a double phase integral, with the integrand containing trigonometric integrals with the argument $\mu\theta/2\eta$ \cite{Dinu:2013hsd}. The calculation of the expectation value of the emitted photon four momentum $\langle \ell^\mu \rangle = \int \ell^\mu \ud \prob$ gives rise to the definition of the so-called Gaunt factor as the ratio of emitted power in the quantum case of NLC compared to the classical emission. The Gaunt factor is often used to phenomenologically introduce some quantum effects in classical radiation reaction calculations \cite{RIDGERS2014273,Niel:PRE2017}, see also Sec.~\ref{sec:higher:classicalRR}.

    \subsubsection{Regularisation}
    \label{sec:first:nlc:reg}
    
    For the numerical evaluation of Compton scattering spectra one often directly performs the phase integrals on the amplitude level, for which it is convenient to expand $\mathcal M= \sum_j T_j I_j$, with constant coefficients $T_j = \bar u_{p'} \Gamma_j u_p$, where $\Gamma_j$ are products of Dirac-matrices and $I_j$ are scalar phase integrals, see e.g.~Refs.~\cite{Seipt:2012tn,Seipt:2013taa,Krajewska:2012gc,Titov:2015tdz}. One of the phase integrals is an integral over a pure phase, $I_0 = \int e^{i\dynamicphase} \ud \varphi $, which requires a careful treatment. 
    {By explicitly writing out convergence factors $e^{- 0^+|\varphi|}$ and integrating by parts \cite{Boca:2009zz,Seipt:2010ya,Mackenroth:2010jr,Dinu:2012tj,Krajewska:2012gc,Ilderton:2019bop} it is possible to separate field-free and field-dependent terms, and to make the amplitudes manifestly gauge invariant with no ambiguous boundary terms \cite{Ilderton:2020rgk}. For Compton scattering we specifically have
    \begin{align} \label{eq:first:regularization}
    I_0 = 2\pi\delta(\nu) + \frac{\nu}{\nu+i0^+} \int \! \ud \varphi \: \Delta(\varphi)\, e^{i\dynamicphase} \,, \qquad \Delta(\varphi) =  1-\frac{\ell \cdot \pi_p(\varphi)}{\ell \cdot p}\,.
    \end{align}
    The second term is the regularised, explicitly background-field dependent part of integral $I_0$, where it can be seen that $\Delta\to0$ as $\xi\to0$. The first term, $\propto \delta(\nu)$, is associated to the background-field independent part of the amplitude \cite{Ilderton:2020rgk}. Here it has support at vanishing lightfront momentum exchange with the background and is thus kinematically forbidden. Hence, this term has no physical significance for calculations of the spectrum of NLC (it would correspond to the emission of zero-frequency photons) \cite{Boca:2009zz,Seipt:2011zz}. Nonetheless, such terms are relevant for higher-order processes \cite{Ilderton:2020rgk}. For instance, it was pointed out in Ref.~\cite{Acosta:2019bvh} that those terms are significant for obtaining the correct weak-field (perturbative QED) limit of trident pair production.}

    \subsubsection{Infinite monochromatic plane waves} The majority of recent developments in NLC have been investigating laser pulses with finite duration. However, in order to better understand short pulse effects, let us briefly consider, for later comparison, the case of infinitely long (monochromatic) plane waves, see for instance Refs.~\cite{landauQED,Ritus1985,2009RPPh...72d6401E,DiPiazza:2011tq} for details. The essential quantity to analyse is $\ell \cdot \pi_p(\varphi) \propto \dynamicphase'$ in the phase of $\mathcal M$, which contains both oscillating and constant terms. The oscillating parts of the dynamic phase $\dynamicphase$ can be treated by means of the Jacobi-Anger expansion \cite{abramowitzStegun},
    \begin{align} \label{eq:first:jacobi-anger}
         e^{-i\alpha\sin\varphi} = \sum_{n=-\infty}^{\infty} J_{n}(\alpha)\:e^{-in\varphi} \,,
    \end{align}
    which is used to separate the Fourier components of the amplitude. This turns $\mathcal M$ into a sum over harmonics $\mathcal M=\sum_{n} \mathcal M_n$. The $n$th Fourier component of the amplitude is often referred to in the literature as the net absorption of $n$ photons from the background field. The constant term in $\dynamicphase'$---its Fourier zero-mode---can be understood in terms of the cycle-averaged electron momentum, also called the quasi-momentum,
    \begin{align} \label{eq:first:quasimomentum}
    \overline{\pi}^\mu_p = \frac{1}{2\pi} \intop_{-\pi}^{\pi} \! \ud \varphi \: \pi^\mu_p(\varphi) = p^\mu + U \, k^\mu \,,
    \end{align}
    where\footnote{\label{foot:LPCP}Note that the explicit result of the cycle average is $\overline{\vec a_\LCperp^2}=m^2 \xi^2/2$ for linearly polarised pulses $\vec a^\LCperp = m \xi (\cos\varphi ,0)$. For circular polarisation, $\vec a^\LCperp = m \xi (\cos\varphi ,\sin\varphi)$, one has to make the replacement $\xi^2/2\to\xi^2$.} $U = \xi^2/4\eta$. The transfer of $P^\LCm$-lightfront momentum between the particles and the background field becomes discrete, $ \nu\to \nu_n = n - \beta$ with $\beta = U\, \frac{\ell\cdot k}{q\cdot k}=\frac{\xi^2 u }{4\eta}$ and $u=s/(1-s)$, but not integer. The $\xi^2$-dependent term stemming from the `ponderomotive potential' $U$ is responsible for the intensity-dependent red-shift of the discrete photon frequencies $\omega'_n = \omega'(\nu_n)$ \cite{Nikishov:1964zza,landauQED}.

    \subsubsection{Pulse shape and interference effects} \label{sec:first:NLC:shape}
    
    We now consider the case of pulsed plane waves with an envelope $\psi(\varphi)$ of finite duration. {For instance the vector potential $\vec a^\LCperp = m\xi \psi(\varphi) (\cos\varphi,0)$ specifies a linearly polarised background. It is convenient to demand that $\text{max}|\psi|=1$.} Often used pulse envelopes include hyperbolic secants, $\psi(\varphi)=1/\cosh(\varphi/\phaselengthparameter)$ \cite{Boca:2009zz,Mackenroth:2010jr,Seipt:2010ya,Seipt:2016rtk}, Gaussians \cite{Boca:2009zz,Seipt:2010ya,Tang:2021qht}, $\sin^K$ envelopes with compact support \cite{Heinzl:2010vg,King:2020hsk,Angioi:2016vir,Meuren:2015mra}, multi-parameter profiles \cite{Seipt:2010ya,Titov:2014usa,Seipt:2016fyu}, and simple flat-top profiles with constant amplitude \cite{King:2020hsk,Tang:2021qht}. For pulses with a non-constant amplitude the differential NLC probability acquires rich structures due to interference effects \cite{1996JETP...83...14N,Boca:2009zz,Seipt:2010ya,Mackenroth:2010jr,Boca:2012pz,Krajewska:2012gc,Krajewska:2013poa,Seipt:2014yga,Krajewska:2014fwa,Krajewska:2015hua,Dinu:2015aci,Seipt:2016rtk}, see for instance Fig.~\ref{fig:first:chirp} (a). Most of the interesting features due to finite pulses have been found in the intermediate intensity regime $\xi\sim 1$ as will be described in the following.

    From the stationary phase condition of the scattering amplitude, $\dynamicphase'(\varphi_\star)=0$, we get $(\vec r_\LCperp-\boldsymbol \pi_\LCperp(\varphi_\star)/m)^2=-1$.
    This shows that the location of the stationary points is determined by the classical transverse kinetic momentum $\boldsymbol \pi^\LCperp = \vec p^\LCperp - \vec a^\LCperp$. The stationary points $\varphi_\star$ have an imaginary part, and if this is large, the amplitude becomes exponentially suppressed \cite{Mackenroth:2010jr}.
    For an oscillatory background field, one usually finds more than one relevant stationary point, which implies that photons with the same momentum can be emitted at different laser phases. Hence, the observed spectra can be seen as the interference pattern due to the interference of multiple paths with the same final state.

\begin{figure}[!th]
        \centering
        \includegraphics[width=0.8\columnwidth]{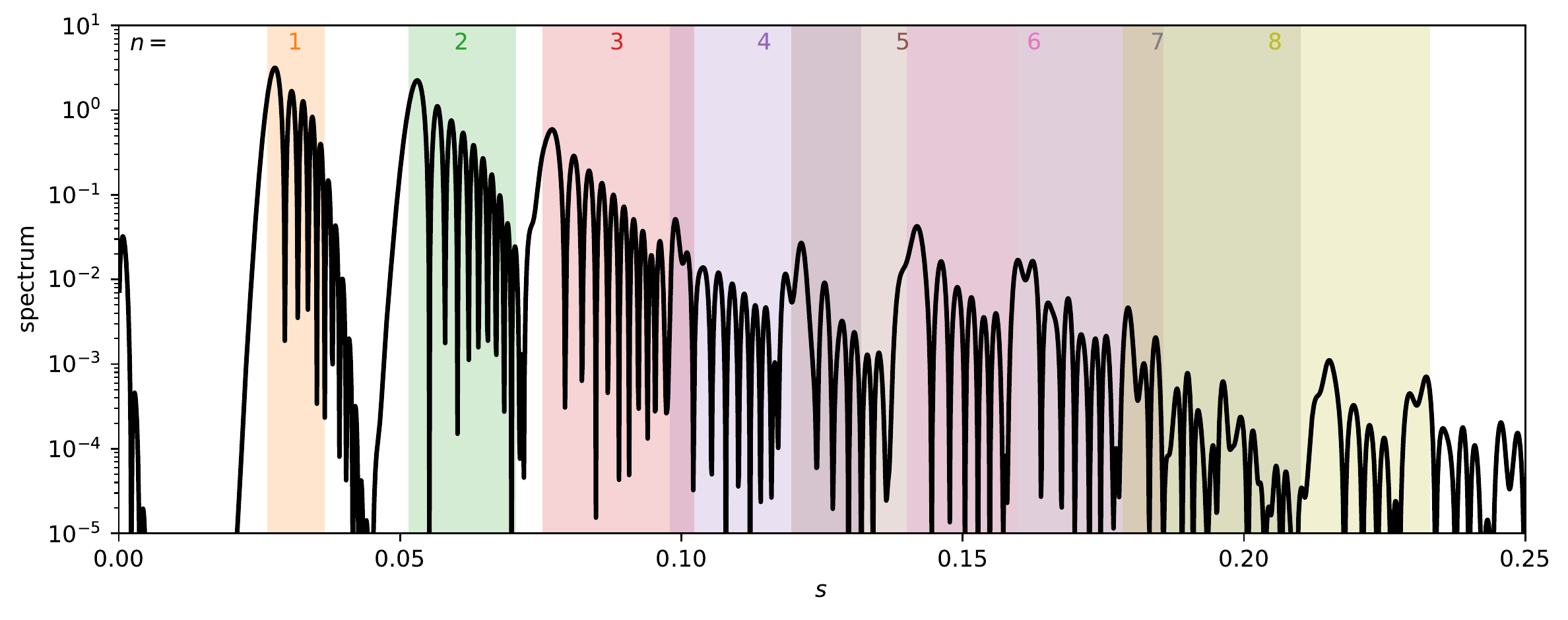}
        \caption{Nonlinear Compton spectrum for $\xi=1$ and $\vec r^\LCperp=(0.5,0)$. The coloured shaded regions depict the spectral range due to ponderomotive broadening (ii), which are overlapping here for $n\geq3$. The harmonic peaks itself exceed the coloured regions due to the laser bandwidth (i), and the substructure is due to interference effects (iii).}
        \label{fig:first:harmonics}
    \end{figure}

    For multi-cycle pulses, $\phaselengthparameter \gg 1$, one can separate the contributions to the phase $\dynamicphase$ due to the fast carrier-wave oscillation and the slow change of the pulse envelope by discarding small terms $\mathcal O(1/\phaselengthparameter)$ \cite{1996JETP...83...14N}. {To} this end, one defines the cycle average as in \eqref{eq:first:quasimomentum} over the floating interval $[\varphi-\pi,\varphi+\pi]$ \cite{1996JETP...83...14N}. Now the quasimomentum and the ponderomotive potential are not constant (as the were for infinitely long plane waves) but instead change on the slow time scale of the pulse envelope $\psi(\varphi)$ by $U\to U \psi^2(\varphi)$. This has profound consequences for the spectra as we discuss below.
    For the oscillating parts of $\dynamicphase$ one can employ the slowly varying envelope approximation, by performing an integration by parts, e.g.
    \begin{align} \label{eq:first:svea}
       \int \! \ud \varphi \: \psi(\varphi) \cos \varphi
       = \psi(\varphi) \sin\varphi 
       - \int \! \ud \varphi  \: \frac{\ud \psi(\varphi)}{\ud\varphi} \sin \varphi
    \end{align}
    and dropping the second term, since derivatives of the pulse envelope are $\mathcal O(1/\phaselengthparameter) \ll 1$. It is now possible to apply the Jacobi-Anger expansion, Eq.~\eqref{eq:first:jacobi-anger}, to the fast oscillating parts of the non-linear phase $e^{i\dynamicphase}$. This amounts to a Fourier series expansion over the floating interval $[\varphi-\pi,\varphi+\pi]$ \cite{1996JETP...83...14N}. The Fourier coefficients here depend on $\varphi$ on the slow timescale of the envelope, e.g.~$\alpha \to \alpha \psi(\varphi)$ in Eq.~\eqref{eq:first:jacobi-anger}. Thus, the arguments of the Bessel functions become phase dependent, $J_n[\alpha \psi(\varphi)]$, but only on the slow envelope time scale. The Jacobi-Anger expansion again transforms the amplitude into a sum over `harmonics', but in contrast to the infinite plane waves case those harmonics are much more complex in a pulse due to an interplay between the frequency bandwidth associated to the finite duration and nonlinear effects. The three main contributions to the harmonic line shape can be identified in the case of multi-cycle pulses as (i) the laser bandwidth, (ii) ponderomotive broadening and (iii) interference effects, as will be discussed in the following, see also Fig.~\ref{fig:first:harmonics}.
    
    (i) The laser bandwidth contributes to the linewidth of the harmonics. Its contribution is inversely proportional to the pulse duration $\propto 1/\phaselengthparameter$ and hence becomes dominant for very short pulses. For weak fields $\xi^2\ll1$ this is the main contribution to the harmonic bandwidth. The shape of the $n$th harmonic line in the perturbative limit is given by the Fourier transform of the $n$th power of the laser pulse envelope $\psi$. {If this were the only effect, one would expect the harmonics in a pulse to converge to narrow lines for long pulses also for $\xi\gtrsim 1$. But this is not the case due to a another effect: ponderomotive broadening.}

    (ii) The ponderomotive broadening effect occurs because the ponderomotive potential $U$ is not constant in a pulse, but instead continuously shifts as the electron travels through the pulse with envelope $\psi(\varphi)$. 
    This yields a contribution to the phase according to $\beta \:\int \!\ud\varphi\: \psi^2(\varphi)$, {where $\beta$ was defined below Eq.~\eqref{eq:first:quasimomentum}}. The harmonics in a pulsed plane wave have spectral support not only on a delta-comb as they have in the infinite plane wave case. Barring the bandwidth, their spectral support is determined by the existence of real stationary points of the phases of the $n$th harmonic amplitude 
    \begin{align} \label{eq:first:spa}
       0 = n - \nu - \beta\, \psi^2(\varphi) = n - \frac{s}{2\eta(1-s)}\left( 1+\rho_\LCperp^2 + \frac{\xi^2}{2} \psi^2(\varphi) \right) \,. 
    \end{align}
    Thus, in a pulse the $n$th harmonic has support between the nonlinearly red-shifted $n$th infinite plane wave harmonic and the linear $n$th harmonic since the envelope $\xi\psi(\varphi) \to 0$ as $\varphi\to\pm\infty$, see the coloured ranges in Fig.~\ref{fig:first:harmonics}. As was pointed out in Ref.~\cite{2010PhRvL.105m0801H}, ponderomotive broadening is a nonlinear effect that does not necessarily require strong fields $\xi\gtrsim1$. The effect could be observed also for $\xi\ll1$ if the laser bandwidth is small enough such that $\xi^2\gtrsim 1/\phaselengthparameter$.

    (iii) The interference of emission from different parts of the pulse evokes an oscillatory structure in each harmonic with multiple sub-peaks. For a simple pulse with a single peak, the same ponderomotive red-shift occurs twice: once at the rising  slope of the pulse and again in the falling slope. The sub-peak multiplicity is determined by the value of the total ponderomotive phase shift, $\sim \beta \int_{-\infty}^\infty \!\ud\varphi \: \psi^2(\varphi) \sim \frac{\xi^2 \nu \phaselengthparameter }{1+\rho_\LCperp^2}$, divided by $2\pi$. It also remains an open question to investigate under which circumstances the fine structures in the spectra can be observed experimentally. {For instance, by considering the scattering of electron wavepackets \cite{Ilderton:2012qe,Angioi:2016vir} (i.e.~including electron momentum spread), it was shown that these features are partly or even completely washed out \cite{Angioi:2016vir}}, see also \cite{Seipt:2011zz,Harvey:2016wte} for similar effects due to laser focusing. Note that in the local monochromatic approximation (LMA, see Sec.~\ref{sec:approx:lma}) only the ponderomotive broadening effect (ii) is described properly, the other two can be considered to be effects that are non-local over length scales longer than the wavelength of the central carrier frequency.

    Due to the ponderomotive broadening in a pulsed field, the individual NLC harmonics are usually distinguishable only for relatively small $\xi$. For $\xi\gg1$ the ponderomotive broadening typically becomes larger than the inter-harmonic separation, $\frac{s\xi^2}{4\eta (1-s)} >1$. In case the harmonics overlap, the differential NLC spectrum appears as a broad continuum with complex peak structure, where $|\mathcal M|^2 = \sum_{n,n'} \mathcal M^*_n \mathcal M_{n'}$ contains interference between different harmonics \cite{2013LaPhy..23g5301S,Heinzl:2020ynb}. While contributions from $n'\neq n$ can be important for differential spectra, they cease to be important for very long pulses and do not affect the total probabilities because the interference averages out by virtue of a finite detector resolution \cite{1996JETP...83...14N,2013LaPhy..23g5301S}.

    While for infinite plane waves analytic results for the NLC probability are well known \cite{Ritus1985,landauQED}, for finite duration laser pulses completely analytical expressions are rare. Explicit expressions for circularly polarised flat-top pulses were discussed recently in \cite{King:2020hsk}. Finite duration pulses with non-flat-top break the symmetry in the longitudinal direction, making the phase integration more difficult. In that case closed form analytic results can only be expected in a few rare cases in terms of special functions. Using classical electrodynamics, Ref.~\cite{Hartemann:1996zz} found an analytic expression for the on-axis spectrum for a circularly polarised hyperbolic secant pulse $g=1/\cosh(\varphi/\phaselengthparameter)$ in terms of degenerate (confluent) hypergeometric functions. More general analytic results for arbitrary scattering angles and a wider range of pulse envelope functions $\psi$ were obtained in Ref.~\cite{Seipt:2016rtk}.

    Interference effects also play an important role in the scattering of multiple laser pulses or pulse trains \cite{Krajewska:2013tla,Twardy:2013jca,Krajewska:2014fwa,Krajewska:2015hua,Ilderton:2020dhs}. The differential NLC probability in two identical pulses is $\ud\prob/\ud s\ud\vec r_\LCperp|_\mathrm{2\,pulses} = 2 (1+\cos\phi_f) \ud \prob/\ud s\ud\vec r_\LCperp|_\mathrm{1\,pulse}$, were the interference phase $\phi_f = \nu \Delta  + const.$ depends linearly on the pulse separation $\Delta$ and lightfront momentum exchange $\nu$, hence on $s$ and $\vec r_\LCperp$~\cite{Ilderton:2020dhs}. A possible experimental signal was proposed as an oscillation of the detected number of photons in a fixed narrow energy range as a function of varying the pulse separation. For a train of $N_\mathrm{rep}$ pulses the photon energy distribution shows a comb of equally spaced peaks with maxima scaling as $N_\mathrm{rep}^2$, indicating temporal coherence. The distance between the peaks can be controlled by a delay of the pulse in the train \cite{Twardy:2013jca,Krajewska:2014fwa}. A generalised (kinematic) Klein-Nishina formula predicting the locations of the teeth of the frequency comb was derived in \cite{Krajewska:2015hua}. It was demonstrated that the broad bandwidth structures of the scattered radiation are temporally coherent, thus form a supercontinuum, which could be synthesised into zeptosecond (or even shorter) MeV radiation pulses \cite{Krajewska:2013tla}. 

    Interference effects in NLC for pulses with several colours were also studied \cite{2000JETP...90..415N,Wistisen:2014ysa}. It was demonstrated that the two-colour spectra noticeably differ from the `incoherent sum' of the spectra of two infinite plane waves, which results from the nonlinearity of the NLC scattering process. 
    
    Partially integrated cross sections were investigated in \cite{Titov:2014usa,Titov:2015pre,Titov:2019kdk} as a sensitive tool to study the dependence on laser pulse properties. They show a step-like dependence as a function of $\omega'$, where the smoothness of the steps depend on the pulse shape and duration. By introducing a low-momentum cutoff as $c<s/(1-s)$, the partially integrated cross section, summed over all harmonics behaves as $\sigma_c\propto \chi^{3/2} f(c) e^{-2c/3\chi}$ for $c\gg \chi$ \cite{HernandezAcosta:2020agu,Kampfer:2020cbx}. A similar exponential behaviour as a function of $\chi$ had been found for the emission of hard photons using a saddle-point approximation \cite{Dinu:2018efz,Blackburn:2017dpn}.
    Similarly, the differential cross section for fixed photon frequency $\omega'$ first increases as function of $\xi$, but eventually drops with increasing the value of $\xi$ \cite{Acosta:2021iyu}. 

    So far the discussion of NLC was based on a single-incident-particle approximation, where the radiation of $N$ particles is incoherently added. From classical electrodynamics it is known, however, that the emitted power of $N$ particles should scale proportional to $N^2$ instead of $N$ if the wavelength of the emitted radiation is larger than the inter-particle distance. In Ref.~\cite{Angioi:2017ygv}, using two-electron wave packets, it was found that quantum effects can partly or even completely suppress the coherence of emission, even if $\chi\ll1$. This result was explained by means of an additional quantum parameter which relates coherence to the quantum electron recoil during emission.

    \subsubsection{Ultrashort pulses and carrier envelope phase effects}

    The angular distribution of the emitted photons shows some symmetries in infinite plane waves or long pulses. For instance, for a circularly polarised laser, the vector potential rotates in the transverse plane with constant amplitude in an infinite plane wave. Consequently, the photon distribution has azimuthal symmetry around the beam axis for circular laser polarisation in head-on collisions. But for ultra-short (subcycle) pulses, it does not rotate in the transverse plane with full length, which distinguishes one particular direction where $\vec a^\LCperp(\varphi)$ is maximised. Hence, an ultra-short pulse duration can introduce pronounced asymmetries of angular distributions, which gradually disappear with increasing number of laser field oscillations \cite{Krajewska:2012gc,Seipt:2013taa}.

    {The emission pattern in ultra-short pulses also depends on the  carrier-envelope-phase (CEP)---the relative phase $\varphi_\mathrm{CE}$ between the pulse envelope and the carrier wave. For a linearly polarised pulse, e.g.~$a^x(\varphi) = m \xi \psi(\varphi) \cos (\varphi + \varphi_\mathrm{CE})$. Focusing for simplicity on head-on collisions, the relevant term in the dynamic phase is $\propto \vec a^\LCperp \cdot \vec r^\LCperp\propto \xi \psi(\varphi) \cos(\varphi+\varphi_\mathrm{CE})$. Thus, the CEP affects the angular emission range in the laser polarisation plane \cite{Mackenroth:2010jk,Seipt:2013taa}. It was also discussed how that angular shift of emitted photons could be employed to detect the CEP of intense two-cycle pulses Ref.~\cite{Mackenroth:2010jk}.

    The specific effect of CEP also strongly depends on the laser polarisation. 
    For a circularly polarised laser for instance the relevant term in the dynamic phase is $\propto m\xi \psi(\varphi) \sin\vartheta  \cos(\varphi +\varphi_\mathrm{CE} - \varpi)$. Since the photon azimuthal angle $\varpi$ only appears in the combination $\varpi - \varphi_\mathrm{CE}$ the case of circular laser polarisation contains an interesting symmetry: A change in the CEP rotates the azimuthal angle spectrum by $\varphi_\mathrm{CE}$ \cite{Seipt:2013taa,Titov:2014usa,Titov:2015tdz,Titov:2015pre}.
    The general case of elliptical polarization was discussed e.g.~in Ref.~\cite{Seipt:2013taa}.}

    \subsubsection{Chirped pulses}
    \label{sec:first:nlc:chirp}
    
    The Compton scattering of high-energy electrons can be employed as a source of bright and ultrashort femtosecond x- and gamma-ray pulses \cite{Corde:RevModPhys2013,Rykovanov:2014ira,Albert:2016}. As discussed above, the ponderomotive broadening effect increases the bandwidth of the scattered photons. For narrow-bandwidth sources this limits the allowed laser intensity and thus the achievable photon yield. It was proposed first in Ref.~\cite{2013PhRvS..16c0705G} to employ optimally chirped laser pulses for a compensation of the nonlinear spectral broadening. The basic idea is to use pulses with a non-constant frequency $\omega( \tilde \varphi)\neq const.$, where $\tilde \varphi=\omega_0 x^\LCp$ with a constant reference frequency $\omega_0$, such that the ponderomotive red-shift in the high-intensity parts of the pulse is compensated by a higher laser frequency at the peak compared to the pulse edges.

    \begin{figure}[bt!] 
        \centering
        \includegraphics[height=6cm]{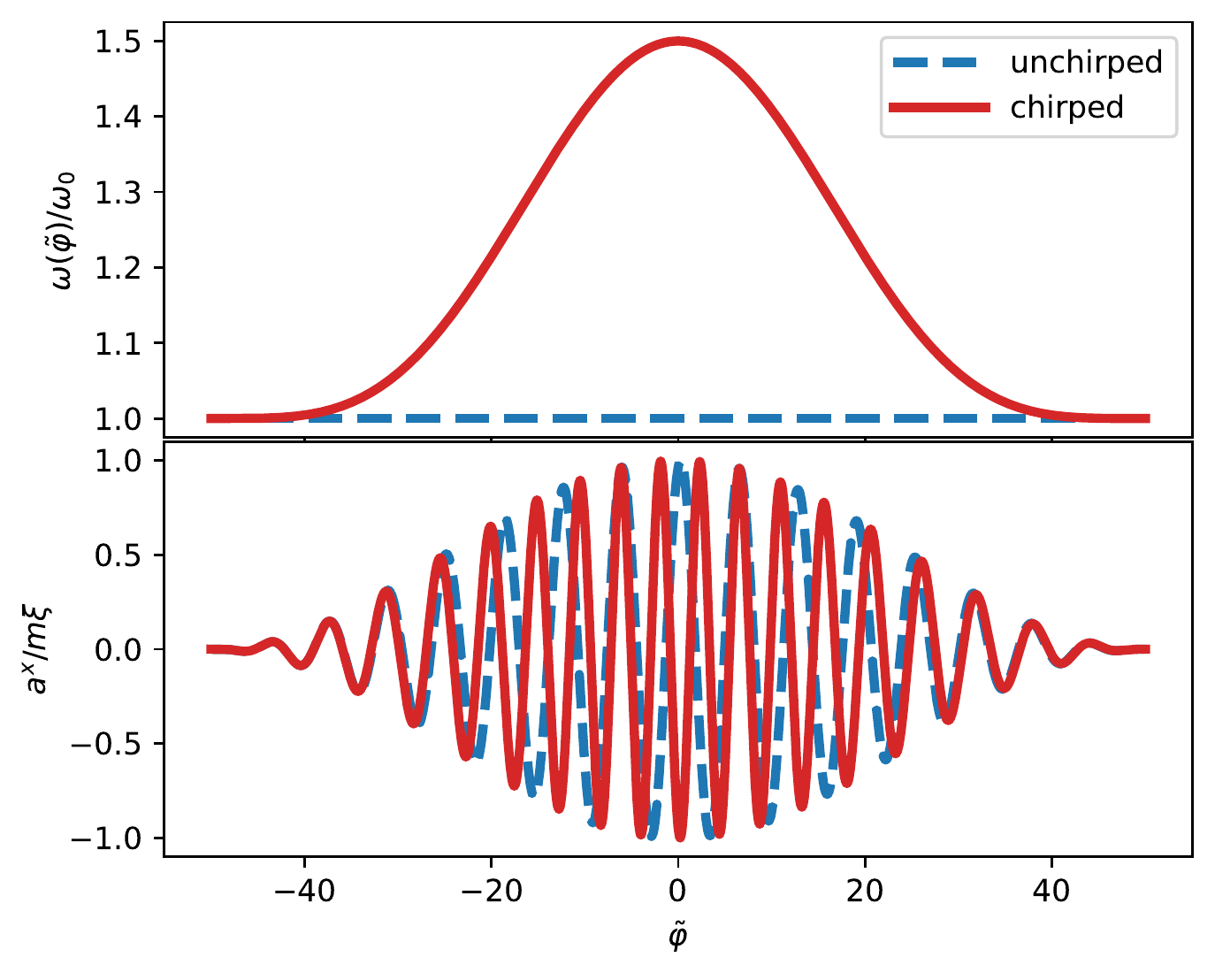}
        \includegraphics[height=6cm]{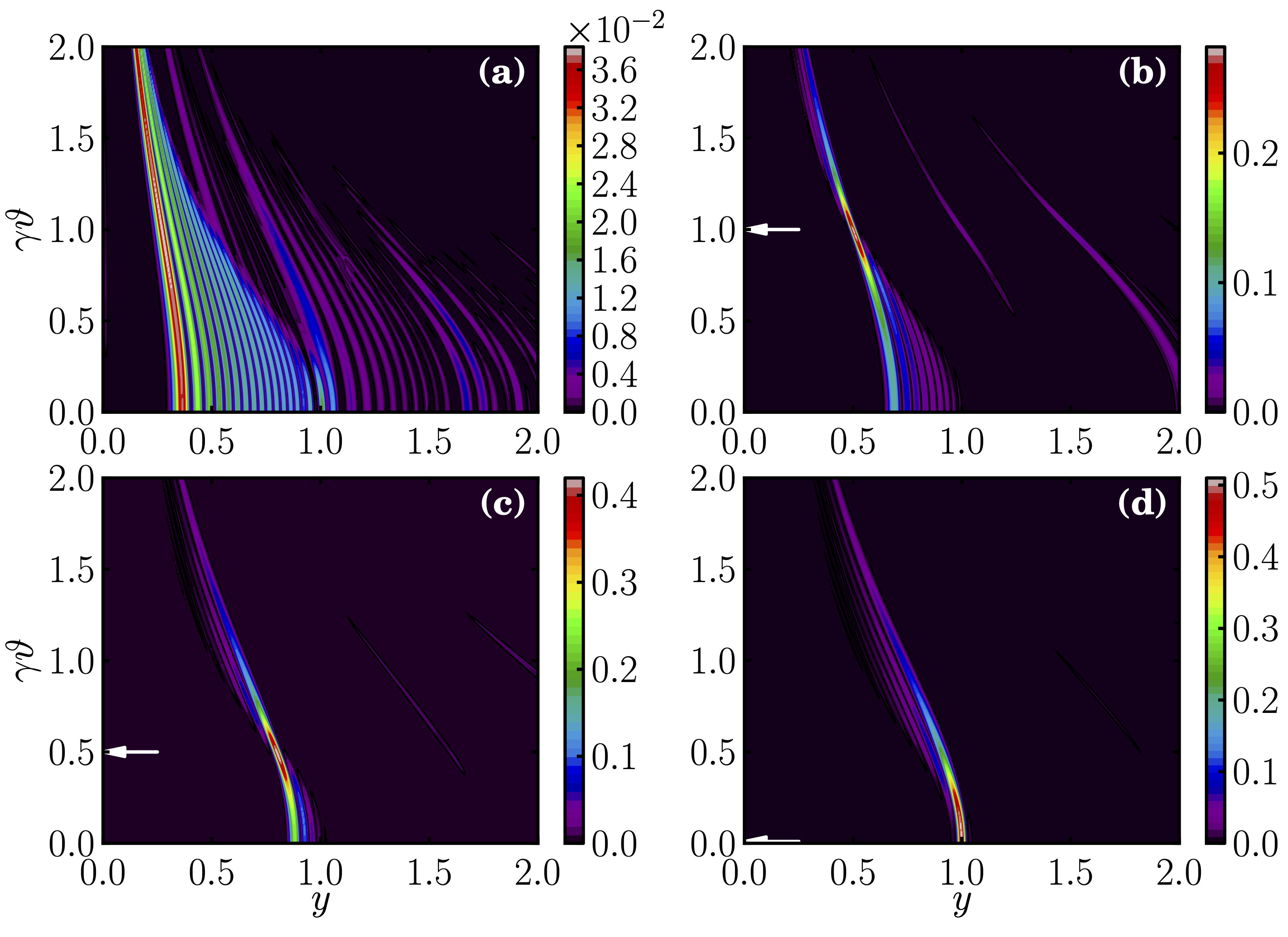}
        \caption{Left panels: Unchirped and chirped laser pulse and non-constant frequency.
        Right panels reproduced from Ref.~\cite{Seipt:2014yga}: Differential NLC photon emission probability as function of normalised frequency $y =\omega'/4\gamma^2\omega_0$ and normalised scattering angle $\gamma\vartheta$ for $\xi=2$. The case of an unchirped pulse (a) vs. optimum chirp to compensate ponderomotive broadening at scattering angle $1/\gamma$ (b), $0.5/\gamma$ (c) and $0$ (d).
        }
        \label{fig:first:chirp}
    \end{figure}
    
    Nonlinear Compton scattering with chirped laser pulses was discussed in Ref.~\cite{Seipt:2014yga}. Since the laser pulse has a non-constant frequency $\omega(\tilde \varphi)$, the phase of the carrier wave is now given by an integral $\varphi(x^\LCp) = \int^{x^\LCp} \! \ud y^\LCp \: \omega(y^\LCp)$, such that $\omega = \ud \varphi/\ud x^\LCp$. Despite having a nonconstant frequency, chirped plane waves are just plane waves. Therefore, the Volkov states are the same and the S-matrix elements can be evaluated just as for the unchirped case. In order to identify the individual harmonics, here we also have to apply a `slowly varying frequency approximation', in addition to the slowly varying envelope approximation, such that $\int \! \ud \tilde \varphi \: \psi(\tilde \varphi) \cos \varphi \approx \psi(x^\LCp) \frac{\omega_0}{\omega(x^\LCp)} \, \sin \varphi$ \cite{Seipt:2014yga}. With this, one can apply the Jacobi-Anger expansion, Eq.~\eqref{eq:first:jacobi-anger}, as before. The stationarity of the phase of the amplitude $\mathcal M_n$, together with the requirement $\omega'(x^\LCp) = const$. allows one to find the form of the optimal laser pulse chirping as
    \begin{align}
        \omega(\tilde \varphi) = \omega_0 \left( 1   + \frac{\xi^2 \psi^2(\tilde \varphi)}{2 (1 + \rho_\LCperp^2 )} \right) \,.
    \end{align}
    Surprisingly, the optimised laser chirping is capable of removing the ponderomotive broadening from all harmonics simultaneously, but only for one particular scattering angle, see also Fig.~\ref{fig:first:chirp}. Moreover, the optimal form of $\omega(x^\LCp)$ found from the NLC amplitude including quantum recoil effects is exactly the same as previously predicted from classical electrodynamics \cite{Terzic:2013ysa}.

    Several further studies of chirped pulses and other techniques to narrow the harmonic bandwidth at large $\xi$ were performed using classical electrodynamics (i.e.~nonlinear Thomson scattering) \cite{Rykovanov:2016ndn,Seipt:2019yds,Terzic:2019eoe,Valialshchikov:2020dhq,Holkundkar:2015rya,Ruijter:2021scx}. An application of these results to the quantum regime of NLC seems straightforward, either by direct calculation or application of a scaling law {that allows one to approximately obtain the NLC spectra by rescaling the classical spectra $\omega' = (1-s) \omega'_\mathrm{class}$ \cite{Seipt:2010ya,Krajewska:2013poa}.} A different approach for generating narrowband emission for $\xi>1$ was based on finding higher-order stationary points, so-called catastrophes. For a linearly chirped pulse the catastrophe is of cusp-type, evoking an off-axis spectral focusing, with the location of the cusp being controllable by the amount of linear chirp \cite{Kharin:2018dxa}.

    \subsubsection{Infrared behaviour}
    \label{sec:first:nlc:IR}
    
    In addition to the usual $n\geq1$-photon harmonics, a mid-IR peak in the photon spectrum far below the first harmonic arises due the equivalence of optical rectification \cite{Reimann:2007}. The nonlinear mixing of different frequency components of a pulse results in the emission of low-frequency radiation. The effect can also be considered as a long-range interference effect associated with the pulse envelope \cite{King:2020hsk}, hence this feature is not seen in infinite plane waves. The location of the mid-IR peak was found approximately at $s=\frac{1}{1+N(1+\xi^2)/\eta}$ for a sine-squared $N$-cycle pulse. The mid-IR peak is essentially a classical phenomenon, since the photon quantum parameter $\chi_\gamma = s \eta \xi \ll 1$. As the laser strength increases above $\xi \approx 1$, the mid-IR peak in the lightfront spectrum increases as $\ud \prob/\ud s \propto \xi^3$, faster than the first harmonic peak which grows as $\ud \prob/\ud s\propto \xi^2$.

    Nonlinear Compton scattering shows a low-frequency (infrared) divergence for the case of \emph{unipolar} fields that contain a Fourier zero mode, i.e. fields where $0=a^\mu(-\infty) \neq a^\mu(+\infty)=a_\infty$ \cite{Dinu:2012tj,Ilderton:2012qe}. 
    Physically that means the plane wave field with a zero mode is capable of accelerating the particle such that $\pi_p^\mu(+\infty)\neq p^\mu$. (This is the memory effect as discussed in Sec.~\ref{sec:intro:PW}.) In the Furry expansion, a constant gauge field $a_\infty$ appears in the phase of the outgoing Volkov wave function, which eventually appears as a contribution to the momentum conservation law. For soft photon emission, $q + a_\infty - p -\nu k =\ell \to 0$. In the limit $\omega'\to0$, the NLC probability is proportional to the square of the current $J = -ie (\frac{\pi_p(\infty)}{\upsilon \cdot \pi_p(\infty)} - \frac{p}{\upsilon \cdot p})$, where $\upsilon^\mu$ is the direction of the emitted soft photon, times an IR divergent factor $\propto \int_0 \frac{\ud t}{t}$ \cite{Dinu:2012tj}. Hence, the NLC probability is IR divergent if and only if the field is unipolar with $J\neq0$.
    
    For non-unipolar fields the infrared limit for small $s$ is given by the simple finite result \cite{DiPiazza:2017raw,Heinzl:2020ynb}
    \begin{align}
    \lim_{s\to0} \frac{\ud \prob}{\ud s} = \frac{\alpha}{2\eta m^2} \int \! \ud \varphi \: \vec a_\LCperp^2(\varphi) \,.
    \end{align}
    It should be noted though that the limit $s\to0$ in the angularly integrated probabilities is in general \emph{not} equivalent to the limit $\omega'\to 0$ \cite{DiPiazza:2017raw,Edwards:2020npu}; the large angle limit (forward scattering with respect to the background plane wave propagation direction, large values of $\rho_\perp$) at finite $\omega'$ also results in $s\to0$.

    For higher-order processes, such as double Compton scattering, the scattering probabilities show infrared divergences also for non-unipolar fields \cite{Ilderton:2012qe,Seipt:2012tn}, see also Sec.~\ref{sec:second}. This divergence occurs when one of the two emitted photons is soft and one is hard. The infrared behaviour of photon emission in classical electrodynamics was studied in \cite{DiPiazza:2018luu}, taking radiation reaction effects into account.

    \subsubsection{Spin and polarisation effects} 
    \label{sec:first:nlc:spin}
    
    The investigation of spin and polarisation effects in QED in strong background fields has seen major progress in recent years.
    A convenient basis for transverse photon polarisation vectors is given in \cite{Baier:1975ff,King:2013zw} as 
    \begin{align} \label{eq:first:photonbasis}
        \epsilon_{(j)}^\mu = \varkappa^\mu_j - \frac{\ell\cdot\varkappa_j}{\ell\cdot k} \, k^\mu \,,
    \end{align}
    where $\varkappa_j^\mu=\delta^\mu_j$ for $j=1,2$ are two space-like unit vectors in the transverse plane, $\varkappa_j\cdot k=0$. The vectors $\epsilon_{(j)}$ are not only mutually orthogonal, but also fulfil $k\cdot \epsilon_{(j)}=\ell\cdot \epsilon_{(j)}=0$. If the laser is linearly polarised along the $x$-axis one calls $j=1$ ($j=2$) the $E$ ($B$) polarisation \cite{King:2020btz}. For a circularly polarised laser one conveniently also defines the complex left/right-handed basis vectors $\epsilon_{(\pm)} = (\epsilon_{(1)}\pm i \epsilon_{(2)})/\sqrt{2}$.

    The emission of polarised photons (by unpolarised  electrons) in NLC for $\xi= \mathcal O(1)$ in multi-GeV electron collisions was considered in Refs.~\cite{King:2020btz,Tang:2020xlj}. In the limit $s\to0$ the emitted photons are unpolarised for a linearly polarised laser, but completely polarised for a circular laser \cite{King:2020btz,Tang:2020xlj}. A relationship between angular harmonics and photon polarisation degree was investigated, and found to be absent in the locally constant field approximation (constant crossed field case).
    Note that in \cite{Wistisen:2019tgu} the circular polarisation degree for $\omega'\to0$ depends on the pulse model and the angular collimation of the emitted photons. For larger $s$ the emitted photons are predominantly $E$-polarised, and it was found that $E$-polarised photons are much more tightly collimated than $B$-polarised photons, which also show a different emission pattern. Since photons close to the edge of the first harmonic are polarised to a very high degree this could serve as a bright source of polarised GeV photons \cite{Tang:2020xlj}.

    The spin 4-vector $\alpha^\mu$ of an electron in a pure state is related to the spinors 
    \begin{align}
    u\bar{u}=\frac{1}{2}(\slashed{p}+m)(1+\gamma^5\slashed{\alpha}) \,,
    \end{align}
    where $\alpha^\mu$ refers to the asymptotic spin state of the particle (with $\alpha^2=-1$, $p\cdot \alpha=0$). A convenient spin-basis was given by~\cite{Seipt:2018adi,Dinu:2018efz,Dinu:2019pau,Torgrimsson:2020gws}
    \begin{align} \label{eq:first:spinbasis}
        \alpha^\mu_{(1,2)} = \varkappa^\mu_{1,2}-\frac{p\cdot \varkappa_{1,2}}{p\cdot k} \,k^\mu \,,
        \qquad
        \alpha^\mu_{(3)} = \frac{p^\mu}{m} - \frac{m}{k\cdot p}\, k^\mu \,,
    \end{align}
    $\alpha_{(i)}\cdot \alpha_{(j)} = -\delta_{ij}$, where $\alpha_{(3)}$ is called the lightfront helicity vector \cite{Ilderton:2020gno}. If the laser is linearly polarised along the $x$-axis, the vector $\alpha_{(1)}$ points along the rest frame electric field, and $\alpha_{(2)}$ is along the direction of the magnetic field in the particle rest frame and is non-precessing. However, in general for an arbitrary polarisation one may not even be able to define a locally-constant polarisation direction.
    {This does not pose a serious problem because it is not necessary to choose the vectors $\varkappa^\mu_{j}$ in \eqref{eq:first:spinbasis} to be along the electric and magnetic field, they just have to be mutually orthogonal and perpendicular to $k^\mu$.}
    The evolution of the spin basis vectors as the electron propagates in a plane wave is given by $\alpha^\mu_{(j)}(p\to\pi)$, which ensures that the spin vector remains orthogonal to the kinetic momentum for all times \cite{Ilderton:2020gno,Seipt:2020diz}.
    
    The polarisation vector $\alpha^\mu$ of a partially polarised electron in a mixed state, $-1 < \alpha^2 \leq 0$ can be expanded in the basis \eqref{eq:first:spinbasis} as
    \begin{align} 
        \alpha^\mu =\sum_i n_i \alpha^\mu_{(i)}
    \end{align}
    which defines the Stokes vector $\stokes$ \cite{Dinu:2018efz,Dinu:2019pau,Torgrimsson:2020gws}. For an electron at rest, ${\bf n}$ points in the spin direction,
    \be
    {\stokes}=\frac{1}{2}u^\dagger{\bf \Sigma}u({\bf p}=0) \;,
    \ee
    where ${\bf\Sigma}=i\{\gamma^2\gamma^3,\gamma^3\gamma^1,\gamma^1\gamma^2\}$. It is convenient to put this into a 4D Stokes vector\footnote{Note that the 4D Stokes vector---despite having four components---is \emph{not} a Minkowski-space four-vector; ${\bf N}$ is not $\alpha^\mu$.}~\cite{Dinu:2019pau,Torgrimsson:2020gws}, ${\bf N}=\{1,{\stokes} \}$. A fully polarised electron has $\stokes^2=1$, an unpolarised has $\stokes={\bf 0}$, and $0<\stokes^2<1$ represents a partially polarised state. The Stokes vectors are used for instance in Sec.~\ref{Spin sums in the two-step} in conjunction with strong-field QED Mueller matrices for treating spin sums in higher order processes.

    An alternative way to describe an arbitrary polarisation state of an electron employs the spin density matrix $\rho = \frac{1}{2} (1 + \boldsymbol \sigma \cdot\vec \Xi)$, where $\boldsymbol \sigma$ are the Pauli matrices, and $\vec \Xi = \mathrm{tr} (\boldsymbol \sigma \rho)$. The polarisation vector $\vec \Xi$ (sometimes called a Bloch vector) determines the polarisation state on the Bloch sphere, with its poles corresponding to pure spin states along an arbitrarily chosen spin quantisation axis. By choosing e.g. $\alpha^\mu_{(2)}$ as the spin quantisation axis, $\vec \Xi = (n_3,n_1,n_2)$. The density matrix of the final state particles is related to the initial state density matrix formally as $\rho_f = S \rho_i(\vec \Xi_i) S^\dagger$, where $S$ is the S-matrix \cite{book:densitymatrix,Seipt:2018adi}.  The final-state density matrix $\rho_f$ contains the complete information on all particles in the final state. Tracing out unobserved quantities allows to calculate the (reduced) spin density matrix of the scattered electrons \cite{Seipt:2018adi}. The probability and polarisation vector of the final electrons are given by
    $\prob(\vec \Xi_i) = \mathrm{tr} [\rho_f(\vec \Xi_i)]$ 
    and $\vec \Xi_f = \mathrm{tr} [\boldsymbol \sigma \rho_f(\vec \Xi_i)]/\prob(\vec \Xi_i)$ respectively.
    
\begin{figure}
        \centering
        \includegraphics[height=4.5cm]{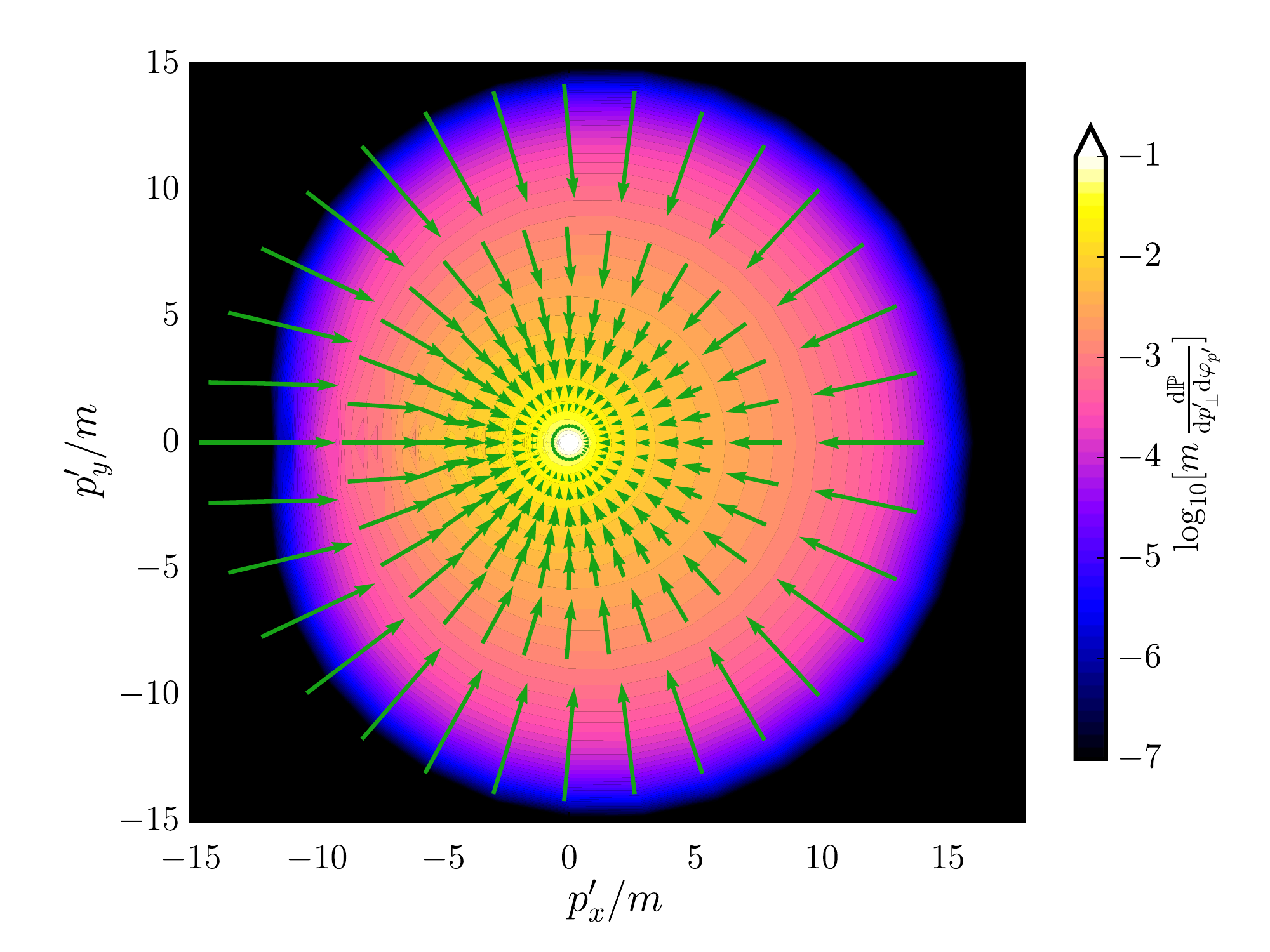}
        \includegraphics[height=4.5cm]{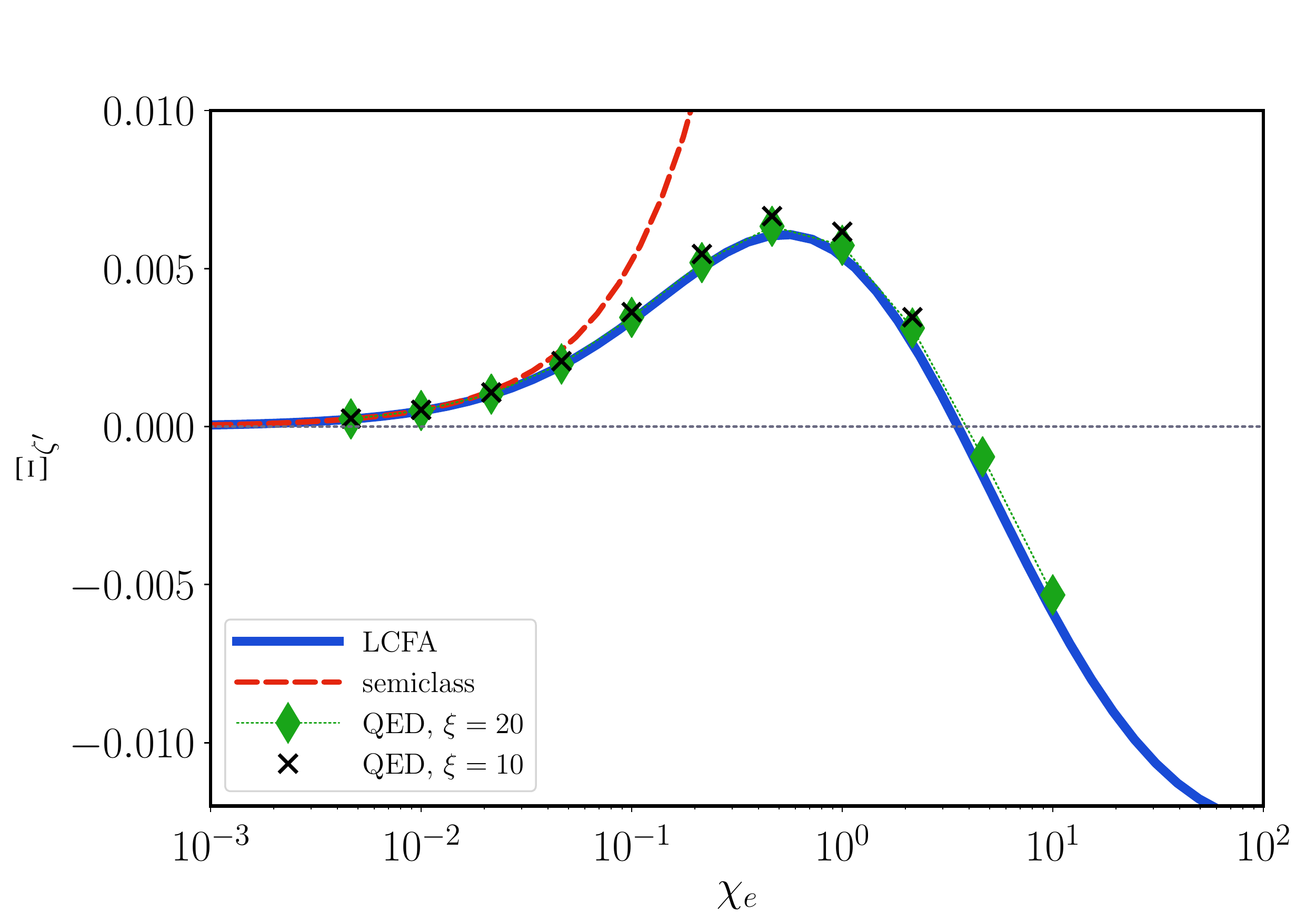}
        \caption{
        Left: Transverse momentum distribution of the scattered electrons (for $\xi=25$, $\Phi=2\pi$, $\gamma=5000$) and the polarisation of the scattered electrons transverse to the beam axis (green arrows). The length of the arrows indicates the magnitude of the polarisation for given $\vec p_\LCperp'=\vec q_\LCperp$. Right: Scattered electron polarisation parameter $\Xi_{\zeta'} (= n_2)$ for an ultra-short linearly polarised laser pulse. Reproduced from Ref.~\cite{Seipt:2018adi}. 
        }
        \label{fig:polarizaion}
    \end{figure}

    For circularly polarised pulses, the scattered electrons show a radial polarisation pattern with a strong correlation between the momenta of the scattered electrons and their polarisation \cite{Seipt:2018adi}, see Fig.~\ref{fig:polarizaion} (left). For a linearly polarised sub-cycle pulse with a large asymmetry of the magnetic field a small transverse polarisation of the final electrons was found for initially unpolarised electrons after just a single photon emission, see Fig.~\ref{fig:polarizaion} (right). Simulations of multiple photon emissions show that larger polarisation degrees can be achieved by using asymmetric bichromatic laser pulses \cite{Seipt:2019ddd,Chen:2019vly}.  The spin-flip and non-flip rates were calculated for an arbitrary direction of the initial spin \cite{Seipt:2018adi}, generalising results of \cite{Baier:1967}. In particular it should be noted that radiative spin flips in the lightfront helicity basis have a different rate than in the transverse basis $\alpha_{(2)}^\mu$ \cite{Seipt:2018adi,Ilderton:2020gno}.

    The general probabilities for the emission of a polarised photon off a spin-polarised electron, with the final electron spin taken into account have been calculated for the specific transverse spin-basis $\alpha_{(2)}$ in Ref.~\cite{Seipt:2020diz}, and for arbitrary spin orientation in Ref.~\cite{Torgrimsson:2020gws}. {(The results in the Locally constant field approximation (LCFA) and Locally monochromatic approximation (LMA) in \cite{Torgrimsson:2020gws} were obtained by approximating the general results in \cite{Dinu:2019pau}, see \secref{sec:approx} for details on these approximation frameworks).}
    {The asymptotic scalings of the completely polarised NLC rates for small and large $\chi$ were presented in \cite{Seipt:2020diz}.} The rates for polarised photon emission from polarised electrons have been independently derived in the quasiclassical method of Baier, Katkov and Strakhovenko (BKS) \cite{baier98} in  \cite{BaierKatkovFadin,Li:2018fcz,Li:2019oxr,Chen:2022dgo}. As far as we can tell the first complete results for all particles polarised were presented just very recently in Ref.~\cite{Li:2019oxr}, where the results were obtained using the BKS. For an infinite monochromatic plane wave the completely spin-and polarisation dependent cross sections had been obtained already in \cite{Ivanov:2004fi} (see Ref.~\cite{Seipt:2020diz} for earlier references). A numerical method for calculating spin and polarisation dependent NLC rates was presented in Ref.~\cite{Wistisen:2019tgu} based on BKS, where the numerical results were in good agreement with the Volkov state approach. Radiative positron polarisation in a bent crystal was calculated \cite{Wistisen:2019tgu}, and NLC was proposed as a method for electron beam polarimetry \cite{Li:2019wkl}. The spin dynamics in Kapitza-Dirac scattering has also attracted some interest recently \cite{Ahrens:2021kww}.

    Ordinary Volkov wavefunctions represent electrons without an anomalous magnetic moment. The latter can be taken into account by introducing an explicit Pauli interaction term as an effective field theory, which yields both a modification to the Volkov states as well as an additional Pauli-vertex in the NLC S-matrix \cite{Ekman:2020fdp}. In a constant crossed field the Pauli term has a different high-intensity scaling $\propto \kappa^2\chi^{4/3}$, where $\kappa$ is the Pauli coupling, compared to $e^2\chi^{2/3}$ for the ordinary QED contribution \cite{Ekman:2020fdp}. {The Pauli-term shows the same high-intensity scaling for $\chi\gg1$ as the one-loop vertex correction in QED. For a detailed discussion of the behaviour of strong-field QED when $\alpha\chi^{2/3}\sim 1$ we refer to Sec.~\ref{sec:RN} of this review.}

    \subsection{Nonlinear Breit-Wheeler pair production}
    \label{sec:first:nbw}

    Nonlinear Breit-Wheeler (NBW) pair production is characterised by the reaction $\gamma(\ell)\to e^+(p) \: e^-(q)$. The amplitude for this process is given by crossing symmetry from the NLC amplitude, Eq.~\eqref{eq:first:MNLC}, by means of the replacements $\mathcal M_\mathrm{NBW} (p,\ell,\epsilon) = \mathcal M_\mathrm{NLC} (-p,-\ell,\epsilon^*)$. Explicitly the amplitude reads
    \begin{align}
        \mathcal M_\mathrm{NBW} = \int \! \ud\varphi\: \bar u_{\pi_q} \slashed \epsilon v_{\hat \pi_p} e^{i\Psi(\varphi)} \,,
    \end{align}
    where the NBW dynamic phase $\dynamicphase = \int^{\varphi} \! \ud \varphi' \:  \Psi'$, $\Psi' = \ell \cdot \hat \pi_p(\varphi) / k\cdot q $, with the kinetic momentum of the positron $\hat{\pi}_p^\mu(\varphi) = - \pi^\mu_{-p}(\varphi)$. Moreover, the outgoing positron spinor $v_{\hat{\pi}_p} = u_{\pi_{-p}}$ fulfills $(\hat{\slashed{\pi}}_p +m )v_{\hat \pi_p} = 0 $, see also Sec.~\ref{sec:intro:PW}. For pair production, the normalised lightfront variables of the produced positron are $s=k\cdot p/k\cdot \ell$ and $\vec r^\LCperp = \vec p^\LCperp/ms$, and the momentum absorbed from the background is $\nu = (\ell\cdot p)/(k\cdot q) = \frac{1}{2\eta_\gamma s (1-s)} (1 + \rho_\LCperp^2)$ with $\boldsymbol \rho_\LCperp  = s(\vec r_\LCperp - \boldsymbol \ell_\LCperp/m )$ and $\eta_\gamma = \ell \cdot k/m^2$.  For a numerical evaluation of the spectra one once again expands the amplitude into a sum of Dirac structures not dependent on the laser phase, and scalar integrals over the laser phase, see for instance \cite{Krajewska:2012eb,Titov:2012rd}
    The appearing pure phase integral $\int \! \ud \varphi \: e^{i\dynamicphase}$ can be regularised \cite{Krajewska:2012eb} analogously to NLC (discussed in Sec.~\ref{sec:first:nlc:reg}).
    
    In comparison to NLC, the NBW process has a threshold energy, which has profound consequences for the probabilities. From momentum conservation in a {plane wave background, $\ell +\nu k = p+q$}, one can derive that the lightfront momentum absorbed from the field has a lower limit $\nu \geq 2/\eta_\gamma$, in contrast to $\nu>0$ for NLC. For pair production to be efficient in a strong field, the quantum parameter of the incident gamma photon $\chi=\xi\eta_\gamma$ must not be small. If $\chi\ll1$, the total pair production probability is exponentially suppressed $\prob \sim e^{-f(\xi)/\chi}$.

    One can distinguish three main regimes of NBW pair production: (i) the perturbative linear regime ($\xi \to 0$) where the energy threshold is already satisfied by a single laser photon colliding with the probe. This situation was first discussed within perturbative QED by Breit and Wheeler \cite{Breit:1934zz} as the collision of two gamma quanta; (ii) the multi-photon regime ($\xi\lesssim 1$, with $\eta_\gamma$ sufficiently large) where $n\geq1$ photons from the laser may contribute to meeting the threshold condition; and (iii) the non-perturbative quasi-static strong-field regime ($\xi\gg1$) where typically a very large number of laser photons are required to overcome the threshold. For $\chi\ll1$ the NBW pair production rate in a plane wave scales as $\sim \alpha \chi^{3/2} e^{-8/3\chi}$, with an exponential suppression akin to a tunneling process. For $\chi\gg 1$ the rate scales as $\sim \alpha \chi^{2/3}$ which resembles over-the-barrier ionisation \cite{Nikishov:1964zza,Ritus1985}.
    These scalings are obtained after integrating the pair production rates in a constant crossed field 
    \begin{align} \label{eq:first:NBWrate}
        \mathsf{R} \sim \begin{cases}
            \: \alpha \chi e^{-8/3\chi} \,, & \chi\ll1 \,, \\
            \: \alpha \chi^{2/3} \,, & \chi \gg 1 \,,
        \end{cases}
    \end{align}
    over the oscillating field amplitude in a plane wave \cite{Nikishov:1964zza,Ritus1985}. The same scalings are found for gamma photons in a constant electric or magnetic field, see e.g.~\cite{Karbstein:2013ufa} and references therein. Several studies have investigated the prospects of using Bremsstrahlung $\gamma$-rays for an experimental observation of NBW pair production providing scaling laws for the expected positron yield \cite{Blackburn:2018ghi,Eckey:2021rgn,2021NJPh...23j5002S}. The NBW pair production rates in a plane wave background have also been calculated in QED$_{2+1}$ \cite{Golub:2021nhj}. The reduced dimensionality manifests itself in a different $\chi$ dependence of the rate at $\xi\gg1$, which scales $\sim \alpha \chi e^{-8/3\chi}$ for $\chi\ll1$, and $\sim \alpha \chi^{1/3}$ for $\chi\gg1$.

    The NBW pair production in linearly or circularly polarised infinite plane waves and constant crossed fields was studied for a long time and details can be found in reviews, e.g.~\cite{Ritus1985}. A more recent derivation for generally elliptically polarised plane waves can be found in \cite{He:2012un}. In the multi-photon regime $\xi\lesssim 1$ it is possible to identify $n$-photon channels by applying the Jacobi-Anger identity, Eq.~\eqref{eq:first:jacobi-anger}. In a linearly-polarised\footnote{As usual, for circular polarisation $\xi^2/2\to \xi^2$.} infinite monochromatic plane wave, the squared cycle-averaged momentum yields the intensity dependent effective mass $m^2_\star=\overline{\pi_p(\varphi)}^2=m^2(1+\xi^2/2)$. Pairs are produced with this effective mass (encompassing the average quiver momentum), which affects the effective threshold and hence the open $n$-photon channels, $n\geq n_\star= \lceil (2+\xi^2)/\eta_\gamma \rceil$. For small $\xi$ the pair production probability of the $n$-photon channel scales as $\propto \xi^{2n}$, hence the overall probability scales as $\propto \xi^{2n_\star}$ in the multi-photon regime for infinite plane waves. {The production probability exactly at the threshold always vanishes, which is a pure final particle phase space effect. This is different in QED$_{2+1}$, where the pair production rate at the threshold can be non-zero for even multi-photons channels, but not for odd ones \cite{Golub:2021nhj}.}

    \subsubsection{Short pulses and threshold effects}

    The total pair production probability is related to the imaginary part of photon polarisation tensor according to $\prob(\ell) = \mathrm{Im} (\epsilon \cdot \Pi(\ell,\ell)\cdot \epsilon^*)/(k\cdot \ell)$. Using this cutting rule, it was found that the total pair production probability in a short pulse grows approximately proportional to $\xi$ in the range $1\leq \xi\leq 10$ for $\chi=1$ \cite{Meuren:2014uia}. Moreover, for $\xi=10$ the probability increases roughly linearly with pulse length for $\chi\sim 1$. The influence of a finite temporal and spatial extent of the laser field on the NBW for $\xi\gg1$ was determined in \cite{Mercuri-Baron:2021waq} using the scaling relations \eqref{eq:first:NBWrate} and computer simulations.
    
    However, in a pulsed background field, in general if $\xi \not \gg 1$, there arise novel effects near threshold: 
    
    (i) A short pulse laser has a finite bandwidth $\propto 1/\phaselengthparameter$, and this `softens' the threshold. Pair production can be possible from the high-frequency components of the laser spectrum, even if the central frequency would suggest that pair production should not be allowed. This has been called `sub-threshold' pair production \cite{Titov:2012rd,Nousch:2012xe} and can lead to orders-of-magnitude increase of the pair production probability especially for small $\xi\lesssim 1$. In the perturbative limit this effect can be understood for the first few harmonics as a convolution of the pair production cross section with an appropriate power of the laser spectrum. The softening of the pair production threshold means that pair production depends less strongly on the photon energy in a short pulse compared to infinite plane waves. In a delta pulse which contains all frequency modes with equal amplitude, the NBW probability is even completely independent of the probe photon energy \cite{Ilderton:2019vot}.

\begin{figure}
        \centering
        \includegraphics[width=0.8\columnwidth]{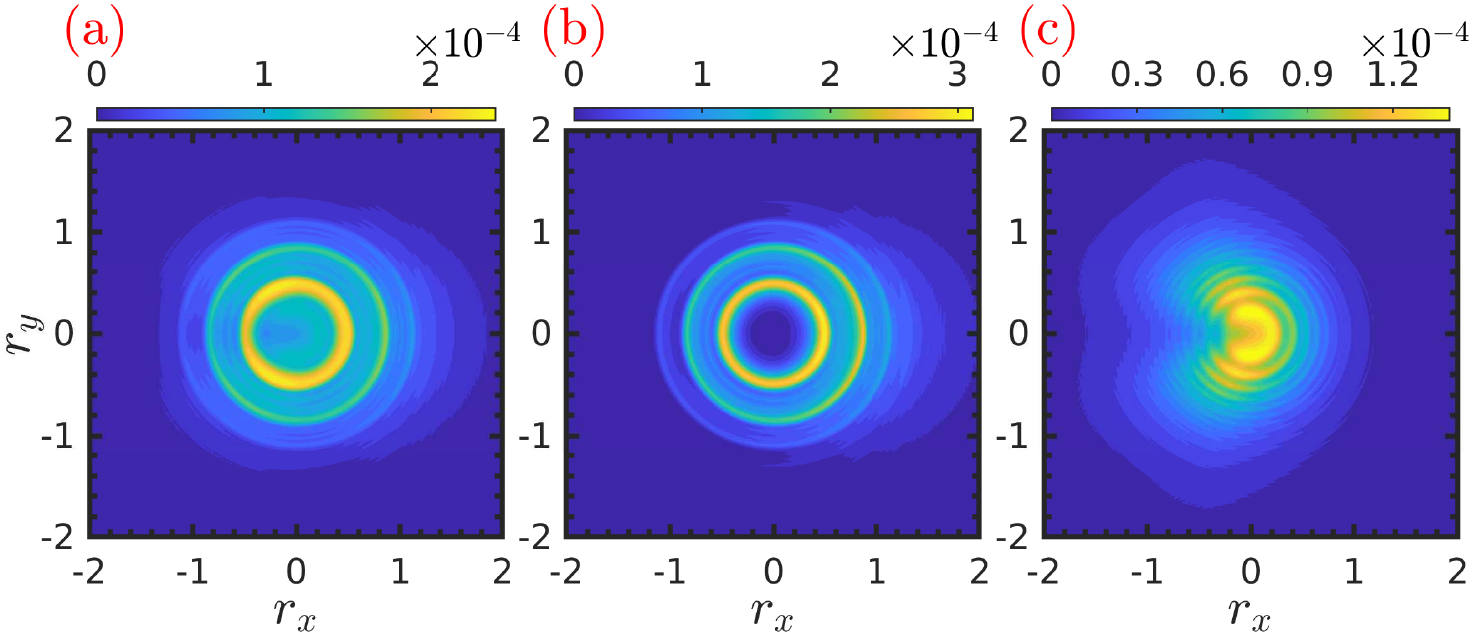}
        \caption{Double differential positron probability $\ud^2\prob/\ud r_x \ud r_y$ for (a) all the harmonic channels open in the flat-top pulse; (b) only the harmonics accessible in a local monochromatic approach i.e.~ignoring the bandwidth effects; and (c) harmonics inaccessible to the local approach. Note that $r_j$ in this Figure corresponds to $\rho_j$ in the general notation of this Review. Reproduced from Ref.~\cite{Tang:2021qht}.
        }
        \label{fig:first:pairs}
    \end{figure}

    (ii) In a pulse the ponderomotive potential experienced by the produced pair is not constant, but proportional to the pulse intensity $\propto \xi^2 \psi^2(\varphi)$. This implies that in a pulse with $\phaselengthparameter\gg1$ that is long enough for a meaningful separation of fast and slow timescales the effective mass $m^2_\star=m^2(1+\xi^2\psi^2(\varphi)/2)$ changes with the pulse envelope $\psi$. Thus, the effective thresholds for the different $n$-photon channels are variable in a pulse. This makes the analysis of the pair production rate near threshold quite intricate: on the one hand, the pair production probability is higher in the high-intensity parts of the pulse, but on the other hand fewer $n$-photon channels are open in the high-intensity parts due to the increased mass threshold. A recent discussion of harmonic channel opening in a pulse can be found in Ref.~\cite{Tang:2021qht}, see also Figure~\ref{fig:first:pairs}.
    
    A semiclassical picture on the amplitude level was established in Ref.~\cite{Meuren:2015mra}. It is based on the observation that for $\xi\gg1$ the $\varphi$-integrals are highly oscillatory and hence can be evaluated using stationary phase approximation. The dominant contributions come from coalescing pairs of complex conjugate stationary points with imaginary part $\sim 1/\xi$. This allows one to understand many features of the momentum distribution of the produced pair in terms of the classical evolution of the created particles in the background field.

    The ultimate limit of an ultra-short pulse, a delta pulse of the electric field, was studied in \cite{Ilderton:2019vot}. In this case, all final state integrals can be performed analytically, and closed form expressions for the total scattering probabilities can be derived, which are a function of $\xi$ alone. The probability is in particular completely independent of the incident photon energy and hence its corresponding $\chi$ parameter. In the large intensity limit $\xi\gg 1$, the pair production probability in a delta pulse has a logarithmic dependence on $\xi$, not a power-law one as in a constant crossed field \cite{Ilderton:2019vot} (see also the discussion on the Ritus-Narozhny conjecture in Sec.~\ref{sec:RN}).

    The energy-angular distributions of produced positrons have been calculated numerically for pulses of various duration and $\xi=1$ \cite{Krajewska:2012eb}. For fixed angles, the positron energy spectra shows distinct harmonic peaks (i.e. the multi-photon channels) with substructure as discussed for NLC. The peaks broaden and merge for decreasing pulse duration, i.e.~the bandwidth effect becomes more important than the ponderomotive broadening effect. The typical line broadening with substructures was also found in transverse momentum spectra \cite{Heinzl:2010vg}. The positron spectra for short pulses with various shape functions $\psi$ have been investigated in \cite{Titov:2013kya} over a wide range of $\xi$, where special emphasis was given to asymmetries in the angular spectra.

    \subsubsection{Interference effects}

\begin{figure}
    \centering
    \includegraphics[width=0.4\columnwidth]{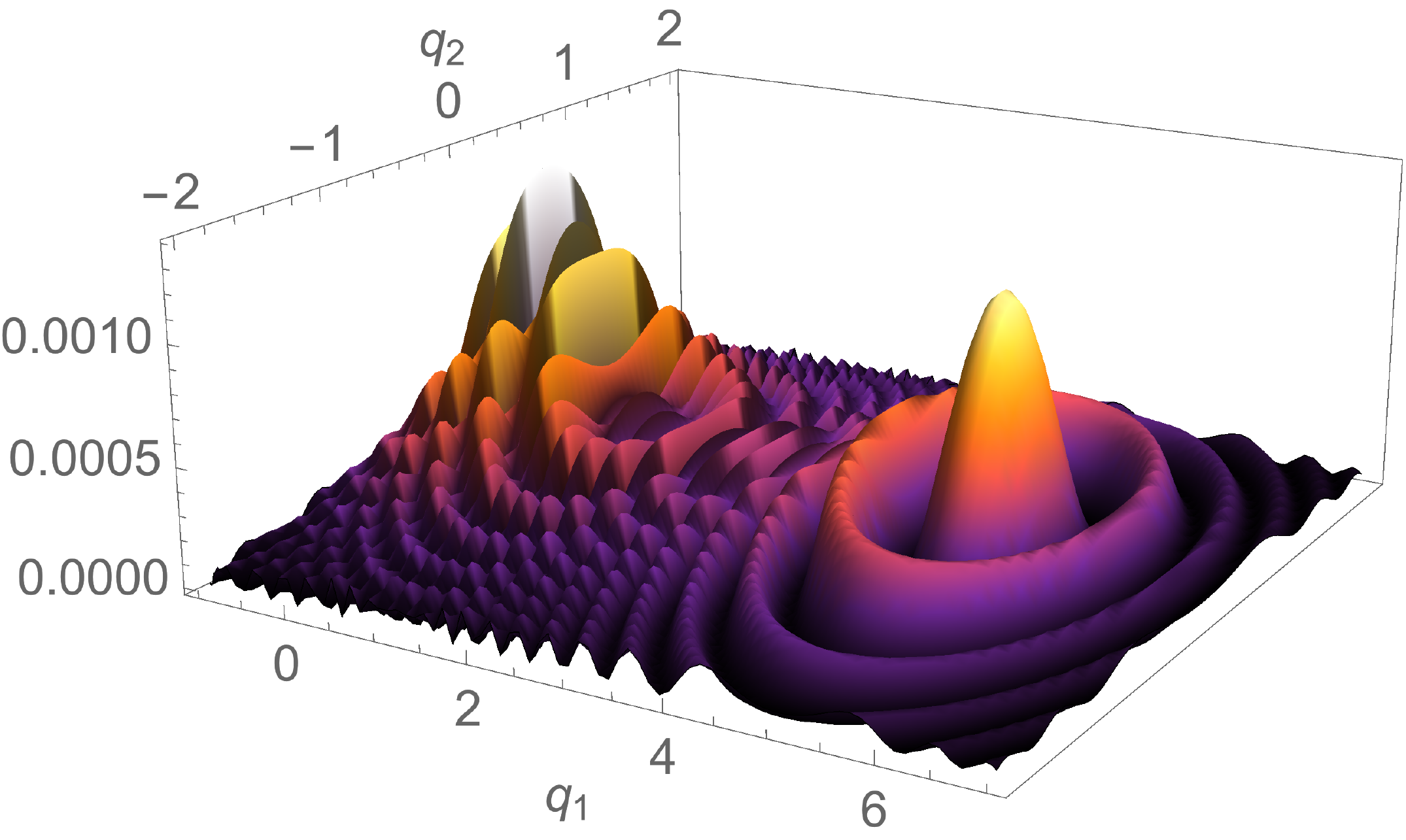}
    \hspace*{0.05\columnwidth}
    \includegraphics[width=0.4\columnwidth]{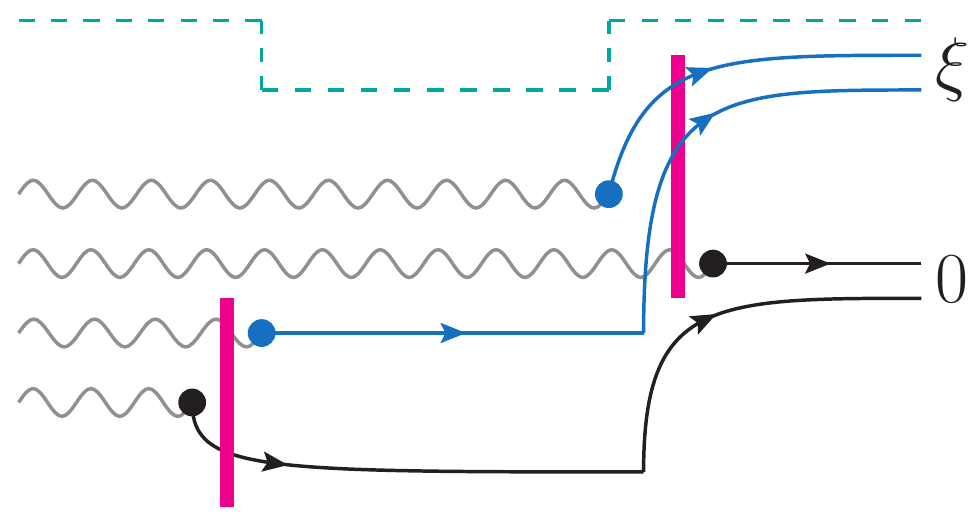}
    \caption{Coherent quantum interference in the positron spectrum produced in two delta-pulses of opposite sign (left), with coherently interfering quantum paths illustrated on the right. In this plot, $q_{1,2}$ refers to the transverse momentum components of the produced positron. Reproduced from Ref.~\cite{Ilderton:2019ceq}.}
    \label{fig:first:delta}
    \end{figure}

    Further quantum interference effects occur during the NBW process when the initial photon interacts with several laser pulses. The interfering quantum paths can be most clearly identified in double delta-pulses \cite{Ilderton:2019vot,Ilderton:2019ceq}, see Fig.~\ref{fig:first:delta}. The interference pattern in the differential probability is controlled by both the separation of the delta pulses $\Delta\varphi$ and initial photon energy in the dimensionless combination $\Delta\varphi/\eta_\gamma$. In the large $\xi$ limit, interference effects drop out of the integrated probability, which becomes approximately equal to twice that in a single delta pulse \cite{Ilderton:2019vot}. {A similar phenomenon has been observed for Schwinger pair production in time dependent electric fields \cite{Akkermans:2011yn}, see also Sec.~\ref{sec:Schwinger}.}    
    
    The production of scalar particles via the NBW process in few-cycle double pulses was considered in \cite{Jansen:2016crq} in multiphoton regime $\xi\lesssim 1$. Here, the time delay strongly influences not only the energy spectrum of created particles but also the total production probability, which is an oscillating function of the gap distance \cite{Jansen:2016crq}. In the case of non-identical pulses (different $\omega$ and $\xi$) the total probability oscillation depends not only on the pulse distance, but also on the temporal order of the pulse arrival \cite{Jansen:2016crq}. (Similar effects have been seen also for the non-perturbative pair production in oscillating electric fields \cite{Granz:2019sxb}.)
    
    Pair production in double pulses with circular polarisation and same frequency was studied in \cite{Titov:2018bgy} for $\xi < 1$. The interplay between pulse separation and carrier envelope phase has a strong impact on the azimuthal positron distribution, especially for very short pulse duration. An enhancement of the pair production cross section was found for very small pulse separation such that the two pulses started to overlap, effectively increasing peak $\xi$.

    For trains composed of $N_\mathrm{rep}$ identical pulses without delay, the appearance of comb-like structures in the energy spectra of produced positrons were found \cite{Krajewska:2014ssa}. The pair production amplitude for a pulse train equals, up to a phase, $\mathcal M_\mathrm{train} \sim \frac{\sin \pi \nu N_\mathrm{osc}N_\mathrm{rep}}{\sin \pi \nu N_\mathrm{osc}}\: \mathcal M_\mathrm{single}$, where $N_\mathrm{osc}$ is the number of carrier wave oscillations per single pulse. When mod-squared, the ratio of sines defines the locations of comb's teeth, while the single-pulse amplitude yields the relative height of the teeth. In cases where the teeth are equidistant {in energy} it was suggested that their coherent superposition would form ultra-short (anti-)matter wave packets \cite{Krajewska:2014ssa}.

    \subsubsection{Multi-colour laser fields}

    If pair production takes place in background fields with multiple distinct frequency modes, further interference and enhancement effects occur, which had been first investigated in Ref.~\cite{2000JETP...90..415N}. An enhanced production probability in two-colour fields was found for the combination of a weak high-frequency mode ($\omega_1\sim m$, $\xi_1\ll1 $) and a strong low-frequency mode ($\omega_2\ll m$,  $\xi_2\sim 1$) \cite{Jansen:2013dea} of linear polarisation interacting with the gamma-ray photon with frequency $\omega_\gamma$. The threshold condition in this case turns into $(n_1\omega_1 +n_2\omega_2)\omega_\gamma \geq m_\star^2$, where the dominant channels have $n_1=1$ and $n_2 \gg 1$. A strong enhancement of the pair production is found if $\omega_1$ alone is below threshold for perturbative pair production by a gap $\Delta = 1 - \eta_{1\gamma}/2$, which is then overcome by absorption of a large number of $\omega_2$ photons if $\xi_2\gg1$ such that the interaction with the low-frequency field commences in the quasi-static tunneling regime. This picture is quite similar to the dynamically assisted Schwinger effect \cite{Schutzhold:2008pz, Dunne:2009gi}, see also Section~\ref{sec:Schwinger}. The modification of the positron spectrum in a similar set-up, but when the perturbative one-photon channel of the high-frequency field is open was investigated in \cite{Nousch:2015pja,Otto:2016fdo}. Features in the spectra could be explained in terms of spectral caustics due to the classical dynamics of the produced particles with the strong low-frequency field.
    
    The production of scalar particles by high-energy photons interacting with bichromatic fields is studied in \cite{Jansen:2015loa} for commensurate frequencies, $\omega_2 = h \omega_1$, with integer $h$. For multiple commensurate frequencies interference exists in the multiphoton regime if there are several channels with the same value of $\Omega = \sum_j n_j \omega_j$ \cite{Jansen:2015idl}, where $n_j$ is the number of absorbed laser photons form mode $j$. For the particular case of two frequency components the interference affects the angular distributions, and in some cases also total pair yield. For the perturbative regime $\xi\ll 1$ it was shown that the magnitude of interference effects is maximised if the intensities of the two waves are balanced $\xi_1=\xi_2^h$ \cite{Jansen:2015loa}. This opens the possibility of coherent phase control of quantum interferences between different multiphoton pathways in pair production in strong fields. The idea was advanced further in Ref.~\cite{Brass:2019pzr}, where phase-of-the-phase spectroscopy \cite{2015PhRvL.115d3001S} was discussed for non-perturbative Schwinger electron-positron pair production in strong two-colour oscillating electric fields (cf. also \cite{Akal:2014eua}).

    \subsubsection{Spin and polarisation effects}

    Pair production from linearly polarised photons in ultrashort linearly polarised laser pulses was studied numerically in Ref.~\cite{Krajewska:2012eb}. The oscillation patterns in the angular positron distributions are much more pronounced in the case of collinear polarisations. The coupling of polarisation to laser pulse shape effects was studied in Ref.~\cite{Tang:2021qht}.

    The influence of the incident photon polarisation being parallel or perpendicular to the linear background field polarisation was investigated in Ref.~\cite{Titov:2020taw} in terms of the asymmetry of cross sections $\mathfrak A = (\sigma_\perp - \sigma_\parallel)/(\sigma_\perp + \sigma_\parallel)$. In the quasi-static regime for $\xi\gg1$, the asymmetry behaves, according to the asymptotic scaling \cite{Ritus1985}, as $\mathfrak A=1/3$ for $\chi\ll1$ and as $\mathfrak A=1/5$ for $\chi\gg1$. In the multiphoton regime, $\xi\sim1$, the asymmetry $\mathfrak A$ has sharp features at integer values of the threshold parameter $\zeta=2/\eta_\gamma$ which are more pronounced for smaller $\xi$, and the asymmetry parameter gets smoother for shorter pulse duration.

    The relevance of spin effects in NBW pair production in few-cycle laser pulses is investigated by comparing the production of spinor and scalar particles for unpolarised incident photons~\cite{Villalba-Chavez:2012kko,Jansen:2016gvt}. For linear laser polarisation and for small $\xi$, the leading contribution to the pair production rate near threshold has a total spin of $S=1$ ($S=0$) (in the spinor case), depending on whether the leading multi-photon channel is even (odd). The rate for scalar particle production is about half that of the spinor rate for odd multiphoton channels. For even multiphoton channels the scalar rate is subleading, which can be understood by angular momentum balance. In short pulses, the broad laser spectrum can either enhance or reduce the spin effect depending on the specific conditions \cite{Jansen:2016gvt}. In circularly polarised pulse at large $\xi\gg1$, the spinor rate is always larger than the scalar one. Specifically, for $\chi\ll1$ this ratio is 6 and for $\chi\gg1$ the ratio is 5 \cite{Villalba-Chavez:2012kko}.
    
    General expressions for the probabilities of the production of a spin-polarised pair from a polarised photon have been calculated for the non-precessing transverse spin-basis $\alpha_{(2)}$, see Eq.~\eqref{eq:first:spinbasis}, in Ref.~\cite{Seipt:2020diz}, and for arbitrary spin orientation in Ref.~\cite{Torgrimsson:2020gws}. {The asymptotic scalings of the completely polarised NBW rates for small and large $\chi$ were given in \cite{Seipt:2020diz}.}
    {The pair production rates from the quasiclassical method of Baier, Katkov and Strakhovenko (BKS) \cite{baier98}, with all particles polarised were given in Ref.~\cite{Dai:2021vgl,Chen:2022dgo}. For an infinite monochromatic plane wave the completely spin-and polarisation dependent cross sections had been obtained already in \cite{Ivanov:2004vh} (cf.~also Ref.~\cite{Seipt:2020diz} for earlier references). A numerical method for calculating polarised pair production probabilities from  polarised photons has been presented in Ref.~\cite{Wistisen:2020rsq} using the BKS method. For plane waves, the numerical results agree well with the Volkov state approach.}

    \subsubsection{Carrier envelope phase effects and chirped pulses}

    \paragraph{CEP effects}For ultra-short pulse duration the background field has asymmetries that can be controlled by the carrier envelope phase. Thus, the latter plays an important role for the differential positron spectra for ultra-short pulses. For linearly polarised two-cycle pulses the CEP was shown to shift the transverse momentum spectra \cite{Meuren:2015mra} or the angular positron distribution \cite{Krajewska:2012eb}. These shifts occur both for parallel and perpendicular linear polarisation of the gamma photon \cite{Krajewska:2012eb}. 
    
\begin{figure}
    \centering
    \includegraphics[width=0.45\columnwidth]{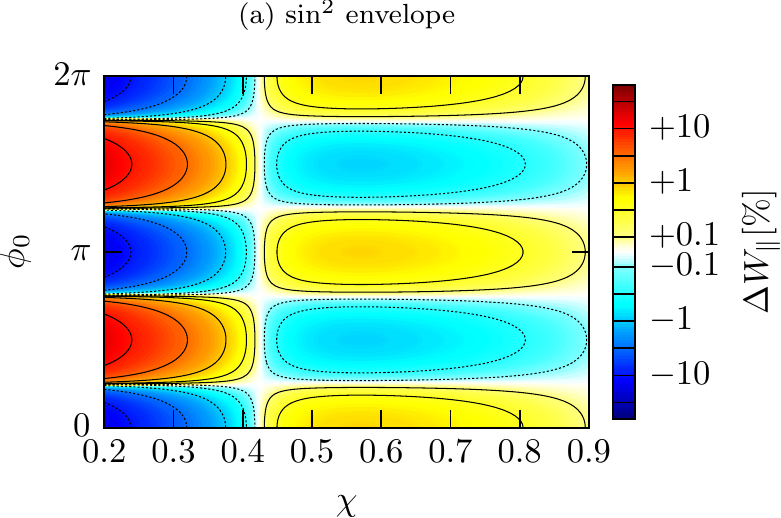}
    \includegraphics[width=0.45\columnwidth]{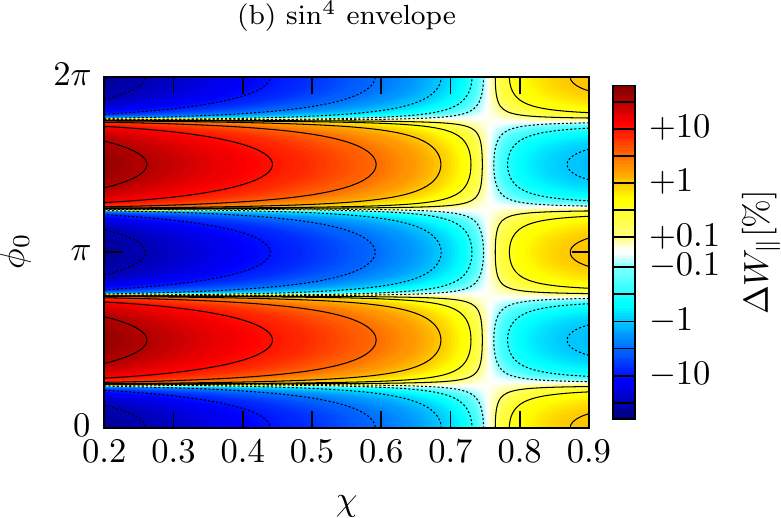}
    \caption{
    The relative pair-creation probability $\Delta W_\parallel$ as a function of $\chi$ and the CEP $\phi_0$ for $\xi = 10$ and $N = 3$ cycle pulses. The dependence on the CEP is quite pronounced for $\chi\ll1$ where the total probability is strongly suppressed. Reproduced from Ref.~\cite{Meuren:2014uia}.}
    \label{fig:first:cep}
    \end{figure}

    In the quasi-static regime for $\xi\gg1$, large CEP effects have been found for $\chi\ll1$ where the pair-creation probability is exponentially suppressed and thus depends sensitively on the maximum field strength, see Fig.~\ref{fig:first:cep}. In the regime $\chi\sim 1$, the CEP effect for the total pair-creation probability is very small \cite{Meuren:2014uia}. For circularly polarised pulses the azimuthal positron spectra become highly asymmetric for single-cycle or sub-cycle pulses over a wide range of $\xi$ \cite{Titov:2015pre}. Similar to NLC, the spectra are invariant when considered as a function of the difference $\varpi - \varphi_{CE}$ between azimuthal angle and CEP. 
    
    For multichromatic laser fields, the relative phases between the different frequency components can be related to the spectral phase \cite{Jansen:2015idl}, where the CEP is associated to a constant spectral phase. (Higher order spectral phase terms are in fact related to the chirp of the pulse \cite{book:optics}.) In the numerical results the CEP modifies the particle spectra locally, while their general multiphoton peak structure is preserved \cite{Jansen:2015idl}.

    \paragraph{Chirped pulses}
    NBW pair production in chirped pulses was investigated in Ref.~\cite{Tang:2021vfh} with the goal to produce quasi-monoenergetic positron beams. The underlying idea is to find a tailored chirping prescription that compensates the ponderomotive broadening of the multi-photon channels, similar to what was discussed for NLC in Sec.~\ref{sec:first:nlc:chirp}. The stationary phase condition for the $n$-photon channel amplitudes, 
    \begin{align}
        \frac{\boldsymbol \rho_\LCperp^2 +m^2_\star(x^\LCp)/m^2}{2\eta_\gamma s(1-s)} - n\frac{\omega(x^\LCp)}{\omega_0}=0 \,,
    \end{align}
   shows that $s$ changes with $x^\LCp$ for a pulse with constant frequency by means of the shift of the effective mass (for circular polarisation)  $m_\star(x^\LCp)=m\sqrt{1+\xi^2 \psi^2(x^\LCp)}$, where $\psi(x^\LCp)$ is the pulse envelope. The optimum chirping prescription is then derived by requiring that all positrons are produced with the same $s$ throughout the pulse, yielding $\omega(x^\LCp)/\omega_0 = 1 + \xi^2 \psi^2(x^\LCp)/ (1 + \boldsymbol \rho_\LCperp^2)$, which clearly depends on the transverse momentum. By optimally chirping also NLC scattering to produce quasi-monoenergetic gamma-rays, the energy spread of a positron beam produced in the two-step NLC+NBW process can controlled to be less than 5~\% at $\xi=1$ \cite{Tang:2021vfh}.

    \subsection{Inverse processes}
    \label{sec:first:absorption}

    In vacuum, momentum conservation implies that an electron-positron pair must annihilate to (at least) two photons. A photon can only be absorbed by an electron (positron) if some other particle, e.g. a nucleus, is involved absorbing some of the momentum. In the presence of a plane wave background, the $2\to 1$ channels of one-photon annihilation and photon absorption are open, where the background field plays the role of the nucleus or additional photon absorbing excess momentum. Contrary to the $1\to 2$ processes discussed so far, light-front momentum conservation for the $2\to1$ processes reads
    \begin{align} \label{eq:emc2to1}
     p_1^\LCp + p_2^\LCp = p_\mathrm{out}^\LCp \,, \qquad 
     \vec p_1^\LCperp + \vec p_2^\LCperp = \vec p_\mathrm{out}^\LCperp \,.
    \end{align}
    The normalised light-front momentum transfer from one arbitrarily chosen initial particle to the final particle is 
    $s=p_\mathrm{out}^\LCp/p_1^\LCp > 1$, i.e.~exceeds unity. This is because the final particle has to take up the light-front momentum of both incident particles.
    Moreover, $P^-$-momentum is deposited into the background field during the interaction, thus $\nu < 0$. The $2\to1$ processes have a rather peculiar kinematics: since there is only one particle in the final state, momentum conservation \eqref{eq:emc2to1} alone determines the scattering products, hence, the values of $s$ and $\nu$ are completely determined by the initial state.

    \subsubsection{Photon absorption}
    
    Due to crossing symmetry, the S-matrix element of photon absorption $\gamma(\ell)\: e(p)\to e(q)$ is related to that for NLC by $S_\mathrm{absorption}(\ell,\epsilon) = S_\mathrm{NLC}(-\ell,\epsilon^*)$. The fact that there is only one outgoing particle has a significant impact on the regularisation of the amplitude as has been detailed in \cite{Ilderton:2019bop}. A perturbative expansion of the absorption S-matrix element for $\xi\ll1$ explicitly demonstrated that the lowest-order contribution to the probability is indeed coming from a negative frequency components of the background field, i.e.~$\nu<0$, which means that the perturbative expression is equivalent to absorbing the probe photon and emitting a photon into the laser field. The relevance for the one-photon absorption process in laser-plasma interactions was studied in detail in \cite{Blackburn:2020fqo}, where it was demonstrated that a consistent description of photon absorption processes in laser-plasma simulations also requires the inclusion of stimulated emission.

    \subsubsection{One-photon pair annihilation}
    
    For pair annihilation into a single photon, $e^+(p) \: e^-(q)\to \gamma(\ell)$, {the strong field S-matrix is related to NBW by time inversion}.
    Due to the restricted $2\to 1$ kinematics, the photons are emitted into a very narrow angular region, yet with rich oscillating substructure reminiscent of the NLC and NBW spectra. Similar to NBW, one-photon pair annihilation has a lower bound on the harmonic number which increases with intensity \cite{Tang:2019ffe}. However, for increasing $\xi$ the possible angular range widens since multiple photons can be emitted into the laser mode.
    
    An initial momentum distribution was included by treating the incoming positron as a wavepacket in Ref.~\cite{Tang:2019ffe}. The optimal conditions for pair annihilation are such that the particles should have similar energy, and their kinetic momenta $\boldsymbol \pi$ should be parallel. For large incident collision angles the process is strongly suppressed, unless $\xi$ is large enough to make the local momenta of the pair particles parallel to one another.
    One-photon pair annihilation has the largest probability if the quantum nonlinearity parameter $\chi=4/3$ \cite{Tang:2019ffe}. It was concluded that one-photon annihilation will have a negligible effect on QED cascades \cite{Tang:2019ffe}, as the number of annihilation events was estimated at less than six orders of magnitude smaller than the positron number at the relativistic critical density.

    
\subsection*{Discussion}

In conclusion, in this section we have given an overview of new insights into first-order (one-vertex) processes in plane waves, which are the simplest processes in SFQED. While lots of work had already been completed before the last decade, this early work was mainly focused on monochromatic and constant crossed fields. 
    In contrast, the research in the last decade has brought a significant refinement of the predictions by considering more realistic finite pulse configurations, and by investigating more detailed observables for laser-particle experiments.
    This included the identification of novel spectral signatures due to different properties of the laser pulses such as their short duration, chirp, carrier envelope phase, interference of multiple pulses or different frequency components, as well as the study of the behaviour and control of particle polarisation.
    
    At this point we emphasise again that all observables calculated from one-vertex processes are strictly meaningful only for `thin' targets, for which all probabilities $\prob <1$. In the general case one has to extend the studies to multi-vertex processes, see Sec.~\ref{sec:second} and \ref{sec:higher}, which can under certain conditions be approximated by using one-vertex diagrams as building blocks. Despite the considerable progress in multi-vertex processes in recent years, much less detailed calculations have been performed than for single-vertex processes. One might therefore  expect a similar richness of phenomena at higher orders as found already at first order.
    
    This section focused on plane waves, where lightfront momentum is conserved. In case one is not dealing with plane waves, explicit calculations become much harder. Some of the known results beyond plane waves are presented in Sec.~\ref{sec:beyondPW}. Approximation frameworks that can serve as basis for simulations are discussed in Sec.~\ref{sec:approx}.


\section{Second-order processes}\label{sec:second}
Even in QED without a background field, it can be difficult to calculate at higher orders in $\alpha$, e.g.~for higher point amplitudes and/or when including many loops. For example, the complete cross-section for ordinary Compton scattering at $\mathcal{O}(\alpha^3)$ was calculated only recently~\cite{Lee:2021iid}. By adding a background field one may expect the calculations to become even more difficult. For processes in plane-wave backgrounds, already the second-order processes are challenging to calculate, which is why we devote an entire section to this.
Most studies in the last 10 years have focused on two specific $\mathcal{O}(\alpha^2)$ processes: the nonlinear trident process $e^\LCm\to e^\LCm e^\LCm e^\LCp$ and double nonlinear Compton scattering $e^\LCm\to e^\LCm \gamma \gamma$. We will therefore focus on these processes, both of which only have one particle in the initial state. We note though that historically there have been a larger number of papers on $2\to2$ processes, $e^\LCm e^\LCm\to e^\LCm e^\LCm$ and $e^\LCm\gamma\to e^\LCm\gamma$, following the seminal papers~\cite{1967JETP...25..697O,1968JETP...26.1132O} which showed that these processes have resonances at the pole of the propagator of the intermediate particle, which gives another example where one has to take higher-order-in-$\alpha$ corrections into account. The literature on these processes is reviewed in~\cite{Hartin:2006qnk}. For more recent papers on $2\to2$ processes see~\cite{Roshchupkin:2021yhd} and references therein.

Before we turn to $1\to3$ processes, we will briefly mention some further studies of other $\mathcal{O}(\alpha^2)$ processes.
As for $2\to1$ processes (see Sec.~\ref{sec:first}), $2\to2$ processes are conceptually quite different from $1\to3$. For example, the $2\to2$ processes are nonzero even in the limit $\xi\to0$.
Pair annihilation $e^\LCm e^\LCp\to\gamma\gamma$ was studied in~\cite{Bragin:2020akq}, which explained the more nontrivial role that particle wave packets play in such processes, as the two initial particles have to meet while a process with only one initial particle can happen anywhere in the field. 

The opposite process, $\gamma\gamma\to e^\LCm e^\LCp$, have been studied recently in electric fields~\cite{Satunin:2018rdw,Torgrimsson:2019sjn}. Pair production in an electric field at finite temperature (with thermal fermions and/or thermal photons) has a long history with many conflicting papers. Recently~\cite{Brown:2015kgj,Gould:2017fve} the exponential part of the pair-production probability was calculated {as a function of $\gamma_T:=2mT/eE\sim1$}, where $T$ is the temperature in the imaginary time formalism. In~\cite{Torgrimsson:2019sjn} it was shown that this can be studied by treating the thermal photons perturbatively, and that the dominant contribution comes from the absorption of two thermal photons, i.e. $\gamma\gamma\to e^\LCm e^\LCp$. This approach also allowed the pre-exponential factor to be calculated.
The process $\gamma\gamma\to e^\LCm e^\LCp$ has also been studied for a weak plane wave, where the process only involves one photon from the laser at each vertex~\cite{Serov:2020ots}, with focus on resonance where the intermediate photon goes on shell.
{Pair production by thermal photons in a constant-crossed field was studied in~\cite{King:2012kd}.}

We now turn to $1\to3$ processes. We will consider plane-wave background fields, since these are simple and experimentally relevant. In this section we use lightfront notation as in Sec.~\ref{A primer on lightfront coordinates}, which in some places is slightly different from~\ref{sec:first}.

As we will explain below, for fields that are sufficiently strong (large $\xi$) or are  sufficiently long (i.e. have pulse length characterised by some large, dimensionless $\mathcal{T}$), one can in general expect important contributions from higher-order processes, since the $n$-th order scales (to leading order in $\xi\gg1$ and/or $\mathcal{T}\gg1$) as $(\alpha\xi\mathcal{T})^n$, which can exceed unity for modern laser pulses. While large $\xi$ and/or long pulse length makes higher orders in $\alpha$ (in the Furry expansion) important, it also allows us to calculate them in a certain approximation, as the dominant contribution comes from incoherent products of first-order processes\footnote{How large $\xi$ or $\mathcal{T}$ has to be for the corrections to be negligible depends of course on the other parameters. If one keeps the expansion parameter ($\xi$ and/or $\mathcal{T}$ in this case) large but fixed then one can in general not expect the approximation to be valid if one makes some of the other parameters very small or large. One example, see below, where this approximation breaks down is in the very-high-energy limit.}. This is still not trivial, and in the last couple of years much progress has been made in how to actually construct this approximation and include general spin and polarisation, what to do when $\xi$ is not very large, and how to take loops into account.
These problems occur already at $\mathcal{O}(\alpha^2)$, so we will review them in this section. (In Sec.~\ref{sec:higher} we will review some methods for what to do with the $\alpha$ series obtained with this approximation.)
At $\mathcal{O}(\alpha^2)$ one can, at least for simple cases such as constant-crossed fields or more general plane waves, and with considerable effort, calculate the exact results. This means that one can check that any corrections to the incoherent-product approximation are small, or determine in which parameter regime corrections are important.

We are mainly reviewing studies where $\xi$ is not small and where the fields are produced by high-intensity lasers, but we note that trident in a weak field has recently been studied in~\cite{Hu:2014ooa,Acosta:2019bvh,Sizykh:2021ywt}, and the two-step/one-step separation (see below) has also been studied in double Compton scattering by an electron in the field of a crystal~\cite{Wistisen:2019pwo}. As many ideas are similar for these two $\mathcal{O}(\alpha^2)$ processes, we will initially focus on trident. (The two-step part alone has been studied in a CCF in \cite{King:2013zw} and in a pulse using the LMA in \cite{Titov:2021kbj}.)

\subsection{On-shell/off-shell vs. one-step/two-step separation}\label{one-step/two-step separation}

\begin{figure}
     \centering
     \includegraphics[width=.8\textwidth]{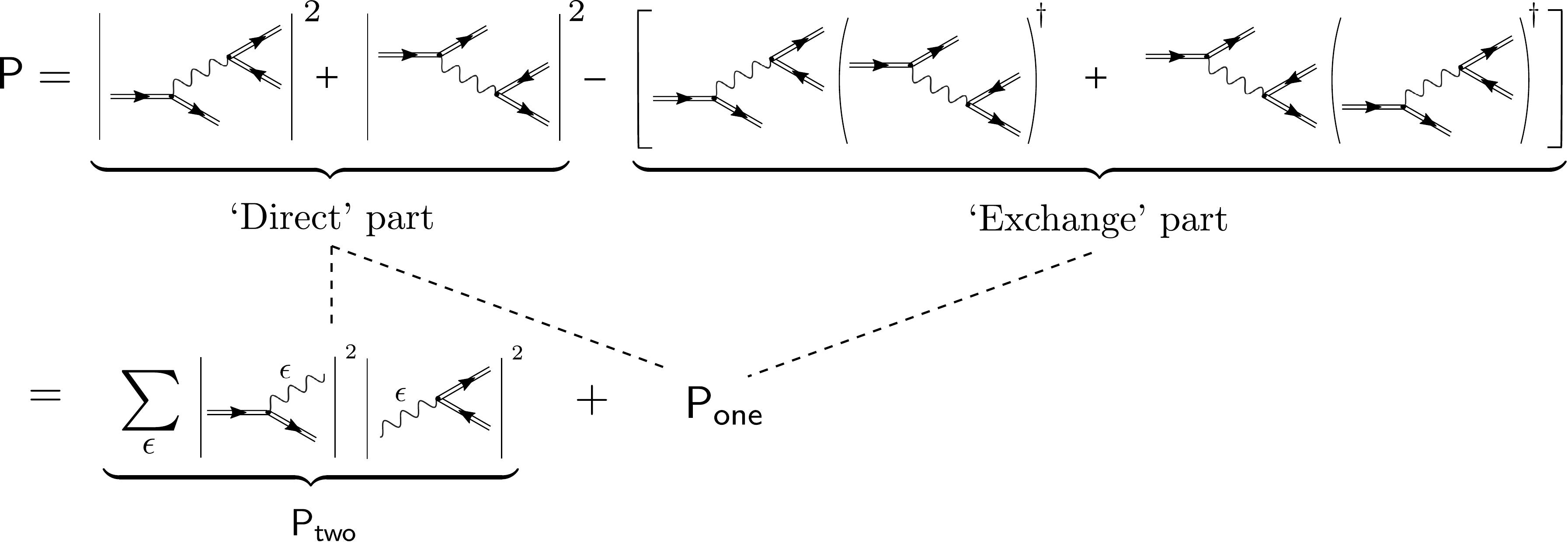}
     \caption{Overview of the one-step/two-step separation for trident, and diagrams corresponding to direct and exchange parts.}
     \label{fig:onetwostepfig}
\end{figure}

In the standard covariant formalism the Feynman diagram for trident contains an intermediate photon that can be both on or off shell. 
The amplitude is given by $\mathcal{M}=\mathcal{M}^{12}-\mathcal{M}^{21}$, where $\mathcal{M}^{21}$ is obtained from $\mathcal{M}^{12}$ by swapping the two electrons in the final state, and
\be\label{S12}
\frac{1}{k_\LCp}(2\pi)^3\delta^3([p_1+p_2+p_3-p]_{\LCm,\LCperp})\mathcal{M}^{12}
=e^2\int\ud^4x \, \ud^4y\; \bar{\psi}_{p_2}(y)\gamma^\nu\psi_{p_3}^{(\LCp)}(y) D_{\nu\mu}(y-x)\bar{\psi}_{p_1}(x)\gamma^\mu\psi_p(x) \;,
\ee
where the states are given by~\eqref{volkov-e-in}, \eqref{volkov-e-out} and~\eqref{volkov-e-out} but without the $a_\infty$ term (i.e. we label also the outgoing states with the initial momentum that the particle would need to have in order for $\pi_\mu(\infty)$ to be the final momentum) and
with the photon propagator 
\be
D_{\nu\mu}(y-x)=-i\int\frac{\ud^4l}{(2\pi)^4}L_{\nu\mu}\frac{e^{-il\cdot (y-x)}}{l^2+i0} \;.
\ee
Rather than e.g. the Feynman gauge $L_{\nu\mu}\to g_{\nu\mu}$, when dealing with plane-wave backgrounds it is convenient to use the lightfront gauge
\be\label{DLFgauge}
L_{\mu\nu}:=g_{\mu\nu}-\frac{k_\mu l_\nu+l_\mu k_\nu}{k\cdot l} \;.
\ee

A first idea of how to split this process into two parts is to split the photon propagator in the amplitude into an on-shell and off-shell part by applying the Sokhotsky-Plemelj formula~\cite{Ritus:1972nf,Ilderton:2010wr}
\be\label{deltaPrincipal}
\frac{1}{r+i0}=-i\pi\delta(r)+{\rm PV}\frac{1}{r} \;,
\ee
where ${\rm PV}$ is the principal value and $r$ some appropriate combination of momenta with $r=0$ corresponding to the intermediate photon going on-shell. However, this is not a unique separation~\cite{King:2013osa}, as one can e.g. perform a partial integration in the principal value part to subtract some arbitrary term with $r=0$. As noted in~\cite{King:2013osa}, a causal step function can be obtained by instead performing the $r$ integral with
\be\label{thetaInstead}
\int\ud r\frac{e^{-ir(\varphi_2-\varphi_1)}}{r+i0}=-2\pi i\Theta(\varphi_2-\varphi_1)
\ee
which only receives contribution from the on-shell pole.
Also, if one applies the lightfront-time Hamiltonian ($H_{\rm LF}$) formalism~\cite{Heinzl:2000ht,Brodsky:1997de} to trident~\cite{Dinu:2017uoj} then all particles, in particular intermediate photons, are on shell from the start and instead of an off-shell intermediate photon one has an instantaneous term in the lightfront-Hamiltonian with 4-fermion direct interaction. 
The lightfront-Hamiltonian approach suggests a split which, in the standard approach, corresponds to a separation of the photon propagator in~\eqref{DLFgauge} as~\cite{Dinu:2017uoj}
\be\label{Lsplit}
L_{\mu\nu}=L_{\mu\nu}\bigg|_{l_\LCp=l_\LCp^{\text{on}}}-\frac{l^2}{(k\cdot l)^2}k_\mu k_\nu=:L_{\mu\nu}^{(\Theta)}+L_{\mu\nu}^{(\delta)} \;,
\ee
where $l_\LCp$ in the first term has been replaced by its on-shell value $l_\LCp^\text{on}=l_\LCperp^2/(4l_\LCm)$. Performing the $l_\LCp$ integral in the photon propagator, with this decomposition, gives a step function $\Theta(\varphi_2-\varphi_1)$ for the $L_{\mu\nu}^{(\Theta)}$ term and a delta function $\delta(\varphi_2-\varphi_1)$ for the $L_{\mu\nu}^{(\delta)}$ term. 
We emphasise that this lightfront separation is just one separation that happens to be convenient. {Different separations on the amplitude level can be useful in various contexts. The separation used in~\cite{Mackenroth:2018smh}, for example, was chosen to explicitly separate direct and cascade channels in such a way that gauge invariance was manifest.

The closely related separation in~\cite{Ilderton:2020rgk} (applied to $2\to2$ scattering) exposed underlying analytic structures, and related these to factorisation of the amplitude into hard/soft contributions.}

What we are actually interested in is having a separation on the probability level, because the dominant contribution to higher-order processes comes from (sums of) incoherent products. Any separation on the amplitude level (such as between on-shell and off-shell) will give cross terms on the probability level~\cite{King:2013osa,King:2014wfa}.
For trident the dominant part of the total probability scales quadratically with the pulse length (at $\mathcal{O}(\alpha^2)$, i.e. neglecting higher-order radiative corrections and assuming that $\mathcal{T}$ {or $\xi$ are} not too large). The quadratic scaling comes from the fact that the initial electron can emit a real intermediate photon anywhere in the field, and then the photon can decay into a pair anywhere in the field, i.e. the intermediate photon can propagate an arbitrary macroscopic distance before it decays. This is called the `two-step' part. Its contribution to the total probability can be obtained from the incoherent product of the probabilities of (real) Compton scattering and pair production. The rest of the trident probability is referred to as the `one-step' part. This is illustrated in Fig.~\ref{fig:onetwostepfig}.
While only real particles contribute to the two-step, both real and off-shell particles contribute to the one-step~\cite{King:2013osa}. 

However, even this separation is not completely unique, because there is still a possibility to choose different step functions to ensure that the intermediate photon is emitted before it decays into a pair. The reason this is not unique is because for each $\mathcal{O}(\alpha)$ process there are two $\varphi$ variables for the probability. Let $\varphi_{\rm C}$ and $\varphi'_{\rm C}$ be the lightfront time variable for the Compton-scattering amplitude and its complex conjugate, and $\varphi_{\rm BW}$ and $\varphi'_{\rm BW}$ likewise for the pair-production step, i.e. $\varphi_{\rm C}=k\cdot x$ and $\varphi_{\rm BW}=k\cdot y$ in~\eqref{S12} (cf.~\eqref{eqn:phidefs1} in Sec.~\ref{sec:first}). The simplest choice is then 
\be\label{stepfunction1}
\Theta(\phi_{\rm BW}-\phi_{\rm C}) \;,
\ee
where $\phi_{\rm C}=(\varphi'_{\rm C}+\varphi_{\rm C})/2$ and $\phi_{\rm BW}=(\varphi'_{\rm BW}+\varphi_{\rm BW})/2$.
However, having lightfront time ordering on the amplitude level means a factor of $\Theta(\varphi_{\rm BW}-\varphi_{\rm C})$, which gives on the probability level~\cite{Dinu:2017uoj} {(this has also been used in~\cite{Mackenroth:2018smh})}
\be\label{stepfunction2}
\Theta(\varphi_{\rm BW}-\varphi_{\rm C})\Theta(\varphi'_{\rm BW}-\varphi'_{\rm C})= \Theta(\phi_{\rm BW}-\phi_{\rm C})\left[1-\Theta\left(\frac{|\theta_{\rm BW}-\theta_{\rm C}|}{2}-[\phi_{\rm BW}-\phi_{\rm C}]\right)\right] \;,
\ee
where $\theta_{\rm C,BW}=\varphi_{\rm C,BW}-\varphi'_{\rm C,BW}$. While the two choices~\eqref{stepfunction1} and~\eqref{stepfunction2} give different separations, they give the same two-step to leading order in $\xi\gg1$ or the pulse length. One reason to choose~\eqref{stepfunction1} is because then the $\theta_{\rm C}$ and $\theta_{\rm BW}$ integrals are independent, which makes the calculation much simpler. Indeed, in the locally-constant-field (LCF) approximation (LCFA) and the locally-monochromatic-field approximation (LMA) one can perform the $\theta$ integrals in terms of Airy or Bessel functions, see Sec.~\ref{sec:approx}. 

The choice~\eqref{stepfunction1} is also natural for the following reason. 
The one-step/two-step separation is mostly useful if the two-step dominates, which it does for either a long pulse or large $\xi$. For large $\xi$ (and with $\chi$ considered an independent variable) one can expand the probability for a general inhomogeneous plane wave as a series in $1/\xi$. The leading term scales as $\xi^2$ (so this is a Laurent series) and can be obtained from the two-step by treating the field as locally constant, i.e. it has one constant value at the Compton step and another constant value at the pair-production step. The one-step scales to leading order as $\xi$. With the choice in~\eqref{stepfunction1}, the next term in the expansion {of the two-step} is $\mathcal{O}(\xi^0)$, i.e. the one-step includes the entire next-to-leading order ($\mathcal{O}(\xi)$) correction. In contrast, with the choice in~\eqref{stepfunction2}, both the two-step and the one-step contribute to $\mathcal{O}(\xi)$.
Note that with the definition we have considered, both the two-step and the one-step also include higher-order terms in the $1/\xi$ expansion. However, in the LCF regime it is common to use two-step to refer to the leading order $\mathcal{O}(\xi^2)$ and the one-step for the next-to-leading order $\mathcal{O}(\xi)$. Hence, to leading and next-to-leading order, the choice in~\eqref{stepfunction1} corresponds to the separation that was considered in the first papers~\cite{BaierTrident,Ritus:1972nf} on trident from the 1970s (which only considered a constant field).   

Thus, the separation that is the most useful for generalisation to higher orders in $\alpha$ is based on having terms that scale differently with respect to the pulse length and/or $1/\xi$. 
Moreover, the two-step part of an $\mathcal{O}(\alpha^2)$ process and in general the $N$-step part of $\mathcal{O}(\alpha^N)$ processes are ingredients that one can implement in PIC codes, as will be discussed in Sec.~\ref{sec:higher}.   

Before we discuss in detail how to obtain the dominant two-step terms, let us briefly discuss another important way to separate the probability, which is related to the fact that the process has more than one Feynman diagram.

\subsection{Direct vs. exchange terms}\label{direct vs exchange}

\begin{figure}
    \centering
\includegraphics[width=0.5\linewidth]{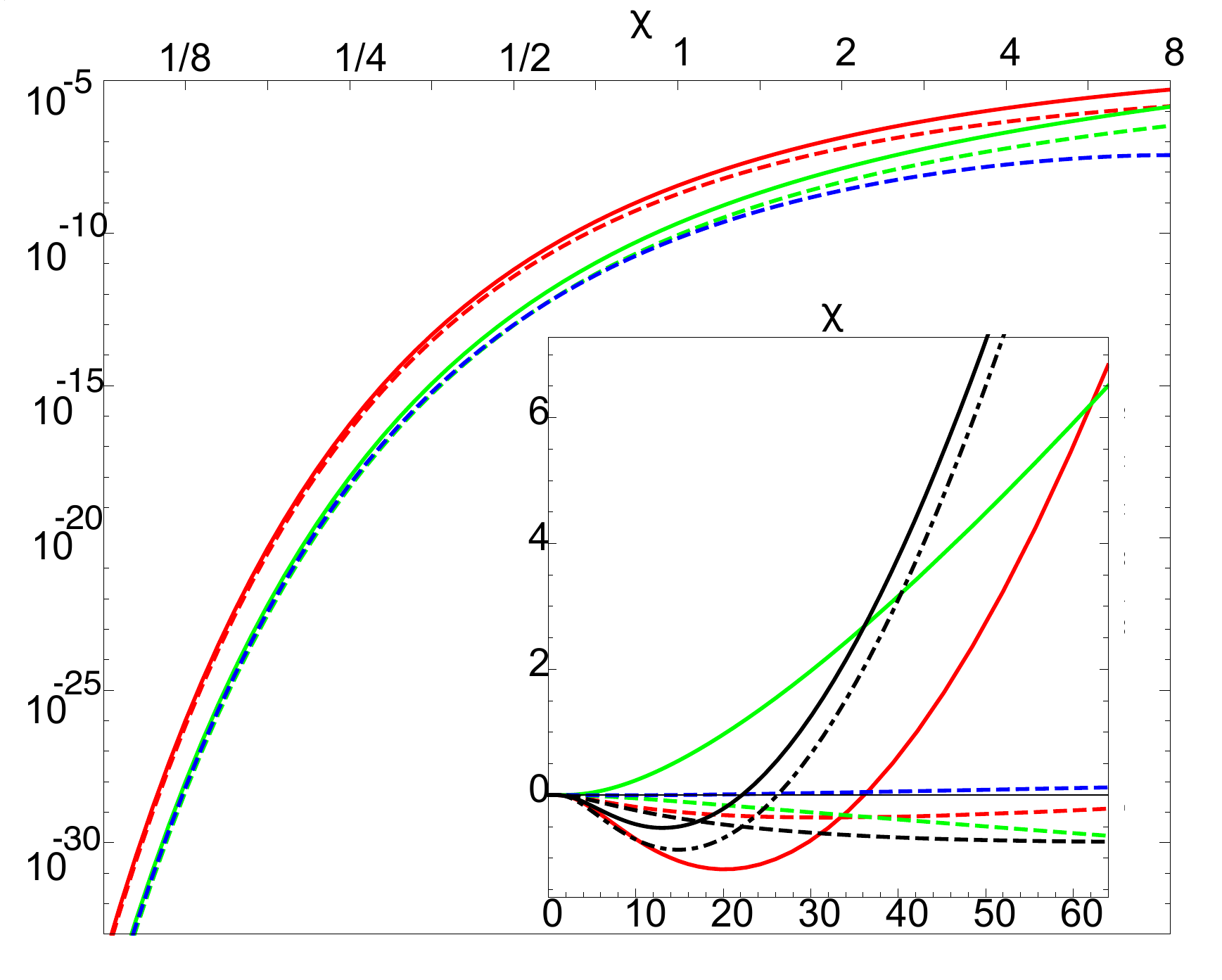}
    \caption{Plot adapted from~\cite{Dinu:2017uoj} showing the exchange and direct parts of trident. The different lines show the different terms that~\eqref{Lsplit} leads to on the probability level. The black (solid, dashed, dot-dashed) lines give the (direct, exchange, total) one-step. (The other lines are not discussed in this review.)}
    \label{exchangeVsDirectFig}
\end{figure}

Apart from the one-step/two-step separation, there is another separation that is useful, perhaps mostly because it separates out a rather complicated term. This separation is due to the fact that the final state has two identical particles, in the trident case two identical electrons. The amplitude therefore has two terms, $M=M_{12}-M_{21}$, where $M_{21}$ is obtained from $M_{12}$ by swapping the momenta and spin of the two final-state electrons. On the probability level this gives a cross-term $2\text{Re }M_{12}^*M_{21}$ which is referred to as the exchange part of the probability. The rest of the probability is called the direct part. This is illustrated in Fig.~\ref{fig:onetwostepfig}. It turns out that the exchange part only contributes to the one-step, while the direct part contributes to both the one-step and to the two-step, i.e.
\be
\prob_{\rm two}=\prob_{\rm two}^{\rm dir}
\qquad
\prob_{\rm one}=\prob_{\rm one}^{\rm dir}+\prob_{\rm one}^{\rm ex} \;.
\ee
The reason it is relevant to talk about the exchange term is because it is much harder to calculate than the direct term. In fact, the exchange term was neglected for decades until very recently when it was finally calculated in~\cite{Dinu:2017uoj,King:2018ibi}. Of course, studies where the amplitude is computed numerically~\cite{Hu:2010ye,Ilderton:2010wr,Seipt:2012tn,Mackenroth:2012rb,2015JPhCS.594a2024K,Mackenroth:2018smh} include everything, since then there is not much motivation for neglecting the exchange term. 
One can expect the exchange term to be negligible at very high energies, although this was only proven recently~\cite{Dinu:2019wdw,Dinu:2019pau,Torgrimsson:2020wlz}. However, in general, the exchange term is not negligible compared to the direct part of the one-step; see Fig.~\ref{exchangeVsDirectFig} and similar plots in~\cite{Dinu:2017uoj,King:2018ibi} and the analytical approximations below, cf. also~\cite{Dinu:2017uoj}. Of course, in a regime where the two-step dominates one can neglect the exchange term to leading order, but then one should for consistency also neglect the direct part of the one-step. One of the main reasons for calculating the one-step in such a regime is to determine how good the two-step approximation is, which is important also for estimating the size of corrections to the $N$-step approximation of higher-order processes. Thus, the exchange term is important in order to answer this fundamental question. 

This discussion also applies to nonlinear double Compton scattering ($e^\LCm\to e^\LCm\gamma\gamma$)~\cite{Morozov:1975uah,King:2014wfa,Dinu:2018efz} and photon trident ($\gamma\to e^\LCm e^\LCp\gamma$)~\cite{MorozovNarozhnyiPhTr,Torgrimsson:2020mto} (resonance conditions for photon trident have been studied for weak fields in~\cite{Yelatontsev:2020zfk}). For double Compton scattering there are two diagrams because the two emitted photons are identical, and for photon trident there are two diagrams because the final photon can be emitted by either the electron or the positron. It turns out that, from an analytical point of view, the exchange term is very similar in these three processes. In double Compton scattering it turns out to be even more important. First, the two individual diagrams for this process are not separately gauge invariant~\cite{Dinu:2018efz}. Second, in the low energy regime the exchange term even cancels the direct part of the one-step to leading order~\cite{Dinu:2018efz}. In other words, the exchange term changes the order of magnitude of the one-step, and so considering only the direct part of the one-step would give a wrong estimate on the size of the corrections to the two-step approximation.

\subsection{Intermediate particle polarization sums in the two-step}\label{Spin sums in the two-step}

While the momentum of the intermediate photon is determined by energy-momentum conservation, there is nothing that determines its polarisation, i.e. there is a sum over its polarisation on the amplitude level. Hence, the two-step part is given by a sum of \emph{incoherent} products.
On the amplitude level one can sum over the two orthogonal polarisation vectors from any basis. Some bases might be simpler, but there is always just a single sum. However, on the probability level this gives a double sum, because the modulus square of the amplitude receives also contributions from off-diagonal terms characterised by different polarisations. It has been known since~\cite{Ritus:1972nf} that for a constant field there is a certain simple basis ($\epsilon_\mu^{(1)}$ and $\epsilon_\mu^{(2)}$) for which this double polarisation sum reduces to a single sum, where the two vectors correspond to photons polarized parallel or perpendicular to the field, i.e $\epsilon_\LCperp^{(1)}=\{1,0\}$ and $\epsilon_\LCperp^{(2)}=\{0,1\}$ with $\epsilon_\LCp=\epsilon_\LCperp l_\LCperp/(2l_\LCm)$ (sum over $\perp$ components implied) and $\epsilon_\LCm=0$ (cf.~\eqref{eq:first:photonbasis}). In other words, the two-step part of the trident probability is then given by the probability to emit a photon with $\epsilon_\mu^{(1)}$ times the probability for such a photon to decay into a pair, plus the same quantity but with $\epsilon_\mu^{(2)}$. This also works for a general inhomogeneous plane wave field with linear polarisation~\cite{Dinu:2017uoj}, and there is another basis that works in a similar way for a circularly polarised field~\cite{Torgrimsson:2020gws}. However, in general there is no such simple basis for which the off-diagonal terms in the double sum vanish.

There is though another way of treating these double spin/polarisation sums which is simple and works for any field and spin~\cite{Dinu:2018efz,Dinu:2019pau,Torgrimsson:2020gws}. In this approach both fermion spin and photon polarization are treated in the same way.
The spin dependence in this approach is expressed in terms of Stokes vectors (see Sec.~\ref{sec:first:nlc:spin}) and "strong-field QED Mueller matrices". 
This approach resembles the formalism used to describe changes in the Stokes vector of a light beam when it passes an optical element. (Similar matrix representations have been used in QED without a background field in~\cite{RevModPhys.33.8}.) In the optics analogy the Stokes vector of a beam that has passed several optical elements is obtained by multiplying a sequence of Mueller matrices. Something similar happens in QED. 

For linear polarization we could choose the two vectors $\epsilon_{(1,2)}^\mu$ to be along the electric and magnetic field, but for a generally polarised field we just choose some fixed orthogonal vectors $\epsilon_{(1,2)}^\mu$.

\begin{figure}
    \centering
    \includegraphics{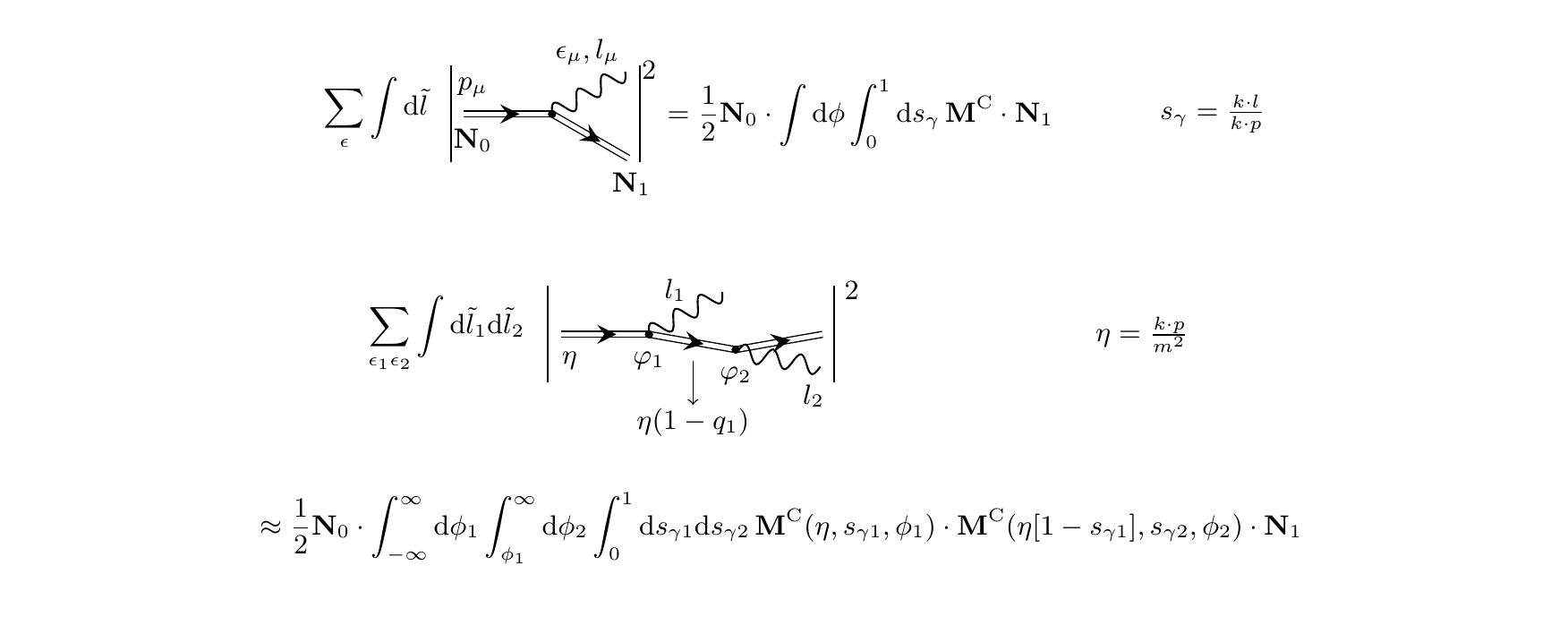}
    \caption{Definition of the Compton Mueller matrix and how it is used to to construct the two-step part of double Compton scattering~\cite{Dinu:2019pau}. Figure adapted from~\cite{Torgrimsson:2021wcj}.}
    \label{fig:ComptonMuellerMatrixReview}
\end{figure}

\begin{figure}
    \centering
    \includegraphics[width=.8\linewidth]{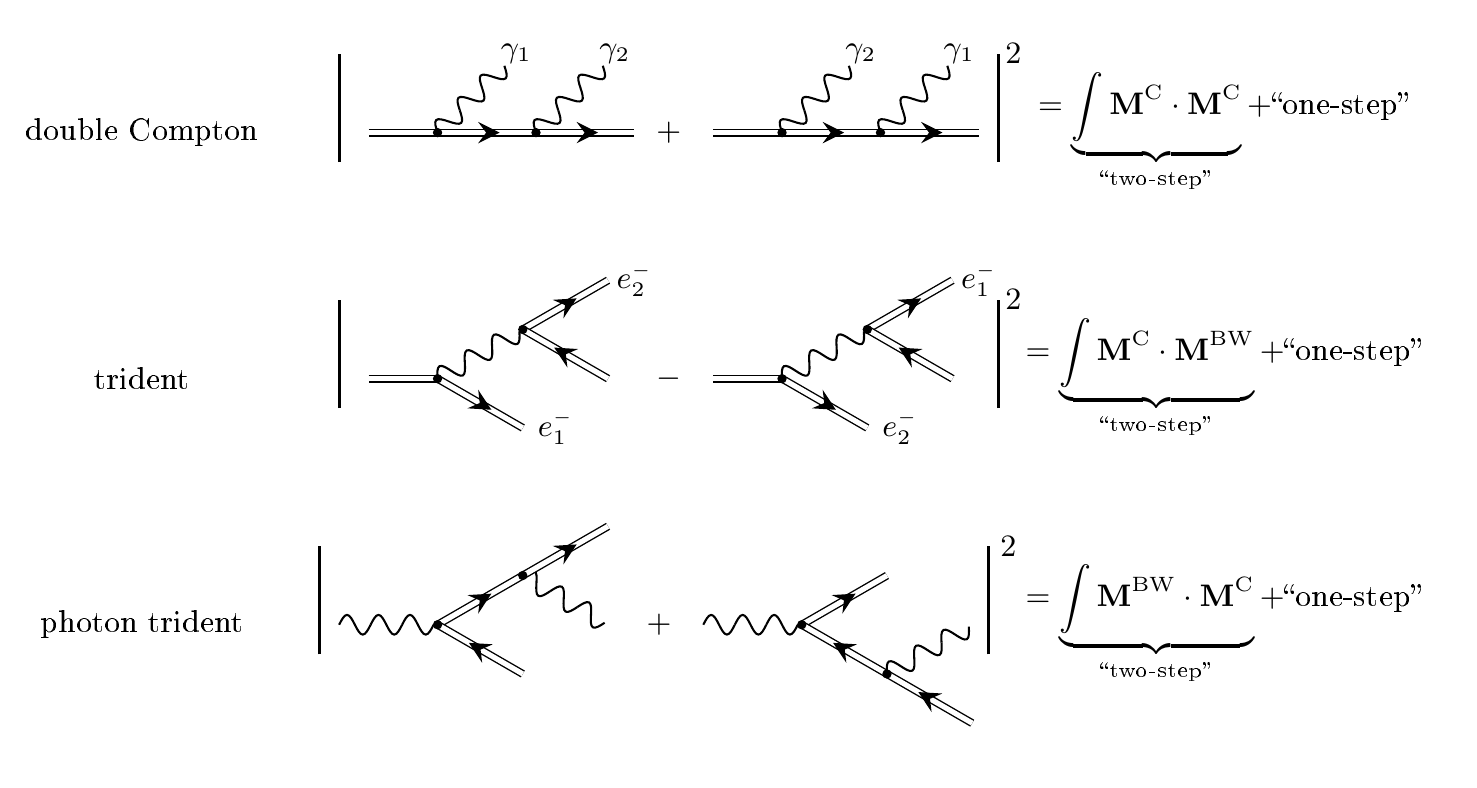}
    \caption{One-step/two-step separation for trident, double Compton and photon trident, with the two-step expressed in terms of of Mueller matrices. Figure taken from~\cite{Torgrimsson:2021wcj}.}
    \label{fig:MuellerPlusOneStep}
\end{figure}

Consider for example the probability of nonlinear Compton scattering. The initial electron has a spin described by ${\bf N}_0$. The probability that the electron emerges with a Stokes vector ${\bf N}_1$ after emitting a photon can then be expressed as (illustrated in Fig.~\ref{fig:ComptonMuellerMatrixReview})
\be\label{MCdefinition}
\prob=\frac{1}{2}{\bf N}_0\cdot\int\ud\phi\int_0^1\ud s_\gamma\,{\bf M}^{\rm C}\cdot{\bf N}_1 \;,
\ee
where $s_\gamma=k\cdot l/k\cdot p$ is the ratio of the longitudinal momentum of the emitted photon and the initial electron, and {${\bf M}^{\rm C}$ is independent of the electron spin.
Given the comparison with Mueller matrices in optics (see e.g.~\cite{1985AmJPh..53..468B}), it is natural to call ${\bf M}^{\rm C}$ a strong-field-QED Mueller matrix.}
It might be more common to choose a spin basis and only consider spin up and down in that basis, but considering general spin in terms of Stokes vectors has been done in e.g.~\cite{baier98,2000PThPh.104..769B,Ivanov:2004fi,Ivanov:2004vh,1983JETP...57..935G,Li:2018fcz}. 
Such treatments of the spin and polarisation have also become popular in recent PIC codes~\cite{Li:2018fcz}, see Sec.~\ref{sec:higher}.
The two-step part of double nonlinear Compton scattering is given by
\be
\prob_{\rm two}=\frac{1}{2}{\bf N}_0\cdot\int_{-\infty}^\infty\ud\phi_1\int_{\phi_1}^\infty\ud\phi_2\int_0^1\ud s_{\gamma1}\ud s_{\gamma2}\,{\bf M}^{\rm C}(\eta,s_{\gamma1},\phi_1)\cdot{\bf M}^{\rm C}(\eta[1-s_{\gamma1}],s_{\gamma2},\phi_2)\cdot{\bf N}_1 \;,
\ee
where $\eta=k\cdot p/m^2$, see Fig.~\ref{fig:ComptonMuellerMatrixReview}.
Note that the two Mueller matrices are lightfront-time ordered and the second step depends on how much longitudinal momentum was emitted in the first step, while the transverse momentum integrals $l^{(1)}_\LCperp$ and $l^{(2)}_\LCperp$ ($l_\mu^{(1,2)}$ being the photon momenta) can be performed for each Mueller matrix separately. This continues to higher orders: The $N$-step part of the probability of emitting $N$ photons is obtained from a lightfront-time and longitudinal-momentum ordered product of $N$ Mueller matrices. There is another Mueller matrix for a pair-production step, ${\bf M}^{\rm BW}$, which can be used e.g. for trident and photon trident as illustrated in Fig.~\ref{fig:MuellerPlusOneStep}. 
The advantage of this approach is that it works for any plane-wave field, of arbitrary polarisation and pulse profile,
and for arbitrary initial and final particle spin. If $\xi$ is large then one can use an LCF approximation of the Mueller matrices. However, $\xi$ does not have to be large as long as the pulse is sufficiently long so that the incoherent product is still dominant.

If one is also interested in the polarisation of the emitted photon then the probability of emitting one photon is given schematically by (with integrals left implicit)
\be\label{NLCMueller3}
\prob=N^{(\gamma)}_i N^{(e,{\rm in})}_j N^{(e,{\rm out})}_k M^{\rm C}_{ijk} \;,
\ee
where $M^C_{ijk}$ are the components of a $4\times4\times4$ matrix, $N_i$ are the components of ${\bf N}$, ${\bf N}^{(\gamma)}$ is the Stokes vector for the photon, which can be defined from the polarisation vector $\epsilon_\mu$ as follows~\cite{Dinu:2019pau}. In the lightfront gauge we have
$\epsilon_\LCp=l_\LCperp\epsilon_\LCperp/(2l_\LCm)$, for a photon with momentum $l_\mu$, and the transverse components are given in terms of two real constants, $\rho$ and $\lambda$, by
\be\label{epsilonDefinition}
\epsilon_\LCperp=\left\{\cos\left(\frac{\rho}{2}\right),\sin\left(\frac{\rho}{2}\right)e^{i\lambda}\right\} \;.
\ee
The Stokes vector ${\bf N}=\{1,{\bf n}\}$ is given by\footnote{${\bf N}$ can be obtained from $\epsilon_\mu$ without using this explicit expression with $\rho$ and $\lambda$, see~\cite{Torgrimsson:2020gws}.}
\be\label{Stokes3D}
{\bf n}=\{\cos\lambda\sin\rho,\sin\lambda\sin\rho,\cos\rho\} \;.
\ee
This vector is similar to the Stokes vector of a light beam, but for a single photon, so ${\bf N}=\{1,0,\pm1,0\}$ would for example correspond to circular polarisation.
One can also obtain Mueller matrices for the other $\mathcal{O}(\alpha)$ processes. For nonlinear Breit-Wheeler pair production one has the same structure as in~\eqref{NLCMueller3}, but with a Stokes vector for the positron instead of ${\bf N}^{(e,{\rm in})}$ (and obviously, the same is true for photon emission by a positron.)


\begin{figure}
    \centering
    \includegraphics{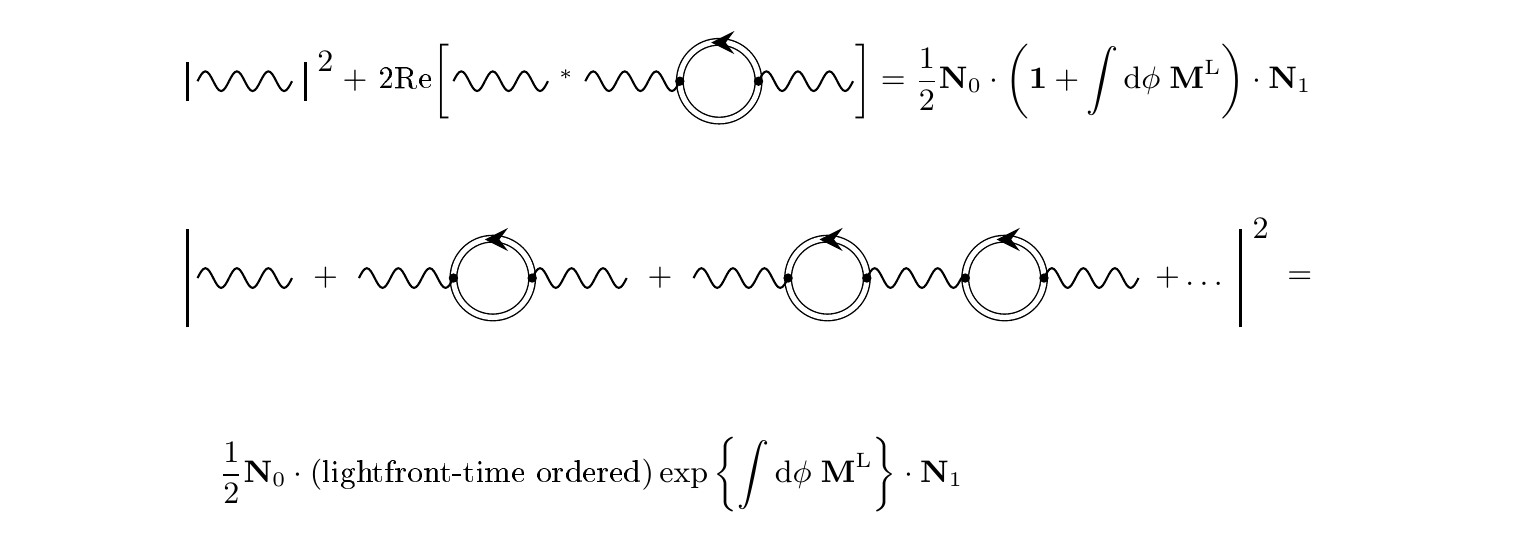}
    \caption{Definition of the $\gamma\to\gamma$-loop Mueller matrix and how it is used to obtain the dominant contribution at higher orders~\cite{Torgrimsson:2020gws}. {A similar figure for the fermion-mass loops can be found in~\cite{Torgrimsson:2021wcj}.}}
    \label{fig:polarizationLoopSum}
\end{figure}

However, one also needs to include loops in an incoherent-product approach, as certain loops have the same scaling with respect to $\alpha\xi\mathcal{T}$ as the tree-level processes, which one should expect from unitarity. The relevant loops can be obtained from the following first-order loop terms~\cite{Torgrimsson:2020gws}. For the one-loop electron mass operator we have
\be\label{eMLdefinition}
\prob_{e^\LCm\to e^\LCm}=\frac{1}{2}{\bf N}_0\cdot\left({\bf 1}+\int\ud\phi\int_0^1\ud s_\gamma\,{\bf M}^{\rm L}\right)\cdot{\bf N}_1+\mathcal{O}(\alpha^2) \;,
\ee
where $s_\gamma=k\cdot l/k\cdot p$ and $l_\mu$ is the momentum of the photon in the loop. This $s_\gamma$ integral could in principle be performed for each ${\bf M}^{\rm L}$ separately, i.e. one could absorb it into the definition of ${\bf M}^{\rm L}$. The reason for not doing so is that it is useful to write it in similar form as the Compton Mueller matrix~\eqref{MCdefinition}, so that IR and soft-photon problems cancel directly in the sum ${\bf M}^{\rm C}+{\bf M}^{\rm L}$, see~\cite{Torgrimsson:2021wcj} and Sec.~\ref{sec:higher}. The Mueller matrix for the ``polarisation tensor" loop is given by (this time absorbing the momentum integrals into the definition of ${\rm M}_\gamma$)
\be\label{gammaMLdefinition}
\prob_{\gamma\to\gamma}=\frac{1}{2}{\bf N}_1\cdot\left({\bf 1}+\int\ud\phi\;{\bf M}_\gamma^{\rm L}\right)\cdot{\bf N}_0+\mathcal{O}(\alpha^2) \;.
\ee
{Fig.~\ref{fig:polarizationLoopSum} shows how ${\bf M}_\gamma^{\rm L}$ can be used to obtain the dominant contribution to higher orders processes with loops.}
With these Mueller matrices for the tree-level and one-loop $\mathcal{O}(\alpha)$ processes one can construct an approximation for general higher-order processes~\cite{Torgrimsson:2020gws,Dinu:2019pau,Dinu:2018efz}. 
Loops give for example the spin precession~\cite{Ilderton:2020gno,Torgrimsson:2020gws}, but are also needed to remove IR divergences and are needed to even be able to take the classical limit, see Sec.~\ref{sec:higher}.

Parallel to the development of this Mueller matrix approach, some PIC codes started taking spin and polarisation of particles into account also using Stokes vectors, see e.g.~\cite{Li:2018fcz,Li:2019oxr,Li:2020bwo,Tang:2021azl} and further discussion in Sec.~\ref{sec:higher}. This is implemented in PIC codes in a very different way. The particles have some Stokes vectors and then at each time step random numbers are generated to determine whether a photon is emitted or a photon decays into a pair as well as how to update the Stokes vectors. This approach has already been used in several papers to study the generation of polarised particle beams, see Sec.~\ref{sec:higher}.
No comparison between the Mueller-matrix and the PIC approach has so far been made. This partly illustrates the fact that this is still an active research field where the formalism is still being developed.

\subsection{Numerical computation of the amplitude for momentum spectra}

\begin{figure}
    \centering
    \includegraphics[width=.5\linewidth]{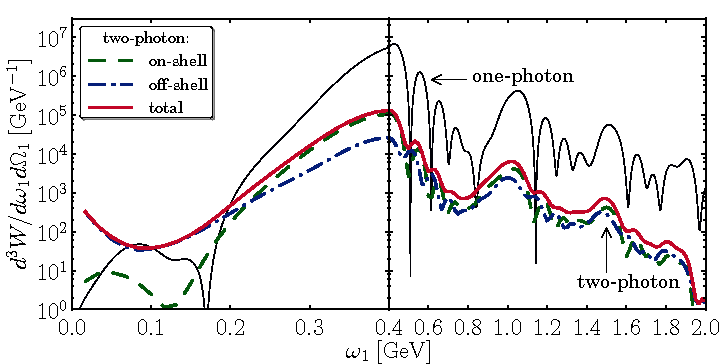}
    \caption{Double vs single Compton scattering as a function of the emitted photon frequency, where the other photon in the double Compton case has been integrated over. Figure taken from~\cite{Seipt:2012tn}.}
    \label{fig:doubleComptonSeiptFig}
\end{figure}

\begin{figure}
    \centering
    \includegraphics[width=0.24\linewidth]{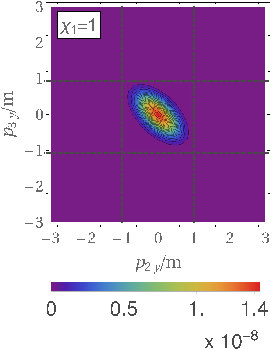}
    \includegraphics[width=0.24\linewidth]{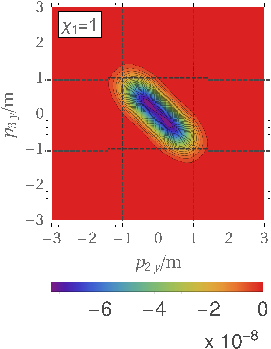}
    \includegraphics[width=0.24\linewidth]{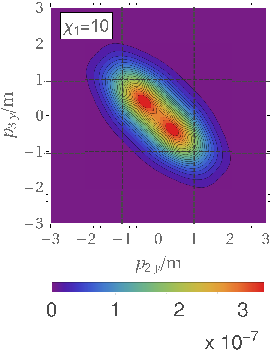}
    \includegraphics[width=0.24\linewidth]{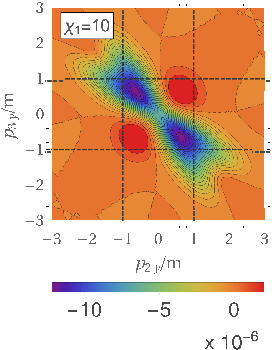}
    \caption{Transverse differential rate of the nonlinear trident process in a constant crossed field for the two-step process (first and third plots) and the one-step contribution (second and fourth plots). Taken from~\cite{King:2018ibi}.}
    \label{fig:KingTransverseTridentFigs}
\end{figure}

\begin{figure}
    \centering
    \includegraphics[width=.24\linewidth]{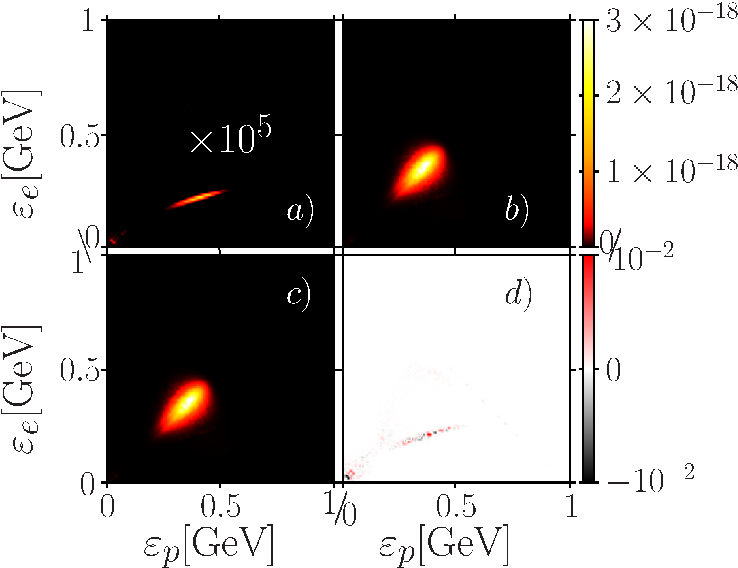}
    \includegraphics[width=.24\linewidth]{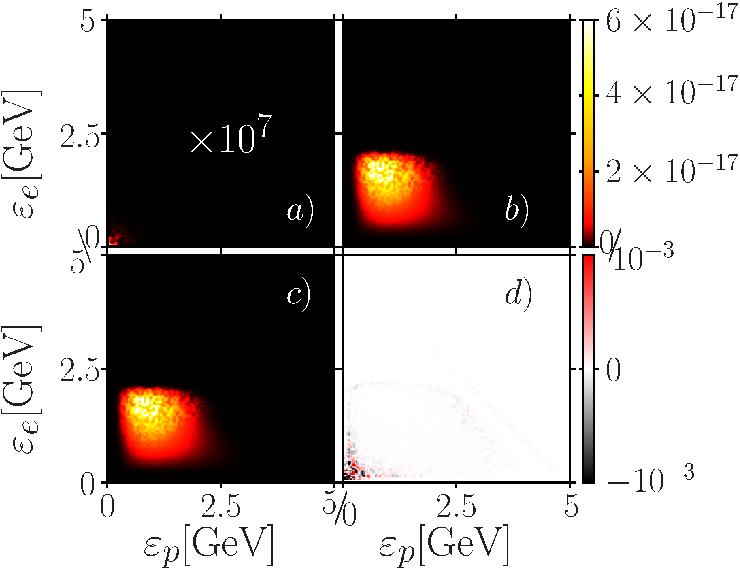}
    \includegraphics[width=.24\linewidth]{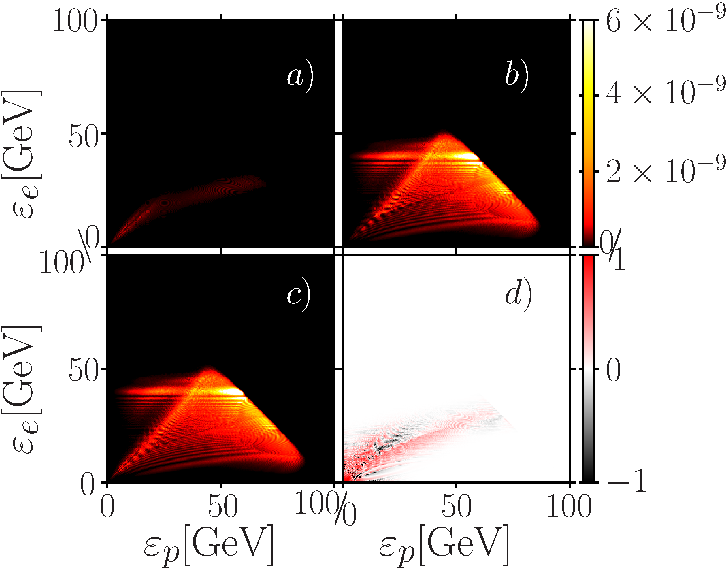}
    \includegraphics[width=.24\linewidth]{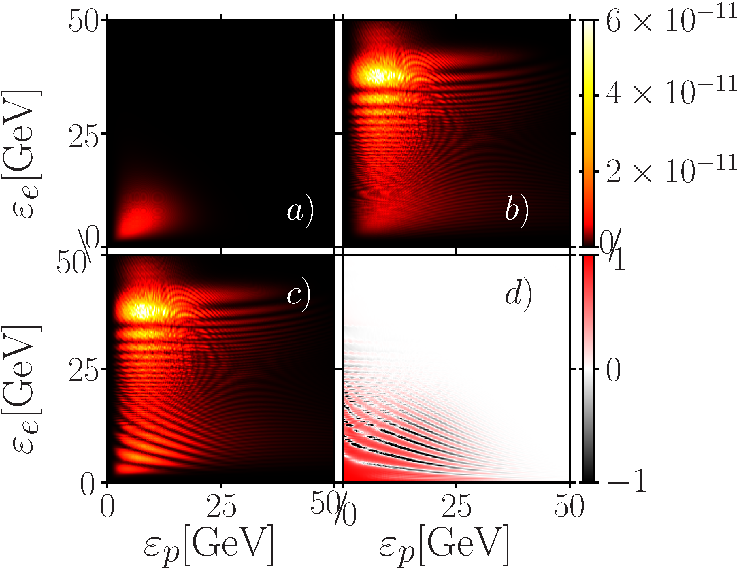}
    \caption{Trident spectrum for ($\xi=22$,$\chi=0.25$), ($\xi=50$,$\chi=3$), ($\xi=22$,$\chi=26$) and ($\xi=11$,$\chi=13$), from left to right. One-step and two-step contributions are shown in a) and b), and the full result in c). d) shows the relative error for the two-step approximation. Positron and electron energies are shown on the x and y axes.  Figure taken from~\cite{Mackenroth:2018smh}.}
    \label{fig:MacDiPiFigs}
\end{figure}

\begin{figure}
    \centering
    \includegraphics[width=.4\linewidth]{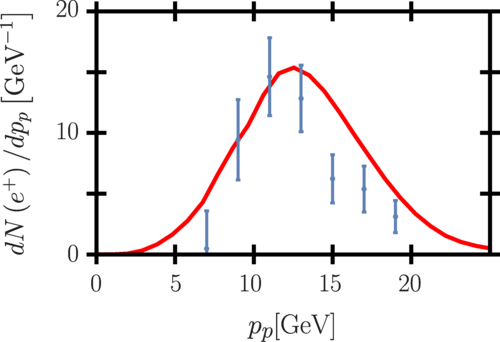}
    \caption{{Plot from~\cite{Mackenroth:2018smh} showing a comparison between theory and the experimental results from~\cite{Burke:1997ew} for the positron spectrum in the nonlinear trident process.
    }}
    \label{fig:MacDiPiSLAC}
\end{figure}

The discussion so far has been focused on separations that are useful as a guide for how to generalise methods and approximations at $\mathcal{O}(\alpha^2)$ to higher-order processes. However, the $\mathcal{O}(\alpha^2)$ processes have also been studied numerically without separations or approximations. In the studies mentioned so far, the amplitude was squared before performing the lightfront time integrals. One reason why this can be a good idea is that one can then perform the integrals over the transverse components of the momenta of external particles, because they are Gaussian, see e.g.~\cite{Dinu:2013hsd}. However, this means that, unless one is in a regime where one can approximate some of the $\varphi$ integrals, there are then twice as many $\varphi$ integrals than on the amplitude level. For example, the exchange term involves four $\varphi$ integrals that are intertwined in a complicated way. Even the exponential part of the integrand is complicated, which is an obstacle for using numerical methods that otherwise works well for the direct terms, where the simple exponentials appearing can be used as integration variables~\cite{Dinu:2013hsd,Dinu:2019wdw}, see also Sec.~\ref{sec:approx}. This is the reason why the exchange term is very difficult to compute, even for the simplest case of a constant field, and more so in a pulse.

This approach of performing the $\varphi$ integrals on the probability level is well suited for studying the longitudinal momentum spectra. One may wish to also study transverse spectra and hence not perform these Gaussian integrals. This has been done in~\cite{King:2013osa,King:2018ibi} in the LCF regime, see Fig.~\ref{fig:KingTransverseTridentFigs}.
However, to study the full momentum spectrum/phase space, it may be advantageous to perform the $\varphi$ integrals numerically on the amplitude level. This is the approach taken in~\cite{Seipt:2012tn,Mackenroth:2012rb} for double nonlinear Compton scattering, see Fig.~\ref{fig:doubleComptonSeiptFig}, and in~\cite{Mackenroth:2018smh} for trident, see Fig.~\ref{fig:MacDiPiFigs}. Then usually the $\varphi$ integrals are performed numerically along the real $\varphi$ axis, which require regularisation in a pulse ~\cite{Ilderton:2010wr,Seipt:2012tn,Mackenroth:2012rb,Mackenroth:2018smh} (see also~\cite{Ilderton:2020rgk}). 
Similar numerical integrals on the amplitude level have been studied in the first-order processes, see Sec.~\ref{sec:first:nlc:reg}.

In~\cite{Seipt:2012tn,Mackenroth:2012rb} this approach was used to numerically compare the differential probabilities of single and double nonlinear Compton scattering, where~\cite{Seipt:2012tn} focused on $\xi\sim1$ and~\cite{Mackenroth:2012rb} on $\xi\gg1$.
In~\cite{Seipt:2012tn} the covariant approach was used to calculate the Feynman diagrams, but it was shown that the fermion propagator for the intermediate electron state naturally separates into two terms with one of them being instantaneous in the sense of the lightfront Hamiltonian formalism. 
In \cite{Seipt:2012tn}, the probability was plotted as a function of the two photon frequencies and peaks in regions around Oleinik resonances were observed, where the intermediate particle goes on shell. 
The probability as a function of one of the photon frequencies after integrating over the other is shown in Fig.~\ref{fig:doubleComptonSeiptFig}.
In~\cite{Mackenroth:2012rb} the numerical spectra were interpreted using a saddle-point treatment of the $\varphi$ integrals (on the amplitude level), and the parts of phase space outside the emission cone for single Compton scattering were observed to be favourable for detecting double rather than single Compton scattering.

A connection between these two approaches was made in~\cite{Mackenroth:2018smh}, which studied the dominance of the two-step\footnote{{Note that the two-step was called the cascade channel, while the remainder of the amplitude was referred to as ``direct'' (whereas here we use ``direct'' for a part that includes the two-step or the cascade channel).}} in selected parts of the phase space (rather than for integrated probabilities) for different values of  $\xi$ and $\chi$. {The last two plots in Fig.~\ref{fig:MacDiPiFigs}, panel d) in particular, illustrate cases where the two-step alone is not enough. These are, in other words, parameter regions that one may want to study in order to be able to observe the one-step in experiments. In the first experimental observation of nonlinear trident in the SLAC-E144 experiment~\cite{Burke:1997ew}, the two-step gave the dominant contribution. The experimental result from~\cite{Burke:1997ew} was compared with theory in~\cite{Hu:2010ye,Mackenroth:2018smh,Titov:2021kbj}, see Fig.~\ref{fig:MacDiPiSLAC}. A two-step approximation was also used in~\cite{Titov:2021kbj} to numerically study the positron spectrum for energies and intensities typical for the upcoming experiments at LUXE~\cite{Abramowicz:2021zja} and FACET-II (E-320, at SLAC).}

Trident in a bichromatic plane wave pulse (comprising co-propagating optical and X-ray pulses) has been considered in~\cite{2015JPhCS.594a2024K}\footnote{Note that \cite{2015JPhCS.594a2024K} has a different notation to us, using instead ``direct" and ``exchange" to refer to the two terms in the \emph{amplitude}.}. The results were obtained by Fourier transforming the two currents ($\bar{\psi}\gamma^\mu\psi$), similar to the LCFA treatment in~\cite{King:2013osa,King:2018ibi}.

\subsection{High-energy limit}

As mentioned, the probabilities of higher-order processes can be approximated by incoherent products of first-order processes if $\xi$ is large or if the pulse is long. This does not simply mean that this approximation is good as long as $\xi\gg1$ (or $\mathcal{T}\gg1$). How large $\xi$ has to be depends on the other parameters of the process. For example, the LCFA can be applied to e.g. trident if $\xi$ is large compared to some function of $\chi$. Expressed in terms of $\eta=\chi/\xi$ instead of $\chi$, the condition for the applicability of the LCFA is (see e.g.~\cite{Ritus:1972nf,Torgrimsson:2020wlz}) in the high-energy regime:
$\xi\gg\eta^2\gg1$ and in the low-energy regime: $\xi^3\gg\frac{1}{\eta}\gg1$,
or for $\eta\sim1$ simply $\xi\gg1$ (see also \secref{sec:approx:lcfa} for a further discussion of the LCFA). This is an example of the fact that in general one can expect the LCFA to be valid if $\xi$ is much larger than any other large parameters.  
See also~\eqref{eqn:lcfawhen1} for a condition on the validity of the LCFA for Compton scattering. One expects, in particular, that the one-step part of trident may become important if $\xi$ is kept at a constant value and $\chi$ increases, even if the constant value of $\xi$ is large. This has been shown for the momentum spectrum in~\cite{Mackenroth:2018smh} and for a constant crossed field in \cite{King:2013osa}. In~\cite{Dinu:2019wdw} this was demonstrated for the longitudinal momentum spectrum, and it was shown for $\xi=1,2,4$ that the spectrum develops a sharp peak (where the initial electron keeps most of its momentum) as $\eta$ increases, and that this peak is due to the direct part of the one-step. In other words, the incoherent-product approximation breaks down at high energies, where the dominant contribution comes from the one-step rather than the two-step part~\cite{Dinu:2019wdw,Torgrimsson:2020wlz}.

One may wonder whether a Weizs\"acker-Williams (WW) approximation may be used instead. After all, such approximations are common in other high-energy processes without a background field. Such an approximation was in fact used in the SLAC experiment, since at the time there did not exist any complete treatment of trident (the SLAC experiment had $\xi<1$ so the constant-field results by Ritus~\cite{Ritus:1972nf} and Baier et al.~\cite{BaierTrident} could not be used). However, it was shown in~\cite{King:2013osa} that the WW approximation does not agree with the large $\chi$ limit of the constant-crossed field result, even though larger $\chi$ can be obtained by increasing the energy. It was shown in~\cite{Torgrimsson:2020wlz} that this is a consequence of the fact that the limit of large $\xi$ and then large $\eta$ does not commute with the limit of large $\eta$ and then large $\xi$ (the fact that these two limits do not commute has also been studied in other processes in~\cite{Podszus:2018hnz,Ilderton:2019kqp}). It is in the latter, high-energy limit that one can use the WW approximation. In other words, WW gives the correct result when $\eta$ is the largest parameter, while $\xi$ should be the largest parameter for the constant-field/LCF approximation (here by the LCF approximation we mean taking the large-$\xi$ limit before any other limit, which should not be confused with a large-$\xi$/LCF approximation of the WW approximation). 
But, in the $\eta\gg1$ limit the probability scales as $\prob\approx\text{const.}\ln\eta+\text{const.}$ and the WW approximation only gives the log term. Both terms can be obtained by treating the initial electron as a Coulomb field, i.e. in this limit trident agrees with pair production in the combination of a plane wave and a Coulomb field~\cite{Torgrimsson:2020wlz}. In other words, this limit is independent of the mass of the initial particle, and the high-energy limit agrees with the limit where the initial particle is treated as an infinitely massive particle, which is unaffected by the laser field and can effectively be treated as a Coulomb field. This also happens for perturbative trident~\cite{Borsellino,1959PhRv..115..672S,1976tper.book.....J}.
However, having $\eta$ much larger than $\xi>1$ will probably not be particularly relevant for upcoming experiments at e.g. FACET-II and LUXE. In other words, other approximations will probably be of more immediate practical use. 

In the high-energy limit of trident the initial electron emits the intermediate (off-shell) photon effectively without interaction with the laser~\cite{Torgrimsson:2020wlz}, i.e. the Compton vertex is effectively free/not dressed by the external field. This can be compared with the high-energy limit of the probabilities of the first-order processes, which agree with the leading order in a perturbative $\xi\ll1$ expansion of the field~\cite{Podszus:2018hnz,Ilderton:2019kqp}. In trident, the Breit-Wheeler vertex still depends nontrivially on $\xi$.

\subsection{Low-energy approximations}

For sufficiently small\footnote{As always, for problems with more than one parameter, the condition for how small $\chi$ has to be depends on the other parameters. Note that $\xi$ does not have to be large. These methods works even if $\xi\sim1$, provided $\chi$ is small enough. However, if $\chi$ is small but fixed, then the saddle-point approximation eventually breaks down if one decreases $\xi$~\cite{Dinu:2017uoj}.} $\chi$ one can use the saddle-point method (see Sec.~\ref{Saddle-point methods} for details) to calculate all terms of the probability~\cite{Dinu:2017uoj,Dinu:2018efz,Dinu:2019pau,Torgrimsson:2020wlz,Torgrimsson:2020mto}. This can be a very powerful method for calculating complicated one-step terms. The main step is to find the saddle point for the $\varphi$ integrals (there are in general four of them for the one-step probability) and, if one is interested in the total probability rather than the spectrum, the two independent longitudinal momentum variables. The exponential factor in the exchange part is a nontrivial function of the $\varphi$ variables, so the saddle-point equation for this term is also complicated. However, one can often find the saddle points by making educated guesses. After finding the saddle point, one just has to make an appropriate change of variables, e.g. $\varphi\to\varphi_{\rm saddle}+\sqrt{\chi}\delta\varphi$, and then expand the integrand in a series in $\chi$ (see Sec.~\ref{Saddle-point methods}). This leads to elementary Gaussian integrals or integrals of the form of a Gamma function: $\Gamma(n+1)=\int_0^\infty\ud x\; x^n e^{-x}$. This allows one to calculate even the exchange term, which has in general a very big and complicated integrand. Prior to~\cite{Dinu:2017uoj,King:2018ibi} there were no analytical results for the exchange term, not even for a constant-crossed field. With the saddle-point method~\cite{Dinu:2017uoj}, though, it was possible to obtain $\chi<1$ approximations for both constant fields as well as inhomogeneous fields. All terms share the same exponential part. For example, for trident we have e.g.~\cite{Dinu:2017uoj} (see~\cite{Dinu:2017uoj} for the prefactor, and compare with the well-known exponential scaling of NBW~\eqref{eq:first:NBWrate})
\be\label{tridentExpContant}
\prob\sim\exp\left\{-\frac{16}{3\chi}\right\}
\ee
for a (locally) constant field (the direct part was calculated in~\cite{Ritus:1972nf}), or
\be\label{tridentExpSin}
\prob\sim\exp\left\{-\frac{4\xi}{\chi}\left([2+\xi^2]\text{arcsinh}\left[\frac{1}{\xi}\right]-\sqrt{1+\xi^2}\right)\right\}
\ee
for a linearly polarised sinusoidal field, $a(\varphi)=m\xi\sin\varphi$. For the pre-exponential part of the probability one finds a power series in $\chi$. To leading order it turns out that~\cite{Dinu:2017uoj}
\be
\text{trident:}\quad \prob_{\rm one}^{\rm ex}\approx\frac{13}{18}\prob_{\rm one}^{\rm dir} \;,
\ee
so the exchange part is on the same order of magnitude as the direct part of the one-step.  

After having obtained saddle points for trident, one can reuse most of them directly also for double Compton scattering~\cite{Dinu:2018efz} and photon trident~\cite{Torgrimsson:2020mto}. For double Compton it turns out that the exchange term is even more important, because~\cite{Dinu:2018efz}
\be
\text{double Compton:}\quad \prob_{\rm one}^{\rm ex}\approx-\prob_{\rm one}^{\rm dir}
\ee
to leading order in $\chi$. Hence, for this process the exchange term even cancels the direct part of the one-step to leading order. 

Thus, terms that were neglected for decades turn out to be crucial. In Sec.~\ref{sec:LBL} we will discuss a similar situation for reducible diagrams, which were assumed to vanish for a long time but were shown just recently to be nonzero~\cite{Gies:2016yaa}.   

It was realised in the SLAC experiment that in certain regimes one can expect the trident probability to have a Schwinger-pair-production-like exponential scaling. However as already remarked, at that time no such results for trident with $\xi\sim1$ existed (\eqref{tridentExpContant} works for $\xi\gg1$). Instead of~\eqref{tridentExpSin}, a somewhat similar exponential was used, which was obtained from a result for Schwinger pair production in a sinusoidal electric field. However, also in upcoming experiments it is relevant to know whether one observes e.g. perturbative particle production or non-perturbative/Schwinger-like production. Results such as~\eqref{tridentExpSin} can then be used to distinguish these different regimes.

Note that the exponential part of the integrands of these processes is a pure phase for real integration variables, so these saddle points are necessarily complex, since the probability should scale as $e^{-(\text{real and positive})/\chi}$ (with some terms potentially multiplied by some oscillating terms). One can also understand this by noting that $\chi\ll1$ takes us to the semi-classical regime, but these processes are classically forbidden, which means complex rather than real saddle points.

\subsection{Resummation of $\chi$ expansions}\label{resummation of chi expansions}

The power series in $\chi$ in the pre-exponential part of the probability are in general asymptotic. One would in general only expect the leading order to give a high precision for $\chi\ll1$, i.e. in a regime where the probability is very small due to the exponential suppression. It is therefore natural to ask whether one can obtain a higher precision also at larger $\chi\lesssim1$ by including more terms from the $\chi$ expansion. The asymptotic nature of these expansions, though, means that a simple direct summation of the power series does not work\footnote{Of course, for smaller $\chi$ one can improve the approximation by summing the first couple of terms~\cite{Dinu:2017uoj}.}. However, at least the first $\sim10$ terms in these expansions are usually straightforward to obtain, because after finding the saddle points and a suitable change of integration variables (e.g. from $\varphi=\varphi_{\rm saddle}+\sqrt{\chi}\delta\varphi$ to $\delta\varphi$, where $\sqrt{\chi}$ is included so that $\chi$ drops out of the quadratic terms in the exponent, $(\varphi-\varphi_{\rm saddle})^2/\chi$, i.e. so that the integrand can be expanded in powers of $\chi$), the rest of the calculation is ``simply'' performed using a symbolic computation program~\cite{Dinu:2017uoj,Torgrimsson:2020wlz,Torgrimsson:2020mto}. Contrast this with e.g. the high-energy expansion, for which one may need to find different points to expand around and/or a different change of integration variables for each order~\cite{Torgrimsson:2020wlz}. It would therefore be unfortunate if one would not be able to use the first terms in these asymptotic $\chi\ll1$ series. Fortunately this is possible thanks to resummation methods.

As a typical example, consider for simplicity a constant field. Then the probability of e.g. nonlinear Breit-Wheeler or trident can be expanded as
\be\label{expTimesPowerSeries}
\prob=\chi^d e^{-\text{const.}/\chi}\sum_{n=0}^\infty c_n\chi^n \;,
\ee
where $d$ and $c_n$ are some constants.  The series coefficients $c_n$ grow factorially as $n\to\infty$, i.e. these series have zero radius of convergence and therefore has to be resummed. Note that the particular transseries structure in~\eqref{expTimesPowerSeries} is not an ansatz or a guess, it is simply what one finds by applying the saddle-point method. In particular, there cannot be any terms without an exponential $e^{-\text{const.}/\chi}$ because we are expanding around a point where the probability is exponentially suppressed.    

Before describing how these expansions have actually been resummed, for the reader familiar with the Stokes phenomenon, see Sec.~\ref{sec:Schwinger}, it might be useful to recall Berry's treatment of it~\cite{Barry:1989zz}. In that case one also has an asymptotic series times an exponential. He took the late terms starting with the smallest summand, and then approximated these terms with their asymptotic limit, which gives an analytical expression for each $c_n$ for arbitrary large $n$. This means that the Borel transform of this estimate of the late part is also given by analytical expressions for arbitrary large $n$ and can therefore be resummed by recognizing it as the expansion of a simple function.

However, in general, for the $\chi$ expansions as well as e.g. the $\alpha$ expansion, one will not find a series which can be recognised as the expansion of some known/simple function. Hence, the goal is to obtain high precision resummations when only a finite number of terms are available. It has become standard to resum truncated asymptotic series (the ones studied are usually not multiplied by an exponential factor such as $e^{-\text{const.}/\chi}$ though) by resumming the corresponding truncated Borel transform using Pad\'e approximants and/or conformal mappings~\cite{PhysRev.124.768,CiulliFischer,LeGuillou:1979ixc,BenderOrszagBook,Kleinert:2001ax,ZinnJustinBook,Caliceti:2007ra,Costin:2019xql,Costin:2020hwg}, which have been recently used for strong-field QED in~\cite{Florio:2019hzn,Dunne:2021acr,Torgrimsson:2020wlz,Torgrimsson:2020mto,Ekman:2021eqc}. 
To use the Borel-Pad\'e method to resum a factorially divergent series $\sum_{n=0}^\infty a_m z^m$ ($a_m\sim m!$ at large $m$) one follows these steps: 1) Calculate the (truncated) Borel transform: 
\be
B_M(t)=\sum_{m=0}^M\frac{1}{m!}a_m t^m \;,
\ee
where $M$ is the number of terms in the original $z$ expansion that one has access to.
2) The full Borel transform has a finite radius of convergence and can therefore be resummed with a Pad\'e approximant by demanding 
\be
PB[I/J](t)=\frac{\sum_{i=0}^{I}c_i t^i}{1+\sum_{j=1}^J d_j t^j}=B_N(t)+\mathcal{O}(t^{N+1}) \;.
\ee
It is often a good idea to choose diagonal or near-diagonal Pad\'e approximants, with $I=J$ or $I=J\pm1$.
3) The last step is to do an inverse Borel transform, i.e. perform a Laplace integral 
\be
\int_0^\infty\frac{\ud t}{z}e^{-t/z}PB[I/J](t) \;,
\ee
which gives a finite function of $z$.

It was shown in~\cite{Torgrimsson:2020wlz} that for nonlinear Breit-Wheeler pair production or the two-step part of trident one can obtain a large number of terms and then Borel-Pad\'e (-conformal) resummation gives a very good precision up to very large $\chi$, in fact into the regime where $\alpha\chi^{2/3}>1$ where the result anyway ceases to be relevant. In other words, even though the input data comes from a $\chi\ll1$ expansion, the resummation is good for any practical value of $\chi$. 

However, it can be difficult to obtain a large number of terms, so it is very useful to try to find better resummation methods which only need few terms. 
There are various resummation methods that already exist. For example, in~\cite{Torgrimsson:2020wlz} it was shown that for the particular examples considered there, a new resummation method from~\cite{Mera:2018qte}, based on Meijer-G functions, allowed for a high precision using relatively few terms. In general, one may expect to find better resummation methods if one has access to some extra information in addition to the asymptotic series coefficients. For example, when conformal mappings are used in combination with Borel and Pad\'e one uses as extra information the position of the singularity in the Borel transform that is closest to the origin, i.e. the convergence limiting singularity. However, this information is usually obtained from the asymptotic series coefficients.

Another new resummation was introduced in~\cite{Alvarez:2017sza}, which uses the leading scaling for large argument (large $\chi$ for us) as additional information. It was shown in~\cite{Torgrimsson:2020mto,Torgrimsson:2020wlz} that this resummation works very well for trident, double Compton and photon trident, including the complicated exchange term. This is a resummation that works for many different cases. Moreover, it was shown in~\cite{Ekman:2021eqc} that this resummation method is also very useful for resumming the $\alpha$ expansion of LAD (we will return to resummations of RR in Sec.~\ref{sec:higher}).

In some cases one has access to more than just the leading scaling at large $\chi$. In~\cite{Torgrimsson:2020mto} it was shown that for the longitudinal momentum spectrum in trident, double Compton and photon trident, one can obtain many terms in the $\chi\gg1$ expansion. Since the $\chi\gg1$ expansions are not asymptotic, this therefore provides another way to obtain high precision results by just a direct summation (at least within some radius of convergence). However, knowing the first couple of terms in the $\chi\ll1$ as well as the first couple of terms in the $\chi\gg1$ expansion, means that one can obtain precise resummations with very few terms. Indeed, just knowing that the $\chi\gg1$ expansion is given by a series in $1/\chi^{2/3}$ with integer powers is very useful. This is because by knowing the structure of both the $\chi\ll1$ and the $\chi\gg1$ expansions, one can find a set of special functions that have the same structure that can be used as basis functions for the resummation. Note that a Borel-Pad\'e resummation, or even the resummation in~\cite{Alvarez:2017sza} which uses the leading scaling for large argument, does in general not give a function with a correct large argument expansion; for the $\chi$ expansions this would mean that both the coefficients and the powers of $1/\chi$ would be wrong. Thus, if one can find a set of special functions with the correct structure for both the $\chi\ll1$ and the $\chi\gg1$ expansions, then one can expect to find precise resummation with relatively few terms. In~\cite{Torgrimsson:2020mto} this was done using Meijer-G or Fox-H functions (not the same type of Meijer-G as in~\cite{Mera:2018qte}) or sums of products of Airy functions.   

The fact that this works for the complicated exchange term, which is nontrivial to compute with a direct numerical approach, shows that resummation methods are a useful alternative to fully numerical approaches. In any case, it is always good to have more than one way of calculating the results. We will come back to these types of resummation methods in Sec.~\ref{sec:higher}.


\section{Approximation Frameworks}\label{sec:approx}
There are various approaches to obtaining the total probability for a given process to occur in an intense background field. Whatever method is used, one is faced with performing numerically expensive integrals over highly oscillating functions. These originate from complicated spacetime dependencies in the intense background, which are not present in vacuum. Computation of such integrals can be approximated directly, or, particularly for plane-wave backgrounds, spacetime-dependent rates can be inferred and used in Monte Carlo generators employed in numerical simulation codes.  The main part of this chapter will concentrate on approximations used for processes in plane-wave backgrounds; the final section will review approaches to calculating the Schwinger effect and non plane-wave backgrounds. Whilst rates are often applied in numerical simulations of plane-wave processes, we will focus on how these rates are derived from QED. First, we will consider developments in understanding and extending these rates, then we will discuss approaches used in evaluation of the integrals appearing in QED: using asymptotic methods and using numerical methods.

Although the scattering approach provides no information about real-time quantities, nevertheless it is common to define a rate,
$\rate$, satisfying the two conditions: i) the total probability is obtained by integrating the rate over some evolution variable, e.g. the phase\footnote{Typically in numerical simulations, the rate is defined as a probability per unit co-ordinate time, but our description here will focus on plane-wave backgrounds (which are useful approximations to processes in laser pulses), for which a natural covariant definition is in terms of probability per unit lightfront time (equivalently: plane-wave/laser phase, $\phi$).} $\phi$. $\prob = \int {\rm d}\phi~ \rate(\phi)$ and ii) a positivity constraint: $\rate \geq 0$.  

The rate contains integrals over outgoing particle momenta. For a typical tree-level, one-vertex `$1\to2$' strong-field QED process, $\rate$ can be reduced to a single integral over a lightfront momentum (or energy), which is the usual form employed in numerical simulations, (for a recent review of simulational approaches, see \cite{Gonoskov:2021hwf}). 
In this case, simulations sample the lightfront momentum distribution of the QED process, but generally make some extra assumptions like forward emission of particles, which neglects the transverse momentum spectrum (which is often not measurable, e.g. for high energy probes).  However, recent work has investigated more exclusive forms of the rate, in which other variables, such as the angle of emission \cite{Blackburn:2019lgk}, are left in the final form of $\rate$. When employed in simulations, this allows for a more complete resolution of the produced particle spectra.

\subsection{Locally constant field approximation}
\label{sec:approx:lcfa}
Although various approximation schemes have been considered for QED processes in intense backgrounds, the most widely employed is the locally constant field approximation (LCFA): this approximates the probability for a process in an inhomogeneous background by taking the \emph{rate} of the process occurring in a constant crossed field, and integrating it over the field traversed by a point-like probe\footnote{The zeroth order derivative approximation to the Heisenberg-Euler effective Lagrangian is also sometimes referred to as the locally constant field approximation \cite{Gavrilov:2016tuq}.}. This approximation is sometimes referred to as the ``constant crossed field approximation'' and for nonlinear Compton scattering, it is equivalent to the ``synchrotron approximation'' \cite{PhysRevLett.89.094801} or the ``Sokolov-Ternov radiation formula'' \cite{Chen:1988ec}. A similar type of approximation is found in other fields, such as in the calculation of bremsstrahlung \cite{Baier:1990pi}, in beamstrahlung \cite{Yokoya:1991qz}, exotic astrophysical objects such as magnetars \cite{Cruz:2020vfm} and in collisions of high energy charged particles with oriented crystals \cite{Wistisen:2019eza}.

The LCFA involves an integral over a constant crossed field, which is the zero-frequency limit of a plane wave. In a plane-wave background, the semi-classical approach yields the exact solution \cite{Wolkow:1935zz}.
Formally, a semi-classical approximation can be  defined as some expansion in $\hbar$ (see \secref{sec:Schwinger:semiclassical}); practically, such an $\hbar$ expansion is achieved by, e.g. a WKB expansion (see \secref{sec:AI-WKB}), replacing quantum operators by their classical values (see \secref{sec:approx:BKmethod}) or expanding with respect to some variable proportional to $\hbar$ e.g. $\chi$ (see \secref{sec:approx:saddle}).
Calculations in plane wave backgrounds explicitly exhibit a dependence on the classical kinematic momentum (solution of the Lorentz force equation). Therefore, when the LCFA is a good approximation, the interaction of charges with a given EM background can be calculated by using a hybrid approach: step-wise solving the classical dynamics of point-like particles and the background and overlaying Monte-Carlo sampling of the LCFA rate for strong-field QED processes \cite{Gonoskov:2021hwf}. In simulations, it is assumed that chains of first-order processes, between which intermediate particles follow a classical trajectory, are a good approximation to higher-order processes (the justification for this is that the pulses are sufficiently long, that the contribution of coherent processes can be neglected see \secref{sec:second}). The main differences to purely classical propagation are: i) photon emission from a charge (or photon decay to a pair) is \emph{stochastic}~\cite{2018PhRvE..97d3209N}, which allows for e.g. deeper penetration of pulses that would be achieved purely classically (`straggling' \cite{Blackburn:2014cig}) or a suppression of the rate of emitted radiation in very short pulses (`quenching' \cite{Harvey:2016uiy}); ii) the classical trajectory for charges is `corrected' by including the recoil from each Compton emission \cite{Burton:2014wsa,Blackburn:2019rfv,Gonoskov:2021hwf}. The LCFA is therefore a versatile method for including strong-field QED processes when the background field is unknown \emph{a priori}, for example in situations involving a large number of charges such as in intense laser-plasma collisions, and therefore finds widespread use in laser-plasma particle-in-cell (PIC) codes \cite{Gonoskov:2021hwf}. 
The increase in the usage and capability of these codes has led to the LCFA recently being analysed in more detail.

In this section, we outline the arguments for how the LCFA can be obtained for a general background. This is typically a two-step process: \begin{enumerate}[label=\roman*)]
    \item one must make some assumptions about the invariants so that the process in a general background can be well-approximated as occurring in a plane wave; 
    \item in that plane wave, some conditions must be fulfilled such that the probability can be well-approximated by an integral over a local rate.
\end{enumerate}
We begin by outlining an argument for i), by building upon similar reasoning in \cite{Ritus1985}.

Consider a probe particle undergoing some process in a general background field. There are three dimensionless invariants we can define immediately, without reference to the type of background: $\chi$, 
$\bar{\mathfrak{E}}=\mathfrak{E}/\Eschwinger$ and $\bar{\mathfrak{B}}=\mathfrak{B}/\Eschwinger$, where the secular invariants are:{
\be 
\mathfrak{E}=\sqrt{\sqrt{\mathcal{S}^{2}+\mathcal{P}^{2}}+\mathcal{S}}; \quad \mathfrak{B}=\sqrt{\sqrt{\mathcal{S}^{2}+\mathcal{P}^{2}}-\mathcal{S}}, \label{eqn:approx:sec}
\ee}
(the EM invariants $\mathcal{S}$ and $\mathcal{P}$ are defined in \eqnref{eq:invariants}). 
Then we can write the probability, $\prob$, for the process, as $\prob = \prob(\cdots, \chi, \bar{\mathfrak{E}},\bar{\mathfrak{B}})$ where `$\cdots$' indicates the dependency on other invariants related to the background and the probe momentum.  For the LCFA to be applicable, we must assume that $\bar{\mathfrak{E}}, \bar{\mathfrak{B}} \ll 1$ ($\bar{\mathfrak{E}}, \bar{\mathfrak{B}}$ are identically zero in a plane wave). We note that this condition is just on the background field: it is independent of the probe particle's momentum. Another necessary assumption to use the LCFA in a general background, is that the particle is ultra-relativistic such that the set of remaining invariants can be well-described by a single, plane-wave invariant, which we choose to be $\eta$. The invariant $\eta$ is the scalar product of the probe's momentum and background field's Fourier momentum in units of the particle mass squared. (One physical interpretation of $\eta$ is when the probe particle is a photon, $\eta=k\cdot k'/m^{2}$, which is twice the centre-of-mass energy squared ($4\,k\cdot k'$) in units of the pair-creation threshold ($4\,m^{2}$); i.e. $\eta \geq 2$ for linear Breit-Wheeler to proceed.) Then assuming the field invariants can be ignored if sufficiently small, the probability can be written as $\prob\approx\prob(0,\cdots ,0,\eta,\chi,0,0)$, where all other invariants have been set to zero, which is the form of a process occurring in a plane-wave.

To illustrate these arguments, consider the example of a probe electron in a standing wave formed from two counter-propagating, monochromatic, circularly polarised plane waves of the form $a_{j}=m\xi_{j}\left(\eps\cos k_{j}\cdot x+ \beta \sin k_{j}\cdot x\right)$ (where $j \in \{1,2\}$) and $a=a_{1}+a_{2}$ with $k_{j}^{2}=0$, $k_{j}\cdot\eps=k_{j}\cdot \beta=\eps\cdot\beta=0$ and $k_{1}\cdot k_{2}\neq 0$. (For simplicity, consider the lab frame when $\xi_{1}=\xi_{2}$, and $\mbf{k}_{1}=-\mbf{k}_{2}$.) Then the probability in this background can be written: $\prob=\prob(\eta_{1},\chi_{1},\eta_{2},\chi_{2},\bar{\mathfrak{E}}, \bar{\mathfrak{B}})$, where $\eta_{j}=k_{j}\cdot p / m^{2}$ and $\chi_{j} = \xi_{j}\eta_{j}$. For simplicity, we choose the electron's momentum, $p$, to satisfy $p\cdot \eps = p\cdot \beta = 0$, then $\chi=\chi_{1}+\chi_{2}$. Suppose the electron's momentum is such that $\eta_{1}>\eta_{2}$, $\chi_{1}>\chi_{2}$. In this example, the argument that the electron `sees a plane wave' when it is sufficiently high energy, is the condition $\eta_{1}\gg \eta_{2}$, $\chi_{1} \gg \chi_{2}$. Then since we have demanded $\bar{\mathfrak{E}},\bar{\mathfrak{B}}\ll 1$ (we note that here, $\bar{\mathfrak{E}}=0,\bar{\mathfrak{B}}\propto|k_{1}\cdot k_{2}/m^{2}|$), we can write the probability as $\prob\approx\prob(\eta,\chi,0,0,0,0)$, where $\eta=\eta_{1}$ and $\chi=\chi_{1}$. Hence the probability is approximately the form of a plane-wave probability.

In order to define conditions that must be fulfilled such that the plane-wave probability can be well-approximated by a `local' rate, we consider the specific case of a one-vertex first-order $1\to 2$ process. Suppose the plane-wave background is described by the potential $a(\phi)=m\sum_{j=1}^{2}\xi_{j} \Psi_{j}(\phi)\varepsilon_{j}$, where $\varepsilon_{i}\cdot\varepsilon_{j}=-\delta_{ij}$. Suppose we simplify the situation and choose the plane-wave to be either linearly or circularly polarised so that we only have one intensity parameter, $\xi$. Then we start from the total probability before any non-trivial outgoing momentum integrals have been performed (i.e. following \secref{sec:first:nlc}, mod-squaring the matrix element, such as \eqnref{eq:first:MNLC} for nonlinear Compton, and integrating over one outgoing particle's momentum using the phase-space delta-function that conserves momentum in the form \eqnref{eq:emc-lf}). This can be written as:
\be
\prob = \frac{\alpha}{(2\pi\eta)^{2}}\int {\rm d}\vphi\,{\rm d}\vphi'\,{\rm d}s\,{\rm d}^{2}\bs{\rho}^{\perp}~\frac{s}{1-s}\,\bar{F}(\xi,s,\bs{\rho}^{\perp},\vphi,\vphi')~\exp\left[\frac{i u(s)}{2\eta}\bar{G}(\xi,\bs{\rho}^{\perp},\vphi,\vphi')\right], \label{eqn:PaI0}
\ee
where $\eta$ is the energy parameter of the incident probe particle and where, for example for nonlinear Compton:
\be
\bar{F}=-\Delta(\vphi)\Delta(\vphi')+\frac{g(s)}{2}\left[a^{2}(\vphi)\Delta(\vphi')+a^{2}(\vphi')\Delta(\vphi)-a(\vphi)\cdot a(\vphi')\right]
\ee 
\be
\bar{G} = \int_{\vphi'}^{\vphi} {\rm d}x\left[1+\left(\boldsymbol{\rho}^{\perp}+\frac{\mbf{a}(x)}{m}\right)^{2}\right]; \quad
\boldsymbol{\rho}^{\perp}=\frac{\pmb{\ell}^{\perp}-s\mbf{p}^{\perp}}{ms};
\quad \Delta(\vphi) = 1-\frac{\ell\cdot\pi_{p}(\vphi)
}{\ell\cdot p},
\ee
$g(s)=g_{\tiny \textsf{NLC}}(s) = 1+s^{2}/[2(1-s)]$ and $u(s)=u_{\tiny \textsf{NLC}}(s)=s/(1-s)$ (for nonlinear Breit-Wheeler: $g(s)=g_{\tiny \textsf{NBW}}(s) = 1-u_{\tiny \textsf{NBW}}/2$, $u(s)=u_{\tiny \textsf{NBW}}(s)=1/[s(1-s)]$). In order to progress to a local expansion, one defines `average phase' $\phi=(\vphi+\vphi')/2$ and `interference phase' variables $\theta = \vphi-\vphi'$ (which we recall from \eqnref{eqn:phidefs1}, and whose relation is sketched in \figref{fig:lcfafig1}.). 
It is possible at this point to derive an angularly-resolved LCFA, by expanding in $\theta$, retaining lowest-order terms and integrating out $\theta$ to obtain Airy functions see \cite{Blackburn:2019lgk}. However, for clarity of exposition, we will concentrate on the more standard lightfront momentum spectrum.
After integrating out the transverse momentum variables, $\bs{\rho}^{\perp}$, which appear in the integrand in the form of a Gaussian (as explained in the text around \eqnref{eq:first:rpGauss1}) one arrives at the probability as an integral over the lightfront momentum spectrum:
\be
\prob = \frac{\alpha}{\eta}\int {\rm d}\phi\,{\rm d}\theta\,{\rm d}s~f(\phi,\theta,\xi,s)~\exp\left[\frac{iu(s)}{2\eta}\theta \mu (\xi,\theta,\phi)\right], \label{eqn:PaI1}
\ee 
where, for nonlinear Compton $s=s_{\tiny \textsf{NLC}}=k\cdot \ell/k\cdot p=\eta_{\gamma}/\eta_{e^{-}}$, and for nonlinear Breit-Wheeler $s=s_{\tiny \textsf{NBW}}=k\cdot q/k\cdot \ell=\eta_{e^{+}}/\eta_{\gamma}$ and where $\eta_{\gamma}$ is the energy parameter of the photon and $\eta_{e^{\pm}}$ for the positron/electron respectively.

For unpolarised nonlinear Compton, the pre-exponent $f$, can be written in the form $f_{\tiny \textsf{NLC}}$ where (see e.g. \cite{Dinu:2013hsd,Seipt:2020diz}): 
\be 
f_{\tiny \textsf{NLC}} = \frac{i}{8\pi\theta}\left\{1+\frac{g_{\tiny \textsf{NLC}}(s)}{2}\left[\mbf{a}\left(\phi+\frac{\theta}{2}\right)-\mbf{a}\left(\phi-\frac{\theta}{2}\right)\right]^{2}\right\}, \label{eqn:fnlc1}
\ee 
(the equivalent expression for nonlinear Breit-Wheeler, $f_{\tiny \textsf{NBW}}$, can be obtained by substituting $g_{\tiny \textsf{NLC}}(s)$ with the factor $g_{\tiny \textsf{NBW}}(s)$).

The quantity $\mu(\xi,\theta,\phi)$ is the Kibble mass squared, which we recap here from \eqnref{eq:first:kibble} in Sec. \ref{sec:first:nlc}:
\be
\mu(\xi,\theta,\phi) = 1 - \frac{\langle a^{2} \rangle}{m^{2}} + \frac{\langle a \rangle^{2}}{m^{2}}; \qquad \langle f \rangle := \frac{1}{\theta}\int^{\phi+\theta/2}_{\phi-\theta/2} f(x) {\rm d}x,
\label{eqn:approx:kibble}
\ee
where $\langle f \rangle$ is a phase window average. The Kibble mass
encapsulates interference effects in the amplitude for the process to occur in different parts of the plane wave, which is responsible for determining structure in the outgoing particle momentum spectrum. To interpret the probability as the integral over a `rate', it is desirable that the interference phase integral be evaluated. This can in general only be done in approximate form. The range of $\theta$ that must be integrated over, in order that the integral be well-approximated, is sometimes referred to as the ``formation length'' of a process (see e.g. \cite{Ritus1985,baier98,Baier:2003hf}). If one chooses to neglect interference effects at the scale of the wavelength and larger (e.g. assuming the process has a sub-wavelength formation length), a Taylor expansion of exponent and pre-exponent in $\theta$ can be performed \cite{Schwinger:1949ym,Baier1968}. This can be justified when $\xi\gg1$, where the dominant contribution to the $\theta$ integral originates from the region $|\theta|<1/\xi$ \cite{Ritus1985,baier98}, justifying such an expansion in this parameter regime. In this case the substitution $\theta\to \bar{\theta}/\xi$ can be performed, where the contributing region of $\bar{\theta}$ is at most $\bar{\theta}\sim {\mathcal O}(1)$. Expanding the exponent from \eqnref{eqn:PaI1} in $\bar{\theta}/\xi$ gives:
\begin{align}
\frac{h(s)}{\eta}\frac{\bar{\theta}}{\xi} \mu \left(\xi,\frac{\bar{\theta}}{\xi},\phi\right) = \frac{h(s)}{\chi}\left[ \bar{\theta}   + \frac{\Psi_{j}'(\phi)\Psi_{j}'(\phi)}{12}\bar{\theta}^{3}+ \frac{\Psi''_{j}(\phi)\Psi''_{j}(\phi)+3\, \Psi'_{j}(\phi)\Psi_{j}'''(\phi)}{720 }\bar{\theta}^{3}\left(\frac{\bar{\theta}}{\xi}\right)^{2} + \ldots 
\right] \label{eqn:KibExp1}
\end{align}
(the index $j$ is summed over, where repeated).
Performing the same expansion in the pre-exponent, and neglecting terms of order $(\bar{\theta}/\xi)^{2}$ in both the exponent and pre-exponent, gives typical Airy function kernels. The integration in $\theta$ then generates $\Ai'$ and $\Ai_{1}$ functions where $\Ai_{1}(x) = \int_{x}^{\infty}\Ai(y){\rm d}y$ (sometimes the Airy functions are written as modified Bessel functions, $\trm{K}_{n}(\cdot)$, using $\Ai'(x) = -(\pi\sqrt{3})^{-1}x\,\trm{K}_{2/3}[(2/3)x^{3/2}]$, and $\Ai_{1}(x)=(\pi\sqrt{3})^{-1}\int_{x}^{\infty} \sqrt{y}\,\trm{K}_{1/3}[(2/3)y^{3/2}]{\rm d}y$ \cite{vallee2010airy}), from which it follows the LCFA rate can be written as:
\be
\rate_{\tsfs{LCFA}} = -\frac{\alpha}{\eta} \int {\rm d}s \left[\bar{c}\,\Ai_{1}(z) + \frac{2g(s)}{z}\Ai'(z)\right].
\label{eqn:lcfaQED}
\ee
We note that this result can also be acquired by simply expanding the total probability in $1/\xi$ and keeping the leading order term.
For QED processes, $0<s<1$. For unpolarised nonlinear Compton scattering, $z=[s/\chi(1-s)]^{2/3}$, $\bar{c}=1$ and $g(s)=g_{\tiny \textsf{NLC}}(s)=1+s^{2}/[2(1-s)]$, for nonlinear Breit-Wheeler pair-creation, $z =[\chi s(1-s)]^{-2/3}$, $\bar{c}=-1$ and $g(s)=g_{\tiny \textsf{NBW}}(s)=-1+[2s(1-s)]^{-1}$. 
(For the classical process of nonlinear Thomson scattering  $z=(s/\chi)^{2/3}$, $\bar{c}=1$ and $g(s)=g_{\tiny \textsf{NLT}}(s)=1$ and $s>0$.)
To adapt the rate to a non-plane-wave background, the lightfront momentum is first re-written in terms of the instantaneous classical momentum (e.g. for fermions, the kinetic momentum $\pi$, which arises from solving the Lorentz equation in a plane wave background), by recognising $k\cdot p = k \cdot \pi$ (we recall $p$ is the momentum of an incident electron and $p=\pi_{p}(\vphi=-\infty)$), and then the plane-wave momentum $\pi$ is replaced with the numerically-solved instantaneous momentum in the simulation. 

To ascertain the correct expansion parameter for acquiring the LCFA, we note that to obtain the Airy functions one performs a change of variables:
\be 
\theta = \frac{\tilde{\theta}}{\xi|\Psi'(\phi)|}\left(\frac{8\chi|\Psi'(\phi)|}{u(s)}\right)^{1/3},
\ee
and integrates in $\tilde{\theta}$. Then for the LCFA to be a good approximation, it follows that higher-order terms of the expansion in this parameter can be neglected. This implies that \cite{DiPiazza:2017raw,Ilderton:2018nws}:
\be
\left(\frac{\xi^{2}[\Psi'(\phi)]^{2}u(s)}{8\eta}\right)^{1/3} \gg 1, \label{eqn:lcfawhen1}
\ee
where we recall that $\xi \gg1$ is assumed, which provides an approximate region of validity for the LCFA (a similar condition to \eqnref{eqn:lcfawhen1} was found much earlier for the synchrotron approximation \cite{PhysRevLett.89.094801}). We note that \eqnref{eqn:lcfawhen1} is a \emph{local} condition, which is violated whenever the field is weak enough, which happens in the tails of a plane-wave pulse. However, when the validity condition is violated like this, the contribution from the LCFA is suppressed, so that, at worst, it underestimates the QED result (which is safer to use for predictions of experiment than if the LCFA would overpredict these regions). We also note that \eqnref{eqn:lcfawhen1} shows that the validity of the LCFA depends on what part of the spectrum is being calculated. For the Compton process, we recall $u(s)=s/(1-s)$, and so if $s$ is small enough, the LCFA validity condition is violated (this is illustrated in \figref{fig:lcfa1}). 

Comparing the LCFA to an exact calculation of Compton scattering in a circularly-polarised monochromatic field, it was shown in \cite{Harvey:2014qla}, that in the spectrum ${\rm d}\prob/{\rm d}s$, the LCFA misses harmonic structure arising from interference effects over a wavelength of the background and indeed diverges in the infra-red limit, $s\to0$. Due to progress made at the end of the 2000s in exact calculations of processes in plane wave pulses of finite duration (see Sec.~\ref{sec:first:nlc}), there has been an interest in ``benchmarking'' exact solutions with the LCFA and investigating the LCFA's regime of applicability.  Several such analyses of comparing the LCFA to pulsed plane wave results have been undertaken, e.g. \cite{DiPiazza:2017raw,Ilderton:2018nws,DiPiazza:2018bfu,Blackburn:2018sfn,King:2019igt}.

The condition in \eqnref{eqn:lcfawhen1} is useful in understanding where the LCFA fails. For different processes, it is natural to have different conditions on the applicability of approximations. For example, for the nonlinear trident process in a constant crossed field, it was found in \cite{Ritus:1972nf} that the incoherent (two-step) probability scales as $\xi^{2}\mathcal{T}^{2}$, whereas the coherent (one-step) part scales as $\xi\mathcal{T}$. It was suggested in \cite{King:2013osa} that, as a result, an additional condition on the LCFA is that $\xi \mathcal{T} \gg 1$ must be fulfilled, because if the external field is too brief in duration, the `one-step' process (which, in a constant crossed field, had been found for some regions of $\chi$ to lower the total probability due to interference effects), is no longer negligible and the total probability for trident could not be well estimated by the standard LCFA approach of approximating the two-step process using the LCFA for Compton followed by the LCFA for Breit-Wheeler. Exactly how large $\xi \mathcal{T}$ has to be for the two-step term to dominate depends on $\chi$, because the two-step and the one-step scale differently at large or small $\chi$. In the high-$\chi$ limit of the LCFA~\cite{Ritus:1972nf,Torgrimsson:2020wlz}, the one-step term scales as $\sim \mathcal{T}\xi\ln\chi$, and the two-step as $\sim \mathcal{T}^2\xi^{2}(1/\chi^{2/3})\ln\chi$, which means $\chi^{2/3} \ll \mathcal{T}\xi$ must also be satisfied. For $\chi\ll1$ one finds a different condition.

Various LCFAs have recently been developed: LCFA rates that take into account the polarisation (photons) and or spin (charges) in first-order processes \cite{Li:2019oxr,Seipt:2020diz,Torgrimsson:2020gws} (polarised rates in a constant crossed field are given in \cite{baier98}), with the LCFA having been derived for the polarisation density matrix describing fermion spin evolution in a background \cite{Seipt:2018adi}. The LCFA has been derived for axion-like-particle (ALP) production from an electron \cite{Dillon:2018ypt} and ALP decay to a pair \cite{King:2019cpj} and LCFA rates for the `$2\to1$' processes of photon absorption by an electron \cite{Ilderton:2019bop,Blackburn:2020fqo} and pair-annihilation to a single photon \cite{Tang:2019ffe}. In this last case, an additional shortcoming of the LCFA was identified when benchmarked with the full QED probability: the LCFA result cannot reproduce the physics of narrow wave packets annihilating, which manifests as a highly oscillatory structure in the high-energy region. 
The LCFA has also been applied when evaluating a WKB approximation of first-order processes in a focussed background \cite{DiPiazza:2016maj}.
A type of LCFA has also been used at the level of the Dyson-Schwinger equation to include the `decay' of photon and fermion states in an external field \cite{Podszus:2021lms} (also in a constant crossed field \cite{Meuren:2011hv}). 
In addition, the LCFA has been applied to classical radiation formulas which include radiation reaction modelled via the LL equation~\cite{Heinzl:2021mji,Piazza:2021vxi} and this has been extended to all orders, by resumming mass-operator and Compton processes in the limit of large $\xi\mathcal{T}$ \cite{Torgrimsson:2020gws,Torgrimsson:2021wcj,Torgrimsson:2021zob}. The concept of the LCFA can also be extended to non plane-wave processes such as the Schwinger effect (pair creation directly from a quasi-constant field)
\cite{Sevostyanov:2020dhs}. In \cite{Aleksandrov:2018zso}, an LCFA-like approximation for Schwinger pair creation was compared to several exact solutions, such as for a Sauter pulse as well as spacetime-dependent fields (see also Sec.~\ref{sec:Schwinger}). {In \cite{Gavrilov:2012jk}, the mean current and energy-momentum tensor in QED were calculated in arbitrary dimensions for a flat-top electric field of finite duration, which in the long pulse limit, gives an LCFA-type approximation.}

For one-loop diagrams, the LCFA can be applied straightforwardly. The two phase variables (or equivalent variables for the evolution of the field, e.g. proper time) that occur at the level of the amplitude are the same as the two phase variables $\vphi$ and $\vphi'$ occurring in the amplitude and its Hermitian conjugate of the $1\to 2$ processes discussed above. This can be clearly seen for example, using the optical theorem to `cut' the polarisation operator and relate it to nonlinear Breit-Wheeler pair-creation. In the context of loops, an LCFA-equivalent expression was derived for a polarisation tensor in \cite{Karbstein:2013ufa} by expanding in proper time, and in a plane wave for the mass-operator \cite{Ilderton:2020gno} and the vertex-correction \cite{DiPiazza:2020kze}.

\subsubsection{Extensions to the LCFA}
{In this section, we consider extensions to the LCFA presented above.  First we consider single-vertex processes in a plane wave background. A systematic way to extend the LCFA is to go beyond the leading order in $1/\xi$.}
One line of enquiry focusses on including derivative terms of \eqnref{eqn:PaI1} that come from higher orders in the expansion in $\bar{\theta}/\xi$ (analogous investigations {were performed decades ago for} 
beamstrahlung \cite{Chen:1988ec} and synchrotron radiation \cite{PhysRevLett.89.094801}).  {These corrections are treated as small, which excludes treating} ultra-short pulses, where coherent and incoherent parts of higher-order processes can be comparable. If including higher derivative corrections, an immediate difficulty is faced by the fact that, for some pulse shapes, the quantity defined above as the `rate', i.e. $d\prob/d\phi$, can become negative in places, violating the positivity constraint given in the introduction, which makes it unsuitable for e.g. numerical simulation. In \cite{Ilderton:2018nws}, a solution was suggested, in the form of applying filters to achieve the desired behaviour. 
\begin{figure}[h!!]
    \centering
    \includegraphics[width=16cm]{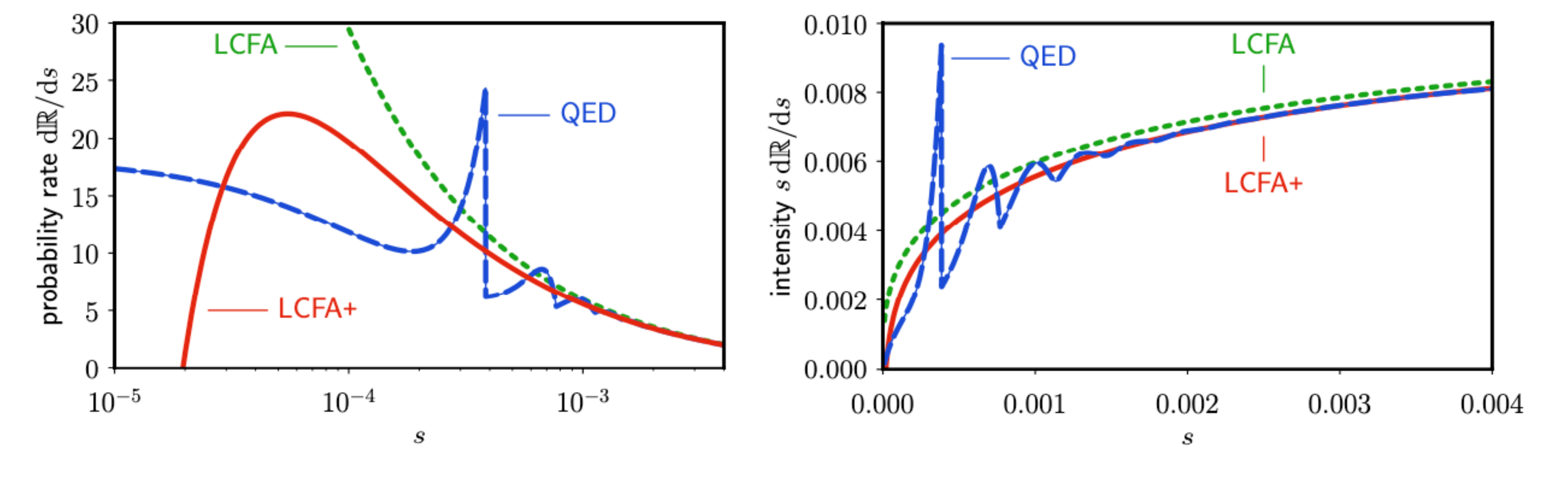}
    \caption{Spectrum for nonlinear Compton scattering (here $\mathbb{R}\equiv \rate = d\prob/d\phi$) for an electron with relativistic gamma factor $\gamma=1250$ colliding head-on with a monochromatic field with $\xi=5$ and frequency $1\,\trm{eV}$. `LCFA+' denotes an extension to the LCFA including higher derivatives of the field. Taken from \cite{Ilderton:2018nws}.}
    \label{fig:lcfa1}
\end{figure}
The higher-derivative corrections from \eqnref{eqn:KibExp1} were included in an improved LCFA (``LCFA+'') by perturbatively expanding the exponential and integrating the new terms by parts to express them in terms of Airy functions. The new expression required two filters: one for the corrections, and one overall positivity filter. The resulting rate was found to be more accurate for nonlinear Compton scattering in the infra-red and at intermediate electron energies, as illustrated in \figref{fig:lcfa1} (where, also, the harmonic structure, missed by the LCFA, is clearly evident in the QED result). In \cite{King:2019igt}, an alternative method (applied to the Breit-Wheeler process) to include higher derivative corrections was shown  by retaining them in the exponent, but casting the exponential in the form of an Airy kernel using a uniform Airy approximation \cite{vallee2010airy}. The result is that the derivative corrections then appear only in the Airy arguments:
\be
z \to z\left[1 + \frac{(\Psi_{j}''(\phi))^{2}+3\Psi_{j}'(\phi)\Psi_{j}'''(\phi)}{30\,\xi^{2} |\Psi'(\phi)|^{4}}\right]^{2/3} \label{eqn:AiryPlus1}
\ee
(the index $j$ is summed over, where repeated) and hence they are manifestly positive. However, a filter was still required, because if the corrections are too large, the LCFA becomes less accurate than without the corrections. 

{An alternative approach to extending the LCFA for nonlinear Compton scattering focusses on the fact that the LCFA rate diverges in the infra-red part of the spectrum. In \cite{Harvey:2014qla}, the LCFA was benchmarked with a monochromatic wave, and the condition $s < \eta/(1+\xi^2)$ given as a range for where the LCFA becomes inaccurate. In \cite{DiPiazza:2017raw}, it was pointed out that, in the $\xi\gg 1$ regime, where the condition becomes $s \ll \eta/\xi^2$, the correct limit of $\prob$ as $s\to 0$, is the perturbative, linear Compton result of the plane-wave background. In \cite{DiPiazza:2017raw} it was emphasised that, for e.g. a $10\,\trm{GeV}$ electron colliding with a plane wave pulse of $\xi\approx 10$ and frequency $1.55\,\trm{eV}$, the LCFA becomes inaccurate for photon energies as high as $30\,\trm{MeV}$. Therefore, in \cite{DiPiazza:2017raw}, it was suggested to modify the LCFA by introducing a type of low-frequency prescription so that the correct infra-red limit is approximated. This was initially achieved by defining a `formation length' $\Phi_{\trm{form.}}$ of the process, to be when the expansion of the LCFA exponent (i.e. the expansion in $\bar{\theta}$ in \eqnref{eqn:KibExp1} up to order $\bar{\theta}^{3}$) equals $\pi$, giving
\be 
\Phi_{\trm{form.}} \sim \frac{8}{\xi_{0}}\sinh\left[\frac{1}{3}\sinh^{-1}\left(\frac{3\pi\chi_{0}}{4}\frac{1-s}{s}\right)\right], \label{eqn:ADPphform}
\ee
where $\xi_{0}$ and $\chi_{0}$ are the global values (i.e. maximum values) of the $\xi$ and $\chi$ parameters.
Then a lightfront momentum cutoff is applied to nonlinear Compton spectrum so that when $s>s_{\tiny\tsf{lcfa}}$, the LCFA is used, otherwise the linear Compton scattering formula is used. The cutoff $s_{\tiny\tsf{lcfa}}$ is defined as being the lightfront momentum fraction where $\Phi_{\trm{form.}}$ is equal to $2\pi$. When this was implemented in a numerical simulation code in \cite{DiPiazza:2017raw}, the total probability of the process was indeed closer to the correct plane-wave probability. However, as was commented in \cite{DiPiazza:2017raw}, the ad-hoc prescription introduced an unphysical discontinuity in the spectrum at $s=s_{\tiny\tsf{lcfa}}$ and, as was remarked in \cite{DiPiazza:2018bfu}, the cutoff requires knowledge of $\xi_{0}$ and $\chi_{0}$ which are not local quantities. One way to improve this approximation is to find a way to define a `local' timescale for the background field $\tau_{\tiny\tsf{bg}}(t)$ in terms of kinematic quantities and use this to define a local $\xi_{0}$ and $\chi_{0}$. In \cite{DiPiazza:2018bfu}, this local timescale was defined as 
\be
\tau_{\tiny\tsf{bg}}(t) = 2\pi \sqrt{\frac{|\mbf{F}_{\trm{L},\perp}^{2}(t)|^{2}}{|\mbf{\dot{F}}_{\trm{L},\perp}(t)|^{2}+|\mbf{F}_{\trm{L},\perp}(t)\cdot\mbf{\ddot{F}}_{\trm{L},\perp}(t)|}}, \label{eqn:taubgADP}
\ee
where $\mbf{F}_{\trm{L},\perp}$ is the transverse Lorentz force (note the similarity of the terms occurring in \eqnref{eqn:taubgADP} and the correction from higher orders of $1/\xi$ in \eqnref{eqn:AiryPlus1}, albeit with different coefficients). Then \eqnref{eqn:ADPphform} is rewritten replacing the formation phase with a local formation time $\Phi_{\tiny\tsf{form.}} \to t_{\tiny\tsf{form.}}(t)$, with $\chi_{0} \to q^{0}(t) \chi(t)/\omega$ and the prefactor $8/\xi_{0}\to 8\gamma(t)/m\chi(t)$, where $q^{0}(t)$ and $\gamma(t)$ are the local values for the scattered electron energy and the Lorentz factor of the electron before scattering respectively. A time-dependent threshold energy cutoff, $\omega^{\ast}_{\tiny\tsf{lcfa}}(t)$, below which the LCFA is replaced with its cutoff value $d\prob_{\tiny\tsf{lcfa}}(\omega^{\ast}_{\tiny\tsf{lcfa}}(t),t)/d\omega\,dt$, is defined via $\omega^{\ast}_{\tiny\tsf{lcfa}}(t)=\rho_{ft}\omega_{\tiny\tsf{lcfa}}(t)$, where $\omega^{\ast}_{\tiny\tsf{lcfa}}$ is the value at which the formation length equals the local field length scale, i.e. $t_{\tiny\tsf{form.}}(t)=\tau_{\tiny\tsf{bg}}(t)$ and $\rho_{ft}$ is an $O(1)$ `fine-tuning constant'. 
The energy $\omega_{\tiny\tsf{lcfa}}(t)$ is then:
\be 
\omega_{\tiny\tsf{lcfa}}(t) = \frac{p^{0}(t)}{1+\frac{4}{3\pi\chi(t)}\sinh\left\{3\sinh^{-1}\left[\frac{\chi(t)}{8\gamma(t)}\,m\tau(t)\right]\right\}},
\ee 
where $p^{0}(t)$ is the electron energy before scattering.
This scheme was implemented in a numerical simulation in \cite{DiPiazza:2018bfu} and a better agreement found with the full plane-wave result, in the infra-red.
}

In terms of corrections to higher-order processes, an expansion of the probability of the trident process in $1/\xi$ has been studied in~\cite{Dinu:2017uoj,Dinu:2019wdw}, in which it was found:
\be
    \prob=\xi^2\Big(P_{-2}+ \frac{1}{\xi}P_{-1} + \frac{1}{\xi^2} P_0 + \ldots\Big) \label{eqn:tridentLargeXi1}
\ee
where the coefficients $P_j$ depend on all other parameters, but not $\xi$. For Compton scattering the corresponding expansion reads
\be
    \prob_{\tiny\tsf{NLC}}=\xi \Big( P_{-1,C}+\frac{1}{\xi^2}P_{1,C} + \ldots \Big)
\ee
in which the next-to-leading (NLO) terms, i.e.~derivative corrections to the LCFA, scale as $1/\xi^2$ with respect to the leading term (e.g. consider expanding out the extra terms in \eqnref{eqn:KibExp1}). The same holds for the probability of nonlinear Breit-Wheeler $\prob_{\tiny\tsf{NBW}}$. If these are combined to give the two-step part of trident, we see (not writing out the relevant integrals and spin sums explicitly) that
\be 
    \prob_{\tiny\tsf{NLC}} \prob_{\tiny\tsf{NBW}} = \xi^2P_{-1,C}P_{-1,B}+P_{-1,C}P_{1,B}+P_{1,C}P_{-1,B}+\ldots \sim \xi^{2}P_{-2} + \delta P_{-2}+\ldots, \label{eqn:corrections1}
\ee 
where $\delta P_{-2}$ are the corrections which scale as $\sim \xi^{0}$ and therefore contribute to $P_{0}$ in \eqnref{eqn:tridentLargeXi1}. Therefore the corrections are suppressed for large $\xi$ with respect to some coherent parts of trident, $P_{-1}$. However, this is not the end of the story: one expects $P_{-2}$ and the corrections in \eqnref{eqn:corrections1} to scale with pulse duration as $\sim \mathcal{T}^{2}$ and $P_{-1}$ only as $\sim \mathcal{T}$. Therefore, if $\mathcal{T}\gg\xi$, it is consistent to include corrections to the two-step part of trident and neglect the one-step part. However, if $\mathcal{T}\lesssim\xi$, corrections to the two-step part of trident can only be consistently included if the one-step part is also included.

The attraction and power of the LCFA is that it can in principle be applied to general (i.e. non-plane-wave) EM backgrounds when the formation length of the process is much smaller than the typical field inhomogeneity (as explained in the introduction to this section). This is exploited regularly, for example in particle-in-cell simulations of laser-matter interactions. However, recently the applicability of the LCFA in non-plane-wave backgrounds has been scrutinised more closely, for example, it has recently been compared to particular solutions in a standing-wave background \cite{Raicher:2020nkq,Lv:2021ayt}.

\subsection{Locally monochromatic approximation}
\label{sec:approx:lma}
The LCFA is very useful for studying strong-field effects in laser-matter interactions, where a plasma is formed and the strong field is not known \emph{a priori}\footnote{The LCFA's applicability still requires, in general, that $\xi\gg1$ and that particles are ultrarelativistic so that the fields can be approximated as plane waves as outlined in the introduction to this chapter.}. However, in contrast to laser-plasma set-ups, experiments have also been performed \cite{E144:1996enr,Burke:1997ew,Bamber:1999zt} and are planned \cite{Abramowicz:2019gvx,Abramowicz:2021zja,Meuren:2020nbw}, using well-characterised particle and laser beams. In these experiments, often $\xi\sim O(1)$, which is outside the region of applicability of the LCFA and therefore requires an alternative approximation framework. It is possible to develop this, by exchanging the versatility of the LCFA with an assumption that the background is well-described as a many-cycle laser pulse. If a high-energy probe is collided almost head-on with a focussed laser pulse, then by e.g. a WKB analysis (see \secref{sec:beyondPW}) it can be shown that processes are well-approximated using the probabilities in a plane-wave background. Assuming the timescale of the carrier frequency is much shorter than the duration of the pulse envelope for the plane wave, treating the former exactly and ignoring derivatives of the latter, one arrives at a `slowly-varying-envelope' approximation (SVEA) (see \secref{sec:first:nlc}). The SVEA has been used for the nonlinear Compton process for example to calculate harmonic \cite{Seipt:2016rtk} and pulse shape effects \cite{Seipt:2010ya} and frequency modulation \cite{Seipt:2015rda} (see also Sec.\,\ref{sec:first:NLC:shape}). When this is combined with a local phase approximation, in a similar vein to the LCFA (but at the level of the triple differential probability, which includes transverse momentum integrals), then one arrives at a `locally monochromatic approximation' (LMA) \cite{Heinzl:2020ynb}. The LMA approximates a process occurring in a laser pulse, by taking the rate for the process to occur in a monochromatic wave, and integrating it over the pulse envelope sampled by a point-like particle traversing the pulse. The LMA has been implemented in numerical simulations in this form for higher-order processes \cite{Bamber:1999zt,cain1,Abramowicz:2019gvx,Blackburn:2020fqo,Blackburn:2021cuq} and explored analytically looking at how spin and polarisation enter the incoherent product in the LMA \cite{Torgrimsson:2020gws}.

To introduce the LMA we again consider the simple case of a $1\to2$ first-order tree-level process. {The simplest way to acquire the LMA is to rescale the average phase variable $\phi=\Phi\bar{\phi}$}, and include the leading order expansion of the probability in $1/\Phi$~\cite{Torgrimsson:2020gws}. 
However, to understand the LMA in more depth, it is useful to consider it as combining two approximations: the slowly varying envelope approximation (SVEA), followed by a local expansion in the interference phase. First, we will look at the intervening approximation: the SVEA. The main idea is to separate fast and slow timescales, approximating the slow timescale and but treating the fast timescale exactly. A simple case to demonstrate this is a circularly-polarised plane wave pulse, with a potential of the form:
\be
\mbf{a}^{\perp}(\vphi) = m \xi \psi\left(\frac{\vphi}{\Phi}\right)(\cos\vphi, \sin \vphi), \label{eqn:aperpLMA1}
\ee
where $\psi$ is the pulse envelope with phase length-scale $\Phi \gg 1$ (the `slow' timescale). From the general form of the probability \eqnref{eqn:PaI0}, it can be seen that products $\bs{\rho}^{\perp}\cdot \mbf{a}^{\perp}(\vphi)$ and $[\mbf{a}^{\perp}(\vphi)]^{2}$ occur in the exponent. Using integration by parts in the exponent, one can separate out the fast and slow timescales as already explained in \eqnref{eq:first:svea} from \secref{sec:first}, for example, for a term in the $\bs{\rho}^{\perp}\cdot \mbf{a}^{\perp}(\vphi)$ part of the exponent:
\be
\int^{\vphi} \psi\left(\frac{y}{\Phi}\right)\cos y \,{\rm d}y = \psi\left(\frac{\vphi}{\Phi}\right)\sin\vphi + O\left[\frac{1}{\Phi}\psi'\left(\frac{\vphi}{\Phi}\right)\right].
\ee
If $\Phi \gg 1$, then the length scale associated with the pulse envelope is much larger than the wavelength of the carrier wave, and the remainder term in the above integral can be neglected. We see from \eqnref{eqn:PaI0} there are still two phase integrals remaining, but this step has simplified the exponent so that the fast timescale occurs as simple cosine and sine factors multiplied by functions of the slow timescale. Using the Jacobi-Anger expansion \eqnref{eq:first:jacobi-anger} allows the probability to be written as a sum over partial probabilities for each of the harmonics of the fast timescale (in which the functions of the slow timescale from the exponent, appear in the arguments of the resulting Bessel functions). Applying this to \eqnref{eqn:PaI0} for nonlinear Compton scattering, one can integrate over the transverse momentum azimuthal angle, which takes a simple form due to the symmetry of a circularly-polarised background, and gives a Kroenecker delta function that reduces the double-harmonic sum (originating from the Jacobi-Anger expansion to functions in $\vphi$ and $\vphi'$) to a single harmonic sum. Then the probability can be written as $\prob=\sum_{n=n_{\ast}}^{\infty}\prob_{n}$:
\begin{align}
\prob_{n} =& \frac{\alpha}{2\pi\eta^{2}}\int {\rm d}\vphi\,{\rm d}\tilde{\vphi}\,{\rm d}s\,{\rm d}(\rho^{2})\,\frac{s}{1-s}~\exp\left\{i\left[\frac{u(s)}{2\eta}(1+\rho^{2})-n\right](\vphi-\tilde{\vphi})+i\frac{\xi^{2}u(s)}{2\eta}\int_{\tilde{\vphi}}^{\vphi}\psi^{2}\left(\frac{x}{\Phi}\right){\rm d}x \right\}, \nonumber\\
& \times \left\{w\tilde{w}J_{n}(\zeta)J_{n}(\tilde{\zeta})+\frac{\xi^{2}g_{\tiny\tsf{NLC}}}{2}\left[2yJ_{n}(\zeta)J_{n}(\tilde{\zeta})-\psi\tilde{\psi}\left(J_{n+1}(\zeta)J_{n+1}(\tilde{\zeta})+J_{n-1}(\zeta)J_{n-1}(\tilde{\zeta})\right)\right]\right\} \nonumber
\end{align}
    \be 
\zeta(\vphi) = \frac{1}{\eta}u(s)\rho\xi\psi\left(\frac{\vphi}{\Phi}\right); \quad w = \frac{1+\rho_{\infty}^{2}(\vphi)}{1+\rho^{2}}; \quad \rho^{2}_{\infty}(\vphi) = 2n\eta (1-s) - s\left[1+\xi^{2}\psi^{2}\left(\frac{\vphi}{\Phi}\right)\right];\nonumber
\ee
\be
 y = \frac{1+(\psi^{2}+\tilde{\psi}^{2})n\eta (1-s)/s - (1+\xi^{2}\psi^{2}\tilde{\psi}^{2})}{1+\rho^{2}}\label{eqn:sveaBK}
\ee
where $\tilde{\vphi}$ is the phase variable in the Hermitian conjugate of the scattering amplitude, $\rho=|\bs{\rho}^{\perp}|$, $J_{n}$ are Bessel functions of the first kind \cite{abramowitzStegun}, and unless otherwise stated, variables with a tilde ($\tilde{w},\tilde{\psi},\tilde{\zeta}$), have argument $\tilde{\vphi}$, and those variables without a tilde indicate an argument of $\vphi$. By rewriting the fast timescale in terms of Bessel functions, \eqnref{eqn:sveaBK} casts the SVEA in a new way, and retains interference due to the pulse envelope, on scales as long as the envelope itself, but with higher derivatives having been neglected. This can be seen e.g. by the harmonic sum having no threshold harmonic (i.e. $n_{\ast}\to-\infty$), allowing the bandwidth of the pulse to activate harmonic channels that are kinematically forbidden in a local approach. If the envelope $\psi$ is chosen to be a top-hat function, the phase integrals can be performed exactly, and bandwidth effects made manifest \cite{King:2020hsk,Tang:2021qht}. If instead, the envelope $\psi$ is chosen to be constant over all $\vphi$ (i.e. a monochromatic wave), then using $\phi=(\vphi+\vphi')/2$ and $\theta=\vphi-\vphi'$ again, the $\theta$ integral yields a delta-function which can be evaluated to give $\rho=\rho_{\infty}$, which leads to all bandwidth factors, $w,\tilde{w},y$ becoming unity, and \eqnref{eqn:sveaBK} reducing to the monochromatic formula (the integrand becomes independent of $\phi$, and $\int d\phi$ is then a divergent phase length pre-factor).
\begin{figure}[h!!]
    \centering
    \includegraphics[width=14cm]{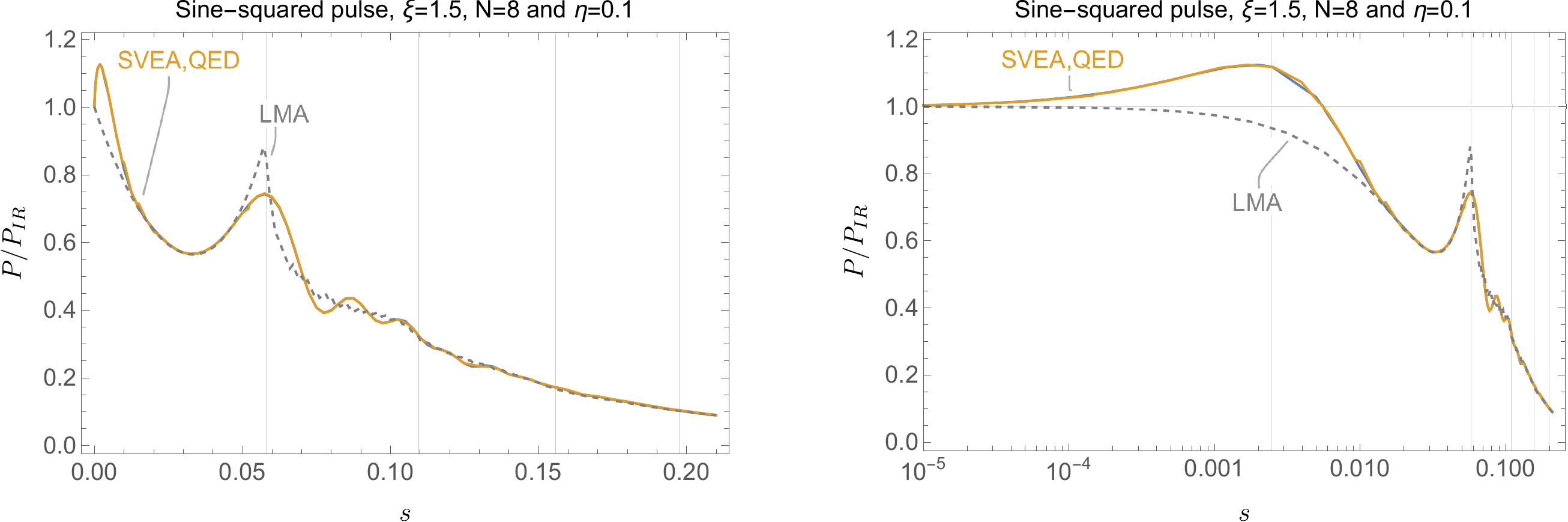}
    \caption{The photon lightfront momentum spectrum for an electron with $\eta=0.1$ colliding with an 8-cycle, sine-squared pulse with $\xi=1.5$, with a full QED calculation (blue solid line), the SVEA as given by \eqnref{eqn:sveaBK} (yellow solid line), and the LMA \eqnref{eqn:lmaQED} (grey dashed line). (The spectrum is normalised by the infra-red limit, $P_{IR}$ The SVEA and QED results are almost indistinguishable, whereas the LMA misses subharmonic structure and the mid-IR peak.}
    \label{fig:svea1}
\end{figure}
The SVEA is in general, a very good, but perhaps less versatile, approximation. The excellent agreement is demonstrated in \figref{fig:svea1}, in which a comparison is made between QED and the SVEA for the potential \eqnref{eqn:aperpLMA1} with a pulse envelope $\psi(\vphi/\Phi) = \sin^{2}(\vphi/2N)$ for $0<\vphi<2N$ and $\psi(\vphi/\Phi)=0$ otherwise.

To obtain the LMA from the SVEA, one can make a local expansion of the pulse envelope in $\theta$. This can also be justified by rescaling the average phase $\bar \phi = \phi/\Phi$, and then making an expansion in $1/\Phi$, retaining the leading order. When the pre-exponent of \eqnref{eqn:sveaBK} is expanded, at the leading order one simply finds $\psi(\bar \vphi')\approx \psi(\bar\vphi) \approx \psi(\bar \phi)$.
Expanding the exponent brings it into the form:
\be
i\left\{\left[\frac{u(s)}{2\eta}\left(1+\xi^{2}\psi^{2}\left(\bar \phi\right)+\rho^{2}\right) - n \right] \theta  
+
\frac{1}{\Phi^{2}}
\frac{\xi^2 u(s)}{24\eta }\left[\psi'^{2}(\bar \phi)+\psi''(\bar \phi)\psi(\bar \phi)\right] \theta^{3} + \ldots\right\}. \label{eqn:LMAexpo1}
\ee
The leading order expansion of the probability in $1/\Phi$ means that the cubic term in $\theta$ is neglected here, unlike in the LCFA. The delta function generated by the integral over $\theta$, $\delta\left(u [1+\xi^{2}\psi^{2}(\bar \phi)+\rho^{2}]/2\eta - n \right)$, has an argument which is exactly the stationary phase condition that was found for the scattering amplitude, Eq.~\eqref{eq:first:spa}. (We note the curiosity that the delta function is evaluated at exactly the point that the linear term in \eqnref{eqn:LMAexpo1} disappears, while all other neglected terms do not in fact not vanish.)
When integrating over the remaining phase variable $\phi$, the argument of the delta function can become stationary itself, which corresponds to a fold caustic due to two coalescing stationary points. This occurs at the peak of the pulse envelope, $\psi'=0$, and has the consequence that the double-differential probability ${\rm d}\tsf{P}_{\tsfs{LMA}}/{\rm d}s\, {\rm d}\rho$ \emph{diverges} at these points, and the LMA formally breaks down. However, practically this is seldom a problem, because: i) when using the LMA in simulations, the double-differential spectrum is sampled with a finite resolution and the delta function is replaced with a non-divergent regularised delta function; ii) often one is only interested in the single-differential lightfront momentum spectra ${\rm d}\tsf{P}_{\tsfs{LMA}}/{\rm d}s$, in which case $\rho$ is integrated over first upon which the delta function is consumed and yields a finite result.
{An alternative derivation of the LMA was provided in \cite{Torgrimsson:2020gws}, in which the transverse momentum integrals ($\rho$) were integrated analytically before any approximation was made (i.e.~using the exact results from~\cite{Dinu:2019pau} as a starting point). In this case, when one rescales the average phase as $\phi=\Phi\bar{\phi}$ and expands the integrand to leading order in $1/\Phi$, one does not encounter the delta function above. The resulting $\theta$ integrals are convergent and give an equivalent representation of the LMA that does not involve Bessel functions.
}

Finally, then, for a circularly-polarised plane-wave pulse, using the shorthand $\nphi=\phi/\Phi$, the LMA rate is $\rate_{\tsfs{LMA}}(\nphi) = \sum_{n=\lceil n_{\ast}(\nphi)\rceil}^{\infty}\rate_{\tsfs{LMA},n}(\nphi)$,
\begin{align}
\rate_{\tsfs{LMA},n}(\nphi) = -\frac{\alpha}{\eta}\int_{s_{-,n}(\nphi)}^{s_{+,n}(\nphi)} {\rm d}s \left\{\tilde{c}J_{n}^{2}[z(\nphi)] - \frac{\xi^{2}\psi^{2}(\nphi)g(s)}{2}\left(J_{n+1}^{2}[z(\nphi)]+J_{n-1}^{2}[z(\nphi)]-2J_{n}^{2}[z(\nphi)] 
 \right)\right\},
\label{eqn:lmaQED}
\end{align}
where the Bessel functions have arguments:
\be
z(\nphi) = \frac{2n\xi|\psi(\nphi)|}{\sqrt{1+\xi^{2}\psi^{2}(\nphi)}}\left[\frac{u(s)}{u_{n}}\left(1-\frac{u(s)}{u_{n}}\right)\right]^{1/2},
\ee
where $\bar{c}$ and $g(s)$ are the same as for the LCFA: for the Compton process, $\tilde{c}=1$, $g(s)=1+s^{2}/[2(1-s)]$, $u(s)=s/(1-s)$, $u_{n}=2n\eta_{e}/[1+\xi^{2}\psi^{2}(\nphi)]$ for Breit-Wheeler, $\tilde{c}=-1$ $g(s)=1/[2s(1-s)]-1$, $u(s)=1/[s(1-s)]$, $u_{n}=2n\eta_{\gamma}/[1+\xi^{2}\psi^{2}(\nphi)]$
the threshold harmonic, $\lceil n_{\ast}(\nphi) \rceil$ and the integration limits are those from the monochromatic probabilities, but where the field strength now depends on phase, i.e. for Compton: $n_{\ast}(\nphi)=1$, $s_{-,n}(\nphi)=0$, $s_{+,n}(\nphi)=2n\eta/[1+2n\eta +\xi^{2}\psi^{2}(\nphi)]$, for Breit-Wheeler: $n_{\ast}(\nphi)=2[1+\xi^{2}\psi^{2}(\nphi)]/\eta$, $s_{\pm,n}(\nphi)=(1\pm\sqrt{1-n_{\ast}(\nphi)/n})/2$. i.e. the threshold harmonic is local, and changes depending on where the electron is in the pulse. For a linearly-polarised plane-wave background, there is an extra integration over azimuthal angle, and, formally, a double sum over harmonic order. However, it has been commented that, to a good approximation, the double harmonic sum can be replaced with a single sum \cite{Heinzl:2020ynb}. An example comparison of the Compton spectrum as predicted by the LMA compared to direct evaluation from QED, is given in \figref{fig:svea1}.

The LMA has been benchmarked against direct evaluation of the QED result for finite pulses and its accuracy assessed. In general, as one would expect, the accuracy increases with pulse duration. It was shown analytically in \cite{Heinzl:2020ynb} that at high intensities, the LMA tends to the LCFA, and this was also seen in the results of numerical calculations \cite{Blackburn:2021rqm,Blackburn:2021cuq}.
Small discrepancies between the LMA and QED have been found in a few areas. First, at channel openings, where the parameters are such that a small change of intensity or energy allows for an extra harmonic to become accessible, a discrepancy was noted for nonlinear Breit-Wheeler, with the larger errors made close to the opening of lower harmonic channels \cite{Blackburn:2021cuq}. 
 Second, at small $\xi$, the LMA can drastically underestimate the rate of nonlinear Breit-Wheeler, due to the `enhancement' from the bandwidth of the pulse envelope  \cite{Titov:2012rd,Nousch:2012xe,Blackburn:2021cuq,Tang:2021qht}.
Third, in the lightfront momentum spectrum of nonlinear Compton scattering, the mid-IR peak (see Sec.\,\ref{sec:first:nlc:IR} for more details), is due to the pulse envelope bandwidth and is missed by the LMA \cite{Heinzl:2020ynb,King:2020hsk}. The position of this peak is indicated by the gridline in the right-hand plot of \figref{fig:svea1} (it is located approximately where the first harmonic would be, if the energy parameter were scaled to correspond to the wavelength of the pulse envelope rather than the carrier frequency, i.e. $\eta\to\eta/2N$, where $N$ is the number of cycles of a finite sine-squared pulse). Other bandwidth effects missed by the LMA, such as the presence of subharmonics and softening of harmonic edges, but that \emph{are} captured by the SVEA, are shown in \figref{fig:svea1}.

While the LCFA is included in simulations by writing the rates using the instantaneous (kinematic) momentum, $\pi^\mu$, the LMA is included by writing the rates in terms of the quasimomentum $\overline{\pi}^\mu$ (introduced in \eqnref{eq:first:quasimomentum} and further discussed around \eqnref{eq:first:svea}), which corresponds to an average of the fast, carrier frequency, timescale. In a plane wave, we recall this is $\overline{\pi}^\mu  = p^\mu + k^\mu \, \overline{\mbf{a}_{\LCperp}^{2}}/(2k\cdot p)$, and the replacement is made in \eqnref{eqn:lmaQED} and below as 
$\xi^{2}\psi^{2}(\phi)\to \overline{ \mbf{a}_{\LCperp}^{2} }/m^{2}$.

\begin{figure}[h!!]
    \centering
    \includegraphics[width=16cm]{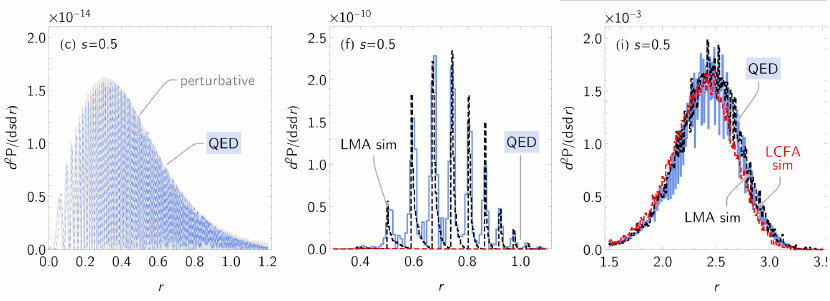}
    \caption{Lineout of the double-differential spectrum in transverse momentum (here, $r\equiv|\mbf{\rho}_{\perp}|$ defined in \secref{sec:first:nbw}) and lightfront fraction for the nonlinear Breit-Wheeler process when $\xi=0.2$ (left) $\xi=0.5$ (centre) and $\xi=2.5$ (right) for $\eta=0.2$ and $N=16$ for a sine-squared plane-wave pulse. At sufficiently low $\xi$ ($\xi=0.2$), pulse envelope effects dominate and the LMA underpredicts the yield. For $\xi=0.5$, the LMA reproduces the main structure, but sub-harmonics can be seen in the QED result, which are missed. At $\xi \gg1$ (here $\xi=2.5$), the LMA and LCFA tend to the same result. Taken from \cite{Blackburn:2021cuq}.}
    \label{fig:lma1}
\end{figure}

The accuracy of the LMA has been evaluated by comparing its predictions from simulation, to those from a numerical calculation of exact QED formulas in a plane-wave background. Throughout this chapter, we have not discussed regularisation of the $\theta$ integral that must be performed in exact solutions of finite pulses. This is because, in the `rate' approach, the approximation of the $\theta$ integral means that no contributions are picked up from outside the pulse, and hence there are no issues with constant phase terms needing to be regularised (see \secref{sec:first:nlc}). (As already commented, all the `width' factors of \eqnref{eqn:sveaBK}, which originate from regularisation terms, cancel in the monochromatic or locally-monochromatic limit.) Therefore, one can expect that the contribution from regularisation corresponds to effects that are missing in the local approach. Total yields as well as lightfront and transverse momentum spectra have been compared for nonlinear Compton \cite{Blackburn:2021rqm} and Breit-Wheeler \cite{Blackburn:2021cuq}, for regular and chirped plane waves. A comparison was also made for a focussed background using the high-energy approximation of integrating the plane-wave QED result over the transverse structure of the focused background. An example of the benchmarking is given in \figref{fig:lma1}, where it can be seen that integer harmonics are reproduced by the LMA, but in general sub-harmonics (as well as interference effects on the scale of the pulse length), are missed. For the range of applicability, it was argued in \cite{Blackburn:2021rqm, Blackburn:2021cuq} that in addition to the condition $\Phi\gg1$, the error of the LMA increases around harmonic channel openings and when pulse-interference effects begin to dominate (when $s\ll s_{+,1}(\phi)$ for Compton and when $\xi\psi(\phi) \ll 1$ for Breit-Wheeler). In \cite{Torgrimsson:2020gws,Abramowicz:2021zja}, the LMA was also found to approximate well the longitudinal momentum spectrum of the full trident process well (when $\xi=1$ and $\eta=1/2$ and when $\xi=4$ and $\eta=1/8$, example cases where the LCFA underestimated the yield). 

    In the E144 experiment, where nonlinear Compton and Breit-Wheeler were measured for the first time, the LMA was used \cite{bula97} to model the interaction between the electron beam and the long laser pulse. Although the LMA was used without derivation, the long laser pulse of full-width-half-max duration of $1.5\,$ps (of order ${\mathcal O}(100)$ cycles), comfortably fulfills the criterion $\Phi\gg 1$ for using the approximation. (Due to the $17$ degree collision angle, the effective interaction time with the laser pulse would be smaller than $1.5\,$ps.)
    The LMA has since been implemented in simulation codes CAIN \cite{etde_74290,cain1}, which models beam-beam interactions between electrons, positrons and photons; IPStrong \cite{Hartin:2018egj}, which has been used in the initial design of the proposed LUXE experiment \cite{Abramowicz:2019gvx}; and the open-source code Ptarmigan \cite{ptarmigan}, which is the current simulation code of the LUXE experiment and has been benchmarked against QED  \cite{Blackburn:2021rqm,Blackburn:2021cuq}.

\subsection{The Quasiclassical Approach} \label{sec:approx:BKmethod}
So far, we have focussed on the LCFA and LMA, as they have seen significant development in the last decade. However, there are also some established methods that have been employed in new situations in strong-field QED.

A notable example is a semi-classical approach that can be used to write a radiation formula that resembles the classical Li\'enard-Wiechert radiation formula, but that takes into account the recoil of the electron due to emitting photons. In this approach, sometimes referred to as the `quasiclassical' or `Baier-Katkov' \cite{Baier1968} method, a two-step method is employed to calculate the emitted radiation from an electron traversing a given background field. First, the classical trajectory is solved for: either there exists an analytical solution, or else the Lorentz equation can be employed in a numerical calculation and the trajectory found. Once the trajectory is know, it can then be integrated over to obtain the radiation spectrum. The radiated energy,  ${\rm d}\mathcal{I}$, of an unpolarised electron can be written as (see e.g.  \cite{Baier:1990pi}, Eq. (2.46)):
\be
{\rm d}\mathcal{I} = \frac{i\alpha}{8\pi^{2}}\frac{m^{2}}{(p^{0})^{2}} \ell_{0}\,{\rm d}\ell_{0} \int \frac{{\rm d}t\,{\rm d}\tau}{\tau - i0} \left\{1+\frac{1}{4}\left(\frac{p_{0}}{q_{0}}+\frac{q_{0}}{p_{0}}\right)\gamma^{2}\left[\Delta\mbf{v}(t+\tau/2)-\Delta\mbf{v}(t-\tau/2)\right]^{2}\right\}e^{-i\Upsilon}; \nonumber
\ee
\be 
\Upsilon = \frac{p_{0}\ell_{0}\tau}{2q_{0}\gamma^{2}}\left\{1 + \frac{\gamma^{2}}{\tau}\int^{\tau/2}_{-\tau/2}{\rm d}u [\Delta \mbf{v}(t+u)]^{2} - \left(\frac{\gamma}{\tau}\int^{\tau/2}_{-\tau/2}{\rm d}u\,\Delta\mbf{v}(t+u)\right)^{2}\right\}, \label{eqn:BKmeth1}
\ee 
where the photon emission angle has already been integrated over, $\gamma$ is the electron's Lorentz factor and $q_{0}=p_{0}-\ell_{0}$ is the outgoing electron energy ($\ell_{0}$ is the emitted photon's energy). To explain the terms in this formula, we note that the particle velocity can be written $\mbf{v}(t) = \mbf{v}_{0} + \Delta \mbf{v}(t,\mbf{v}_{0})$, where $\mbf{v}_{0}$ is the average velocity, and that the integration variables are related to the original time variables $t_{1}$ and $t_{2}$ that occur in the amplitude and its Hermitian conjugate respectively, via $t_{1}=t-\tau/2$, $t_{2}=t+\tau/2$. We note the analogous form of \eqnref{eqn:BKmeth1} to the QED formula for nonlinear Compton in a plane-wave background (where no approximations are used) \eqnref{eqn:fnlc1}. We can make this comparison more explicit by setting: $s=\ell_{0}/p_{0}$ and recalling in a plane-wave that ${\rm d}\vphi/{\rm d}t = \eta/\gamma$, then by setting $\gamma \Delta \mbf{v} = \mbf{a}$, and recalling that $1-s+(1-s)^{-1} = 2g_{\tiny \textsf{NLC}}(s)$ we see the pre-exponents take the same form. Using these replacements, it can be seen that $\Upsilon$ contains exactly the squared Kibble mass factor \eqnref{eqn:approx:kibble}. The formula \eqnref{eqn:BKmeth1} is derived assuming $\gamma \gg 1$ and using the `operator method' of writing S-matrix elements with operator-valued functions. To proceed, one `disentangles' the operators and drops terms that result from commutators that scale with $\mathcal{O}(1/\gamma^{2})$ and then replaces operators with their classical expectation values. This involves writing \cite{Wistisen:2016bry}:
\be 
q_{0} = \sqrt{(\mbf{p}-\bs{\ell})^{2}+m^{2}} = p_{0}-\ell_{0} + \frac{p_{0}\ell_{0}-\mbf{p}\cdot\bs{\ell}}{p_{0}-\ell_{0}} + {\mathcal O}\left(\frac{1}{\gamma^{2}}\right),
\ee 
and neglecting terms of order ${\mathcal O}(1/\gamma^{2})$. However if the background is a plane wave, the kinematics are such that the neglected terms are zero. Therefore the Baier-Katkov approach is exact in a plane wave, which is logical when one considers the solution to the Dirac equation in a plane-wave background (the Volkov solution), is identical to the semi-classical solution. The power of the semi-classical approach, however, is that it can be applied to non-plane-wave backgrounds. 
{Indeed, an approach based on using WKB to approximate solutions of the the Dirac equation in non plane-wave fields} has been used and extended recently, in a series of papers using semi-classical solutions of the Dirac equation in focussed background fields to go beyond the plane-wave approach \cite{DiPiazza:2013vra,DiPiazza:2015xva,DiPiazza:2016tdf,DiPiazza:2016maj}, {with a result that the probability for nonlinear Compton is obtained by averaging \eqnref{eqn:BKmeth1} over the initial electron position.} 
see \secref{subsec:highE}.

If the angular integral is kept and not integrated over to obtain \eqnref{eqn:BKmeth1}, the differential energy radiated into an angle ${\rm d}\Omega$, can be written in terms of two `radiation integrals' \cite{Wistisen:2014ysa}:
\begin{align}
\frac{{\rm d}^{3}\mathcal{I}}{{\rm d}\ell^{0}\,{\rm d}\Omega} = \frac{e^{2}}{(4\pi)^2}\left(\frac{q_{0}^{2}+p_{0}^{2}}{2 p_{0}^{2}}|\mathbf{I}|^{2} + \frac{\ell_{0}^{2}m^{2}}{2p_{0}^{4}}|J|^{2}\right),
\end{align}
where the integrals $\mbf{I}$ and $J$ are given by:
\be
\mbf{I}=\int \frac{\mbf{n} \times [(\mbf{n}-\mbf{v})\times\mbf{\dot{v}}]}{(1-\mbf{n}\cdot\mbf{v})^{2}}\,e^{i\ell'\cdot x} {\rm d}t; \qquad J = \int \frac{\mbf{n}\cdot \mbf{\dot{v}}}{(1-\mbf{n}\cdot\mbf{v})^{2}}\,e^{i\ell'\cdot x} {\rm d}t,  \label{eqn:BKradintegrals1}
\ee
where $\ell' = (p_{0}/q_{0})\ell$ (this is a manifestation of photon recoil \cite{1991PhRvA..43.6032L}), $\times$ is the cross-product and $\mbf{n}$ is the direction of emission i.e. for the photon momentum $\mbf{\ell}=\ell_{0}\mbf{n}$.

The quasiclassical approach has a history of being employed to calculate strong-field QED effects in beam-crystal experiments (recent examples include \cite{Wistisen:2014twa,Wistisen:2017pgr,Wistisen:2019eza,DiPiazza:2019vwb}). It has recently been applied to electron-laser collisions. In \cite{Wistisen:2014ysa} an extended version of the radiation integrals \eqnref{eqn:BKradintegrals1} that includes electron polarisation was derived, and the method was used to investigate interference in two-color pulses, in \cite{Wistisen:2015rua} the method was used to calculate Compton scattering in a constant field of finite extent; in \cite{Wistisen:2020rsq} the semi-classical method was used to calculate pair production in a plane wave and compared with the directly evaluated but QED expression and the LCFA; in \cite{Raicher:2018cih,Nielsen:2021nuo} it was used to calculate nonlinear Compton scattering in a plane wave and benchmarked with the LCFA and simulations employing a linearly-polarised version of the LMA and in \cite{Raicher:2020nkq} it was used to investigate the validity of the LCFA in a standing wave background.

\subsection{Saddle-point methods}\label{Saddle-point methods} \label{sec:approx:saddle}

A useful analytical method for approximating probabilities is to apply saddle-point methods. For example, for $\chi\ll1$ and $\xi\gg1$ one can obtain the probability of nonlinear Breit-Wheeler pair production $\prob\sim e^{-8/(3\chi)}$ by performing the lightfront time and momentum integrals with the saddle-point method. However, in this simple case one can simply use well-known expansions of Airy functions to obtain this probability. The saddle-point method is instead most useful in cases when one cannot find such special-function representations, e.g. in cases where one cannot use (the leading order) LCFA or LMA. Having probabilities with exponential scaling and thus being able to use saddle-point methods is sometimes associated with $\chi\ll1$ {\it and} $\xi\gg1$, but, we emphasize, $\xi\gg1$ is {\it not} a necessary condition. While it is true that the saddle-point approximations in general break down if one keeps $\chi\ll1$ fixed and decreases $\xi$, these approximations work even for $\xi\sim \mathcal{O}(1)$ as long as $\chi$ (or $\eta$) is sufficiently small. The expansion parameter for saddle-point approximations is $\chi$. As for other expansions we have discussed, one can in general expect the $\chi\ll1$ expansion to be good as long as no other parameter is very small or large. In this case $\xi$ cannot be taken too small (with $\chi\ll1$ kept fixed), but $\xi$ \emph{can} be taken large.

Consider for example nonlinear Breit-Wheeler in a Sauter pulse, $a_1(\varphi)=\xi\tanh(\varphi)$ and $a_2=0$. The saddle points are given by $\phi=0$, $\theta=2i\text{arctan}(1/\xi)$ and the electron and positron share the absorbed momentum equally. One finds~\cite{Esposti:2021wsh}
\be\label{BWSauterPulse}
\prob\approx\frac{\alpha\sqrt{\pi \xi\chi}}{32\sqrt{(1+\xi^2)\text{arccot}(\xi)}}
\frac{\exp\left\{-\frac{4\xi}{\chi}[(1+\xi^2)\text{arccot}(\xi)-\xi]\right\}}{(1+\xi^2)\text{arccot}(\xi)-\xi} \;.
\ee
In the limit $\xi\gg1$, \eqref{BWSauterPulse} gives $\prob\propto e^{-8/(3\chi)}$ in agreement with the expansion of the LCFA approximation, i.e. the limits $\chi\ll1$ and $\xi\gg1$ commute. But~\eqref{BWSauterPulse} also works for $\xi\sim1$ as long as $\chi\ll1$. The same method can be used for more complicated processes, like trident~\cite{Dinu:2017uoj}, where both the direct and exchange terms, $\prob_{\rm dir}$ and $\prob_{\rm ex}$, have a very similar form as in~\eqref{BWSauterPulse}. 
By expanding these in $1/\xi$, one finds, to leading order in $1/\xi$, $\prob_{\rm ex}\approx c_{\rm ex}\xi e^{-16/(3\chi)}$, while $\prob_{\rm dir}\approx(c_{\rm dir}^{(-2)}\xi^2+c_{\rm dir}^{(-1)}\xi)e^{-16/(3\chi)}$, where the $c$'s are independent of $\xi$. The term that is quadratic in $\xi$ gives (the leading order contribution of) the two-step part, and the terms that are linear in $\xi$ give (the leading order contribution of) the one-step part (definitions of the two-step/one-step and direct/exchange separations can be found in~\ref{one-step/two-step separation} and~\ref{direct vs exchange}). These terms in the expansion in $1/\xi$ agree with what one finds by first expanding the exact probability in $1/\xi$ and then making a $\chi\ll1$ approximation of the coefficients. However, $\prob_{\rm ex}$ had never been calculated analytically before, not even for a constant-crossed field. This is therefore an example of the fact that the saddle-point methods can allow one to calculate even complicated processes. The most nontrivial part of the calculation is to find the saddle point(s). After that one just has to make a suitable change of integration variables, e.g. $\theta\to\theta_{\rm saddle}+\sqrt{\chi}\delta\theta$\footnote{The reason for the factor of $\sqrt{\chi}$ is to have all the terms in the exponent that are quadratic in the perturbation around the saddle point independent of $\chi$, e.g. $x=x_{\rm saddle}+\sqrt{\chi}\delta x$ so that $\exp\{-[1/\chi]\text{const.}(x-x_{\rm saddle})^2\}\to\exp\{-\text{const.}\delta x^2\}$ and so that one can expand the integrand in a series in $\chi$.}, and expand the integrand in $\chi\ll1$. Then it does not matter much that the original integrand for $\prob_{\rm ex}$ is complicated, because after expanding the integrand around the saddle point one just finds simple (Gaussian) integrals.

In other strong-field cases, a Sauter pulse is often chosen because one can find simple analytical solutions of the Dirac equation in e.g. a purely time-dependent electric field, and in such cases it can be very difficult to obtain similar results for other field shapes. This is not the case here. There are many other field shapes that lead to simple compact approximations similar to the ones above for a Sauter pulse~\cite{Dinu:2018efz}. In fact, for a general symmetric field with one field maximum, $a_1(\varphi)=m\xi f(\varphi)$ with $f(-\varphi)=-f(\varphi)$, one finds a saddle point at $\theta=2iz$ with $z=-if^{-1}(i/\xi)>0$. This was used in~\cite{Dinu:2018efz} to obtain approximations for single and double Compton scattering. The probability that an electron has longitudinal momentum fraction $s_\gamma=k\cdot l/k\cdot p$ is given by
\be\label{PCsaddleAntiSym}
\prob_{\rm C}(s_\gamma)\approx\frac{\alpha}{2u_{\rm NLC}}\frac{(2g_{\rm NLC}-1)\exp\left\{\!-\frac{u_{\rm NLC}}{\chi}\xi z\left[1+\xi^2\langle f^2\rangle\right]\right\}}{z\xi f'(iz)\sqrt{1-\frac{1}{z\xi f'(iz)}}} 
\qquad
\langle f^2\rangle=\frac{1}{2iz}\int_{-iz}^{iz}\!\ud u\; f^2(u) \;,
\ee
where $u_{\rm NLC}$ and $g_{\rm NLC}$ are defined below~\eqref{eqn:PaI0}. Similar expressions were derived for double Compton scattering (including all terms, i.e. two-step/one-step, direct/exchange). Note that very similar expressions can be derived for other processes, e.g. nonlinear Breit-Wheeler~\cite{Esposti:2021wsh}. The main difference between e.g. Compton and Breit-Wheeler is that for Compton one needs to avoid soft photons (because the probability to emit soft photons does not have an exponential scaling). In~\eqref{PCsaddleAntiSym} this is done by considering the longitudinal momentum spectrum with photon momentum fraction $s_\gamma$ not small. One could call this hard photons, but does not mean that one has to assume that the photon takes away most of the momentum (which would be $1-s_\gamma\ll1$); one can have, say, $s_\gamma=0.5$, as long as $\chi$ is sufficiently small. If one wants to integrate over all the photon momentum components then one needs to include a cut-off to avoid integrating over soft photons~\cite{HernandezAcosta:2020agu}.

For fields defined implicitly via $f'(\varphi)=(1-f^2(\varphi))^c$ one can express the probability in~\eqref{PCsaddleAntiSym} in terms of hypergeometric functions ${}_2F_1$ with $c$ and $\xi$ in the arguments~\cite{Dinu:2018efz}. The hypergeometric functions can be expressed in terms of more elementary functions for $c=j/2>0$, with integer $j$. For $c=1$ one recovers the results for a Sauter pulse. For $c=1/2$ one finds the results for one peak of a sinusoidal field, $f(\varphi)=\sin\varphi$.

\begin{figure}[h!!]
    \centering
    \includegraphics[width=15cm]{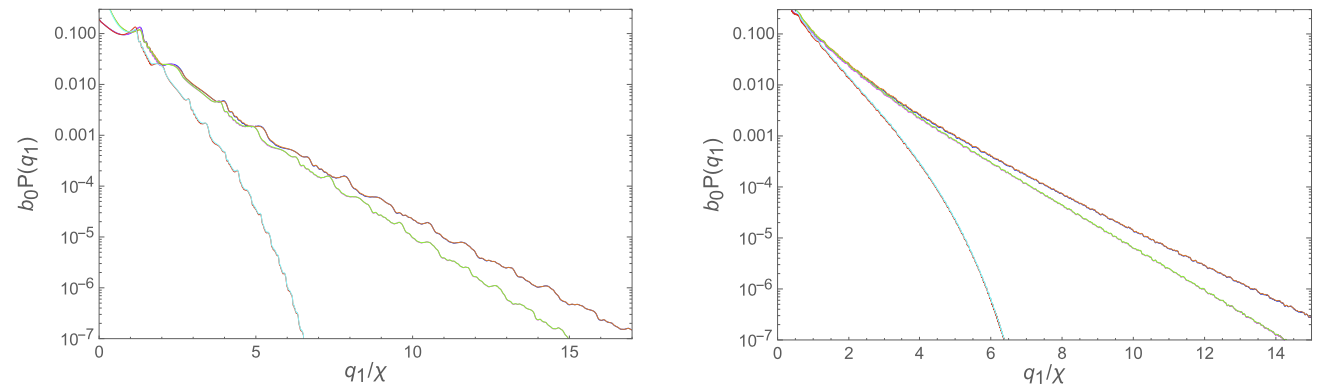}
    \caption{Spectrum for single nonlinear scattering for a Gaussian plane wave $a=m\xi \sin \vphi \,e^{-(\vphi/\Phi)^{2}}$, with $\Phi=80$ for $\chi=0.001$ (blue and orange curves), $\chi=0.01$ (magenta and green curves) and $\chi=0.1$ (red and cyan curves) for $\xi=1$ (left plot) and $\xi=2$ (right plot). The $\chi=0.1$ lines intersect the x-axis first, and the $\chi=0.001$ lines last/extend outside the plotted range. For each case there are two very close lines: one is for the exact numerical result and the other is for the saddle-point approximation. $b_0$ and $q_1$ are what we, in this review, denote $\eta$ and $s$ or $s_\gamma$. Adapted from~\cite{Dinu:2018efz}.}
    \label{fig:SpectrumManySaddlePoints}
\end{figure}

However, for a field with many oscillations one also finds many saddle points. Since these lie in the complex plane with both real and imaginary parts, it is not trivial to find all the relevant saddle points. This problem was considered in~\cite{Dinu:2018efz} for $a_1(\varphi)=m\xi\sin\xi e^{-(\varphi/\mathcal{T})^2}$. For large $\mathcal{T}$ there are many contributing saddle points. Finding these directly for a given $\mathcal{T}$ can be difficult. Instead one can start with the monochromatic limit $\mathcal{T}\to\infty$, where it is much simpler to find the saddle points. If $\mathcal{T}$ is finite but very large then one can find the saddle points numerically by using the saddle points from the monochromatic limit as starting points. For a moderately large $\mathcal{T}$ one can obtain the saddle points by following their movement in the complex plane as one gradually decreases $\mathcal{T}$ from $\infty$. 
The saddle-point approximation is compared with the exact result in Fig.~\ref{fig:SpectrumManySaddlePoints}. As seen, the approximation agrees very well with the exact result, including the small, fast and seemingly irregular patterns in the spectrum (in fact, one might have to zoom in on the plots to actually see the small difference). The same method was used in~\cite{Dinu:2019pau} for the longitudinal momentum spectrum of trident. Also in this case the approximation captures fine details of the spectrum.

\subsection{Numerical calculation of QED expressions}
Direct evaluation of QED probabilities for plane wave pulses requires numerical evaluation of nonlinearly oscillating exponentials. This task can be made more efficient by employing well-suited quadrature. For some field shapes, such as constant crossed or monochromatic, probabilities reduce to evaluating well-known special functions and they will not be considered here. However, if the field is a pulse, the amplitude must, in general, be regularised (see \secref{sec:first:nlc}) and this method of regularisation affects the numerical approach chosen. An advantage of the `$i\epsilon$' prescription is that transverse momentum integrals can be performed analytically, so that, if one is only interested in lightfront spectra or the total yield, the dimension of numerical integration is reduced. Also, if one is using the `gluing' approach to calculate higher order processes, only the lightfront spectrum of intermediate blocks is required \cite{Dinu:2019pau}. A disadvantage of this method of regularisation is that the (two-dimensional) oscillating phase integration is over an infinite interval. This means that in order to capture long-range interference effects, a phase range much longer than the pulse must be integrated over (although some parts of this phase integration may be performed analytically \cite{Dinu:2013hsd,Heinzl:2020ynb}). Therefore, this method of regularisation is less popular for numerical evaluation. It has been used for nonlinear Compton and Breit-Wheeler \cite{Heinzl:2020ynb}, as well as ALP decay to an electron-positron pair \cite{King:2019cpj}. A form of regularisation more amenable to numerical evaluation is the Boca-Florescu method \cite{Boca:2009zz}, discussed in \secref{sec:first:nlc:reg}, applied at the level of the amplitude. This has the advantage that, for a laser pulse with support on a finite interval the phase integration region is finite. The rest of this section describes numerical schemes to evaluate expressions with this form of regularisation. (Other forms of regularisation are also possible, such as transverse dimensional regularisation used in \cite{Dinu:2013gaa}.)

A standard method is to perform an integral substitution to linearise the exponent \cite{Dinu:2013gaa} and expand the resulting pre-exponent in a sum over an orthogonal function basis set. For example:
\be
\int_{0}^{\theta} {\rm d}t \,f(t)e^{ig(t)} = \int_{0}^{1} {\rm d}y \,H(y)e^{iKy}, \label{eqn:linexp1}
\ee
where the old and new integration variables are related by $g(t) = g(\theta)y$,
$K=g(\theta)$ and  $H(y) = Kf[g^{-1}(Ky)]/g'[g^{-1}(Ky)]$. Typical choices for expanding $H(y)$ include the set of Chebyshev \cite{Meuren:2014uia} and Legendre polynomials \cite{King:2019igt}, for which the right-hand side of \eqnref{eqn:linexp1} can be evaluated analytically. If the integral over the linear exponential cannot be evaluated analytically, it may be evaluated numerically by using quadrature suited to linear but highly-oscillating functions. One example is the Filon method \cite{filon_1930}, which has been applied to nonlinear Compton \cite{Angioi:2016vir} and photon-photon scattering \cite{King:2012aw}. Another possibility is to perform a fast Fourier transform which has been used for nonlinear Breit-Wheeler \cite{Meuren:2015mra}.

An alternative method to linearising the exponent is a deformation of the integration contour in the complex plane. Using Cauchy's theorem, an integration path can be found, over which the nonlinear oscillations are removed. If the exponent has stationary points, this approach is closely related to the saddle point method, in which contributions from regions away from the stationary points are exponentially suppressed. This has been exploited in calculations of Compton \cite{Dinu:2019pau} and Breit-Wheeler \cite{Dinu:2019pau}, double Compton \cite{Dinu:2018efz} and trident \cite{Dinu:2017uoj,Dinu:2019wdw}. As an alternative to a complex contour, one can use the fact that in some cases, the imaginary part of the stationary point scales as $1/\xi$, which can be neglected to a good approximation when $\xi \gg 1$, so that the method of stationary phase can be employed (which uses real stationary points), as was exploited for the Compton \cite{Mackenroth:2010jr,mackenroth2014quantum} and double Compton \cite{Mackenroth:2012rb} processes.

Another approach is to numerically integrate already at the level of the amplitude, which is treated as a matrix with spin and polarisation indices, and then mod-square to calculate the probability. Although less sophisticated, this method has the advantage of halving the number of oscillating phase integrals, with the disadvantage that the outgoing momenta must also be integrated numerically (whereas with e.g. the ``$i\epsilon$'' prescription, the transverse momenta can be integrated out analytically). However, since the integrand has a well-defined width in outgoing transverse particle momenta, this straightforward method leads to an overall reduction in numerical evaluation time. This method has been used widely, and is specifically referred to in \cite{Krajewska:2012gc,2013LPB....31..503K,Mackenroth:2018smh,Tang:2019ffe,King:2020hsk,Tang:2021vfh,Blackburn:2021rqm,Blackburn:2021cuq}.

\subsection{Alternative numerical approaches}
We review here other numerical methods in strong field QED. These methods provide alternative means of tackling problems in realistic background fields with complicated spacetime inhomogenities, including the Schwinger effect.

\subsubsection{DHW formalism}\label{sec:approx:DHW}
The Schwinger effect, as outlined in \secref{sec:intro}, is an example of spontaneous electron-positron pair creation from a strong field. Within the mean-field approximation, i.e.~neglecting back-reaction, extracting observables such as the number of pairs created in a given electromagnetic (background) field ${\mathcal A}_\mu$ requires, essentially, solving the Dirac equation in that background. Since it is impossible to do so analytically for arbitrary or even very realistic field configurations, numerical methods are often required.

In practice however, directly solving the Dirac equation in spacetime-dependent fields may be numerically expensive. It can be more convenient to instead solve another equation, equivalent to the Dirac equation or an approximation to it, to discuss the Schwinger effect. This is the situation which arises in the widely-used Dirac-Heisenberg-Wigner (DHW) formalism~\cite{Vasak:1987um, Bialynicki-Birula:1991jwl, Zhuang:1995pd, Hebenstreit:2010vz,Li:2014nua}. (See~\cite{Gelis:2015kya} and references therein for real-time lattice techniques, which provide another powerful numerical method for investigating the Schwinger effect; this will also be discussed in Sec.~\ref{sec:beyondPW:back:schwing}).  In the limit of spatially homogeneous fields, the DHW formalism includes the quantum Vlasov approach \cite{Kluger:1998bm, Schmidt:1998vi} and can be solved analytically for some special cases (where the Dirac equation is also solvable \cite{Bialynicki-Birula:2011cln, Sheng:2018jwf}).  The equivalence of kinetic approaches with S-matrix methods is demonstrated in~\cite{Dumlu:2009rr,Fedotov:2010ue}.  See also~\cite{Aleksandrov:2019ddt} for numerical comparison between the DHW formalism and results from directly solving the Dirac equation.

In the DHW formalism one calculates the spacetime evolution of a (quasi-)phase-space distribution function.  A number of studies have been made within the DHW formalism to numerically study spacetime inhomogeneous-field effects (see Sec.~\ref{sec:ht3}).  A key ingredient is the equal-time Wigner function ${\mathcal W}$ \cite{Bialynicki-Birula:1991jwl}:
\begin{align}
	{\mathcal W}(\mbf{x},\mbf{p};t) 
	= \frac{-1}{2} \bra{{\rm 0;in}} \int {\rm d}^3\mbf{y}\,e^{-i\mbf{p}\cdot \mbf{y}} e^{+ie\int^{\mbf{x}+\mbf{y}/2}_{\mbf{x}-\mbf{y}/2}{\rm d}\mbf{z}\cdot {\bm {\mathcal A}}(t,\mbf{z})} [ \psi(t,\mbf{x}+\mbf{y}/2) , \bar{\psi}(t,\mbf{x}-\mbf{y}/2) ] \ket{{\rm 0;in}} , \label{eq:ht5---25}
\end{align}
in which $\mbf{x}$ and $\mbf{p}$ are to be interpreted as particle positions and (kinetic) momenta, while the Wilson line guarantees gauge invariance of ${\mathcal W}$.  Using the Dirac equation, one can derive an evolution equation for the Wigner function ${\mathcal W}$: 
\begin{align}
\label{eq:ht2--28}
	D_t {\mathcal W} = -\frac{1}{2} \mbf{D}_\mbf{x} \cdot [\gamma^0\mbf{\gamma}, {\mathcal W}] - im[\gamma^0, {\mathcal W}] - i\mbf{P} \cdot \{ \gamma^0\mbf{\gamma},{\mathcal W} \} \;,
\end{align}
where
\be\begin{split}
    D_t = \partial_t + e\int^{1/2}_{-1/2}{\rm d}\lambda\,\mbf{E}(t,\mbf{x}&+\lambda i\mbf{\partial}_\mbf{p}) \cdot \mbf{\partial}_\mbf{p} \;,
    \qquad
    D_\mbf{x} = \partial_\mbf{x} + e\int^{1/2}_{-1/2}{\rm d}\lambda\,\mbf{B}(t,\mbf{x}+\lambda i\mbf{\partial}_\mbf{p}) \times \mbf{\partial}_\mbf{p} \;, \\
    \mbf{P} &= \mbf{p} - ie\int^{1/2}_{-1/2}{\rm d}\lambda\,\lambda \mbf{B}(t,\mbf{x}+\lambda i\mbf{\partial}_\mbf{p}) \times \mbf{\partial}_\mbf{p} \;.
\end{split}
\ee
The initial condition for the differential equation (\ref{eq:ht2--28}) is given by considering times before the background electromagnetic field switches on at $t=t_{\rm in}$, in which case $\mathcal{W}$ can be calculated directly; one finds ${\mathcal W}(t_{\rm in}) = {\mathcal W}_{\rm vac} = \frac{1}{2p_0} [\mbf{\gamma}\cdot\mbf{p} - m] $.  
The Wigner function ${\mathcal W}$ carries, from (\ref{eq:ht2--28}), bispinor indices and has 16 independent components. These may be parametrised in terms of gamma matrices as
\begin{align}
	{\mathcal W} = \frac{1}{4} \left[ {\mathbb s} + i\gamma_5{\mathbb p} + \gamma^\mu {\mathbb v}_\mu + \gamma^\mu \gamma_5 {\mathbb a}_\mu + \frac{i}{2}[\gamma^\mu, \gamma^\nu] {\mathbb t}_{\mu\nu} \right] . 
\end{align}
This basis can be used to express various physical quantities once $\mathcal{W}$ is obtained by solving (\ref{eq:ht2--28}) numerically (or analytically in special cases). For example, the unregularised energy density is given by
\be
    {\mathcal E}(t,\mbf{x}) = \bra{0;{\rm in}}  \bar{\psi} \gamma^0 (i/2)\overset{\leftrightarrow}{\partial}_t \psi \ket{0;{\rm in}} = \int {\rm d}^3\mbf{p}\, [ m{\mathbb s} + \mbf{p}\cdot {\mathbb v}] \;.
\ee
One can infer from the \emph{integrand} of this expression a phase-space distribution function for created ``particles'', with the caveat that it is only strictly safe to do so at asymptotic times where interactions are turned off, see Sec.~\ref{sec:Schwinger}, while there always exists an ambiguity in how to define a distribution function for particles at intermediate times.  As a consequence, the construction of a distribution function is not unique.  A common choice is 
\begin{align}
	 f(\mbf{x},\mbf{p},t) 
		= \frac{ m{\mathbb s} + \mbf{p}\cdot {\mathbb v} }{p_0} - \frac{ m{\mathbb s}_{\rm vac} + \mbf{p}\cdot {\mathbb v}_{\rm vac} }{p_0} , \label{eq:ht---30}
\end{align}
which is obtained by regularising the energy density by subtracting the vacuum contributions at each instant of time and supposing that the energy density ${\mathcal E}$ is given by a momentum integral of the one-particle energy $p_0$ times the distribution function $f$.  Note that $f(\mbf{x},\mbf{p},t)$ includes contributions from both electrons and positrons and that spin degrees of freedom are already summed over.  As alternatives/improvements, one may for example construct spin-resolved phase-space distribution functions by inserting appropriate spin projection operators~\cite{Blinne:2015zpa}, or consider coarse-graining~\cite{Rafelski:1993uh, Kohlfurst:2017git}.  A good property of the choice (\ref{eq:ht---30}) is that, for spatially homogeneous and linearly-polarized electric fields, it coincides with quantum Vlasov approach~\cite{Dumlu:2009rr,Hebenstreit:2010vz}, see also~\cite{Li:2019rex}. Note that there are (infinitely) many other possible choices, reflecting the infinite ambiguity in defining  particle number at non-asymptotic times -- see Sec.~\ref{sec:Schwinger} for a discussion. (There is a further ambiguity in the Wigner function itself: the Wilson line could be replaced by any other function which guarantees gauge invariance, see e.g.~\cite{Lavelle:2011yc} and references therein. However, it is only for the straight Wilson line used above that ${\bf p}$ has the interpretation of a particle momentum.)

\subsubsection{Basis lightfront quantisation}
In `basis light-front quantisation' one begins in lightfront field theory~\cite{Bakker:2013cea}, see also Sec.~\ref{one-step/two-step separation}, so that $x^\LCp$ is again time, and then reduces the infinite number of degrees of freedom in QED to a finite number by truncating the state space in particle number, (discretised) momentum, and so on, in a particular basis inspired by lightfront holography~\cite{Brodsky:2014yha}. One then diagonalises the \emph{full} QED Hamiltonian in this finite basis to obtain non-perturbative eigenstates describing excitations of electrons, positrons, and photons. These are then evolved in time by numerically solving the Schr\"odinger equation to study their interactions~\cite{Vary:2009gt}. BLFQ inherits some of the useful properties of lightfront field theory; being quantised on a lightlike rather than spacelike hypersurface, the theory is explicitly causal. Lightfront field theories are ghost-free and, while fully relativistic, much of their structure, including dispersion relations and the Schr\"odinger equation to be solved, appears non-relativistic, which offers various simplifications. For more information and references see~\cite{Heinzl:2000ht,Brodsky:1997de,Brodsky:2014yha,Bakker:2013cea}.

Time-dependent basis lightfront quantisation, or `tBLFQ', extends the BLFQ formalism to include background fields in both QED~\cite{Zhao:2013cma,Zhao:2013jia} and QCD~\cite{Li:2020uhl,Li:2021zaw}. tBLFQ has been used to study nonlinear Compton scattering in backgrounds beyond plane waves, i.e.~those having transverse structure~\cite{Hu:2019hjx}. An advantage of tBLFQ is that it is fully quantum and explicitly real time, but a disadvantage is that available computing power severely limits the size of the truncated, finite system and thus the number of particles which can contribute in processes.

While non-perturbative, tBLFQ may, being based on lightfront quantisation, struggle to fully capture zero mode physics, which can be essential for non-perturbative contributions to the Schwinger effect~\cite{Tomaras:2000ag,Ilderton:2014mla}, hence which numerical method is most appropriate depends on the physical situation of interest. Perturbative contributions to pair creation from external fields can though be captured by tBLFQ~\cite{Lei:2022nsk}.


\section{Higher-order processes and resummation}\label{sec:higher}
In this section we will consider processes at higher orders in $\alpha$. One reason for considering higher orders is simply because a particle in a strong electromagnetic field can cause the production of many particles. For example, an electron emitting several photons is a higher-order process. Also, truncating the perturbation series in $\alpha$ at $O(\alpha)$ gives ``probabilities'' that can become larger than 1, if e.g. $\xi$ or $\mathcal{T}$ are ``too" large or in case of IR divergences. This just means that one has to include higher orders. Note that one has to include both tree-level processes {\it and} loops. Including all the relevant diagrams gives probabilities that are automatically less than 1.  

As mentioned, higher orders become important if, for example, $\xi$ or the (dimensionless) pulse length $\mathcal{T}$ are large enough, which is also a regime where higher orders can be approximated by suitable (sums) of incoherent products of $\mathcal{O}(\alpha)$ processes, see Sec.~\ref{sec:second}.
In Sec.~\ref{sec:second} we discussed some new methods for how to construct the two-step part of $\mathcal{O}(\alpha^2)$ processes using $\mathcal{O}(\alpha)$ strong-field-QED Mueller matrices. These Mueller matrices can also be used as building blocks to obtain the $N$-step part of general $\mathcal{O}(\alpha^N)$ processes, where the $N$-step part gives the dominant contribution for large $\xi$ and/or a long pulse and is given by a product of $N$ Mueller matrices. Apart from the Mueller matrices for the tree-level processes nonlinear Compton, see~\eqref{MCdefinition}, and Breit-Wheeler, we also have Mueller matrices for the $\mathcal{O}(\alpha)$ term in the probability of $e^\LCm\to e^\LCm$ and $\gamma\to\gamma$, see~\eqref{eMLdefinition} and~\eqref{gammaMLdefinition} and Fig.~\ref{fig:polarizationLoopSum}, i.e. for loops. Each of these building blocks in general involves two $\varphi$ integrals ($\varphi$ for the amplitude $M$ and $\varphi'$ for $M^*$), giving $2N$ integration variables $\varphi_i$ and $\varphi'_i$ with $i=1,\dots,N$. However, the lightfront-time ordering of the $N$-step can be enforced using step functions with only $\phi_i=(\varphi_i+\varphi'_i)/2$ as arguments ($\Theta(\phi_N-\phi_{N-1})\dots\Theta(\phi_2-\phi_1)$), and each $\theta_i=\varphi_i-\varphi'_i$ integral can be performed in terms of Airy functions in the LCFA regime or Bessel functions for a circularly polarised field in the LMA regime (for linear polarisation one finds more complicated generalised Bessel functions~\cite{1962JMP.....3...59R,2006JPhA...3914947K,2009PhRvE..79b6707L}). And even if one cannot find some pre-defined special functions, the $\theta_i$ integrals can anyway be performed {numerically} for each building block separately. This leaves one $\phi$ integral for each building block and they are nontrivially connected because of lightfront time ordering. Thanks to the simplicity of plane waves, all the transverse momentum integrals can be performed for each building block separately. This leaves one longitudinal momentum integral for each building block (e.g. the longitudinal momentum of the emitted photon in a Compton-scattering step). One therefore has $N$ integrals in $\phi_i$ and $N$ momentum integrals for each $\mathcal{O}(\alpha^N)$ diagram. For the second-order processes discussed in Sec.~\ref{sec:second} one can just perform these integrals with brute force, but at higher orders one is faced with many inter-connected integrals. Also, one is usually interested in inclusive processes (e.g. the expectation value of the electron momentum), which means that there will in general be many diagrams that contribute to each $\mathcal{O}(\alpha^N)$ for large $N$. And even after having obtained each order, one probably needs to find some way of resumming the $\alpha$ series. Indeed, if higher orders are important, then the effective expansion parameter is likely large and may lie beyond the radius of convergence (the radius can be finite for an approximation, even though the full/exact $\alpha$ series is expected to be asymptotic). Thus, developing a way to write down compact expressions for general higher-order diagrams is only half the work. In this section we will review resummation methods and various applications and practical studies of higher-order processes.

Recall that so far almost every calculation in strong-field QED is based on perturbation theory in $\alpha$ in the Furry picture. The worldline formalism offers another approach to both higher-order and non-perturbative calculations. It has been used in~\cite{Edwards:2021vhg} to calculate compact expressions for the $N$-photon amplitude at one-loop level in a general plane-wave background. In~\cite{Affleck:1981bma} the effective action is calculated using worldline methods to all orders in $\alpha$, for a weak electric field (see also Sec.~\ref{sec:ht6.1}). (The form of the result has also been independently guessed, though, based on the first two terms in the $\alpha$ expansion and physical intuition in~\cite{Lebedev:1984mei}.)

Often resummations of the $\alpha$ expansion have involved cases where one can, in some approximation, find explicit expressions for arbitrary orders. A text-book example is the exponentiation of IR divergences (without a background field). IR divergences are canceled when summing over indistinguishable soft processes, see~\cite{Ilderton:2012qe} for this cancellation in plane-wave backgrounds. Processes can also be indistinguishable when particles propagate collinearly. In~\cite{Edwards:2020npu} it was shown that non-laser photons that are collinear with the laser also allow for a resummation as a multiplicative exponential factor, including both real photon emissions and loops, giving a probability that scales like
\be
    P\sim \alpha\xi^2\exp\left\{-\alpha\xi^2\text{const.}\right\} \;.
\ee

Another type of series that often appears in cases where one can find explicit coefficients to all orders is a geometric series. 
In~\cite{Karbstein:2019wmj} it was shown that in the limit of very strong magnetic or electric fields, the reducible diagrams give the dominant contribution to the effective action and they give geometric series. We will come back to this in Sec.~\ref{sec:LBL}, see~\eqref{eq:alpha1loop}. Another geometric series appears in the $\alpha$ expansion of the solution to the Landau-Lifshitz equation~\cite{Heinzl:2021mji}. 

The study of the Ritus-Narozhny conjecture is an example where resummations will be particularly important. We will devote Sec.~\ref{sec:RN} to this topic.

In this section we will instead focus on cases where only a finite number of orders in $\alpha$ are available and where one cannot identify the coefficients with the expansion of some simple function. This is relevant because, although one might naively expect that the product of Mueller matrices might simply give expansions of the form  $z^n/n!$ or $z^n$ in which $z$ is a suitable expansion parameter proportional to $\alpha$, this is not the case; the lightfront-time and momentum ordering of the integrals in these products gives much less trivial expansions. 
However, before we continue we note that if one only considers the loop terms alone, i.e. without sums over the number of final state particles, then one does have an expansion of the form $z^n/n!$ which can hence be summed into an exponential form~\cite{Meuren:2011hv,Bragin:2017yau,Torgrimsson:2020gws,Podszus:2021lms}. Such loop sums are relevant e.g. for vacuum birefringence, where a photon changes its polarisation (both the type and degree) via sequences of any number of fermion loops~\cite{King:2016jnl,Meuren:2011hv}.
However, as emphasised in~\cite{Heinzl:2021mji}, unitarity implies certain cancellations between loops and other diagrams, which is important for e.g.~expectation values in RR.

Regarding light-by-light interactions, in~\cite{Bohl:2015uba} the field of a probe plane wave passing through a stronger pump plane wave was studied (using the Euler-Heisenberg effective action) by resumming a series involving a Bessel function, which is an example of cases where one knows all the coefficients in the series and can therefore find a unique resummation, but where the coefficients are more complicated than simply $(-1)^n$ or $1/n!$.

\subsection{Dyson-Schwinger equations}

The self-energy diagrams shown in the second lines of {Fig.~\ref{fig:polarizationLoopSum} as well as in Fig.~4 of the supplementary in~\cite{Torgrimsson:2021wcj}} can also be resummed by solving the Dyson-Schwinger equations, see e.g.~\cite{Podszus:2021lms},
\be
\begin{split}
(i\slashed{\mathcal D}-m)\Psi(x)&=\int\ud^4y\, M(x,y)\Psi(y) \\
-\partial_\mu\partial^\mu A^\nu(x)&=\int\ud^4y\, \Pi^{\nu\lambda}(x,y)A_\lambda(y) \;,
\end{split}
\ee
where ${\mathcal D}_\mu=\partial_\mu+ia_\mu$, $\Psi$ is the fermion/Dirac field including both the background field (assumed to be a plane wave) $a_\mu$ to all orders as well as all orders in radiative corrections $\alpha$, $A_\mu$ is the photon field, $M(x,y)$ is the electron mass ``operator" (i.e. the loop shown in {(1b) in Table~\ref{tab:RN_studies1980}}), and $\Pi^{\mu\nu}(x,y)$ is the polarisation tensor ({(1a) in Table~\ref{tab:RN_studies1980}}). At $\mathcal{O}(\alpha^0)$, but to all orders in $a_\mu$, $\Psi$ is given by the usual Volkov solution. $M(x,y)$ contains all irreducible diagrams\footnote{In general background fields (beyond plane waves) these quantities no longer need to be exclusively 1PI. Beyond PW there would e.g. be the tadpole on the electron line which is reducible with respect to cutting the photon line. What is relevant is a specific irreducibility, namely the diagrams to be included are irreducible such that in- and out-state are not becoming disconnected by cutting a line, cf. the tadpole correction in~\cite{Karbstein:2017gsb}.}. However, for a long pulse and/or large $\xi$ the dominant contribution comes from the diagrams in {Fig.~\ref{fig:polarizationLoopSum} and Fig.~4 of the supplementary in~\cite{Torgrimsson:2021wcj}}. In this regime, the Dyson-Schwinger equations can be solved~\cite{Meuren:2011hv,Meuren:2014uia,Bragin:2017yau,Podszus:2021lms,DiPiazza:2021szp}. As function of $\alpha$, the solutions take the form
\be
\Psi(p,s,x)\propto\exp\left\{i(\text{real Volkov terms})+i\alpha \tilde{M}_1(p,s,x^\LCp)\right\} \;,
\ee
and
\be
A(q,j)\propto\exp\left\{-il\cdot x+i\alpha\tilde{\Pi}_1(q,j,x^\LCp)\right\} \;,
\ee
where $\tilde{M}_1$ and $\tilde{\Pi}_1$ are derived from the one-loop terms in $M(x,y)$ and $\Pi(x,y)$ as shown in~\cite{Podszus:2021lms}, and depend on the spin $s$ or polarization $j$ (and momentum, $p$ or $q$). $\tilde{M}_1$ has a nonzero imaginary part, since this solution describes the propagation of a fermion without photon emission, which is an unstable state, i.e. the probability that the electron propagates through the laser without having emitted any photons is less than one. Similarly, $\tilde{\Pi}_1(x^\LCp)$ has an imaginary part because the photon can decay into a pair. These states have been used in~\cite{Meuren:2011hv} for the spin-transition probability without photon emission and in nonlinear Compton scattering and Breit-Wheeler pair production in~\cite{Podszus:2021lms}.

The ingredients, $M(x,y)$ and $\Pi^{\mu\nu}(x,y)$ to $\mathcal{O}(\alpha)$  are shown in {(1a) and (1b) in Table~\ref{tab:RN_studies1980}}. The electron mass operator loop has recently been calculated in~\cite{DiPiazza:2021szp,1975JETP...42..400B} for a general plane wave and used and discussed in~\cite{DiPiazza:2021szp} as an ingredient in the Dyson-Schwinger equation. The polarisation tensor has also been calculated in an arbitrary plane wave, see references in Sec.~\ref{sec:LBL}. Resummed expressions on the form $e^{\trm{const.} \times \alpha}$ have been used to study vacuum birefringence in~\cite{Dinu:2013gaa,Meuren:2014uia,King:2016jnl,Bragin:2017yau}. We will discuss vacuum birefringence in detail in Sec.~\ref{sec:LBL}.

The one-loop vertex correction has also been studied recently~\cite{DiPiazza:2020kze}. It was calculated in a general plane wave, focusing on UV/IR divergences and gauge dependence. This is relevant for the Ritus-Narozhny conjecture, see Sec.~\ref{sec:RN}.

\subsection{Resummation of incoherent-product approximation}

We come back now to the question of how to use the incoherent-product approximation (i.e. products of $\mathcal{O}(\alpha)$ Mueller matrices, see Sec.~\ref{Spin sums in the two-step}) for those higher-order processes where summation of loops alone is not enough. Perhaps the simplest and experimentally most relevant case is RR. It should be noted that RR does not automatically mean that one has to consider higher-order processes. Even in the classical limit, RR is nonzero already at $\mathcal{O}(\alpha)$, which is obtained from the emission of a single photon and one loop diagram (the scaling of the photon momentum with respect to $\hbar$ is cancelled by the factor of $1/\hbar$ from $\alpha$, see e.g.~\cite{Ilderton:2013tb}). 
However, in regimes relevant for e.g. upcoming laser experiments (see \secref{sec:intro:experiments}), there is an interest in measuring processes with multiple emissions. In other words, higher-order processes will be important and therefore studying their resummation is important. 

At $\mathcal{O}(\alpha^N)$, $N$ photons can be emitted, which in the Mueller-matrix approach (see Sec.~\ref{sec:second}) is described by the product of $N$ Compton Mueller matrices, ${\bf M}^{\rm C}\cdot{\bf M}^{\rm C}\dots{\bf M}^{\rm C}$, which are integrated with lightfront-time and longitudinal-momentum ordering. However, these are not the only building blocks that are relevant. It is perhaps not always realised, or at least not explicitly mentioned, but, as noted already above, it can also be important to consider loops. Loops provide important cancellations of e.g. infra-red divergences and are needed in order to be able to take the classical limit of RR~\cite{Ilderton:2013tb,Ilderton:2013dba} (see also~\cite{Krivitsky:1991vt,Higuchi:2004pr,Higuchi:2005an} and~\cite{Holstein:2004dn} for other processes). Loops (sequences of $\mathcal{O}(\alpha)$ loops) are also responsible for the nontrivial, anomalous magnetic moment part of the spin precession (see e.g.~\cite{BaierSokolovTernov,Ilderton:2020gno,Torgrimsson:2020gws}), see also discussion in Sec.~\ref{spin in codes} for how spin is treated in recent numerical codes. In the Mueller-matrix approach these loops can be obtained from the $\mathcal{O}(\alpha)$ building block ${\bf M}^{\rm L}$ mentioned in Sec.~\ref{sec:second}, see~\cite{Torgrimsson:2020gws}. At $\mathcal{O}(\alpha^N)$ there are therefore $2^N$ diagrams, e.g. ${\bf M}^{\rm L}\cdot{\bf M}^{\rm C}\cdot{\bf M}^{\rm C}\dots {\bf M}^{\rm L}$. Because of unitarity, many elements (but not all) of ${\bf M}^{\rm L}$ are similar or can be obtained from ${\bf M}^{\rm C}$. It is natural to group these $2^N$ diagrams together as~\cite{Torgrimsson:2021wcj} $({\bf M}^{\rm L}+{\bf M}^{\rm C})\cdot({\bf M}^{\rm L}+{\bf M}^{\rm C})\dots({\bf M}^{\rm L}+{\bf M}^{\rm C})$. Because of the combination ${\bf M}^{\rm L}+{\bf M}^{\rm C}$, many soft-photon problems are either absent or less severe in this inclusive quantity than what one might otherwise expect by considering only e.g. photon emission without loops. Thus, loops need to be included simply because they are numerically important, and omitting them does not actually make things simpler but rather introduces unnecessary problems.      

These products are lightfront-time and longitudinal-momentum ordered. In~\cite{Torgrimsson:2021wcj} a recursive formula was derived: Consider the expectation value of the longitudinal momentum of the electron, which is obtained from an expansion in $\alpha$, $\langle k\cdot P\rangle=\sum_{n=0}^\infty\langle k\cdot P\rangle^{(n)}$ where $\langle k\cdot P\rangle^{(n)}=(1/2){\bf N}_0\cdot{\bf M}^{(n)}\cdot{\bf N}_1=\mathcal{O}(\alpha^N)$ and ${\bf M}^{(n)}$ is essentially $n$ factors of $({\bf M}^{\rm L}+{\bf M}^{\rm C})$ with different arguments. By prepending a factor of ${\bf M}^{\rm L}+{\bf M}^{\rm C}$ at the beginning of the previous product, a recursive formula was obtained which gives ${\bf M}^{(n)}$ from ${\bf M}^{(n-1)}$,
\be\label{generalRecursive}
{\bf M}^{(n)}(\eta,\sigma)= 
\int_\sigma^\infty\!\frac{\ud\sigma'}{\eta}\!\int_0^1\!\!\ud s_\gamma({\bf M}^{\rm L}(\eta,\sigma',s_\gamma)\cdot{\bf M}^{(n-1)}(\eta,\sigma') 
+{\bf M}^{\rm C}(\eta,\sigma',s_\gamma)\cdot{\bf M}^{(n-1)}(\eta(1-s_\gamma),\sigma')) \;,
\ee 
where $\eta=k\cdot p/m^2$ and $s_\gamma=k\cdot l/k\cdot p$ is the ratio of the longitudinal momentum of the photon and the initial electron. The initial condition for $\langle k.P\rangle$ is ${\bf M}^{(0)}=\eta{\bf 1}$. The same equation but with ${\bf M}^{(0)}={\bf 1}$ gives the resummed total Mueller matrix for the probability of a general spin transition. The lower integration limit $\sigma$ ensures lightfront-time ordering, and the factor of $1-s_\gamma$ takes the recoil due to photon emission into account.

It was shown that the recursive formula~\eqref{generalRecursive} can be resummed right from the start, i.e. without choosing a specific field or electron parameters, into a matrix integro-differential equation for ${\bf M}=\sum_{n=0}^\infty{\bf M}^{(n)}$,
\be\label{integro-diff-eq}
\frac{\partial{\bf M}}{\partial\sigma}=-\int_0^1\frac{\ud s_\gamma}{\eta}({\bf M}^{\rm L}\cdot{\bf M}(\eta)+{\bf M}^{\rm C}\cdot{\bf M}(\eta[1-s_\gamma])) \;.
\ee
This equation is valid for arbitrary pulse shape and polarisation of the plane wave and, as long as the pulse is sufficiently long (see discussion in Sec.~\ref{sec:second}), even if $\xi$ is not large. In other words, it goes beyond standard LCFA treatments. Since it is a matrix equation with no reference to any specific spin or polarisation basis, it works for general spin/polarisation. If one omits the spin and considers LCFA then it resembles the structure of older kinetic equations that have been used previously to study RR~\cite{Neitz:2013qba,2018PhRvE..97d3209N}. In those studies no resummation was mentioned.

Having obtained the integro-differential equation one no longer has to think about resummation. This seems therefore like another example where the $\alpha$ expansion can be resummed thanks to the fact that one has an expression for each higher order. However, instead of resumming the $\alpha$ expansion from the start into an integro-differential equation, one can instead approach the problem in a similar way to which the $\chi$ expansions have been resummed in Sec.~\ref{sec:second}. (This method is also closer to how one would approach the problem when, in the future, one has figured out how to calculate the $\alpha$ expansion beyond the incoherent-product approximation). In this approach one starts with ${\bf M}^{(0)}$ and uses the recursive formula to obtain ${\bf M}^{(1)}$, which is then used to obtain ${\bf M}^{(2)}$ etc. This procedure gives explicit results for a finite number of terms in the $\alpha$ expansion. One might for example obtain $\langle k\cdot P\rangle^{(n)}$ for $n=0,...,10$. In the classical limit one finds a geometric series to leading order with an obvious resummation, $1-z+z^2-\cdots=1/(1+z)$, where $z\propto\alpha$. For the difference in the final momentum for an electron that initially has spin up or down, $\langle k\cdot P\rangle(\uparrow)-\langle k\cdot P\rangle(\downarrow)$, one can also find explicit expressions for all orders~\cite{Torgrimsson:2021wcj,Torgrimsson:2021zob}. This, though, is much less trivial; the coefficients are not simply $(-1)^n$ but rather a more complicated expression involving the harmonic number. However, using techniques to solve recursive formulas, one can again find a unique resummation~\cite{Torgrimsson:2021wcj,Torgrimsson:2021zob}, which happens to involve logarithms (it is unique in the sense that it reproduces the expansion coefficients to all orders, rather than just the first e.g. 10 orders which would be the case for any resummation of a series of which only the first 10 orders are known). However, away from the leading low-energy limit, one will not find a series with some obvious resummation such as $(-z)^n$ or $(-z)^n/n!$ and one will only have access to a finite number of terms. 

\begin{figure}
    \centering
    \includegraphics[width=0.49\linewidth]{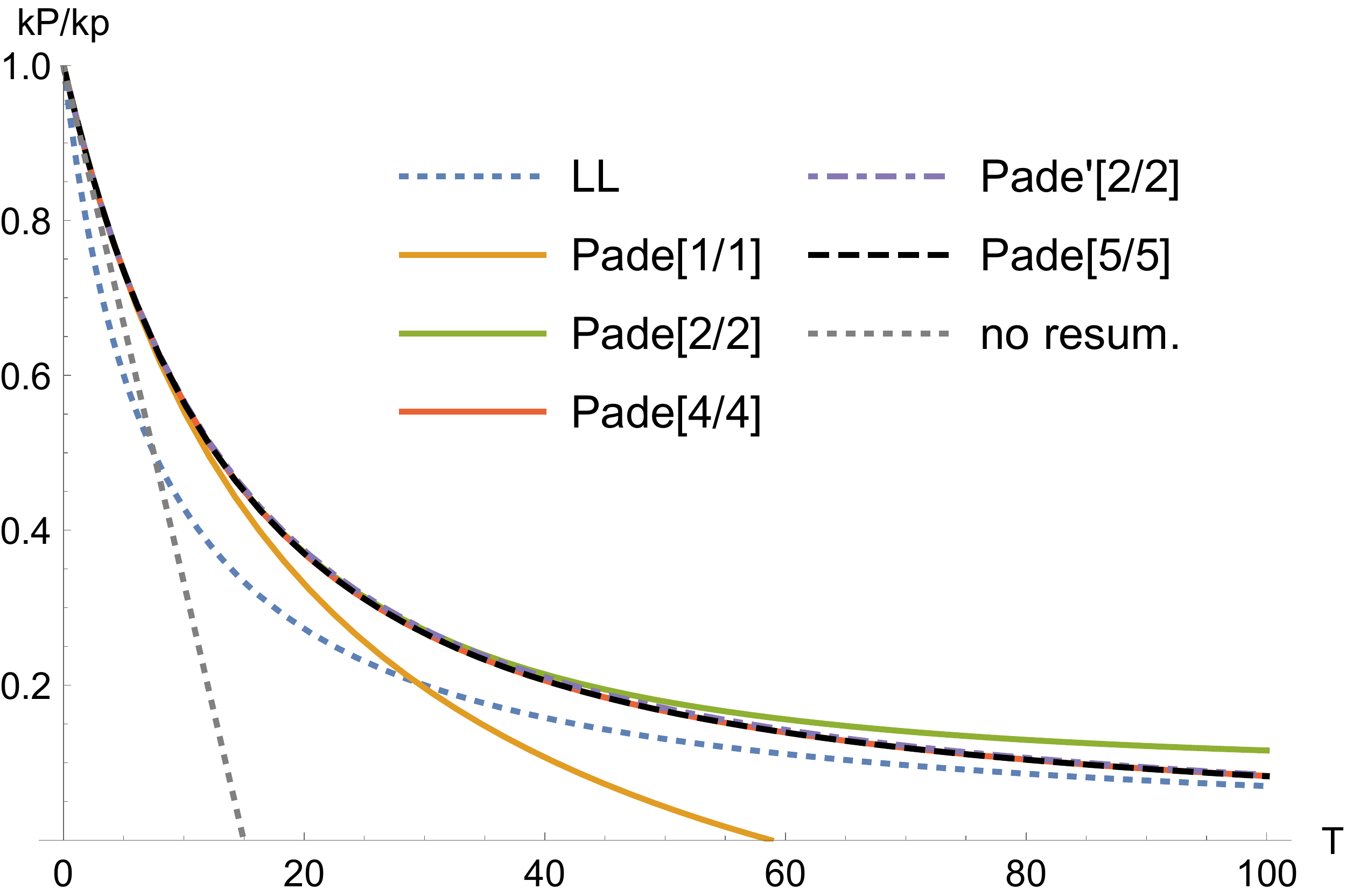}
    \includegraphics[width=0.49\linewidth]{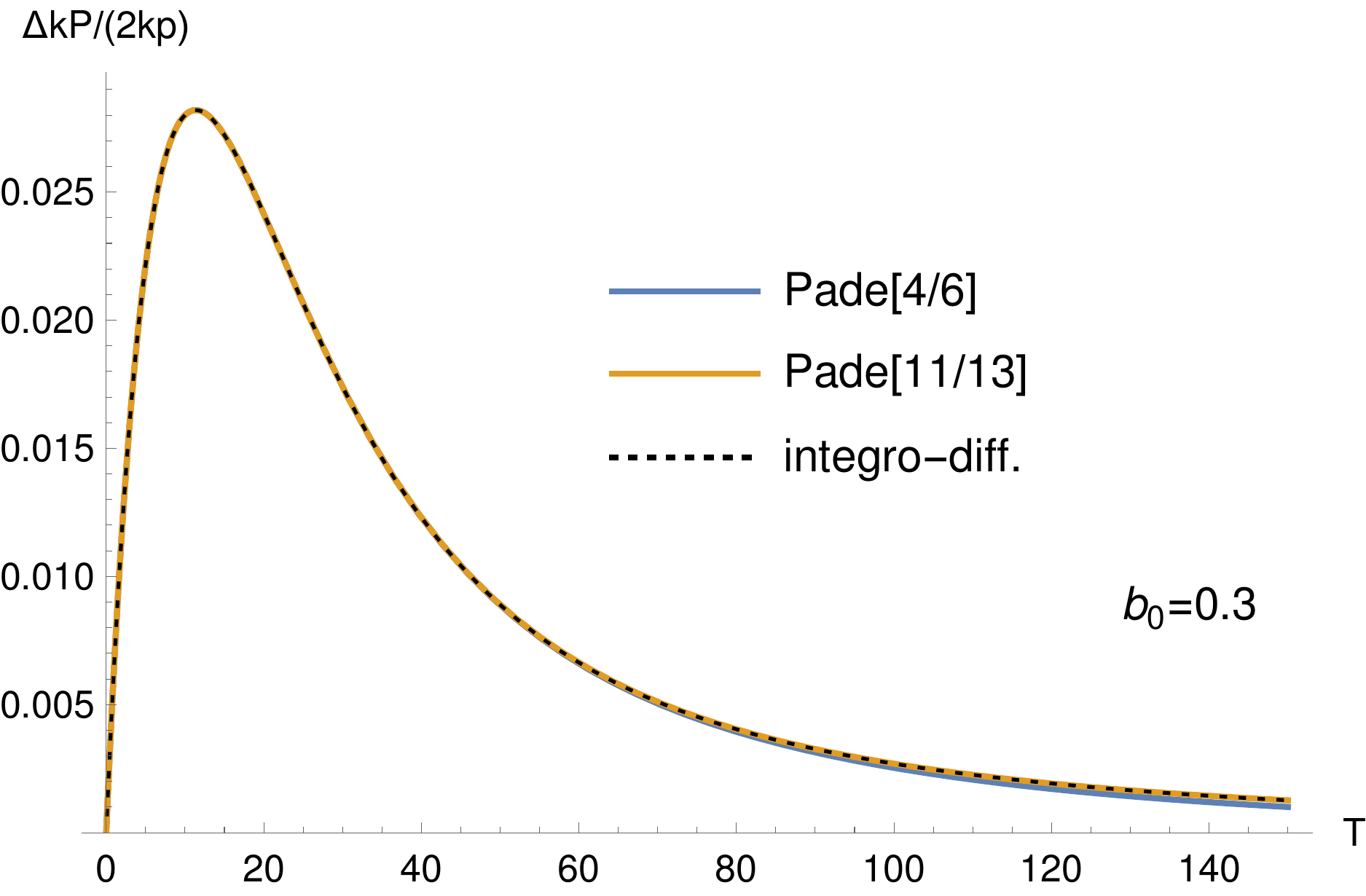}
    \includegraphics[width=0.49\linewidth]{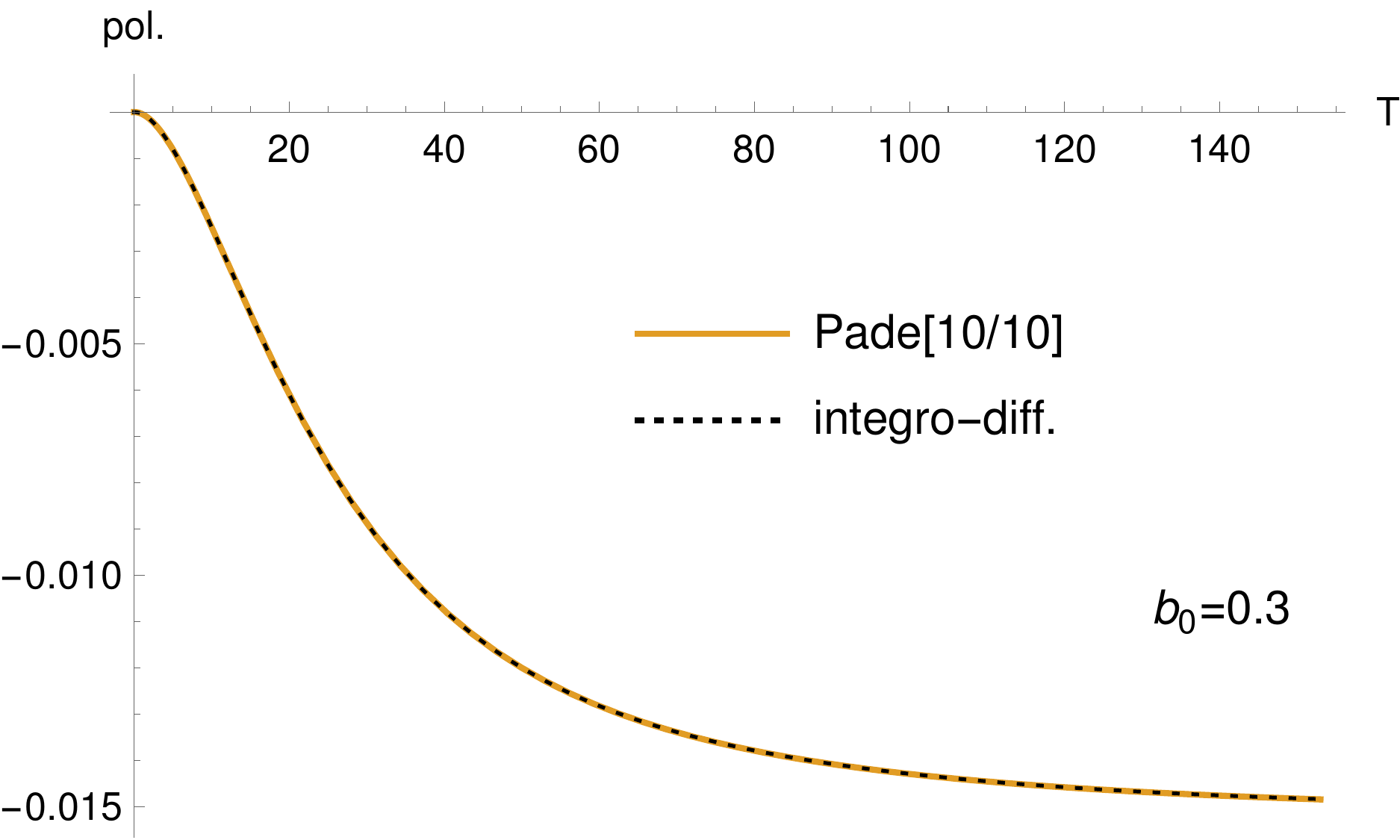}
    \includegraphics[width=0.49\linewidth]{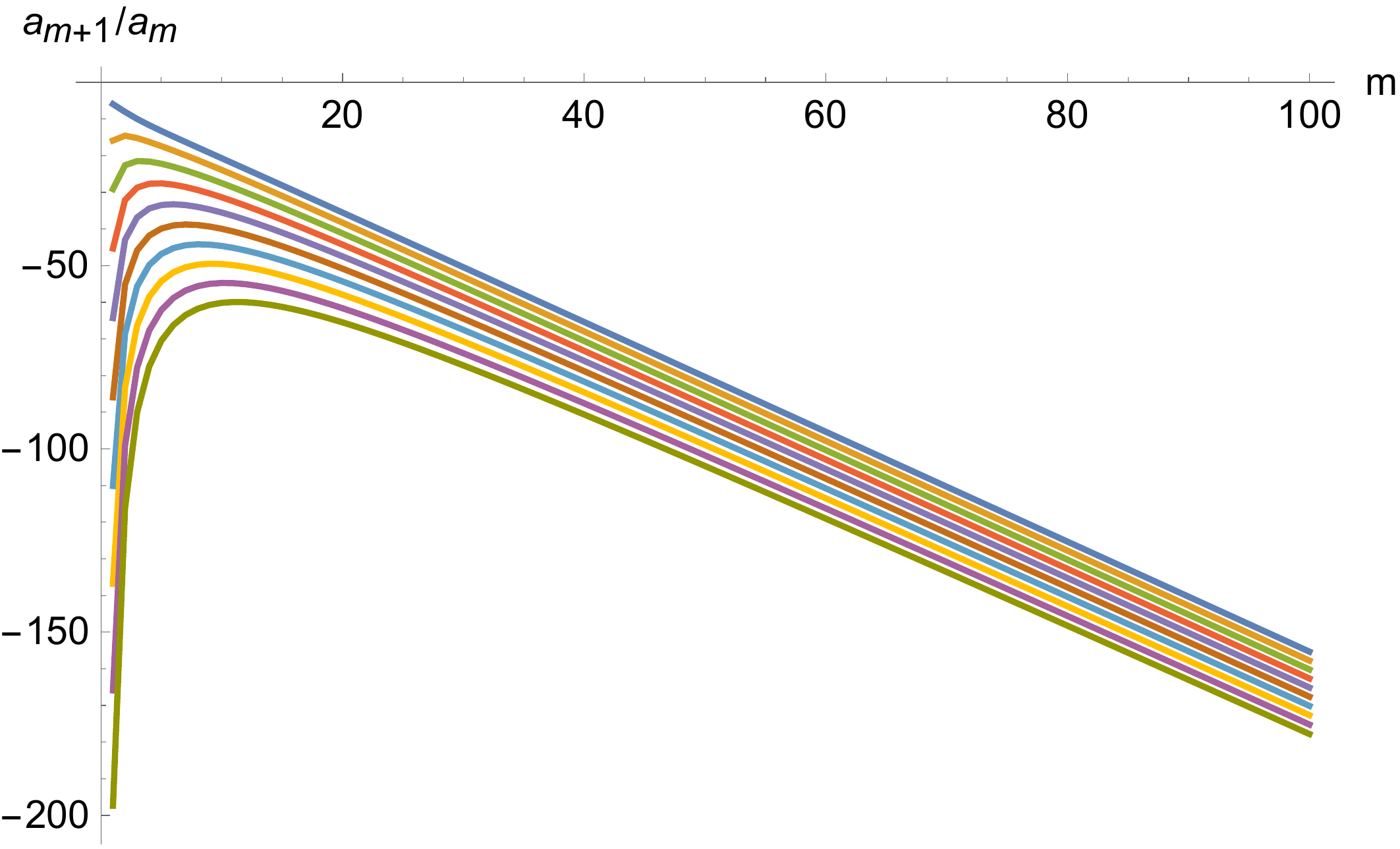}
    \caption{The upper left plot, taken from from~\cite{Torgrimsson:2021wcj}, shows the longitudinal momentum of an initially unpolarised electron as a function of an effective pulse-length parameter for a constant-crossed field. The different Pad\'e curves show the convergence of the Pad\'e resummation in~\eqref{kPPade}. Pad\'e' shows that a faster convergence is achieved by using the anticipated large-$T$ scaling as an extra condition. The upper right plot, taken from~\cite{Torgrimsson:2021zob}, shows the difference in the final momentum due to the initial spin, for a circularly polarised monochromatic field with $\xi=1$, and $\eta=b_0$. "integro-diff" is the result obtained by solving the integro-differential equation in~\eqref{integro-diff-eq}. The lower left plot, taken from~\cite{Torgrimsson:2021zob}, shows the final degree of electron polarisation for an initially unpolarised electron, for a circularly polarised field, which is zero at $\mathcal{O}(\alpha)$. The lower right plot, taken from~\cite{Torgrimsson:2021wcj}, shows the ratios of neighboring coefficients in the $\chi$ expansion of the first ten orders in the $\alpha$ expansion of $k\cdot P$ for an initially unpolarised electron in a constant-crossed field, where the uppermost line is for $\mathcal{O}(\alpha)$ and the last line for $\mathcal{O}(\alpha^{10})$.}
    \label{fig:quantumRRplots}
\end{figure}

Thus, in the general case one needs to choose some resummation method. For the RR expansions considered in~\cite{Torgrimsson:2021wcj,Torgrimsson:2021zob} the first terms in the $\alpha$ expansion indicate a finite radius of convergence. No Borel transform is therefore needed, but a simple direct sum only works (in principle) within the radius of convergence. However, a direct sum may not be the best approach, because if one is close to the radius of convergence, for example, then a very large number of terms may be required, meaning difficult or impractical calculations. A finite radius of convergence suggests that one could resum the series with Pad\'e approximants. In other words, with the recursive formula one obtains a truncated expansion (considering a constant field for simplicity and writing `tru' for truncated)
\be\label{kPexpansion}
\xi\langle k\cdot P\rangle_{\rm tru}=\chi+c^{(1)}T+c^{(2)}T^2+\cdots+ c^{(n)}T^n \;,
\ee
where $T=\alpha\xi\Delta\phi$ is an effective expansion parameter, with $\Delta\phi$ the pulse length, $c^{(n)}(\chi)$ are nontrivial functions of $\chi$, and $n$ is a finite, not very large, order (e.g. $n=10$). 
In the classical limit we have $c^{(n)}\propto\chi^n$, and then it would be natural to also include a factor of $\chi$ in the effective expansion parameter.
Note that, even though $\alpha\ll1$, for sufficiently large $\xi\Delta\phi$, $T$ can be large and higher orders become important. Obviously, the truncated series will eventually break down since it goes as $T^n$ as $T\to\infty$. The idea is to decipher the information that is encoded in $c_i$ in a way that preferably does not require one to keep calculating more and more terms (increasing $n$) as $T$ increases. A Pad\'e resummation is achieved by matching $\xi\langle k\cdot P\rangle_{\rm tru}$ with the $\alpha$ expansion of
\be\label{kPPade}
\xi\langle k\cdot P\rangle_{\rm resum}=\chi+\frac{\sum_{i=1}^IA_i(\chi)T^i}{1+\sum_{j=1}^J B_j(\chi)T^j} \;,
\ee
which determines the coefficients $A_j$ and $B_j$. In general one finds higher precision by increasing $I$ and $J$ (and consequently $n$), but the relative size of $I-J$ is not determined by $\xi\langle k\cdot P\rangle_{\rm tru}$ alone. If one does not know how the exact result should scale in the limit of large argument ($T\gg1$ here), then one often finds good results with diagonal or near diagonal approximants ($I=J$ or $I\sim J$). In this case one expects $\xi\langle k\cdot P\rangle$ to decrease as $T$ increases, which leads one to choose $I=J$ so that the Pad\'e approximant can cancel the leading-order term ($\chi$).    
It was shown in~\cite{Torgrimsson:2021wcj,Torgrimsson:2021zob} that such Pad\'e resummations tend to converge quickly, i.e. one only needs to calculate relatively few terms in the $\alpha$ expansion. This means that this approach can be competitive or faster than the integro-differential approach.

In this case one can obtain an even faster convergence since one can guess that the quantum result should converge to the classical in the limit of a very long pulse, $T\gg1$, see~\cite{Torgrimsson:2021wcj}. This is based on the fact that $\chi$ decreases over time due to RR, i.e. the dynamics becomes more and more classical, and the fact that the classical solution becomes independent of the initial momentum in the long-pulse limit, 
so one can expect that this also happens in the quantum case and then the result for a general initial momentum will agree with that for a low momentum, for which the classical result applies. This gives not just $I=J$ but also fixes $A_I$ and $B_I$, so that given $n$ terms from $\xi\langle k\cdot P\rangle_{\rm tru}$ one can go to one Pad\'e order higher, $[m/m]\to[m+1/m+1]$, which moreover will converge to the exact result in the limit $T\gg1$ even at low orders. 
An important point is that it was not necessary to impose this extra condition on the approximant; one still finds the same results. However, doing so means a faster convergence, since forcing the resummation to the expected result at both small and large $T$ does not leave much room for large discrepancies at intermediate values of $T$. This holds also for other resummations: the more additional information one has (or can guess) the fewer terms are needed.

For $\langle k\cdot P\rangle(\uparrow)-\langle k\cdot P\rangle(\downarrow)$ one can show explicitly in the classical limit that the large-$T$ limit involves a factor of $\ln T$~\cite{Torgrimsson:2021wcj,Torgrimsson:2021zob}. One can therefore also expect such log terms in the quantum case. For this spin difference there is therefore no choice of $I$, $J$ and $A_j$ and $B_j$ that gives a resummation that reproduces the exact scaling at $T\gg1$, since a Pad\'e approximant can never give log terms. However, this is not a problem for reasonably large $T$ because the factor of $\ln T$ grows slowly. Hence, one can still speed up the convergence by choosing a Pad\'e approximant with the same scaling as the factor of $1/T^n$ that multiplies $\ln T$. Hence one can still find fast convergence with Pad\'e approximants. Thus, even though a Pad\'e approximant cannot reproduce the exact large-$T$ scaling in this case, due to the $\ln T$ part, this resummation method still works very well. 

Final results for these $\alpha$ resummations are shown in Fig.~\ref{fig:quantumRRplots} for a constant-crossed and a circularly-polarised monochromatic field.  

Two different methods for calculating the $\alpha$ expansion (without the integro-differential equation) were presented in~\cite{Torgrimsson:2021wcj,Torgrimsson:2021zob}. In one of them one first creates a numerical interpolation function of ${\bf M}^{(1)}(\chi)$ for $0<\chi<\chi_{\rm max}$ and plugs that into the recursive formula to create an interpolation function of ${\bf M}^{(2)}(\chi)$ and so on. In the other approach, one first expands ${\bf M}^{(1)}$ in a power series in $\chi$, and then plugs that series into the recursive formula to obtain a power series expansion of ${\bf M}^{(2)}$ etc. Each of these $\chi$ expansions are asymptotic and therefore need to be resummed with e.g. the Borel-Pad\'e method. But ${\bf M}^{(2)}$ is obtained from ${\bf M}^{(1)}$ before resumming ${\bf M}^{(1)}$ etc. For the constant field and circularly monochromatic field considered in~\cite{Torgrimsson:2021wcj,Torgrimsson:2021zob}, this double resummation approach was the fastest. 

In the double-resummation approach one has for each $n$ in the expansion in~\eqref{kPexpansion}
\be
c^{(n)}=\chi^{1+n}\sum_{m=0}^{M}a^{(n)}_m\chi^m \;,
\ee
where $a_m\sim(-1)^m m!$ for large $m$. The ratios of neighboring coefficients for the first $\sim100$ terms are plotted in Fig.~\ref{fig:quantumRRplots}. Note that we are again dealing with a truncated series, i.e. we only have access to a finite number ($M$) of terms. We can resum this expansion using the methods mentioned in Sec.~\ref{resummation of chi expansions}, e.g. the Borel-Pad\'e method. 
After the $\chi$ expansion has been resummed the $\alpha$ expansion can be resummed.

A different type of recursive formula was obtained in~\cite{Tamburini:2019tzo} for the probability $\prob_n(\varepsilon',t)$ that an electron has energy $\varepsilon'$ and has emitted $n$ photons at time $t$,
\be
\frac{\ud\prob_n}{\ud\varepsilon'}(\varepsilon',t)=\int_{-\infty}^t\ud\tau S(t,\tau;\varepsilon')\int_{\varepsilon'}^{\varepsilon_i}\frac{\ud^2W}{\ud\tau\ud\varepsilon'}(\varepsilon',\varepsilon,\tau)\frac{\ud\prob_{n-1}}{\ud\varepsilon}(\varepsilon,\tau) \;,
\ee
where 
\be
S(t,t';\varepsilon)=\exp\left\{-\int_{t'}^t\ud\tau\frac{\ud W}{\ud\tau}(\varepsilon,\tau)\right\}
\ee
is the probability that the electron does not emit between $t$ and $t'$, and $W$ is the probability of photon emission in a constant-crossed field (see \secref{sec:first:nlc} for more detail on nonlinear Compton and \secref{sec:approx:lcfa} for the LCFA). Apart from the fact that this is a different quantity, this formula uses ordinary time and energy rather than lightfront time and longitudinal momentum, and spin is omitted. Also, it corresponds to a different ordering of terms, as one can see by noting that $W=\mathcal{O}(\alpha)$ so the exponential part of $S$ is linear in $\alpha$ and hence $\prob_n$ is not $\mathcal{O}(\alpha^n)$ but is rather given by a partial resummation (of what corresponds to loops).

\subsection{Classical limit of quantum radiation reaction}\label{sec:higher:classicalRR}

So far we have discussed quantum RR, but even in classical RR there are still unsolved and debated problems. A number of alternative classical RR equations have been proposed (several of which have been compared with the classical limit of QED in~\cite{Ilderton:2013dba}). We will discuss the two most common: the Lorentz-Abraham-Dirac (LAD) and the Landau-Lifshitz (LL) equation. Both can be written
\be\label{eq:allRR}
m\ddot{x}^\mu=e\mathcal{F}^{\mu\nu}\dot{x}_\nu+\frac{2}{3}\frac{e^2}{4\pi}R^\mu \;,
\ee
where the first term is the Lorentz force due to the background field $\mathcal{F}_{\mu\nu}$ and the RR term is given by (suppressing the indices)
\be
R^{\rm LAD}=\dddot{x}+\ddot{x}^2\dot{x} 
\qquad
R^{\rm LL}=e\dot{\mathcal{F}}\dot{x}+e^2\mathcal{F}\mathcal{F}\dot{x}+(e\mathcal{F}\dot{x})^2\dot{x} \;.
\ee
As is well known, the standard equation of motion describing classical RR, LAD, has unphysical solutions with acausal preacceleration or runaway solutions with diverging momentum (see e.g.~\cite{Burton:2014wsa,Blackburn:2019reaching,Gonoskov:2021hwf} for a recent review of RR). A reduction of order, essentially by substituting the solution into itself and omitting higher order terms (though see~\cite{Zhang:2013ria,Ekman:2021eqc} for studies of higher orders and their resummation), leads to LL. (See Sec.~\ref{sec:reduction} for reduction of order applied to the Klein-Gordon and the Dirac equations.) LL is free from the unphysical solutions of LAD but gives predictions that are ``close'' to the preaccelerating solution of LAD. The difference, including the preacceleration part, is small in the classical regime, i.e. where one can expect these classical solutions to be valid and where quantum effects can be neglected, see e.g.~\cite{PhysRevE.84.056605}. LL is therefore widely accepted as a classical equation for practical purposes. 

A QED approach to RR gives us a series in $\alpha$. Calculating higher orders exactly is of course extremely difficult, but even the classical limit can be challenging. Several papers~\cite{Krivitsky:1991vt,Higuchi:2004pr,Higuchi:2005an,Ilderton:2013tb,Ilderton:2013dba} have studied the classical limit of quantum RR to $\mathcal{O}(\alpha)$ and compared with the prediction of classical equations\footnote{The classical limit of non-relativistic QED was studied in~\cite{Moniz:1976kr} without making such an expansion in $\alpha$.}. {In~\cite{Krivitsky:1991vt} both the coupling to the quantized photon field as well as the background field were treated to lowest order in perturbation theory, i.e. to $\mathcal{O}(e^3)$ ($e^2$ from the quantized field and linear in the background field). {This was justified by arguing that RR effects are small in the rest frame of the electron, and then working in that frame.} It was shown that the classical limit of QED agrees with LL to this order. The calculation was done for an arbitrary background field, but the RR contribution to the force (the time-derivative of the momentum expectation value, $\ud\langle {\bf P}\rangle/\ud t$) at $\mathcal{O}(e^3)$ only involves a term proportional to $\ud{\bf E}/\ud t$, which is linear in the field strength and vanishes for a constant field or tends to average out for an oscillating field. A Furry-picture expansion was used in~\cite{Higuchi:2004pr,Higuchi:2005an,Ilderton:2013dba}, i.e. the coupling to the quantized photon field was treated to leading order but without making an expansion in the background field $eE$. It was shown in~\cite{Ilderton:2013dba} (for a plane-wave background)} that both LAD and LL agrees with QED at $\mathcal{O}(\alpha)$, while already at this order some of the other proposed classical equations could be ruled out. LAD and LL can be distinguished by going to higher orders in $\alpha$. If one expects the $\alpha$ expansion to be asymptotic even in the classical limit, then one would also expect the classical limit may agree with LAD rather than LL, since LAD gives asymptotic series while LL has a finite radius of convergence. Regardless of which equation agrees with QED, LAD and LL of course still agree well in the classical limit, but if it turns out that higher orders agree with LAD, then one would also have confirmed both LAD \emph{per se} and as the starting point for the derivation of LL as an approximate equation. 
{The first order quantum correction to the ($\mathcal{O}(\alpha)$) Larmor formula was calculated in~\cite{Higuchi:2009ms} in a non-relativistic approximation for fields that depend on either time or one spatial coordinate. While~\cite{Krivitsky:1991vt,Higuchi:2009ms,Ilderton:2013dba} considered the expectation value of the momentum operator, one can also consider the expectation value of a position operator~\cite{Higuchi:2004pr,Higuchi:2005an,Ilderton:2013dba}.}

However, from a phenomenological point of view, one may ask in which parameter regime higher orders are actually important. As already discussed, one can expect them to be important for example at large $\xi$ or a long pulse, when a large number of photons are emitted. In the classical regime we can see this explicitly from the exact solution to LL in a plane wave~\cite{2008LMaPh..83..305P}
\be\label{LLexactSol}
P_{\LCm,\LCperp}=\pi_{\LCm,\LCperp}+\frac{\Delta}{1+\Delta\int\ud\phi\,{\bf a}'^2}\left[\pi'-\int\ud\phi\,{\bf a}'^2\pi\right]_{\LCm,\LCperp} \,,
\ee
where $\pi_\mu=p_\mu-a_\mu+(2a\cdot p-a^2)k_\mu/(2k\cdot p)$ is the zeroth order, i.e. the solution to the Lorentz-force equation without RR, $\Delta=(2/3)\alpha\eta$ ($\eta=k\cdot p/m^2$), and the remaining component follows from the on-shell condition, $P_\LCp=(m^2+P_\LCperp^2)/(4P_\LCm)$.
{The $\pi'$ term in~\eqref{LLexactSol} corresponds to the term that was reproduced in~\cite{Krivitsky:1991vt}, while all terms in the square brackets were reproduced in~\cite{Ilderton:2013dba}.}
From~\eqref{LLexactSol} we see that higher orders are important if
\be
\Delta\int\ud\phi\,{\bf a}'^2\sim\alpha\xi\mathcal{T}\chi\gtrsim1 \;,
\ee
i.e. if $\xi$ and/or $\mathcal{T}$ are sufficiently large\footnote{As emphasised, what is sufficiently large depends on other large or small parameters. In this classical regime, for example, we see explicitly that $\xi$ and/or $\mathcal{T}$ has to be large enough to compensate for the factor of $\chi\ll1$ in addition to $\alpha\ll1$.}.
As we have already discussed, in such regimes one can approximate higher orders by suitable incoherent products. Thus, with the methods for quantum RR developed in~\cite{Torgrimsson:2021wcj,Torgrimsson:2021zob} one can derive the classical limit of QED to leading order in $\xi\gg1$ and/or a long pulse. It was shown in~\cite{Torgrimsson:2021wcj} that this agrees exactly to all orders with the solution to LL. And, since LL and LAD converges in the limit of a long pulse~\cite{Kazinski:2010ce,Kazinski:2013vga,Ekman:2021eqc,Torgrimsson:2021zob}, this is also what one would expect if LAD is the equation that agrees with exact classical limit of QED (i.e. beyond the leading order in $\xi\gg1$ and/or long pulse length). See also~\cite{Elkina:2010up,Neitz:2013qba,2018PhRvE..97d3209N} for comparisons between kinetic RR equations and LL.

Since LL is the equation of choice for practical computations and since it has now been confirmed to agree with the classical limit of QED to all orders, is there then any motivation for calculating the classical limit of higher orders beyond the leading order in $\xi$ and/or long pulse? The following is one motivation. If LAD is indeed the correct equation then one would expect it to be possible to calculate these higher orders from QED, because then the result should agree with the expansion of LAD which does not appear too complicated~\cite{Ekman:2021eqc,Torgrimsson:2021zob}. This would then be an example where one can study the higher-order structure, e.g. convergence or asymptotic nature, of the QED $\alpha$ expansion.   
If we assume for the moment that LAD is correct, then we might be able to learn something about the general QED expansion by studying how to resum the asymptotic $\alpha$ expansion of LAD, to which we now turn.

\subsection{Resummation of LAD}\label{sec:resum_LAD}

The incoherent product approach seems to give a RR series with a non-zero radius of convergence, see~\cite{Torgrimsson:2021wcj,Torgrimsson:2021zob}. In general one expects the $\alpha$ expansion of QED to be asymptotic. This is usually associated with a factorially growing number of Feynman diagram, which explains why the incoherent-product approximation of RR does not lead to an asymptotic $\alpha$ series (recall that there are only $2^N$ diagrams at $\mathcal{O}(\alpha^N)$). There are currently no tractable methods for obtaining all-order quantum RR beyond the incoherent-product approach. However, as noted above, in the classical limit one can study the LAD equation. 

\begin{figure}
    \centering
    \includegraphics[width=.49\linewidth]{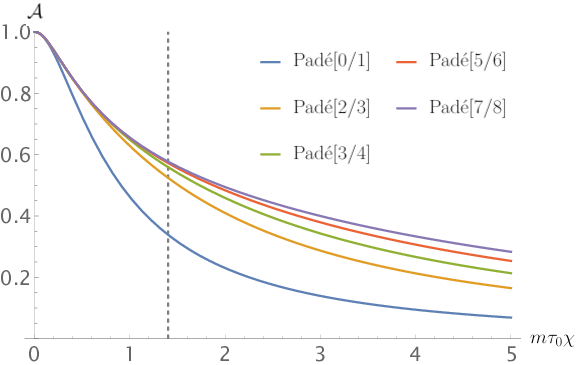}
    \includegraphics[width=.49\linewidth]{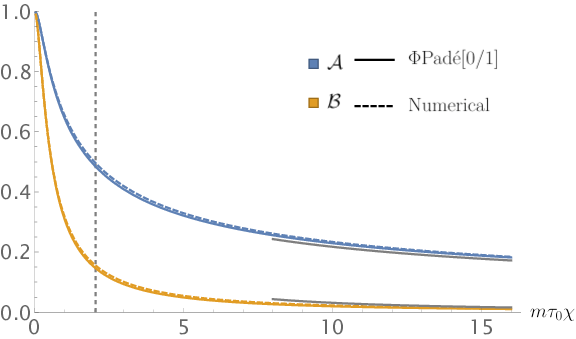}
    \caption{First plot shows $\mathsf{A}$ in~\eqref{ABinLAD} resummed using the standard Borel-Pad\'e method. The second plot shows $\mathsf{A}$ and $\mathsf{B}$ resummed using the new resummation method in~\cite{Alvarez:2017sza}, which allows one to incorporate the scaling at large argument into the resummation, thus significantly reducing the number of terms needed from the small-argument expansion. Figures taken from~\cite{Ekman:2021eqc}.}
    \label{fig:ABinLADresum}
\end{figure}

As mentioned, LL is obtained from LAD by reduction of order and truncating at first order in $\alpha$. This procedure can be continued by repeatedly inserting the equation into itself to remove higher derivatives, which gives an equation with terms proportional to $\alpha$, $\alpha^2$, $\alpha^3$ and so on. In the nonrelativistic limit this simplifies considerably and one finds an asymptotic series in $\alpha \ud/\ud t$, where $\ud/\ud t$ is the time derivative~\cite{Zhang:2013ria}. Resumming this series with the Borel method gives the preaccelerating solution of LAD~\cite{Zhang:2013ria}. This explains how preacceleration can appear even though each order separately does not give preacceleration. In the relativistic case one in general finds many different types of terms which makes the corresponding expansion and resummation much harder. In~\cite{Ekman:2021eqc,Torgrimsson:2021zob} the LAD expansion was resummed in the relativistic case for a constant crossed field, and in~\cite{Ekman:2021czy} for a circularly polarised monochromatic field. Since the field is constant and since $(f^3)_{\mu\nu}=0$ there are only a couple of different terms. In~\cite{Ekman:2021eqc} a reduction of order was performed and then the $\alpha$ expansion was resummed on the level of the differential equation, i.e. 
\be\label{ABinLAD}
\dot{u}^\mu=\mathsf{A}f^{\mu\nu}u_\nu+\tau_0\mathsf{B}(\mathbb{P}f^2)^{\mu\nu}u_\nu \;,
\ee
where $\tau_0=2\alpha/(3m)$, $\mathbb{P}^{\mu\nu}=g^{\mu\nu}-u^\mu u^\nu$ and $\mathsf{A},\mathsf{B}$ are functions of $\tau_0\chi$ which can be obtained by resumming the corresponding asymptotic series. In~\cite{Torgrimsson:2021zob} the $\alpha$ expansion was instead resummed on the level of the solution, by writing it as 
\be
P^\mu(\tau)=m\dot{u}^\mu=\left(g_0\delta^\mu_\nu+g_1\frac{{f^\mu}_\nu}{\chi_0}+g_2\frac{{(f^2)^\mu}_\nu}{\chi_0^2}\right)P^{(0)\nu} \;,
\ee
where $P^{(0)\nu}$ is a constant reference momentum and $g_i$ are functions found by resumming the corresponding asymptotic series (one of them is obtained from the other two by an algebraic equation due to the mass-shell condition $P^2=m^2$). Runaway solutions can be obtained by expanding the $g_i$ functions as trans-series~\cite{Torgrimsson:2021zob,Ekman:2021czy}, but those solutions are anyway unphysical. For times when the electron is inside the constant field, an expansion in powers of $\alpha$ only (and no exponentials) gives an asymptotic series which after a resummation agrees with a numerical solution of the original version of LAD with no runaways but with pre-acceleration shortly before the field turns on.  
The asymptotic $\alpha$ expansion can be resummed either with standard Borel-Pad\'e (using a suitable choice of variable)~\cite{Torgrimsson:2021zob} or~\cite{Ekman:2021eqc} using the new method proposed in~\cite{Alvarez:2017sza}. The result for $\mathsf{A}$ and $\mathsf{B}$ in~\eqref{ABinLAD} is shown in Fig.~\ref{fig:ABinLADresum}.

A suitable expansion parameter for this expansion of LAD is
\be\label{LL_LAD_exppar}
\delta=\frac{2}{3}\alpha\chi_0 \;,
\ee
where $\chi_0$ is some fixed, e.g. initial, $\chi$ value. {The importance of this parameter (and a second parameter involving the wave length of the field) was identified in~\cite{Landau:1975pou}.} If one chooses a rescaled lightfront-time variable $u$ such that the LL solution~\eqref{LLexactSol} can be written as $g_0=1/(1+\delta u)$,
then for LAD $g_0$ can be expanded in a power series $g_0=\sum_{n=0}^\infty\delta^n g_0^{(n)}(u)$, where the coefficients $g_0^{(n)}(u)$ are polynomials in $u$. However, if one instead absorbs a factor of $\delta$ into a new lightfront-time variable $v=\delta u$ then the zeroth order is given by $1/(1+v)$ and the coefficients at higher order are also rational functions with some $\log$ terms~\cite{Torgrimsson:2021zob}. Thus, the parameter $\delta$ characterises the difference between LAD and its leading order approximation, LL. Of course, for a classical description to be valid we need $\chi_0\ll1$, which makes $\delta\ll\chi_0\ll1$, i.e. for physical values of $\alpha$ and $\chi$ the difference between LL and LAD is actually very small.

\subsection{Spin and polarisation in numerical codes for higher-order processes}\label{spin in codes}

In this section we will discuss recent developments in the use of spin and polarisation dependent rates in numerical codes, e.g. particle-in-cell (PIC) codes, for higher order QED processes. No resummation is mentioned in these works, but what is calculated corresponds to quantities that one could analyse using a top-down QFT approach and resummation.
Developments in spin/polarisation in PIC codes have, in the last couple of years, been made in parallel, but with little exchange between the two approaches, or at least without much detailed comparison.

However, we can in fact already compare the new ideas and developments in these two approaches, e.g. with respect to the questions of how to treat spin/polarisation sums of intermediate particles or how to include the "no-photon emission" or loop contributions.

Including spin and polarisation in codes is motivated by the fact that it contributes on the same order of magnitude in $\xi$ and/or $\mathcal{T}$ as the spin-averaged terms and in the last couple of years there have been several promising studies that suggest that high-intensity lasers can be used to generate polarised particles beams.
Some of the first papers to initiate this include~\cite{DelSorbo:2017fod,2018PPCF...60f4003D}, where it was shown that fermions can obtain, in a short amount of time, a high degree of polarisation in intense fields, by considering fermions in a rotating electric field (the magnetic field nodes of two rotating colliding lasers). For this case one can find a non-precessing spin basis, but it was noted in~\cite{2018PPCF...60f4003D} that away from these magnetic nodes it is in general not possible to find such a basis. Recall our discussion in Sec.~\ref{sec:second} about spin sums for intermediate particles. The next papers we will discuss use Stokes vectors instead, which can allow for more general cases. (See Sec.~\ref{sec:first} for an equivalent way of representing the spin dependence in terms of a $2\times2$ polarisation density matrix.)

It is only very recently that spin and polarisation resolved rates have been included in PIC codes, motivated by e.g.~ prospects of generating polarised particle beams using high-intensity lasers. 
However, spin and polarization has previously been included in the CAIN code~\cite{YokoyaCAIN}.
As we have explained in Sec.~\ref{sec:second}, in some cases it is enough to consider fermions that have either spin up or down with respect to a certain direction, or photons that have either $\epsilon_\mu^{(1)}$ or $\epsilon_\mu^{(2)}$, where $\epsilon_\mu^{(i)}$ are some basis vectors. However, on the probability level this is in general not enough, and one is instead led to consider general spin/polarisation using Stokes vectors, see Sec.~\ref{sec:second}. Since PIC codes {work} on the probability level, one would naturally expect Stokes vectors to be useful in PIC codes. 

In~\cite{Li:2018fcz} a PIC code was developed which uses rates that include the dependence on the fermion spins via Stokes vectors (which are called ``spin polarisation vectors'' there), but no polarisation for the photons yet. The code worked as follows. As other PIC codes, random numbers are generated to determine e.g. whether or not a photon should be emitted. At each such event the electron spin ``collapses'' to parallel/antiparallel to some ``instantaneous spin quantisation axis'' (SQA), which was chosen to be the magnetic field in the electron's rest frame. Between photon emissions the spin precesses according to the T-BMT equation. As a first application, it was found that an initially unpolarised electron beam can be split into two parts by a laser with small ellipticity, where the two parts have opposite polarisation with tens of milliradians separation and up to 70\% degree of polarisation~\cite{Li:2018fcz}.     

The same method was used in~\cite{Wan:2019gow} to study the polarisation of positrons created via nonlinear Breit-Wheeler (neglecting the polarisation of the intermediate photon), again for elliptical polarisation, which led to positron beams with 90\% degree of polarisation. 
The method was used in~\cite{Chen:2019vly} to study the production of polarised positrons from unpolarised electrons (with the intermediate photon treated as unpolarised), and a two-color field was used to counteract the fact that the spin polarisation induced on a fermion in one cycle tends to average out in a more symmetrically oscillating field as the magnetic field direction flips sign. This gave positron beams with a 60\% degree of polarisation. (The positron is polarised at creation slightly depolarises due to the subsequent photon emissions~\cite{Chen:2019vly}.)
Electron polarisation by a two-color field has also been studied in \cite{Seipt:2019ddd}, using another Monte-Carlo algorithm and a kinetic approach. \cite{Seipt:2019ddd} also studied spin dependent RR.
Apart from e.g. two-color or elliptical beams, another solution to avoid the averaging out of the induced polarisation was proposed in~\cite{Song:2021wou}, by replacing the laser with the strong field given by a second electron beam, e.g. FACET-II, which gives a field without such oscillations. 

A PIC code with Stokes vectors for both fermion spin and photon polarisation in nonlinear Compton scattering was developed in~\cite{Li:2019oxr}, which found that electrons which initially has longitudinal or transverse polarisation lead to photons that are circularly or linearly polarised, respectively. This gave high-energy photon beams with high degree of polarisation, 95\%, and a flux suitable for vacuum birefringence experiments. 
A PIC code was used in~\cite{Wan:2020zet} to study the role of the polarisation of the intermediate photon in two-step trident with an initially unpolarised electron. It was found that including the polarisation gives a $>10\%$ difference, see also~\cite{King:2013zw}.

In~\cite{Li:2020bwo} it was shown how the polarisation can be transferred to the produced positrons by initially longitudinally polarised electrons and a circularly polarised laser, again resulting in high degrees of polarisation. 
They chose the SQA to be $\pm{\bf b}/{|{\bf b}|}$, where ${\bf b}$ is a vector that gives the dependence of the photon-emission probability $\prob=a+{\bf S}_f\cdot{\bf b}$ on the outgoing (3D) spin Stokes vector ${\bf S}_f$. The sign is chosen with the help of a random number. The spin collapses even if a photon is not emitted. If a photon is not emitted then the spin collapses onto $\pm{\bf d}$, where ${\bf d}$ is \emph{defined} from the no-photon-emission probability $\prob_{\rm no\gamma}=(1/2)(c+{\bf S}_f\cdot{\bf d})$, see also user's manual for the CAIN code~\cite{YokoyaCAIN}.
The no-photon-emission contribution was obtained in~\cite{Li:2020bwo,YokoyaCAIN} using an indirect approach, essentially by appealing to unitarity and using the spin-dependent results for photon emission. Although it was not mentioned in~\cite{Li:2020bwo,YokoyaCAIN}, the no-photon-emission contribution considered there is given by the loop. In other words, by considering the loop one obtains the no-photon-emission contribution directly. Between photon-emission or no-photon-emission events, these PIC codes include spin precession by using the T-BMT equation.
The nontrivial, anomalous magnetic moment part of the T-BMT equation also comes from the loop~\cite{BaierSokolovTernov,Ilderton:2020gno,Torgrimsson:2020gws}.

It was emphasised in~\cite{Tang:2021azl} that different choices of SQA will in general give different results, and that collapsing an electron spin onto up and down a SQA effectively corresponds to making a measurement after each photon emission. An approach was proposed in~\cite{Tang:2021azl} where the spin vector of the electron after emission is again determined by an equation like $\prob=a+{\bf S}_f\cdot{\bf b}$, but where the spin vector is allowed to be partially polarised $|{\bf S}|<1$. It was also emphasised that not including the no-photon-emission part can give incorrect results. The approach in~\cite{Tang:2021azl} was found to give results consistent with the approach in~\cite{Li:2020bwo}.
The results of~\cite{Tang:2021azl} are also compared with those of~\cite{Li:2018fcz} and found to be similar.
In one approach in~\cite{Tang:2021azl} the no-photon-emission part combined with the T-BMT equation. Although it is not mentioned, both contributions do in fact come from the loop.

\subsection{Cascades}\label{subsec:cascades}

QED cascades have received increasing attention in the last decade~\cite{Bell:2008zzb,Fedotov:2010ja}. These can arise in a background when a single probe `seed' particle generates, through successive nonlinear Compton and Breit-Wheeler events, a higher-order process. (Cascades can also be initiated in intense fields in the absence of a probe, by spontaneous pair creation from the Schwinger effect, which will be discussed in more detail in \secref{sec:Schwinger}.) Although the dynamics of the cascade can sensitively depend on initial field and probe configuration (see, e.g., \cite{Tang:2013vot}), two main types of dynamics have been identified. In the first type, a high-energy electron or photon probe can initiate a cascade in a background, with each further generation of particles being produced with a lower lightfront momentum \cite{sokolov:2010pair,bulanov:2013el}. This is sometimes referred to as a `free field' \cite{King:2013zw} or `shower' type cascade \cite{Mironov:2014xba} due to their similarity to cosmic showers. The second type is illustrated by the seed particle being initially at rest, but once illuminated by the intense background is accelerated to sufficiently high lightfront momenta as to begin cascading. This has been dubbed a `field driven' \cite{King:2013zw} or `avalanche' type cascade \cite{Mironov:2014xba}. The principle difference between the two types of cascades is the source of energy for particle production. An avalanche draws the energy from the field via particle acceleration, while a shower requires high energy seed particles rather than a strong field. However, in a strong field after the shower has faded out it can also turn into an avalanche \cite{Mironov:2014xba}. Note that the dynamics of carrier multiplication due to impact ionization, which had been observed in semiconductors pumped by THz pulses (see, e.g., \cite{wen:2008ultrafast,hoffmann:2009impact,hirori:2011extraordinary}), is very similar to the dynamics of avalanche cascades that we consider here.

\begin{figure}
    \centering
    \includegraphics[width=0.8\linewidth]{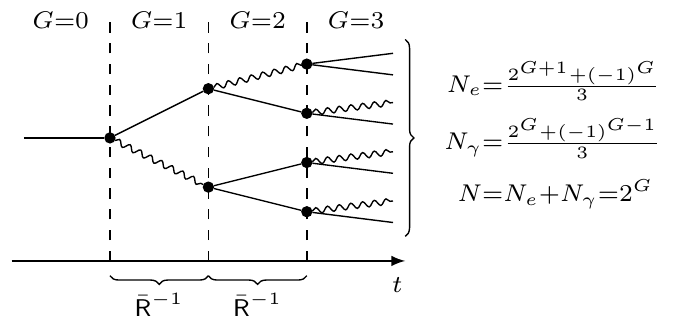}
    \caption{A 'regular branching tree' model of a cascade, for which the successive generations $G=0,1,2,\ldots$ of cascade particles are well separated. Applies literally if the difference of the rates for Compton and Breit-Wheeler processes is ignored and at each vertex the momenta are distributed equally, but remains useful in order-of-magnitude estimates of the actual cascade dynamics.}
    \label{fig:reg_branch_tree}
\end{figure}

The onset and dynamics of a cascade are a complicated nonlinear many-particle phenomenon with an interesting interplay between the QED processes responsible for the production of pairs and photons, and the plasma fields governing the dynamics of radiating electrons and positrons \cite{Nerush:2010fe,Elkina:2010up}. As such, cascades are commonly described using the plasma kinetic theory, with quantum collision operators describing the SFQED processes. {These (quantum) transport equations have been recently derived from first principles using the formalism of out-of-equilibrium quantum field theory \cite{Fauth:2021nwe}.} To the full extent, the electron and positron motion is in general extremely complicated and even stochastic \cite{esirkepov:2015ac,jirka:2016el}. The main tool employed to accurately simulate cascades is PIC (particle-in-cell) codes supplemented with LCFA-based event generators for QED processes. (For a recent review of cascade simulations and their possible applications see~\cite{Gonoskov:2021hwf}.) Still, some insights can be gained if the growth rate and duration of a cascade are sufficiently high that to discuss the physics in terms of average quantities. 

Since an ultrarelativistic particle radiates in a narrow forward cone, a shower type cascade can be well approximated as one dimensional. Then, assuming the validity of LCFA and hence conservation of the parameter $\chi$, and assuming also $\chi\gg1$, the rates for the nonlinear Compton and Breit-Wheeler processes scale the same way $\propto\chi^{2/3}$ (see the lower lines in \eqref{eq:first:NLCrate} and \eqref{eq:first:NBWrate}). If the difference in numerical coefficients in these scalings is ignored, then the shower type cascade would look like a regular branching tree, see Fig.~\ref{fig:reg_branch_tree}. In such a model the total multiplicity would be proportional to the $\chi$-parameter $\chi_0$ of the initial seed particle (electron, positron or photon), as it is clear from the figure that the $\chi$-parameter of the $G^\text{th}$ generation of {$N=2^G$} particles equals $\chi_G=\chi_0/N$, and emission of hard photons capable for pair production effectively stops when $\chi_G\lesssim1$. The same remains true on average even when accounting for the difference in the coefficients of the scalings of the rates, which only changes the overall coefficient of proportionality. Thus the overall multiplicity of such cascades can be notable, but not macroscopically large. The details of the actual cascade dynamics can be studied semi-analytically following the standard approach of solving 1D kinetic equations by a suitable integral transform (see, e.g., \cite{akhiezer:1994ki} and references therein).

In contrast, the dynamics of an avalanche-type cascade is essentially determined by the field-invoked acceleration of the particles in between the emission events. The key point is that in a \textit{general} background {of electric type ($E>B$)} and for slow particles such acceleration results in a short-term growth of their energy and parameter $\chi$, thus eventually restoring them to about the same values as before the emission. (This does not apply to a few \textit{specific} over-idealised backgrounds (e.g., constant fields and plane waves), for which the lightfront momentum is conserved between the emissions and hence avalanche-like cascades cannot set in). It is worth stressing that we assume the parameters are such that the acceleration takes place on a scale exceeding the formation scales of the QED processes, on which LCFA still remains valid. This mechanism maintains energy and parameter $\chi$ of the majority of the particles on average at about the same level, which can be controlled by the shape of the driving laser pulse 
\cite{Tamburini:2017sxg}. In such conditions the number of particles in a cascade initially grows exponentially fast ($\propto e^{\Gamma t}$) and can rapidly achieve macroscopic values. The exponential growth saturates due to either escaping of the particles from the strong field region, or relaxation of the plasma when achieving a balance between the splitting and inverse (merging) processes, or due to the depletion of the field \cite{Fedotov:2010ja,Nerush:2010fe,Gris2016absorption}. The two last scenarios result in the formation of an extremely dense, macroscopically populated region of electron-positron plasma (with density of the order of relativistic critical plasma density), which intensively emits hard $\gamma$-rays. The scenario involving field depletion also questions the possibility to attain field strength beyond or even close to the Schwinger field $E_{\rm S}=m^2/e$ with optical lasers \cite{Fedotov:2010ja}. 

\begin{figure}
    \centering
    \includegraphics[width=0.5\linewidth]{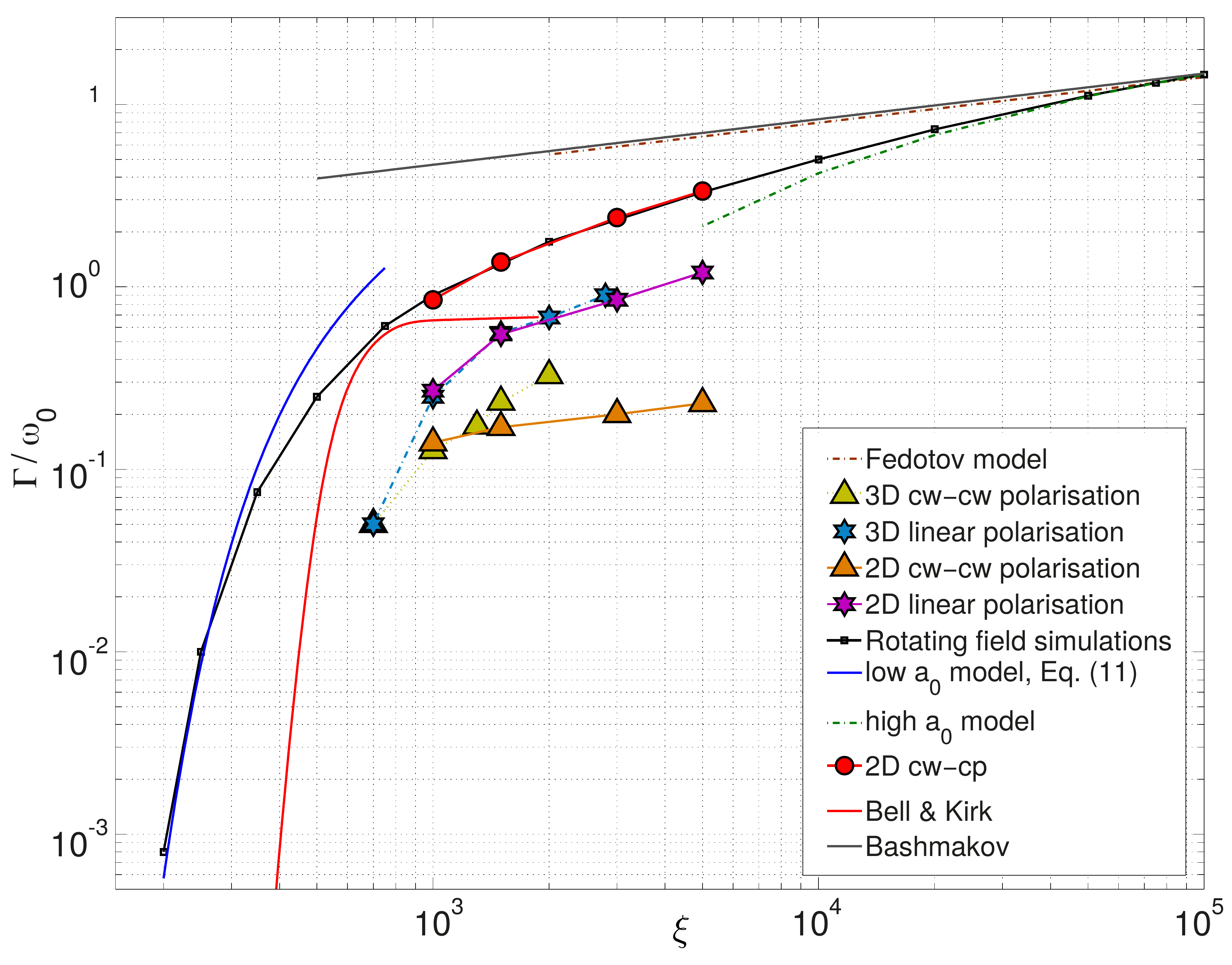}
    \caption{A comparison of the cascade growth rate in several analytical models \cite{Elkina:2010up,Bell:2008zzb,2014PhPl...21a3105B,grismayer:2017se} to the results of 2D and 3D simulations in a uniformly rotating electric field and in standing waves of different [linear, clockwise-clockwise (cw-cw) and clockwise-counterclockwise (cw-cp)] polarisations. Adapted from \cite{grismayer:2017se}.}
    \label{fig:cascade_models}
\end{figure}

A test bed to study avalanche-type cascades, both analytically and in simulations, is a homogeneous uniformly rotating electric field. The advantages of using this model are (i) acceleration of particles entails the growth of the $\chi$-parameter, as required, (ii) the classical evolution of the energy and $\chi$ between emissions can be derived exactly, making some analytical progress possible. Also, a homogeneous uniformly rotating electric field mimics some features of the antinodes of a standing wave formed by two counterpropagating, circularly polarised laser pulses. By taking also into account that one needs $\chi\gtrsim1$ at the moment of photon emission in order for the emitted photon to later produce a pair, one can use the $\chi\gg1$ asymptotics in~\eqref{eq:first:NLCrate} for the NLC rate. With that, by equating the total emission probability to unity, one can estimate the average duration $t_r$ of acceleration between recurrent emissions, hence also the average $\chi$ and Lorentz factor of the particles in a cascade~\cite{Fedotov:2010ja,Elkina:2010up}. 
Specifically, one finds that\footnote{The form of~\eqref{cascade_est} has been chosen so there appears the ratio of the characteristic atomic energy $\alpha^2m$ to the optical frequency $\omega$, which is slightly above unity.}
 \begin{align}\label{cascade_est}
t_r\simeq \frac1{\omega}\sqrt{\frac{\omega}{\alpha^2m}}\left(\frac{\alpha E_{\rm S}}{E}\right)^{1/4},\quad 
\gamma(t_r)\simeq \frac1{\alpha}\sqrt{\frac{\alpha^2 m}{\omega}}\left(\frac{E}{\alpha E_{\rm S}}\right)^{3/4},\quad
\chi(t_r)\simeq \left(\frac{E}{\alpha E_{\rm S}}\right)^{3/2},
\end{align}
where $\omega$ is the field rotation frequency. The cascade growth rate can be then estimated as $\Gamma\sim t_r^{-1}$. By requiring $\chi(t_r)\gtrsim1$ one estimates an effective threshold for cascade production as $E\gtrsim\alpha E_{\rm S}$ \cite{Fedotov:2010ja,Elkina:2010up}. Despite the number of explicit and implicit approximations assumed in the derivation, the scalings \eqref{cascade_est} agree surprisingly well with the simulation results in the strong field limit (i.e., where asymptotics in use for the emission rate are formally valid) \cite{Elkina:2010up,nerush:2011an,grismayer:2017se}. The scalings \eqref{cascade_est} were re-derived explicitly by passing to dimensionless variables in the kinetic equations \cite{nerush:2011an}. Their further analysis also revealed an analytical parametrisation of the high-energy tails of particle energy distributions in a cascade, in terms of the growth rate. This rate could not, however, be determined analytically and was instead used in numerics as a fitting parameter. However, \eqref{cascade_est} overestimates the growth rate for moderate and weaker fields \cite{grismayer:2017se}, see Fig.~\ref{fig:cascade_models}. The figure also demonstrates the sensitivity of the cascade growth rate $\Gamma$ to the choice of field model, which cannot be captured by using simple order-of-magnitude estimates. The position of the effective threshold for cascade production is also notably overestimated with respect to the simulation results, according to which cascades in counterpropagating laser pulses set in at an intensity of about $I\simeq 10^{24}\text{W/cm}^2$ \cite{grismayer:2017se}.
 
Two ways of refining and extending the above estimates have been suggested. First, in order to fix the numerical coefficients in \eqref{cascade_est} and extend these scalings to weaker fields, in which the NBW rate is much smaller than the NLC one, it was proposed to replace the relevant kinetic equations for phase space distributions with the simplified rate equations involving solely the total numbers of particles of each species \cite{2014PhPl...21a3105B,grismayer:2017se}. This proved it possible to correct the analytical model by fixing the proper numerical coefficients in \eqref{cascade_est} for strong fields and moreover to extend it in order to also match the weak field simulation results \cite{grismayer:2017se}, see Fig.~\ref{fig:cascade_models}. The original approach of \cite{Bell:2008zzb} based on the Zeldovich model of electron motion in a uniformly rotating electric field with classical radiation reaction \cite{zeldovich:1975in} is also reliable for estimating the effective threshold of cascade production \cite{grismayer:2017se}. Another improvement was extending the estimates of cascade growth, to fields more general than a uniformly rotating electric field. This is especially demanding if the field possesses multiple incommensurable scales and can be done along the same line as above but with a properly refined short-time dependence of $\chi(t)$. The latter was done for a standing wave by solving electron equation of motion in terms of a Taylor {expansion near the antinodes of the electric field
\cite{2014PhPl...21a3105B}. The same paper also demonstrated that electrons and positrons can first accelerate in an electric region and later radiate their energy in magnetic regions, thus explaining how an intensive avalanche production was possible in a linearly polarised standing wave,} which was initially a subject of debates \cite{bulanov:2010sch}. Very recently, this approach was generalised to an arbitrary field {of electric type} by means of its derivative expansion and a general formula for $\chi(t)$ was obtained. The latter was applied to explain the simulation results for avalanche production in a single focused laser pulse \cite{mironov:2021on}.

An emerging topic in the study of cascades is the evolution and transport of the spin and polarisation of the cascade particles \cite{King:2013zw,Seipt:2020uxv}. This links to recent studies of the spin- and polarisation dependence of the fundamental SFQED processes discussed in~\secref{sec:first}, in particular the issue of chaining together first order processes to approximate higher orders. As has been shown above, reliable predictions require carefully taking into account the polarisation of the intermediate states when resumming. However, investigations of cascade formation usually employed unpolarised particle distributions and particle production rates. Only a few publications have so far taken into account at least photon polarisation \cite{King:2013zw} or both photon polarisation and lepton spin \cite{Seipt:2020uxv} in studies of cascade formation. In both cited cases, numerical simulations showed a reduced cascade growth rate \cite{King:2013zw,Seipt:2020uxv}.

\section{Light-by-light scattering}\label{sec:LBL}
This chapter reviews recent developments with regard to quantum vacuum nonlinearities and light-by-light scattering phenomena driven by macroscopic electromagnetic fields. This includes progress in fundamental theory such as new insights into the strong-field behavior of the Heisenberg-Euler effective action in constant fields, and progress in the study of phenomenological consequences such as a refined modelling of experimental scenarios allowing for the detection of light-by-light scattering effects with current and near future high-intensity laser technology.

As opposed to the classical notion of the vacuum, the quantum vacuum can no longer be considered as an empty and inert ground state, but amounts to a non-trivial quantum state which encodes information about the full particle content of the underlying fundamental quantum theory (QFT) in the form of {virtual processes}.
These can be probed with electromagnetic fields.
Upon subjecting the microscopic QFT to a macroscopic classical electromagnetic field (gauge potential ${\cal A}$) and integrating out the quantum fields one arrives at a low energy effective field theory governing the dynamics of the macroscopic field in the quantum vacuum.
The latter encodes vacuum fluctuations in effective nonlinear self-interactions of the prescribed field, and generically accounts for couplings to all orders in ${\cal A}$.

\subsection{Low energy effective action}\label{sec:6.1}

\begin{figure}
 \centering
 \includegraphics[width=0.65\columnwidth]{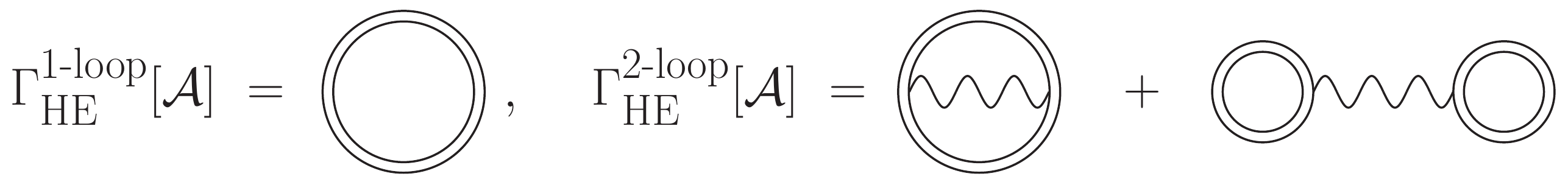}
\caption{Diagrammatic representation of the Heisenberg-Euler effective action up to two loops. The double line denotes the dressed fermion propagator accounting for arbitrarily many couplings to the macroscopic field ${\cal A}$.}
\label{fig:fig_GammaHE}
\end{figure}

Within the Standard Model of particle physics the leading effective self-interactions of $\cal A$ mediated by vacuum fluctuations are governed by QED, whose quantum degrees of freedom are the dynamical photon field $A$ and the electron-positron fields $\psi$ and $\bar\psi$.\footnote{Contributions of other particle sectors of the Standard Model are highly suppressed as their masses $M$ effectively fulfill $M\gg m$ while their charges are of the same order as $e$.
While this statement does not seem to hold for $u$ and $d$ quarks, the latter directly couple to gluons and their physics is governed by quantum chromodynamics (QCD). Central properties of QCD are color confinement and infrared slavery, which implies strong coupling at low energies. Hence, it is to be expected that upon integrating out the gluons, the spectrum of QCD at low energies is characterised by colorless composite states made up of quarks and gluons. Their lightest representatives are the pions fulfilling $M\gg m$.}
The corresponding effective field theory is governed by the {\it Heisenberg-Euler} effective action \cite{Euler:1935qgl,Heisenberg:1936nmg}
\begin{equation}
 \Gamma_{\rm HE}[{\cal A}]=\int{\rm d}^4x\,\left(-\frac{1}{4}{\cal F}_{\mu\nu}{\cal F}^{\mu\nu}\right)+\Gamma_{\rm int}[{\cal A}]  \,, \label{eq:HE}
\end{equation}
where $\Gamma_{\rm int}$ encodes the effective interactions between electromagnetic fields induced by quantum vacuum fluctuations.  {We emphasize that a fully self-consistent theoretical description requires the classical field $
\cal A$ to solve the nonlinear equations of motions $\frac{\delta\Gamma_{\rm HE}[{\cal A}]}{\delta{\cal A}^\mu}=0$ associated with Eq.~\eqref{eq:HE} \cite{Gies:2016yaa}.
Also note} that here we refer to the general low energy effective action for $\cal A$ in the QED vacuum as the Heisenberg-Euler effective action; note, however, that in the literature this term is sometimes used exclusively for the effective action in a constant electromagnetic field, or even just the one-loop constant-field result given in Eq.~\eqref{eq:L_1loop} below.
The interaction part in Eq.~\eqref{eq:HE} follows formally from the Lagrangian given in Eq.~\eqref{action-with-background} by integrating out the quantum fields as \cite{Gies:2016yaa}
\begin{equation}
	{e}^{i\Gamma_{\rm int}[\cal A]}=\int{D}A\int{D}\psi\int{D}\bar\psi\,{e}^{i\int{\rm d}^4x\,{\cal L}}\,. \label{eq:iiintdefGamma}
\end{equation}
It admits a perturbative loop expansion in powers of the fine structure constant $\alpha$, i.e., $\Gamma_{\rm int}=\sum_{l=1}^\infty\Gamma_{\rm HE}^{l\text{-loop}}$ with $\Gamma_{\rm HE}^{l\text{-loop}}=\int{\rm d}^4x\,{\cal L}_{\rm HE}^{l\text{-loop}}\sim\alpha^{l-1}$. Beyond one loop, it moreover receives contributions from both one-particle irreducible (1PI) and one-particle reducible (1PR) diagrams, such that generically $\Gamma_{\rm HE}=\Gamma_{\rm HE}|_{1{\rm PI}}+\Gamma_{\rm HE}|_{1{\rm PR}}$ \cite{Gies:2016yaa}.
See Fig.~\ref{fig:fig_GammaHE} for a graphical representation of the one- and two-loop contributions. We emphasise that until recently all 1PR diagrams in constant fields were believed to vanish identically and were thus completely neglected \cite{Gies:2016yaa}.
Gauge invariance of QED in $d=3+1$ space-time dimensions implies that $\Gamma_{\rm HE}$ depends on ${\cal A}$ only in terms of the field strength tensor ${\cal F}^{\mu\nu}=\partial^\mu{\cal A}^\nu-\partial^\nu{\cal A}^\mu$ and its dual $\tilde{\cal F}_{\rho\sigma}=\frac{1}{2}\epsilon_{\rho\sigma\mu\nu}{\cal F}^{\mu\nu}$.\footnote{ {We note that this does not hold in generic space-time dimensions as can be easily illustrated by the possibility of a gauge-invariant Chern-Simons term $\sim\epsilon^{\mu\nu\rho}A_\mu F_{\nu\rho}$ in $d=2+1$.}}
In turn, it can be schematically expressed as
\begin{equation}
\Gamma_{\rm HE}[{\cal A}]=\int{\rm d}^4x\, {\cal L}_{\rm HE}({\cal F},\partial{\cal F},\partial^2{\cal F},\ldots)\,, \label{eq:GammaHE}
\end{equation}
with Lagrangian ${\cal L}_{\rm HE}$ which is a function of ${\cal F}(x)$ and arbitrary powers of derivatives $\partial$ thereof. As the quantity $\exp\{i\Gamma_{\rm HE}[{\cal A}]\}=\langle0;{\rm out}|0;{\rm in}\rangle$ can be interpreted as the vacuum persistence amplitude in the presence of the
prescribed electromagnetic field ${\cal A}$ \cite{Schwinger:1951nm}, an imaginary part of $\Gamma_{\rm HE}[{\cal A}]$ can be related to the phenomenon of vacuum decay or electron-positron pair creation; see Sec.~\ref{sec:Schwinger}.
Due to {\it Furry's theorem} \cite{Furry:1937zz} ${\cal L}_{\rm HE}$ is even in the elementary charge $e$, and because of CP invariance (charge conjugation parity symmetry) of QED it is also even in both ${\cal F}$ and $\tilde{\cal F}$. For recent studies of CP violating effective Lagrangians quartic in $\cal F$, see \cite{Fan:2017sxk,Ghasemkhani:2021kzf,Gorghetto:2021luj,Neves:2021tbt}.
Because it is a Lorentz scalar, all Minkowski indices are to be contracted and ${\cal L}_{\rm HE}$ is in addition even in the derivative operator.
This directly implies that $\Gamma_{\rm HE}[{\cal A}]$ vanishes identically in transverse null fields ${\cal F}^{\mu\nu}={\cal F}^{\mu\nu}(\kappa\cdot x)$  with $\kappa^2=0$ and $\kappa_\mu{\cal F}^{\mu\nu}=0$ fulfilling ${\cal S}={\cal P}=0$, such as a single plane wave \cite{Schwinger:1951nm}.
Hence, the quantum vacuum is stable in such fields and no electron-positron pairs are created.

The effective Lagrangian ${\cal L}_{\rm HE}$ can be formally expanded in the number of derivatives characterising a given contribution: the zeroth order term in this derivative expansion formally agrees with the result for ${\cal L}_{\rm HE}({\cal F})$ evaluated in a constant field ${\cal F}$ \cite{Heisenberg:1936nmg,Weisskopf:406571,Schwinger:1951nm} with the replacement ${\cal F}\to{\cal F}(x)$.
Derivatives are rendered dimensionless by the electron mass $m$, constituting the only physical scale parameter of QED at zero field. Hence, particularly for fields with typical frequency scales of variation $\Omega\ll m$ the contribution containing a total number of $2n$ derivatives with $n\in\mathbb{N}$ is parametrically suppressed by a factor of $(\Omega/m)^{2n}$ relatively to that with zero derivatives and can be neglected \cite{King:2015tba,Karbstein:2019oej}.
Also note that the contribution containing no derivatives depends on the macroscopic field only via the scalar invariants ${\cal S}$ and ${\cal P}^2$ defined in Eq.~\eqref{eq:invariants}; ${\cal P}$ is CP odd. 
On the other hand, derivative corrections do not only involve derivatives of $\cal S$ and $\cal P$, but also new scalar invariants formed by contractions of the derivative operator and field strength tensors; cf. \cite{Karbstein:2021obd}.
 {This derivative expansion of ${\cal L}_{\rm HE}$ constitutes a very important approach: at least in principle, it allows to ensure the above requirement on $\cal A$ to be a solution of $\frac{\delta\Gamma_{\rm HE}[{\cal A}]}{\delta{\cal A}^\mu}=0$ at each order of the expansion without needing to evaluate Eq.~\eqref{eq:iiintdefGamma} for generic space-time dependent $\cal A$ from the outset. A direct evaluation of Eq.~\eqref{eq:iiintdefGamma} for arbitrary field profiles is impossible {in practice}.}

Presently, in $d=3+1$ space-time dimensions ${\cal L}_{\rm HE}$ is known explicitly up to two loops (one loop) at zeroth (quadratic) order in a derivative expansion \cite{Heisenberg:1936nmg,Ritus:1975pcc,Gies:2016yaa} (\cite{Gusynin:1995bc,Gusynin:1998bt,Karbstein:2021obd}). These results can be conveniently derived employing propertime representations \cite{Dittrich:1985yb}, in the in-out formalism \cite{Kim:2012vr,Kim:2014iia}, and using the wordline formalism \cite{Schubert:2001he}.
Apart from this, higher-loop results in constant fields and lower space-time dimensions \cite{Huet:2017ydx,Huet:2018ksz}, as well as one-loop results for specific purely electric or magnetic (one-dimensional) field inhomogeneities are available; the latter are reviewed in \cite{Dunne:2004nc}. See also \cite{Dunne:2021acr} for a recent investigation of the higher-loop Heisenberg-Euler trans-series structure, \cite{Navarro-Salas:2020oew} for an adiabatic proper time expansion of ${\cal L}_{\rm HE}^{1\text{-loop}}$, and \cite{Pegoraro:2021whz} for a study of nonlinear waves in a dispersive vacuum described with a high-order derivative Lagrangian.
The all-order Landau-level structures of ${\cal L}_{\rm HE}^{1\text{-loop}}$ for QED and QCD were recently investigated by \cite{Hattori:2020guh}.

\subsection{Weak and strong field limits}

Recently it has in particular been demonstrated that the renowned one-loop result for the Heisenberg-Euler effective Lagrangian in a constant field \cite{Heisenberg:1936nmg},
\begin{equation}
 {\cal L}_{\rm HE}^{1\text{-loop}}({\cal F})=-\frac{1}{8\pi^2}\int_0^\infty\frac{{\rm d}s}{s^3}\,{e}^{-m^2s}\left\{\frac{(e\mathfrak{E}s)(e\mathfrak{B}s)}{\tan(e\mathfrak{E}s)\tanh(e\mathfrak{B}s)}+\frac{1}{3}(es)^2(\mathfrak{E}^2-\mathfrak{B}^2)-1\right\}, \label{eq:L_1loop}
\end{equation}
acts as the generator of the previously missed 1PR contribution to the constant-field result for ${\cal L}_{\rm HE}^{2\text{-loop}}$ \cite{Gies:2016yaa}.
Here, $\mathfrak{E}$ and $\mathfrak{B}$ denote the secular invariants of
the electromagnetic field introduced in Eq.~\eqref{eqn:approx:sec}, and the integration contour is implicitly assumed to lie slightly above the positive real $s$ axis.
This results in the following compact representation,
\begin{equation}
    {\cal L}_{\rm HE}^{2\text{-loop}}({\cal F})\big|_{1{\rm PR}}=\frac{1}{2}\frac{\partial{\cal L}_{\rm HE}^{1\text{-loop}}}{\partial{\cal F}^{\mu\nu}}\frac{\partial{\cal L}_{\rm HE}^{1\text{-loop}}}{\partial{\cal F}_{\mu\nu}}=\frac{1}{4}\left[\left(\frac{\partial{\cal L}_{\rm HE}^{1\text{-loop}}}{\partial \mathfrak{B}}\right)^2-\left(\frac{\partial{\cal L}_{\rm HE}^{1\text{-loop}}}{\partial \mathfrak{E}}\right)^2\right]\,. \label{eq:L_2poop_1PR}
\end{equation}
See \cite{Karbstein:2017gsb} for a generalization of this construction procedure allowing to generate generic 1PR contributions from lower-loop 1PI diagrams.
Equation~\eqref{eq:L_1loop} accounts for couplings to $e{\cal F}$ to all orders and admits a perturbative expansion in powers of $\mathfrak{E}^2$ and $\mathfrak{B}^2$, or equivalently $\cal S$ and ${\cal P}^2$. This yields a divergent asymptotic series. Alternatively, Eq.~\eqref{eq:L_1loop} can be represented in terms of a non-perturbative but convergent infinite series \cite{Cho:2000ei,Jentschura:2001qr}; cf. also the paragraph detailing the resurgence approach in Sec.~\ref{sec:ht2}.
For a pertinent review of Heisenberg-Euler Lagrangians see \cite{Dunne:2004nc}; a pedagogical derivation of Eq.~\eqref{eq:L_1loop} is presented in \cite{Karbstein:2016hlj}.

For perturbatively weak fields fulfilling $e{\cal F}/m^2={\cal F}/E_{\rm S}\ll1$, the leading contribution to ${\cal L}_{\rm int}={\cal L}_{\rm HE}+\frac{1}{4}{\cal F}_{\mu\nu}{\cal F}^{\mu\nu}$ at zeroth order in a derivative expansion accounting for contributions up to two loops reads
\begin{equation}
 {\cal L}_{\rm int}({\cal F})\simeq
 \left(\frac{e}{m}\right)^4\left(c_1{\cal S}^2 + c_2{\cal P}^2\right)=
 m^4\left(\frac{e}{m^2}\right)^4\left(c_1{\cal S}^2 + c_2{\cal P}^2\right), \label{eq:Lint_pert}
\end{equation}
with coefficients
\begin{equation}
    c_1=\frac{4}{360\pi^2}\left(1+\frac{40}{9}\frac{\alpha}{\pi}\right)
    \quad\text{and}\quad
    c_2=\frac{7}{360\pi^2}\left(1+\frac{1315}{252}\frac{\alpha}{\pi}\right).
    \label{eq:c12}
\end{equation}
In the last step of Eq.~\eqref{eq:Lint_pert} we recast the overall prefactor depending on the fundamental QED parameters $e$ and $m$ to make the scaling with the parameter $e{\cal F}/m^2$ explicit.
Equation~\eqref{eq:Lint_pert} scales as $\sim m^4(e{\cal F}/m^2)^4$ and amounts to an effective four-field interaction. The overall factor of $m^4$ ensures the Lagrangian to have the correct mass dimension. This expression can be readily obtained by combining the leading term in a perturbative weak-field expansion of Eq.~\eqref{eq:L_1loop} with the analogous contribution from ${\cal L}_{\rm HE}^{2\text{-loop}}|_{1{\rm PI}}$ \cite{Ritus:1975pcc}; the leading contribution of Eq.~\eqref{eq:L_2poop_1PR} in the weak field limit scales as $\sim \alpha m^4(e{\cal F}/m^2)^6$ and is subleading. Obviously the one and two loop contributions in Eq.~\eqref{eq:Lint_pert} have the same structure. The latter result in corrections of the one-loop expressions for $c_{1,2}$ on the $1\%$ level.
Higher-order interaction terms coupling six or more fields are parametrically suppressed by at least another factor of $(e{\cal F}/m^2)^2$. 

On the other hand, the analogous contribution in the strong-field limit characterised by $e\sqrt{|{\cal S}|}/m^2\gg1$ while $e\sqrt{|{\cal P}|}/m^2\ll1$ is given by \cite{Gies:2016yaa}
\begin{equation}
 {\cal L}_{\rm int}({\cal F})\simeq-\frac{e^2{\cal S}}{12\pi^2}\ln\left(\frac{e\sqrt{-{\cal S}}}{m^2}\right)\left\{1+\frac{\alpha}{\pi}\left[\frac{1}{6}\ln\left(\frac{e\sqrt{-{\cal S}}}{m^2}\right)+\frac{1}{4}+4\zeta'(-1)\right]\right\}\,. \label{eq:Lint_sf}
\end{equation}
Here, we neglected subleading terms which scale at most linearly in $\cal S$ or are parametrically suppressed with powers of $(e\sqrt{|{\cal P}|}/m^2)^2\ll1$; $\zeta'(-1)\simeq-0.165$ is the first derivative of the Riemann zeta function evaluated at $-1$.
In this case the two-loop contribution is logarithmically enhanced, and thus can even surpass the one-loop contribution for exponentially large fields $|{\cal S}|\sim(m^2/e)^2\exp(12\pi/\alpha)$.
This signalises that the perturbative loop expansion of ${\cal L}_{\rm HE}$ breaks down for large values of $|{\cal S}|$.

We emphasise that this logarithmic enhancement arises from the previously neglected 1PR diagram, which was only recently found to yield the non-vanishing contribution~\eqref{eq:L_2poop_1PR} \cite{Gies:2016yaa}.
For completeness, we note that subsequently analogous 1PR tadpole corrections have been evaluated for the charged particle propagator \cite{Ahmad:2016vvw,Edwards:2017bte,Ahmadiniaz:2017rrk} and photon-graviton conversion \cite{Ahmadiniaz:2021ltn} in a constant field at one loop, the low-energy limit of the QED $N$-photon amplitudes at two loops \cite{Edwards:2018vjd}, and arbitrary processes in constant fields \cite{Karbstein:2017gsb}; see also \cite{Voskresensky:2021okp}.
Interestingly, all physical tadpole contributions vanish identically in constant crossed and plane-wave fields due to the special field and momentum structure of these backgrounds \cite{Gies:2016yaa,Ahmadiniaz:2019nhk,DiPiazza:2022lij}.

In fact, it can be shown that the leading contribution to ${\cal L}_{\rm int}$ at $l$ loops in the strong-field limit ($e\sqrt{|{\cal S}|}/m^2\gg1$, $e\sqrt{|{\cal P}|}/m^2\ll1$) arises from 1PR diagrams and scales as \cite{Karbstein:2019wmj}
\begin{equation}
{\cal L}_{\rm int}^{l\text{-loop}}({\cal F})\sim -e^2{\cal S}\ln\left(\frac{e\sqrt{-{\cal S}}}{m^2}\right)\left[\frac{\alpha}{\pi}\ln\left(\frac{e\sqrt{-{\cal S}}}{m^2}\right)\right]^{l-1}\,.
\end{equation}
Interestingly, its resummation has the same effect as replacing the factor of $\alpha$ multiplying the logarithm in Eq.~\eqref{eq:Lint_sf} by
\begin{equation}
    \alpha^{1\text{-loop}}(e\sqrt{-{\cal S}})=\frac{\alpha}{1-\frac{\alpha}{3\pi}\ln\left(\frac{e\sqrt{-{\cal S}}}{m^2}\right)}\,.
    \label{eq:alpha1loop}
\end{equation} This matches the expectations of an effective field theory point of view: the latter naturally assumes that the couplings are evaluated at the relevant momentum scale, which -- in the considered limit -- is the renormalisation group invariant combination $e\sqrt{-{\cal S}}$. See \cite{Karbstein:2021gdi} for a prospective route towards insights into the manifestly non-perturbative parameter regime of ${\cal L}_{\rm HE}$ beyond a perturbative loop expansion, and field strengths beyond the Landau pole for which Eq.~\eqref{eq:alpha1loop} diverges. The approach put forward in \cite{Karbstein:2021gdi} resorts to a large $N_f$ expansion.

Equation~\eqref{eq:Lint_pert} comprises the effective nonlinear interactions most relevant for potential experimental verification of QED vacuum nonlinearities in controlled laboratory experiments using macroscopic electromagnetic fields which fulfill $e{\cal F}/m^2\ll1$ and $\Omega\ll m$, i.e., are weak and slowly varying.

\subsection{Probe photon propagation}
\label{sec:LBL:probeprop}
\begin{figure}
 \centering
 \includegraphics[width=0.65\columnwidth]{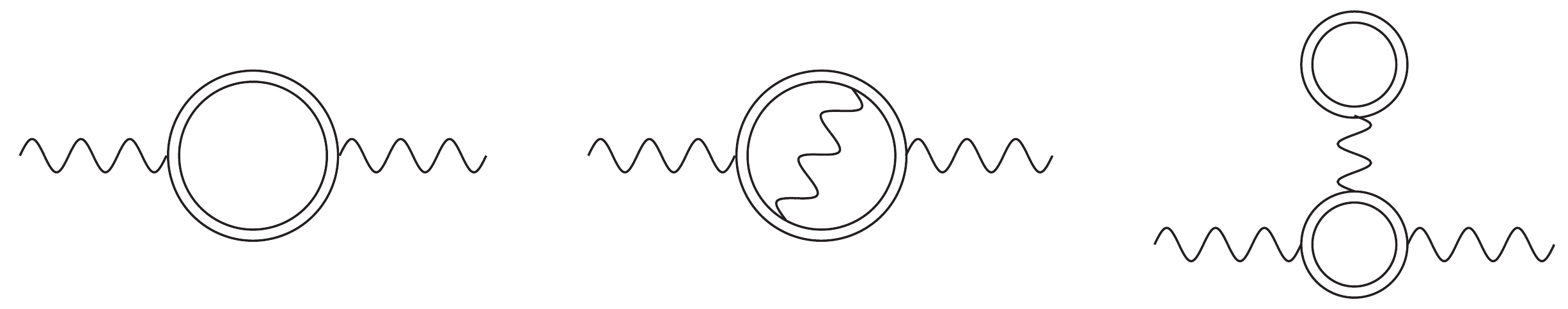}
\caption{Representative contributions to the photon polarisation tensor $\Pi^{\mu\nu}(\ell,\ell'|{\cal F})$ up to two loops. The wiggly line is the photon propagator; for the definition of the double line, cf. Fig.~\ref{fig:fig_GammaHE}.}
\label{fig:fig_Pi}
\end{figure}

The vacuum fluctuation mediated impact of a macroscopic electromagnetic field $\cal F$ on probe photon propagation is encoded in the photon polarisation tensor $\Pi^{\mu\nu}$, being the second-order correlation function of the photon field; cf., e.g., \cite{Dittrich:2000zu,Karbstein:2011ja}.
To determine its low frequency limit, we decompose the field ${\cal F}$ in ${\cal L}_{\rm int}$ as ${\cal F}\to {\cal F}+{\mathfrak f}$, where ${\mathfrak f}^{\mu\nu}=\partial^\mu {\mathfrak a}^\nu-\partial^\nu {\mathfrak a}^\mu$ denotes the field strength tensor of the probe photon field $\mathfrak a$. In this section we use all-incoming conventions, i.e., all external momenta in a given Feynman diagram are formally considered as incoming and thus come with the same sign.
 {At one loop, t}he associated polarisation tensor in momentum space, which effectively mediates the transition from an incident ${\mathfrak a}_\mu(\ell)$ to an outgoing ${\mathfrak a}_\nu(\ell')$ probe photon field, then follows as
$\Pi^{\mu\nu}(\ell,\ell'|{\cal F},\partial{\cal F},\ldots)=\frac{\delta}{\delta {\mathfrak a}_\mu(\ell)}\frac{\delta}{\delta {\mathfrak a}_\nu(\ell')}\Gamma^{ {1\text{-loop}}}_{\rm int}[{\cal A}+{\mathfrak a}]|_{{\mathfrak a}=0}$; cf. Fig.~\ref{fig:fig_Pi} for a graphical representation  {of the photon polarization tensor up to two loops}.
Specifically for ${\cal P}=0$, which encompasses the cases of a purely magnetic (electric) and a crossed field, at zeroth order in a derivative expansion the latter can be expressed as \cite{Karbstein:2015cpa}
\begin{align}
 \Pi^{\mu\nu}(\ell,\ell'|{\cal F})
 \simeq  -\int{\rm d}^4x\,&{e}^{i(\ell+\ell')\cdot x} \left[
 \bigl((\ell\cdot\ell')g^{\mu\nu} - \ell'^\mu \ell^\nu \bigr)\frac{\partial{\cal L}^{ {1\text{-loop}}}_{\rm int}({\cal F})}{\partial{\cal S}}\right.
  \nonumber\\
 &\quad+\left. (\ell\cdot{\cal F})^\mu  (\ell'\cdot{\cal F})^\nu \frac{\partial^2{\cal L}^{ {1\text{-loop}}}_{\rm int}({\cal F})}{\partial{\cal S}^2}
 + (\ell\cdot\tilde{\cal F})^\mu (\ell'\cdot\tilde{\cal F})^\nu \frac{\partial^2{\cal L}^{ {1\text{-loop}}}_{\rm int}({\cal F})}{\partial{\cal P}^2}\right]\Bigg|_{{\cal P}=0}, \label{eq:Pi_kk}
\end{align}
where $(\ell\cdot{\cal F})^\mu=\ell_\rho{\cal F}^{\rho\mu}$, etc. For the more general case with $\{{\cal S},{\cal P}\}\neq0$, cf. \cite{Karbstein:2015cpa}.
The leading contribution to Eq.~\eqref{eq:Pi_kk} in the perturbative small field limit $e{\cal F}/m^2\ll1$ scales quadratically in ${\cal F}$.
See \cite{Gies:2016czm} for a generalization of Eq.~\eqref{eq:Pi_kk} to an effective three photon interaction, and \cite{Adler:1970gg,Adler:1971wn,Papanyan:1971cv,Papanyan:1973xa} for the seminal studies of photon splitting in constant fields from the 1970s. 
We emphasise that Eq.~\eqref{eq:Pi_kk} comprises the full result at ${\cal O}(\Omega^2)$.
It allows for the reliable study of low-frequency probe photon propagation in slowly-varying inhomogeneous fields of arbitrary field strengths.  {The determination of higher-loop contributions to the photon polarization tensor in Eq.~\eqref{eq:Pi_kk} requires knowledge of ${\cal L}_{\rm int}({\cal F})$ at higher-loop orders.} Recall, however, that presently ${\cal L}_{\rm int}({\cal F})$ is only known up to two loops and that the loop expansion may break down at very large field strengths; cf. the discussion above. Via the above definition of the one-loop photon polarization tensor in terms of a functional derivative of $\Gamma_{\rm int}=\int{\rm d}^4x\,{\cal L}_{\rm int}$, higher order derivative corrections to ${\cal L}_{\rm int}$ translate to corrections to the polarisation tensor.
In constant \cite{Batalin:1971au} and plane-wave \cite{Baier:1975ff,Becker:1974en} backgrounds these corrections to the one-loop polarisation tensor can even be accounted for exactly as the corresponding Feynman diagram with the charged-particle propagator dressed to all orders in the respective background field has been evaluated explicitly for generic momentum transfers.
See also the more recent studies \cite{Meuren:2013oya,Karbstein:2013ufa} and references therein.
Moreover, the full result for the 1PR contribution to the polarisation tensor in constant fields at two loops (right-most diagram in Fig.~\ref{fig:fig_Pi}) has been determined by \cite{Karbstein:2017gsb}.
For instance the explicit result for the one-loop polarisation tensor in a constant crossed field \cite{Narozhny:1968} can be represented as 
\begin{equation}
  \Pi^{\mu\nu}(\ell,\ell'|{\cal F})
 =(2\pi)^4\delta(\ell+\ell')\left[
 \bigl(\ell^2g^{\mu\nu} - \ell^\mu \ell^\nu \bigr)\pi_T
 + \frac{(\ell\cdot{\cal F})^\mu  (\ell\cdot{\cal F})^\nu}{(\ell\cdot{\cal F})^2} \pi_{{\cal F} {\cal F}}
 + \frac{(\ell\cdot\tilde{\cal F})^\mu (\ell\cdot\tilde{\cal F})^\nu}{(\ell\cdot{\cal F})^2}  \pi_{\tilde{\cal F} \tilde{\cal F}}\right]\,, \label{eq:Pi_CCF}
\end{equation}
with scalar coefficients
\begin{align}\label{poltensor_CCF2}
 \left\{\begin{array}{c}
  \pi_T \\ \pi_{{\cal F} {\cal F}} \\ \pi_{\tilde{\cal F} \tilde{\cal F}}
 \end{array}\right\}=\frac{\alpha}{24\pi}\int_0^\infty{\rm d}s\int_0^1\,{\rm d}\nu\,&{e}^{-i\left[\left(m^2-\frac{1-\nu^2}{4}\ell^2\right)s-\frac{e^2(\ell\cdot{\cal F})^2}{48}(1-\nu^2)^2s^3\right]} \nonumber\\
 &\times
 \left\{\begin{array}{c}
  -i\nu^2(3-\nu^2)[2\ell^2+e^2(\ell\cdot{\cal F})^2(1-\nu^2)s^2] \\ 2e^2(\ell\cdot {\cal F})^2(1-\nu^2)(3-\nu^2)s \\ e^2(\ell\cdot{\cal F})^2(1-\nu^2)(3+\nu^2)s 
 \end{array}\right\},
\end{align}
The particular representation given here follows from the expression given in \cite{Dittrich:2000zu} with the help of the identity (A.1) of \cite{Karbstein:2013ufa}.
We emphasise that the scalar coefficients $\pi_T$, $\pi_{{\cal F} {\cal F}}$ and $\pi_{\tilde{\cal F} \tilde{\cal F}}$ depend on the constant crossed field only in terms of the combination $e^2(l\cdot{\cal F})^2$. The entire momentum dependence is encoded in $\ell^2$ and $e^2(\ell\cdot{\cal F})^2$. Also note that $e^2(\ell\cdot{\cal F})^2=-m^6\chi_{\ell}^2$.

In constant fields, the expression in the square brackets in Eq.~\eqref{eq:Pi_kk} does not depend on $x$ and the integral over position space results in an overall momentum conserving delta function, such that Eq.~\eqref{eq:Pi_kk} can be cast in the form of Eq.~\eqref{eq:Pi_CCF}. This reflects the fact that due to translational invariance in constant fields the four-momentum of traversing photons is conserved. Analogously, upon performing the integration over $x$ in a monochromatic plane-wave background ${\cal F}\sim\cos(\kappa\cdot x)$, a term $\sim{\cal F}^{2n}$ in Eq.~\eqref{eq:Pi_kk} generically decomposes into contributions $\sim\delta(\ell+\ell'+ 2\mathfrak{n}\kappa)$ with integer index $-n\leq\mathfrak{n}\leq n$ 
associated with the transfer of $2\mathfrak{n}$ fundamental-frequency quanta from the plane wave to the probe photon field \cite{Baier:1975ff}.
On the other hand, for generic inhomogeneous fields ${\cal F}(x)$ the four-momentum transfer involves all Fourier modes of the field, typically resulting in diffraction effects for traversing probe photons.
However, for kinematic reasons the signature of quantum vacuum nonlinearity is typically dominated by quasi-elastically scattered signal photons characterised by energies and propagation directions close to those of the incident probe photons \cite{King:2012aw}. See \cite{Karbstein:2017jgh,Karbstein:2017uzq} for explicit expressions of Eq.~\eqref{eq:Pi_kk} evaluated in Laguerre- and Hermite-Gaussian beams of arbitrary mode composition.

\subsection{Vacuum birefringence and diffraction}

By contraction of the Minkowski indices of the polarisation tensor~\eqref{eq:Pi_kk} with polarisation vectors $\epsilon_\mu(\ell)$ and $\epsilon_\nu(\ell')$ representing the incident and outgoing photons, respectively, it can be easily checked that the quantum vacuum subjected to ${\cal F}$ can give rise to a polarisation/helicity-flip phenomenon \cite{Dinu:2013gaa,Karbstein:2015xra}, which can be related to vacuum birefringence \cite{Toll:1952rq}.
Particularly in constant fields it can be straightforwardly shown that the quantum vacuum subjected to the macroscopic field gives rise to two distinct dispersion relations for probe photons with linear polarisation directions ${\mathfrak a}_1^\mu(\ell)\sim(\ell\cdot{\cal F})^\mu$ and ${\mathfrak a}_2^\mu(\ell)\sim(\ell\cdot\tilde{\cal F})^\mu$.
The associated indices of refraction $n_{1,2}$ are \cite{Bialynicka-Birula:1970nlh,Brezin:1971nd}
\begin{align}
	n_i &\simeq1- c_i e^2\left(\frac{e}{m^2}\right)^2\frac{(\ell\cdot{\cal F})^2}{\pmb{\ell}^2} \nonumber\\
	&\simeq1+c_i e^2 \left(\frac{e}{m^2}\right)^2\left[(\hat{\pmb{\ell}}\times{\bf B})^2+(\hat{\pmb{\ell}}\times{\bf E})^2+2\hat{\pmb{\ell}}\cdot({\bf B}\times{\bf E})\right]\,,
	\label{eq:dispers}
\end{align}
where $\hat{\pmb{\ell}}=\pmb{\ell}/|\pmb{\ell}|$ is the normalized probe wave vector and we neglected higher order corrections scaling at least quartic with $e{\cal F}/m^2$. See Eq.~\eqref{eq:c12} for the explicit values of the coefficients $c_{1,2}$ at two-loop accuracy. Correspondingly, probe light polarised such as to have a non-zero overlap with both polarisation eigenmodes $i\in\{1,2\}$ in the background field ${\cal F}$ experiences a vacuum birefringence phenomenon. 
Analogous considerations are possible beyond the perturbative weak field limit, e.g., in strong magnetic fields; cf. \cite{Hattori:2012ny,Hattori:2012je,Karbstein:2013ufa,Ishikawa:2013fxa,Karbstein:2015cpa,Denisov:2016pfu,Kim:2022fkt} for recent studies in this parameter regime. The phase difference accumulated by the probe light of wavelength $\lambda$ (frequency $\omega=2\pi/\lambda$) after traversing a field extending over a length $L$ is $\Phi=2\pi(L/\lambda)\Delta n$ with $\Delta n=n_2-n_1$. In the limit where $\Phi\ll1$ the number of photons scattered into a perpendicularly polarised mode is given by $N_\perp\simeq(\Phi/2)^2N$, where $N$ is the number of probe photons available.

Especially for slowly varying pump and probe fields featuring no structure transverse to the propagation direction of the probe $\hat{\pmb{\ell}}$ the constant-field result for $\Phi$ can be generalised to inhomogeneous pump fields by the substitution $L\Delta n\to\int{\rm d}(\hat{\pmb{\ell}}\cdot{\bf x})\,\Delta n(x)$.
However, due to the possibility of a finite momentum transfer $\Delta\omega$ from the inhomogeneous pump to the probe field, the birefringence signal along $+\hat{\pmb{\ell}}$ is generically accompanied by a quantum reflection signal in the $-\hat{\pmb{\ell}}$ direction \cite{2011PhRvL.107e3604K,Gies:2013yxa,Gies:2014wsa}. Typically, the latter receives an exponential suppression $\sim\exp\{-\#w\Delta\omega\}$ with $\Delta\omega$ made dimensionless by multiplication with the characteristic length scale of variation $w$ of the pump field.
Intuitively, the reflection signal is a direct consequence of $n(x)\neq1$ in the localised space-time region polarised by the background field: probe light entering this region characterised by $n(x)>1$ from the field-free region where $n=1$ effectively experiences an attractive potential resulting in above-barrier reflection.
Both phenomena can be consistently accounted for along the lines of \cite{Dinu:2013gaa,Dinu:2014tsa,Karbstein:2021otv}.

For pump and probe fields depending on all space-time coordinates the signal can be determined with Green's functions methods (cf., e.g., \cite{DiPiazza:2006pr,King:2010nka,King:2010kvw}), within the S-matrix formalism (cf., e.g., \cite{Dinu:2013gaa,Dinu:2014tsa}), and by a direct numerical solution of the non-linear wave equation (cf., e.g., \cite{domenech2017implicit,grismayer2021,Lindner:2021krv}).
A particularly convenient and easy to handle approach which was put forward in the past few years is the vacuum emission picture \cite{Galtsov:1971xm,Karbstein:2014fva,Gies:2017ygp} which does not distinguish between pump and probe fields, but treats all driving electromagnetic fields $\cal A$ on the same footing.
The latter is based on the S-matrix formalism and provides direct access to the photonic signal of quantum vacuum nonlinearity sourced by the driving field configuration. This signal is assumed to be detected far outside the interaction region of the driving laser fields. In turn, the evaluation of the signal boils down to the determination of the zero-to-single signal photon amplitude ${S}_{(\epsilon)}(\pmb{\ell})=\langle \gamma_\epsilon(\pmb{\ell})|\Gamma_{\rm int}[{\cal A}+{\mathfrak a}]|0\rangle$ in the quantum vacuum sourced by the prescribed electromagnetic field configuration ${\cal A}$.
In this particular context, $\mathfrak{a}$ is to be understood as canonically quantised field operator, the $|0\rangle$ denotes the vacuum state with zero signal photons and $| \gamma_\epsilon(\pmb{\ell})\rangle=a_{\pmb{\ell},\epsilon}^\dagger|0\rangle$ a state containing a single on-shell signal photon of wave vector $\pmb{\ell}$ and transverse polarisation $\epsilon$. The associated differential number of $\epsilon$-polarised signal photons is obtained from this matrix element by Fermi's golden rule, ${\rm d}^3N_{\text{signal},\epsilon}=\frac{{\rm d}^3\ell}{(2\pi)^3}|{S}_{(\epsilon)}(\pmb{\ell})|^2$. The differential number of signal photons ${\rm d}^3N_{\rm signal}$ attainable in a polarisation-insensitive measurement follows upon summation over two transverse signal-photon polarisations $\epsilon$.
Also note that the generalization of this approach towards out-states containing more than one signal photon is straightforward.
As only the contribution of $\Gamma_{\rm HE}[{\cal A}+{\mathfrak a}]$ linear in the signal photon field $a$ contributes to ${S}_{(\epsilon)}$, the latter can equivalently be expressed as
\begin{equation}
    {S}_{(\epsilon)}(\pmb{\ell})=\langle \gamma_\epsilon(\pmb{\ell})|\int\frac{{\rm d}^4\ell'}{(2\pi)^4} {\mathfrak a}_\mu(\ell')\frac{\delta \Gamma_{\rm int}[{\cal A}+{\mathfrak a}]}{\delta a_\mu(l')} |0\rangle\,.
\end{equation}
 {For completeness, we note that for the field strengths presently achieved in the laboratory fulfilling $|e{\cal F}/m^2|\ll1$, in the present context it is typically sufficient to model the driving fields by solutions of the linear Maxwell equations in vacuum.}
In addition decomposing the driving field into pump and probe fields ${\cal A}={\cal A}_{\rm pump}+{\cal A}_{\rm probe}$ and linearising in ${\cal A}_{\rm probe}$,  {on the one-loop level} we obtain
\begin{equation}
    {S}_{(\epsilon)}(\pmb{\ell})\simeq\int\frac{{\rm d}^4\ell'}{(2\pi)^4}\langle \gamma_p(\pmb{\ell})|{\mathfrak a}_\mu(\ell') |0\rangle\int\frac{{\rm d}^4\ell''}{(2\pi)^4}\, \Pi^{\mu\nu}(\ell',\ell''|{\cal F}_{\rm pump})\,{\cal A}_{\text{probe,}\nu}(\ell'')\,.
\end{equation}
The latter expression encodes the vacuum birefringence signal, which is linear in the probe field on the level of the amplitude, and highlights the importance of the photon polarisation tensor introduced in the previous section for its theoretical analysis.

Alternatively the signal can be analyzed by directly solving the non-linear Maxwell equations for the macroscopic electromagnetic fields \cite{Heisenberg:1936nmg}:
\begin{equation}
\pmb{\nabla} \times \mbf{H} = \frac{\partial\mbf{D}}{\partial t}, \quad \pmb{\nabla} \cdot \mbf{D}=\pmb{0}, \qquad \mbf{D} = \frac{\partial \mathcal{L}_{\trm{int}}}{\partial \mbf{E}}, \quad \mbf{H} = -\frac{\partial \mathcal{L}_{\trm{int}}}{\partial \mbf{B}}; \label{eq:modMW}
\end{equation}
the source-free Maxwell equations remain unaltered. This can be achieved e.g. analytically using Green's function methods. In this `quantum-vacuum-modified' classical equation approach, reviewed in \cite{Marklund:2006my} and pursued in the last decade in many papers, e.g.
\cite{King:2012aw,Hu:2014kda,Briscese:2017htx,Briscese:2017wuh}, the nonlinear terms in the wave-equation are conventionally interpreted as a current sourcing outgoing signal fields. This provides access to both near-field and far-field scattering effects. As detailed in \cite{Karbstein:2019oej}, the far-field signal from this approach can be mapped onto the `vacuum emission' signal analyzed above. Progress has been made recently in particular in solving the coupled set of equations in Eq.~\eqref{eq:modMW} numerically using nonlinear Maxwell equation solvers both in $d=1+1$ \cite{King:2014vha,Bohl:2015uba} and in $d=3+1$ \cite{domenech2017implicit,grismayer2021,Lindner:2021krv} space-time dimensions. Although computationally demanding, this approach has the advantage that it captures the full time evolution of the electromagnetic fields and thereby self-consistently accounts for back-reaction effects as well as the depletion of the pump fields by signal emission.

Especially in the last decade, it has been emphasised that the relevant quantity to be maximised for experiment is not the integrated number of signal photons $N_{\rm signal}$ but rather the fraction of signal photons fulfilling a {\it visibility} or {\it discernibility} (cf., e.g., \cite{King:2010nka,Karbstein:2019oej}) criterion, such as \cite{Karbstein:2016lby,Karbstein:2019dxo}
\begin{equation}
 {\rm d}^3 N_{\rm signal}(\pmb{\ell})\geq C{\rm d}^3{\cal N}(\pmb{\ell})\,, \label{eq:discernibility}
\end{equation}
where the constant $C>0$ quantifies the sensitivity of the considered experiment and ${\rm d}^3{\cal N}(\pmb{\ell})$ is the differential number of background photons of energy $|\pmb{\ell}|$ emitted in the direction $\hat{\pmb{\ell}}$ far outside the strong field region where the driving laser fields overlap. Of course, a similar criterion can be given for the number of $\epsilon$-polarised signal photons, e.g., the number of signal photons polarised perpendicularly to the incident probe photons constituting the birefringence signal.

Ongoing experimental activities have mainly focused on the signature of vacuum birefringence induced by quasi-static magnetic fields of several Tesla in combination with continuous wave lasers and high-finesse/high-Q cavities to increase the effective path length of the probe light in the magnetic field \cite{Battesti:2018bgc,Ejlli:2020yhk,Agil:2021fiq,Fan:2017fnd}.
All these experiments resort to the measurement scheme devised by \cite{Iacopini:1979ci}. {See~\cite{Fouche:2016qqj} for a more general discussion of the limits on the low-energy effective coupling constants $c_{1,2}$ attainable from experiment.}

Recent progress in x-ray and high-intensity laser technology as well as x-ray high-definition polarimetry \cite{Marx:2013xwa,Bernhardt:2020vxa} have substantiated the perspectives of detecting vacuum birefringence in laser pulse collisions \cite{Dinu:2013gaa,Dinu:2014tsa,Karbstein:2015xra,Karbstein:2016lby,Ataman:2018ucl,Shen:2018lbq,Karbstein:2018omb,Tzenov:2019ovq,Ahmadiniaz:2020lbg} following the proposal of \cite{Aleksandrov:1985,Heinzl:2006xc,DiPiazza:2006pr}. Brilliant x-ray pulses delivered by an XFEL look particularly promising as probes of vacuum birefringence because the birefringence signal scales with a positive power of the probe frequency and linearly with the number of probe photons. 
Modeling the high-intensity pump as constant crossed field, Eq.~\eqref{eq:dispers} predicts the birefringence effect to become maximal for a head-on collisions of the pump and probe laser pulses.
This prediction persists for the collision of manifestly inhomogeneous laser pulses. Such vacuum birefringence measurements, employing probe photon energies of $\omega={\cal O}(10){\rm keV}$, are envisioned for the near future, e.g., at the Helmholtz International Beamline for Intense Fields (HIBEF) at the European XFEL  \cite{Schlenvoigt:2016jrd}, SACLA in Japan \cite{Inada:2017lop}, and the station of extreme light (SEL) at the Shanghai Coherent Light Facility in China \cite{2020NIMPA.98264553X}. 
First steps towards taking into account experimentally realistic constraints and performing detailed parameter scans to optimise the vacuum birefringence signal for HIBEF parameters were taken in \cite{Schlenvoigt:2016jrd,Mosman:2021vua}.
See also \cite{Seino:2019wkb,Karbstein:2019bhp,Karbstein:2021hwc} for activities aiming at the detection of x-ray vacuum diffraction in a polarisation insensitive measurement at SACLA, and a recent theoretical proposal of Coulomb-assisted vacuum birefringence \cite{Ahmadiniaz:2020kpl}.
Experimental bounds on the closely related signature of elastic photon-photon scattering (cf. \cite{Ahmadiniaz:2020jgo} and references therein) from direct searches with optical and x-ray beams are discussed in \cite{Bernard:2000ovj,Inada:2014srv}.
A recent proposal envisions the detection of light-by-light scattering in the collision of XUV pulses with broadband gamma-ray radiation \cite{Sangal:2021qeg}.
See also \cite{dEnterria:2013zqi,ATLAS:2017fur,CMS:2018erd,ATLAS:2019azn} for recent experimental evidences of light-by-light scattering with almost real photons in the ATLAS and CMS experiments at CERN, and \cite{STAR:2019wlg}
for an attempt to study vacuum birefringence indirectly by relating experimental results for the Breit-Wheeler process in ultra-peripheral Au+Au collisions to the light-by-light scattering process via the optical theorem.

The scaling with the probe frequency has moreover motivated studies concerning the perspectives of detecting vacuum birefringence (and dichroism) with high-energy probes beyond the hard x-ray regime. In scenarios where the pump field is weak and slowly varying, i.e., fulfills $e{\cal F}/m^2\ll1$ and $\bar\omega/m\ll1$, with typical frequency scale of variation $\bar\omega$, the leading contribution to the Heisenberg-Euler Lagrangian~\eqref{eq:Lint_pert} allows for analytical insights up to fairly high probe photon energies $\omega$; the probe photon energy is delimited by $\omega/m\ll{\rm min}\{(e{\cal F}/m^2)^{-1},(\bar\omega/m)^{-1}\}$, such that $\omega/m$ can be much larger than unity \cite{Karbstein:2021otv}.

On the other hand, resorting to the analytic results for the one-loop photon polarisation tensor in constant and plane-wave backgrounds, e.g., Eq.~\eqref{eq:Pi_CCF}, studies at arbitrary probe energies are possible.
Solving the Dyson equation for the photon propagator in a given background field, in position space these vacuum polarisation corrections are effectively resummed into the exponential which gives rise to characteristic phase shifts for the two different transverse photon polarisation eigenmodes.
See \cite{Kotkin:1996nf,Ilderton:2016khs,Nakamiya:2015pde,King:2016jnl,Bragin:2017yau} for theoretical proposals and studies aiming at probing vacuum birefringence with synchrotron or gamma radiation and high-intensity lasers as pump, and in macroscopic magnetic fields \cite{Cantatore:1991sq,Wistisen:2013waa}. 

Here we would like to draw the attention to an important point: 
in the slowly-varying regime, Eq.~\eqref{eq:Pi_kk} provides a method to calculate in arbitrary backgrounds, whereas calculations based on the exact results for the polarisation tensor are inevitably restricted to a fixed background with a high degree of symmetry (such as a constant or plane-wave field).
It can be readily seen that the polarisation tensor~\eqref{eq:Pi_kk} determined via functional derivatives of the low energy effective action~\eqref{eq:GammaHE} accounts for the full contribution and momentum structure at quadratic order in $\ell$ and $\ell'$ \cite{Karbstein:2015cpa}: the reason is that $\Gamma_{\rm int}$ depends on the electromagnetic field only in terms of ${\cal F}(x)$, whereas the polarisation tensor mediates between two $\cal A$ fields featuring different space-time coordinates or momenta; the Fourier transform ${\cal F}(\ell)$ of the field strength tensor scales linearly with the momentum ${\cal F}(\ell)\sim\ell{\cal A}(\ell)$.
This allows for studies of generic diffraction phenomena in focused laser fields.
Contrarily, an explicit limitation to either constant or plane-wave backgrounds from the outset, such as naturally invoked in the cases for which exact analytic results for the polarisation tensor are available, immediately comes with a less general tensor structure. This inhibits the study of generic diffraction phenomena with the latter results. Hence, presently no first-principles studies of generic diffraction phenomena in focused laser fields at arbitrarily large probe photon momenta are possible; worldline numerics may constitute a prospective route in this direction \cite{Gies:2011he}.

Also note that due to the explicit restriction on specific backgrounds involved in their construction, the low-energy limits of the analytically known one-loop polarisation tensors accounting for terms quadratic in $\ell$ and $\ell'$ can never reproduce the generic tensor structures in Eq.~\eqref{eq:Pi_kk}.
This can be nicely demonstrated on the example of quantum reflection: a study based on the full result for the polarisation tensor in a constant \cite{Batalin:1971au} (monochromatic plane-wave \cite{Baier:1975ff,Becker:1974en}) background, which is {\it a posteriori} adopted to the study of photon propagation in a slowly varying inhomogeneous background comes with restrictions on specific probe photon polarisation modes which can be reliably studied \cite{Gies:2013yxa} (\cite{Gies:2014wsa}). On the other hand, no such restriction appear when instead basing the analysis on Eq.~\eqref{eq:Pi_kk} \cite{Karbstein:2015qwa}.
Similar arguments hold for higher-order correlation functions, such as, e.g., the three-photon polarisation tensor. Extracting the latter by functional derivatives of $\Gamma_{\rm int}$ yields an expression comprising the full momentum dependence at cubic order in the three momenta $\ell$, $\ell'$ and $\ell''$ \cite{Gies:2016czm}.

\subsection{All-optical signatures of quantum vacuum nonlinearity}

Using the same approaches and techniques as for the analysis of vacuum birefringence discussed in the preceding section, in the past years many theoretical works have studied the perspectives of detecting the effective nonlinear couplings between electromagnetic fields in an all-optical high-intensity laser experiment; see also the recent reviews~\cite{Battesti:2012hf,King:2015tba,Karbstein:2019oej}.
For instance, in the vacuum emission picture the analysis of all-optical quantum vacuum signatures typically boils down to the evaluation of the signal transition amplitude ${S}_{(\epsilon)}(\pmb{\ell})$ in the prescribed driving field configuration ${\cal F}$ without utilizing the additional linearisation in the probe field used for the analysis of vacuum birefringence and diffraction.
Because high-intensity laser fields are slowly varying and fulfill $e{\cal F}/m^2\ll1$, these studies are typically based on the weak-field limit of ${\cal L}_{\rm int}$ at zeroth order in a derivative expansion, at most supplemented with a number of additional low-order interaction terms, and potentially lowest-order derivative corrections.

Explicit analytical insights in photonic quantum vacuum signals induced by focused paraxial Gaussian laser fields are often possible resorting to an infinite Rayleigh range approximation; cf., e.g., \cite{King:2010nka,Gies:2017ygp,King:2018wtn,Karbstein:2021ldz}.
The infinite Rayleigh range approximation  is characterised by formally sending the Rayleigh range to infinity, while retaining a finite beam radius in transverse direction.
Because the signal photons predominately originate from the strong-field region where the driving laser beams overlap, typically only local information about the driving fields is needed for determining the signal.
Hence, an infinite Rayleigh range approximation is well-justified as long as the field profiles in the overlap region are essentially insensitive to variations of the Rayleigh ranges of the driving laser beams. This requires the existence of an additional physical length scale which limits the spatial extent of the interaction region along the beam axes of the driving laser fields to scales much smaller than their Rayleigh ranges. For instance, a sufficiently small pulse duration can constitute such a scale.

On the other hand, in \cite{Blinne:2018nbd,Sainte-Marie:2022efx} it has been explicitly demonstrated that in combination with a numerical solver of the linear Maxwell equations \cite{2018arXiv180104812B}, which self-consistently propagates any initially given laser field configuration, the vacuum emission picture allows for accurate theoretical predictions of photonic signatures of vacuum nonlinearity in high-intensity laser experiments from first principles. A convenient laser field model for exploratory studies of prospective nonlinear QED signatures in laser pulse collisions has been put forward by \cite{Waters:2017tgl}.
While a single laser pulse focused so tightly that the field invariants $\cal S$ and $\cal P$ differ substantially from zero in principle suffices to induce a quantum vacuum signal \cite{Fedotov:2006ii,2011PhRvL.107g3602M,Paredes:2014oxa,Blinne:2018nbd,Sainte-Marie:2022efx},
the associated signal photon yield is typically very small, rendering a detection with state-of-the-art and near-future high-intensity laser technology highly unlikely.
Correspondingly, most studies have focused on prospective photonic quantum vacuum signals in the collision of several high-intensity laser pulses.

Various proposals have considered the perspectives of detecting quasi-elastic photon-photon scattering \cite{2012PhRvA..86c3810M,King:2012aw,King:2014vha,Briscese:2017wuh,Briscese:2017htx,Gies:2017ygp,Kadlecova:2018vty,Kadlecova:2019dxv,Huang:2019ojh,Pegoraro:2019obj,Robertson:2020nnc,Karbstein:2021ldz,Roso:2021hfo}, frequency up/down conversion \cite{Gies:2014jia,Gies:2016czm,King:2018wtn}, as well as higher-harmonic generation \cite{PhysRevD.72.085005,Fedotov:2006ii,Bohl:2015uba,Bohl:2016ddu,Sasorov:2021anc} and soliton formation \cite{Bulanov:2019wfw} in the collision of two monochromatic laser fields.
The approach adopted in \cite{Gies:2014jia} is to be highlighted as it constitutes an example where one can go beyond the low-energy approximation and study arbitrary-order higher-harmonic generation from a general pulsed plane wave probe impinging on a electromagnetic pump field of generic space-time dependency but fulfilling $e{\cal F}/m^2\ll1$, such as provided by a state-of-the-art focused high-intensity laser pulse. The key ingredient to this calculation is the exact analytical result for the one-loop photon polarisation tensor in a plane-wave background \cite{Baier:1975ff,Becker:1974en}: to this end the arbitrary-frequency background field is identified with the incident probe, one of the indices of the polarisation tensor is contracted with an arbitrary-frequency pump, and the other one with the outgoing signal photon field. Because only the contribution linearly in the pump field (on the amplitude level) can be accounted for in this way, only pump fields fulfilling $e{\cal F}/m^2\ll1$ can be reliably considered.
Moreover, also relying on the availability of two driving laser pulses, recent works have studied the perspectives of enhancing photonic quantum vacuum signals by employing strongly peaked pump field configurations such as generated by high numerical aperture parabolic mirror \cite{Fillion-Gourdeau:2014uua}, a $4\pi$-spherically-focused electromagnetic wave \cite{Jeong:2020vpy} or coherent harmonic focusing (CHF) \cite{Karbstein:2019dxo,Sainte-Marie:2022efx}, and using tailored probe laser fields with far-field characteristics modified such as to substantially improve the signal-to-background separation \cite{Karbstein:2020gzg}.
See \cite{Doyle:2021mdt} for recent experimental estimates of the photon background in a potential light-by-light scattering study.

\begin{figure}
 \centering
 \includegraphics[width=0.4\columnwidth]{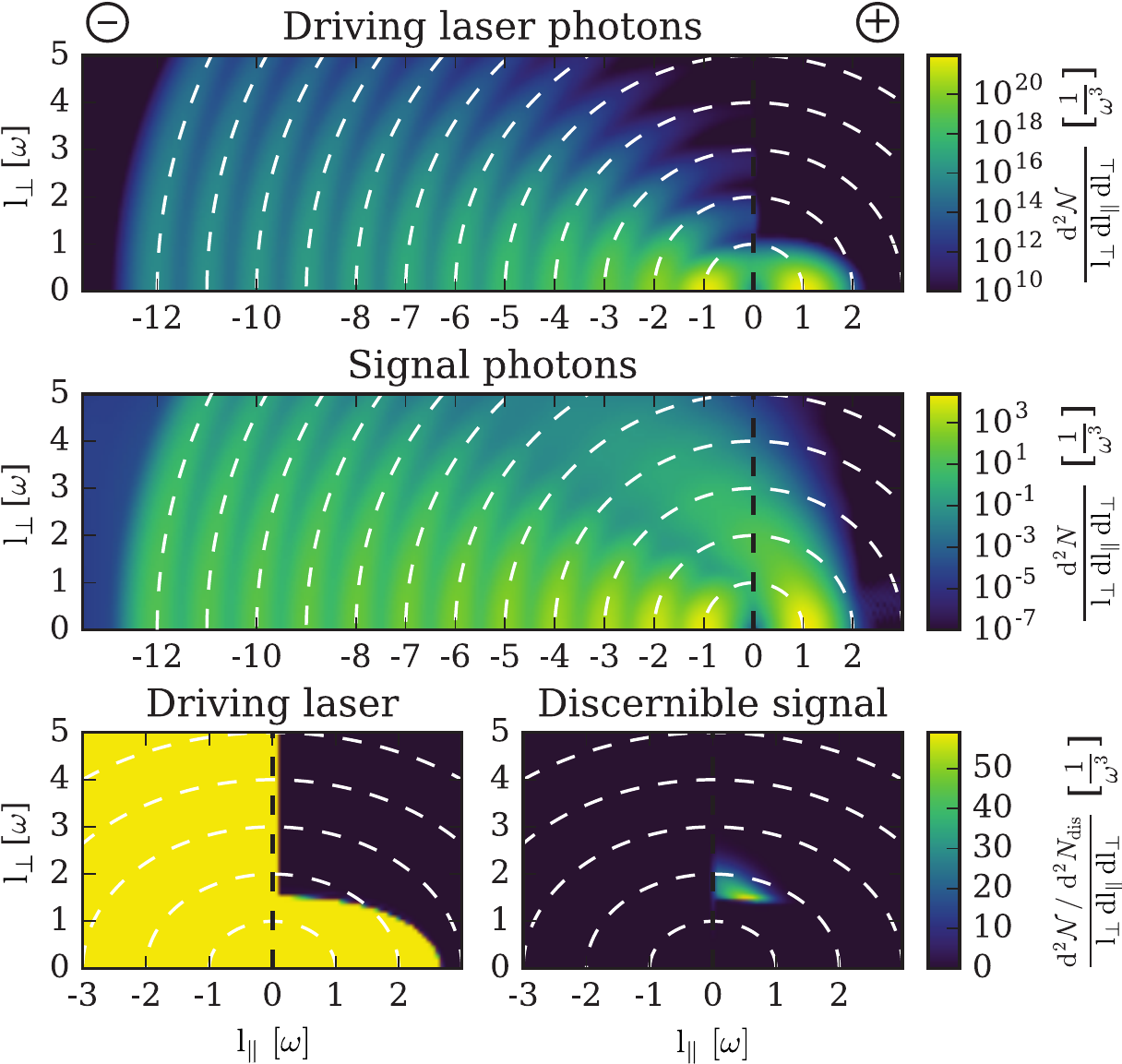}
\caption{Spectra of the driving laser photons and signal photons attainable in a polarisation insensitive measurement. In the highlighted example, the fundamental mode laser propagates in ``$+$" ($\hat{\bf z}$) direction and collides head-on with a CHF pulse made up of $12$ harmonics. White dashed lines indicate constant photon energy $|\pmb{\ell}|$, which is decomposed into parallel and perpendicular components as ${\rm l}_\parallel=\hat{\bf z}\cdot\pmb{\ell}$ and ${\rm l}_\perp=\sqrt{\pmb{\ell}^2-{\rm l}_\parallel^2}$. The bottom panels focus on the spectral domain where the differential number of signal photons surpasses the differential number of driving laser photons; adapted from \cite{Karbstein:2019dxo}.}
\label{fig:fig_exampleQVacSignature}
\end{figure}

While quasi-elastically scattered signal photons induced in the forward cone of the driving laser beams are hardly discerned from the photons comprising the driving laser fields, those with sufficiently different emission characteristics may be. Figure~\ref{fig:fig_exampleQVacSignature} compares of the spectra of the driving laser photons and the associated signal photons on the example of a fundamental-mode laser pulse colliding with a CHF pulse: in this case the discernible signal (fulfilling the criterion~\eqref{eq:discernibility} with $C=1$) is scattered outside the forward cones of the driving lasers and arises from an inelastic scattering process \cite{Karbstein:2019dxo}.

Strategies aiming at a clear directional as well as spectral separation typically require the use of more than two laser fields; cf., e.g., \cite{Moulin:1999hwj} and references therein.
This especially allows inducing a discernible signal at a frequency not contained in the spectrum of the driving laser fields.
Recent studies have in particular refined the theoretical modeling of the three-beam scenarios originally devised by \cite{Varfolomeev:1966,Lundstrom:2005za,Lundin:2006wu} describing the driving laser fields as plane-waves. In these scenarios, the collision of one beam of frequency $\omega$ and two beams of frequency $2\omega$ results in a signal of frequency $3\omega$ outside the spectra of the driving laser fields. Especially if the three beams collide at right angles, the signal of vacuum polarisation is in addition produced in a direction well-separated from the propagation axes of these beams. While \cite{Gies:2016czm,Gies:2017ezf,King:2018wtn} reanalyzed the effect in focused laser pulses, \cite{Aboushelbaya:2019ncg} studied the effect of orbital angular momentum on elastic photon-photon scattering.
See also \cite{Mendonca:2017tdw} for another recent study considering the emission of twisted photons from the laser-driven quantum vacuum.
Additional setups involving the scattering of three or more laser pulses of different frequencies were analyzed in \cite{2016PhRvD..93l5032T,King:2018wtn,Klar:2020ych,Gies:2021ymf}, and the potential of inducing interference effects in quantum vacuum signals was highlighted in \cite{King:2010nka,King:2010kvw,2011PhRvL.107e3604K}.

\subsection{Indirect signatures of quantum vacuum nonlinearity}

In a medium of refractive index $n$, a particle moving at speed $\beta$ greater than the phase velocity of light $1/n$  emits `Cherenkov' radiation at a typical angle $\cos\theta = 1/(n\beta)$. As reviewed above, the vacuum exposed to a macroscopic electromagnetic field behaves as a medium with nontrivial refractive indices $n_{1,2}$; see~Eq.~(\ref{eq:dispers}). Hence, an electron impinging on a strong field can emit Cherenkov radiation when its velocity obeys $\beta n_{1,2} > 1$~\cite{Erber:1966vv,Dremin:2002eb}. The advent of intense lasers has generated renewed interest in this area~\cite{Macleod:2018zcb}, and it has been estimated that Cherenkov radiation could be observable in laser-electron collisions at intensities of $\sim 10^{24}\,$W/cm$^2$ (at $\lambda=0.25\,\mu{\rm m}$) using $14\,{\rm GeV}$ electrons~\cite{Jirka:2020vih}. As a high energy is required, one should consider effects beyond the Heisenberg-Euler approach of~\cite{Macleod:2018zcb}; cf. \cite{Lee:2020tay} for a recent discussion of the magnetic field case. For example, exact solutions afforded by constant crossed fields or plane waves show that very high-energy Cherenkov radiation cannot be generated because the vacuum refractive indices return to unity at high energies~\cite{Bulanov:2019gdh} (as well as changes sign at intermediate energies; cf., e.g., \cite{Shore:2007um,Dinu:2013hsd}). See~\cite{Shabad:2021ran} for kinematic limitations in the case of ultra-strong magnetic fields.
As emphasised in~\cite{Artemenko:2020uxk}, Cherenkov radiation from e.g.~an electron in a strong laser field is not something separate from nonlinear Compton scattering. The electron enters the field, and it emits; Cherenkov radiation is automatically included in any change to the tree-level phase space of radiation caused by fermion loop corrections, which are required for the process to become sensitive to vacuum polarisation. See also~\cite{Bufalo:2021myy} and references therein for Cherenkov radiation in Lorentz-violating scenarios.

Another indirect signal of quantum vacuum nonlinearity is the recollision contribution to lepton-antilepton pair creation by a gamma photon impinging on an intense laser pulse \cite{Meuren:2014kla}. The underlying process proceeds in two steps: first, the incident high-energy photon decays in an electron-positron pair, which absorbs laser photons and subsequently annihilates back into a photon; this process is described by the photon polarisation tensor evaluated in a plane-wave background  \cite{Baier:1975ff,Becker:1974en}. Second, this virtual photon creates an asymptotic lepton-antilepton pair. Due to the macroscopic separation of the creation and annihilation points set by the laser wavelength, the energy absorbed in the recollision process corresponds to a large number of laser photons and can be exploited to prime high-energy reactions. Clearly, this recollision process manifestly requires accounting for the laser-photon dressed electron-positron loop.


\section{The Schwinger effect and spontaneous pair creation}\label{sec:Schwinger}
We turn to the Schwinger effect, as introduced in Sec.~\ref{sec:ht1}, that is the spontaneous creation of charged particle pairs by electromagnetic fields.  We review recent developments in theoretical frameworks for Schwinger-effect calculations in Sec.~\ref{sec:ht2}, inhomogeneous-field effects in Sec.~\ref{sec:ht3}, advances beyond linearly-polarised electric fields in Sec.~\ref{sec:ht4}, higher-order radiative corrections in Sec.~\ref{sec:ht6}, the intermediate particle picture in Sec.~\ref{sec:ht5}, and analogue experiments in condensed-matter systems in Sec.~\ref{sec:Schwinger:newCondMatt}.  For simplicity, we neglect backreaction from created pairs on the driving field (this will be reviewed in Sec.~\ref{sec:beyondPW:backreaction}) and focus on zero-temperature limit (see \cite{Brown:2015kgj, Medina:2015qzc, Gould:2017fve, Gould:2018ovk, Gould:2018efv, Korwar:2018euc, Wang:2019bjp, Torgrimsson:2019sjn} for finite temperature).  See also Sec.~\ref{sec:beyondQED} for the Schwinger effect beyond QED, and Sec.~\ref{sec:approx:DHW} for the DHW formalism.

\subsection{Theoretical frameworks} \label{sec:ht2}

\subsubsection{Semi-classical methods and the Stokes phenomenon} \label{sec:Schwinger:semiclassical}

Semi-classical methods have been used widely in Schwinger-effect calculations.  The idea of the semi-classical treatment is to expand observables $O$ in powers of the Planck constant $\hbar$ (re-introduced here for clarity), including not only perturbative corrections but also non-perturbative instanton corrections.  In modern language, we expand $O$ in a trans-series \cite{2008arXiv0801.4877E, Marino:2012zq, Dorigoni:2014hea, Aniceto:2018bis} as
\begin{align}
	O = \hbar^a \sum_{n=0}^\infty \sum_{m=0}^\infty e^{-I_n /\hbar} \hbar^m O_{n,m}  \;, \label{eq:ht2--10}
\end{align}
where $I_n$ are instanton actions, obeying $I_0=0$ (no instantons), $I_{n+1}>I_{n}>0$ for $n>0$, $O_{n,m}$ are perturbations around the $n$-th instanton, and $a$ is a constant that determines the leading power behaviour of $O$ at small $\hbar$.  Non-perturbative information, hence the Schwinger effect, is encoded in the instanton actions $I_{n>0}$. These can be evaluated conveniently with semi-classical techniques such as the worldline instanton method (see Ref.~\cite{Schneider_2019} and references therein), WKB analysis of the Stokes phenomenon~\cite{Li:2019ves,Hashiba:2020rsi,Taya:2020dco,Enomoto:2020xlf,Taya:2021dcz,Enomoto:2021hfv,Hashiba:2021npn,Sou:2021juh}, steepest-descent analysis of the Bogoliubov coefficients~\cite{Brezin:1970xf,Dumlu:2011rr,Strobel:2013vza, Fukushima:2019iiq,Taya:2020dco}, and imaginary-time method~\cite{Popov:2005rp}.  In these semi-classical methods, which are essentially equivalent to each other~\cite{Kim:2019yts,Taya:2020dco}, the problem reduces to determining the analytic structure of some action \cite{Ilderton:2015lsa}, or, in spatially homogeneous cases, the instantaneous on-shell energy $\pi_0(w)=\sqrt{m^2+(\mbf{p}-e{\bm {\mathcal A}(w)})^2}$, in the complexified time plane $t \in {\mathbb R} \to w \in {\mathbb C}$.  Such an analysis is sometimes easier than directly solving the Dirac equation in inhomogeneous fields, and hence semi-classical methods are frequently used to discuss effects of inhomogeneities, see Sec.~\ref{sec:ht3}.  

WKB analysis of the Stokes phenomenon is one of the common semi-classical methods.  There are several variants of this method, e.g., the phase-integral method~\cite{Kim:2000un,Kim:2003qp,Kim:2007pm,Kim:2018dsp} and the divergent asymptotic series method~\cite{Barry:1989zz,Dingle1973,Li:2019ves}.  We shall review this method in the language of exact WKB~\cite{Taya:2020dco, Enomoto:2020xlf, Enomoto:2021hfv, Taya:2021dcz}.  To illustrate the exact WKB method, let us for simplicity assume spatially homogeneous electric fields and consider scalar QED.  The starting point of exact WKB is the fact that the na\"ive WKB expansion is a {\it formal} asymptotic expansion.  Namely, we first make a WKB ansatz for the mode function $\phi_{\pm}$, 
\begin{align}
    \phi_{\pm}(x)
    = \frac{e^{\frac{i}{\hbar}(\mbf{p}\cdot\mbf{x}\mp i\int^t{\rm d}t'\,S(t'))}}{\sqrt{2S(2\pi \hbar)^3}} \;, \label{eq:ht2--13}
\end{align}
and expand $S$ and the exponential in terms of the Planck constant $\hbar$ to express $\phi_{\pm}$ as a formal power series in $\hbar$ as
\begin{align}
	\phi_{\pm}(x) 
	 \simeq \frac{e^{\frac{i}{\hbar} \left( \mbf{p}\cdot\mbf{x} \mp \int^t{\rm d}t\,\pi_0  \right)}}{\sqrt{2\pi_0 (2\pi \hbar)^3}} \sum_{n=0}^\infty \hbar^n \phi_{\pm}^{(n)}(t) \;. \label{eq:ht2-13}
\end{align}
(Here we used ``$\simeq$", instead of ``$=$", to make sure that the expression is formal.)  The series coefficients $\phi_{\pm}^{(n)}$ are in general factorially divergent $\phi_{\pm}^{(n)} \sim n!$ and thus the radius of convergence is zero, meaning that the $\hbar$ expansion (\ref{eq:ht2-13}) is mathematically ill-defined.  The exact WKB method resolves the factorial-divergence problem via Borel resummation, which is achieved by computing a Laplace transformation of a Borel transform $B[\phi_{\pm}]$.  The resulting Borel-summed $\phi_{\pm}$ is a well-defined function, and its asymptotic expansion coincides with the na\"ive WKB expansion (\ref{eq:ht2-13}).  The Borel sum $\phi_{\pm}$ is, however, defined only locally and cannot be continued analytically to the whole complexified time space $t\in{\mathbb R}\to w\in {\mathbb C}$ without experiencing discontinuous jumps, which is the Stokes phenomenon.  The jumps are rigorously and quantitatively formulated, mathematically, under suitable conditions: a jump occurs at the so-called Stokes lines ${\mathcal C}_j = \{ w \in {\mathbb C}\, |\, 0 = {\rm Im}\, [ i \int^w_{w^{\rm tp}_j} {\rm d}w'\,\pi_0(w') ] \}$, with $w^{\rm tp}_j \in {\mathbb C}$ being turning points defined by the condition $\pi_0(w^{\rm tp}_j)=0$, and the associated discontinuity can also be identified (see the mathematical literature for more details, e.g., \cite{CNP, AIHPA_1983__39_3_211_0, AIF_1993__43_1_163_0, doi:10.1063/1.532206, AIHPA_1999__71_1_1_0, AKT1, AKT2, Aoki:1993ra, Takei2008}).    By analyzing the analytic structure of the energy $\pi_0$, one can determine the topology of the Stokes lines ${\mathcal C}_j$ for the Schwinger effect \cite{Taya:2020dco}.  Then, by considering an analytic continuation of a Borel sum at $t=-\infty$ to $t=+\infty$, one obtains 
\begin{align}
	\begin{pmatrix} \phi_{+}^{\rm out} \\ \phi_{-}^{\rm out} \end{pmatrix} 
		&\approx \begin{pmatrix} 1 & -i\sum_j \Theta(t-t^{\rm cr}_j) e^{\sigma^*(w^{\rm tp}_j)/\hbar} \\ +i\sum_j \Theta(t-t^{\rm cr}_j)  e^{\sigma(w^{\rm tp}_j)/\hbar} & 1 \end{pmatrix} \begin{pmatrix} \phi_{+}^{\rm in} \\ \phi_{-}^{\rm in} \end{pmatrix}  \label{eq:ht2-21}
\end{align}
at the leading order in $\hbar$\footnote{We note some more details on Eq.~(\ref{eq:ht2-21}).  For second-order ordinary differential equations with potentials having linear dependencies around turning points (called simple turning points) and if Stokes lines do not degenerate with each other, the problem can be reduced to analyzing an Airy equation (cf. WKB-theoretic transformation \cite{AKT1,AKT2}).  For an Airy equation, all the higher-order WKB terms can be computed and resummed exactly and hence the corresponding connection matrix can be obtained without using any approximations (see, e.g., Appendix~A in Ref.~\cite{Taya:2020dco} for the explicit expressions).  In general, however, turning points are not necessarily simple and Stokes lines can degenerate, for which the connection matrix cannot necessarily be obtained exactly (due to, e.g., unavailability of higher-order WKB terms and non-Borel summability).  For the case of the Schwinger effect, what is relevant is connection matrices for degenerated Stokes lines, which suffer from the problem of the so-called fixed singularity when doing the Borel resummation \cite{Takei2008, AKT2} and the exact expressions for the connection matrices are unknown in general.  It was proposed in Ref.~\cite{Taya:2020dco} that, if one neglects higher order $\hbar$ corrections, the degenerated Stokes lines can safely be de-degenerated into two Stokes lines that emanate from two distinct simple turning points and then the Airy-type argument can be applied, which yields Eq.~(\ref{eq:ht2-21}).  }.  Here, $\phi^{\rm in/out}_{\pm}$ are solutions to the Klein-Gordon equation with plane-wave boundary conditions at the asymptotic initial and final times such that $\phi^{\rm in}_{\pm}(t=-\infty), \phi^{\rm out}_{\pm}(t=+\infty) \propto e^{+i \mbf{p}\cdot\mbf{x}} e^{\mp i \pi_0 t}$.  The times $t^{\rm cr}_j$ specify when the Stokes phenomena occur and are identified as crossings between the Stokes lines and the real $t$ axis.  According to the Bogoliubov transformation technique (cf. Refs.~\cite{Fradkin:1991zq, Tanji:2008ku}), the overlap between $\phi^{\rm in}_{+}$ and $\phi^{\rm out}_{-}$ gives the number of created pairs in the out-state as
\begin{align}
	\frac{{\rm d}^3N}{{\rm d}^3{\mbf p}} 
    = \int {\rm d}^3{\mbf p}' \left| \left(\phi_{-}^{{\rm out}}({\mbf p})|\phi_{+}^{{\rm in}}({\mbf p}') \right) \right|^2 \approx \frac{V}{(2\pi \hbar)^3} \Biggl| \sum_j e^{\sigma(w^{\rm tp}_j)/\hbar} \Biggl|^2  \ {\rm with}\ \sigma(w) = +2i\int^w {\rm d}w' \,\pi_0(w') \;, \label{eq:ht2-11}
\end{align}
where the conserved inner product is defined and normalised as 
$(\phi^{\rm in}_{\pm,{\mbf p}}|\phi^{\rm in}_{\pm', {\mbf p}'}) = \pm i\int {\rm d}^3{\mbf x}\,\phi_{\pm,{\mbf p}}^{{\rm in}\dagger} \overset{\leftrightarrow}{\partial}_t \phi^{\rm in}_{\pm',{\mbf p}'} =  \delta_{\pm,\pm'}\delta^3(\mbf{p}-\mbf{p}')$
for bosons [for fermions, $(\psi^{\rm in}_{\pm,{\mbf p},s}|\psi^{\rm in}_{\pm',{\mbf p}',s'}) = \int {\rm d}^3{\mbf x}\,\psi_{\pm,{\mbf p},s}^{{\rm in}\dagger} \psi_{\pm',{\mbf p}',s'}^{\rm in}=\delta_{\pm,\pm'}\delta_{s,s'}\delta^3(\mbf{p}-\mbf{p}')$] and the same for the out-mode function $\phi^{\rm out}_{\pm,{\mbf p}}$.  For spinor QED, Eq.~(\ref{eq:ht2-11}) acquires an additional factor of $(-1)^j$ in the sum due to spin statistics \cite{Dumlu:2011cc,Dumlu:2011rr,Taya:2021dcz}.  The formula (\ref{eq:ht2-11}) is valid in the formal limit of $\hbar \to 0$.  This is equivalent to situations where fields are varying slowly, since $\hbar$ always appears together with a derivative $\partial_\mu$ and hence $\hbar$ essentially controls the magnitude of gradients.  See Sec.~\ref{sec:ht3} for more details on the validity of the semi-classical approximation.  Note that one can derive the formula (\ref{eq:ht2-11}) with various equivalent semi-classical methods (e.g., the worldline instanton method and steepest-descent analysis of the Bogoliubov coefficients) \cite{Dumlu:2011cc,Dumlu:2011rr, Kim:2019yts,Taya:2020dco}.  

Finally, we mention attempts to extend the semi-classical analyses to spatially inhomogeneous fields.  Except for special spacetime inhomogeneities that essentially reduce to single-parameter inhomogeneities in light-front time $E=E(t+z)$ \cite{Ilderton:2014mla} and ``interpolating'' coordinates $E=E(t\,\cos \theta + z\,\sin \theta )$ \cite{Ilderton:2015qda}, the semi-classical analyses for single-parameter inhomogeneities cannot be applied directly.  Among the semi-classical methods, the worldline instanton method is rather easy to extend to multi-parameter inhomogeneities \cite{Dunne:2006ur, Dumlu:2015paa, Akal:2017sbs,Schneider:2018huk,Torgrimsson:2017cyb}.  The problem reduces to finding classical solutions (worldline instantons) to the worldline action $S_{\rm wl} = -\int^\tau_0( \dot{x}^2 + m^2 + ie{\mathcal A} \cdot x )$.  Analytical solutions are not available in general.  As numerical methods, shooting method \cite{Dunne:2006st} and discretisation method \cite{Schneider:2018huk} have been proposed.  Another attempt was made, based on a steepest-descent analysis of the Bogoliubov coefficients, in~\cite{Oertel:2016vsg}.

\subsubsection{Resurgence approach}

The resurgence approach \cite{Dunne:1999uy, Dunne:1999vd, Florio:2019hzn,Dunne:2021acr} enables us to access non-perturbative information of QED, including the Schwinger effect, only using perturbative data.  The idea is to apply Borel resummation to the QED perturbation series and then to obtain information on the Schwinger effect through analysis of the singularities of the Borel transform.  

For example, the one-loop Heisenberg-Euler effective Lagrangian in a constant electric field ${\mathcal L}^{\rm 1\mathchar`-loop}_{\rm HE}$ (\ref{eq:L_1loop}) can be expressed formally as an asymptotic series by perturbatively expanding it in powers of the field as ${\mathcal L}^{\rm 1\mathchar`-loop}_{\rm HE} = \sum_{k=0}^\infty {\mathcal L}^{{\rm 1\mathchar`-loop}(k)}_{\rm HE} (eE/m^2)^k$ \cite{Dunne:2004nc}.  The series coefficients ${\mathcal L}^{{\rm 1\mathchar`-loop}(k)}_{\rm HE}$ are found to be factorially divergent ${\mathcal L}^{{\rm 1\mathchar`-loop}(k)}_{\rm HE} \propto \Gamma(2k+2)$, which can be resummed via Borel resummation.  The resulting Borel transform $B[{\mathcal L}^{\rm 1\mathchar`-loop}_{\rm HE}]$ has singularities on the positive real axis in the Borel plane.  The Laplace transformation picks up the contribution from those poles, which determine the imaginary part of the Heisenberg-Euler effective Lagrangian ${\mathcal L}^{\rm 1\mathchar`-loop}_{\rm HE}$ and exactly reproduce Schwinger's classic result (\ref{eq:ht3}) \cite{Dunne:1998ni, Dunne:1999uy, Dunne:1999vd}.  

For general electric fields and/or higher loops, the perturbation-series coefficients are typically not known to all orders, except for a few special cases, e.g.~for the Sauter pulse at one loop~\cite{Dunne:1999uy}.  For practical use, several approximation schemes have been proposed to carry out Borel resummation approximately, using a finite number of coefficients~\cite{2007PhR...446....1C, Mera:2018qte, Costin:2020hwg}.  The widely-used is the so-called Borel-Pad\'{e} method, in which the Pad\'{e} approximation is applied to a truncated version of Borel transform, constructed with a finite number of coefficients.  The Borel-Pad\'{e} method was applied recently to the Schwinger effect \cite{Florio:2019hzn, Dunne:2021acr} (see also Secs.~\ref{sec:second} and \ref{sec:higher} for application to second- and higher-order processes), and it was demonstrated that ${\mathcal O}(10)$ terms are sufficient to reproduce the exact result {to good precision, see~\cite{Dunne:2021acr} for details. The method can be further improved} by using a conformal mapping technique \cite{2007PhR...446....1C, Costin:2020hwg, Dunne:2021acr} and an optimised order of Pad\'{e} approximation, which is determined by matching the expected behaviour of observables (e.g., logarithmic behaviour in strong-field limit)~\cite{Dunne:2021acr}.

\subsubsection{Perturbation theory in a Furry expansion}

Consider a classical field ${\mathcal A}$ having both strong and weak components, ${\mathcal A}={\mathcal A}_{\rm strong}+{\mathcal A}_{\rm weak}$.  In such a field the Schwinger effect can, following~\cite{Torgrimsson:2017cyb,Torgrimsson:2017pzs,Torgrimsson:2018xdf,Taya:2018eng,Huang:2019uhf,Taya:2020bcd}, be approached in a Furry expansion very similar to that described in Sec.~\ref{sec:intro}; one computes the mode function $\psi_{\pm}$ perturbatively with respect to $e{\mathcal A}_{\rm weak}$ while treating the interaction with $e{\mathcal A}_{\rm strong}$ exactly:
\begin{equation}
    \psi_{\pm}[{\mathcal A}_{\rm strong},{\mathcal A}_{\rm weak}] = \psi_{\pm}^{(0)}[{\mathcal A}_{\rm strong}] + \psi_{\pm}^{(1)}[{\mathcal A}_{\rm strong}] e{\mathcal A}_{\rm weak} + {\mathcal O}((e{\mathcal A}_{\rm weak})^2) \;.
\end{equation}
Once the mode function $\psi_{\pm}^{\rm in/out}$ is obtained, one can compute physical observables of interest based on, e.g., the Bogoliubov transformation technique (cf. Refs.~\cite{Fradkin:1991zq, Tanji:2008ku}).  Perturbation theory in such a Furry expansion is very powerful in situations where there is a clear scale separation ${\mathcal A}={\mathcal A}_{\rm strong}+{\mathcal A}_{\rm weak}$, such as in the dynamically assisted Schwinger effect (see Sec.~\ref{sec:ht3}).  Note that perturbation theory in a Furry expansion can also be extended to other types of perturbation such as thermal photons~\cite{Torgrimsson:2019sjn} and vibrations~\cite{Taya:2020pkm}.  The inclusion of quantum effects in this expansion is also straightforward, see e.g.~\cite{Otto:2016xpn}.

Of particular interest is the number of created pairs at the asymptotic out-state, which is given by (see~\cite{Torgrimsson:2017pzs} for inclusion of higher-order ${\mathcal A}_{\rm weak}$ corrections)
\begin{align}
	\frac{{\rm d}^3N}{{\rm d}^3 {\mbf p}} = \sum_{r'} \int {\rm d}^3{\mbf p}' \left| ( \psi_{+}^{(0){\rm out}}({\mbf p},r)|\psi_{-}^{(0){\rm in}}({\mbf p}',r') ) - ie \int {\rm d}^4 x \, \bar{\psi}_{+}^{(0){\rm out}}({\mbf p},r) \slashed{\mathcal A}_{\rm weak} \psi_{-}^{(0){\rm in}}({\mbf p}',r') + {\mathcal O}((e{\mathcal A}_{\rm weak})^2) \right|^2  \;. \label{eq:ht2--30}
\end{align} 
The formula (\ref{eq:ht2--30}) describes the smooth interplay between the non-perturbative Schwinger effect and perturbative multi-photon processes (including a non-linear Breit-Wheeler process by a single ${\mathcal A}_{\rm weak}$ assisted by ${\mathcal A}_{\rm strong}$), which are represented by the first and second terms, respectively.  The interference between the two terms become important in the intermediate regime where both of the Schwinger effect and the non-linear Breit-Wheeler process contribute to the pair creation \cite{Torgrimsson:2017pzs, Huang:2019uhf}.  The zeroth order solution $\psi_{\pm}^{(0){\rm in/out}}$ and the associated integrals can be evaluated analytically, as can the number of created pairs (\ref{eq:ht2--30}), by using WKB mode functions~\cite{Torgrimsson:2017pzs, Torgrimsson:2017cyb, Torgrimsson:2018xdf, Torgrimsson:2019sjn} or by studying solvable electric field configurations such as a constant electric field~\cite{Nikishov:1971, Taya:2018eng, Huang:2019uhf, Taya:2020bcd}.  Such analytical calculations are available for a wide class of perturbations, e.g., with spatial inhomogeneities~\cite{Torgrimsson:2017pzs, Torgrimsson:2017cyb} and transverse electric-field orientation~\cite{Huang:2019uhf}.

\subsection{Inhomogeneous electric fields} \label{sec:ht3}

Electromagnetic fields realised in physical situations are inhomogeneous in both space and time, and these inhomogeneities can significantly impact the Schwinger effect.  We first focus on temporal effects, and discuss spatial effects in Sec.~\ref{sec:schwinger:spatial}.  We also focus on purely electric fields throughout this section\footnote{Purely time-dependent electric fields do not satisfy the vacuum Maxwell equation without sources.  The vacuum state here should thus be understood in an extended sense, i.e., the ground state with a source.}; see Sec.~\ref{sec:ht4} for magnetic-field effects.

\begin{figure}[t]
\begin{flushleft}
\mbox{(a)}
\hspace*{48mm}
\mbox{(b)}
\hspace*{60mm}
\mbox{(c)}
\vspace*{-8mm}
\hspace*{52mm}
\end{flushleft}
\begin{center}
\hspace*{-15mm}
\includegraphics[trim=0 -4 0 0, clip, width=0.4\textwidth]{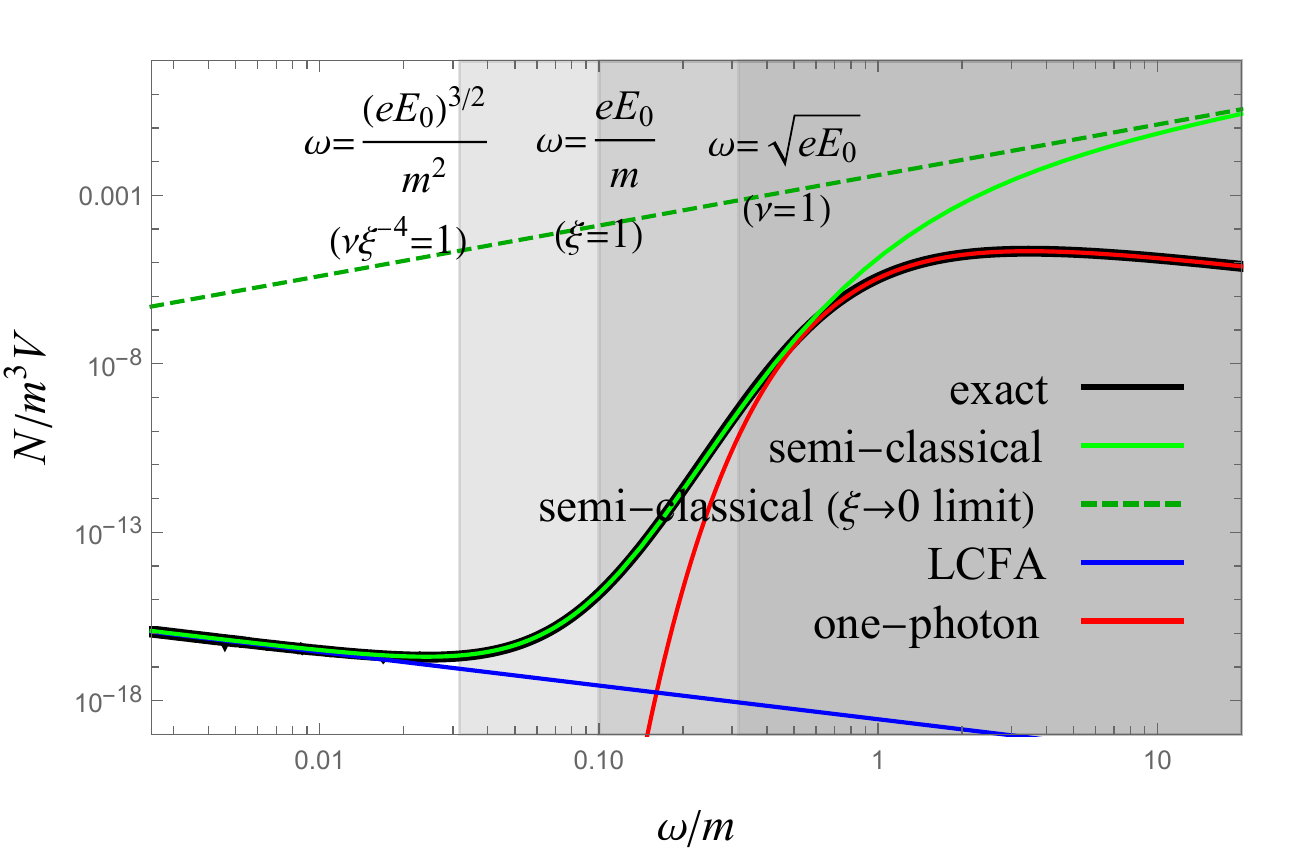}\hspace*{-2mm}
\includegraphics[clip, width=0.4\textwidth]{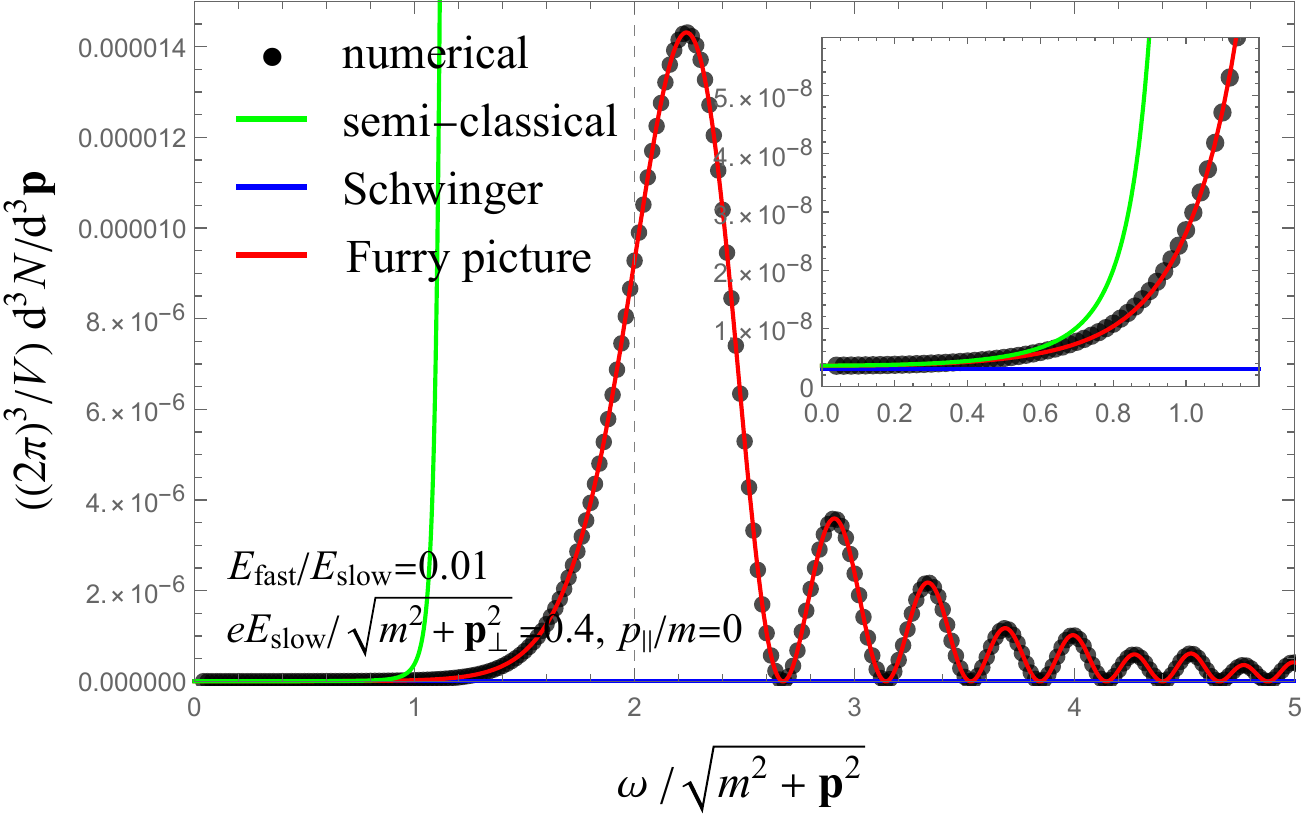}\hspace*{0.5mm}
\includegraphics[trim=0 -24.5 0 0, clip, width=0.36\textwidth]{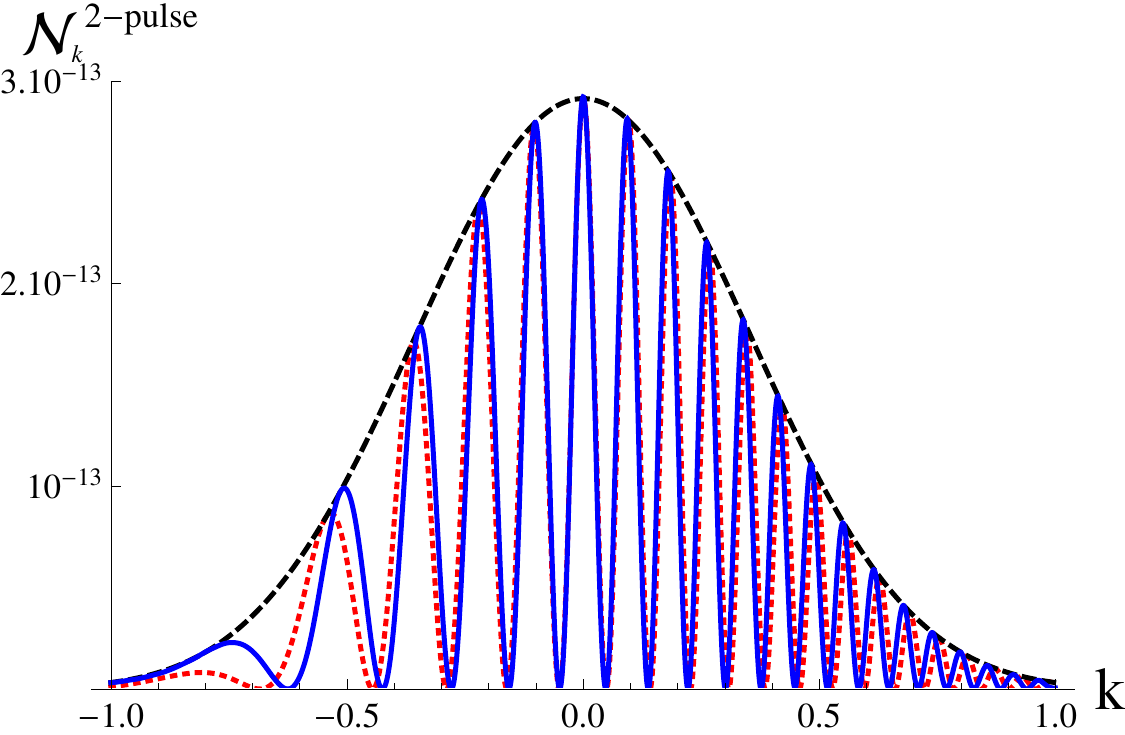}\hspace*{-2mm}
\end{center}
\caption{\label{fig-ht1} Effects of temporal inhomogeneities to the Schwinger effect: (a) The interplay between non-perturbative and perturbative pair creation for Sauter's pulsed electric field $E(t)=E_0/\cosh^2(\omega t)$.  (b) The dynamically assisted Schwinger effect for a constant electric field superimposed by a cosine electric field $E=E_{\rm slow}+E_{\rm fast}\cos(\omega t) $.  (c) Quantum interference effects in parallel-momentum (momentum in the electric-field direction) spectrum for two pulses with alternating signs.  An exact numerical result (blue) is compared with the semi-classical formula (\ref{eq:ht2-11}) (red dotted) and a single pulse result multiplied by a factor of $({\rm pulse\ number})^2=2^2$ (black dashed).  Note that (a) and (b) are adapted from Ref.~\cite{Taya:2020dco} and Ref.~\cite{Taya:2018eng}, respectively, and (c) is taken from Ref.~\cite{Akkermans:2011yn}.  }\vspace*{-5mm}
\end{figure}

\subsubsection{Non-perturbative vs perturbative pair creation}

Temporal inhomogeneities significantly affect the ``non-perturbativity'' of pair creation [see Fig.~\ref{fig-ht1} (a)].  An intuitive understanding is as follows.  In order for the non-perturbative quantum-tunneling picture of the Schwinger effect [see Eq.~(\ref{eq:ht5}) below] to be valid, the typical lifetime or inverse frequency $\omega^{-1}$ of electric fields must be sufficiently larger than the tunneling time, which can be estimated roughly using the width of the gap $m/eE$.  Therefore, the ratio $\omega^{-1}/(m/eE) = eE/m\omega$, which is nothing but the classical nonlinearity parameter $\xi$ for a time-like wavevector $k_\mu=\omega(1,0,0,0)$, should be larger than unity $\xi \gtrsim 1$ for non-perturbative tunneling pair creation to occur.  Note that it is standard, in literature on the Schwinger effect, to call $\xi^{-1}$ the Keldysh parameter~\cite{Keldysh:1965ojf}.  In the opposite limit $\xi \lesssim 1$, the tunneling picture may no longer be valid.  There, the frequency of electric fields becomes sizeable so that perturbative pair creation with a finite number $\sim 2m/\omega$ of photons can occur.  The qualitative features of pair creation change drastically depending on whether non-perturbative or perturbative mechanisms are relevant.  Non-perturbative pair creation is exponentially suppressed, and the momentum spectrum of the produced pairs is `soft', being a Gaussian distribution around zero momentum. In contrast, perturbative pair creation is only weakly suppressed by a power law, and the momentum spectrum acquires a harder, higher-energy, component.  High-frequency components of electric fields, which are more relevant to the perturbative mechanism, enhance pair creation, compared to the na\"ive expectation by the non-perturbative mechanism \cite{Levai:2009mn, Taya:2014taa, Adorno:2014bsa, Aleksandrov:2016lxd, Taya:2016ovo}.  Such ``perturbative'' enhancement becomes significant due to resonance when fields contain threshold frequencies such that $2m/\omega \in {\mathbb N}$ \cite{Kohlfurst:2013ura, Abdukerim:2013vsa, Taya:2014taa, Gelis:2015kya, Sitiwaldi:2017mfh, Krajewska:2018lwe}.  This is important for pulse shaping to maximise the Schwinger effect for experiments (see also the dynamically assisted Schwinger effect discussed next) \cite{Orthaber:2011cm, Nuriman:2012hn, Abdukerim:2013vsa, Nuriman:2013vba, Hebenstreit:2014lra, Hebenstreit:2015jaa, Otto:2015gla, Panferov:2015yda, Sitiwaldi:2018wad, Banerjee:2018gyt,Wang:2021tmo}.

Semi-classical methods have been widely used to study the interplay between non-perturbative and perturbative pair creation.  According to the semi-classical formula (\ref{eq:ht2-11}), the number of created pairs is controlled by the real part of the instanton action $\sigma(w^{\rm tp})$.  For linearly-polarised electric fields with general time-dependence $E(t) = E_0f'(\omega t)$, the instanton action $\sigma(w^{\rm tp})$ (\ref{eq:ht2-11}) is expressed as \cite{Dunne:2006st}
\begin{align}
	{\rm Re}\,\sigma(w^{\rm tp}) = \frac{\pi}{2}\frac{m^2}{eE_0} g(\xi) \ \ {\rm with}\ \ g(\xi) = \frac{2i}{\pi}  \xi \int^{w^{\rm tp}}_{w^{{\rm tp}*}} {\rm d}w\, \sqrt{1 + (\xi f(w))^2 } \;, \label{eq:ht3-30}
\end{align}
where we have set $\mbf{p}=\mbf{0}$ for simplicity and normalised $g$ so that $g=1$ corresponds to the constant-electric-field result (\ref{eq:ht5}).  Note that the prefactor in the number of created pairs can also be expressed in terms of~$g$~\cite{Popov:1971iga, Dunne:2006st}.  Thus, {\it within semi-classical methods}, the number of created pairs is determined by a single function $g$, which depends only on the classical nonlinearity parameter $\xi$ (neglecting the $\mbf{p}$ dependence).  One can explicitly evaluate the function $g$ for some particular field configurations.  For example, it reads $g(\xi) = 2\xi ( \sqrt{1 + \xi^2} - \xi ) = 1 - \xi^{-2}/4 + \cdots$ for Sauter's pulsed electric field $E=E_0/\cosh^{2}(\omega t)$.  For large $\xi$, the function $g$ reduces to the tunneling formula for constant electric fields (\ref{eq:ht5}), meaning that pair creation becomes dominated by the non-perturbative mechanism.  Temporal inhomogeneities typically decrease $g<1$, i.e., enhance pair creation.  Intuitively, this is because the time dependence supplies energy to the Dirac sea and thereby effectively reduces the gap size.  For small $\xi$, the number of created pairs deviates significantly from the tunneling formula $g=1$, implying that the na\"ive tunneling picture needs to be modified.  

Recently, more detailed analyses have been made specifically for Sauter's pulsed electric field, for which an exact expression for the number of created pairs is available [see Fig.~\ref{fig-ht1} (a)] \cite{Taya:2014taa, Taya:2021dcz, Gelis:2015kya}; see also relevant numerical works \cite{Hebenstreit:2008ae, Blaschke:2013ip}.  It was found that the semi-classical methods cannot fully capture temporal effects and that there exists another dimensionless parameter that controls the interplay, in addition to $\xi$; one needs two dimensionless parameters to completely characterise the system, since the system has in total three parameters with dimensions: $m, eE$, and $\omega$.  Indeed, the exact expression for the number of created pairs reads $((2\pi)^3/V) {\rm d}^3N/{\rm d}^3{\mbf p} = \sinh^2 (\pi\nu) / \sinh^2(\pi \nu \sqrt{1+\xi^{-2}})$ at $\mbf{p}=\mbf{0}$ \cite{Sauter:1932gsa, Narozhnyi:1970uv, Nikishov:1970br}, which is characterised not only by $\xi$ but also by $\nu = eE_0/\omega^2$ \cite{Taya:2014taa, Taya:2021dcz}.  The leading order semi-classical approximation overlooks the existence of $\nu$, which is responsible for higher-order instanton effects.  In the formal limit $\omega \to \infty$ such that $\nu, \xi \to 0$ one finds precise agreement with the one-photon perturbative process ${\mathcal A} \to e^+e^-$, with ${\mathcal A}$ being the Sauter electric field, while in the opposite limit $\omega\to 0$ such that $\nu, \xi \to \infty$, one reproduces the tunneling formula~(\ref{eq:ht5}).

For sufficiently slow electric fields, pair creation may occur ``instantaneously" compared to the typical variation scale of the field.  This implies that the pair creation is driven by ``constant" electric fields at each spacetime point, and the LCFA based on Nikishov's formula (\ref{eq:ht5}) [or the Heisenberg-Euler effective Lagrangian (\ref{eq:ht3})] may provide a good estimate of the number of created pairs.  Namely, one may estimate the {\it total} number of created pairs by na\"ively replacing the constant electric field in Nikishov's result (\ref{eq:ht5}) with a time- (or spacetime-) dependent one $E \to E(t)$ and carrying out the spacetime integration via replacement $VT \to \int {\rm d}^4x$ \cite{Bulanov:2004de, Gies:2016yaa, Gavrilov:2016tuq, Karbstein:2017pbf, Sevostyanov:2020dhs}.  Recently, another LCFA prescription is proposed to estimate the {\it momentum spectrum} ${\rm d}^3N/{\rm d}^3\mbf{p}$ \cite{Aleksandrov:2018zso}.  The idea is to assume that a pair creation event occurs instantaneously at a time $t_*$, which is speculated to be the time when the gap, i.e., the instantaneous energy $\pi_0$, takes the minimum.  One may then estimate the momentum spectrum from Eq.~(\ref{eq:ht5}) with the instantaneous value of the electric field $E=E(t_*)$ as 
\begin{align}
    \frac{(2\pi)^3}{V} \frac{{\rm d}^3 N_{\rm LCFA}}{{\rm d}^3\mbf{p}} = \exp\left[  - \pi \frac{m^2+\mbf{p}_\perp^2}{|eE(t_*)|} \right] \Theta\left(-p_\parallel \left( p_\parallel + \int^{+\infty}_{-\infty} {\rm d}t\,eE(t)  \right) \right) \;,
\end{align}
where a step function is introduced to approximate the distribution of parallel (canonical) momentum in the electric-field direction $p_\parallel$, which incorporates the fact that the maximum parallel momentum that a created particle can obtain from the applied electric field is $\int {\rm d}t\,eE$.  The validity condition for the LCFA was investigated in Refs.~\cite{Aleksandrov:2018zso, Sevostyanov:2020dhs} by comparing LCFA with exact and semi-classical results [see Fig.~\ref{fig-ht1} (a)].  For subcritical field strength $m^2 \ll eE$, it was shown that the LCFA is valid when $(eE_0)^{3/2}/m^2 \omega \gg 1$, with $\omega$ being the typical frequency of the field.  Notice that this condition is stronger than the na\"ive condition based on the classical nonlinearity parameter $\xi \gg 1$.  The quantity $(eE_0)^{3/2}/m^2\omega$ can be understood as a ratio between the tunneling time $\sim m/eE_0$ and the effective variation time-scale of the LCFA rate $\tau_{\rm eff} = \sqrt{eE_0/m^2}/\omega$ \cite{Sevostyanov:2020dhs}.

\subsubsection{Dynamically assisted Schwinger effect}\label{sec:dase}

The dynamically assisted Schwinger effect \cite{Schutzhold:2008pz, Dunne:2009gi} is an idea to enhance the Schwinger effect by utilizing perturbative effects [see Fig.~\ref{fig-ht1} (b)].  This is an analog of Franz-Keldysh effect in materials \cite{Taya:2018eng,Torgrimsson:2018xdf}.  One may also understand this mechanism as a variant of a non-linear Breit-Wheeler process, as discussed in Sec~\ref{sec:first}, in a strong slow electric field with a perturbative fast {\it classical off-shell} photon~\cite{Gelis:2015kya, Torgrimsson:2017pzs, Torgrimsson:2017cyb, Torgrimsson:2018xdf, Taya:2018eng, Huang:2019uhf, Taya:2020bcd}.  The idea is to superimpose a fast perturbative electric field $\xi \lesssim 1$ onto a slow one $\xi \gtrsim 1$.  An intuitive picture is that the fast perturbative electric field assists the non-perturbative tunneling by the slow electric field.  The perturbative field first kicks up an electron in the Dirac sea into the gap.  Then, the quantum tunneling occurs from the inside of the gap and hence the tunneling length and the potential height are reduced compared to the original situation without the perturbative kick  {(see, e.g., Fig.~2 of Ref.~\cite{DiPiazza:2009py} and Fig.~1 of Ref.~\cite{Schutzhold:2008pz})}.  Thus, the tunneling probability is enhanced, so is the Schwinger effect.  One may also understand the dynamically assisted Schwinger effect from a reverse point of view: the non-perturbative tunneling by the slow electric field assists the perturbative pair creation by the fast electric field \cite{Gelis:2015kya, Taya:2018eng, Huang:2019uhf, Taya:2020bcd, Torgrimsson:2017pzs, Torgrimsson:2017cyb, Torgrimsson:2018xdf}.  Quantum tunneling allows electrons in the Dirac sea to exist even inside the gap.  Then, the threshold energy for perturbative pair creation is reduced.  In particular, one-photon pair creation can occur even below the threshold $\omega<2m$ (threshold $\omega=2m$ for purely time-dependent electric fields that do not have momentum).  The magnitude of the one-photon pair creation is suppressed only weakly by square of the field strength, which gives more abundant creation compared to the exponentially-suppressed non-perturbative tunneling.  Consequently, the Schwinger effect is enhanced because of the reminiscent of the perturbative peak extended to the low-frequency regime.  (An analogous effect occurs in nonlinear Breit-Wheeler pair-creation in a plane-wave pulse where linear Breit-Wheeler from photons in the slow timescale of the envelope become, with decreasing field strength, less suppressed than the exponentially suppressed nonlinear Breit-Wheeler from the fast timescale of the carrier frequency \cite{Titov:2012rd,Nousch:2012xe}.)  The one-photon picture tells us that the enhancement becomes the most significant at around the one-photon threshold $\omega \sim 2m$ \cite{Gelis:2015kya, Otto:2016fdo, Sitiwaldi:2018wad, Taya:2018eng, Huang:2019uhf, Taya:2020bcd}.  It also indicates that momentum spectrum becomes very different from the soft Gaussian distribution, which is na\"ively expected from the non-perturbative Schwinger effect (\ref{eq:ht5}).  This is because one-photon pair creation has harder spectrum and is sensitive to the Fourier spectrum of the field \cite{Linder:2015vta, Torgrimsson:2017pzs, Torgrimsson:2017cyb, Torgrimsson:2018xdf}.  

Below, we focus on the simplest but the most essential situation $E(t) = E_{\rm slow} + E_{\rm fast} f'(\omega t)$ and assume $E_{\rm slow}/E_{\rm fast} \ll 1$, i.e., a constant strong electric field is superimposed by a time-dependent weak electric field with arbitrary time dependence in the same direction.  Note that, since the essence of the dynamically assisted Schwinger effect is the perturbative energy supplied from fast fields, the polarisation of the superposition is not essential.  The dynamical enhancement can equally occur for transverse superpositions \cite{Huang:2019uhf}.  One may also consider general field configurations such as weak fields with spatial inhomogeneities \cite{Schneider:2014mla, Torgrimsson:2017pzs, Torgrimsson:2017cyb, Aleksandrov:2018uqb} and magnetic components \cite{Copinger:2016llk}.  

An analytical approach to the dynamically assisted Schwinger effect is provided by semi-classical methods (e.g., the worldline instanton method \cite{Schutzhold:2008pz, Linder:2015vta, Schneider:2016vrl,Torgrimsson:2017cyb}, WKB analysis \cite{Fey:2011if, Taya:2020dco}, phase-integral method \cite{Kim:2021dcw}).  Interplay of dominant turning points is the essence of the dynamically assisted Schwinger effect within the semi-classical methods.  Namely, a dominant turning point that gives the largest contribution in the semi-classical number formula (\ref{eq:ht2-11}) is changed from the original one, responsible for the na\"ive Schwinger effect by the slow field $E_{\rm slow}$, to another one, describing (perturbative) pair creation by the fast field $E_{\rm fast}$, once the so-called combined Keldysh parameter $\tilde{\gamma} = m \omega /eE_{\rm slow}$ becomes larger than a certain value $\tilde{\gamma}\gtrsim\tilde{\gamma}_{\rm cr}$.  The critical value $\tilde{\gamma}_{\rm cr}$ is determined by the analytic structure of the fast field $f(w)$ and hence depends on details of $f(w)$ \cite{Schutzhold:2008pz, Linder:2015vta, Schneider:2016vrl}.  The number of created pairs (\ref{eq:ht2-11}) behaves differently below and above the critical value $\tilde{\gamma}_{\rm cr}$.  Note that this is a smooth change and is not a ``phase transition,'' where the physics changes discontinuously.  Below the critical value $\tilde{\gamma} \lesssim \tilde{\gamma}_{\rm cr}$, it is justified to compute the instanton action $\sigma(w^{\rm tp})$ as a perturbation series in terms of the ratio $E_{\rm fast}/E_{\rm slow}$ and one can obtain a closed expression for general $f$ \cite{Schneider:2016vrl, Taya:2020dco}.  On the other hand, above the critical value $\tilde{\gamma} \gtrsim \tilde{\gamma}_{\rm cr}$, the perturbative expansion in $E_{\rm fast}/E_{\rm slow}$ is unjustified.  This means that the na\"ive tunneling formula (\ref{eq:ht5}) is significantly modified even by infinitesimally small fields $E_{\rm fast}/E_{\rm slow} \to 0$.  An analytical expression for general $f$ is not available within the semi-classical methods, and can be computed explicitly only for specific field configurations (see, e.g., Refs.~\cite{Linder:2015vta, Schneider:2016vrl}).  For example, for Sauter's  pulsed electric field, it was shown that the number of created pairs monotonically increases for large $\tilde{\gamma}$ as $N \propto e^{-({\rm const.})/\tilde{\gamma}}$ \cite{Schutzhold:2008pz, Fey:2011if, Linder:2015vta, Schneider:2016vrl}.  

Another analytical approach is the perturbation theory in a Furry expansion \cite{Torgrimsson:2017pzs, Torgrimsson:2017cyb, Torgrimsson:2018xdf, Taya:2018eng, Huang:2019uhf, Taya:2020bcd}.  An advantage of this approach is that one can obtain a closed expression for the number of created pairs for fast fields $f$ with arbitrary time dependence for any $\tilde{\gamma}$.  For example, the formal expression for the spectrum (\ref{eq:ht2--30}) was evaluated analytically when the slow field is a constant electric field \cite{Taya:2018eng, Huang:2019uhf, Taya:2020bcd} (see Ref.~\cite{Taya:2020dco} for scalar QED).  It reproduces exact numerical results very accurately; see Fig.~\ref{fig-ht1} (b).  As shown in the figure, the perturbation theory in a Furry expansion agrees with the semi-classical results in the low-frequency limit $\omega \to 0$, while they disagree for large $\omega$, where the semi-classical approximation is invalidated \cite{Torgrimsson:2017pzs, Taya:2018eng}.

\subsubsection{Quantum interference}

If pair creation occurs not at a single time but at more than one time, the pair creation events at different times interfere with each other and modify momentum spectra [see Fig.~\ref{fig-ht1} (c)].  This is an analog of St\"{u}ckelberg-phase interference in materials \cite{stuckelberg1933theorie}, and can also be understood as a multiple-slit interferometry in the temporal direction.  Namely, suppose we have two pair creations at different times $t=t_1 < t_2$.  A wavefunction of an electron produced at the time $t_1$ obtains a phase $\sim \exp[+i\int^{t_2}_{t_1}{\rm d}t\,\pi_0]$ at the time $t_2$.  On the other hand, a wavefunction for an electron produced at the time $t_2$ has a phase $\sim \exp[-i\int^{t_2}_{t_1}{\rm d}t\,\pi_0]$ because this electron sits in the Dirac sea and hence has negative energy $-\pi_0$ from $t=t_1$ to $t_2$.  Thus, the two electrons created at $t=t_1, t_2$ have a non-vanishing relative phase $\sim \exp[+2i\int^{t_2}_{t_1}{\rm d}t\,\pi_0]$ (this is the St\"{u}ckelberg phase; see also Fig.~4 in Ref.~\cite{Shevchenko_2010}), which leads to constructive or destructive interference in the pair creation.  The relative phase is obviously momentum dependent.  Hence, by varying $\mbf{p}$, the interference changes from constructive to destructive and vice versa, which results in characteristic oscillating patterns in momentum spectra \cite{Hebenstreit:2009km, Li:2014xga, Li:2014psw, Kaminski:2018ywj, Krajewska:2019vqd}.  Conversely, one can change the locations of the pair creation times and thereby the interference pattern by designing fields, e.g., sub-cycle structures of a laser pulse \cite{Hebenstreit:2009km, Abdukerim:2013vsa, Abdukerim:2015dsa, Abdukerim:2017hkh, Olugh:2018seh, Gong:2019sbw, Ababekri:2019qiw, Li:2021vjf, Mohamedsedik:2021pzb}.  At particular momenta where the interference becomes fully constructive, the total probability amplitude for pair creation is enhanced by the number of pair-creation times $n$ and hence the peak heights of the momentum spectrum scale as $n^2$ \cite{Akkermans:2011yn, Li:2014xga, Li:2014psw, Kaminski:2018ywj, Krajewska:2019vqd}.  The quantum interference also plays an important role in the realtime dynamics \cite{2005PhRvL..94j0602O, Taya:2021dcz}.  Without the quantum interference, the total creation number monotonically increases whenever pair creation occurs.  The number of created pairs increases or decreases at a pair creation, depending on the relative phase at each pair-creation time.  It eventually exhibits complicated time dependence, which is very distinct from that of the applied electric field (e.g., high-harmonic generation \cite{Taya:2021dcz}).  The semi-classical formula (\ref{eq:ht2-11}) gives a theoretical grounding for the intuitive argument for the quantum interference effects \cite{Dumlu:2010ua, Dumlu:2010vv, Dumlu:2011rr, Dumlu:2011rr, Taya:2020dco}.  One may extend the semi-classical formula (\ref{eq:ht2-11}), which counts the asymptotic particle number, to intermediate times by assuming some intermediate particle picture and confirm that quantum interference actually takes place at the pair creation times \cite{Tanji:2008ku, Tanji:2010eu, Dabrowski:2014ica, Dabrowski:2016tsx}.  Recent studies have shown that the interference effects can survive even beyond the semi-classical regime but there may appear several new features such as coherent enhancement with pulses of the same sign and deviations from the $n^2$ scaling \cite{Kaminski:2018ywj, Ilderton:2019ceq}.

\subsubsection{Spatial inhomogeneity}\label{sec:schwinger:spatial}

There are  {three }
important effects due to spatial inhomogeneities.   {One is that magnetic fields come into play for realistic electromagnetic fields obeying the Maxwell equation (i.e., the Faraday law $\partial_t \mbf{B} = -{\bf \nabla} \times \mbf{E}$), which shall be discussed in Sec.~\ref{sec:ht4.1}.  }
 {Another } is that spatial inhomogeneities typically suppress the Schwinger effect \cite{Nikishov:1970br, Dunne:2005sx, Dunne:2006st, Dunne:2006ur, Ruf:2008ahs, Ilderton:2014mla, Gies:2015hia, Gies:2016coz}.  An intuitive way of understanding is that spatial inhomogeneities supply momentum  {$\Delta {\mbf p}$} to the Dirac sea and thereby effectively increase the mass gap  {$2m \to 2\pi_0 = 2\sqrt{m^2 + (\Delta {\mbf p})^2}$}.  
This is in contrast to temporal inhomogeneities, which typically enhance the Schwinger effect by supplying energy to the Dirac sea.   {See also Ref.~\cite{Ruf:2008ahs} for multi-photon pair creation in realistic counter-propagating laser beams and how the momentum supply modifies the kinematics and the resulting spectrum of the pair creation.  }
 {The last one } is an effect on post pair creation.  After creation, created pairs are accelerated by the electric field.  If the spatial extent of the electric field is small (large), pairs can (cannot) go outside of the electric field after the creation and hence can acquire small (large) momentum from the field.  Thus, spatial inhomogeneities, through modifying the dynamics of post pair creation, affect momentum spectra, leading to self-bunching \cite{Hebenstreit:2011wk, Ababekri:2019dkl} and ponderomotive effects \cite{Kohlfurst:2017hbd}.  The temporal effects such as the dynamically assisted Schwinger effect and the quantum interference are also affected.  For the dynamically assisted Schwinger effect with a spatially inhomogeneous slow electric field, the suppression of the Schwinger effect effectively enhances the degree of the dynamical assistance \cite{Schneider:2014mla, Ababekri:2019dkl}.  Also, the dynamical assistance becomes manifest for lower combined Keldysh parameter $\tilde{\gamma}$ \cite{Schneider:2014mla}.  Contrarily, when fast electric fields to be superimposed are inhomogeneous in space, the dynamical assistance is reduced because the momentum supply by the fast fields increases the gap \cite{Torgrimsson:2017cyb, Aleksandrov:2018uqb}.  The dynamical assistance is thus basically maximised when perturbations are purely time dependent.  On the other hand, the quantum-interference effects also tend to be suppressed by spatial inhomogeneities \cite{Li:2021vjf, Mohamedsedik:2021pzb}.  Indeed, it becomes difficult for pairs created at different spacetime points to overlap the same phase-space after the inhomogeneous acceleration.  Therefore, the different pairs rarely interfere with each other, resulting in suppression of, e.g., the oscillating structures and $n^2$ enhancement in momentum spectra \cite{Dong:2017vse, Ababekri:2019dkl, Kohlfurst:2019mag}.

The worldline instanton method \cite{Dunne:2005sx, Dunne:2006st, Dunne:2006ur, Schneider:2014mla, Ilderton:2014mla,Gies:2015hia, Copinger:2016llk} is a possible analytical approach for spatially inhomogeneous effects.  The worldline instanton method can cover a wide parameter region that the LCFA cannot cover \cite{Gies:2016yaa, Karbstein:2017pbf, Aleksandrov:2018zso, Gavrilov:2019sbt, Aleksandrov:2019ddt}, while it cannot go beyond the semi-classical regime, for which one basically needs to rely on numerical methods (e.g., the DHW formalism \cite{Hebenstreit:2011wk, Kohlfurst:2017hbd, Kohlfurst:2019mag, Ababekri:2019dkl}, direct solving of the Dirac equation \cite{Aleksandrov:2019ddt}).  Within the worldline instanton method, the magnitude of pair creation is controlled by the same function $g$ as the purely time-dependent case (\ref{eq:ht3-30}) but with a different argument $\xi \to -i\xi$, i.e., ${\rm Re}\,\sigma(w^{\rm tp}) = \frac{\pi}{2}\frac{m^2}{eE_0} g(-i\xi)$, where $\xi  {= \frac{eE_0}{m\omega}}$ is the classical nonlinearity parameter for a space-like wavevector $k_\mu=\omega(0,0,0,1)$ \cite{Dunne:2005sx, Dunne:2006st}.  See also Ref.~\cite{Ilderton:2014mla} for how the replacement $\xi \to -i\xi$ is smoothly connected when spatial and temporal inhomogeneities coexist $k_\mu=\omega(\cos \theta,0,0,\sin \theta)$.  As discussed below Eq.~(\ref{eq:ht3-30}), $g(\xi) = 1 - ({\rm positive\ const.}) \times \xi^{-2}$ for $\xi \gg 1$ for a wide class of electric fields and thus the change of the argument $\xi \to -i\xi$ means that spatial inhomogeneities typically suppress pair creation $g(-i\xi) = 1 + ({\rm positive\ const.}) \times \xi^{-2}> 1$.  Note that the enhancement and suppression by temporal and spatial inhomogeneities, respectively, cancel with each other for light-like fields $k_\mu =\omega(1,0,0,1)$ and the number of created pairs is given exactly by the LCFA~\cite{Tomaras:2000ag, Tomaras:2001vs,Fried:2001ur,Ilderton:2014mla}.  The worldline instanton method has also been used to show, for Sauter and monochromatic fields~\cite{Dunne:2006st}, that due to the prefactor contribution, which is also described by the function $g$ as above, pair creation does not occur for $\xi \geq 1$ (or smaller $\xi$ when temporal inhomogeneities coexist \cite{Ilderton:2015qda}).  This is consistent with our intuition that spatially small electric fields cannot supply enough energy to create pairs because virtual pairs escape from the electric field before becoming real.  In other words, for pair production one needs $2m<\int_{-\infty}^{+\infty} {\rm d}z\, eE_3(z)=:2m\xi$, where the final equality \emph{defines} $\xi$ for these fields, and is chosen such that the critical point is at $\xi=1$ rather than some other value.  The onset of no-pair creation was investigated in Refs.~\cite{Gies:2015hia, Gies:2016coz} in and beyond the semi-classical regimes, respectively; it was found that the vacuum decay probability ${\rm Im}\,\Gamma^{\rm 1\mathchar`-loop}$ obeys universal scaling behaviours around the critical point $\xi \sim 1$.  The universal behaviours are insensitive to details of electric fields but are determined solely by asymptotic power dependence of the fields $E \sim |x|^{-d}$ as
\begin{align}
	{\rm Im}\,\Gamma^{\rm 1\mathchar`-loop} \propto (1 - \xi^{-2} )^\beta\ {\rm with}\ \beta = \left\{ \begin{array}{ll} \frac{5d+1}{4(d-1)} & ( (eE)^2 \ll 1 - \xi^{-2}) \\ 3 & ( 1 - \xi^{-2} \ll (eE)^2 ) \end{array}
 \right. 	
\end{align}
for sufficiently rapidly decaying fields $d>3$.

\subsection{Beyond linearly-polarised electric fields} \label{sec:ht4}

We review here the Schwinger effect in fields other than linearly-polarised electric fields, e.g., electro{\it magnetic} and rotating/multi-component fields, with an emphasis on spin/chirality creation.

\subsubsection{Magnetic fields} \label{sec:ht4.1}

Magnetic fields are important to discuss the parameter region ${\mathcal S} < 0, {\mathcal P} \neq 0$, which purely electric configurations cannot cover.  The role of magnetic fields in the Schwinger effect is, however, somewhat indirect: magnetic fields cannot supply energy and hence cannot create particles out of the vacuum without the help of electric fields (in fact, the Schwinger effect is prohibited for a purely magnetic case ${\mathcal S}<0, {\mathcal P}=0$ \cite{Schwinger:1951nm, Ilderton:2015qda}).  Magnetic fields can affect the Schwinger effect through modifying the energy $\pi_0$.  Inhomogeneous electric-field effects (e.g., the dynamically assisted Schwinger effect with magnetic fields \cite{Copinger:2016llk, Taya:2020bcd}) are also modified accordingly.  

To be concrete, let us consider scalar ($r=0$), fermion ($r=1/2$), and vector ($r=1$) particles with electric charge $q$, in the presence of a constant parallel electromagnetic field $\mbf{E} \parallel \mbf{B}$.  For constant fields, it is sufficient to consider the parallel setup because it can cover all the possible values of the Lorentz invariants ${\mathcal S}$ and ${\mathcal P}$.  Perpendicular electromagnetic fields can always be transformed into purely electric ones (if ${\mathcal S}>0$) via Lorentz transformations and hence the effect of perpendicular magnetic fields is to effectively reduce the electric-field strength as $E \to \sqrt{E^2-B^2}$.  In the parallel setup, the energy $\pi_0$ is quantised via the Landau quantisation $\pi_0 \to \sqrt{m^2 + (2n+1)|qB| - 2r_{\parallel}qB}$, with $n=0,1,2,\cdots$ and $r_{\parallel}=-r,-r+1,\cdots,r$ being Landau levels and spin in the magnetic-field direction, respectively.  The number of created particles $N$ can be obtained by substituting the Landau-quantised energy into the momentum spectrum for constant electric fields (\ref{eq:ht5}) and reads \cite{Marinov:1972nx, Tanji:2008ku, Tanji:2010eu, Tanji:2011di, Sheng:2018jwf} (cf. see Refs.~\cite{Karbstein:2019oej, Hattori:2020guh} for corresponding expressions for the Heisenberg-Euler effective action): 
\begin{align}
\begin{split}
	\frac{N}{VT} 
		&= \frac{1}{T} \sum_{r_{\parallel}=-r}^{r} 2\pi |qB| \sum_{n=0}^\infty \int {\rm d}p_\parallel \frac{1}{(2\pi)^3} \exp \left[ -\pi \frac{m^2 + (2n+1)|qB| - 2r_{\parallel}qB}{|qE|} \right] \\
		&= (2r+1) \frac{(qE)^2}{(2\pi)^3} \exp \left[ -\pi \frac{m^2}{eE} \right] \times \underbrace{ \frac{1}{2r+1} \left|\pi \frac{B}{E} \right| \frac{\sinh \left[ (2r+1) \left|\pi \frac{B}{E} \right| \right] }{ \sinh^2 \left|\pi \frac{B}{E} \right|} }_{=N(B)/N(B=0)} \;.  \label{eq:ht3-41}
\end{split}
\end{align}
Note that one can express Eq.~(\ref{eq:ht3-41}) in a manifestly covariant manner by using Eq.~(\ref{eqn:approx:sec}): $E \to \mathfrak{E}=\sqrt{\sqrt{{\mathcal S}^2+{\mathcal P}^2}+{\mathcal S}}$ and $B \to \mathfrak{B}=\sqrt{\sqrt{{\mathcal S}^2+{\mathcal P}^2}-{\mathcal S}}$.  Magnetic fields become particularly important when they are stronger than electric ones $|B/E| \gg 1$.  In this limit, the number of created particles $N$ is dominated by the lowest energy mode ($n=0, r_{\parallel}=r\,{\rm sgn}\,q$) and other modes are exponentially suppressed, i.e., created particles are spin  {aligned }
in the magnetic-field direction.  Also, for scalar ($r=0$) and vector particles ($r=1$), the pair creation is exponentially suppressed and enhanced, respectively, because the corresponding lowest energy modes $\pi_0^2 \to m^2 + (1-2r)|qB|$ are decreased and increased by the magnetic field, respectively.  Notably, for vector particles ($r=1$), the enhancement can result in positive exponent for the number of created particles with very strong magnetic fields $|qB|> m^2$, which is reminiscent of the Nielsen-Olesen instability \cite{Marinov:1972nx, Tanji:2011di, Karabali:2019ucc, Hattori:2020guh}.  Note that particle interpretation is available for asymptotic states, where unstable modes are stabilised by the infinite acceleration by the electric field \cite{Tanji:2011di}.  For fermions ($r=1/2$), the lowest energy level is unchanged $\pi_0^2 \to m^2$ and hence there is no exponential modification in the number of created particles.  Nevertheless, the number is enhanced roughly linearly in $qB$ because the transverse phase-space $\int {\rm d}^2\mbf{p}_\perp \to 2\pi |qB| \sum_n$ increases linearly with $qB$ due to the Landau quantisation.

We recall that the formula (\ref{eq:ht3-41}) is limited to constant electromagnetic fields.  For general spacetime-dependent electromagnetic fields, there exist extra Lorentz invariants, other than ${\mathcal S}$ and ${\mathcal P}$, and also it is in general not sufficient to focus on the parallel setup.  Due to the complexity, inhomogeneous magnetic-field effects are less understood.  Nevertheless, there exist some attempts from various viewpoints/frameworks, e.g., gradient-expansion approach \cite{Karbstein:2021obd}, worldline instanton method \cite{ Ilderton:2014mla,Copinger:2016llk}, and numerical approaches \cite{PhysRevLett.109.253202, Kohlfurst:2017git,Aleksandrov:2019ddt}.  An immediate consequence of magnetic inhomogeneities is that the energy $\pi_0$ deviates from the Landau-quantised one, which in turn affects the Schwinger effect.  For example, it was demonstrated in Ref.~\cite{Copinger:2016llk} that the energy $\pi_0$ increases/decreases for magnetic fields with modulations having negative/positive curvature (i.e., decaying/growing profile in space), leading to suppression/enhancement of the Schwinger effect.  Another interesting feature of magnetic inhomogeneities is appearance of spin-dependent force (Stern-Gerlach force) because force is roughly given by the spatial derivative of the energy $\sim \mbf{\partial} \pi_0 \propto r_{\parallel} \mbf{\partial} B $.  The spin-dependent force affects the post dynamics of the Schwinger effect, leaving spin-dependent momentum spectra \cite{Kohlfurst:2017git}.

\subsubsection{Chirality}

In spinor QED, parallel electromagnetic fields have another physical consequence due to the chiral anomaly \cite{Adler:1969gk, Bell:1969ts}: the numbers of right-handed and left-handed electrons created by the Schwinger effect are unequal.  The chiral imbalance $N_5$, defined to be the difference between the number of right- and left-handed electrons, $N_5=N_{\rm R}-N_{\rm L}=\int{\rm d}^3\mbf{x}\langle \bar{\psi} \gamma^0 \gamma_5 \psi \rangle$, induces intriguing physical phenomena such as the generation of an anomalous current along magnetic fields, known as the chiral magnetic effect~\cite{Kharzeev:2007jp,Fukushima:2010vw, Warringa:2012bq,Fukushima:2015tza}.  The chiral imbalance $N_5$ may be expressed in terms of the number of electrons in the lowest energy mode ($n=0, r_\parallel=1/2$) as $N_5 = 2N_{e^-}(n=0,r_\parallel=1/2)$ in the out-state \cite{Tanji:2010eu,Warringa:2012bq,Taya:2020bcd,Aoi:2021azo}: this is because up and down spin degrees of freedom are degenerated in higher energy levels and hence their contributions cancel with each other, while in the lowest energy level only up spin, along the magnetic field, exists and hence no cancellation occurs.  The lowest energy mode can move only in the magnetic-field direction due to the dimensional reduction and hence the spin and momentum are parallel to each other.  The sign of the momentum is determined by that of the electric field, and, if it is taken to be plus, one understands that non-zero +1 chirality is produced (assuming that electrons are infinitely accelerated by the electric field to be ultra-relativistic, for which helicity is the same as chirality).  A similar argument holds for the created anti-particles, yielding the same amount of +1 chirality creation.  Hence, $N_5 = N_{e^-}(n=0,r_\parallel=1/2)+N_{e^+}(n=0,r_\parallel=-1/2)=2N_{e^-}(n=0,r_\parallel=1/2)$ in total.  Therefore, for a constant parallel electromagnetic field, one can immediately read off from the formula for the number of created pairs for parallel electromagnetic fields (\ref{eq:ht3-41}) that the chiral imbalance $N_5$ is given by \cite{Warringa:2012bq, Copinger:2018ftr}
\begin{align}
	\frac{N_5}{VT} = \frac{e^2}{2\pi^2} EB e^{-\pi \frac{m^2}{eE}} \;. \label{eq:ht3-43}
\end{align}
The chiral imbalance $N_5$ can be enhanced/suppressed by changing the electromagnetic-field configuration~\cite{Aoi:2021azo} and through dynamical assistance by fast electric fields~\cite{Taya:2020bcd}.  Note that once the chiral imbalance $N_5$ is changed, the pseudo-scalar condensate $2m \langle \bar{\psi} i\gamma_5\psi \rangle$, responsible for the explicit breaking of chiral symmetry by finite Dirac mass, must change as well so that the chiral anomaly relation $N_5 = 2m \int {\rm d}^3\mbf{x} \langle \bar{\psi} i\gamma_5\psi \rangle + \frac{e^2}{2\pi} \int {\rm d}^3\mbf{x} \, \mbf{E}\cdot\mbf{B}$ is satisfied \cite{Warringa:2012bq,Copinger:2018ftr}.

\subsubsection{Rotating/multi-component electric fields}

The Schwinger effect is modified if orientation of electric fields changes in time.  Similarly to the linearly-polarised case, the time dependence results in enhancement of pair creation \cite{2012ChPhL..29b1102X,Blinne:2013via, Ilderton:2015qda,Huang:2019uhf} and quantum interference that distorts momentum spectra \cite{Blinne:2013via,Li:2015cea,Li:2017qwd,Olugh:2018seh,Li:2019rex}.  A novel feature is that the Schwinger effect becomes spin (or chirality) dependent even without magnetic fields \cite{Strobel:2014tha,Wollert:2015kra,Blinne:2015zpa,Ebihara:2015aca,Kim:2016yxx,Kohlfurst:2018kxg,Huang:2019uhf, Li:2019rex,Takayoshi:2020afs} and can produce spin current out of the vacuum~\cite{Huang:2019szw, Chu:2021eae}.  The spin dependence arises due to the spin-orbit coupling $\mbf{s}\cdot (\mbf{p} \times \mbf{E})$, which is inherent in the Dirac equation as a relativistic effect \cite{Wollert:2015kra, Huang:2019uhf,Huang:2019szw,Chu:2021eae} (mathematically, it can also be understood in terms of a geometrical effect, related to the Berry phase \cite{Takayoshi:2020afs, 2020CmPhy...3...63K}).  Intuitively, an electron, moving with velocity $\mbf{v}$ under an electric field $\mbf{E}$, effectively feels a magnetic field in its rest frame $\mbf{B}_{\rm eff} \sim \mbf{v} \times \mbf{E}$ and hence the one-particle energy becomes spin dependent due to the Zeeman coupling $\sim \mbf{s}\cdot \mbf{B}_{\rm eff}$.  If the electric-field orientation varies in time, the velocity $\mbf{v}$ is no longer parallel to $\mbf{E}$ and hence spin is  {aligned} 
in the direction of $\mbf{v} \times \mbf{E}$.  Note that the 
 {spin alignment}
can modify momentum spectra in such a way that the directions of particles' momentum $\mbf{p}$ and spin $\mbf{s}$ become parallel to each other.  This is a relativistic effect, originating from (approximate) conservation of helicity in the relativistic limit \cite{Wollert:2015kra, Huang:2019szw}.

\subsection{Radiative corrections} \label{sec:ht6}

\subsubsection{Beyond one-loop}\label{sec:ht6.1}

Ritus was the first who went beyond the one-loop result (\ref{eq:ht3}) by evaluating the two-loop effective action for constant fields~ \cite{Ritus:1975pcc,Lebedev:1984mei}.  The result is a complicated double-integral representation, which cannot be simplified into a closed form (except for special cases, e.g., Euclidean self-dual fields ${\mathcal F}^{\mu\nu}=\tilde{\mathcal F}^{\mu\nu}$ \cite{Dunne:2004nc}).  For weak electric fields $eE/m^2 \ll 1$, it can be expanded as   
\begin{align}
	2\, {\rm Im}\, \left[ {\mathcal L}^{\rm 1\mathchar`-loop}_{\rm HE} + {\mathcal L}^{\rm 2\mathchar`-loop}_{\rm HE} \right]
	&= 2\frac{(eE)^2}{(2\pi)^3}  \sum_{n=1}^\infty \left[ \frac{1}{n^2} + \frac{e^2}{4} \left( - \frac{m}{\sqrt{eE}} c_n + 1 + {\mathcal O}\left( \frac{\sqrt{eE}}{m} \right) \right) \right] e^{-n\frac{\pi m^2}{eE}}  \;, \label{eq:ht5-43}
\end{align}
where $c_1=0$ and $c_n = \sum_{k=1}^{n-1} 1/2\sqrt{nk(n-k)} \sim \pi/2\sqrt{n}$ for $n>1$.  Note that Eq.~(\ref{eq:ht5-43}) is independent of spin at this order, and it holds equally for scalar particles after including the statistical factor $(-1)^{n-1}$ and replacing the overall spin degeneracy factor $2 \to 1$.  The radiative correction can be interpreted as an effective mass shift $m \to m_{\rm eff}$, and it was conjectured that summation of higher-order radiative corrections beyond two loops amounts to exponentiating Eq.~(\ref{eq:ht5-43}) as (the exponentiation conjecture~\cite{Affleck:1981bma,Lebedev:1984mei})
\begin{align}
	2 \,{\rm Im}\, \sum_{l=1}^\infty {\mathcal L}^{l{\rm \mathchar`-loop}}_{\rm HE} \overset{?}{=} 2\frac{(eE)^2}{(2\pi)^3}  \sum_{n=1}^\infty \frac{e^{-n\frac{\pi m^2_{\rm eff}}{eE}}}{n^2} \ {\rm with}\ m_{\rm eff} = m + \frac{e^2}{4\pi} \frac{n c_n}{2}\sqrt{eE} - \frac{e^2}{4\pi} \frac{neE}{2m} + {\mathcal O}\left( \left| \frac{\sqrt{eE}}{m} \right|^3 \right) \;.  \label{eq:ht111}
\end{align}
The effective mass has a simple physical interpretation in terms of Coulomb interactions, if one interprets the label $n$ as a number of coherently created pairs~\cite{Lebedev:1984mei}.  The negative mass shift in the third term of $m_{\rm eff}$ is due to an attractive Coulomb potential between the created electrons and positrons (see also~\cite{Evans:2019zyk}).  Namely, an electron in a bunch of $n$ coherently created electrons is attracted by the Coulomb force from $n$ created positrons at a distance typically given by the length of the gap $2m/eE$, and therefore $\delta m = -ne^2/4\pi r$ with $r \sim 2m/eE$.  Similarly, the positive mass shift in the second term is interpreted as due to a repulsive Coulomb potential between an electron and the other $n-1$ electrons in the bunch {: As the momentum spectrum of the produced particles in a coherent $n$ electron-production event can be estimated as $\propto \exp[-n \pi (m^2+\mbf{p}_\perp^2)/eE]$, the typical momentum of an electron in the bunch is $|\mbf{p}_\perp| \sim \sqrt{eE/n}$, implying that the typical distance among the electrons is $r \sim 1/|\mbf{p}_\perp| \sim \sqrt{n/eE}$.  Therefore, the repulsive Coulomb potential among the electrons gives $\delta m = +(n-1)e^2/4\pi r \propto n c_n \sqrt{eE}$}.  The exponentiation conjecture seems to be supported by the worldline instanton method in the weak-field limit, where multi-pair creations ($n > 1$) and the associated positive mass shift are neglected \cite{Affleck:1981bma} [see Eq.~(\ref{eq:ht3-44})].  If the exponentiation conjecture is true,  resurgence theory predicts a constraint on a large-order behaviour of QED perturbation series.  Namely, so as to yield the same dominant non-perturbative factor for any order of the coupling $e^{2l}$, the large-order behaviour of perturbative ($k \gg 1$)-photon diagrams containing $l$-loops should all have the same magnitude of factorial divergence, regardless of values of $l$, as ${\mathcal L}^{l{\rm \mathchar`-loop}}_{\rm HE} \propto {\mathcal L}^{{\rm 1\mathchar`-loop}}_{\rm HE} \propto \Gamma(2k+2)$.  Such a large-order perturbative behaviour was numerically confirmed for the two-loop case \cite{Dunne:2021acr}.  

It is still an open issue if the exponentiation conjecture is correct.  At the present, even a three loop result is available neither for scalar nor spinor QED in (3+1) dimensions.  Instead of ${\rm QED}_{3+1}$, more tractable ${\rm QED}_{1+1}$ was considered by several authors.  Two-loop calculations in ${\rm QED}_{1+1}$ were carried out in~\cite{Dunne:2006sx, Krasnansky:2006sx,Huet:2010nt} (see also Sec.~\ref{sec:6.1}).  Weak-field expansion coefficients were obtained in closed forms, confirming a large-order behaviour consistent with the exponentiation conjecture and the resurgence theory.  A three-loop calculation in ${\rm QED}_{1+1}$ was performed recently in Ref.~\cite{Huet:2018ksz}.  Unlike the two-loop case, the expansion coefficients were obtained only numerically up to the first six terms (or seven \cite{Huet:2020awq}).  The number of the coefficients turned out to be insufficient to conclude the validity of the exponentiation conjecture.    

Effects of temporal inhomogeneities on radiative corrections are discussed in Refs.~\cite{Lan:2018xnq,Gould:2019myj}, which show, using the worldline instanton method, that in the weak-field limit the effective negative mass shift is enhanced, i.e.~the Schwinger effect is enhanced by radiative corrections.  Intuitively, temporal inhomogeneities supply finite energy to electrons in the Dirac sea and hence the tunneling length is reduced.  Thus, the attractive Coulomb potential between a pair is increased.  

Holographic approach, a radically different approach based on gauge/gravity duality, is also applied to study radiative corrections to the Schwinger effect \cite{Gorsky:2001up, Semenoff:2011ng}.  A suggested advantage is that the holographic approach may be applicable even to strong fields $eE>m^2$ with a strong coupling, for which the semi-classical worldline instanton method is invalidated.  A disadvantage is that it cannot deal with real electrons, but just with an analogous particle with a different microscopic theory.  A typical setup is ``$W$ bosons'' (sometimes called ``quarks'' in the holographic literature) on the Coulomb branch of ${\mathcal N}=4$ supersymmetric $SU(N)$ Yang-Mills theory in a large `t Hooft coupling limit.  In other words, charged vector bosons exposed to a $U(1)$ gauge field that are interacting with each other by mediating $SU(N)$ bosons in the planar limit.  This is realised in type IIB string theory, a gravity dual to ${\mathcal N}=4$ supersymmetric $SU(N)$ Yang-Mills theory, by preparing a stack of $N$ $D_3$-branes at the horizon $r=0$ and a probe $D_3$-brane at an intermediate position $r=r_0<\infty$.  A remarkable prediction of the holographic approach is that there exists an upper limit for electric-field strength $E_{\rm max}$ at which the probability of pair creation becomes unity and hence the field decays immediately.  Namely, Semenoff-Zarembo \cite{Semenoff:2011ng} proposed that the number of created pairs can be obtained holographically by computing a circular Wilson loop on the probe $D_3$ brane, which leads to 
\begin{align}
	2\, {\rm Im}\,\Gamma_{\rm hol} 
		\propto \exp\left[ -\frac{\sqrt{\lambda}}{2}\left(  \sqrt{\frac{E_{\rm max}}{E}}  -  \sqrt{\frac{E}{E_{\rm max}}} \right)^2 \right] \ {\rm with}\ E_{\rm max} = \frac{2\pi m^2}{\sqrt{\lambda}} \;, \label{eq:ht5--45}
\end{align}
where $\lambda = g_{\rm YM}^2 N \gg 1$ is an `t Hooft coupling, with $g_{\rm YM}$ being the Yang-Mills coupling.  The existence of the upper limit $E_{\rm max}$ is natural from a holographic or string-theory point of view: the Dirac-Born-Infeld action, describing the probe $D_3$ brane becomes unstable above the upper limit $E>E_{\rm max}$.  More intuitively, the attractive Coulomb potential between electron-position pairs become more important for stronger electric fields as the length of the gap $2m/eE$ is reduced.  The attractive Coulomb potential is negatively divergent for $r \to 0$ and hence the effective mass can be vanishing for sufficiently strong electric fields \cite{Sato:2013iua, Sato:2013hyw}.

\subsubsection{Photon creation}

One of the observable consequences of radiative corrections is the production of real photons accompanying the Schwinger effect.  This effect has been discussed using perturbation theory in a Furry expansion~\cite{Otto:2016xpn, Aleksandrov:2022rgg}, semi-classical methods~\cite{Taya:2021dcz}, lattice techniques~\cite{Tanji:2015ata}, and kinetic approaches~\cite{Blaschke:2008wf,Blaschke:2010vs, Blaschke:2011is,Smolyansky:2019yma,Aleksandrov:2021ylw, Fauth:2021nwe}.  All these approaches rely on low-order perturbation theory to compute the photon spectra, meaning that the possibility of QED cascades, and resulting modifications to the photon spectrum, are neglected (along with potential infrared divergence in the very soft regime).

It has been suggested that sizeable numbers of soft photons may be created even for subcritical fields~\cite{Blaschke:2010vs, Otto:2016xpn, Aleksandrov:2019irn}.  The created photon spectrum is, like the pair spectrum, affected by field inhomogeneities~\cite{Tanji:2015ata, Kuchiev:2015qua, Aleksandrov:2021ylw, Taya:2021dcz} and can be modified by dynamical assistance~\cite{Villalba-Chavez:2019jqp, Otto:2018jbs}.  The effect of an additional probe photon beam was investigated in~\cite{Aleksandrov:2022rgg}.  A proposed characteristic signature is the generation of high harmonics from the vacuum~\cite{PhysRevD.72.085005, Fedotov:2006ii, Kuchiev:2015qua, Aleksandrov:2021ylw, Taya:2021dcz}.  Namely, due to the interplay of creation, acceleration, annihilation and interference of pairs (cf. three step model in materials \cite{Corkum:1993zz, Vampa2014}), the electric current induced in the Schwinger effect acquires a complicated time-dependence.  This leads to generation of photons with odd high-harmonics $\omega = \Omega, 3\Omega, 5\Omega, \cdots$ having an upper cutoff at $\omega \sim eE/\Omega$, with $\Omega$ being frequency of the field \cite{Taya:2021dcz}.

\subsection{Intermediate particle picture} \label{sec:ht5}

\begin{figure}[t]
\begin{center}
\includegraphics[clip, width=0.7\textwidth]{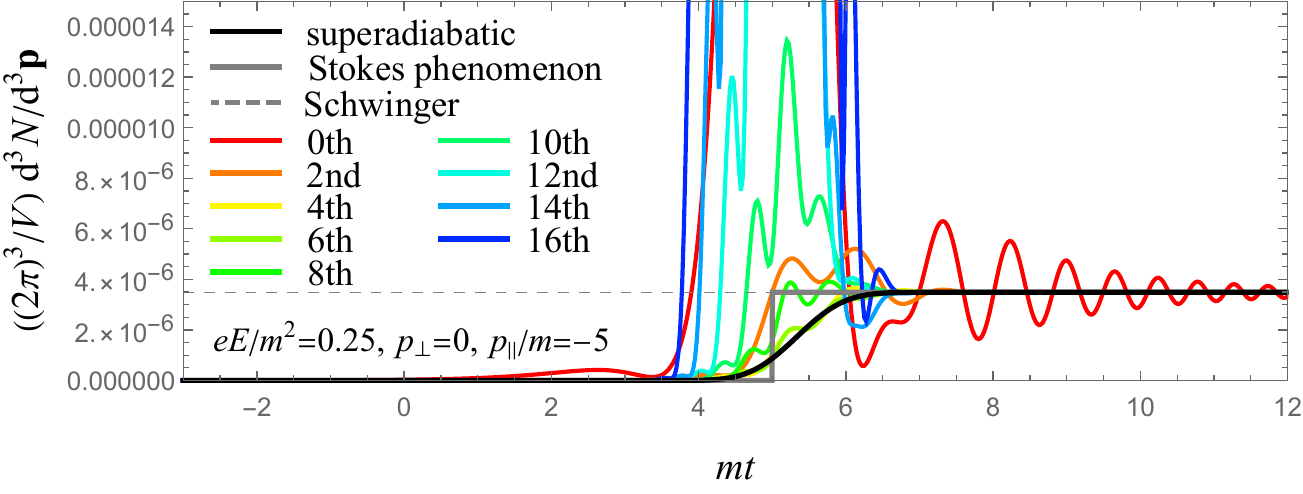}
\end{center}
\caption{\label{fig-ht2} Superadiabatic particle number vs intermediate particle number with the $n$-th order WKB expansions for a constant and linearly polarised electric field $E(t)=E$.  For the given parameter choice, the optimal order of truncation is $n=6$.  }\vspace*{-5mm}
\end{figure}
 
Since the Schwinger effect is a non-equilibrium process, it is tempting to consider the number of pairs created at intermediate, or non-asymptotic, times.  One should, however, bear in mind that the concept of intermediate particle number is highly ambiguous~\cite{Parker:1968mv,Tanji:2008ku, Fedotov:2010ue,Dabrowski:2014ica,Dabrowski:2016tsx}.  In fact, the presence of an electric fields breaks time-translation symmetry, so that ``energy,'' which is usually defined as an eigenvalue of the time-translation operator, is no longer a good quantum number.  This means that the notion of electrons and positrons, usually thought of as one-particle states having positive and negative energy eigenvalues respectively, becomes ill-defined.  One can unambiguously define electrons and positrons only when there are no electric fields so that the system exhibits time-translation symmetry.  There is, therefore, no rigorous definition for intermediate particles, and one has to make an ansatz to define them.  The intermediate number of created pairs is thus ambiguous and is sensitive to the choice of the ansatz.

To be concrete, let us consider scalar QED and assume spatial homogeneity for simplicity.  The polarisation of the electric field is left arbitrary~\cite{Pervushin:2006vh, Blaschke:2011is, Huang:2019szw, Aleksandrov:2020mez}.  To discuss the intermediate number, one needs to define creation/annihilation operators at time $t$.  This may be achieved by expanding the field operator $\phi$ with an ``intermediate" mode function $\phi^{\rm int}_{\pm}$ at each instant of time $t$ as
\begin{align}
	\phi(x) 
	= \int {\rm d}^3{\mbf p} \left[  \phi_{+}^{{\rm int}}(x;{\mbf p}) a^{\rm int}(t;{\mbf p}) + \phi_{-}^{{\rm int}}(x;{\mbf p}) b^{{\rm int}\dagger}(t;-{\mbf p}) \right] .  
\end{align}
The choice of $\phi^{\rm int}_{\pm}$ is where one has to make an ansatz.  Once the interpolating mode function is specified, one may compute the intermediate number of created electrons at time $t$:
\begin{align}
	\frac{{\rm d}^3N(t)}{{\rm d}^3{\mbf p}} 
	= \bra{{\rm 0;in}} a^{{\rm int}\dagger}(t;{\mbf p}) a^{{\rm int}}(t;{\mbf p}) \ket{{\rm 0;in}}
	= \int {\rm d}^3{\mbf p}' \left| \left( \phi_{-}^{{\rm int}}({\mbf p})|\phi_{+}^{{\rm in}}({\mbf p}') \right) \right|^2 \;.   \label{eq:ht5-45}
\end{align}
Differentiation of Eq.~(\ref{eq:ht5-45}) with respect to time $t$ leads to a closed set of differential equations for the intermediate number $N$.  The forms of the differential equations depend on the choice of the intermediate mode function $\phi_{\pm}^{{\rm int}}$ \cite{Dabrowski:2014ica, Dabrowski:2016tsx}, which is a reflection of the ambiguity of intermediate particle states.

A widely-used ansatz for the intermediate mode function $\phi^{\rm int}_{\pm} $ is to use the $n$-th order WKB mode function $\phi^{{\rm WKB},(n)}_{\pm}$, which is sometimes called adiabatic basis~\cite{Tanji:2008ku,Tanji:2010eu, Dabrowski:2014ica,Dabrowski:2016tsx,Ilderton:2021zej}.  The quantum Vlasov approach is a special case of $n=0$ \cite{Dabrowski:2014ica, Dabrowski:2016tsx}.  The $n$-th order WKB mode function $\phi^{{\rm WKB},(n)}_{\pm}$ is defined by truncating the function $S$ in the WKB ansatz (\ref{eq:ht2--13}) up to the order of $\hbar^n$\footnote{The series coefficients $\phi^{(n)}_{\pm}$ in Eq.~(\ref{eq:ht2-13}) are obtained by further expanding the exponential in the $n$-th order WKB mode function $\phi^{{\rm WKB},(n)}_{\pm}$ (\ref{eq:ht2---13}).  In general, $\phi^{{\rm WKB},(n)}_{\pm}$ is asymptotic for large $n$, as was the case in $\phi^{(n)}_{\pm}$.  }:  
\begin{align}
	\phi^{{\rm WKB},(n)}_{\pm} = \frac{e^{\mp \frac{i}{\hbar}\int^t{\rm d}t'\,S_n(t')}}{\sqrt{2S_n}}   \frac{e^{+i\mbf{p}\cdot\mbf{x}}}{(2\pi \hbar)^{3/2}} \ \ {\rm with}\ S_n\ {\rm s.t.}\ S = S_n + {\mathcal O}(\hbar^{n+1}) \;. \label{eq:ht2---13}
\end{align}
The function $S_n$ can be determined by solving the Klein-Gordon equation order by order in $\hbar$.  The WKB mode function seems to be a natural ansatz; in particular, when electric fields are not so strong nor fast, the system is adiabatic and higher-order ${\mathcal O}(\hbar^{n+1})$ corrections may be neglected.  In such a case, the WKB mode function becomes an approximate eigenfunction for the time-translation operator and may distinguish the positive and negative energy states.  In practice (e.g., in backreaction problem), it is typical to adopt the zeroth- (or low-) order WKB mode function because it is simple and sufficient to eliminate the ultraviolet divergences from vacua at each instant of time \cite{Tanji:2008ku}.  It is also appealing that one may assign a physical meaning to the intermediate particle number with the zeroth-order WKB mode function: it coincides with the number at the out-state for the same electric field with a sudden switch-off at time $t$ \cite{Tanji:2008ku, Ilderton:2021zej}.  Nonetheless, there is no reason forbidding us to use higher-order ($n\geq 1$) WKB mode functions (which may also have a physical interpretation~\cite{Ilderton:2021zej}).  As emphasised, the intermediate particle number (\ref{eq:ht5-45}) is sensitive to the choice of intermediate mode functions and hence the order of the WKB expansion $n$ (see Fig.~\ref{fig-ht2}).  The sensitivity was investigated in detail by Dabrowski-Dunne  \cite{Dabrowski:2014ica, Dabrowski:2016tsx}.  It was shown that there exists an optimal truncation order $n$, giving the smoothest intermediate particle number, and that it is approximated well by the so-called superadiabatic particle number.  The key point is that the WKB expansion is a divergent asymptotic series (see Sec.~\ref{sec:ht2}), and hence one has to resum it to make it well-defined.  The idea of the superadiabatic particle number is to apply Borel resummation only to the leading-order factorial divergence of the series coefficients $\phi_{\pm}^{(n)}$ in Eq.~(\ref{eq:ht2-13}) and carry out the Laplace integral using the steepest descent method to obtain a Borel sum in an approximate manner.  One can then obtain a connection matrix between $\phi^{\rm in}_\pm$ and $\phi^{\rm out}_\pm$, similar to the exact WKB method (\ref{eq:ht2-21}), which can be used to compute the intermediate particle number.  The result is the superadiabatic particle number \cite{Dabrowski:2014ica, Dabrowski:2016tsx}: 
\begin{align}
	\frac{1}{V} \frac{{\rm d}^3N(t)}{{\rm d}^3\mbf{p}} \approx \frac{1}{(2\pi \hbar)^3} \left| \sum_{j} \frac{1}{2} {\rm Erfc}\left( - \frac{{\rm Im}\, F_j(t)}{\sqrt{2\,{\rm Re}\,F_j(t)}} \right) e^{\sigma(w^{\rm tp}_j)/\hbar} \right|^2\ {\rm with}\ F_j(w) = \sigma(w) - \sigma(w^{\rm tp}_j) \;.  
\end{align}
The function $F_j$ is called singulant \cite{Barry:1989zz, Dingle1973}.  The advantage of the superadiabatic particle number is that it connects the asymptotic numbers of created pairs very smoothly and does not show unpleasant oscillations and peaks observed with the WKB mode functions (see Fig.~\ref{fig-ht2}).  The optimal truncation order of WKB is defined as the one giving the closest number of created pairs to the superadiabatic one.  For electric fields having a single pair of turning points, it was identified that $n$ is the closest integer to $ {\rm Re}\,\sigma(w^{{\rm tp}*}) - 1$ \cite{Dabrowski:2014ica, Dabrowski:2016tsx}.  Note that the reason why the superadiabatic basis has a finite ``tunneling time" is that, due to the approximations, the Stokes phenomenon at each Stokes-line crossing acquires a finite time width compared to the exact treatment (\ref{eq:ht2-21}) as $\Theta(t-t^{\rm cr}_j) \to (1/2) {\rm Erfc}( - {\rm Im}\, F_j(t)/\sqrt{2\,{\rm Re}\,F_j(t)} )$.  In general, the finiteness of ``tunneling time" is attributed to the difference between the basis adopted and the exact Borel sum.  This means that the concept of ``tunneling time" is highly dependent on the choice of a basis and hence is ambiguous \cite{Landsman:2014auv,2016PhRvL.116w3603Z}.

\subsection{Analogue Schwinger effect}\label{sec:Schwinger:newCondMatt}

The past decade has seen a large amount of work in the pursuit of analogue models of strong-field QED effects in condensed-matter systems.  While these analogues can be contaminated by non-QED effects such as impurities, lattice structures, dimensionality, temperature, and so on, they may provide fruitful insights to strong-field QED and vice versa.  This research direction has largely been pursued in the context of the Schwinger effect.  While it remains difficult to create strong fields exceeding the Schwinger field $eE > eE_{\rm S} = m^2$ with e.g.~current laser technology, it is possible for condensed-matter materials, where the band-gap size $\Delta$ is much smaller than the electron mass [$\Delta = {\mathcal O}(1\;{\rm eV}) \sim 10^{-5}m$ for typical solids] to create ``strong'' fields such that $eE > \Delta^2$ in actual experiments.  We highlight some results below; for comprehensive reviews on strong-field condensed-matter phenomena, see, e.g., Refs.~\cite{Oka:2011kf, 2014JPhB...47t4030G, Basov2017, 2018RvMP...90b1002K, 2019ARCMP..10..387O}.

There are a number of theoretical proposals for condensed-matter analogues of the Schwinger effect, including the dielectric breakdown of materials (the Landau-Zener effect), e.g., graphene \cite{Allor:2007ei, Gavrilov:2012jk, Zubkov:2012ht, Katsnelson:2012tp, Katsnelson:2012cz, Fillion-Gourdeau:2015dga, Akal:2016stu, Akal:2018txb}, cold atom \cite{Szpak:2011jj, Kasper:2015cca}, semi-conductor \cite{Linder:2015fba}, Dirac/Weyl materials \cite{Vajna:2015qra, Abramchuk:2016afc, Chernodub:2019blw}.  The Schwinger effect has also been used to propose novel mechanisms in condensed-matter processes including Mott breakdown~\cite{2010PhRvL.105n6404E, Oka:2011ct, PhysRevB.81.033103, 2012PhRvA..85c3625Q}, magnon creation~\cite{Hongo:2020xaw}, spin-current generation \cite{Huang:2019szw}, and a superconductor analog of the Schwinger effect~\cite{Solinas:2020woq}.  

Intriguing results have been reported from the experimental side as well, including the consistency of ``electron-hole pair creation'' mimicked through cold atoms with the QED formula (\ref{eq:ht5})~\cite{Pineiro:2019uzb}.  A realtime study of the vacuum persistence probability of ${\rm QED}_{1+1}$ was carried out with a few-qubit quantum simulator realised through trapped ions~\cite{Martinez:2016yna} (see also a recent quantum simulation on ${\rm QED}_{3+1}$ for a spatially homogeneous electric field with an IBM quantum computer \cite{Xu:2021tey}).  Doublon-holon pair creation in strongly-coupled materials~\cite{2015PhRvB..91w5113M, 2017NatMa..16.1100Y} has also been studied, and an exponential dependence similar to Eq.~(\ref{eq:ht5}) was found, but with an exponent modified by the strong correlation.  The static/dynamical Franz-Keldysh effect, which is an analogue of the dynamically assisted Schwinger effect (as well as non-linear Breit-Wheeler), was studied by measuring field-dependent photo-absorption rates under strong THz lasers or infrared pulses~\cite{2010ApPhL..97u1902S, 2013NatSR...3E1227N, 2013Natur.493...75S, 2016Sci...353..916L}.  Quantum-interference in the Schwinger effect has been studied under the name of the St\"{u}ckelberg-phase-interference effect in condensed-matter physics~\cite{Shevchenko_2010}, and this has been confirmed experimentally in, e.g., an optimal lattice~\cite{2010PhRvL.105u5301K} and in graphene~\cite{2017Natur.550..224H}.  A graphene analogue of vacuum decay due to a strong Coulomb field was observed in~\cite{2013Sci...340..734W} and was consistent with the prediction of QED~\cite{Rafelski:2016ixr} that the vacuum starts to decay for $Z \gtrsim 1/\alpha$, with $Z$ being atomic number. 

\section{Ritus-Narozhny Conjecture}\label{sec:RN}
We have now seen many Furry picture calculations, where the charge-field coupling $e \ll1$ is enhanced in a strong background field, $e\to\xi\gg1$ that must be treated exactly, and the coupling to dynamical fields $e\ll1$, is treated perturbatively as usual. It is, however, possible for strong fields to also enhance the coupling between dynamical fields as $e\to e\xi^n$ where the (positive) power $n$ varies from case to case~\cite{Heinzl:2021mji}. This implies that the Furry expansion breaks down for sufficiently large $\xi$. Here by a breakdown of a perturbative expansion we mean that the magnitude of the successive terms of the expansion grows from the start, so that the expansion is meaningless even as an asymptotic one (for which the magnitude of the successive terms initially decreases but at some point starts growing) unless an all-order resummation is performed. There are many examples of such a breakdown in both quantum and classical physics; let us use the latter, simpler, case to illustrate.

Consider the classical dynamics of a charged particle. Treating the Lorentz force exactly, but radiation reaction effects in perturbation theory, is the classical analogue of the Furry expansion in QED, see e.g.~\cite{Ilderton:2013tb}. One example is an electron in a rotating electric field, strength $E_0$, frequency $\omega$. The electron has a stable orbit for a particular energy~\cite{PhysRevE.84.056605} which is, according to the Lorentz force, determined by the condition $\gamma^2-1 = \xi^2$, where $\xi = eE_0/(m\omega)$. Corrections calculated in the Furry expansion are expected to scale with powers of $\epsilon = (2/3)(e^2)(\omega/m)$, which should be small for a classical analysis to be valid. However, one finds that these corrections in fact scale with powers of $\epsilon \xi^3$, with leading behaviour $\gamma^2 -1 \sim \xi^2(1-\epsilon^2\xi^6)$. For sufficiently large $\xi$ the correction becomes large, and the energy becomes imaginary, which signals the breakdown of the Furry expansion.
    
This is an example system in which the Furry expansion can be effectively resummed, as the exact solution to the LAD equation, which describes radiation reaction effects, is available (see, e.g., \cite{zeldovich:1975in}); this implies that $\gamma$ satisfies an equation which recovers the perturbative result for $\epsilon \xi^3$ small, but behaves as $\gamma^2-1 \sim \xi^2 (\epsilon \xi^3)^{-1/2}$ for large $\epsilon\xi^3$, and the energy stays real. Hence resummation (achieved here through an exact solution rather than direct resummation methods) corrects the unphysical behaviour seen in perturbation theory. For other examples see~\cite{Heinzl:2021mji}.

This section reviews the yet fragmentary and incomplete results on a specific example of the phenomena described above: the breakdown of the QED Furry expansion in extremely strong fields for which the (locally) constant field approximation (see \secref{sec:approx:lcfa}) holds. As discussed below, it appears that in this case the actual emergent expansion parameter is $g_q=\alpha\chi^{2/3}$, where $\chi$ is the quantum nonlinear parameter \eqref{eq:chi:def}. This is the Ritus-Narozhny (RN) conjecture on the behaviour of QED in a constant crossed field (CCF), to which we restrict our attention from here on. Though formulated by the beginning of 1980's, the conjecture remained almost unnoticed 
until very recently. A notable revival of the interest in this topic was accompanied by recent experimental proposals for attaining the breakdown threshold $g_q\sim1$ in the near future. Our goal here is to review all the relevant results, both old and recent, in order to clarify the state-of-the-art, the main directions of current research, and the future prospects.

\subsection{Early evidence and the criterion proposed for breakdown of the Furry expansion}
\label{sec:N-R_prev}

A perturbation theory breakdown in a CCF occurs already at a classical level in a transition from the LL to LAD equation, see \secref{sec:resum_LAD} for more details on this specific case and the rest of \secref{sec:higher} for the discussion of breaking down and resummation of perturbative expansions from a broader perspective. The LL equation emerges at the leading order of an expansion of LAD that develops assuming the radiation reaction force is smaller than the Lorentz force in the electron rest frame. The relevant expansion parameter in classical theory is $g_c=\alpha\chi$, see \eqref{LL_LAD_exppar}. As explained in \secref{sec:resum_LAD}, within the scope of the classical theory this parameter is always small and LL represents an approximation to LAD that is well-suited and sufficient for practical purposes. However, if this parameter is formally considered large, then the expansion in powers of $g_c$ breaks down and an all-order resummation is required, by which one eventually arrives back at LAD. As is well known, LAD possesses unphysical runaway or pre-accelerating solutions, signalizing that classical electrodynamics becomes {inconsistent} in strong fields $\chi\gtrsim\alpha^{-1}$, though this is revealed only for parameters beyond its regime of validity, when it should be replaced by QED.

In contrast, QED remains self-consistent at least as long as it stays well below the scales associated with the Landau pole, which can be viewed as infinite for any practical purposes. (QED is in any case modified at much lower energy scales when embedded into the unified electroweak theory of the Standard Model and after that possibly into an as-yet hypothetical Grand Unified Theory.) Radiation processes are considered in the framework of a perturbative approach, typically well justified by the smallness of the QED coupling $\alpha$. Action of a strong external field is taken into account exactly by means of working in the Furry picture, see \secref{sec:prelim}. The field strength is characterised by several dimensionless Lorentz invariant parameters, among which let us here focus on $\chi$ and $\xi$ (see \secref{sec:intro:experiments}). The zero frequency limit, in which $\xi\to\infty$ while $\chi={\rm const}$, formally corresponds to the case of a constant field. A feasibility of the corresponding approximation when applied to realistic fields is formalised by the LCFA, as discussed earlier in \secref{sec:approx:lcfa}.   
 
From the very beginning, studies of the simplest tree-level first-order processes of photon emission (nonlinear Compton) and pair photoproduction (nonlinear Breit-Wheeler) in a CCF revealed that for $\chi\gg1$ their rates (probability per unit time) scale as $\propto\alpha\chi^{2/3}$, see Eqs.~\eqref{eq:first:NLCrate} and \eqref{eq:first:NBWrate}, respectively. The appearance of the (large by assumption) dynamical factor $\chi^{2/3}$, common for both processes, was originally called a `universal enhancement of radiation processes in intense field' \cite{Ritus:1970radiative}. However, in a lack of genuine insights into the higher-order behaviour, the term `universal' at the time referred only to the same behaviour of electrons and photons at lowest order. The consequence of such an enhancement is that if $\alpha\chi^{2/3}\gtrsim1$, then these processes occur on a sub-Compton proper time scale. As the Compton time in many aspects emerges as a minimal duration on which a particle could be localised in a relativistic quantum theory, this already implies that in a strong enough constant crossed field the concept that a particle can be driven purely by the field without the accompanying radiative processes, which serves a zeroth-order approximation for the Furry expansion, becomes somehow limited when $\alpha\chi^{2/3}\gtrsim1$. 

Furthermore, by virtue of the optical theorem, the first order processes under discussion are related to the one-loop radiative corrections to the electron and photon self-energies, called the electron mass operator and the photon polarisation tensor, respectively. To be specific, the renormalised exact one-loop contribution to the polarisation tensor in a CCF is given by Eq.~\eqref{eq:Pi_CCF}, where for $\chi_{\ell}=e\sqrt{-(F.\ell)^2}/m^3\gg1$ and
{$\ell^2\lesssim m^2\chi_{\ell}^{2/3}$} the scalar coefficients \eqref{poltensor_CCF2} asymptotically scale as \cite{Narozhny:1968,Mironov:2020resummation}
\begin{align}\label{poltensor_large_chi}
\pi_{{\cal F}\cal{F}}(\ell^2,\chi_\ell),\,\pi_{\tilde{{\cal F}}\tilde{\cal{F}}}(\ell^2,\chi_\ell)\propto \alpha\chi_{\ell}^{2/3},
\end{align}
(the coefficient $\pi_T(\ell^2,\chi_\ell)$ scales weaker as $\propto \alpha\log{\chi_{\ell}}$). Similarly, the renormalised one-loop mass operator for $\chi\gg 1$ and
{$p^2\lesssim m^2\chi^{2/3}$} 
is dominated by the term \cite{Ritus:1972radiative,Narozhny:1980expansion,Mironov:2021structure}
\begin{align}\label{massop_large_chi}
M(p)\propto \frac{\alpha \gamma^\mu F_{\mu\nu} F^\nu_{\;\;\lambda} p^\lambda}{m^4\chi^2}V(p^2,\chi),\quad V(p^2,\chi)\propto\alpha\chi^{2/3}.
\end{align}
It turns out, for $\chi\gg1$ both real and imaginary parts of these corrections (the former on-shell giving the field-induced dynamical electron and photon squared mass shifts and the latter proportional to their above mentioned decay rates) also grow as $\propto \alpha\chi^{2/3}$ and for $\alpha\chi^{2/3}\gtrsim1$ exceed the free electron squared mass $m^2$ (which is a tree-level reference value). This also implies that at this point the field-enhanced radiative corrections can no longer be regarded small, thus giving another indication of a possible Furry expansion breakdown.

\begin{table*}[ttt!]
	\caption{\label{tab:RN_studies1980}Known asymptotic scaling for the radiative corrections in a CCF to the polarisation operator (left) and to the mass operator (right). For each diagram the row specifies the $\chi\gg1$ asymptotic behaviour together with the corresponding source. The dominant scaling in $\chi$ is highlighted in bold for each loop order. (Reproduced from the introduction of \cite{Mironov:2020resummation}).}\vspace{0.5ex}
	\renewcommand*{\arraystretch}{1.6}
	\newcolumntype{A}{>{\hsize=.1\hsize\linewidth=\hsize\raggedleft\arraybackslash}X}
    \newcolumntype{B}{>{\hsize=1.8\hsize\linewidth=\hsize\centering\arraybackslash}X}
    \newcolumntype{C}{>{\hsize=1.1\hsize\linewidth=\hsize}X}
    \newcolumntype{D}{>{\hsize=1.0\hsize\linewidth=\hsize\raggedleft\arraybackslash}X}
		\begin{tabularx}{\textwidth}{lccr}
		    \hline\hline
			\multicolumn{4}{c}{\bf 1 loop} \\ [0.5ex]
			$\,$ &
			\multicolumn{1}{c}{
				\begin{tabularx}{0.45\textwidth}{ABCD}
					(1a) & \includegraphics[valign=c,width=2.5cm]{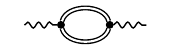}& $\alpha\bm{\chi^{2/3}}$ & \cite{Narozhny:1968} \\ [2ex]
				\end{tabularx} 
			} &
			\multicolumn{1}{c}{
				\begin{tabularx}{0.45\textwidth}{ABCD}
					(1b) & \includegraphics[valign=c,width=2.5cm]{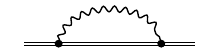}& $\alpha\bm{\chi^{2/3}}$ & \cite{Ritus:1970radiative} \\ [2ex]
				\end{tabularx}
			} &
			$\,$ \\ [1ex] \hline
			\multicolumn{4}{c}{\bf 2 loops} \\ [0.5ex]
			$\,$ &
			\multicolumn{1}{c}{
				\begin{tabularx}{0.45\textwidth}{ABCD}
					(2a) & \includegraphics[valign=c,width=2.5cm]{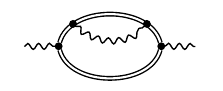}& \hspace*{-0pt}$\alpha^2\bm{\chi^{2/3}}\log{\chi}$ & \cite{MorozovNarozhnyiPhTr} \\ [2ex]
				\end{tabularx}
			} &
			\multicolumn{1}{c}{
				\begin{tabularx}{0.45\textwidth}{ABCD}
					(2b) & \includegraphics[valign=c,width=2.5cm]{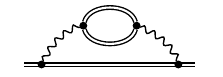} & $\alpha^2\bm{\chi}\log{\chi}$ & \cite{Ritus:1972radiative,Ritus:1972nf}\\
					(2c) & \includegraphics[valign=c,width=2.5cm]{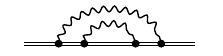} & $\alpha^2\chi^{2/3}\log{\chi}$ & \cite{Morozov:1975uah} \\ [2ex]
				\end{tabularx}
			} &
			$\,$ \\ \hline
			\multicolumn{4}{c}{\bf 3 loops} \\ [0.5ex]
			$\,$ &
			\multicolumn{1}{c}{
				\begin{tabularx}{0.45\textwidth}{ABCD}
					(3a) & \includegraphics[valign=c,width=2.5cm]{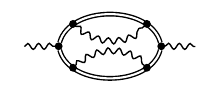}& $\alpha^3\chi^{2/3}\log{\chi}$ & \cite{Narozhny:1979radiation} \\
					(3b) & \includegraphics[valign=c,width=2.5cm]{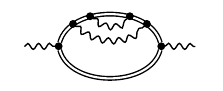}& $\alpha^3\chi^{2/3}\log{\chi}$ & \cite{Narozhny:1979radiation} \\
					(3c) & \includegraphics[valign=c,width=2.5cm]{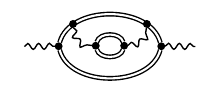}& $\alpha^3\bm{\chi}\log^2{\chi}$ & \cite{Narozhny:1980expansion} \\ [2ex]
				\end{tabularx}
			} &
			\multicolumn{1}{c}{
				\begin{tabularx}{0.45\textwidth}{ABCD}
					(3d) & \includegraphics[valign=c,width=2.5cm]{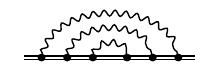}& $\alpha^3\chi^{2/3}\log^2{\chi}$ & \cite{Narozhny:1979radiation} \\
					(3e) & \includegraphics[valign=c,width=2.5cm]{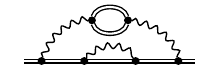}& $\alpha^3\chi^{4/3}$ & \cite{Narozhny:1979radiation} \\
					(3f) & \includegraphics[valign=c,width=2.5cm]{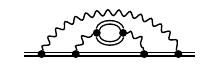}& $\alpha^3\chi\log^2{\chi}$ & \cite{Narozhny:1980expansion}\\
					(3g) & \includegraphics[valign=c,width=2.5cm]{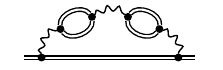}& $\alpha^3\bm{\chi^{5/3}}$ & \cite{Narozhny:1980expansion} \\ [2ex]
				\end{tabularx}
			} &
			$\,$ \\ \hline\hline
		\end{tabularx}	
\end{table*}

Subsequently, the Ritus group put a lot of effort in calculations of the higher-order SFQED effects. Unlike the others, who had studied mostly the processes in a plane wave or magnetic field as certainly more practically important ones, they had focused on  the simplest case of a constant crossed field. Its simplicity, however, made it possible to substantially advance calculations of the high-order processes. Another notable point of their strategy was to consider the self-energy corrections, from which the probabilities of real radiation processes could be deduced by means of the optical theorem. The results thus obtained in 1969--1980 are sketched in Table~\ref{tab:RN_studies1980}. While the one- and two- loop corrections indicated in the table were computed exactly, the high-$\chi$ behaviour of the three-loop ones could be only roughly estimated. Note that the contributions containing 1PI-vertex corrections could not be evaluated at the time, hence are missing in the table (see further discussion below). However, they were always believed \cite{Ritus:1972nf,Narozhny:1979radiation} to be at high energies smaller than, or at most of the same order as the corrections shown in Table~\ref{tab:RN_studies1980}; it would be interesting to investigate this with explicit calculations.

The left column in the table lists the results on the photon self-energy (polarisation tensor) and the right column shows the same for the electron self-energy (mass operator). In each loop order the fastest growing corrections are highlighted in bold. Clearly, it looks as in each order shown they are always given by the bubble-type diagrams (i.e. those including photon propagator maximally saturated with one-loop vacuum polarisation insertions). One can see that, ignoring the weakly growing log-factors, the next-to-current order ratios of the successive fastest growing corrections in the two cases form the sequences
\begin{equation}\label{secN-R:ratios}
\begin{aligned}
&& &\ilfrac{\text{(2b)}}{\text{(1b)}}\simeq\alpha\chi^{1/3},& &\ilfrac{\text{(3g)}}{\text{(2b)}}\simeq\alpha\chi^{2/3},& &\ldots&\\
&\ilfrac{\text{(2a)}}{\text{(1a)}}\simeq\alpha,& &\ilfrac{\text{(3c)}}{\text{(2a)}}\simeq\alpha\chi^{1/3},& &\ldots&
\end{aligned}
\end{equation}
As each photon self-energy correction necessarily contains an extra electron loop outside the inner photon propagator, the number of bubble insertions it contains in each order is always less by one than the number of such insertions in the electron self-energy corrections of the same order. This explains the one-order shift observed in comparing the ratios of the fastest growing successive corrections for the two cases. In particular, it can be expected that the fourth-to-third loop order ratio of the leading corrections to the photon self-energy would be of the same order as the third-to-second order ratio in the right column, i.e. $\simeq\alpha\chi^{2/3}$. Once the second-loop order results became available, it was initially claimed \cite{Ritus:1972radiative,Ritus:1972nf,Morozov:1975uah,MorozovNarozhnyiPhTr} that the expansion parameter is $\alpha\chi^{1/3}$. But later, after having obtained the third-loop results, it was finally conjectured by Narozhny in \cite{Narozhny:1980expansion} that to higher orders (not shown in the table) all the ratios should stabilise at the universal value $g_q=\alpha\chi^{2/3}$, thus representing the actual expansion parameter of the Furry expansion for self-energy corrections in a CCF for $\chi\gg1$. If so then this would also imply a breakdown of the Furry expansion for $g_q\gtrsim1$ (i.e., for $\chi\gtrsim\alpha^{-3/2}\simeq 1600$). As mentioned, however, this conjecture for quite long remained almost unnoticed. It was suggested in \cite{Fedotov:2017conjecture} to call it the Ritus-Narozhny (RN) conjecture. 

So far, we have not discussed the field-induced modifications of radiative corrections to a QED vertex. In fact, calculation of such type of corrections is most challenging, therefore they remain least studied (hence are missing in Table~\ref{tab:RN_studies1980}). In the past the one-loop vertex correction in a CCF was discussed only in \cite{Morozov:1981vertex} (see also \cite{Morozov:1981preprint} presenting the relevant calculation details), where it was first estimated that for $\chi\gg1$ it scales the same way $\propto\alpha\chi^{2/3}$ as all the other one-loop corrections. However, no attempt was made since then to analyse this result in a more general context, in particular to consider the impact of inserting vertex corrections into high order diagrams on their strong field scaling (moreover, it was always implied without proper justification in \cite{Ritus:1972radiative,Ritus:1972nf,Morozov:1975uah,MorozovNarozhnyiPhTr,Narozhny:1979radiation,Narozhny:1980expansion} that such type of corrections should always remain subleading). Only recently the findings of \cite{Morozov:1981vertex,Morozov:1981preprint} have been revisited and generalised in \cite{DiPiazza:2020kze} to the case of a plane wave background. By specialising to the CCF case the scaling $\propto\alpha\chi^{2/3}$ was independently confirmed. Besides, an additional important insight was made that the only terms of such leading scaling are precisely those which exactly cancel on-shell due to gauge invariance when the one-loop vertex correction is combined with the one-loop self-energy corrections to the electron lines attached to the vertex. This might imply that in a certain special gauge both electron mass and vertex corrections would remain subleading with respect to photon mass (vacuum polarisation) corrections, thus singling out the bubble chain corrections as most enhanced. Existence of such a gauge, however, has not been demonstrated so far and might require extremely challenging calculations involving vertex corrections to higher orders. This is one reason that an unambiguous, rigorous identification of which diagrams provide the leading scaling in $g_q$, to all orders, is still lacking.

\subsection{Feasibility of reaching the regime \texorpdfstring{$\alpha\chi^{2/3}\gtrsim1$}{alpha*chi2/3>1}}\label{sec:RN:feasibility}

The current interest in the RN conjecture is in part motivated by the experimental proposals that appeared recently \cite{Blackburn:2019reaching,Yakimenko:2018kih,Baumann:2019probing,Fedeli:2021probing,Baumann:2019laser,DiPiazza:2019vwb}. Although the same applies to hard photons as probes, let us follow the proposals that assume the probe particles are electrons. To start with, attaining the values $\chi\simeq 1600$ is deemed extremely challenging by itself, it is worth mentioning that all the planned  experimental campaigns with high power lasers will operate at best at $\chi\lesssim10$ (see Fig.~\ref{FIG:PARAMPLOT}, where the required level $\alpha\chi^{2/3}\simeq1$ is shown in the top right corner). Going far beyond that will require increase of the energies of probe electrons or field strength to substantially higher values than readily available and thus imply serious further advances in accelerator and/or laser technologies. 

Another challenge, as specifically stressed in the proposals, is the necessity to mitigate the unsolicited radiative losses of the incoming probe particles before they reach the strong field region (see also \cite{Bradley:2021dnu}). Indeed, as they approach the region of the strongest field with $\chi\gg1$ by assumption, the field along the particle trajectory gradually ramps up. But already as soon as the quantum regime $\chi\gtrsim1$ is attained, the energy transfer to the emitted photons becomes on average of the order of the energy of the emitting particle, hence as a rule photon emission results in significant drop of the $\chi$ parameter, thus obstructing its further growth to the demanded higher values. The only way to overcome this is to make the field rising sharper than the average radiation path, in which case a notable fraction of probe electrons could avoid hard photon emissions before reaching the field core (this is in essence the same effect as quantum quenching of radiation losses as discussed for short laser pulses in \cite{Harvey:2016uiy}, but here demanded solely in the transient region). Assuming the initial electron energies do not exceed hundreds GeVs, the average radiation path would be less than about ten nanometres. As shown in \cite{Blackburn:2019reaching}, under such conditions the oblique collisions are advantageous over the normal ones and it might be still possible to attain $\chi\simeq 10^2$ with high-power optical lasers by using tight focusing and injecting probe electrons into the focal spot sideways. However, an even sharper turn-on of the field can hardly be realised directly using optical lasers. Therefore it has been suggested to probe the regime $g_q=\alpha\chi^{2/3}\gtrsim1$ using several alternative setups.

One of the proposed options relied on a future electron-electron or electron-positron collider. Namely, motivated by the successive demonstration of beam compression at FACET-II it was proposed \cite{Yakimenko:2018kih} to tune the collider design in such a way to make the produced bunches of total charge $\sim 0.14$nC round and tightly focused in transverse directions ($\sim10$nm), and longitudinally compressed (up to $\sim 10-100$nm), thus enabling their peak current $\sim1.7$MA. The suggested parameters were obtained by trading the demanded enhancement of $\chi$-parameter at the impact against reducing the radiative losses while also keeping the disruption of the collided bunches under a reasonable control. According to simulations, with energy $125$GeV per electron (or positron) in the bunches, the $\chi$-parameter of a substantial part ($\sim 40\%$) of the particles acquires the demanded values $\chi\simeq1600$ during the overlap of the colliding bunches. 

Most of other discussed setups also relied on using $\simeq 100$GeV electrons, but rather collided with frequency-upshifted laser pulses. It was proposed in \cite{Baumann:2019probing} to use tightly focused ultraintense ($\xi\sim1450$) attosecond pulses (duration $\sim150\,$as) obtained from moderately intense ($\xi\sim350$) optical laser pulses via their reflection off a solid target. As a side remark in \cite{Fedeli:2021probing}, the demanded values of the $\chi$ parameter could be achieved also by reflecting petawatt-class laser pulses off a counterpropagating relativistic plasma mirror (according to the estimates in \cite{Fedeli:2021probing} in such a case it might even be sufficient to use $16\,\trm{GeV}$ electrons). Another idea \cite{Baumann:2019laser} was to inject $100\,$GeV electrons from the rear side of a sharp ($\sim10\,$nm) laser-plasma interface of hole boring by an ultraintense\footnote{Note that in plasma physics the parameter $\xi$ is commonly denoted by $a_0$.} ($\xi\sim1400$) circularly polarised optical laser pulse in an overdense ($n\sim 1000 n_{cr}$, where $n_{cr}=m\omega^2/4\pi e^2$ is the critical plasma density) plasma target. Finally, it was argued that the desired values of $\chi$ could be also attained by channeling multi-TeV electrons/positrons in aligned crystals \cite{DiPiazza:2019vwb}. {Note that one advantage of the crystal setup is that the length of a crystal can be chosen such that at most one photon is emitted by the electron.}

Neither one of the above proposals was capable of specifying definite experimental signatures of the nonperturbative regime { $\alpha\chi^{2/3}\gtrsim1$} due to the lack of precise, relevant, theory  predictions. In fact, they all only demonstrated a principle feasibility of attaining the values $\chi\sim1600$ with the focus on the required advances in the near-future technology. However, it was stressed and explicitly demonstrated in \cite{Ilderton:2019note} for photon emission and spin flip rates and in \cite{Podszus:2019high} for one-loop mass and polarisation operators in a pulsed plane wave background that it is not enough just to reach the values $\alpha\chi^{2/3}\gtrsim1$, as one needs also to ensure the validity of LCFA, as has been explicitly done in \cite{Yakimenko:2018kih}. For tree-level first order radiation processes and one-loop corrections the corresponding condition reads $\xi\gg\chi^{1/3}$ (cf. \eqref{eqn:lcfawhen1} and the discussion after it) and implies that, being sharp, the strong field region should be also reasonably long. Otherwise, e.g. the rate of a first-order process instead of $\alpha\chi^{2/3}/\eta$ turns into $(\alpha\xi^3/\chi)\log{\chi}$ and no considerable enhancement can take place \cite{Ilderton:2019note}. The same applies to spin flip rates and one-loop radiative corrections as well. The latter is in agreement with exactly solvable, though opposite to assumed in the experimental proposals,  cases of $\delta$-function longitudinal field profile considered in the context of beam-beam collision in \cite{Adamo:2021jxz} (see more detailed discussion in Sec.~\ref{sec:null_vs_nonnull}) and in the context of electron collision with laser pulses in \cite{Ilderton:2019vot}, which as expected demonstrated absence of power-low scaling of radiative corrections. Furthermore, as we discuss below, the condition for LCFA applicability is likely modified in higher orders, getting even more restrictive than in lowest order.

\subsection{Bubble-chain diagrams to all orders and their resummation}

\begin{figure}[t]
\begin{center}\includegraphics[width=0.95\textwidth]{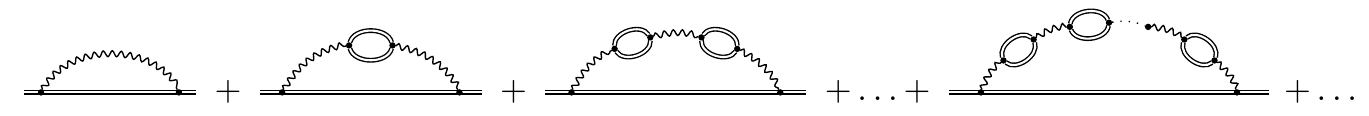}\end{center}
\caption{Bubble-type polarisation corrections to the electron mass operator (double lines denote electron
propagators dressed by a CCF).}
\label{fig:mass_bubble_corr}
\end{figure}

As we have seen earlier in Sec.~\ref{sec:N-R_prev}, older calculations of mass and polarisation operators in a CCF in the few lowest (up to three-loops) orders were giving a strong hint that the contribution growing quickest with $\chi$ at each order, might be those associated with bubble-chain corrections. Recall that by this we mean the diagrams that are maximally saturated with one-loop vacuum polarisation insertions, like 2b, 2c or 3g in Table~ \ref{tab:RN_studies1980}. Maximal saturation means they contain a single chain of photon propagators sandwiched by one-loop vacuum polarisation insertions (bubbles). 

As the next step, it is instructive to examine such diagrams to an arbitrary order and consider their resummation. For bubble chain contributions to the mass operator shown in Fig.~\ref{fig:mass_bubble_corr} this task was implemented in Ref.~\cite{Mironov:2020resummation} for an incoming on-shell electron. Instead of describing the calculations of \cite{Mironov:2020resummation} in depth, let us explain the main essence of the arguments of this work in plain terms. Let $p^\mu$, $q^\mu$ and $\ell^\mu$ be the 4-momenta of the incoming electron, virtual electron in the external loop, and of virtual photons (all of which have the same momentum, due to the field homogeneity and overall momentum conservation) in the bubble chain, respectively. We assume the incoming electron is on-shell, $p^2=m^2$. It is convenient to use the gauge \eqref{eqn:this-is-the-potential}, which for CCF just reads $A^\mu=\{0,-\vec{E}x^\LCp,0\}$, and perform calculations in the basis of electron Volkov states \eqref{volkov-e-in} (in the context also often called the Ritus $E_p$-representation or `the Ritus basis') \cite{Ritus:1972radiative}. Then the transverse and lightfront plus-component of generalised momentum are conserved at the vertices, $\vec{p}^\LCperp=\vec{q}^\LCperp+\pmb{\ell}^\LCperp$, $p^\LCp=q^\LCp+\ell^\LCp$, while the minus-component is not due to the exchange of energy and momentum with the field (see the discussion in \secref{sec:first_order_general_props}). Then it is easy to perform all the integrals in the external loop but the ones over $\ell^2$, $\chi_\ell=e\sqrt{-(F\cdot\ell)^2}/m^3$ and over the lightfront extension $x^\LCp$ of the external loop (the latter, if also done, reduces to Airy functions; at this stage the loop integrals in the bubbles also remain to be evaluated). 

Both the lightfront time extent $\Delta x^\LCp$ of the external loop (the formation scale of the process, similar in meaning to the characteristic interference phase $\theta$ which we introduced in \secref{sec:first_order_general_props} and discussed in detail in \secref{sec:approx:lcfa}) and the photon virtuality $\ell^2$ can be estimated by identifying a non-oscillatory domain of the exponential composed from the electron Volkov functions and the photon plane wave involved in the effective field-dressed vertex (this argument is essentially equivalent to applying the uncertainty principle \cite{Bassompierre:1995search,Fedotov:2019qualitative}):
\begin{align}\label{secN-R:uncrt_cond}
\int dx^\LCp\left(p^\LCm(x^\LCp)-q^\LCm(x^\LCp)-\ell^\LCm\right)\lesssim1,
\end{align}
where
\begin{align}\label{secN-R:mass_shells}
p^\LCm(x^\LCp)=\frac{m^2+\left(\vec{p}^\LCperp+e\vec{E}x^\LCp\right)^2}{p^\LCp},\quad
q^\LCm(x^\LCp)=\frac{q^2+\left(\vec{q}^\LCperp+e\vec{E}x^\LCp\right)^2}{q^\LCp},\quad
\ell^\LCm=\frac{\ell^2+(\pmb{\ell}^\LCperp)^2}{\ell^\LCp}.
\end{align}
In the strong field limit, by retaining only the terms quadratic in field strength, whilst accounting for conservation laws, one obtains from \eqref{secN-R:uncrt_cond} 
\begin{align}\label{secN-R:x_plus}
\frac{e^2E^2\ell^\LCp}{p^\LCp q^\LCp}\left(\Delta x^\LCp\right)^3\lesssim1,\quad\Delta x^\LCp\simeq\frac{m}{eE}\left(\frac{\chi\chi'}{\chi_\ell}\right)^{1/3},
\end{align}
where the result is expressed in terms of the parameters $\chi=eEp^\LCp/m^3$, $\chi_\ell=eE\ell^\LCp/m^3$ and $\chi'=eEq^\LCp/m^3=\chi-\chi_\ell$ of the corresponding particles. Likewise, with $\Delta x^\LCp$ known, the effective virtuality $\ell^2$ is estimated from $\left(\ell^2/\ell^\LCp\right)\Delta x^\LCp\lesssim1$ as
\begin{align}\label{secN-R:l**2}
\ell^2\simeq m^2\frac{\chi_\ell^{4/3}}{(\chi\chi')^{1/3}}.
\end{align}

The one-loop polarisation tensor in a CCF (dubbed here a bubble for shortness) was presented previously in Sec.~\ref{sec:LBL}, see Eq.~\eqref{eq:Pi_CCF}. In a CCF its different components $\pi_\Omega$ with $\Omega\in\{T,\mathcal{FF},\mathcal{\tilde{F}\tilde{F}}\}$ depend exclusively on $\ell^2$ and $\chi_\ell$ and never intertwine. The only effect of $\pi_T$ is a finite renormalisation, transforming the fine structure constant into a running coupling. The corresponding one-loop $\beta$-function grows as $\propto\alpha\log{\chi_\ell}$ and can be neglected for our purposes. The remaining components $\pi_{\mathcal{FF}}$, $\pi_{\mathcal{\tilde{F}\tilde{F}}}$ {contribute bubble corrections to the electron elastic scattering amplitude.}
For $\ell^2\lesssim m^2\chi_\ell^{2/3}$ they can be written as $\pi_\Omega(\ell^2,\chi_\ell)\approx \pi_\Omega(0,\chi_\ell)=\alpha m^2 f_\Omega(\chi_\ell)$, where both $f_\Omega(\chi_\ell)$ for $\chi_\ell\ll1$ are almost real-valued ($\text{Im}\,f_\Omega(\chi_\ell)\propto - e^{-8/3\chi_\ell}$), negative and quadratic in $\chi_\ell$, whereas for $\chi_\ell\gg1$ their (positive) real and (negative) imaginary parts are commensurable and $\propto\chi_\ell^{2/3}$. A chain with $r\ge1$ bubbles contributes to the diagrams in Fig.~\ref{fig:mass_bubble_corr} by a factor $(-1/\ell^2)g_\Omega^r(\ell^2,\chi_\ell)$, where
\begin{align}\label{secN-R:g}
g_\Omega(\ell^2,\chi_\ell)=\frac{\pi_\Omega(\ell^2,\chi_\ell)}{\ell^2}\simeq \frac{\alpha(\chi\chi')^{1/3} f_\Omega(\chi_\ell)}{\chi_\ell^{4/3}}
\end{align}
can be thus regarded as an expansion parameter of the expansion in the number of bubbles. Besides this factor, the integrand also contains other factors originating from the electron propagator and the Jacobian of passing to the new integration variables as described. However, for sufficiently large $r$ (it turns out practically for $r\ge2$) the effective values of $\chi_\ell$ are those maximizing the ratio \eqref{secN-R:g}, i.e. $\chi_\ell\simeq1\ll\chi$ according to the above discussed behaviour of $f_\Omega$. With this, the ratio \eqref{secN-R:g} indeed reduces to $\simeq\alpha\chi^{2/3}$, justifying this combination as the expansion parameter for the set of bubble chain diagrams. 

It is worth emphasizing that it turns out the photons dominant in a loop with bubble chain (in contrast to the case $r=0$ of no bubbles) are much softer than the incoming electron and that the non-smallness of the effective expansion parameter \eqref{secN-R:g} emerges as just a consequence of their small virtuality $\ell^2\simeq m^2\chi^{-2/3}\ll m^2$ (see Eq.~\eqref{secN-R:l**2}). In particular, this implies that for bubble chain diagrams the actual loop extension $\Delta x^\LCp$ is larger than for leading order one-loop corrections, thus potentially imposing more severe restrictions on the applicability of LCFA. 

Upon interchanging the order of integration over photon momenta variables in the loop and of summation over the number of bubbles (by putting the latter first), it is possible to resum the contribution of the whole set of bubble chain diagrams. Such resummation is effectively reduced to a geometric series
\begin{align}\label{secN-R:resum}
\sum\limits_{r=1}^\infty \left(\frac{\pi_\Omega(\ell^2,\chi_l)}{\ell^2}\right)^r=\frac{\pi_\Omega(\ell^2,\chi_l)}{\ell^2-\pi_\Omega(\ell^2,\chi_l)}.
\end{align}
Obviously, appearance of the pole structure in the resummed expression \eqref{secN-R:resum} means dynamical generation of a field-induced effective photon mass (with gauge invariance remaining preserved due to preserving the longitudinal photon modes massless). This is a standard way of how resummation of the 1PI corrections takes effect on the whole amplitude by dressing the propagators.

The expression for the resummed mass operator can be rearranged in terms of the Fourier transform of \eqref{secN-R:resum} with respect to virtuality $\ell^2$ (with the remaining weighted integration over $\chi_\ell$ performed afterwards). In a CCF $\pi_\Omega(\ell^2,\chi_l)$ has a complicated analytical structure in both variables, in particular has an essential singularity at $\ell^2=\infty$ \cite{Ritus:1972radiative}. Accordingly, by means of Picard's theorem the resummed expression \eqref{secN-R:resum} possesses an infinite number of poles accumulated at infinity in the lower plane of $\ell^2$. Though under such conditions the Fourier transform cannot be evaluated straightforwardly by enclosing integration path, it was argued and confirmed numerically in Ref.~\cite{Mironov:2020resummation} that asymptotically it is still determined by contribution of the main (closest to the origin) pole $\ell^2\approx \pi_\Omega(0,\chi_l)$. This is enough to evaluate the whole resummed amplitude asymptotically for $\chi\gg1$. The remaining integrations can be performed analytically for $\alpha\chi^{2/3}\gg1$ by taking into account that also $\chi_\ell\ll\chi$. This way it was shown in Ref.~\cite{Mironov:2020resummation} that for $\alpha\chi^{2/3}\gg1$ the resummed all-order bubble chain contribution to the on-shell mass operator $\mathcal{M}(\chi)$ can be separated into the two terms
\begin{align}\label{secN-R:M_comp}
\mathcal{M}(\chi)-\mathcal{M}^{(r=0)}(\chi)=\mathcal{M}^{(\text{II})}(\chi)+\mathcal{M}^{(\text{III})}(\chi),
\end{align}
depending differently on $\chi$. Namely,
\begin{align}
\label{secN-R:M(II)} 
\mathcal{M}^{(\text{II})}(\chi)=(-0.995 + 1.72i)\alpha^{3/2}\chi^{2/3}m^2,\\
\label{secN-R:M(III)} 
\mathcal{M}^{(\text{III})}(\chi)=-(0.103 + 1.18i)\alpha^2\chi m^2.
\end{align}
This should be compared to the leading order no-bubble one-loop contribution \cite{Ritus:1970radiative,Ritus:1972radiative}
\begin{align}\label{secN-R:M(r=0)} 
\mathcal{M}^{(r=0)}(\chi)=0.843(1-i\sqrt{3})\alpha\chi^{2/3}m^2.
\end{align}
 Different dependencies of the terms \eqref{secN-R:M(II)} and \eqref{secN-R:M(III)} on $\chi$ were attributed to different formation regions of the integrals involved; for example, the effective values of $\chi_\ell$ contributing to  \eqref{secN-R:M(II)} and \eqref{secN-R:M(III)} are $\chi_\ell\simeq\alpha^{3/2}\chi$ and $\chi_\ell\simeq1$, respectively. Whereas \eqref{secN-R:M(II)} always remains smaller than \eqref{secN-R:M(r=0)}, \eqref{secN-R:M(III)} becomes dominant for extremely large values $\chi\gg\alpha^{-3}$.

However, these results can serve only as a proof of concept: the implications for potentially observable features that could characterise the non-perturbative regime $\alpha\chi^{2/3}\gtrsim1$, such as modification of the emission rates, still remain unclear. At first sight, the two possible cuts of a bubble-chain diagram in Fig.~\ref{fig:mass_bubble_corr} (with a cut crossing either a photon line or a bubble) could be associated via the optical theorem with corrections to either photon emission or trident the process~\cite{Mironov:2020resummation}. However, application of the optical theorem requires much more caution than expected. First, assuming no restrictions on $\chi$, the imaginary parts of both terms \eqref{secN-R:M(II)} and \eqref{secN-R:M(III)} are unbounded, indicating a conflict with unitarity~\cite{Heinzl:2021mji}. As already discussed in \secref{sec:higher}, to reconcile with unitarity one needs to consistently take into account all the relevant loop contributions. {This suggests that} it is sufficient to take into account, in the same bubble-chain approximation, dressing of the external electron legs, meaning that the optical theorem should be applied to 1PR rather than 1PI diagrams. Second, the foundations of the $S$-matrix approach (in the Furry picture in particular) explicitly require that both the initial and final states for a process are always chosen to correspond to stable, or at worst, weakly unstable particles. According to the general consideration of \cite{Veltman:1963unitarity}, unitarity also requires that in the presence of unstable particles only stable intermediate states show up in the optical theorem. However, in the case $\chi\gg1$ both photon and electron states (whether real or virtual) are obviously highly unstable. From this perspective one may expect that in such a regime meaningful rates can be formulated only for such processes that involve massive production of soft (hence almost stable) particles. At the moment of writing the derivation of the emission rates along these lines remains in progress. 

\section{Beyond the plane wave and background field approximations}\label{sec:beyondPW}
{We have seen in previous sections that discussions of many laser-particle scattering phenomena in strong-field QED are based on constant crossed field (CCF) results. However, such fields fail completely to correctly describe infra-red physics~\cite{Harvey:2014qla,Dinu:2012tj,DiPiazza:2017raw,Ilderton:2018nws,DiPiazza:2018bfu}, high-energy physics~\cite{Podszus:2018hnz,Ilderton:2019kqp}, harmonics~\cite{Harvey:2014qla}, mid-IR physics~\cite{King:2020hsk,Tang:2021qht}, and collinear physics~\cite{Dinu:2012tj,Edwards:2020npu}. This is all seen by comparing CCFs with the more realistic case of pulsed plane waves, demonstrating the importance of going beyond the most simple, and unphysical, fields.}

{However, it is then very natural to ask where the plane wave model fails, in comparison with an more realistic descriptions of a laser field (which include e.g.~focussing effects), and how one can go beyond this.} The Dirac equation cannot, however, be solved exactly in a general background field; many backgrounds for which it (and the Klein-Gordon equation) can be solved are listed in~\cite{BagrovGitman,Bagrov:2014rss}. For backgrounds describing a realistic, high-intensity, focussed laser pulse, however, no exact solutions exist. This limits the practical use of the Furry expansion, because the analogues of the Volkov wavefunctions and propagator cannot be found.  In this Section we discuss attempts to overcome this difficulty and to solve the Dirac equation in backgrounds more complex than plane waves and constant fields.

There are, very broadly, two approaches; i)  identify methods by which to construct the wavefunctions approximately, but without using perturbation theory in $\xi$, ii) identify special cases for which one can find, either by construction or by accident, exact solutions. We describe below several examples of both approaches, all of which have the common goal of extending the methods of strong-field QED to more physically realistic background fields. We will also review methods by which to go beyond the approximation of strong fields as fixed backgrounds.

\subsection{Reduction of order} \label{sec:reduction}
A commonly considered background in which to develop and study approximate solutions of the Klein-Gordon and Dirac equations is that of a field $e\mathcal{A}_\mu(x) = a_\mu(\phi)$ depending on (like a plane wave) a single-variable $\phi=k_\mu x^\mu$, but where $k_\mu$ is now spacelike, $k^2<0$ or timelike, $k^2>0$. The spacelike case may describe (boosting to an appropriate frame) the magnetic field of an undulator, or a plane wave in a medium of refractive index $n_r>1$. The timelike case may describe a time-dependent electric field, or a plane wave in a medium with $n_r<1$.  The Klein-Gordon equation for a scalar field $\Phi(x)$ in this background is, making the ansatz $\Phi(x)=e^{-ip\cdot x}F(k\cdot x)$,
\be\label{KG4F}
    k^2 F''-2ik\cdot p F'+(2p\cdot a-a\cdot a)F = 0\;,
\ee
in which the gauge $k\cdot a=0$ is used. In the particular case of a circularly polarised plane wave $a$, with $p$ chosen such that $p\cdot a=0$, the equation is `constant coefficients' and hence immediately solvable~\cite{Raicher:2013cja}, but there is no completely general solution to (\ref{KG4F}). However, as the difference to the plane-wave case is seemingly slight -- only the first term of (\ref{KG4F}) is new -- one is encouraged to seek an approximate solution.

The slowly varying approximation $|k^2 F''| \ll |k\cdot p F'|$ is used in~\cite{MendoncaPRE2011} to ignore the double-derivative term in (\ref{KG4F}). The equation then becomes first order, is trivially integrated, and the solution is formally identical to the scalar Volkov solution, i.e.~the exponential part in e.g.~(\ref{volkov-e-in}), but with the $k_\mu$ appearing no longer lightlike. (The same approximation is assumed in~\cite{Mackenroth:2018rtp}.) 

The implied expansion parameter is $\epsilon := k^2/(2k\cdot p)$. A warning is that this parameter is not \textit{automatically} small: for $k_\mu$ spacelike $k\cdot  p$ and hence $\epsilon$ can be of any sign and size, for example. This puts kinematic restrictions on the validity of the approximations considered in this section~\cite{Heinzl:2016kzb}. Under the assumption that parameters are chosen such that $\epsilon$ is small, though, we can ask how to improve upon the slowly varying envelope approximation. Formally this is achieved by `reduction of order'; one substitutes (\ref{KG4F}) into itself and trades higher derivative terms for powers of $\epsilon$, which are then dropped at a chosen truncation order. For example, writing $V(\phi)=(2p\cdot a-a\cdot a)/(2k\cdot p)$, (\ref{KG4F}) becomes
\begin{equation}
\begin{split}
    i F' &= V F +\epsilon F'' \\
    &= V F-i\epsilon(V F + \epsilon F'')' \\
    &= (V - i \epsilon V'-\epsilon V^2) F +\mathcal{O}(\epsilon^2) \;.
\end{split}
\end{equation}
Dropping the $\mathcal{O}(\epsilon^2)$ terms one has again a first-order equation which is trivially solved. The process can be iterated to higher orders, or in some cases all orders~\cite{Ekman:2021eqc}. However, a comparison of results derived from reductive methods with those derived from exact solutions or other approaches~\cite{Heinzl:2016kzb}, shows that reduction of order typically yields `over the barrier' approximations. As such they cannot capture `under the barrier' physics of quantum tunneling (barrier penetration). Current conservation and unitarity are also violated~\cite{Heinzl:2016kzb}. As an example, (\ref{KG4F}) can in the case of a circularly polarised monochromatic field with $k^2>0$ be transformed to the Mathieu equation (or the Hill equation for general polarisation)~\cite{1977PhLA...60..137C,Becker:1977,2014LaPhL..11a6001V}, the solutions of which exhibit a rich band-gap structure. The approximate solutions obtained through reduction of order, though, do not access this structure ~\cite{Raicher:2015kaa}. See also~\cite{Raicher:2016bbx} in which reductive solutions of (\ref{KG4F}) were used to calculate nonlinear Compton scattering in rotating electric fields. {Note that the same reduction of order is performed to convert the LAD equation into the LL equation, as discussed in Sec.~\ref{sec:higher:classicalRR}, with a truncation at lowest order. For discussions of the equations obtained by truncating at higher orders see~\cite{DiPiazza:2018luu,Ekman:2021vwg}.}

The case of two counter-propagating plane waves with momenta $k_\mu$ and $k'_\mu$ is similar. The Klein-Gordon equation again contains a double-derivative term proportional to $k\cdot k'\not=0$. There is again no general solution\footnote{Claims to the contrary in the literature are incorrect. Many of them `choose' $k'_\mu$ such that $k\cdot k'=0$, but this implies $k'_\mu \propto k_\mu$ for null vectors, so one has simply a single plane wave with a different profile.}, but there are two analytically solvable cases for circularly polarised monochromatic fields; if the particle has zero initial transverse canonical momentum, or if it is confined to a magnetic node, then the Klein-Gordon equation again reduces to a Mathieu equation and can so be solved~\cite{King:2016oei}. The classical dynamics is also solvable for these cases but is in general chaotic, for a review see~\cite{Gonoskov:2021hwf}. Turning to approximations, if one assumes that $k\cdot k'$ (suitably normalised) is small, then reduction of order can again be performed. Simply neglecting $k\cdot k'$ terms again leaves a first-order equation which is solved by a product of Volkov-like solutions for the individual waves~\cite{Hu:2015qya}. Various approximations are compared in~\cite{King:2016oei}.

\subsection{WKB approximations}\label{sec:AI-WKB}
In WKB, one makes an expansion  {in powers of $\hbar$. At leading order in the expansion,} the resulting wavefunctions are closely related to classical results (see below). WKB has recently been applied to calculate electron wavefunctions in pulsed electric and magnetic fields~\cite{Heinzl:2016kzb}, rotating electric fields~\cite{Raicher:2018cih} (where it was compared to, see below, high-energy approximations), Gaussian beams~\cite{DiPiazza:2021rum}, and in the context of higher harmonic generation, in 1+1 dimensional systems~\cite{Taya:2021dcz}.

While the general method of WKB is well-known, see also Sec.~\ref{sec:Schwinger}, we summarise it here with an emphasis on its application to the Furry expansion, more specifically the solution of the Klein-Gordon and Dirac equations in strong fields. Beginning with the former case, i.e.~scalar QED, we look for a solution $\phi$ to the Klein-Gordon equation
$(\hbar^2 D^2+m^2)\phi=0$ with $D_\mu = \partial_\mu +i e\mathcal{A}_\mu/\hbar$ the covariant derivative in a now \textit{arbitrary} background $\mathcal{A}_\mu(x)$, and we have reinstated $\hbar$. We make the standard WKB ansatz
\be\label{WKBexpansion}
    \phi = \exp\Big(-\frac{i}{\hbar} S_0 -i S_1 -i \hbar S_2 + \ldots\Big) \;,
\ee
and solve perturbatively in $\hbar$, treating it as a small parameter. Working to lowest order one immediately finds that the $S_n$ for $n>0$ can be dropped, while the leading term $S_0$ must obey
\be\label{classical1}
    (\partial S_0- e\mathcal{A})^2 - m^2 = 0 \;.
\ee
This is just the equation for the classical Hamilton-Jacobi action; in other words, and has long been known, if one can solve the classical problem to find an $S_0$ obeying $\partial_\mu S_0 - e\mathcal{A}_\mu = \pi_\mu$, with $\pi_\mu$ being the on-shell classical momentum (in general a function of $x^\mu$), then one has a semi-classical approximation to the quantum wavefunction, $\phi \simeq \exp iS_0/\hbar$. The appearance of the Hamilton-Jacobi action emphasises that the leading-order WKB approximation is semiclassical~\cite{DiPiazza:2021rum}. One can proceed to higher orders in $\hbar$. Reinstating $\hbar$ in the Volkov solutions (\ref{volkov-e-in})--(\ref{volkov-e-out}), one sees again that they are semiclassical-exact -- higher order WKB corrections are exactly zero in a plane wave background.

This lowest order WKB approach is easily extended to approximate solutions of the Dirac equation. We make the long-known ansatz~\cite{Pauli:1932,PhysRev.131.2789,Rosen1964ATW}
\be
    \psi = e^{-\frac{i}{\hbar}S_0}\big( u_0 + \hbar u_1 +\ldots)
\ee
in which $S_0$ is taken to be the classical action and we may combine higher-order exponential terms with the spinor parts labelled $u_n$. Acting with the Dirac operator $i\hbar\slashed{\partial}-e\slashed{\mathcal{A}}$ and again retaining only lowest order terms in $\hbar$, we obtain the equation of motion
\be\label{classical2}
    \big(\slashed{\pi} - m \big)u_0 = 0
\ee
so that $u_0$ is simply a `free' spinor for the classical momentum $\pi$, with boundary conditions dictated by the free-theory limit; this is exactly as for the Volkov solution.

    In summary, if one can solve the classical problem, an approximate quantum solution follows (see for example~\cite{Raicher:2018cih}), in which the wavefunctions share several characteristics with Volkov solutions. Solving the classical problem is still hard in general, and even when WKB wavefunctions are available, they may not be simple enough to allow easy calculation of observables. Progress can be made, though, {by supplementing WKB with additional approximations; for example~\cite{DiPiazza:2021rum} imposes the additional condition that the electron is ultrarelativistic such that its energy $\mathcal{E}$ obeys $\mathcal{E}\gg \text{max}(m, m\xi)$, which allows a further expansion and thus simplification of the WKB wavefunctions,} as will be discussed in Sec.~\ref{subsec:highE}. Estimates of the regime of validity for WKB, in the case of a Gaussian laser pulse background, can be found in~\cite{DiPiazza:2021rum}. {In other fields, the general approach remains valid, but one should re-derive the precise conditions of validity specific to those fields.}  

There are several methods by which to improve on leading order WKB approximations. `Uniform WKB'~\cite{cherry1948,MillerGood} for example can provide good approximations through turning points where ordinary WKB breaks down. Heuristically, suppose one wants to obtain the wavefunction $\psi(x)$ (obeying the Schr\"odinger, Dirac, or Klein Gordon equation) in a potential $V(x)$. Suppose also that $\psi_0(x)$ is a \textit{known} solution of the `comparison problem', that is the same equation but in the potential $V_0(x)$. Then one makes the ansatz $\psi(x)=\psi_0(f(x))$ and applies perturbation theory in $\hbar$ to the unknown function $f(x)$. The key step is of course the choice of comparison problem.  To illustrate, consider the Schr\"odinger equation in a periodic cosine potential (equivalent to the Mathieu equation). In the weak-coupling regime, the potential resembles an infinite series of well-separated harmonic (oscillator) wells, as follows from expanding $\cos x \sim 1- x^2/2$. One therefore makes the ansatz that $\psi_0$ should be an eigenstate of the Schr\"odinger equation in the harmonic oscillator, i.e.~a parabolic cylinder function~\cite{Dunne:2014bca}.

Uniform WKB has recently been applied to study trans-series and resurgence in the periodic cosine potential~\cite{Dunne:2014bca,Dunne:2016qix} and to  tunneling through potential barriers, using a linear potential as comparison problem~\cite{Heinzl:2016kzb}. The latter case leads to the appearance of Airy functions -- uniform Airy approximations have been used to extend the LCFA (see Sec.~\ref{sec:approx}), applied to pair production, beyond its usual range of validity~\cite{King:2019igt}. See also Sec.~\ref{sec:Schwinger} for `exact WKB'.

{It is clear from (\ref{classical1}) and (\ref{classical2}) that the standard WKB ansatz (without the Stokes phenomenon) leads to single-particle wavefunctions, i.e. it~does not mix positive and negative energy states, which is an integral part of pair production phenomena. WKB can also be generalised to better capture pair production by mixing the lowest order WKB ansatz with a Bogoliubov transformation. This approach was used in~\cite{Oertel:2016vsg, Kohlfurst:2021skr} to study the impact of magnetic fields on pair production in space-and-time dependent fields $\mathcal{A}(x) = (f(x^\LCm) + f (x^\LCp))\,\mathbf{e}$ (with $\mathbf{e}$ transverse, as in plane waves) modelling colliding laser pulses.}

\subsection{High energy approximations  {and the eikonal}}\label{subsec:highE}
High-energy approximations have a long history of application to strong field systems~\cite{Baier1968,Bjorken:1970ah,Blankenbecler:1987rg} and,  {together with the WKB approach above}, have seen renewed attention in the last few years~\cite{DiPiazza:2013vra,DiPiazza:2015xva,DiPiazza:2016tdf}.

Consider a high-energy scalar particle (for simplicity) colliding with an external field ${\cal A}_\mu(x)$, the form of which is essentially arbitrary for now.  {We assume the leading-order WKB approximation of Sec.~\ref{sec:AI-WKB}. We want to solve (\ref{classical1}) to obtain the analogue of the Volkov wavefunctions in a high-energy approximation.  We present here a compact derivation which is almost trivialised by working in suitable coordinates and in a suitable gauge. We can always say the particle is moving in the $z$-direction, hence its momentum is $p_\mu = p_\LCm {\bar n}_\mu +  m^2 n_\mu /(4 p_\LCm)$ without approximation, for $n_\mu$ as before and ${\bar n}_\mu$ obeying ${\bar n}\cdot {\bar n}=0$ and $n\cdot{\bar n}=2$. We work in the gauge ${\bar n}\cdot {\cal A}=0$.}

 {We now make the high-energy assumption $p_\LCm\gg m$. The particle thus travels \emph{approximately} along a null ray, $p_\mu \sim p_\LCm {\bar n}_\mu$, as a massless particle would.} If $p_\LCm$ is also larger than field-induced energies, so roughly~$p_\LCm \gg m\xi$  {for $\xi$ here simply typifying the field strength}, then $p_\LCm$ sets the dominant energy scale of the interaction, and $1/p_\LCm$ can be used as a small expansion parameter.  {We thus make the ansatz that the leading order WKB exponent $S_0$ in (\ref{WKBexpansion}) takes the form $S_0 = p\cdot x + T_0 +\frac{T_1}{p_\LCm} + \frac{T_2}{p_\LCm^2} + \ldots$ and we solve for the $T_j$ by expanding (\ref{classical1}) in powers of $1/p_\LCm$ (made appropriately dimensionless and noting that a $1/p_\LCm$ term is already present in $p\cdot x$). At leading order in the expansion, the choice of gauge yields the equation $\partial_\LCp T_0=0$; scattering boundary conditions thus imply that $T_0$ vanishes. It is then trivial to solve for $T_1$, and one finds~\cite{Blankenbecler:1970qd,Bjorken:1970ah,Blankenbecler:1987rg,DiPiazza:2013vra}
\be\label{eq:allofprl}
        S_0 = p \cdot x + \frac{1}{2p_\LCm}\int\limits_{-\infty}^{x^\LCp}\!\ud u \, e^2\! {\cal A}_\LCperp^2(u,x^\LCm,x^\LCperp) \;.
\ee
This expresses the classical action as an integral along the null ray which approximates the ultra-relativistic particle's trajectory. The combination of the leading order WKB ansatz (\ref{WKBexpansion}) with the leading-order high-energy result (\ref{eq:allofprl}) is therefore just a relativistic generalisation of the well-known eikonal approximation used in quantum mechanics, as was already pointed out in~\cite{Bjorken:1970ah}. To make the connection to the eikonal completely explicit, we can combine the mass and background into an effective potential $V := m^2 + e^2{\cal A}_\LCperp^2$. The exponent in (\ref{eq:allofprl}) then takes the form $p_\LCm x^\LCm + \int V/p_\LCm$, in direct analogy to the non-relativistic wavefunction of a particle scattering on a potential $V$, see e.g.~\cite{Sakurai,ZinnJustinBook}. We note with interest that there is no need to go via the classical equations of motion to obtain these results. One could proceed to higher order, but  working to the given order in $1/p_\LCm$ is enough} to recover the Volkov solution (\ref{volkov-e-in}) in the plane-wave limit~\cite{DiPiazza:2013vra} which includes terms going like $1/p_\LCm$ in both the exponent and the spin structure. 

The ideas above have been applied to develop {approximations to the} equivalent of Volkov solutions in strong, \emph{focussed} laser pulses~\cite{DiPiazza:2013vra,DiPiazza:2015xva}, something which otherwise escapes an analytic treatment and is of crucial importance to upcoming experiments. (The simplifying assumption of counter-propagation between the particle and field used in those papers can be relaxed, see~\cite{DiPiazza:2016tdf} for details.  {Such assumptions may be more practical, of course, when it comes to explicit calculations, as may other gauge choices, compared to working with the general expression~(\ref{eq:allofprl}).)} Investigation of scattering processes using these wavefunctions shows that focussing effects can lead to significant differences in the angular emission spectrum of  {pairs created via the nonlinear Breit-Wheeler process, compared to that} in plane waves~\cite{DiPiazza:2016maj}. 

Adding structure to the background fields under consideration means that physical processes can be characterised by more invariants than in plane waves; this has been explored using WKB/high-energy methods, see e.g.~\cite{DiPiazza:2020wxp} for the case of `flying focus' beams, which have been suggested as useful for enhancing signals of radiation reaction~\cite{Formanek:2021bpw}  {and for investigating the role of \emph{transverse} formation length~\cite{DiPiazza:2020wxp} which, unlike the longitudinal formation length discussed in \secref{sec:first_order_general_props} and \secref{sec:approx:lcfa}), has a quantum origin}. In another example, that of electron dynamics in counter-propagating beams, small time-scales can arise in the electron motion due to the interaction geometry and impact the radiation spectrum -- while this can be missed by the LCF approximation~\cite{Raicher:2020nkq}, it can be investigated using WKB~\cite{Lv:2021ayt}. High-energy approximations have also been used to investigate signatures of the intensity-dependent `mass shift'~\cite{Harvey:2012ie,Kohlfurst:2013ura} in rotating electric fields~\cite{Raicher:2019flc}, to construct solutions of the Dirac equation in linearly growing fields modelling dense particle bunches~\cite{Wistisen:2018rlg}, in superpositions of laser and atomic fields~\cite{DiPiazza:2014uea,Krachkov:2019ovr}, and to study photon emission from ultra-relativistic electrons moving in periodic potentials modelling aligned crystals~\cite{Wistisen:2019cew}.
    
\subsection{Highly symmetric backgrounds and exact solutions}\label{sec:beyondPW:super}
We turn now to the construction of exact solutions of the Klein-Gordon and Dirac equations~\cite{BagrovGitman}. It is useful to recall that plane waves are highly symmetric fields, being invariant under a five-dimensional group of spacetime transformations (the Carroll group~\cite{Levy1965} with broken rotations~\cite{Duval:2017els,Zhang:2019gdm}). What is crucial is that these symmetries \emph{extend to symmetries of particle dynamics}, in the plane wave background, i.e.~they generate conserved quantities, making the dynamics (both classical and quantum) integrable, and in fact `superintegrable'. As reviewed in~\cite{Miller:2013gxa}, a system with $2n$ dimensional phase space is `superintegrable' if there exists more than $n$ independent integrals of motion. The topic has a long history in non-relativistic mechanics, and is frequently encountered in systems containing background electromagnetic fields. (Our interest here is of course in relativistic theories.)

It is the loss of symmetry, and hence of conserved quantities, which is one of the main challenges in solving the Lorentz, Klein-Gordon or Dirac equations in more realistic, and hence more complicated, backgrounds than plane waves. One method of circumventing this problem, and thus regaining the ability to solve the Dirac equation exactly, is thus to first construct backgrounds with as many symmetries as possible which guarantee conserved quantities of motion.  We discuss the classical case first. Let ${\mathcal A}_\mu$ be a gauge potential for some background field. Let $\zeta$ represent a Poincaré transformation of spacetime, under which ${\mathcal A}_\mu$ is symmetric, i.e. in terms of the Lie derivative $\mathcal{L}_\zeta$, the potential obeys $\mathcal{L}_\zeta {\mathcal A}_\mu = \partial_\mu \Lambda_\zeta$ for some scalar $\Lambda_\zeta$. (In other words the change in ${\mathcal A}_\mu$ under the Poincar\'e transformation is zero up to gauge.) For a particle moving under the Lorentz force in the background, \emph{canonical} momentum $P_\mu$, the quantity 
    \be
        Q = \zeta\cdot P - e \Lambda_\zeta
    \ee
    is then conserved~\cite{Heinzl:2017zsr,Heinzl:2017blq}. If three independent, conserved quantities can be found/imposed (in $3+1$ dimensions), the dynamics becomes integrable. If more can be found (up to a maximum of five, as for plane waves), motion becomes superintegrable. The generalisation to the Dirac equation is as follows; if ${\mathcal A}_\mu$ has a Poincar\'e symmetry then $J^\mu_\zeta = {\bar \psi}\gamma^\mu (i\mathcal{L}_\zeta -e\Lambda_\zeta)\psi$ is a conserved current. A sufficient condition which guarantees this conservation is
    \be\label{eq:2impose}
        i\mathcal{L}_\zeta \psi - e \Lambda_\zeta \psi = \lambda \psi \;,
    \ee
    for some constant $\lambda$. The method of solution is to first impose (\ref{eq:2impose}), for as many symmetries as possible, to determine as much of the structure of $\psi(x)$ as possible, and then use the Dirac equation itself to complete the solution~\cite{Heinzl:2017zsr}. For examples see~\cite{Heinzl:2017blq}. Interestingly, it is often the case that superintegrable systems admit explicit analytic solutions, whereas integrable systems may still only be solvable e.g.~up to quadratures. 

For particles in \textit{scalar} background fields (which may be viewed as a conformally flat spacetime, or a spacetime-dependent mass~\cite{Nikitin:2014cfl}) both Poincar\'e, dilation and special conformal symmetries of the field all automatically generate conserved quantities~\cite{Ansell:2018dro} in classical and quantum particle dynamics. If an \emph{electromagnetic} field is symmetric under a special conformal tranformation, it does not \emph{in general} generate a conserved quantity in particle motion. However, using ideas based on double copy (which allows one to calculate scattering amplitudes in gravity from scattering amplitudes in Yang-Mills), it is possible to find cases where special conformal symmetries of the background do  generate conserved quantities in particle motion~\cite{Andrzejewski:2019hub,Elbistan:2020ffe}. For systematic approaches to the classifications of non-relativistic superintegrable systems in background fields see~\cite{Marchesiello2015,Marchesiello:2018kog}.

\subsection{Inverse approaches for exact solutions}\label{sec:inverse}
The standard starting point for strong-field QED is that one is handed a background field (or, as above, constructs one with desired properties) and must then solve for the particle dynamics in that field. `Inverse' approaches take the opposite path to achieve the same ends~\cite{2012PhRvL.109f0401B,Hebenstreit:2015jaa,Oertel:2015yma,Campos:2017waa}. We illustrate using first the results of~\cite{Oertel:2015yma} in 1+1 dimensions, where the two components of the Dirac equation can be rearranged into the two equations (adapted to our conventions)
\be
    e {\mathcal A}_\LCp = i\frac{\partial_\LCp \psi_2}{\psi_2} - \frac{m}{2}\frac{\psi_1}{\psi_2} \;, \qquad e {\mathcal A}_\LCm = i\frac{\partial_\LCm \psi_1}{\psi_1} - \frac{m}{2}\frac{\psi_2}{\psi_1} \;.
\ee
The inverse approach now specifies the Dirac field $\psi$ and from this reads off the background ${\mathcal A}_\mu$ which generates that $\psi$. By construction, $\psi$ obeys the Dirac equation in the background ${\mathcal A}_\mu$. {There are two caveats; first, and ideally, the background field obtained should be independent of the kinematic properties (e.g.~initial momentum) of $\psi$, as it is only in this way that one can hope to find a complete set of solutions in the presence of a single, fixed background.} Second, one must ensure that ${\mathcal A}_\mu$ is real for a physical solution, which imposes two real constraints on the two complex degrees of freedom in $\psi$.  The method has been used to construct exact analytic solutions of the Dirac equation in counter-propagating \textit{pulses} in 1+1 dimensions~\cite{Oertel:2015yma}, a system which escapes any exact treatment in $3+1$ dimensions (but see~Sec.~\ref{sec:AI-WKB} above for WKB methods).

The complexity of inverse approaches increases with the number of spacetime directions on which the solutions are to depend. In 3+1 dimensions~\cite{Campos:2017waa} represents the 4-component Dirac spinor $\psi$ as a matrix $\Psi$,
\be
    \psi \to \Psi =
    \begin{pmatrix}
    \psi_1 + \psi_3 & -\psi_2^*+\psi_4^* \\
    \psi_2 + \psi_4 & \psi_1^*-\psi_3^*
    \end{pmatrix} \;,
 \ee
upon which the Dirac equation becomes, writing ${\tilde X} = \sigma_\mu X^\mu$ for any vector $X^\mu$ and $\sigma_\mu$ the Pauli spin matrices,
 \be
    i \tilde{\partial} \Psi \sigma_3 - e \tilde{{\mathcal A}}\Psi - m {\tilde\Psi}^\dagger = 0
    \implies 
    e {\tilde {\mathcal A}} = (i\tilde\partial \Psi\sigma_3- m {\tilde\Psi}^\dagger)\Psi^{-1} \;.
 \ee
 Upon specifying $\Psi$, one again reads off the background field generating the implied motion from ${\mathcal A}_\mu = \text{Tr} (\tilde{{\mathcal A}} \sigma_\mu)/2$. In this approach, called relativistic dynamical inversion (RDI), one specifies the spinor in terms of recognisable dynamics, writing $\Psi = \sqrt{\rho}L$ in which $\rho$ is a scalar and the invertible matrix $L$ is a product of (spacetime-dependent) rotations, boosts and internal transformations. (It is interesting to note that, as in the superintegrable approach~\cite{Heinzl:2017blq}, Poincar\'e symmetries again play a key roll.)

The constraints which ensure a real solution for ${\mathcal A}_\mu$ are far more complicated than in 1+1 dimensions~\cite{Oertel:2015yma}, and instead of solving them~\cite{Campos:2017waa} advises a trial and error approach in order to find a suitable parametrisation of the spinor which gives desired particle dynamics and a physical background. RDI has provided both stationary and time-dependent exact solutions in two and three dimensions~\cite{Oertel:2015yma}, including dispersionless circular, elliptic and linear motion, ~\cite{Campos:2017waa,Campos:2020ked}, as well as solutions carrying orbital angular momentum~\cite{Campos:2020cza}. Inverse methods have also been applied to optimise pair production yields through pulse shaping in~\cite{Hebenstreit:2015jaa}.  {Finally, we normally think of external, or background, fields as being classical; it is interesting to note, then, that the fields constructed by RDI can carry an $\hbar$-dependence. On the one hand, this might imply that the constructed fields are excluded from being used in classical systems. On the other hand $\hbar\not=0$ in our world. Furthermore, a background field is really a (strong) coherent state of photons, and there is nothing which formally prevents the wave profiles in these states from being $\hbar$-dependent. It would thus be interesting to further explore this $\hbar$-dependence.}

\subsection*{Discussion}

 {It is unlikely that one will be able to find an exactly solvable model which captures all aspects of physics in a given, realistic, strong field. Of course, it is not possible to find an approximation which could do this either, by definition. It is however possible to capture and explore \emph{aspects} of a system which may not be seen in the most common solvable models, namely constant fields and plane waves. For example, a model of exact dynamics in counter-propagating beams is provided by inverse methods~\cite{Oertel:2015yma}, while symmetry methods provide a toy model of particle focussing by structured beams~\cite{Heinzl:2017zsr}.}

 {Just as with approximations, one should take care in generalising results of exactly solvable systems beyond their regime of validity.  For example, results in 1+1 dimensions may well not generalise to higher dimensions. However, even if toy models are not of \emph{direct} experimental relevance, they can still be of indirect relevance, providing e.g.~evidence to support numerical simulations. Quite aside from this, exact solutions are very useful for theory. In future, it is likely that a \emph{combination} of approximations and the other methods of Sec.~\ref{sec:reduction}--Sec.~\ref{sec:inverse} will be necessary to provide more comprehensive models of physics in strong fields.}

\subsection{Null vs. non-null fields}\label{sec:null_vs_nonnull}
Fields for which the invariants (\ref{eq:invariants}) are non-zero may spontaneously produce pairs via the Schwinger effect. Exactly solvable cases investigated in this context include constant electric fields~\cite{Nikishov:1969tt, Tanji:2008ku}, Sauter fields~\cite{Sauter:1932gsa, Narozhnyi:1970uv, Nikishov:1970br}, exponentially increasing/decreasing fields $E\propto e^{-\omega t}, e^{-\omega x}$~\cite{Adorno:2014bsa, Adorno:2015ibo, Adorno:2016bjx}, inverse-square fields~\cite{Gies:2016coz, Adorno:2019ffg, Adorno:2018jbp}, longitudinal fields depending on lightfront time, $E=E(x^\pm)$ \cite{Tomaras:2001vs,Ilderton:2014mla}, and asymmetric pulses~$E\propto \sqrt{1+e^{t/\sigma}}\cosh^{-2}(t/2\sigma)$~\cite{Breev:2021lpn}. Plane waves, on the other hand, are examples of `null' fields for which both invariants vanish and thus there is no pair production. This simplification prompts the question: does insisting on nullity also simplify the identification of fields in which classical and quantum equations of motion can be solved? There does not seem to be, in the literature, an answer to this question or a systematic approach to including the restriction in the construction of fields and dynamics in them. Here we simply review a class of null backgrounds which do allow progress. These are a class of `plane fronted waves with parallel propagation', or pp waves for short.

Let $n_\mu$ be, as for plane waves, a lightlike direction, and define
\be
    {\mathcal A}_\mu = -n_\mu \delta(n\cdot x)\Phi(x^\LCperp) \;,
\ee
in which $\Phi$ is an arbitrary function of the transverse coordinates $x^\LCperp$. These fields are in general \textit{sourced} by charge distributions `ultra' boosted to the speed of light -- as such, the fields are approximations of those for very high energy sources. {(Note that the form of such sources, typically $J\sim \delta(x^\LCm)\rho(x^\LCperp)$, is that used in the colour glass condensate formalism for heavy ion collisions, see Sec.~\ref{sec:beyondQED:QCD}.)} The case $\Phi = \tfrac{Q}{4\pi} \log (\mu^2 r^2)$, for constant $\mu$, is the famous `shockwave' describing the ultra-boosted Coulomb field of a point charge $Q$~\cite{Jackiw:1991ck}. The gravitational and colour-electromagnetic counterparts of the shockwave are well-studied~\cite{Aichelburg:1970dh,tHooft:1987vrq}, playing important roles in high-energy hadron physics~\cite{Caron-Huot:2013fea} and transplanckian scattering~\cite{Lodone:2009qe}. The case $\Phi \sim r^2$, representing a uniform, space-filling distribution of charge boosted to the speed of light, is studied in~\cite{Hollowood:2015elj} in the context of causality violation and UV completion, in both QED and gravity. Combining these two cases gives Bonnor's solution describing the electromagnetic fields of an ultra-relativistic beam, of finite radius $r_0$ and total charge $Q$~\cite{Bonnor:1969gr,Bonnor:1969rb}:
    \be\label{eqn:shock}
    \Phi(x^{\LCperp})
    =
    \frac{Q}{4\pi}
        \begin{cases}
            r^{2}/r_{0}^{2} & r \le r_{0} \\
            1 + \log(r^{2}/r_{0}^{2}) & r \ge r_{0} \;.
        \end{cases}
    \,.
\ee
{Shockwave methods have long been a staple of high-energy physics. Motivated by interest in using beam-beam collisions to access the strong field regime of QED~\cite{DelGaudio:2018lfm,Yakimenko:2018kih} (see Sec.~\ref{sec:RN}), Bonnor's beam (\ref{eqn:shock}) was therefore used in~\cite{Adamo:2021jxz} to investigate the scattering of particles on high-energy particle beams. It was found that the locally constant field approximation (Sec.~\ref{sec:approx:lcfa}) and RN conjectures (Sec.~\ref{sec:RN}) do not apply in this system.}

Finally, we note that $\Phi(x^\LCperp) = c_\LCperp x^\LCperp$ with $c_\LCperp$ constant is an `impulsive' plane wave, that is one of infinitely short duration (as is perhaps more easily seen from by the gauge-equivalent potential ${\mathcal A}_\mu = \delta_\mu^\LCperp c_\LCperp\theta(n\cdot x)$). Impulsive plane waves have the advantage that more progress can be made, analytically, in evaluating scattering rates and probabilities than in general plane waves. They have been studied in the context of the Schwinger effect~\cite{Fedotov:2013uja}, nonlinear Compton and Breit-Wheeler~\cite{Ilderton:2019vot} and quantum interference~\cite{Ilderton:2019ceq}.

\subsection{Back-reaction on background fields}\label{sec:beyondPW:backreaction}

We have thus far treated strong electromagnetic fields as fixed backgrounds. This itself is an approximation, as it neglects back-reaction on the field from any given physical process (scattering, particle creation, and so on) occurring within it~\cite{Gavrilov:2008fv}. Natural questions to ask are then i) when does back-reaction on the field become important, ii) what are its consequences, and iii) how is it modelled? (See also~\cite{Yang:2017xyh} for subtleties in the modelling of quantum sub-systems using classical backgrounds, and implications for unitarity.) We address some of these questions in this section.

\subsubsection{Back-reaction in the Schwinger effect}\label{sec:beyondPW:back:schwing}

Consider first the Schwinger effect in a given electromagnetic field. Backreaction on the field due to the created particles is crucial for overall energy conservation, and can be expected to be important when, roughly, the energy of the created particles becomes comparable to the energy in the field. Heuristically, backreaction can be taken into account by simultaneously solving the Dirac equation (or phenomenological equations that determine the time-evolution of pair creation) and the Maxwell equation, 
\begin{align}
	\partial_\mu {\mathcal F}^{\mu\nu} = e{\mathcal J}^\nu ,
\end{align}
in which the source term $e{\mathcal J}^\nu$ describing the charge current of the created particles may contain nonlinear quantum corrections. Early studies of the backreaction problem in the Schwinger effect (which began in the context of heavy-ion collisions~\cite{Glendenning:1983qq}) were limited to spatially homogeneous fields and the weak coupling limit such that radiative/collisional effects of created particles could be neglected.  For purely electric fields, it was found that backreaction induces plasma oscillations~\cite{Kluger:1992gb,Bloch:1999eu}: created particles are accelerated by the field, driving a current in the electric-field direction.  The field then decays according to Amper\'{e}'s law, $\partial_t \mbf{E} = -{\bm {\mathcal J}}$ until it reaches zero; at this point, though, ${\mathcal J}$ is still flowing in the positive direction and therefore $E$ changes direction, decelerating the particles.  Eventually, ${\mathcal J}$ also changes sign, and $E$ begins to increase again. These processes occur repeatedly, yielding oscillations of $E$ and ${\mathcal J}$ in time, until $E$ decays completely.  Quantum interference can be captured in this picture, and it impacts the spectrum of created pairs as reviewed in Sec.~\ref{sec:ht3}.

While a complete account of backreaction in the Schwinger effect {(as well as its analogue in, e.g., graphene \cite{Gavrilov:2012jk})} remains an open problem, there have been steady efforts to overcome the limitations of early studies. Some analytic and numerical progress can be made in, for example, ${\rm QED}_{1+1}$ using numerical solution of quantum kinetic equations~\cite{Otto:2018hya} and lattice methods. {The real-time lattice approach using Monte-Carlo sampling techniques provides a numerical scheme in which to compute physical observables in background fields $\mathcal{A}_\mu$~with complicated spacetime inhomogeneities, as well as include backreaction on those fields from pairs created via the Schwinger effect~\cite{Aarts:1998td, Hebenstreit:2013qxa, Hebenstreit:2013baa, Gelis:2013oca,Kasper:2014uaa,Gelis:2015eua,Mueller:2016aao,Buyens:2016hhu}. Realtime lattice techniques have also been applied to quark production~\cite{Tanji:2015ata,Gelfand:2016prm,Spitz:2018eps}, relativistic quantum plasmas~\cite{Shi:2018sxm} and chiral plasma instability~\cite{Mace:2019cqo}. Further applications can be found in the reviews~\cite{Gelis:2015kya,Berges:2020fwq}.} We outline some particular advances which have followed from the inclusion of (i) magnetic fields, (ii) spatial inhomogeneities, and (iii) radiative/collisional effects.

(i) The effect of magnetic fields was discussed in~\cite{Tanji:2008ku, Tanji:2010eu, Mueller:2016aao}.  If spatially homogeneous, magnetic fields do not receive backreaction corrections by Faraday's law $\partial_t \mbf{B} = -\nabla\times\mbf{E} = 0$ and hence stay constant in time.  Magnetic fields then enter the Schwinger effect indirectly via modifying the energy $\pi_0$, as was the case without backreaction, see Sec.~\ref{sec:ht4}.  For parallel electric and magnetic fields, pair creation is enhanced, hence the created charge current becomes larger, which in turn speeds up the plasma oscillations~\cite{Tanji:2008ku, Tanji:2010eu}.  For magnetic fields having both parallel and perpendicular components, it has been proposed that the chiral anomaly leads to a chiral imbalance $N_5$ (the difference between the numbers of created right- and left-handed particles, see Sec.~\ref{sec:ht4}), induced by the parallel magnetic component, generating an anomalous current in the total magnetic-field direction via the chiral magnetic effect \cite{Kharzeev:2007jp,Fukushima:2010vw, Warringa:2012bq,Fukushima:2015tza}.  Therefore, the total charge current ${\bm {\mathcal J}}$ is not parallel to the electric field $\mbf{E}$, and so the electric-field direction becomes rotated in time according to Amper\'{e}'s law $\partial_t \mbf{E} = -{\bm {\mathcal J}}$ \cite{Mueller:2016aao}.  

(ii) Advances in realtime lattice approaches have made it feasible to study backreaction in the Schwinger effect with {\it space}time dependent fields.  The backreaction problem in ${\rm QED}_{1+1}$ with spatially finite electric fields was investigated numerically in~\cite{Hebenstreit:2013qxa, Hebenstreit:2013baa, Kasper:2014uaa,Gold:2020qzr} (see~\cite{Chu:2010xc} for an analytical study in the massless limit and also~\cite{Spitz:2018eps} for a numerical study of massive ${\rm QCD}_{1+1}$). 

In (1+1) dimensions, the homogeneous plasma oscillations described above are  modified as follows~\cite{Hebenstreit:2013qxa, Hebenstreit:2013baa, Spitz:2018eps}: an initial electric field creates two spatially-localized bunches of electrons and positrons and accelerates them apart. If the bunches have sufficiently large charge then a strong, localised, electric field is formed between them, with orientation precisely opposite to that of the original field.  The secondary electric field also decays via the Schwinger effect and creates another two bunches, and a field between them, with opposite charges and orientation, respectively. These processes repeat, again leading to plasma oscillations in time around the peak of the initial electric field, as well as oscillating charge distributions in the spatial direction, which continue until the electric field is sufficiently weakened such that sizable pair creation no longer occurs. In the massless limit, the electric field decays in time $t$ as $t^{-1/2}$~\cite{Chu:2010xc}.

(iii) Radiative/collisional effects are important for describing equilibriation processes (and also QED cascades discussed in Sec.~\ref{sec:higher}). So far, such effects have been typically treated within Boltzmann equations with phenomenological collision (e.g., relaxation-type kernel and LCFA) and source terms~\cite{Kluger:1991ib,Ruffini:2003cr}. 
There have been recent attempts to obtain self-consistent strong-field QED Boltzmann equations from quantum-field theory \cite{Smolyansky:2019yma, Fauth:2021nwe}. Ref.~\cite{Smolyansky:2019yma} straightforwardly generalised the intermediate particle picture of the spatially homogeneous Schwinger effect by adding a coupling to dynamical photons and truncating higher-order correlation functions in the Bogoliubov-Born-Green-Kirkwood-Yvon (BBGKY) hierarchy.  Ref.~\cite{Fauth:2021nwe} is based on a non-equilibrium quantum-field theoretic approach, called the two-particle irreducible effective action technique (2PI formalism) \cite{Luttinger:1960ua, Cornwall:1974vz, Berges:2004yj}, and systematic power countings in the coupling $e$ and gradients.  The 2PI formalism consistently solves a coupled equations for the one-point function, i.e., the classical field ${\mathcal A}^\mu$, and two-point functions defined on a closed-time path contour, which contain all the statistical and spectral information of quantum fields out-of-equilibrium.  This formalism provides a powerful framework to describe far-from-equilibrium dynamics from classical fields to quantum particles in a unified manner.  
{Although in general 2PI formalism has problems with ensuring gauge invariance due to the absence of e.g.~vertex corrections~\cite{Arrizabalaga:2002hn}, it was found that gauge invariance can be recovered in the kinetic limit~\cite{Carrington:2007fp}, reproducing the gauge-invariant strong-field QED kinetic equation (see e.g.~\cite{Elkina:2010up}) under suitable conditions~\cite{Fauth:2021nwe}.}
A few attempts have been made to numerically solve the 2PI formalism to trace non-equilibrium dynamics of gauge theories directly without assuming the kinetic limit, but this appears to be challenging as it requires, for example, a heavy numerical cost due to memory integrals and unphysical contributions~\cite{Nishiyama:2010mn, Zoller:2013lfa}.  

\subsubsection{Beam depletion}
Depletion of a strong field is due not only to pair creation, but also to direct absorption of photons. The impact of beam depletion on observables such as the photon emission angle in the nonlinear Compton scattering of electron bunches against intense lasers has been investigated in~\cite{Seipt:2016fyu}; an analysis of the energy absorbed from the laser {leads to a criterion for the onset of depletion as follows. First one equates the number of absorbed photons with that in the laser (normalised per $\lambda^3$): $N_{abs}\sim N_L$, with $N_L=2\times 10^{14} \xi^2$. To estimate the number of absorbed photons itself one writes $N_{abs} \sim N_e \bar n {\lambda}/{L_C}$, in which $N_e$ is the particle density, $\bar n$ is the average number of photons absorbed per emission, and $L_C\sim 1/\prob$ is the radiation length in terms of the emission probability $\prob$. Finally, the number of photons per absorption scales classically as $\bar n\sim \xi^3$, and was determined in~\cite{Seipt:2016fyu} by a best fit to numerical calculations of nonlinear Compton for $\chi > 2$. The} results indicate that depletion becomes important for intensities of $\xi \sim 10^3$, and for particle densities of order
\begin{equation}
    N_e \sim 6.8 \times 10^{11} \gamma^{0.92} \xi^{-1.08} \quad \text{per laser wavelength cubed.}
\end{equation}
As such, depletion effects must be considered when new ultra-high  intensity scenarios are envisaged for the study of strong field phenomena~\cite{Magnusson:2019qop}. Depletion through laser energy absorption will be both caused by and have an impact on cascades~\cite{Fedotov:2010ja,Gris2016absorption,Jirka:2017cvr,Slade-Lowther:2018kgv,Seipt:2020uxv}, see Sec.~\ref{subsec:cascades}. (Note that `quantum' depletion is not included in PIC simulations, whereas classical depletion is. To illustrate, the energy required to create a pair via the Schwinger effect is at least $2m$, but this is negligible compared to the energy lost in the accelerating the pair, which goes as $m\gamma\sim m\xi  \gg 2m$ for $\xi$ large, which must hold for the LCFA approximations underpinning PIC methods to hold~\cite{Gonoskov:2014mda}.

Recall that scattering in a background field $\mathcal{A}_{\mu}$ is described by the Lagrangian~(\ref{Furry-action}). Let $S[A+\mathcal{A}]$ be the corresponding $S$-matrix, in which $A$ represents the dynamical, quantised degrees of freedom. If $\mathcal{A}_\mu$ obeys Maxwell's equations in vacuum, then it has long been known that~\cite{Kibble:1965zza,Frantz:1965}
\be\label{S-relate}
	S[A+\mathcal{A}] = D^\dagger(z)S[A]D(z) \;,
\ee
in which $D(z)$ is, in complete analogy to optics, the displacement operator for the photon modes $a_\mu$, $a^\dagger_\mu$,
\be
 {\qquad D(z) = \exp \int\!\frac{\ud^3\ell}{(2\pi)^32\ell_0}\, z^\mu(\ell) a_\mu^\dagger(\ell) + z_\mu^\dagger(\ell) a^\mu(\ell)}
\ee
with the profile $z_\mu(\ell)$ being the positive frequency component of the background $\mathcal{A}_\mu$:
\be
	\mathcal{A}_\mu(x) = \int\!\frac{\ud^3\ell}{(2\pi)^32\ell_0}\,
	e^{-i\ell.x} z_\mu(\ell) + e^{i\ell.x} {\bar z}_\mu(\ell) \;.
\ee
Taking matrix-elements of (\ref{S-relate}) shows that scattering in the background is equivalent to scattering in vacuum, described by the usual QED $S[A]$, but with states which contain the \textit{same} incoming and outgoing coherent state created from the vacuum by the action of $D(z)$. That the initial coherent state is unchanged by the scattering process simply reflects that depletion is neglected when considering a background, i.e.~fixed, field. (See~\cite{Frantz:1965} for going beyond the case of coherent states.)

{Expanding coherent states in terms of photon number states provides a link between background field amplitudes and ordinary amplitudes in vacuum; the link may equivalently be studied by expanding amplitudes in background fields directly in powers of the background. This is of course equivalent to treating the background perturbatively from the beginning, but expanding e.g.~plane wave amplitudes can have advantages over purely perturbative calculations. For example, the method has been used to study processes in weak fields or superpositions of weak and strong fields, including nonlinear Compton scattering~\cite{Seipt:2013hda,Seipt:2015rda}, nonlinear and linear Breit-Wheeler \cite{Nousch:2015pja,Golub:2020kkc} and trident~\cite{HernandezAcosta:2019vok}. The method has also been used to study high-multiplicity, multi-collinear limits of QED and Yang-Mills amplitudes~\cite{Adamo:2021hno} which would be difficult to obtain using purely perturbative methods, due to both the large number of legs involved, and the subtlety of colinear limits.}

Coherent states are a key ingredient in the extraction of classical observables from quantum scattering amplitudes, as has long been studied in the context of radiation reaction~\cite{Krivitsky:1991vt,Higuchi:2004pr,Higuchi:2005an,Ilderton:2013dba,Ilderton:2013tb,Torgrimsson:2021wcj} and which has recently received renewed attention in particle-particle scattering with applications to the extraction of classical gravitational wave observables from QFT calculations~\cite{Kosower:2018adc,Cristofoli:2021vyo}.
 
The appearance of quantum states prompts us to recall~\cite{Berson1969}, in which the Dirac equation is solved in the background of a \textit{quantised} rather than classical plane wave, i.e.~in a single quantised mode of the electromagnetic field. The solution has structural similarity with the Volkov solution. As such, its matrix elements have been interpreted in terms of quantum depletion of a highly populated intense laser mode~\cite{Bergou:1980cp}. This interpretation is questioned in~\cite{Heinzl:2018xnv}, simply because there is no laser or intense field in the theory. A top-down approach suggests instead that the obtained solutions of the Dirac equation in a quantised background describe an electron together with the single-mode limit of the virtual photon cloud which always surrounds it and generates its Coulomb field of~\cite{Lavelle:1995ty}. Thus, rather than gaining access to quantum depletion, one is instead considering a greatly simplified, but seemingly exactly solvable, version of QED (and one which is surprisingly subtle from the viewpoint of lightfront field theory~\cite{Bakker:2013cea}, as one must explicitly retain zero modes).  Non-relativistic interactions of light and matter in this single-mode theory are investigated in~\cite{Mati:2017zyc}.

 These observations prompt the following method to include in calculations back-reaction on background fields; one calculates amplitudes between asymptotic states containing \textit{different} coherent photon states. Generalising (\ref{S-relate}), the properties of the displacement operator give (assuming that the particles in `in' and `out' are distinct, in momentum space, from those in the coherent states) 
    \be\label{A2B}
        \bra{\text{out}}D^\dagger(z^f) S[A] D(z^i)\ket{\text{in}} = N(z_f,z_i) \bra{\text{out}}S[A+\mathcal{B}] \ket{\text{in}}\;,
    \ee
    in which $N$ is a normalisation constant and the new background field $\mathcal{B}$ has positive (negative) frequency mode $z_i$ ($\bar{z}_f$).  The use of different coherent states, as in the left hand side of (\ref{A2B}) was used in~\cite{Endlich:2016jgc} to study superradiance, a mechanism for extracting energy from spinning compact objects such as black holes. The right hand side of (\ref{A2B}) says that changes in coherent states may equivalently be captured by performing the Furry expansion around the new background $\mathcal{B}$ which is in general \textit{complex} valued. This was applied to beam depletion in~\cite{Ilderton:2017xbj}. While this approach may be enough to capture exclusive depletion amplitudes, such as e.g.~complete depletion in which $z_f=0$, it is not enough to fully capture back-reaction on an initial coherent state. This can be seen by exploring the same approach in the far simpler Jaynes-Cummings model (for a recent review see~\cite{Kasper:2020akk}). There one sees that an initial coherent state will typically evolve into a  `cat' state~\cite{Ekman:2020vsc}, exhibiting a quantum aspect of back-reaction. Hence the coherent state methods described here could be extended by including, in the final state, a non-classical superposition of coherent states.
 
 We note that for light-by-light scattering phenomena, see Sec.~\ref{sec:LBL}, beam depletion is \textit{automatically} taken into account by standard numerical solvers of the Heisenberg-Euler equations of motion that account for the full nonlinear dynamics of the electromagnetic fields~\cite{domenech2017implicit,grismayer2021,Lindner:2021krv}.

\section{Beyond QED}\label{sec:beyondQED}
 {A sufficiently strong external field can modify the weak, strong, and gravitational interactions of elementary particles, as well as their electromagnetic interactions.}
In this section we review how strong-field methods, in particular those developed initially or mainly in QED, have found application in other sectors of the Standard Model (SM), and beyond.

\subsection{Nuclear physics}
%
 {The impact of a strong field on nuclear decays is characterised by an intensity parameter $\tilde\xi$ and a quantum parameter $\tilde\chi$ defined with respect to the \emph{parent} particle. The `tilde' notation on these parameters is used to emphasise their mass-dependence: since $\tilde\xi$ is normalised by the, usually very large, parent particle mass, it remains small for all currently available and near-future field strengths. The quantum parameter, which is similarly suppressed, governs the modification of the total (fully integrated) decay rate.}

 {Several $\beta$-decays of particles in a constant crossed field or periodic plane wave (including $\mu\to e\nu\tilde{\nu}$, $\pi^\pm\to\pi^0 e^\pm\nu$, $\pi\to\mu\nu$, $\pi\to e\nu$ and $e\to e\nu\tilde{\nu}$), were studied in 60s~\cite{ritus-jetp69} and nuclear decays were studied in the 80s~\cite{Becker83,Nikishov83,Akhmedov:1983gts}. For decays which can occur in vacuum, the impact of the field is typically proportional to $\tilde{\chi}^2$, whereas for field-induced decays the rate is exponentially suppressed as $\exp(-\mathrm{const}/\tilde{\chi})$. In all practical cases and assuming that the parent particle is non-relativistic, $\tilde{\chi}$ can approach unity only for fields about or higher than the Schwinger value (still defined with respect to the electron). Hence the total rate remains almost unaffected by weaker fields. In contrast, the spectral distribution of decay products is controlled by composite parameters such as $\tilde{\xi}^2/\tilde{\chi}$ and $\tilde{\xi}^3/\tilde{\chi}$, and can thus be substantially modified even by weak fields. A comprehensive discussion can be found in~\cite{Akhmedov:2010ee}. Proton transmutation $p\to ne^+\nu$ has recently been reconsidered in strong, pulsed plane waves~\cite{Wistisen:2020czu} and, in agreement with the above, it is shown to be extremely challenging to observe with current technology.}

 {The impact of external fields on $\alpha$-decay is usually considered by incorporating the external field into the Gamov precluster tunneling model. The broad results are the same as for $\beta$-decay: whereas even relatively weak fields can substantially modify the spectral distribution of the decay products, in particular inducing recollisions~\cite{Cortes:2012tu}, the total tunneling rate remains almost unaffected by the field, unless the field strength is of the order of the Schwinger limit or higher~\cite{Qi:2018fwx} -- it is found in~\cite{Qi:2020yhj}, for example, that intensities of the order of $10^{24}$W/cm$^2$ can only induce sub-percent level effects on fission processes.}

 {A clear discussion of the extent to which $\alpha$-decays may be controlled with  intense electromagnetic fields, highlighting relevant dimensionless parameters and identifying the limitations of  is given by~\cite{Palffy:2019scn}, along with illustrative estimates of the lifetime shifts of several typical isotopes. That paper also identifies shortcomings of earlier works which claim \emph{extreme} modifications of decay rates.}

 {One possible mechanism for enhancing nuclear transitions is dynamical assistance, as discussed earlier in the context of the Schwinger effect, see Sec.~\ref{sec:dase}. According to the estimates of~\cite{Queisser:2019nuh,Queisser:2020mns}, for example, deuterium-tritium fusion could be sped up by an X-ray laser of keV frequency, within the reach of present or near-future technology. See~\cite{Wang:2020woc} for a discussion of the enhancement in near-IR frequency, monochromatic fields, and~\cite{Kohlfurst:2021dfk} for the non-adiabatic, or `impulsive' pulsed fields.}

\subsection{Electroweak and Higgs physics}

 {Consider now the full electroweak sector of the Standard model; in comparison to QED there are, along with electrons and positrons, additional} charged and neutral particles, and new weak interactions between them. 
The only really new ingredient is that the charged $W^\LCpm$  {bosons become, unlike the photon, dressed by an electromagnetic background $\mathcal{A}$}; however, the high mass of the $W$ means that,  {as above}, its coupling ${\tilde \xi}_W\sim e \mathcal{A}/m_W$ is around $6\times 10^{-6}$ times smaller than the coupling $\xi\sim e\mathcal{A}/m$ to the electron. If we consider laser backgrounds, then even for future facilities with $\xi=1000$ (see~Fig.~\ref{FIG:PARAMPLOT} in Sec.~\ref{sec:intro}), $\xi_W\ll 1$ and the $W$ only couples very weakly to the laser. {Note that the effective coupling ${\tilde \xi}_\mu$ of muons is ${\tilde \xi}_\mu \sim \xi/200$ due to the difference in mass  {between the muon and electron}, and so ${\tilde \xi}_\mu\sim 1$ may be reached in the coming years.}

Higgs production from laser-assisted lepton-lepton collisions has recently been studied~\cite{Muller:2013xaa} with the idea being to use the laser fields to give an additional boost to the colliding particles, i.e.~to raise their energies, giving better access to Higgs physics~\cite{Muller:2014iza}.

Neutrinos  {are not charged, but} can couple to electromagnetic fields through charged particle loops. Effective actions for low-energy neutrino-photon interactions can be derived by integrating out the (heavy) particles running in the loop; for the corresponding Heisenberg-Euler action see~\cite{Gies:2000tc} and for an application to neutrino-pair creation from external fields see~\cite{Gies:2000wc}.

 {The electroweak sector of the Standard Model, unlike QED, violates parity conservation. Consider neutrinos; due to being only left-handed (while anti-neutrinos are only right-handed, maximally violating parity symmetry), it is possible for interactions with neutrinos} to convert linearly polarised photons to circular polarisation~\cite{Mohammadi:2013ksa}. This could lead to birefringence-like signals (see Sec.~\ref{sec:LBL}) in lasers colliding with neutrino beams. The axial-vector–vector current allows neutrino to photon conversion at one loop in strong backgrounds~\cite{Gies:2000wc}; this is investigated for massless neutrinos and arbitrary plane waves in~\cite{Meuren:2015iha}.

The very small neutrino mass is crucial in solving the solar neutrino problem; a flavour state  {(i.e. an electron, muon, or tau neutrino)} is then a superposition of mass eigenstates, and a neutrino produced with a definite flavour can oscillate between other flavours in flight as the superposition evolves in time. Massive (Dirac) neutrinos can couple to electromagnetic fields through their magnetic moment; it was suggested in~\cite{Formanek:2017mbv} that the dynamics of ultra-relativistic neutrinos in strong fields could allow a measurement of the moment, and thus a test of whether neutrinos are of Dirac or Majorana type,  {something which remains unresolved}.  {(Recall that all other spin 1/2 particles in the Standard Model are known to be Dirac fermions.)} Other processes which can be triggered by strong fields include spin oscillations (conversion of left-handed neutrinos to right-handed) and spin-flavour oscillations (e.g.~left-handed electron neutrino to right-handed muon neutrino), as has been studied in  monochromatic plane waves~\cite{Dvornikov:2018tmm,Dvornikov:2019pxd}, as well as combinations of plane waves and curved spacetimes~\cite{Dvornikov:2019sfo}.

\subsection{QCD and heavy-ion collisions}\label{sec:beyondQED:QCD}
There appear transient, strong, colour (chromoelectromagnetic) and electromagnetic fields in relativistic heavy-ion collisions at RHIC and the LHC.  The typical strength of the created fields is of order $(100\;{\rm MeV})^2$ to $(1\;{\rm GeV})^2$, which is beyond the critical field strength $E_{\rm S}$ of QED. Heavy-ion collisions may thus provide a unique opportunity to explore strong-field physics in a complementary regime than that covered by intense lasers (in which the field strength is still subcritical but the characteristics of the fields themselves are better understood). The lifetime of the produced fields is, though, very short, {${\mathcal O}(10^{-1})\;{\rm fm}/c$ at most},
which may affect the non-perturbativity of processes under investigation, see below. {Heavy-ion physics is a broad research area, and we will not attempt to review it comprehensively here, instead directing the reader to the reviews~\cite{Kharzeev:2013jha,Huang:2015oca,Hattori:2016emy,Fukushima:2016xgg,Fukushima:2018grm,Berges:2020fwq,Gelis:2021zmx}. We will briefly highlight, though, some similarities and differences to ongoing work in strong-field QED.}

Strong \emph{colour} fields play crucial roles in the early-stage dynamics of relativistic heavy-ion collisions.  In the high-energy limit, the colliding ions are essentially composed of high-density gluonic matter, and can be effectively described as classical colour charges. This is a non-Abelian analogue of the Weizs\"{a}cker-Williams approximation in QED, and is conveniently described by the colour-glass condensate (CGC) framework~\cite{Kovchegov:2012mbw,Gelis:2012ri,Albacete:2014fwa}.  The two classical colour charges act like a ``parallel-plate capacitor'' just after a collision, and form colour flux tubes in between (which is often called {\it glasma}~\cite{Lappi:2006fp}).  Reflecting the largeness of the gluon density, or the classical colour charges, the created colour flux tubes become strong and can be parametrised as $E_{\rm colour},B_{\rm colour} \sim Q^2_{\rm s}/g$, where $g$ is the QCD coupling constant
and $Q_{\rm s} = {\mathcal O}(1\;{\rm GeV})$ is the so-called saturation scale of the CGC that characterises the density of the gluons.  Importantly, while the QCD coupling $g$ is small at high energies due to asymptotic freedom, it is compensated by the strength of the field $E_{\rm colour},B_{\rm colour} \propto 1/g$ and hence the interaction with the colour field must be treated non-perturbatively.  This is the same situation as in strong-field QED, where the smallness of the QED coupling $e$ is compensated by the intensity of the coherent electromagnetic fields, such that $\xi>1$ becomes the relevant coupling parameter. While it is widely accepted that colour flux tubes sourced by the highly dense gluons provide the initial conditions for heavy-ion collisions, it is an active research topic to address how the flux tubes evolve and eventually decay, via e.g. the Schwinger effect, into the quark-gluon plasma -- matter composed of quarks and gluons liberated from confinement. See~\cite{Fukushima:2016xgg,Berges:2020fwq,Gelis:2021zmx} and reference therein for recent progress.

Heavy-ion collisions can also produce strong electromagnetic fields due to, e.g., the strong Lorentz contraction of Coulomb fields. {Estimates of the electromagnetic field-strengths and spacetime distributions in (ultra-)peripheral collisions were made in~\cite{Bzdak:2011yy,Deng:2012pc}.  See also \cite{Hirono:2012rt,Voronyuk:2014rna} for asymmetric collisions.  It was shown that the created fields are very strong $eE,eB ={\mathcal O}(100\;{\rm MeV})^2 -{\mathcal O}(1\;{\rm GeV})^2$ but are very short-lived.  (Ref.~\cite{Baur:2008hn} suggests that the short duration also means that, for ultra-relativistic collisions, the external field approximation itself does not hold. See also~\cite{Peroutka:2017esw} for a discussion of how to model the classical fields, and their charged particle sources, including spin and wavepacket effects.)  }  Ultra-peripheral heavy ion collisions have recently led to the first measurements of both light-by-light scattering~\cite{ATLAS:2017fur,CMS:2018erd,ATLAS:2019azn} and linear Breit-Wheeler pair-creation~\cite{STAR:2019wlg} in QED. {In ultra-peripheral collisions it is consistent to describe the two processes using just {leading} order perturbation theory}, even though the Coulomb fields involved can in some circumstances be considered `strong'. The reason is that the processes occur when two heavily boosted Coulomb fields overlap, and the higher the energy with which the fields collide, the shorter the duration of the appreciable spatio-temporal overlap. This in turn implies a lower `intensity', and the effect of low intensity fields can be studied using perturbation theory. 
{In \cite{BAUR20071}, it is explained that the Keldysh parameter of the interaction is $m \omega / eE \gtrsim 1$ such that processes are perturbative because the effective field frequency is so high, even though field strengths above the Schwinger limit are possible. In a plane-wave background, the Keldysh parameter is $1/\xi$, and so in this sense ultraboosted heavy-ion Coulomb fields correspond to low 'intensity'.} It has been found that the contribution of higher-order interactions with the Coulomb field (so-called `Coulomb corrections') beyond the Born approximation, can be included adequately using perturbation theory, without requiring an all-orders result~\cite{Baur:2008hn}, for experimentally-relevant parameters at current heavy ion colliders. There remain many open questions and directions for further study of QED in ultra-peripheral collisions, for example the transition from `fast' to `slow' collisions~\cite{Obraztsov:2021tip}, creation of pair in bound states, {``bound-free" combinations~\cite{Francener:2021wzx,Francener:2021wzx} and even ``bound-bound'' states \cite{Voitkiv:2009ef}} and the application of tested QED results, such as using linear Breit-Wheeler as a probe of the strong magnetic fields in heavy ion collisions~\cite{Brandenburg:2021lnj}, the generation of which is reviewed in~\cite{Huang:2015oca,Hattori:2016emy}.

\subsection{Yang-Mills, gravity and the double copy}
Another motivation for studying background colour fields in Yang-Mills theory (YM) is colour-kinematic duality, or `double copy'~\cite{Kawai:1985xq,Bern:2008qj,Bern:2010ue,Bern:2010yg}. Double copy shows that scattering amplitudes in gravitational theories can be obtained exactly from scattering amplitudes in gauge theory.  {For example, $n$-graviton scattering amplitudes can be obtained from $n$-gluon scattering amplitudes in gauge theory, by replacing the colour structure of the latter with a copy of their kinematic structure.} 
Double copy may thus provide a whole new way of thinking about gravity as a `squared' gauge theory. Practically, double copy also offers an efficient method of calculating gravitational observables, since perturbative calculations in gauge theories are much simpler than those in gravity. See~\cite{Bern:2019prr,White:2021gvv} for reviews and more details.

If double copy is a fundamental relationship between gravity and gauge theory, though, then it must persist beyond the case of scattering in vacuum, i.e.~on \emph{flat} backgrounds. This prompts the consideration of scattering amplitudes of colour-charged matter on background colour fields, and of matter and graviton scattering on background gravitational fields,  {meaning scattering in}~non flat spacetimes. {Given the progress made in QED, it would seem natural to start with plane wave backgrounds in both cases, and indeed this is a situation in which QED results are informing progress in YM and gravity, as we now discuss.}

 {Background colour fields live, like gluons, in the adjoint representation. They carry both a vector index and a colour index, ${\cal A}_\mu \equiv {\cal A}_\mu^{\mathsf{a}} \mathsf{T}^{\mathsf{a}}$ in which the generators $\mathsf{T}^a$ obey, in terms of the structure constants $f$ of the gauge group obey, $(\mathsf{T}^{\mathsf{a}})_{bc} = i f_{bac}$.} A YM plane wave $\mathcal{A}_\mu$ is a vacuum solution satisfying the same symmetry algebra as electromagnetic and gravitational plane waves\footnote{The `nonabelian plane waves' of~\cite{Coleman:1977ps} do not have the requisite symmetries, and are actually pp-waves -- plane-fronted waves with parallel propagation --  {rather than plane waves.}}.  {A consequence of these symmetries} is that $\mathcal{A}_\mu$ must be valued in the maximally abelian subalgebra of the gauge group, `the Cartan'~\cite{Adamo:2017nia} --  {in other words, to the part of the algebra which is commutative. This presents a significant simplification, as follows:} let $\mathsf{T}$ be the generator of some  {adjoint particle, for example a gluon} scattering on the background. Then we have $[\mathcal{A}_\mu,\mathsf{T}] = e^\mathsf{i} \mathcal{A}_\mu^\mathsf{i} \mathsf{T}$, with the colour indices `$\mathsf{i}$' running over the  {Cartan generators in the background field}, and $e^\mathsf{i}$ being the `root-valued charge' of $\mathsf{T}$  {(essentially a vector). The point is that commutation} with $\mathcal{A}_\mu$  {simplifies to} scalar multiplication, and  {hence} interactions with the  {plane wave} background are  {\emph{effectively abelian\footnote{We illustrate using SU(2), the generators of which are the Pauli matrices. We write these in the Cartan-Weyl basis,
\be
    \mathsf{T}^0 := \frac12 \sigma_3 = \frac12\begin{pmatrix}1 & 0 \\ 0 & -1\end{pmatrix} \;,
    \quad
    \mathsf{T}^+ := \frac12 (\sigma_1 + i\sigma_2) = \begin{pmatrix}0 & 1 \\ 0 & 0\end{pmatrix} \;,
    \quad 
    \mathsf{T}^- := \frac12 (\sigma_1 -i\sigma_2) =\begin{pmatrix}0 & 0 \\ 1 & 0\end{pmatrix} \;.
\ee
The Cartan is spanned by a single element $\mathsf{T}^0$. The remaining two generators $\mathsf{T}^\pm$ obey $[\mathsf{T}^0,\mathsf{T}^\pm] = \pm \mathsf{T}^\pm$ -- because there is only one generator in the Cartan, the roots are in this case scalars equal to $\pm 1$ respectively; these are the charges of the gluons. Any gluons which have $\mathsf{T}^0$ as their generator carry charge $0$; they do not see the background at all. For the extension to matter transforming in the fundamental representation, the charges are replaced by the weights of the fundamental representation, here $\pm 1/2$ in the natural basis:
\be
    \mathsf{T}^0 \begin{pmatrix}1 \\ 0 \end{pmatrix} = \frac{1}{2} \begin{pmatrix}1 \\ 0 \end{pmatrix} \;, \qquad 
    \mathsf{T}^0 \begin{pmatrix}0 \\ 1 \end{pmatrix} = -\frac{1}{2} \begin{pmatrix}1 \\ 0 \end{pmatrix} \;.
\ee
 {Note that while the Cartan-Weyl basis is often just a convenient basis to work in, here it has a particular physical relevance.}}}}.  {As a result of this simplification}, many methods and results of strong-field QED in plane wave backgrounds can be generalised to YM plane waves. For example, the classical momentum of an (adjoint) colour-charged particle, initial momentum $p_\mu$ crossing the plane wave background is (with $\mathcal{A}_\mu$ having only transverse components,  {as in (\ref{eqn:this-is-the-potential}) for QED}),
\be
\label{classical-p-YM}
	\Pi_\mu = p_\mu - e^\mathsf{i} \mathcal{A}^\mathsf{i}_\mu(x^\LCp) + \frac{2e^\mathsf{i}\mathcal{A}^\mathsf{i}(x^\LCp)\cdot p-(e^\mathsf{i} \mathcal{A}^\mathsf{i}(x^\LCp))^2}{2n\cdot p} n_\mu \;,
\ee
in which the similarities to the electromagnetic case~(\ref{pi-def}) are explicit. The same result holds for fields in the fundamental  {e.g.~quarks}, by replacing $e^\mathsf{i}$ with the fundamental weight $\mu^\mathsf{i}$.  The  {underlying} non-abelian nature of  {interactions with the background} can still be seen, though, in that each colour of scattered particle will see a \textit{different} background,  {because} each will have a different charge~$e^\mathsf{i}$.

Turning to the quantum theory,  {consider the calculation of scattering amplitudes in the Furry expansion.} Because gluons are colour charged ,and so interact directly with the background (while photons interact with electromagnetic backgrounds only through charged particles, see Sec.~\ref{sec:LBL}), their propagator and asymptotic wavefunctions are nontrivial. To illustrate, an incoming gluon is represented in scattering amplitudes by the wavefunction
\be\label{Volkov-YM}
	\mathcal{E}_\mu (x^\LCp) \mathsf{T} \exp\bigg[-i p \cdot x  - i \int\limits_{-\infty}^{x^\LCp} \!\ud y \frac{2e^\mathsf{i}\mathcal{A}^\mathsf{i}(y)\cdot p -(e^\mathsf{i} \mathcal{A}^\mathsf{i}(y))^2}{2n\cdot p} \bigg]  \;,
\ee
in which $\mathsf{T}$ is the generator, the exponent is again the Hamilton-Jacobi action for the classical momentum (\ref{classical-p-YM}), and the polarisation vector, here in lightfront gauge $n\cdot \mathcal{E}=n\cdot \mathcal{\epsilon}=0$, is
\be\label{YMpol}
	\mathcal{E}_\mu(x^\LCp) = \epsilon_\mu + \frac{ e^\mathsf{i} \mathcal{A}^\mathsf{i}(x^\LCp)\cdot \epsilon}{2n\cdot p} n_\mu \;.
\ee
 {Comparing to a photon wavefunction $\epsilon_\mu \exp(-ip\cdot x)$ in QED, we see that the polarisation picks up a nontrivial spacetime dependence, like the spin factor in the Volkov solution. Like the spin, though, the structure in (\ref{YMpol}) is only there to maintain the transversality relation $\Pi\cdot\mathcal{E}=0$ between the kinematic momentum of the gluon (\ref{classical-p-YM}) and its polarisation vector}: the helicity of the gluon is not changed by propagation through a YM plane wave.  Three-point scattering amplitudes for gluons on YM plane wave backgrounds have been calculated in~\cite{Adamo:2017nia,Adamo:2017sze} and further studied in~\cite{Adamo:2018mpq} along with higher-point tree amplitudes. At one loop, gluon helicity flip in YM is investigated in~\cite{Adamo:2019zmk}, along with the influence on the process of fundamental quarks  in QCD.

Turning to gravity, a plane wave spacetime may be represented in `Einstein-Rosen' coordinates by the line element~\cite{Einstein:1937qu}  
\be\label{metric-PW}
	\ud s^2 = 4\ud x^\LCp \ud x^\LCm - \gamma_{ij}(x^\LCp)\ud x^i \ud x^j \;, \qquad i,j \in\{1,2\}
\ee
in which the transverse metric $\gamma^{ij}$ depends on $x^\LCp$, just as the transverse gauge field depends on $x^\LCp$ in YM and QED. In the flat space limit, $\gamma_{ij}\to \delta_{ij}$ and we recover the Minkowski metric in lightfront coordinates. Note that the form of $\gamma^{ij}$ is \emph{not} arbitrary, but is constrained by the vacuum equations, which is unlike the situation  {for plane waves in QED and YM, where the funtional dependence of the wave is arbitrary}.  {We note briefly that Einstein-Rosen coordinates are \emph{not} global, so calculations should be performed in e.g.~Brinkmann coordinates~\cite{Brinkmann:1925fr}. We stay with Einstein-Rosen here because the basic structures we will see are more obviously related to those we have already encountered in QED. For example, it is clear from (\ref{metric-PW}) that, just like in gauge theory, the transverse and longitudinal components of overall momentum will be conserved in any scattering process.  To further emphasise} the similarities to gauge theory, we write down the solution $\phi(x)$ of the wave equation on the spacetime (\ref{metric-PW}) which, in scattering, represents an incoming massless scalar particle of momentum $p_\mu$. Writing $\gamma := \det{\gamma_{ij}}$, the wave equation is $g^{-1/2}(\partial_\mu (g^{1/2} g^{\mu\nu}\partial_\nu \phi) = 0$ and its solution is~\cite{Ward1987}
\[
    	\phi(x) = \frac{1}{\gamma^{1/4}(x^\LCp)} \exp\bigg[-ip_\LCm x^\LCm -i p_\LCperp x^\LCperp - \frac{i}{4p_\LCm}\int\limits^{x^\LCp}\!\ud y^\LCp \gamma^{ij}(y^\LCp) p_i p_j\bigg] \;.
    \]
The only nontrivial dependence in the wavefunction is on $x^\LCp$, and is similar to that in the Volkov wavefunction. In the flat space limit $\gamma^{ij} \to \delta^{ij}$ and we recover the  {wavefunction of a free massless scalar.}

Three-point amplitudes for gravitons scattering on gravitational plane waves have also been calculated in~\cite{Adamo:2017nia,Adamo:2017sze}. A double copy relationship was therein established, confirming that graviton amplitudes on the plane wave spacetime (\ref{metric-PW}) are the double copy of gluon scattering amplitudes on a YM plane wave. This prescription was tested and found to hold also for three-point amplitudes describing massless (glue or graviton) radiation from massive particles in plane wave (YM and gravity) backgrounds in~\cite{Adamo:2020qru}. Now, the classical radiation field emitted by colour-charged particles crossing a YM plane wave, and massive particles crossing a gravitational plane wave can be obtained from the classical limit of these three-point amplitudes. They can also, though, be calculated directly in the classical theory, where they are related by a classical double copy~\cite{Monteiro:2014cda,Goldberger:2017frp,Goldberger:2016iau}. The radiation fields were computed using both methods in~\cite{Adamo:2020qru} and consistency of the classical and quantum double copy prescriptions was established explicitly.

Double copy has also been used to construct electromagnetic beam models for which the Lorentz force equation is analytically solvable~\cite{Andrzejewski:2019hub,Andrzejewski:2018zby}, see Section~\ref{sec:beyondPW}.
\subsection{Gravitational waves}
\begin{figure}[t!]
    \centering
    \includegraphics[height=3cm]{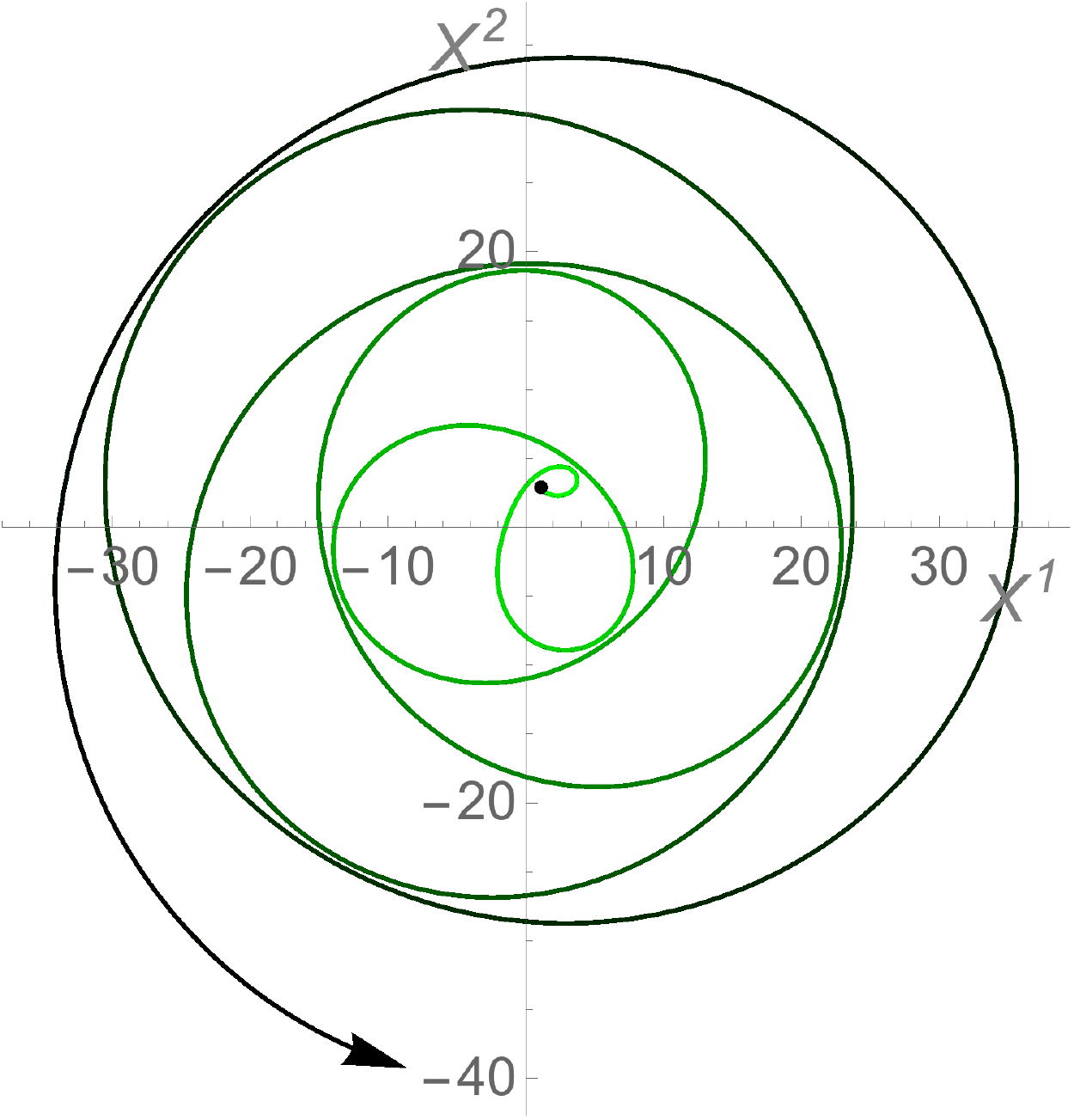}
    \qquad
    \includegraphics[height=3cm]{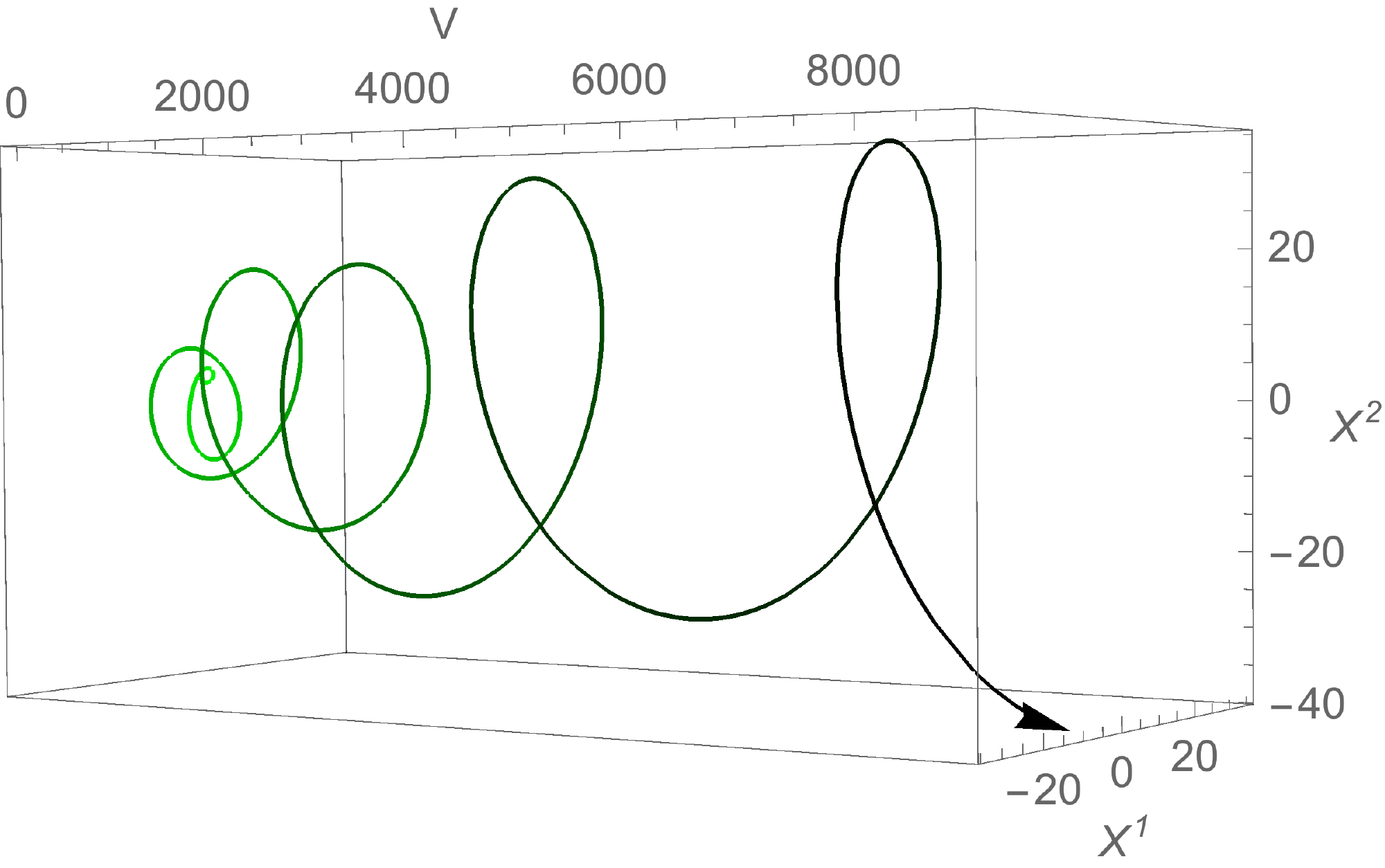}
    \qquad
    \includegraphics[height=3cm]{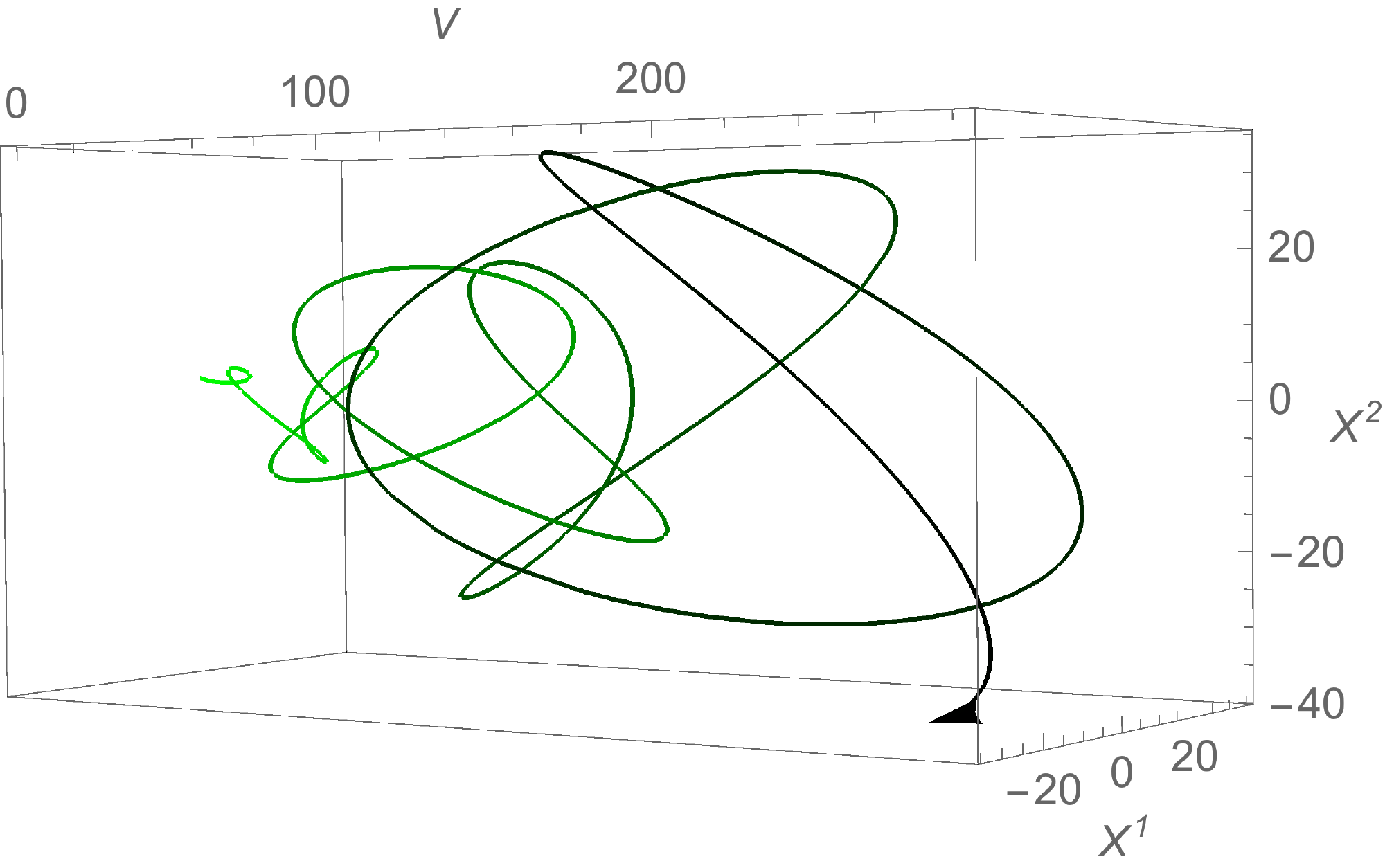}
    \caption{\label{fig:GRwave}
    Classical particle motion in an electromagnetic vortex~\cite{Bialynicki-Birula:2004bvr}, and in its double copy, a screw-symmetric gravitational wave~\cite{Ilderton:2018lsf}. Motion projected into the plane transverse to the wave propagation direction (left hand panel, $(X^1,X^2)$ are our $x^1,x^2$) is identical in the two waves for ``matched parameters'', here $H^{\text{grav}} = H^{U(1)}/(2p_V)$ where $H$ is essentially the field strength and $p_V\sim$ our $p_\LCm$, see~\cite{Ilderton:2018lsf} for details. Motion in the longitudinal direction $V$ (essentially our $x^\LCm$) is however very different in the electromagnetic (middle panel) and gravitational (right hand panel). Figure taken from~\cite{Ilderton:2018lsf}.}
\end{figure}

{Electromagnetic beams with angular momentum can trap particles near the beam centre~\cite{Bialynicki-Birula:2004bvr}. By generalising properties of such fields, e.g.~Bessel beams, to \emph{linearised} gravity~\cite{Bialynicki-Birula:2015mvf,Bialynicki-Birula:2018nnk}, it is possible to find approximate gravitational wave solutions which carry angular momentum, and exhibit very similar particle dynamics and trapping phenomena to the U(1) case. However, such physics exists beyond linearised gravity. Exact solutions of the vacuum Einstein equations, including circularly polarised gravitational waves, carry angular momentum and allow for bound particle orbits, which can also be calculated analytically~\cite{Zhang:2018srn,Zhang:2018msv,Zhang:2021lrw}. See also~\cite{Zhang:2021lrw} for results in Lukash waves.}
    
{An exact correspondence between motion in electromagnetic and gravitational waves is again made by the double copy. The classical double copy of an electromagnetic vortex~\cite{Bialynicki-Birula:2004bvr} carrying angular momentum is a gravitational wave with helical, or screw, symmetry~\cite{Ilderton:2018lsf}, an exact solution of the vacuum Einstein equations. One finds from this matching conditions for which motion in the plane transverse to the two waves is not just similar, but identical, as illustrated in Fig.~\ref{fig:GRwave}. The high degree of symmetry of these waves allows for analytic solutions of the equations of particle motion, see also Sec.~\ref{sec:beyondPW:super}. In general, relations between waves in electromagnetism and gravity are naturally underpinned by symmetries, see~\cite{Zhang:2019gdm} for a comprehensive discussion.}

The observation of gravitational waves~\cite{LIGOScientific:2016aoc,LIGOScientific:2016sjg} has prompted the question of whether it is possible to generate detectable gravitational waves in lab-based experiments, using e.g.~laser-plasma interactions~\cite{Ribeyre1} or optical media~\cite{Pustovoit:2020ksx}. Using linearised gravity (sufficient given the tiny signal sizes), one can straightforwardly calculate metric perturbations due to laser-accelerated relativistic ions~\cite{Gelfer:2015fbj}, standing waves of light~\cite{MorozovGravStanding}, or laser pulses themselves~\cite{Lageyre:2021cxe}.  The latter were found to give slightly better prospects for detection than waves generated by acceleration of matter through laser-driven ablation of foil targets~\cite{Kadlecova:2016fbx}.  However, none of the above currently provide realistic detection scenarios.

\subsection{Monopole pair creation}
The electro-magnetic duality of Maxwell's equations is broken only by the apparent absence of fundamental magnetic charges, or monopoles. It is therefore natural to consider the possibility of a magnetic analogue of the Schwinger effect, i.e., pair creation of monopoles in strong magnetic fields.  The main obstacle is that the monopole coupling $g$ is strong due to the Dirac quantisation condition $eg/2\pi \in {\mathbb N}$, implying that one needs to resum radiative corrections in a non-perturbative manner.  Affleck-Alvarez-Manton applied the worldline instanton method~\cite{Affleck:1981bma} (see also~\cite{Affleck:1981ag}) to this problem and showed for scalar monopoles with arbitrary coupling $g$ that
\begin{align}
	2\,{\rm Im}\,{\mathcal L}^{\rm 1\mathchar`-loop}_{\rm HE} \sim \frac{(gB)^2}{(2\pi)^3}{\rm exp}\left[ -\pi \left( \frac{M^2}{gB} - \frac{g^2}{4\pi } \right) \right] , \label{eq:ht3-44}
\end{align}
provided that the magnetic field $B$ is constant, homogeneous, and sufficiently weak compared to the monopole mass $M$, i.e.~$gB/M^2, g^3B/M^2 \ll 1$.  Note that the same formula holds for electrically charged scalar particles after replacement $B \to E, g\to e$, and $eE/m^2, e^3E/m^2 \ll 1$.  (See also Sec.~\ref{sec:ht6} on radiative corrections.)
The non-perturbative formula (\ref{eq:ht3-44}) exhibits different magnetic-field- and/or coupling-dependencies compared to na\"ive perturbative pair creation formulas for, e.g., the tree-level Drell-Yan process (in which a quark and antiquark from a pair of hadrons annihilate to a pair of leptons).  A proper non-perturbative treatment of monopole pair creation is, thus, important for making a correct interpretation of experimental results and setting bounds on the monopole mass $M$.  

Monopole pair creation via the Schwinger effect has recently received renewed interest~\cite{Gould:2017zwi,Gould:2018ovk, Gould:2019myj, Gould:2021bre, Ho:2019ads, Ho:2021uem}, motivated by ultra-relativistic heavy-ion collision experiments at RHIC and LHC which create strong magnetic fields of order $eB = {\mathcal O}((100\;{\rm MeV})^2\mathchar`-(1\;{\rm GeV})^2)$.  To discuss monopole pair creation in heavy-ion collision experiments, it is important to include strong spatial and temporal inhomogeneities of the generated magnetic fields (see Sec.~\ref{sec:beyondQED:QCD}) and the impact of finite temperature. These effects were included in~\cite{Gould:2019myj, Gould:2021bre} using the LCFA and the worldline instanton {method (the accuracy of which in relevant experimental regimes is discussed in~\cite{Gould:2019myj}). For} realistic magnetic field strength in heavy-ion collision experiments, it was concluded that LHC experiments could test the existence of monopoles with masses of order ${\mathcal O}(100\;{\rm GeV})$. {Details of the first experimental monopole search at the LHC were recently published in~\cite{Acharya:2021ckc}. For monopole production in the primordial magnetic fields of the early universe see~\cite{Kobayashi:2021des}.}

\subsection{Beyond the Standard Model}
The hunt for physics beyond the standard model (BSM) is a pervasive topic throughout physics (for reviews, see \cite{Jaeckel:2010ni,Essig:2013lka,Graham:2015ouw,Irastorza:2018dyq}). Here, we are interested in the connection with intense background fields (for a review in this context, see \cite{Tajima:2012mx}). Typically BSM and dark matter searches are characterised as using: a) direct detection (BSM$\to$SM); b) production-and-detection  SM$\to$BSM$\to$SM; c) collider detection SM$\to$BSM (characterised by e.g. missing transverse momentum). The majority of scenarios considered that would employ intense EM fields are of the direct detection type, with some collider detection examples. One reason is that intense EM fields are typically of very short duration (laser pulses of the order of tens of femtoseconds and ions in heavy-ion collisions contracted to less than a femtometre in the lab frame) and so only provide a small spacetime overlap with any probe beam, making production-and-detection less favourable.

Most BSM searches employing EM fields use weak fields (which are easier to generate and control than intense fields) and we summarise these briefly here (a review of magnetic fields used for BSM searches can be found in \cite{Battesti:2018bgc}). Combining a weak magnetic field with a high-finesse cavity, as in the PVLAS \cite{Ejlli:2020yhk}, BMV \cite{Cadene:2013bva}, Q\&A \cite{Chen:2006cd} and OVAL \cite{Fan:2017fnd} experiments, or by using a long magnet as a helioscope to capture solar axion-like particles (ALPs) as in CAST \cite{CAST:2017uph} and IAXO \cite{Armengaud:2014gea} allows for the weakness of the interaction to be compensated for by the large interaction volume. Weak EM fields are also easier to screen from the detector region, such as in `light shining through the wall' (LSW) experiments \cite{Redondo:2010dp} like ALPS \cite{Bahre:2013ywa}, GammeV \cite{GammeV:2008cqp}, LIPPS\cite{Afanasev:2008jt} and OSQAR \cite{OSQAR:2015qdv}. An alternative to using cavity searches, is to collide probe and pump laser pulses directly, and look for signals of BSM physics in the scattered probe photons, such as frequency conversion due to exchange of virtual ALPs as performed in SAPPHIRES \cite{SAPPHIRES:2021vkz}. (Although intense fields are used in laser pulse collisions, due to the weak photon-ALP interaction, the physics is still perturbative in the coupling.)  In this section, we will concentrate on the physics case explored in the literature in the last decade, for using intense EM fields in BSM searches (an early overview is given \cite{Gies:2008wv}).

ALPs, by which we include (neutral) pseudoscalars and scalars, can play a role in intense fields in different ways. Through the (dimension-5) diphoton coupling, $g_{\phi\gamma\gamma}$, ALPs can contribute to photon-photon scattering. {(In this sub-section we apply a common convention that $\phi$ refers to scalars or pseudoscalars, and not to laser phase.)} In a collision of a probe with a pump laser beam, ALPs can induce a chiral vacuum birefringence and dichroism signal \cite{Villalba-Chavez:2013bda,Villalba-Chavez:2013goa} as well as an ellipticity \cite{Shakeri:2020sin} and the pulse shape can play an important role in the induced rotation \cite{Villalba-Chavez:2016hxw}. When colliding two intense pump laser pulses orthogonally, with frequencies $\omega_{1}=2\omega_{2}$, a probe of the overlap can experience frequency up/down shifts at resonant axion masses
\cite{Dobrich:2010hi}. The diphoton coupling is also predicted to lead to parametric instabilities in the propagation of an intense laser pulse \cite{Beyer:2021xql} and a flux of ALPs being generated by a laser-driven plasma wakefield propagating along a constant strong magnetic field \cite{Burton:2017bxi}. The change in an x-ray beam (produced by a free electron laser) as it probes the quasi-static fields in hollow plasma structures, formed by colliding an intense laser pulse with a low-density target, has also been suggested as a probe of ALPs \cite{Huang:2020lxo}. The diphoton coupling to pseudoscalar ALPs can be compared to QED light-by-light scattering in a background in a simple way in the weak-field limit by a simple rotation of the mass and interaction term $g_{\phi\gamma\gamma}\phi\mathcal{P}$ (recall \eqnref{eq:invariants}). This procedure generates an effective four-photon interaction $g_{\phi\gamma\gamma}^{2}\mathcal{P}^{2}/2m_{\phi}^{2}$, which is as large as the corresponding Heisenberg-Euler QED term  if $g_{\phi\gamma\gamma}^{2} \geq 28\alpha^2 m_{\phi}^2/(45m^{4})$ \cite{Evans:2018qwy}. The production of scalar particles by a photon propagating in a circularly polarised monochromatic wave has also been investigated \cite{Villalba-Chavez:2012kko}.

ALPs can also couple to fermions via a (dimension-4) $ig_{\phi e}\phi\bar{\psi}\gamma^{5}\psi$ pseudoscalar or $g_{\phi e}\phi\bar{\psi}\psi$ scalar term and hence can be radiated by fermions in intense fields via nonlinear Compton scattering, or can decay in an intense field via nonlinear Breit-Wheeler to an electron-positron pair. Allowing the ALPs to have a mass, as is standard, leads to a modification of the Compton and Breit-Wheeler kinematics from the massless photon case. An example is given in \cite{King:2018qbq} for Compton scattering of an ALP by an electron in a circularly-polarised monochromatic background. The total rate can be written, as is usual for the QED case as a sum of partial rates for each harmonic $\rate = \sum_{n=n_{0}}^{\infty}\rate_{n}$, except now that the ALP has a mass (compared to the QED case where the photon is massless), there is a `threshold harmonic' which can be larger than $1$: 
\be
n_{0} = \left\lceil\frac{m_{\phi}}{m\eta_{e}}\left(\frac{m_{\phi}}{2m}+\sqrt{1+\xi^{2}}\right)\right\rceil,
\ee 
where {$\eta_{e}=k \cdot p / m^{2}$ is the energy parameter of the probe electron and $k_\mu$ is the plane wave background wavevector (see also the paragraph preceding \eqnref{eq:chi:def})}. This also affects the harmonic range of each partial rate. It was also shown in the same work, that the production of pseudoscalars is suppressed for $\eta_{e}<1$ as $\sim \eta^{2}$, which is an obstacle to their production in typical beam-laser collisions, in which $\eta \ll 1$. Nonlinear Compton scattering of ALPs has been calculated in the weak-field (perturbative) limit and in a constant crossed field \cite{Dillon:2018ypt}, where the LCFA was used to model emission of ALPs from an electron in an intense Gaussian pulse. In \cite{Dillon:2018ouq} scalar ALP emission from an electron beam colliding with a laser pulse was calculated, and coherent enhancement was suggested as a means by which to significantly increase the ALP signal. The decay of ALPs via Breit-Wheeler to an electron-positron pair has been calculated in weak fields and a monochromatic wave \cite{King:2018qbq}, as well as in a constant crossed field, and a finite pulse \cite{King:2019cpj} where the edge effects of a quasi-constant magnetic field were investigated.

U(1) BSM candidates such {sometimes referred to as `hidden' or `para' or `dark'} photons that couple directly to the photon via a $F_{\mu\nu}G^{\mu\nu}$
`kinetic mixing' term (where $F$ is the standard model U(1) gauge field and $G$ the hidden U(1) gauge), can also be probed in intense fields. In particular, the hidden U(1) sector can interact with photons if there exists a charge that couples both. Such charges must have a correspondingly small coupling in order to have so far evaded experimental detection. The corresponding mini-charged particle (MCP) is therefore itself a BSM candidate that can be accommodated straightforwardly in strong-field QED in a plane-wave by employing Volkov wavefunctions for MCPs instead of electrons/positrons. (Dark radiation constraints on MCPs are reviewed in \cite{Vogel:2013raa}.) MCPs have a fraction of the electron charge (and hence a weak interaction, explaining why they have not so far been detected in experiments), and an unknown mass. Depending on the ratio of the mass squared to the charge,  the `Schwinger field' for MCPs may lie below or above the QED value, and in addition, the MCP intensity parameter may be small, allowing a perturbative treatment in a background field \cite{Gies:2006ca}, or it could be large and hence necessitate the use of dressed states. In particular, it has been shown that MCPs can contribute to the birefringence of the vacuum and induce a signal in the polarisation of a probe beam due to dichroism \cite{Villalba-Chavez:2016hht}, which can be increased close to the two-photon mass threshold \cite{Villalba-Chavez:2013txu,Villalba-Chavez:2014nya,Villalba-Chavez:2015xna}. A similar set-up to an LSW experiment has been suggested, where the wall is traversed by the loop of MCPs, allowing recombination to photons on the other side. This type of tunneling `of the third kind' is enhanced by applying a magnetic field \cite{Dobrich:2012sw,Dobrich:2012jd}.

With regards to the hidden photon in intense fields, the phenomenological consequences of a hidden photon have been explored in a monochromatic background \cite{Villalba-Chavez:2014nya}. The effect on photon-paraphoton oscillations in a circularly-polarised monochromatic wave on having a spin-$1/2$ or scalar MCP has been investigated in \cite{Villalba-Chavez:2015xna}, with an enhancement found in the scalar case. Mixing of photon and dark photons due to the nonlinearities in QED in intense fields, has been calculated in constant electric and magnetic backgrounds in \cite{Fortin:2019npr} and a search employing a $10\,\trm{PW}$ laser suggested.

Physics beyond the SM may also be revealed in intense fields if the photon's interaction with itself is modified by some nonlinearity in Maxwell's equations which appears at a higher energy scale, such as in the Born (B) \cite{Born:1933qff} and Born-Infeld (BI) \cite{Born:1934gh} generalisations of electrodynamics. The interaction Lagrangian for these generalisations can be written:
\be
\mathcal{L}_{\trm{B}} = -b^{2}\left[\left(1-\frac{2\mathcal{S}}{b^{2}}\right)^{1/2}-1\right]; \qquad \mathcal{L}_{\trm{BI}} = -b^{2}\left[\left(1-\frac{2\mathcal{S}}{b^{2}}-\frac{\mathcal{P}^{2}}{b^{4}}\right)^{1/2}-1\right], \label{eqn:BBI}
\ee
and $b$ is a free parameter with units mass squared. (In the original formulation, by equating the EM self-energy of the electron with its rest mass, Born and Infeld arrived at the value $b=1.2 \times 10^{20}\,\trm{Vm}^{-1}$.) Therefore, the nonlinear modification of electromagnetism predicted by the BI model, would modify the signal of photon-photon scattering and so schemes to measure the QED process will also bound the value of $b$ in the BI model. For example, the ATLAS \cite{ATLAS:2017fur,ATLAS:2019azn} and CMS \cite{CMS:2018erd} results for photon-photon scattering in ultra-peripheral collisions of heavy ions were used to bound BI theory over a large energy range \cite{Ellis17} (much higher than typical centre-of-mass energies in intense laser experiments). In the Born-Infeld theory, the vacuum is not birefringent at leading order unlike in QED (see Sec. \ref{sec:LBL}), because the predicted vacuum refractive index is independent of field polarisation. 
This can be seen by performing a ``weak-field'' expansion of \eqnref{eqn:BBI} for small $\mathcal{S}/b^{2}$ and $\mathcal{P}/b^{2}$ and comparing to the prediction of the QED:
\be
\widetilde{\mathcal{L}}_{\trm{B}} \simeq \frac{\mathcal{S}^{2}}{2b^{2}}; \qquad \widetilde{\mathcal{L}}_{\trm{BI}} \simeq \frac{\mathcal{S}^{2}+\mathcal{P}^{2}}{2b^{2}}, \label{eqn:BIwf1}
\ee
where, just in this equation we have used $\widetilde{\mathcal{L}} = \mathcal{L} - \mathcal{S}$ to subtract off the Maxwell term so as to compare with the weak-field expansion of $\mathcal{L}_{\trm{int}}$ in \eqnref{eq:Lint_pert}. A comparison of Eqs.~\eqref{eq:Lint_pert} and \eqref{eqn:BIwf1} unveils that \eqnref{eqn:BIwf1} can be represented in the form of \eqnref{eq:Lint_pert} with the replacement $m^2/e^{2}\to b$. For the Born model we infer $c_1=1/2$, $c_2=0$ and for the Born-Infeld model $c_1=c_2=1/2$. Plugging these results into \eqnref{eq:dispers} (and accounting for the above replacements) we see that while the Born model exhibits vacuum birefringence there is no birefringence in the Born-Infeld model at quartic order in the field. However, it was shown in \cite{Davila:2013wba} that if the standard QED four-photon scattering vertex \emph{and} the tree-level four-photon Born-Infeld interaction are included in calculations, the interference leads to a \emph{modification} of birefringence, i.e. $n^{\parallel,\perp}_{\trm{total}} = n^{\parallel,\perp}_{\trm{QED}}+n_{\trm{BI}}$. Therefore, vacuum birefringence experiments can place a bound on BI theory. 
In general, laser-based light-by-light scattering experiments can be used to probe Born \cite{Kadlecova:2021bmp} and Born-Infeld \cite{Kadlecova:2021jqq} electrodynamics. It has been found, for example, that the polarised angular differential cross-section for scattering from scalar, spinor and supersymmetric QED and also BI electrodynamics shows discernible differences in the weak-field case~\cite{Rebhan:2017zdx}.

We remark that there is no vacuum birefringence in (unbroken) SUSY QED in a plane wave background, because the contribution of loop sfermions to helicity flip precisely cancels that of the fermions. The same is true in SUSY Yang Mills~\cite{Adamo:2021hno}. Interestingly, this holds  despite a background (gauge boson) plane wave not being itself a supersymmetric vacuum solution.

\begin{figure}[t!!]
    \centering
    \includegraphics[width=12cm]{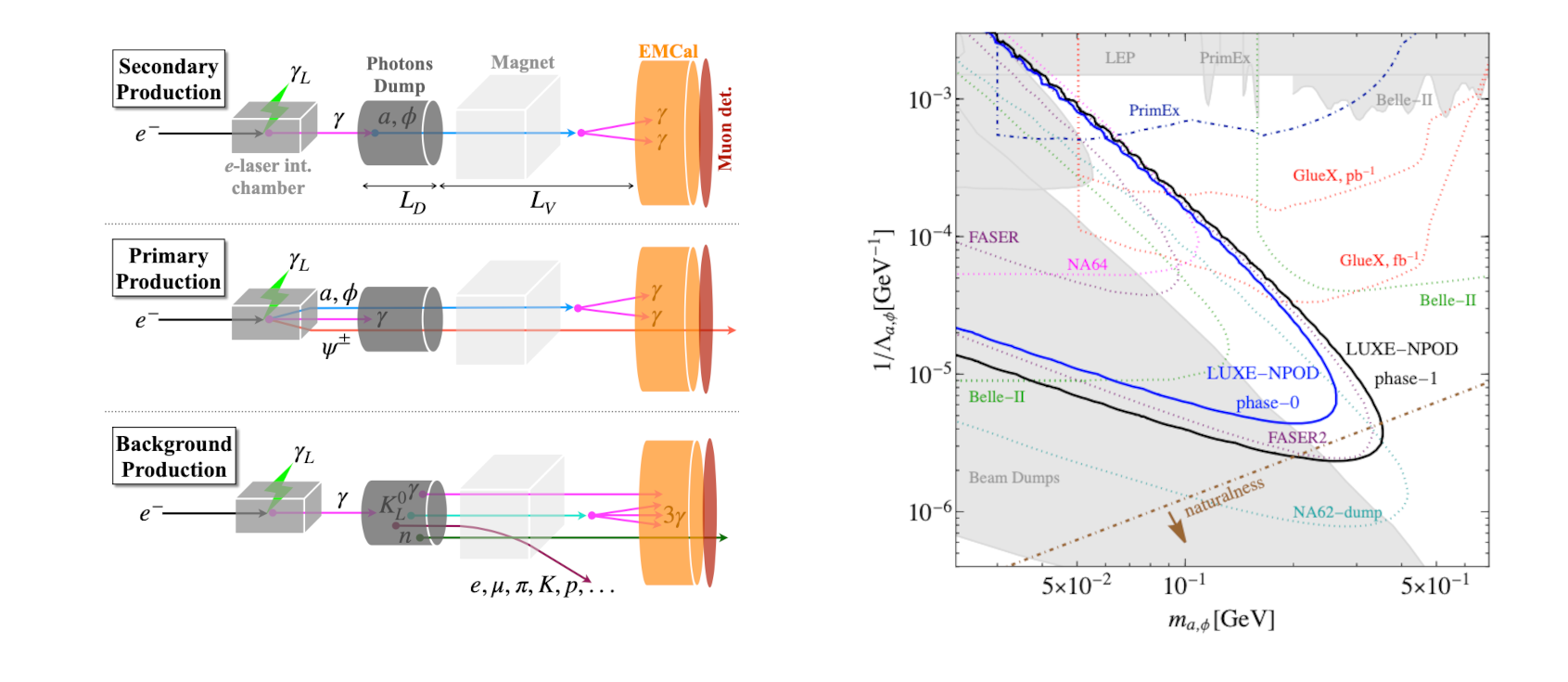}
    \caption{Left: LUXE-NPOD production mechanisms; Right: LUXE-NPOD exclusion region for stage 0 and stage 1 of the LUXE experiment. (Adapted from \cite{Bai:2021dgm}.)}
    \label{fig:luxenpod1}
\end{figure}

In addition to direct detection schemes, an intense EM field may also be used to generate a source of probe particles, which can then be employed in a BSM measurement. For example, the proposed LUXE-NPOD experiment \cite{Bai:2021dgm} would employ an electron-laser collision to generate a bunch of high energy photons which then hit a beam dump, with any ALPs escaping the dump and decaying to two photons on the other side being detected. This `secondary production' mechanism would be sensitive to ALP masses of the order $O(50-300)\,\trm{MeV}$ (see \figref{fig:luxenpod1}.)

\subsubsection{Lorentz-invariance violating scenarios}
{One possible source of physics beyond the Standard Model is the violation of Lorentz invariance. It has been proposed that intense lasers interacting with high-energy lepton beams could probe Lorentz-violating extensions of the Standard Model~\cite{Tizchang:2018mzr}, including noncommutativity in QED interactions~\cite{Tizchang:2016hml}, in which quantised spacetime coordinates obey $[x^\mu,x^\nu] = i\theta^{\mu\nu}$ for $\theta^{\mu\nu}$ a noncommutativity tensor.}

{Properties of photon propagation in noncommutative QED in the presence of constant crossed fields is explored in \cite{Fresneda:2015zya}. The investigations above are ``$\theta$-expanded'', meaning that $\theta$ is treated perturbatively. However, for the case of lightlike noncommutativity (where $\theta^{\LCp\LCperp}$ is the only nonzero component of the noncommutativity tensor), the Volkov solutions in a background plane wave can be found exactly, allowing a more precise treatment: see the review~\cite{Ilderton:2010rx}.}

Strong-field QED can itself be seen as an analog BSM theory, which admits violations of Lorentz invariance. In Very Special Relativity (VSR)~\cite{Cohen:2006ir}, the spacetime symmetry group of nature is taken to be the subgroup SIM(2) of the Lorentz group (plus translations). The resulting theory admits neutrino mass without adding new particles or interactions to the Standard Model~\cite{Cohen:2006ky}. Now consider QED in a circularly polarised, monochromatic plane wave. If the wave is of very high frequency, it will generate (in amplitudes and observables) terms which are rapidly oscillating (e.g.~$a_\mu(\phi)$) leading to cancellations, as well as terms which are slowly varying (e.g.~$a^2(\phi) = m^2\xi^2$, constant). Removing the former terms and carefully retaining all of the latter transforms strong-field QED into exactly QED in VSR~\cite{Ilderton:2016rqk}. In particular, the intensity-dependent `mass shift' in QED~\cite{Harvey:2012ie} becomes the physical mass in VSR: it is therefore VSR which realises, exactly, the old idea of the intensity-dependent mass shift as a genuine effective mass.

\section{Conclusions and open questions}\label{sec:conclusions}
We have reviewed work in the theory and phenomenology of strong-field QED over the past decade.
We conclude by summarising progress made, identifying important lessons learnt, and outlining some open problems and possible avenues of future research, for each of the topics covered in this review.

\subsection*{First order processes}
Tree-level ($\mathcal{O}(\alpha)$) processes in plane wave backgrounds have been studied extensively in the literature for many years; our understanding of nonlinear Compton scattering and nonlinear Breit-Wheeler pair production is very well developed. Many open theory questions in this area are really questions of how to go beyond plane wave backgrounds, beyond first-order processes, beyond tree-level, and so on. However, there are a few theory directions which seem to be worth pursuing, as well as linking the knowledge gained to real-world applications.

One possible direction stems from the fact that closed-form expressions for the probability of nonlinear Compton and nonlinear Breit-Wheeler in a finite pulse are still unknown, aside from the two limiting cases of infinitely long and infinitely short pulses. Analytical results {for even one plane wave pulse of finite duration} would be useful in understanding field shape effects, understanding strong field and high energy limits, benchmarking numerical results, and delineating the limitations of local approaches.

Another possible theory direction is related to the many applications of polarised beams of electrons and photons: in fundamental research, for example at upcoming colliders \cite{Ari:2015tca,Fujii:2018mli} and in applications such as studies of surface physics \cite{2016SurSR..71..547H}. In the intense background context, there is an interest in understanding how the shape of the field (e.g. laser pulse) and parameters of a probe beam can be tuned so as to optimise the desired properties of the scattered beam of photons, electrons and positrons. Detailed studies of spin and polarisation have only just begun and as attainable background field strength and typical colliding particle energy increase, there is still much room for improving understanding and applications.

\subsection*{Second order processes}

Work in recent years has lead to several new insights into tree-level processes at $\mathcal{O}(\alpha^2)$. We now have a much better understanding of not only how to approximate higher order processes with incoherent products of first-order processes, but from these $\mathcal{O}(\alpha^2)$ tree-level processes we also have a better idea of the size of the corrections. Incoherent-product approximations for higher orders including loops have also been developed. However, very little has been done beyond this approximation for loops. It would be worthwhile, for example, to study the probability of $e\to e\gamma$ to $\mathcal{O}(\alpha^2)$. Call this probability $\prob_{\rm C}^{(2)}$; it can be separated into two-step and one-step parts, where in the two-step part the photon is emitted in the first or second step and the other step contains a loop.
This is like in Compton scattering: the loop contributes an $\mathcal{O}(\alpha)$ Mueller matrix building block, given by the interference term between the $\mathcal{O}(\alpha^0)$ and $\mathcal{O}(\alpha)$ parts of the amplitude for scattering without emission. The two-step probability takes the form "${\bf M}_{\rm C}\cdot{\bf M}_{\rm L}+{\bf M}_{\rm L}\cdot{\bf M}_{\rm C}$".
It is already known how to calculate the two-step part. Calculating the one-step part of $\prob_{\rm C}^{(2)}$ would give us a more precise understanding of the size of corrections of $N$-step parts of general $\mathcal{O}(\alpha^N)$ processes and would allow us to study in more detail the cancellation of IR divergences from soft photons in degenerate processes (this process is degenerate to the emission of two photons with one of them soft, $e\to e\gamma\gamma$ at $\mathcal{O}(\alpha^2)$, which is IR divergent and, recall, can also be separated into two-step and one-step parts, where the two-step part is given by the incoherent product of two single-photon emission). Such calculations would, perhaps more interestingly, also allow us to check the contribution of $\prob_{\rm C}^{(2)}$ to RR. Including loops gives many more processes and quantities to consider, beyond the already considered $\mathcal{O}(\alpha^2)$ tree-level processes of trident, double Compton and photon trident. This can potentially provide many research problems for the next decade.

\subsection*{Approximation frameworks}

A general lesson from studying higher-order processes is that, when making approximations (e.g. the locally constant field approximation, LCFA), one should start with the general, all-order result and make approximations to that, before isolating the particular terms of interested. For example, use of the LCFA is normally in the long-pulse limit, where only the $n$-step contribution is included from $n$-vertex processes: in trident only the two-step part is included. In the long-pulse limit the two-step part of trident scales as $\tsf{R}_{\tsf{two}}\sim \xi^{2}\mathcal{T}^{2}$, and the (neglected) one-step part as $\tsf{R}_{\tsf{one}}\sim \xi T$. However, corrections to the LCFA two-step part scale as $\delta \tsf{R}_{\tsf{two}}\sim  \mathcal{T}^{2}$, so if $\xi \gtrsim \mathcal{T}$, it would be inconsistent to neglect the one-step term but include corrections to the LCFA for the two-step term.

As for future research directions, there are two main fronts on which progress can be made: i) improving the precision of existing modelling to meet the demands of improved experimental precision; ii) extending the localisation scheme, which has been used in the local constant crossed field approximation and local monochromatic approximation, to cover `new' effects.

i) One example of a higher-precision approximation is the semi-classical Baier-Katkov approach (see \secref{sec:approx:BKmethod}) to nonlinear Compton scattering. This is exact in a plane wave and therefore for this process, superior in accuracy to the local approach. When $\xi^{2}/\gamma \ll 1$, the Baier-Katkov approach can even be applied to focussed backgrounds. Can the formalism be developed to deal with higher orders, pairs and cascades? And does it maintain an advantage in accuracy when extended in this way? In terms of local approaches, the local monochromatic approximation (LMA) misses pulse bandwidth effects such as harmonic broadening in nonlinear Compton and nonlinear Breit-Wheeler but is straightforward to adapt to higher orders. Can particle polarisation be added to numerical codes, which can be significant when $\xi\sim O(1)$ where the LMA differs substantially from the standard approach of using the locally constant field approximation? (Including polarisation for the trident process using a `gluing' approach has already been demonstrated for the LMA in \cite{Torgrimsson:2020gws}, and polarisation effects were included in the numerical code, CAIN \cite{cain1}, for a locally-monochromatic linearly-polarised background.)

ii) Local approximation schemes are essentially used in the ``long-pulse-limit'' and so only the ``n-step'' (fully incoherent) contribution is included from n-th order tree-level process. However, for certain choices of scattered angle, production from the coherent process can in principle be more probable than the fully incoherent process. Can, therefore, coherent contributions to higher-order processes be included in the local approximation framework? Another possible direction is in density-dependent processes. So far, it has mainly been $1\to2$ processes that have been included in simulations (but see recent results \cite{Blackburn:2020fqo} including the $2\to1$ process of photon absorption). However, in systems where the particle density is sufficiently high, $2\to 2$ and $2\to1$ absorption processes, can play important roles. Can these be included in the approximation programme?

\subsection*{Higher-order processes and resummation}

In general, one must calculate consistently at each perturbative order. For example, if one is looking at nonlinear Compton scattering, one may suppose it is sufficient to calculate the tree-level diagram. Eventually though, {as higher field intensities and pulse duration are used} the associated probability and cross section will exceed unitarity bounds.  This is \emph{not} a sign that strong-field QED is broken, nor are any ad-hoc prescriptions or normalisations needed. Rather, it is a sign that for the chosen parameters one must consider higher-order effects (be they UV (loop corrections) or IR (inclusive corrections) or both), and resummation. 

A new method of resumming the $\alpha$ expansion for radiation reaction quantities has recently been developed for plane-wave backgrounds, see Sec.~\ref{sec:higher}, where the $\alpha$ expansion is calculated by gluing together strong-field QED ``Mueller matrices'', see Sec.~\ref{sec:second}. There are still many applications and generalisations to work out, even within this class of fields. Looking to the future, an obvious goal is to generalise the incoherent-product approximation, with its products of Mueller matrices, to more general space-time dependent fields. One aspect that simplifies such approximations in plane waves is the fact that only the lightfront longitudinal components of the momenta play a nontrivial role when gluing together Mueller matrices, i.e.~the transverse momentum components can all be integrated  for each $\mathcal{O}(\alpha)$ step separately. As lightfront time gives the only nontrivial space-time dependence, one only has to deal with two variables for each step/Mueller matrix.  
For more general fields one can expect more than one momentum component to be important for the gluing process, so that even if one considers a regime in which the probabilities can in principle be approximated with some $N$-step parts, it might still be significantly more complicated than the plane wave case. 
However, it is often well-motivated to consider high-energy particles, for which one can expect significant simplifications since in this limit a (quite) general field behaves essentially like a plane wave, and the solutions to the corresponding Dirac equation look very much like Volkov solutions, see~\cite{DiPiazza:2015xva} and Sec.~\ref{sec:beyondPW}. While this does not make the problem trivial, one can hope to make progress on finding such a generalisation in the coming decade.

\subsection*{Light-by-light scattering}

In the first part of Sec.~\ref{sec:LBL} we focused on the Heisenberg-Euler effective Lagrangian ${\cal L}_{\rm HE}$, highlighted the importance of one-particle reducible contributions, studied the strong field limit of ${\cal L}_{\rm HE}$ and demonstrated that a perturbative loop expansion of ${\cal L}_{\rm HE}$ breaks down for exponentially large fields. It would certainly be very interesting and important to advance the study of higher loop orders of ${\cal L}_{\rm HE}$ from lower \cite{Huet:2018ksz} to $3+1$ space-time dimensions, particularly by using resummation and resurgence techniques \cite{Dunne:2021acr}. Another topical research direction is to put forward strategies to go beyond the loop expansion~\cite{Karbstein:2021gdi} and so obtain insights into the manifestly non-perturbative parameter regime, where such an expansion no longer makes sense.

An important lesson learnt in this context is that a larger Feynman diagram containing a sub-diagram which, considered as an isolated object, vanishes identically because of an overall momentum conserving delta function can still yield a finite result; cf. \cite{Gies:2016yaa,Karbstein:2017gsb,Ahmadiniaz:2019nhk,DiPiazza:2022lij}.
This can be illustrated by the following schematic example: while $g(k)\sim k\delta(k)$ clearly vanishes, upon sewing it to another (finite) contribution $h(k)\sim k$ with a 'propagator' $\sim1/k^2$ which is non-regular at $k=0$, and integrating over $k$, we obtain $\int_k g(k)f(k)/k^2\sim\int_k\delta(k)\neq0$.
Therefore, one should never set such a superficially vanishing contribution to a larger diagram to zero from the outset, but rather keep the full expression and only apply simplifications on the level of the final expression to be calculated.
This subtlety was precisely the reason why, e.g., one-particle reducible contributions to ${\cal L}_{\rm HE}$ were previously erroneously assumed to vanish.

Moreover, we detailed that analytical results for the photon polarization tensor are presently only available in homogeneous constant and generic plane-wave backgrounds.
Remarkably, even in a constant field the one-particle irreducible contribution to the photon-polarization tensor at two loops and arbitrary momentum transfer has not been evaluated to date. Together with the one-particle reducible contribution determined in~\cite{Karbstein:2017gsb}, this would constitute the full result for the two-loop photon polarization tensor. The result for the one-particle irreducible contribution would in particular allow one to assess the relative importance of the irreducible and reducible contributions in the strong field limit for large moment transfers.
Apart from this, it would be worthwhile to go beyond constant and plane wave backgrounds and evaluate the (one-loop) photon polarization tensor in a localized inhomogeneous field. This would for instance allow for the study of vacuum diffraction and quantum reflection phenomena generic to inhomogeneous fields from first principles without needing to employ a slowly varying field approximation.

Finally, to actively assist the discovery of light-by-light scattering phenomena in high-intensity laser experiments, in the upcoming years it will definitely be very important to refine various theoretical estimations of prospective photonic quantum vacuum signals to account for the full details of the actual discovery experiment. This requires an accurate modelling of the precise field configurations available in experiment as well as accounting for any real-world complications such as the inevitable presence of shot-to-shot fluctuations and jitter \cite{Schlenvoigt:2016jrd}. The combination of the vacuum emission picture with a Maxwell solver \cite{Blinne:2018nbd,Sainte-Marie:2022efx} as well as complementary approaches based on a direct numerical solution of the non-linear wave equation in $3+1$ dimensions \cite{domenech2017implicit,grismayer2021,Lindner:2021krv} should allow tackling this challenge.

\subsection*{The Schwinger effect and spontaneous pair production}

Our understanding of field inhomogeneities in the Schwinger effect is essentially limited to one-dimensional cases.  Complete analytical frameworks have not yet been established to go beyond this, though there are attempts based on semi-classical methods and Furry picture expansions, see Sec.~\ref{sec:ht2}.  Numerical simulations also become resource-heavy in the presence of multi-dimensional field inhomogeneities.  Magnetic-field effects become important for multi-dimensional inhomogeneities, and can induce non-trivial phenomena such as Stern-Gerlach forces, spin-current generation, chiral-magnetic effects, anomaly-induced dynamical refringence, etc (see Secs.~\ref{sec:ht4} and \ref{sec:ht5}), all of which require further investigation.

Field inhomogeneities may also significantly affect radiative corrections to the Schwinger effect, a topic in which there remains a great deal to be understand.  Even for the simplest case of a constant electric field, for example, only the two-loop result is available, see Sec.~\ref{sec:ht6}.  The exponentiation conjecture still remains an open question.  It is also interesting that the holographic approach predicts an upper limit for electric fields in the super-critical regime, which is worthwhile to be tested with field-theoretical calculations.   

Photon emission due to radiative corrections is also an interesting subject, as it may provide characteristic experimental signatures such as high-harmonic generation but still requires further theoretical investigation.

There is no established realtime framework for the Schwinger effect that goes beyond the mean-field approximation.  Going beyond this is crucially important for the discussion of equilibriation processes and possible QED cascades.  It is predicted that there exist universal behaviours in far-from-equilibrium quantum processes initiated by overpopulated gauge fields, see e.g.~the review~\cite{Berges:2020fwq}, and it would be interesting to make connections between those processes and the Schwinger effect.

It is envisaged that available laser strengths may reach a few orders of magnitude below the Schwinger field 
{$eE_S = 10^{16}\;{\rm V/cm}$ (corresponding to intensity $10^{29}\;{\rm W/cm}^2$)} in the near future.  It is therefore timely to discuss possible experimental signatures; for laser experiments, it is important to continue investigation into enhancement effects including pulse shape optimisation and/or multiple colliding pulse setups~\cite{Bulanov:2010ei,Gonoskov:2013ada,Torgrimsson:2016ant,Magnusson:2019qop}.

Strong fields can also be realised in other systems, e.g, heavy-ion collisions, which may offer an opportunity to explore the regime of super strong-fields beyond the Schwinger field $E\gg E_{\rm S}$.  This implies that the Schwinger effect may be testable via electromagnetic probes in various collision geometries, but because the associated spacetime volume of the strong fields generated is very small~\cite{Bzdak:2011yy,Deng:2012pc,Hattori:2016emy}, care must be taken to properly account for the impact of finite volume effects which affect the non-perturbativity of the pair creation process.  This, along with quark/gluon production by strong colour flux tubes, remain open questions, but may be analysed in a quantitative manner using the developments reviewed here.

\subsection*{The Ritus-Narozhny Conjecture}

The Ritus-Narozhny (RN) conjecture applies to those backgrounds and parameter regimes where the locally constant field approximation (LCFA) holds. Therefore it is required to conceive of a scenario where the high-$\chi$ region can be accessed experimentally, whilst the LCFA remains valid. This is an active area of research and several suggestions are reviewed in \secref{sec:RN:feasibility}.

However, on the theoretical side the main challenge is to specify what to resum, and to perform that resummation, in the high-$\chi$ region where the conjecture holds. In spite of a certain progress in understanding the nature of bubble chain corrections and the reasons for their enhancement in a CCF, the overall understanding of the possible non-perturbative regime of QED at $\alpha\chi^{2/3}\gtrsim1$ has not been yet achieved. It is  likely that the most important feature of this regime, yet to be accounted for properly, is the severe "instability" of photons and electrons in a strong CCF, with respect to nonlinear Breit-Wheeler and nonlinear Compton scattering. A systematic route to future progress in understanding this regime (including better understanding of the role of vertex corrections) might be based on a more rigorous selection of diagrams with potentially leading scaling for resummation, possibly starting with a general form of the Dyson-Schwinger equations. Examples of successful application of these equations in similar problems include magnetic catalysis of spontaneous chiral symmetry breaking \cite{Gusynin:1999dynamical} and development of non-perturbative results using $1/N_f$ expansion \cite{Coquereaux:1981fermionic,Palanques:1984The} (for application of the latter approach to all-order resummation of reducible contributions to the Heisenberg-Euler effective action see Ref.~\cite{Karbstein:2021gdi}). As a necessary prerequisite for formulating DS equations explicitly in a CCF, a general form of off-shell tensor and gamma-matrix structure of the mass and vacuum polarisation corrections was discussed in \cite{Mironov:2021structure}. A systematic approach to resummation can be useful also in understanding possible interrelations between higher order corrections and backreaction, that were observed in the Jaynes-Cummings model \cite{Ekman:2020vsc}.

It has recently been suggested \cite{Heinzl:2021mji} that the classical domain contributes to the RN scaling of $\chi^{2/3}$, since it arises in the electron radiation spectrum in the form $(\chi/s)^{2/3}$ (where $s\propto \hbar$ is the lightfront momentum fraction), which is independent of the $\hbar \to 0$ limit. Further work may investigate the link between the classical limit, and the RN scaling.

It was also stressed in \cite{Edwards:2020npu,Heinzl:2021mji} that it can be relevant to consider inclusive observables and degenerate processes, as these modify the probability rates in strong fields. An example is the cumulative, damping, impact of IR and background-collinear photon emissions on hard-photon emission rates. Therefore, in order to judge the feasibility of particular experimental proposals, it may not be enough to focus on specific sets of diagrams (chosen for e.g.~simplicity or accessibility) or exclusive observables. This is a topic requiring further investigation.

\subsection*{Beyond the plane wave and background field approximations}
The advances made in both approximate and exact methods for solving the Dirac equation in background fields should be systematically exploited to explore physics beyond the plane wave model. Even considerations which only go a step beyond plane waves (rather than immediately to fully realistic beams) are highly worthwhile if they reveal phenomenological effects missed by the plane wave model. The extension of inverse and superintegrable methods to problems of radiation reaction, even in the classical theory, would be interesting to pursue. {See also~\cite{Bialynicki-Birula:2021yvi} for recent results on generating whole families of solutions to relativistic wave equations using spinorial methods and fractional Fourier transforms.} Much work also remains to be done on the extension of WKB methods (including the Stokes phenomenon) to multi-dimensional inhomogeneities. 

It would be very useful to have a systematic, analytic, and accessible approach to back-reaction on background fields (e.g.~going beyond the mean-field approximation)-- ideally, one would be able to build on the methods and results of the Furry picture, rather than abandon the progress made. Perhaps though, a practical approach to back-reaction will require very different methods instead.

\subsection*{Beyond QED}
It would be intriguing to explore how more of the strong-field methods from QED could be brought to bare on heavy-ion physics and the colour glass condensate, where strong classical fields play a crucial role. Developing methods for going beyond the mean-field approximation is for example important for the thermalisation/hydrodynamisation of the quark-gluon plasma in heavy-ion collisions.

As gravitational wave astronomy becomes an ever more well-established field, new approaches will be needed for calculations of observables in gravity, calculations which are extremely challenging even classically. Motivated by this, recent years have seen the development of efficient methods for extracting the classical limits of observables from quantum scattering amplitudes \emph{without} having to calculate the full amplitude itself~\cite{Kosower:2018adc}. There is clearly scope for adapting such methods to strong-field QED.  First-quantised and worldline approaches have very recently been applied to double copy~\cite{Ahmadiniaz:2021fey}. In gravity, such methods can greatly simplify the calculation of graviton scattering amplitudes compared to standard diagrammatic approaches, see e.g.~\cite{Edwards:2022qiw}; there remains a great deal of potential, and work to be done, in the extension of these methods to include background gravitational fields.

In `new physics' particle searches, weak magnetic fields $B$ are often combined with high-finesse cavities to achieve a long interaction length $L$ and hence increase the production probability by increasing the product $BL$. The equivalent quantity in the interaction of a probe with a weak laser pulse of intensity parameter $\xi$ and frequency $\omega$ is $\xi \omega L$. Using strong fields introduces nonlinearities, which compared to this perturbative scaling, tend to \emph{decrease} the probability. It remains an open question as to how other properties of intense fields phenomenology might be used to enhance signals of new physics. For example in \cite{Dillon:2018ouq} it was suggested to use coherent production of scalars on an electron beam colliding with a laser pulse, which drastically increases the yield of scalars from $\sim N_{e}$ to $\sim N_{e}^{2}$. This mechanism requires that the electron bunch length is around the same order of magnitude as the wavelength of emitted particle. Therefore, for higher energies, as would be required for pseudoscalars, other methods would be required to boost the signal. Depending on which part of parameter space should be probed, an alternative way forward may be to use QED in intense backgrounds to provide the probe, which can then be used to test for new physics in standard ways, such as in \cite{Bai:2021dgm}.

\begin{center}
    ------------------------------------------
\end{center}

The topics reviewed here lie at the confluence of high-power laser, particle, and plasma physics, and of theory, simulation, and experiment. As such they are relevant to the intensity frontier of particle physics, to future colliders such as the International Linear Collider (ILC) \cite{Behnke:2013xla, Baer:2013cma} and Compact LInear Collider (CLIC) \cite{Aicheler:2012bya,Roloff:2018dqu} through strong field effects at the interaction point {(studied by, e.g. the Advanced LinEar collider study GROup (ALEGRO) \cite{Cros:2019tns})}, and also to matter exposed to extreme electromagnetic fields, such as in exotic astrophysical objects \cite{Kaspi:2017fwg} as tested using high intensity lasers \cite{2016RPPh...79d6901M}.

The past decade has seen huge progress in the understanding of the `traditional' processes of strong-field QED, such as nonlinear Compton and Breit-Wheeler, the Schwinger effect, and vacuum birefringence, while access to higher-order processes, such as trident and double nonlinear Compton, has been made possible by a combination of novel methods and appropriate approximations. More recently, calculations have been pushed to higher loop orders, and in some cases to all orders, motivated by results on the very high intensity behaviour of QED in strong fields, and by results on resurgence in quantum field theory. Connections to non-Abelian gauge theories and gravity have also begun to be explored, and there is scope here for a fruitful exchange of ideas and methods.

This review has concentrated on the last decade of progress. At the beginning of this period, the state-of-the-art experimental result for testing QED in intense laser fields, was still dominated by the landmark E144 \cite{E144:1996enr,Burke:1997ew} experiment from the mid-1990s. 
In the last decade, several multi-PW lasers have come or are coming online, that will be able to reach intensities of the order of $10^{23}\,\trm{Wcm}^{-2}$ and higher (the E144 experiment operated with a peak intensity of $0.5\times 10^{18}\,\trm{Wcm}^{-2}$).
The increase in the number of papers published in the last 10 years (Fig.~\ref{FIG:ABSPLOT}), has likely been driven in part by recent experiments that have probed the edge of the nonlinear quantum regime \cite{Cole:2017zca, Poder:2017dpw} and in part by upcoming experiments that will use more energetic probes and higher field strengths. These future experiments will be able to probe the nonlinear quantum regime in depth, and do so at a higher precision. It is a reasonable expectation that ten years from now, there will be papers that feature comparisons of experimental data with the theoretical models and predictions reviewed here.

\section*{Acknowledgments}
We thank Tom Heinzl, Holger Gies, Arseny Mironov and Suo Tang for many useful comments and discussions. The authors are supported by the MEPhI Program Priority 2030 [AF], the Russian Foundation for Basic Research, grant No. 20-52-12046 [AF], the STFC consolidated grant ST/X000494/1 [AI], the Deutsche Forschungsgemeinschaft (DFG, German Research Foundation), grant No.416607684 within the Research Unit FOR2783/1 [FK], 
the Deutsche Forschungsgemeinschaft (DFG, German Research Foundation) under Germany’s Excellence Strategy – EXC 2121 ``Quantum Universe'' – 390833306 [BK],
the RIKEN special postdoctoral researcher program [HT], the Nonequilibrium working group at RIKEN Interdisciplinary Theoretical and Mathematical Sciences [HT], and the Swedish Research Council, contract 2020-04327 [GT].

\clearpage

\section{Conventions, commonly used symbols and abbreviations}
\label{sec:conventions}

\begin{center}
\fbox{
    \centering
    \begin{tabular}{c|l|l}
    $(+,-,-,-)$ & metric convention \\
    $\epsilon_0 = c = \hbar = 1$ & ($\hbar$ may be reinstated when needed) \\
    $x\cdot y$, $\mbf{x}\cdot\mbf{y}$ & inner product (all-plus metric for bold symbols) \\
    $m$ & electron (positron) mass \\
    $\alpha = e^2/(4\pi) = 1/137$ & fine structure constant (at scale $m$)\\
    $\tau_0 = 2\alpha/(3m)$ & radiation reaction parameter & Sec.~\ref{sec:higher:classicalRR}, Eq.~(\ref{ABinLAD}) and (\ref{eq:allRR}) \\
    $\xi$     & classical nonlinearity parameter & Sec.~\ref{sec:intro:experiments} \\
    $\eta$     & quantum energy parameter & Sec.~\ref{sec:intro:experiments} \\
    $\chi$     & quantum nonlinearity parameter & Sec.~\ref{sec:intro:experiments}, Eq.~(\ref{eq:chi:def})\\
    $k_\mu=\omega(1,0,0,1)$     & laser (plane-wave) wavevector & Sec.~\ref{sec:intro:PW}\\
$\varphi = k\cdot x$ & laser (plane-wave) phase &  Sec.~\ref{sec:intro:PW}\\
    $k\cdot x = \omega (t+z) = \omega x^+$ & phase in terms of lightfront time $x^\LCp$ & Sec.~\ref{sec:intro:PW}, Eq.~(\ref{eq:lightfront-conventions})\\
        $\Phi$, $\mathcal{T}$ & pulse duration \\
    $\ell$     & photon momentum \\
    $p$, $q$ & fermion momenta \\
    $\ud{\tilde p}$ & invariant measure for on-shell momentum, $p$ & Sec.~\ref{sec:first_order_general_props}, Eq.~(\ref{eq:onshell:measure}) \\
    $u_p$, $v_p$ & free electron/positron spinors & Sec.~\ref{sec:prelim}, Eq.~(\ref{volkov-e-in}) \\ 
    $s$, $s_\gamma$ & lightfront momentum fraction & Sec.~\ref{sec:first_order_general_props}, Eq.~(\ref{eq:first:srperp}) \& Sec.~\ref{Spin sums in the two-step}, Eq.~(\ref{MCdefinition})\\
    $\phi = (\varphi+\varphi')/2$ & average phase & Sec.~\ref{sec:first_order_general_props}, Eq.~(\ref{eqn:phidefs1}) \\
    $\theta = \varphi-\varphi'$ & interference phase & Sec.~\ref{sec:first_order_general_props}, Eq.~(\ref{eqn:phidefs1})\\
    $i0$ & poles \\
    $\mathcal{A}$, $\mathcal{F}$ & classical/background/macroscopic fields & Sec.~\ref{sec:intro:Furry}\\
    $A$, $F$ & quantum fields & Sec.~\ref{sec:intro:Furry}\\
    $\mathfrak{a}$, $\mathfrak{f}$ & probe fields & Sec.~\ref{sec:LBL:probeprop}\\
    $\mathcal{S}$, $\mathcal{P}$ & field invariants & Sec.~\ref{sec:intro:Furry}, Eq.~(\ref{eq:invariants}) \\
    $\mathfrak{E}$, $\mathfrak{B}$ & secular invariants & Sec.~\ref{sec:approx:lcfa}, Eq.~(\ref{eqn:approx:sec}) \\
    $\psi$ & fermion wavefunction \\
    $\phi$ & scalar wavefunction \\
    $\Theta$ & Heaviside step function & \\[2pt]
    \hline
    \hline
 &&\\[-5pt]
    1PI / 1PR & One-particle irreducible / reducible\\
    CCF & constant crossed field & Sec.~\ref{sec:prelim:plane:quantum}\\
    LAD & Lorentz-Abraham Dirac (equation) & Sec.~\ref{sec:higher:classicalRR}\\
    LCFA & locally constant field approximation & Sec.~\ref{sec:approx:lcfa}\\
    LL & Landau-Lifshitz (equation) & Sec.~\ref{sec:higher:classicalRR}\\
LMA & locally monochromatic approximation & Sec.~\ref{sec:approx:lma} \\
    NBC & nonlinear Breit-Wheeler (pair production) & Sec.~\ref{sec:first:nbw}\\
    NLC & nonlinear Compton scattering & Sec.~\ref{sec:first:nlc} \\
    PW & plane wave \\
    RN & Ritus-Narozhny (conjecture) & Sec.~\ref{sec:RN}\\
    RR & radiation reaction & Sec.~\ref{sec:higher:classicalRR}
    \end{tabular}
    \label{tab:defs}
}
\end{center}

\bibliography{ReviewBib}

\begin{thebibliography}{1000}
\expandafter\ifx\csname url\endcsname\relax
  \def\url#1{\texttt{#1}}\fi
\expandafter\ifx\csname urlprefix\endcsname\relax\def\urlprefix{URL }\fi
\expandafter\ifx\csname href\endcsname\relax
  \def\href#1#2{#2} \def\path#1{#1}\fi

\bibitem{Strickland85}
D.~Strickland, G.~Mourou, Compression of amplified chirped optical pulses, Opt.
  Commun (1985) 219--221.

\bibitem{nobellink}
2018 nobel prize in physics,
  \url{https://www.nobelprize.org/prizes/physics/2018/summary/}.

\bibitem{Proceedings:2012ulb}
{Fundamental Physics at the Intensity Frontier}.
\newblock \href {http://arxiv.org/abs/1205.2671} {\path{arXiv:1205.2671}},
  \href {https://doi.org/10.2172/1042577} {\path{doi:10.2172/1042577}}.

\bibitem{Sturm:2011uow}
S.~Sturm, A.~Wagner, B.~Schabinger, J.~Zatorski, Z.~Harman, W.~Quint, G.~Werth,
  C.~H. Keitel, K.~Blaum, {g Factor of Hydrogenlike Si13+28}, Phys. Rev. Lett.
  107~(2) (2011) 023002.
\newblock \href {https://doi.org/10.1103/PhysRevLett.107.023002}
  {\path{doi:10.1103/PhysRevLett.107.023002}}.

\bibitem{Sailer:2022azt}
T.~Sailer, et~al., {Measurement of the bound-electron $g$-factor difference in
  coupled ions}, Nature 606~(7914) (2022) 479--483.
\newblock \href {http://arxiv.org/abs/2204.12182} {\path{arXiv:2204.12182}},
  \href {https://doi.org/10.1038/s41586-022-04807-w}
  {\path{doi:10.1038/s41586-022-04807-w}}.

\bibitem{2015JPCRD..44c1205S}
V.~M. {Shabaev}, D.~A. {Glazov}, G.~{Plunien}, A.~V. {Volotka}, {Theory of
  Bound-Electron g Factor in Highly Charged Ions *}, Journal of Physical and
  Chemical Reference Data 44~(3) (2015) 031205.
\newblock \href {http://arxiv.org/abs/1508.00392} {\path{arXiv:1508.00392}},
  \href {https://doi.org/10.1063/1.4921299} {\path{doi:10.1063/1.4921299}}.

\bibitem{Yerokhin:2017sfg}
V.~A. Yerokhin, Z.~Harman, {One-loop electron self-energy for the
  bound-electron g factor}, Phys. Rev. A 95~(6) (2017) 060501.
\newblock \href {http://arxiv.org/abs/1704.08080} {\path{arXiv:1704.08080}},
  \href {https://doi.org/10.1103/PhysRevA.95.060501}
  {\path{doi:10.1103/PhysRevA.95.060501}}.

\bibitem{Hanneke:2010au}
D.~Hanneke, S.~F. Hoogerheide, G.~Gabrielse, {Cavity Control of a
  Single-Electron Quantum Cyclotron: Measuring the Electron Magnetic Moment},
  Phys. Rev. A 83 (2011) 052122.
\newblock \href {http://arxiv.org/abs/1009.4831} {\path{arXiv:1009.4831}},
  \href {https://doi.org/10.1103/PhysRevA.83.052122}
  {\path{doi:10.1103/PhysRevA.83.052122}}.

\bibitem{Aoyama:2017uqe}
T.~Aoyama, T.~Kinoshita, M.~Nio, {Revised and Improved Value of the QED
  Tenth-Order Electron Anomalous Magnetic Moment}, Phys. Rev. D 97~(3) (2018)
  036001.
\newblock \href {http://arxiv.org/abs/1712.06060} {\path{arXiv:1712.06060}},
  \href {https://doi.org/10.1103/PhysRevD.97.036001}
  {\path{doi:10.1103/PhysRevD.97.036001}}.

\bibitem{ATLAS:2017fur}
M.~Aaboud, et~al., {Evidence for light-by-light scattering in heavy-ion
  collisions with the ATLAS detector at the LHC}, Nature Phys. 13~(9) (2017)
  852--858.
\newblock \href {http://arxiv.org/abs/1702.01625} {\path{arXiv:1702.01625}},
  \href {https://doi.org/10.1038/nphys4208} {\path{doi:10.1038/nphys4208}}.

\bibitem{ATLAS:2019azn}
G.~Aad, et~al., {Observation of light-by-light scattering in ultraperipheral
  Pb+Pb collisions with the ATLAS detector}, Phys. Rev. Lett. 123~(5) (2019)
  052001.
\newblock \href {http://arxiv.org/abs/1904.03536} {\path{arXiv:1904.03536}},
  \href {https://doi.org/10.1103/PhysRevLett.123.052001}
  {\path{doi:10.1103/PhysRevLett.123.052001}}.

\bibitem{CMS:2018erd}
A.~M. Sirunyan, et~al., {Evidence for light-by-light scattering and searches
  for axion-like particles in ultraperipheral PbPb collisions at
  $\sqrt{s_\mathrm{NN}} =$ 5.02 TeV}, Phys. Lett. B 797 (2019) 134826.
\newblock \href {http://arxiv.org/abs/1810.04602} {\path{arXiv:1810.04602}},
  \href {https://doi.org/10.1016/j.physletb.2019.134826}
  {\path{doi:10.1016/j.physletb.2019.134826}}.

\bibitem{STAR:2019wlg}
J.~Adam, et~al., {Measurement of $e^+e^-$ Momentum and Angular Distributions
  from Linearly Polarized Photon Collisions}, Phys. Rev. Lett. 127~(5) (2021)
  052302.
\newblock \href {http://arxiv.org/abs/1910.12400} {\path{arXiv:1910.12400}},
  \href {https://doi.org/10.1103/PhysRevLett.127.052302}
  {\path{doi:10.1103/PhysRevLett.127.052302}}.

\bibitem{dEnterria:2013zqi}
D.~d'Enterria, G.~G. da~Silveira, {Observing light-by-light scattering at the
  Large Hadron Collider}, Phys. Rev. Lett. 111 (2013) 080405, [Erratum:
  Phys.Rev.Lett. 116, 129901 (2016)].
\newblock \href {http://arxiv.org/abs/1305.7142} {\path{arXiv:1305.7142}},
  \href {https://doi.org/10.1103/PhysRevLett.111.080405}
  {\path{doi:10.1103/PhysRevLett.111.080405}}.

\bibitem{Li:2019yzy}
C.~Li, J.~Zhou, Y.-J. Zhou, {Probing the linear polarization of photons in
  ultraperipheral heavy ion collisions}, Phys. Lett. B 795 (2019) 576--580.
\newblock \href {http://arxiv.org/abs/1903.10084} {\path{arXiv:1903.10084}},
  \href {https://doi.org/10.1016/j.physletb.2019.07.005}
  {\path{doi:10.1016/j.physletb.2019.07.005}}.

\bibitem{Heisenberg:1936nmg}
W.~Heisenberg, H.~Euler, {Consequences of Dirac's theory of positrons}, Z.
  Phys. 98~(11-12) (1936) 714--732.
\newblock \href {http://arxiv.org/abs/physics/0605038}
  {\path{arXiv:physics/0605038}}, \href {https://doi.org/10.1007/BF01343663}
  {\path{doi:10.1007/BF01343663}}.

\bibitem{Schwinger:1951nm}
J.~S. Schwinger, {On gauge invariance and vacuum polarization}, Phys. Rev. 82
  (1951) 664--679.
\newblock \href {https://doi.org/10.1103/PhysRev.82.664}
  {\path{doi:10.1103/PhysRev.82.664}}.

\bibitem{Toll:1952rq}
J.~S. Toll, {The Dispersion relation for light and its application to problems
  involving electron pairs}, Other thesis (1952).

\bibitem{Sauter:1931zz}
F.~Sauter, {Uber das Verhalten eines Elektrons im homogenen elektrischen Feld
  nach der relativistischen Theorie Diracs}, Z. Phys. 69 (1931) 742--764.
\newblock \href {https://doi.org/10.1007/BF01339461}
  {\path{doi:10.1007/BF01339461}}.

\bibitem{Bulanov:2010ei}
S.~S. Bulanov, V.~D. Mur, N.~B. Narozhny, J.~Nees, V.~S. Popov, {Multiple
  colliding electromagnetic pulses: a way to lower the threshold of $e^+e^-$
  pair production from vacuum}, Phys. Rev. Lett. 104 (2010) 220404.
\newblock \href {http://arxiv.org/abs/1003.2623} {\path{arXiv:1003.2623}},
  \href {https://doi.org/10.1103/PhysRevLett.104.220404}
  {\path{doi:10.1103/PhysRevLett.104.220404}}.

\bibitem{Tanji:2008ku}
N.~Tanji, {Dynamical view of pair creation in uniform electric and magnetic
  fields}, Annals Phys. 324 (2009) 1691--1736.
\newblock \href {http://arxiv.org/abs/0810.4429} {\path{arXiv:0810.4429}},
  \href {https://doi.org/10.1016/j.aop.2009.03.012}
  {\path{doi:10.1016/j.aop.2009.03.012}}.

\bibitem{Dabrowski:2014ica}
R.~Dabrowski, G.~V. Dunne, {Superadiabatic particle number in Schwinger and de
  Sitter particle production}, Phys. Rev. D 90~(2) (2014) 025021.
\newblock \href {http://arxiv.org/abs/1405.0302} {\path{arXiv:1405.0302}},
  \href {https://doi.org/10.1103/PhysRevD.90.025021}
  {\path{doi:10.1103/PhysRevD.90.025021}}.

\bibitem{Dabrowski:2016tsx}
R.~Dabrowski, G.~V. Dunne, {Time dependence of adiabatic particle number},
  Phys. Rev. D 94~(6) (2016) 065005.
\newblock \href {http://arxiv.org/abs/1606.00902} {\path{arXiv:1606.00902}},
  \href {https://doi.org/10.1103/PhysRevD.94.065005}
  {\path{doi:10.1103/PhysRevD.94.065005}}.

\bibitem{Landsman:2014auv}
A.~S. Landsman, M.~Weger, J.~Maurer, R.~Boge, A.~Ludwig, S.~Heuser, C.~Cirelli,
  L.~Gallmann, U.~Keller, {Ultrafast resolution of tunneling delay time},
  Optica 1~(5) (2014) 343.
\newblock \href {https://doi.org/10.1364/optica.1.000343}
  {\path{doi:10.1364/optica.1.000343}}.

\bibitem{Parker:2012at}
L.~Parker, {Particle creation and particle number in an expanding universe}, J.
  Phys. A 45 (2012) 374023.
\newblock \href {http://arxiv.org/abs/1205.5616} {\path{arXiv:1205.5616}},
  \href {https://doi.org/10.1088/1751-8113/45/37/374023}
  {\path{doi:10.1088/1751-8113/45/37/374023}}.

\bibitem{Birrell:1982ix}
N.~D. Birrell, P.~C.~W. Davies, {Quantum Fields in Curved Space}, Cambridge
  Monographs on Mathematical Physics, Cambridge Univ. Press, Cambridge, UK,
  1984.
\newblock \href {https://doi.org/10.1017/CBO9780511622632}
  {\path{doi:10.1017/CBO9780511622632}}.

\bibitem{1962JMP.....3...59R}
H.~R. {Reiss}, {Absorption of Light by Light}, Journal of Mathematical Physics
  3~(1) (1962) 59--67.
\newblock \href {https://doi.org/10.1063/1.1703787}
  {\path{doi:10.1063/1.1703787}}.

\bibitem{Nikishov:1964zza}
A.~I. Nikishov, V.~I. Ritus, {Quantum Processes in the Field of a Plane
  Electromagnetic Wave and in a Constant Field 1}, Sov. Phys. JETP 19 (1964)
  529--541.

\bibitem{Nikishov:1964zzab}
A.~I. Nikishov, V.~I. Ritus, {Quantum Processes in the Field of a Plane
  Electromagnetic Wave and in a Constant Field 1}, Sov. Phys. JETP 19 (1964)
  529--541, [Zh. Eksp. Teor. Fiz.46,776(1964)].

\bibitem{Brown:1964zzb}
L.~S. Brown, T.~W.~B. Kibble, {Interaction of Intense Laser Beams with
  Electrons}, Phys. Rev. 133 (1964) A705--A719.
\newblock \href {https://doi.org/10.1103/PhysRev.133.A705}
  {\path{doi:10.1103/PhysRev.133.A705}}.

\bibitem{Kibble:1965zza}
T.~W.~B. Kibble, {Frequency Shift in High-Intensity Compton Scattering}, Phys.
  Rev. 138 (1965) B740--B753.
\newblock \href {https://doi.org/10.1103/PhysRev.138.B740}
  {\path{doi:10.1103/PhysRev.138.B740}}.

\bibitem{Frantz:1965}
L.~M. Frantz, \href{https://link.aps.org/doi/10.1103/PhysRev.139.B1326}{Compton
  scattering of an intense photon beam}, Phys. Rev. 139 (1965) B1326--B1336.
\newblock \href {https://doi.org/10.1103/PhysRev.139.B1326}
  {\path{doi:10.1103/PhysRev.139.B1326}}.
\newline\urlprefix\url{https://link.aps.org/doi/10.1103/PhysRev.139.B1326}

\bibitem{E144:1996enr}
C.~Bula, et~al., {Observation of nonlinear effects in Compton scattering},
  Phys. Rev. Lett. 76 (1996) 3116--3119.
\newblock \href {https://doi.org/10.1103/PhysRevLett.76.3116}
  {\path{doi:10.1103/PhysRevLett.76.3116}}.

\bibitem{Dinu:2017uoj}
V.~Dinu, G.~Torgrimsson, {Trident pair production in plane waves: Coherence,
  exchange, and spacetime inhomogeneity}, Phys. Rev. D 97~(3) (2018) 036021.
\newblock \href {http://arxiv.org/abs/1711.04344} {\path{arXiv:1711.04344}},
  \href {https://doi.org/10.1103/PhysRevD.97.036021}
  {\path{doi:10.1103/PhysRevD.97.036021}}.

\bibitem{King:2018ibi}
B.~King, A.~M. Fedotov, {Effect of interference on the trident process in a
  constant crossed field}, Phys. Rev. D 98~(1) (2018) 016005.
\newblock \href {http://arxiv.org/abs/1801.07300} {\path{arXiv:1801.07300}},
  \href {https://doi.org/10.1103/PhysRevD.98.016005}
  {\path{doi:10.1103/PhysRevD.98.016005}}.

\bibitem{Mackenroth:2018smh}
F.~Mackenroth, A.~Di~Piazza, {Nonlinear trident pair production in an arbitrary
  plane wave: a focus on the properties of the transition amplitude}, Phys.
  Rev. D 98~(11) (2018) 116002.
\newblock \href {http://arxiv.org/abs/1805.01731} {\path{arXiv:1805.01731}},
  \href {https://doi.org/10.1103/PhysRevD.98.116002}
  {\path{doi:10.1103/PhysRevD.98.116002}}.

\bibitem{Harvey:2012ie}
C.~Harvey, T.~Heinzl, A.~Ilderton, M.~Marklund, {Intensity-Dependent Electron
  Mass Shift in a Laser Field: Existence, Universality, and Detection}, Phys.
  Rev. Lett. 109 (2012) 100402.
\newblock \href {http://arxiv.org/abs/1203.6077} {\path{arXiv:1203.6077}},
  \href {https://doi.org/10.1103/PhysRevLett.109.100402}
  {\path{doi:10.1103/PhysRevLett.109.100402}}.

\bibitem{Battesti:2012hf}
R.~Battesti, C.~Rizzo, {Magnetic and electric properties of quantum vacuum},
  Rept. Prog. Phys. 76~(1) (2013) 016401.
\newblock \href {http://arxiv.org/abs/1211.1933} {\path{arXiv:1211.1933}},
  \href {https://doi.org/10.1088/0034-4885/76/1/016401}
  {\path{doi:10.1088/0034-4885/76/1/016401}}.

\bibitem{King:2015tba}
B.~King, T.~Heinzl, {Measuring Vacuum Polarisation with High Power Lasers},
  High Power Laser Sci. Eng. 4 (2016).
\newblock \href {http://arxiv.org/abs/1510.08456} {\path{arXiv:1510.08456}},
  \href {https://doi.org/10.1017/hpl.2016.1} {\path{doi:10.1017/hpl.2016.1}}.

\bibitem{Karbstein:2019oej}
F.~Karbstein, {Probing vacuum polarization effects with high-intensity lasers},
  Particles 3~(1) (2020) 39--61.
\newblock \href {http://arxiv.org/abs/1912.11698} {\path{arXiv:1912.11698}},
  \href {https://doi.org/10.3390/particles3010005}
  {\path{doi:10.3390/particles3010005}}.

\bibitem{Ritus:1970radiative}
V.~I. Ritus,
  \href{http://www.jetp.ras.ru/cgi-bin/dn/e_030_06_1181.pdf}{Radiative effects
  and their enhancement in an intense electromagnetic field}, Sov. Phys. JETP
  30 (1970) 1181.
\newline\urlprefix\url{http://www.jetp.ras.ru/cgi-bin/dn/e_030_06_1181.pdf}

\bibitem{Fedotov:2017conjecture}
A.~M. Fedotov, {Conjecture of perturbative QED breakdown at $\alpha\chi^{2/3}
  \gtrsim 1$}, J. Phys. Conf. Ser. 826~(1) (2017) 012027.
\newblock \href {http://arxiv.org/abs/1608.02261} {\path{arXiv:1608.02261}},
  \href {https://doi.org/10.1088/1742-6596/826/1/012027}
  {\path{doi:10.1088/1742-6596/826/1/012027}}.

\bibitem{Dorigoni:2014hea}
D.~Dorigoni, {An Introduction to Resurgence, Trans-Series and Alien Calculus},
  Annals Phys. 409 (2019) 167914.
\newblock \href {http://arxiv.org/abs/1411.3585} {\path{arXiv:1411.3585}},
  \href {https://doi.org/10.1016/j.aop.2019.167914}
  {\path{doi:10.1016/j.aop.2019.167914}}.

\bibitem{Dunne:2016nmc}
G.~V. Dunne, M.~\"Unsal, {New Nonperturbative Methods in Quantum Field Theory:
  From Large-N Orbifold Equivalence to Bions and Resurgence}, Ann. Rev. Nucl.
  Part. Sci. 66 (2016) 245--272.
\newblock \href {http://arxiv.org/abs/1601.03414} {\path{arXiv:1601.03414}},
  \href {https://doi.org/10.1146/annurev-nucl-102115-044755}
  {\path{doi:10.1146/annurev-nucl-102115-044755}}.

\bibitem{Salgado:2021fgt}
F.~C. Salgado, et~al., {Single particle detection system for strong-field QED
  experiments}, New J. Phys. 24~(1) (2022) 015002.
\newblock \href {http://arxiv.org/abs/2107.03697} {\path{arXiv:2107.03697}},
  \href {https://doi.org/10.1088/1367-2630/ac4283}
  {\path{doi:10.1088/1367-2630/ac4283}}.

\bibitem{Altarelli:2019zea}
M.~Altarelli, R.~Assmann, F.~Burkart, B.~Heinemann, T.~Heinzl, T.~Koffas, A.~R.
  Maier, D.~Reis, A.~Ringwald, M.~Wing, {Summary of strong-field QED Workshop},
  2019.
\newblock \href {http://arxiv.org/abs/1905.00059} {\path{arXiv:1905.00059}}.

\bibitem{Abramowicz:2019gvx}
H.~Abramowicz, et~al., {Letter of Intent for the LUXE Experiment} (9 2019).
\newblock \href {http://arxiv.org/abs/1909.00860} {\path{arXiv:1909.00860}}.

\bibitem{Heinemann2020}
{Heinemann, Beate}, {Heinzl, Tom}, {Ringwald, Andreas},
  \href{https://doi.org/10.1051/epn/2020401}{Luxe: combining high energy and
  intensity to spark the vacuum}, Europhysics News 51~(4) (2020) 14--17.
\newblock \href {https://doi.org/10.1051/epn/2020401}
  {\path{doi:10.1051/epn/2020401}}.
\newline\urlprefix\url{https://doi.org/10.1051/epn/2020401}

\bibitem{Abramowicz:2021zja}
H.~Abramowicz, et~al., {Conceptual design report for the LUXE experiment}, Eur.
  Phys. J. ST 230~(11) (2021) 2445--2560.
\newblock \href {http://arxiv.org/abs/2102.02032} {\path{arXiv:2102.02032}},
  \href {https://doi.org/10.1140/epjs/s11734-021-00249-z}
  {\path{doi:10.1140/epjs/s11734-021-00249-z}}.

\bibitem{2021Optic...8..630Y}
J.~W. {Yoon}, Y.~G. {Kim}, I.~W. {Choi}, J.~H. {Sung}, H.~W. {Lee}, S.~K.
  {Lee}, C.~H. {Nam}, {Realization of laser intensity over 1023 W/cm2}, Optica
  8~(5) (2021) 630.
\newblock \href {https://doi.org/10.1364/OPTICA.420520}
  {\path{doi:10.1364/OPTICA.420520}}.

\bibitem{Turcu:2016dxm}
I.~C.~E. Turcu, et~al., {High field physics and QED experiments at ELI-NP},
  Rom. Rep. Phys. 68~(Supplement) (2016) S145.

\bibitem{sel18}
E.~Cartlidge,
  https://www.science.org/content/article/physicists-are-planning-build-lasers-so-powerful-they-could-rip-apart-empty-space
  (2018).
\newblock \href {https://doi.org/10.1126/science.aat0939}
  {\path{doi:10.1126/science.aat0939}}.

\bibitem{MP3ref}
L.~Rochester, {MP3 Multi-Petawatt Physics Prioritization Workshop},
  https://mp3.lle.rochester.edu/ (2022).

\bibitem{Heinzl:2008rh}
T.~Heinzl, A.~Ilderton, {A Lorentz and gauge invariant measure of laser
  intensity}, Opt. Commun. 282 (2009) 1879--1883.
\newblock \href {http://arxiv.org/abs/0807.1841} {\path{arXiv:0807.1841}},
  \href {https://doi.org/10.1016/j.optcom.2009.01.051}
  {\path{doi:10.1016/j.optcom.2009.01.051}}.

\bibitem{Seipt:2019dnn}
D.~Seipt, A.~G.~R. Thomas, {A Frenet\textendash{}Serret interpretation of
  particle dynamics in high-intensity laser fields}, Plasma Phys. Control.
  Fusion 61~(7) (2019) 074005.
\newblock \href {http://arxiv.org/abs/1903.11463} {\path{arXiv:1903.11463}},
  \href {https://doi.org/10.1088/1361-6587/ab1e77}
  {\path{doi:10.1088/1361-6587/ab1e77}}.

\bibitem{Nielsen:2021ppf}
C.~F. Nielsen, J.~B. Justesen, A.~H. S\o{}rensen, U.~I. Uggerh\o{}j,
  R.~Holtzapple, {Experimental verification of the Landau\textendash{}Lifshitz
  equation}, New J. Phys. 23~(8) (2021) 085001.
\newblock \href {https://doi.org/10.1088/1367-2630/ac1554}
  {\path{doi:10.1088/1367-2630/ac1554}}.

\bibitem{Wistisen:2019eza}
T.~N. Wistisen, A.~Di~Piazza, C.~F. Nielsen, A.~H. S\o{}rensen, U.~I.
  Uggerh\o{}j, {Quantum radiation reaction in aligned crystals beyond the local
  constant field approximation}, Phys. Rev. Research. 1 (2019) 033014.
\newblock \href {http://arxiv.org/abs/1906.09144} {\path{arXiv:1906.09144}},
  \href {https://doi.org/10.1103/PhysRevResearch.1.033014}
  {\path{doi:10.1103/PhysRevResearch.1.033014}}.

\bibitem{CERNNA63:2012zsc}
K.~K. Andersen, J.~Esberg, H.~Knudsen, H.~D. Thomsen, U.~I. Uggerhoj, P.~Sona,
  A.~Mangiarotti, T.~J. Ketel, A.~Dizdar, S.~Ballestrero, {Experimental
  investigations of synchrotron radiation at the onset of the quantum regime},
  Phys. Rev. D 86 (2012) 072001.
\newblock \href {http://arxiv.org/abs/1206.6577} {\path{arXiv:1206.6577}},
  \href {https://doi.org/10.1103/PhysRevD.86.072001}
  {\path{doi:10.1103/PhysRevD.86.072001}}.

\bibitem{CERNNA63:2012hqm}
K.~K. Andersen, S.~L. Andersen, J.~Esberg, H.~Knudsen, R.~Mikkelsen, U.~I.
  Uggerhoj, P.~Sona, A.~Mangiarotti, T.~J. Ketel, S.~Ballestrero, {Direct
  measurement of the formation length of photons}, Phys. Rev. Lett. 108 (2012)
  071802.
\newblock \href {https://doi.org/10.1103/PhysRevLett.108.071802}
  {\path{doi:10.1103/PhysRevLett.108.071802}}.

\bibitem{CERNNA63:2013ahd}
K.~K. Andersen, S.~L. Andersen, J.~Esberg, H.~Knudsen, R.~E. Mikkelsen, U.~I.
  Uggerh\o{}j, T.~N. Wistisen, P.~Sona, A.~Mangiarotti, T.~J. Ketel,
  {Experimental investigation of the Landau-Pomeranchuk-Migdal effect in low-Z
  targets}, Phys. Rev. D 88~(7) (2013) 072007.
\newblock \href {http://arxiv.org/abs/1309.5765} {\path{arXiv:1309.5765}},
  \href {https://doi.org/10.1103/PhysRevD.88.072007}
  {\path{doi:10.1103/PhysRevD.88.072007}}.

\bibitem{DiPiazza:2015oia}
A.~Di~Piazza, T.~N. Wistisen, U.~I. Uggerh\o{}j, {Investigation of classical
  radiation reaction with aligned crystals}, Phys. Lett. B 765 (2017) 1--5.
\newblock \href {http://arxiv.org/abs/1503.05717} {\path{arXiv:1503.05717}},
  \href {https://doi.org/10.1016/j.physletb.2016.10.083}
  {\path{doi:10.1016/j.physletb.2016.10.083}}.

\bibitem{Wistisen:2017pgr}
T.~N. Wistisen, A.~Di~Piazza, H.~V. Knudsen, U.~I. Uggerh\o{}j, {Experimental
  evidence of quantum radiation reaction in aligned crystals}, Nature Commun.
  9~(1) (2018) 795.
\newblock \href {http://arxiv.org/abs/1704.01080} {\path{arXiv:1704.01080}},
  \href {https://doi.org/10.1038/s41467-018-03165-4}
  {\path{doi:10.1038/s41467-018-03165-4}}.

\bibitem{DiPiazza:2019vwb}
A.~Di~Piazza, T.~N. Wistisen, M.~Tamburini, U.~I. Uggerh\o{}j, {Testing Strong
  Field QED Close to the Fully Nonperturbative Regime Using Aligned Crystals},
  Phys. Rev. Lett. 124~(4) (2020) 044801.
\newblock \href {http://arxiv.org/abs/1911.04749} {\path{arXiv:1911.04749}},
  \href {https://doi.org/10.1103/PhysRevLett.124.044801}
  {\path{doi:10.1103/PhysRevLett.124.044801}}.

\bibitem{danson19}
C.~N. Danson, C.~Haefner, J.~Bromage, T.~Butcher, J.-C.~F. Chanteloup, E.~A.
  Chowdhury, A.~Galvanauskas, L.~A. Gizzi, J.~Hein, D.~I. Hillier, et~al.,
  Petawatt and exawatt class lasers worldwide, High Power Laser Science and
  Engineering 7 (2019) e54.
\newblock \href {https://doi.org/10.1017/hpl.2019.36}
  {\path{doi:10.1017/hpl.2019.36}}.

\bibitem{Gonoskov:2021hwf}
A.~Gonoskov, T.~G. Blackburn, M.~Marklund, S.~S. Bulanov, {Charged particle
  motion and radiation in strong electromagnetic fields} (7 2021).
\newblock \href {http://arxiv.org/abs/2107.02161} {\path{arXiv:2107.02161}}.

\bibitem{papadopoulos2016}
D.~Papadopoulos, J.~Zou, C.~Le~Blanc, G.~Chériaux, P.~Georges, F.~Druon,
  G.~Mennerat, P.~Ramirez, L.~Martin, A.~Fréneaux, et~al., The apollon 10 pw
  laser: experimental and theoretical investigation of the temporal
  characteristics, High Power Laser Science and Engineering 4 (2016) e34.
\newblock \href {https://doi.org/10.1017/hpl.2016.34}
  {\path{doi:10.1017/hpl.2016.34}}.

\bibitem{2015PhRvL.114s5003K}
K.~{Khrennikov}, J.~{Wenz}, A.~{Buck}, J.~{Xu}, M.~{Heigoldt}, L.~{Veisz},
  S.~{Karsch}, {Tunable All-Optical Quasimonochromatic Thomson X-Ray Source in
  the Nonlinear Regime}, Phys. Rev. Lett. 114~(19) (2015) 195003.
\newblock \href {https://doi.org/10.1103/PhysRevLett.114.195003}
  {\path{doi:10.1103/PhysRevLett.114.195003}}.

\bibitem{Sakai:2015mra}
Y.~Sakai, et~al., {Observation of redshifting and harmonic radiation in inverse
  Compton scattering}, Phys. Rev. ST Accel. Beams 18~(6) (2015) 060702.
\newblock \href {https://doi.org/10.1103/PhysRevSTAB.18.060702}
  {\path{doi:10.1103/PhysRevSTAB.18.060702}}.

\bibitem{2021NJPh...23j5002S}
F.~C. {Salgado}, K.~{Grafenstein}, A.~{Golub}, A.~{D{\"o}pp}, A.~{Eckey},
  D.~{Hollatz}, C.~{M{\"u}ller}, A.~{Seidel}, D.~{Seipt}, S.~{Karsch},
  M.~{Zepf}, {Towards pair production in the non-perturbative regime}, New
  Journal of Physics 23~(10) (2021) 105002.
\newblock \href {https://doi.org/10.1088/1367-2630/ac2921}
  {\path{doi:10.1088/1367-2630/ac2921}}.

\bibitem{Chen:2013mba}
S.~Chen, et~al., {MeV-Energy X Rays from Inverse Compton Scattering with
  Laser-Wakefield Accelerated Electrons}, Phys. Rev. Lett. 110~(15) (2013)
  155003.
\newblock \href {https://doi.org/10.1103/PhysRevLett.110.155003}
  {\path{doi:10.1103/PhysRevLett.110.155003}}.

\bibitem{Yan2017}
W.~Yan, C.~Fruhling, G.~Golovin, D.~Haden, J.~Luo, P.~Zhang, B.~Zhao, J.~Zhang,
  C.~Liu, M.~Chen, S.~Chen, S.~Banerjee, D.~Umstadter,
  \href{https://doi.org/10.1038/nphoton.2017.100}{High-order multiphoton
  thomson scattering}, Nature Photonics 11~(8) (2017) 514--520.
\newblock \href {https://doi.org/10.1038/nphoton.2017.100}
  {\path{doi:10.1038/nphoton.2017.100}}.
\newline\urlprefix\url{https://doi.org/10.1038/nphoton.2017.100}

\bibitem{Hannasch:2021kyh}
A.~Hannasch, et~al., {Compact spectroscopy of keV to MeV X-rays from a laser
  wakefield accelerator}, Sci. Rep. 11~(1) (2021) 14368.
\newblock \href {http://arxiv.org/abs/2103.01370} {\path{arXiv:2103.01370}},
  \href {https://doi.org/10.1038/s41598-021-93689-5}
  {\path{doi:10.1038/s41598-021-93689-5}}.

\bibitem{Bamber:1999zt}
C.~Bamber, et~al., {Studies of nonlinear QED in collisions of 46.6-GeV
  electrons with intense laser pulses}, Phys. Rev. D 60 (1999) 092004.
\newblock \href {https://doi.org/10.1103/PhysRevD.60.092004}
  {\path{doi:10.1103/PhysRevD.60.092004}}.

\bibitem{meuren2019probing}
S.~Meuren, Probing strong-field qed at facet-ii (slac e-320), in: Third
  Conference on Extremely High Intensity Laser Physics (ExHILP), 2019.

\bibitem{Zuegel:14}
J.~D. Zuegel, S.-W. Bahk, I.~A. Begishev, J.~Bromage, C.~Dorrer, A.~V. Okishev,
  J.~B. Oliver,
  \href{http://opg.optica.org/abstract.cfm?URI=CLEO_SI-2014-JTh4L.4}{Status of
  high-energy opcpa at lle and future prospects}, in: CLEO: 2014, Optica
  Publishing Group, 2014, p. JTh4L.4.
\newline\urlprefix\url{http://opg.optica.org/abstract.cfm?URI=CLEO_SI-2014-JTh4L.4}

\bibitem{Sarri:2014gea}
G.~Sarri, et~al., {Ultrahigh Brilliance Multi-MeV \ensuremath{\gamma} -Ray
  Beams from Nonlinear Relativistic Thomson Scattering}, Phys. Rev. Lett.
  113~(22) (2014) 224801.
\newblock \href {http://arxiv.org/abs/1407.6980} {\path{arXiv:1407.6980}},
  \href {https://doi.org/10.1103/PhysRevLett.113.224801}
  {\path{doi:10.1103/PhysRevLett.113.224801}}.

\bibitem{Cole:2017zca}
J.~M. Cole, et~al., {Experimental evidence of radiation reaction in the
  collision of a high-intensity laser pulse with a laser-wakefield accelerated
  electron beam}, Phys. Rev. X 8~(1) (2018) 011020.
\newblock \href {http://arxiv.org/abs/1707.06821} {\path{arXiv:1707.06821}},
  \href {https://doi.org/10.1103/PhysRevX.8.011020}
  {\path{doi:10.1103/PhysRevX.8.011020}}.

\bibitem{Poder:2017dpw}
K.~Poder, et~al., {Experimental Signatures of the Quantum Nature of Radiation
  Reaction in the Field of an Ultraintense Laser}, Phys. Rev. X 8~(3) (2018)
  031004.
\newblock \href {http://arxiv.org/abs/1709.01861} {\path{arXiv:1709.01861}},
  \href {https://doi.org/10.1103/PhysRevX.8.031004}
  {\path{doi:10.1103/PhysRevX.8.031004}}.

\bibitem{Hernandez_Gomez_2010}
C.~Hernandez-Gomez, S.~P. Blake, O.~Chekhlov, R.~J. Clarke, A.~M. Dunne,
  M.~Galimberti, S.~Hancock, R.~Heathcote, P.~Holligan, A.~Lyachev,
  P.~Matousek, I.~O. Musgrave, D.~Neely, P.~A. Norreys, I.~Ross, Y.~Tang, T.~B.
  Winstone, B.~E. Wyborn, J.~Collier,
  \href{https://doi.org/10.1088/1742-6596/244/3/032006}{The vulcan 10 {PW}
  project}, Journal of Physics: Conference Series 244~(3) (2010) 032006.
\newblock \href {https://doi.org/10.1088/1742-6596/244/3/032006}
  {\path{doi:10.1088/1742-6596/244/3/032006}}.
\newline\urlprefix\url{https://doi.org/10.1088/1742-6596/244/3/032006}

\bibitem{Li:18}
W.~Li, Z.~Gan, L.~Yu, C.~Wang, Y.~Liu, Z.~Guo, L.~Xu, M.~Xu, Y.~Hang, Y.~Xu,
  J.~Wang, P.~Huang, H.~Cao, B.~Yao, X.~Zhang, L.~Chen, Y.~Tang, S.~Li, X.~Liu,
  S.~Li, M.~He, D.~Yin, X.~Liang, Y.~Leng, R.~Li, Z.~Xu,
  \href{http://opg.optica.org/ol/abstract.cfm?URI=ol-43-22-5681}{339\&\#x2009;\&\#x2009;j
  high-energy ti:sapphire chirped-pulse amplifier for 10\&\#x2009;\&\#x2009;pw
  laser facility}, Opt. Lett. 43~(22) (2018) 5681--5684.
\newblock \href {https://doi.org/10.1364/OL.43.005681}
  {\path{doi:10.1364/OL.43.005681}}.
\newline\urlprefix\url{http://opg.optica.org/ol/abstract.cfm?URI=ol-43-22-5681}

\bibitem{Mukhin_2021}
I.~Mukhin, A.~Soloviev, E.~Perevezentsev, A.~Shaykin, V.~Ginzburg, I.~Kuzmin,
  M.~Mart'yanov, I.~Shaikin, A.~Kuzmin, S.~Mironov, I.~Yakovlev, E.~Khazanov,
  \href{https://doi.org/10.1070/qel17620}{Design of the front-end system for a
  subexawatt laser of the {XCELS} facility}, Quantum Electronics 51~(9) (2021)
  759--767.
\newblock \href {https://doi.org/10.1070/qel17620}
  {\path{doi:10.1070/qel17620}}.
\newline\urlprefix\url{https://doi.org/10.1070/qel17620}

\bibitem{Nees21}
J.~Nees, A.~Maksimchuk, G.~Kalinchenko, B.~Hou, Y.~Ma, P.~Campbell,
  A.~McKelvey, L.~Willingale, I.~Jovanovic, C.~Kuranz, A.~Thomas,
  K.~Krushelnick, Zettawatt equivalent ultrashort pulse laser system: An nsf
  mid-scale facility for laser-driven science in the qed regime, in: 2021
  Conference on Lasers and Electro-Optics (CLEO), 2021, pp. 1--2.

\bibitem{Schlenvoigt:2016jrd}
H.-P. Schlenvoigt, T.~Heinzl, U.~Schramm, T.~E. Cowan, R.~Sauerbrey, {Detecting
  vacuum birefringence with x-ray free electron lasers and high-power optical
  lasers: a feasibility study}, Phys. Scripta 91~(2) (2016) 023010.
\newblock \href {https://doi.org/10.1088/0031-8949/91/2/023010}
  {\path{doi:10.1088/0031-8949/91/2/023010}}.

\bibitem{yabashi2015overview}
M.~Yabashi, H.~Tanaka, T.~Ishikawa, Overview of the sacla facility, Journal of
  synchrotron radiation 22~(3) (2015) 477--484.

\bibitem{Inada:2014srv}
T.~Inada, et~al., {Search for Photon-Photon Elastic Scattering in the X-ray
  Region}, Phys. Lett. B 732 (2014) 356--359.
\newblock \href {http://arxiv.org/abs/1403.2547} {\path{arXiv:1403.2547}},
  \href {https://doi.org/10.1016/j.physletb.2014.03.054}
  {\path{doi:10.1016/j.physletb.2014.03.054}}.

\bibitem{Ejlli:2020yhk}
A.~Ejlli, F.~Della~Valle, U.~Gastaldi, G.~Messineo, R.~Pengo, G.~Ruoso,
  G.~Zavattini, {The PVLAS experiment: A 25 year effort to measure vacuum
  magnetic birefringence}, Phys. Rept. 871 (2020) 1--74.
\newblock \href {http://arxiv.org/abs/2005.12913} {\path{arXiv:2005.12913}},
  \href {https://doi.org/10.1016/j.physrep.2020.06.001}
  {\path{doi:10.1016/j.physrep.2020.06.001}}.

\bibitem{Cadene:2013bva}
A.~Cad\`ene, P.~Berceau, M.~Fouch\'e, R.~Battesti, C.~Rizzo, {Vacuum magnetic
  linear birefringence using pulsed fields: status of the BMV experiment}, Eur.
  Phys. J. D 68 (2014) 16.
\newblock \href {http://arxiv.org/abs/1302.5389} {\path{arXiv:1302.5389}},
  \href {https://doi.org/10.1140/epjd/e2013-40725-9}
  {\path{doi:10.1140/epjd/e2013-40725-9}}.

\bibitem{Neitz:2013qba}
N.~Neitz, A.~Di~Piazza, {Stochasticity Effects in Quantum Radiation Reaction},
  Phys. Rev. Lett. 111~(5) (2013) 054802.
\newblock \href {http://arxiv.org/abs/1301.5524} {\path{arXiv:1301.5524}},
  \href {https://doi.org/10.1103/PhysRevLett.111.054802}
  {\path{doi:10.1103/PhysRevLett.111.054802}}.

\bibitem{Yoffe:2015mba}
S.~R. Yoffe, Y.~Kravets, A.~Noble, D.~A. Jaroszynski, {Longitudinal and
  transverse cooling of relativistic electron beams in intense laser pulses},
  New J. Phys. 17~(5) (2015) 053025.
\newblock \href {http://arxiv.org/abs/1504.03480} {\path{arXiv:1504.03480}},
  \href {https://doi.org/10.1088/1367-2630/17/5/053025}
  {\path{doi:10.1088/1367-2630/17/5/053025}}.

\bibitem{Gonoskov:2013aoa}
A.~Gonoskov, A.~Bashinov, I.~Gonoskov, C.~Harvey, A.~Ilderton, A.~Kim,
  M.~Marklund, G.~Mourou, A.~Sergeev, {Anomalous radiative trapping in laser
  fields of extreme intensity}, Phys. Rev. Lett. 113 (2014) 014801.
\newblock \href {http://arxiv.org/abs/1306.5734} {\path{arXiv:1306.5734}},
  \href {https://doi.org/10.1103/PhysRevLett.113.014801}
  {\path{doi:10.1103/PhysRevLett.113.014801}}.

\bibitem{Blackburn:2014cig}
T.~G. Blackburn, C.~P. Ridgers, J.~G. Kirk, A.~R. Bell, {Quantum radiation
  reaction in laser-electron beam collisions}, Phys. Rev. Lett. 112 (2014)
  015001.
\newblock \href {http://arxiv.org/abs/1503.01009} {\path{arXiv:1503.01009}},
  \href {https://doi.org/10.1103/PhysRevLett.112.015001}
  {\path{doi:10.1103/PhysRevLett.112.015001}}.

\bibitem{Harvey:2016uiy}
C.~Harvey, A.~Gonoskov, A.~Ilderton, M.~Marklund, {Quantum quenching of
  radiation losses in short laser pulses}, Phys. Rev. Lett. 118~(10) (2017)
  105004.
\newblock \href {http://arxiv.org/abs/1606.08250} {\path{arXiv:1606.08250}},
  \href {https://doi.org/10.1103/PhysRevLett.118.105004}
  {\path{doi:10.1103/PhysRevLett.118.105004}}.

\bibitem{E320ref}
{E320 Experiment}.

\bibitem{Heinzl:2020ynb}
T.~Heinzl, B.~King, A.~J. Macleod, {The locally monochromatic approximation to
  QED in intense laser fields}, Phys. Rev. A 102 (2020) 063110.
\newblock \href {http://arxiv.org/abs/2004.13035} {\path{arXiv:2004.13035}},
  \href {https://doi.org/10.1103/PhysRevA.102.063110}
  {\path{doi:10.1103/PhysRevA.102.063110}}.

\bibitem{Nielsen:2021nuo}
C.~F. Nielsen, R.~Holtzapple, {Radiation Emission In Strong Electromagnetic
  Fields} (9 2021).
\newblock \href {http://arxiv.org/abs/2109.09490} {\path{arXiv:2109.09490}}.

\bibitem{Ritus1985}
V.~I. Ritus, \href{https://doi.org/10.1007/BF01120220}{{Quantum effects of the
  interaction of elementary particles with an intense electromagnetic field}},
  Journal of Soviet Laser Research 6~(5) (1985) 497--617.
\newblock \href {https://doi.org/10.1007/BF01120220}
  {\path{doi:10.1007/BF01120220}}.
\newline\urlprefix\url{https://doi.org/10.1007/BF01120220}

\bibitem{Dunne:2005sx}
G.~V. Dunne, C.~Schubert, {Worldline instantons and pair production in
  inhomogeneous fields}, Phys. Rev. D 72 (2005) 105004.
\newblock \href {http://arxiv.org/abs/hep-th/0507174}
  {\path{arXiv:hep-th/0507174}}, \href
  {https://doi.org/10.1103/PhysRevD.72.105004}
  {\path{doi:10.1103/PhysRevD.72.105004}}.

\bibitem{Marklund:2006my}
M.~Marklund, P.~K. Shukla, {Nonlinear collective effects in photon-photon and
  photon-plasma interactions}, Rev. Mod. Phys. 78 (2006) 591--640.
\newblock \href {http://arxiv.org/abs/hep-ph/0602123}
  {\path{arXiv:hep-ph/0602123}}, \href
  {https://doi.org/10.1103/RevModPhys.78.591}
  {\path{doi:10.1103/RevModPhys.78.591}}.

\bibitem{2009RPPh...72d6401E}
F.~{Ehlotzky}, K.~{Krajewska}, J.~Z. {Kami{\'n}ski}, {Fundamental processes of
  quantum electrodynamics in laser fields of relativistic power}, Reports on
  Progress in Physics 72~(4) (2009) 046401.
\newblock \href {https://doi.org/10.1088/0034-4885/72/4/046401}
  {\path{doi:10.1088/0034-4885/72/4/046401}}.

\bibitem{DiPiazza:2011tq}
A.~Di~Piazza, C.~Muller, K.~Z. Hatsagortsyan, C.~H. Keitel, {Extremely
  high-intensity laser interactions with fundamental quantum systems}, Rev.
  Mod. Phys. 84 (2012) 1177.
\newblock \href {http://arxiv.org/abs/1111.3886} {\path{arXiv:1111.3886}},
  \href {https://doi.org/10.1103/RevModPhys.84.1177}
  {\path{doi:10.1103/RevModPhys.84.1177}}.

\bibitem{Narozhny:2015vsb}
N.~B. Narozhny, A.~M. Fedotov, {Extreme light physics}, Contemp. Phys. 56~(3)
  (2015) 249--268.
\newblock \href {https://doi.org/10.1080/00107514.2015.1028768}
  {\path{doi:10.1080/00107514.2015.1028768}}.

\bibitem{Seipt:2017ckc}
D.~Seipt, {Volkov States and Non-linear Compton Scattering in Short and Intense
  Laser Pulses}, in: {Quantum Field Theory at the Limits}: {from Strong Fields
  to Heavy Quarks}, 2017.
\newblock \href {http://arxiv.org/abs/1701.03692} {\path{arXiv:1701.03692}},
  \href {https://doi.org/10.3204/DESY-PROC-2016-04/Seipt}
  {\path{doi:10.3204/DESY-PROC-2016-04/Seipt}}.

\bibitem{Zhang:2020lxl}
P.~Zhang, S.~S. Bulanov, D.~Seipt, A.~V. Arefiev, A.~G.~R. Thomas,
  {Relativistic Plasma Physics in Supercritical Fields}, Phys. Plasmas 27~(5)
  (2020) 050601.
\newblock \href {http://arxiv.org/abs/2001.00957} {\path{arXiv:2001.00957}},
  \href {https://doi.org/10.1063/1.5144449} {\path{doi:10.1063/1.5144449}}.

\bibitem{BrodinZamanian2021}
G.~Brodin, J.~Zamanian, Quantum kinetic theory of plasmas (2021).
\newblock \href {http://arxiv.org/abs/2111.00994} {\path{arXiv:2111.00994}}.

\bibitem{Fauth:2021nwe}
G.~Fauth, J.~Berges, A.~Di~Piazza, {Collisional strong-field QED kinetic
  equations from first principles}, Phys. Rev. D 104~(3) (2021) 036007.
\newblock \href {http://arxiv.org/abs/2103.13437} {\path{arXiv:2103.13437}},
  \href {https://doi.org/10.1103/PhysRevD.104.036007}
  {\path{doi:10.1103/PhysRevD.104.036007}}.

\bibitem{2014JPhB...47t4001P}
S.~V. {Popruzhenko}, {Keldysh theory of strong field ionization: history,
  applications, difficulties and perspectives}, Journal of Physics B Atomic
  Molecular Physics 47~(20) (2014) 204001.
\newblock \href {https://doi.org/10.1088/0953-4075/47/20/204001}
  {\path{doi:10.1088/0953-4075/47/20/204001}}.

\bibitem{2019RPPh...82k6001A}
K.~{Amini}, J.~{Biegert}, F.~{Calegari}, A.~{Chac{\'o}n}, M.~F. {Ciappina},
  A.~{Dauphin}, D.~K. {Efimov}, C.~{Figueira de Morisson Faria}, K.~{Giergiel},
  P.~{Gniewek}, A.~S. {Landsman}, M.~{Lesiuk}, M.~{Mandrysz}, A.~S. {Maxwell},
  R.~{Moszy{\'n}ski}, L.~{Ortmann}, J.~{Antonio P{\'e}rez-Hern{\'a}ndez},
  A.~{Pic{\'o}n}, E.~{Pisanty}, J.~{Prauzner-Bechcicki}, K.~{Sacha},
  N.~{Su{\'a}rez}, A.~{Za{\"\i}r}, J.~{Zakrzewski}, M.~{Lewenstein}, {Symphony
  on strong field approximation}, Reports on Progress in Physics 82~(11) (2019)
  116001.
\newblock \href {http://arxiv.org/abs/1812.11447} {\path{arXiv:1812.11447}},
  \href {https://doi.org/10.1088/1361-6633/ab2bb1}
  {\path{doi:10.1088/1361-6633/ab2bb1}}.

\bibitem{Lai:2014nma}
D.~Lai, {Physics in Very Strong Magnetic Fields: Introduction and Overview},
  Space Sci. Rev. 191~(1-4) (2015) 13--25.
\newblock \href {http://arxiv.org/abs/1411.7995} {\path{arXiv:1411.7995}},
  \href {https://doi.org/10.1007/s11214-015-0137-z}
  {\path{doi:10.1007/s11214-015-0137-z}}.

\bibitem{2016RPPh...79d6901M}
A.~{Marcowith}, A.~{Bret}, A.~{Bykov}, M.~E. {Dieckman}, L.~{O'C Drury},
  B.~{Lemb{\`e}ge}, M.~{Lemoine}, G.~{Morlino}, G.~{Murphy}, G.~{Pelletier},
  I.~{Plotnikov}, B.~{Reville}, M.~{Riquelme}, L.~{Sironi}, A.~{Stockem Novo},
  {The microphysics of collisionless shock waves}, Reports on Progress in
  Physics 79~(4) (2016) 046901.
\newblock \href {http://arxiv.org/abs/1604.00318} {\path{arXiv:1604.00318}},
  \href {https://doi.org/10.1088/0034-4885/79/4/046901}
  {\path{doi:10.1088/0034-4885/79/4/046901}}.

\bibitem{Kim:2019joy}
S.~P. Kim, {Astrophysics in Strong Electromagnetic Fields and Laboratory
  Astrophysics}, in: {15th Marcel Grossmann Meeting on Recent Developments in
  Theoretical and Experimental General Relativity, Astrophysics, and
  Relativistic Field Theories}, 2019.
\newblock \href {http://arxiv.org/abs/1905.13439} {\path{arXiv:1905.13439}}.

\bibitem{Turolla:2015mwa}
R.~Turolla, S.~Zane, A.~Watts, {Magnetars: the physics behind observations. A
  review}, Rept. Prog. Phys. 78~(11) (2015) 116901.
\newblock \href {http://arxiv.org/abs/1507.02924} {\path{arXiv:1507.02924}},
  \href {https://doi.org/10.1088/0034-4885/78/11/116901}
  {\path{doi:10.1088/0034-4885/78/11/116901}}.

\bibitem{Kaspi:2017fwg}
V.~M. Kaspi, A.~Beloborodov, {Magnetars}, Ann. Rev. Astron. Astrophys. 55
  (2017) 261--301.
\newblock \href {http://arxiv.org/abs/1703.00068} {\path{arXiv:1703.00068}},
  \href {https://doi.org/10.1146/annurev-astro-081915-023329}
  {\path{doi:10.1146/annurev-astro-081915-023329}}.

\bibitem{Kim:2021kif}
C.~M. Kim, S.~P. Kim, {Magnetars as Laboratories for Strong Field QED}, in:
  {17th Italian-Korean Symposium on Relativistic Astrophysics}, 2021.
\newblock \href {http://arxiv.org/abs/2112.02460} {\path{arXiv:2112.02460}}.

\bibitem{Kuznetsov:2004tb}
A.~Kuznetsov, N.~Mikheev, {Electroweak processes in external electromagnetic
  fields}, Springer Tracts Mod. Phys. 197 (2004) 1--120.
\newblock \href {https://doi.org/10.1007/b97444} {\path{doi:10.1007/b97444}}.

\bibitem{Kharzeev:2013jha}
D.~Kharzeev, K.~Landsteiner, A.~Schmitt, H.-U. Yee (Eds.), {Strongly
  Interacting Matter in Magnetic Fields}, Vol. 871, 2013.
\newblock \href {https://doi.org/10.1007/978-3-642-37305-3}
  {\path{doi:10.1007/978-3-642-37305-3}}.

\bibitem{Hattori:2016emy}
K.~Hattori, X.-G. Huang, {Novel quantum phenomena induced by strong magnetic
  fields in heavy-ion collisions}, Nucl. Sci. Tech. 28~(2) (2017) 26.
\newblock \href {http://arxiv.org/abs/1609.00747} {\path{arXiv:1609.00747}},
  \href {https://doi.org/10.1007/s41365-016-0178-3}
  {\path{doi:10.1007/s41365-016-0178-3}}.

\bibitem{Raicher:2013cja}
E.~Raicher, S.~Eliezer, {Analytical Solutions of the Dirac and the Klein-Gordon
  Equations in Plasma Induced by High Intensity Laser}, Phys. Rev. A 88~(2)
  (2013) 022113.
\newblock \href {http://arxiv.org/abs/1306.0486} {\path{arXiv:1306.0486}},
  \href {https://doi.org/10.1103/PhysRevA.88.022113}
  {\path{doi:10.1103/PhysRevA.88.022113}}.

\bibitem{Gies:2016yaa}
H.~Gies, F.~Karbstein, {An Addendum to the Heisenberg-Euler effective action
  beyond one loop}, JHEP 03 (2016).
\newblock \href {http://arxiv.org/abs/1612.07251} {\path{arXiv:1612.07251}},
  \href {https://doi.org/10.1007/JHEP03(2017)108}
  {\path{doi:10.1007/JHEP03(2017)108}}.

\bibitem{Furry:1951zz}
W.~H. Furry, {On Bound States and Scattering in Positron Theory}, Phys. Rev. 81
  (1951) 115--124.
\newblock \href {https://doi.org/10.1103/PhysRev.81.915}
  {\path{doi:10.1103/PhysRev.81.915}}.

\bibitem{Dunne:2004nc}
G.~V. Dunne, {Heisenberg-Euler effective Lagrangians: Basics and extensions},
  2004, pp. 445--522.
\newblock \href {http://arxiv.org/abs/hep-th/0406216}
  {\path{arXiv:hep-th/0406216}}, \href
  {https://doi.org/10.1142/9789812775344_0014}
  {\path{doi:10.1142/9789812775344_0014}}.

\bibitem{Nikishov:1985}
A.~I. Nikishov, \href{https://doi.org/10.1007/BF01120143}{Problems of intense
  external-field intensity in quantum electrodynamics}, Journal of Soviet Laser
  Research 6~(6) (1985) 619--717.
\newblock \href {https://doi.org/10.1007/BF01120143}
  {\path{doi:10.1007/BF01120143}}.
\newline\urlprefix\url{https://doi.org/10.1007/BF01120143}

\bibitem{Fradkin:1991zq}
E.~S. Fradkin, D.~M. Gitman, S.~M. Shvartsman, {Quantum electrodynamics with
  unstable vacuum}, 1991.

\bibitem{Karbstein:2014fva}
F.~Karbstein, R.~Shaisultanov, {Stimulated photon emission from the vacuum},
  Phys. Rev. D 91~(11) (2015) 113002.
\newblock \href {http://arxiv.org/abs/1412.6050} {\path{arXiv:1412.6050}},
  \href {https://doi.org/10.1103/PhysRevD.91.113002}
  {\path{doi:10.1103/PhysRevD.91.113002}}.

\bibitem{Karbstein:2015qwa}
F.~Karbstein, {Vacuum Birefringence as a Vacuum Emission Process}, in:
  {International Conference on the Structure and Interactions of the Photon and
  21st International Workshop on Photon-Photon Collisions and International
  Workshop on High Energy Photon Linear Colliders}, 2015.
\newblock \href {http://arxiv.org/abs/1510.03178} {\path{arXiv:1510.03178}}.

\bibitem{Gies:2017ygp}
H.~Gies, F.~Karbstein, C.~Kohlf\"urst, {All-optical signatures of Strong-Field
  QED in the vacuum emission picture}, Phys. Rev. D 97~(3) (2018) 036022.
\newblock \href {http://arxiv.org/abs/1712.03232} {\path{arXiv:1712.03232}},
  \href {https://doi.org/10.1103/PhysRevD.97.036022}
  {\path{doi:10.1103/PhysRevD.97.036022}}.

\bibitem{Blinne:2018uyn}
A.~Blinne, H.~Gies, F.~Karbstein, C.~Kohlf\"urst, M.~Zepf, {The Vacuum Emission
  Picture Beyond Paraxial Approximation}, J. Phys. Conf. Ser. 1206~(1) (2019)
  012017.
\newblock \href {http://arxiv.org/abs/1812.04620} {\path{arXiv:1812.04620}},
  \href {https://doi.org/10.1088/1742-6596/1206/1/012017}
  {\path{doi:10.1088/1742-6596/1206/1/012017}}.

\bibitem{Klein:1929zz}
O.~Klein, {Die Reflexion von Elektronen an einem Potentialsprung nach der
  relativistischen Dynamik von Dirac}, Z. Phys. 53 (1929) 157.
\newblock \href {https://doi.org/10.1007/BF01339716}
  {\path{doi:10.1007/BF01339716}}.

\bibitem{Nikishov:1969tt}
A.~I. Nikishov, {Pair production by a constant external field}, Zh. Eksp. Teor.
  Fiz. 57 (1969) 1210--1216.

\bibitem{Weisskopf:406571}
V.~F. Weisskopf, \href{https://cds.cern.ch/record/406571}{{Über die
  Elektrodynamik des Vakuums auf Grund des Quanten-Theorie des Elektrons}},
  Dan. Mat. Fys. Medd. 14 (1936) 1--39.
\newline\urlprefix\url{https://cds.cern.ch/record/406571}

\bibitem{Vanyashin:1965ple}
V.~S. Vanyashin, M.~V. Terentev, {The Vacuum Polarization of a Charged Vector
  Field}, Zh. Eksp. Teor. Fiz. 48~(2) (1965) 565--573.

\bibitem{Marinov:1972nx}
M.~S. Marinov, V.~S. Popov, {Pair production in electromagnetic field (case of
  arbitrary spin)}, Yad. Fiz. 15 (1972) 1271--1285.

\bibitem{Frob:2014zka}
M.~B. Fr\"ob, J.~Garriga, S.~Kanno, M.~Sasaki, J.~Soda, T.~Tanaka, A.~Vilenkin,
  {Schwinger effect in de Sitter space}, JCAP 04 (2014) 009.
\newblock \href {http://arxiv.org/abs/1401.4137} {\path{arXiv:1401.4137}},
  \href {https://doi.org/10.1088/1475-7516/2014/04/009}
  {\path{doi:10.1088/1475-7516/2014/04/009}}.

\bibitem{Cohen:2008wz}
T.~D. Cohen, D.~A. McGady, {The Schwinger mechanism revisited}, Phys. Rev. D 78
  (2008) 036008.
\newblock \href {http://arxiv.org/abs/0807.1117} {\path{arXiv:0807.1117}},
  \href {https://doi.org/10.1103/PhysRevD.78.036008}
  {\path{doi:10.1103/PhysRevD.78.036008}}.

\bibitem{Fukushima:2009er}
K.~Fukushima, F.~Gelis, T.~Lappi, {Multiparticle correlations in the Schwinger
  mechanism}, Nucl. Phys. A 831 (2009) 184--214.
\newblock \href {http://arxiv.org/abs/0907.4793} {\path{arXiv:0907.4793}},
  \href {https://doi.org/10.1016/j.nuclphysa.2009.09.062}
  {\path{doi:10.1016/j.nuclphysa.2009.09.062}}.

\bibitem{Gelis:2015kya}
F.~Gelis, N.~Tanji, {Schwinger mechanism revisited}, Prog. Part. Nucl. Phys. 87
  (2016) 1--49.
\newblock \href {http://arxiv.org/abs/1510.05451} {\path{arXiv:1510.05451}},
  \href {https://doi.org/10.1016/j.ppnp.2015.11.001}
  {\path{doi:10.1016/j.ppnp.2015.11.001}}.

\bibitem{Edwards:2019eby}
J.~P. Edwards, C.~Schubert, {Quantum mechanical path integrals in the first
  quantised approach to quantum field theory}, 2019.
\newblock \href {http://arxiv.org/abs/1912.10004} {\path{arXiv:1912.10004}}.

\bibitem{Bulanov:2004de}
S.~S. Bulanov, N.~B. Narozhny, V.~D. Mur, V.~S. Popov, {On e+ e- pair
  production by a focused laser pulse in vacuum}, Phys. Lett. A 330 (2004)
  1--6.
\newblock \href {http://arxiv.org/abs/hep-ph/0403163}
  {\path{arXiv:hep-ph/0403163}}, \href
  {https://doi.org/10.1016/j.physleta.2004.07.013}
  {\path{doi:10.1016/j.physleta.2004.07.013}}.

\bibitem{Gonoskov:2013ada}
A.~Gonoskov, I.~Gonoskov, C.~Harvey, A.~Ilderton, A.~Kim, M.~Marklund,
  G.~Mourou, A.~M. Sergeev, {Probing nonperturbative QED with optimally focused
  laser pulses}, Phys. Rev. Lett. 111 (2013) 060404.
\newblock \href {http://arxiv.org/abs/1302.4653} {\path{arXiv:1302.4653}},
  \href {https://doi.org/10.1103/PhysRevLett.111.060404}
  {\path{doi:10.1103/PhysRevLett.111.060404}}.

\bibitem{Bakker:2013cea}
B.~L.~G. Bakker, et~al., {Light-Front Quantum Chromodynamics: A framework for
  the analysis of hadron physics}, Nucl. Phys. B Proc. Suppl. 251-252 (2014)
  165--174.
\newblock \href {http://arxiv.org/abs/1309.6333} {\path{arXiv:1309.6333}},
  \href {https://doi.org/10.1016/j.nuclphysbps.2014.05.004}
  {\path{doi:10.1016/j.nuclphysbps.2014.05.004}}.

\bibitem{Dinu:2012tj}
V.~Dinu, T.~Heinzl, A.~Ilderton, {Infra-Red Divergences in Plane Wave
  Backgrounds}, Phys. Rev. D 86 (2012) 085037.
\newblock \href {http://arxiv.org/abs/1206.3957} {\path{arXiv:1206.3957}},
  \href {https://doi.org/10.1103/PhysRevD.86.085037}
  {\path{doi:10.1103/PhysRevD.86.085037}}.

\bibitem{Heinzl:2000ht}
T.~Heinzl, {Light cone quantization: Foundations and applications}, Lect. Notes
  Phys. 572 (2001) 55--142.
\newblock \href {http://arxiv.org/abs/hep-th/0008096}
  {\path{arXiv:hep-th/0008096}}, \href
  {https://doi.org/10.1007/3-540-45114-5_2}
  {\path{doi:10.1007/3-540-45114-5_2}}.

\bibitem{Brodsky:1997de}
S.~J. Brodsky, H.-C. Pauli, S.~S. Pinsky, {Quantum chromodynamics and other
  field theories on the light cone}, Phys. Rept. 301 (1998) 299--486.
\newblock \href {http://arxiv.org/abs/hep-ph/9705477}
  {\path{arXiv:hep-ph/9705477}}, \href
  {https://doi.org/10.1016/S0370-1573(97)00089-6}
  {\path{doi:10.1016/S0370-1573(97)00089-6}}.

\bibitem{Bieri:2013hqa}
L.~Bieri, D.~Garfinkle, {An electromagnetic analogue of gravitational wave
  memory}, Class. Quant. Grav. 30 (2013) 195009.
\newblock \href {http://arxiv.org/abs/1307.5098} {\path{arXiv:1307.5098}},
  \href {https://doi.org/10.1088/0264-9381/30/19/195009}
  {\path{doi:10.1088/0264-9381/30/19/195009}}.

\bibitem{Zhang:2017geq}
P.~M. Zhang, C.~Duval, G.~W. Gibbons, P.~A. Horvathy, {Soft gravitons and the
  memory effect for plane gravitational waves}, Phys. Rev. D 96~(6) (2017)
  064013.
\newblock \href {http://arxiv.org/abs/1705.01378} {\path{arXiv:1705.01378}},
  \href {https://doi.org/10.1103/PhysRevD.96.064013}
  {\path{doi:10.1103/PhysRevD.96.064013}}.

\bibitem{Zhang:2017rno}
P.~M. Zhang, C.~Duval, G.~W. Gibbons, P.~A. Horvathy, {The Memory Effect for
  Plane Gravitational Waves}, Phys. Lett. B 772 (2017) 743--746.
\newblock \href {http://arxiv.org/abs/1704.05997} {\path{arXiv:1704.05997}},
  \href {https://doi.org/10.1016/j.physletb.2017.07.050}
  {\path{doi:10.1016/j.physletb.2017.07.050}}.

\bibitem{Pate:2017vwa}
M.~Pate, A.-M. Raclariu, A.~Strominger, {Color Memory: A Yang-Mills Analog of
  Gravitational Wave Memory}, Phys. Rev. Lett. 119~(26) (2017) 261602.
\newblock \href {http://arxiv.org/abs/1707.08016} {\path{arXiv:1707.08016}},
  \href {https://doi.org/10.1103/PhysRevLett.119.261602}
  {\path{doi:10.1103/PhysRevLett.119.261602}}.

\bibitem{Ilderton:2020rgk}
A.~Ilderton, A.~J. MacLeod, {The analytic structure of amplitudes on
  backgrounds from gauge invariance and the infra-red}, JHEP 04 (2020) 078.
\newblock \href {http://arxiv.org/abs/2001.10553} {\path{arXiv:2001.10553}},
  \href {https://doi.org/10.1007/JHEP04(2020)078}
  {\path{doi:10.1007/JHEP04(2020)078}}.

\bibitem{Strominger:2017zoo}
A.~Strominger, {Lectures on the Infrared Structure of Gravity and Gauge Theory}
  (3 2017).
\newblock \href {http://arxiv.org/abs/1703.05448} {\path{arXiv:1703.05448}}.

\bibitem{Lavelle:2020ijh}
M.~Lavelle, D.~McMullan, {Simplest description of charge propagation in a
  strong background}, Phys. Rev. D 103~(3) (2021) 036015.
\newblock \href {http://arxiv.org/abs/2011.03757} {\path{arXiv:2011.03757}},
  \href {https://doi.org/10.1103/PhysRevD.103.036015}
  {\path{doi:10.1103/PhysRevD.103.036015}}.

\bibitem{Wolkow:1935zz}
D.~M. Wolkow, {Uber eine Klasse von Losungen der Diracschen Gleichung}, Z.
  Phys. 94 (1935) 250--260.
\newblock \href {https://doi.org/10.1007/BF01331022}
  {\path{doi:10.1007/BF01331022}}.

\bibitem{Neville:1971uc}
R.~A. Neville, F.~Rohrlich, {Quantum electrodynamics on null planes and
  applications to lasers}, Phys. Rev. D 3 (1971) 1692--1707.
\newblock \href {https://doi.org/10.1103/PhysRevD.3.1692}
  {\path{doi:10.1103/PhysRevD.3.1692}}.

\bibitem{Ilderton:2012qe}
A.~Ilderton, G.~Torgrimsson, {Scattering in plane-wave backgrounds: infra-red
  effects and pole structure}, Phys. Rev. D 87 (2013) 085040.
\newblock \href {http://arxiv.org/abs/1210.6840} {\path{arXiv:1210.6840}},
  \href {https://doi.org/10.1103/PhysRevD.87.085040}
  {\path{doi:10.1103/PhysRevD.87.085040}}.

\bibitem{Angioi:2016vir}
A.~Angioi, F.~Mackenroth, A.~Di~Piazza, {Nonlinear single Compton scattering of
  an electron wave-packet}, Phys. Rev. A 93~(5) (2016) 052102.
\newblock \href {http://arxiv.org/abs/1602.02639} {\path{arXiv:1602.02639}},
  \href {https://doi.org/10.1103/PhysRevA.93.052102}
  {\path{doi:10.1103/PhysRevA.93.052102}}.

\bibitem{Chiu:2017ycx}
K.~Y.-J. Chiu, S.~J. Brodsky, {Angular Momentum Conservation Law in Light-Front
  Quantum Field Theory}, Phys. Rev. D 95~(6) (2017) 065035.
\newblock \href {http://arxiv.org/abs/1702.01127} {\path{arXiv:1702.01127}},
  \href {https://doi.org/10.1103/PhysRevD.95.065035}
  {\path{doi:10.1103/PhysRevD.95.065035}}.

\bibitem{Ilderton:2020gno}
A.~Ilderton, B.~King, S.~Tang, {Loop spin effects in intense background
  fields}, Phys. Rev. D 102~(7) (2020) 076013.
\newblock \href {http://arxiv.org/abs/2008.08578} {\path{arXiv:2008.08578}},
  \href {https://doi.org/10.1103/PhysRevD.102.076013}
  {\path{doi:10.1103/PhysRevD.102.076013}}.

\bibitem{Adamo:2021hno}
T.~Adamo, A.~Ilderton, A.~J. MacLeod, {One-loop multicollinear limits from
  2-point amplitudes on self-dual backgrounds}, JHEP 12 (2021) 207.
\newblock \href {http://arxiv.org/abs/2103.12850} {\path{arXiv:2103.12850}}.

\bibitem{Bargmann:1959gz}
V.~Bargmann, L.~Michel, V.~L. Telegdi, {Precession of the polarization of
  particles moving in a homogeneous electromagnetic field}, Phys. Rev. Lett. 2
  (1959) 435--436.
\newblock \href {https://doi.org/10.1103/PhysRevLett.2.435}
  {\path{doi:10.1103/PhysRevLett.2.435}}.

\bibitem{DiPiazza:2021rum}
A.~Di~Piazza, {WKB electron wave functions in a tightly focused laser beam},
  Phys. Rev. D 103~(7) (2021) 076011.
\newblock \href {http://arxiv.org/abs/2102.06692} {\path{arXiv:2102.06692}},
  \href {https://doi.org/10.1103/PhysRevD.103.076011}
  {\path{doi:10.1103/PhysRevD.103.076011}}.

\bibitem{DiPiazza:2022lij}
A.~Di~Piazza, F.~P. Fronimos, {Quasiclassical representation of the Volkov
  propagator and the tadpole diagram in a plane wave} (1 2022).
\newblock \href {http://arxiv.org/abs/2201.08101} {\path{arXiv:2201.08101}}.

\bibitem{Heinzl:2010vg}
T.~Heinzl, A.~Ilderton, M.~Marklund, {Finite size effects in stimulated laser
  pair production}, Phys. Lett. B 692 (2010) 250--256.
\newblock \href {http://arxiv.org/abs/1002.4018} {\path{arXiv:1002.4018}},
  \href {https://doi.org/10.1016/j.physletb.2010.07.044}
  {\path{doi:10.1016/j.physletb.2010.07.044}}.

\bibitem{King:2020hsk}
B.~King, {Interference effects in nonlinear Compton scattering due to pulse
  envelope}, Phys. Rev. D 103~(3) (2021) 036018.
\newblock \href {http://arxiv.org/abs/2012.05920} {\path{arXiv:2012.05920}},
  \href {https://doi.org/10.1103/PhysRevD.103.036018}
  {\path{doi:10.1103/PhysRevD.103.036018}}.

\bibitem{Tang:2021qht}
S.~Tang, B.~King, {Pulse envelope effects in nonlinear Breit-Wheeler pair
  creation}, Phys. Rev. D 104~(9) (2021) 096019.
\newblock \href {http://arxiv.org/abs/2109.00555} {\path{arXiv:2109.00555}},
  \href {https://doi.org/10.1103/PhysRevD.104.096019}
  {\path{doi:10.1103/PhysRevD.104.096019}}.

\bibitem{Lavelle:2014mka}
M.~Lavelle, D.~McMullan, {Sideband Mixing in Intense Laser Backgrounds}, Phys.
  Lett. B 739 (2014) 421--424.
\newblock \href {http://arxiv.org/abs/1407.1279} {\path{arXiv:1407.1279}},
  \href {https://doi.org/10.1016/j.physletb.2014.11.014}
  {\path{doi:10.1016/j.physletb.2014.11.014}}.

\bibitem{Lavelle:2015jxa}
M.~Lavelle, D.~McMullan, {Fermionic propagator in an intense background}, Phys.
  Rev. D 91 (2015) 105022.
\newblock \href {http://arxiv.org/abs/1502.06529} {\path{arXiv:1502.06529}},
  \href {https://doi.org/10.1103/PhysRevD.91.105022}
  {\path{doi:10.1103/PhysRevD.91.105022}}.

\bibitem{Lavelle:2017dzx}
M.~Lavelle, D.~McMullan, {Electrons in an eccentric background field}, Phys.
  Rev. D 97~(3) (2018) 036013.
\newblock \href {http://arxiv.org/abs/1711.09046} {\path{arXiv:1711.09046}},
  \href {https://doi.org/10.1103/PhysRevD.97.036013}
  {\path{doi:10.1103/PhysRevD.97.036013}}.

\bibitem{Lavelle:2019lys}
M.~Lavelle, D.~McMullan, {Renormalization of the Volkov propagator}, Phys. Rev.
  D 100~(3) (2019) 036001.
\newblock \href {http://arxiv.org/abs/1905.05551} {\path{arXiv:1905.05551}},
  \href {https://doi.org/10.1103/PhysRevD.100.036001}
  {\path{doi:10.1103/PhysRevD.100.036001}}.

\bibitem{Lavelle:2019vcz}
M.~Lavelle, D.~McMullan, {One loop Volkov propagator in the Lorentz class of
  gauges}, Phys. Lett. B 798 (2019) 135021.
\newblock \href {http://arxiv.org/abs/1908.10228} {\path{arXiv:1908.10228}},
  \href {https://doi.org/10.1016/j.physletb.2019.135021}
  {\path{doi:10.1016/j.physletb.2019.135021}}.

\bibitem{Ilderton:2016qpj}
A.~Ilderton, G.~Torgrimsson, {Worldline approach to helicity flip in plane
  waves}, Phys. Rev. D 93~(8) (2016) 085006.
\newblock \href {http://arxiv.org/abs/1601.05021} {\path{arXiv:1601.05021}},
  \href {https://doi.org/10.1103/PhysRevD.93.085006}
  {\path{doi:10.1103/PhysRevD.93.085006}}.

\bibitem{Edwards:2021uif}
J.~P. Edwards, C.~Schubert, {Plane Wave Backgrounds in the Worldline
  Formalism}, in: {29th International Laser Physics Workshop}, 2021.
\newblock \href {http://arxiv.org/abs/2112.13944} {\path{arXiv:2112.13944}}.

\bibitem{Esposti:2021wsh}
G.~D. Esposti, G.~Torgrimsson, {Worldline instantons for nonlinear
  Breit-Wheeler pair production and Compton scattering} (12 2021).
\newblock \href {http://arxiv.org/abs/2112.11433} {\path{arXiv:2112.11433}}.

\bibitem{Blackburn:2019lgk}
T.~G. Blackburn, D.~Seipt, S.~S. Bulanov, M.~Marklund, {Radiation beaming in
  the quantum regime}, Phys. Rev. A 101~(1) (2020) 012505.
\newblock \href {http://arxiv.org/abs/1904.07745} {\path{arXiv:1904.07745}},
  \href {https://doi.org/10.1103/PhysRevA.101.012505}
  {\path{doi:10.1103/PhysRevA.101.012505}}.

\bibitem{baier98}
V.~N. Baier, V.~M. Katkov, V.~M. Strakhovenko, Electromagnetic processes at
  high energies in oriented single crystals, World Scientific, Singapore, 1998.

\bibitem{Dinu:2013hsd}
V.~Dinu, {Exact final state integrals for strong field QED}, Phys. Rev. A
  87~(5) (2013) 052101.
\newblock \href {http://arxiv.org/abs/1302.1513} {\path{arXiv:1302.1513}},
  \href {https://doi.org/10.1103/PhysRevA.87.052101}
  {\path{doi:10.1103/PhysRevA.87.052101}}.

\bibitem{Kibble:1975vz}
T.~W.~B. Kibble, A.~Salam, J.~A. Strathdee, {Intensity Dependent Mass Shift and
  Symmetry Breaking}, Nucl. Phys. B 96 (1975) 255--263.
\newblock \href {https://doi.org/10.1016/0550-3213(75)90581-7}
  {\path{doi:10.1016/0550-3213(75)90581-7}}.

\bibitem{RIDGERS2014273}
C.~Ridgers, J.~Kirk, R.~Duclous, T.~Blackburn, C.~Brady, K.~Bennett, T.~Arber,
  A.~Bell,
  \href{https://www.sciencedirect.com/science/article/pii/S0021999113008061}{Modelling
  gamma-ray photon emission and pair production in high-intensity
  laser–matter interactions}, Journal of Computational Physics 260 (2014)
  273--285.
\newblock \href {https://doi.org/https://doi.org/10.1016/j.jcp.2013.12.007}
  {\path{doi:https://doi.org/10.1016/j.jcp.2013.12.007}}.
\newline\urlprefix\url{https://www.sciencedirect.com/science/article/pii/S0021999113008061}

\bibitem{Niel:PRE2017}
F.~Niel, C.~Riconda, F.~Amiranoff, R.~Duclous, M.~Grech,
  \href{http://arxiv.org/abs/1707.02618
  https://link.aps.org/doi/10.1103/PhysRevE.97.043209}{{From quantum to
  classical modeling of radiation reaction: A focus on stochasticity effects}},
  Phys. Rev. E 97~(4) (2018) 043209.
\newblock \href {http://arxiv.org/abs/1707.02618} {\path{arXiv:1707.02618}},
  \href {https://doi.org/10.1103/PhysRevE.97.043209}
  {\path{doi:10.1103/PhysRevE.97.043209}}.
\newline\urlprefix\url{http://arxiv.org/abs/1707.02618
  https://link.aps.org/doi/10.1103/PhysRevE.97.043209}

\bibitem{Seipt:2012tn}
D.~Seipt, B.~Kämpfer, {Two-photon Compton process in pulsed intense laser
  fields}, Phys. Rev. D 85 (2012) 101701.
\newblock \href {http://arxiv.org/abs/1201.4045} {\path{arXiv:1201.4045}},
  \href {https://doi.org/10.1103/PhysRevD.85.101701}
  {\path{doi:10.1103/PhysRevD.85.101701}}.

\bibitem{Seipt:2013taa}
D.~Seipt, B.~Kämpfer, {Asymmetries of azimuthal photon distributions in
  non-linear Compton scattering in ultra-short intense laser pulses}, Phys.
  Rev. A 88 (2013) 012127.
\newblock \href {http://arxiv.org/abs/1305.3837} {\path{arXiv:1305.3837}},
  \href {https://doi.org/10.1103/PhysRevA.88.012127}
  {\path{doi:10.1103/PhysRevA.88.012127}}.

\bibitem{Krajewska:2012gc}
K.~Krajewska, J.~Z. Kaminski, {Compton Process in Intense Short Laser Pulses},
  Phys. Rev. A 85 (2012) 062102.
\newblock \href {http://arxiv.org/abs/1203.6022} {\path{arXiv:1203.6022}},
  \href {https://doi.org/10.1103/PhysRevA.85.062102}
  {\path{doi:10.1103/PhysRevA.85.062102}}.

\bibitem{Titov:2015tdz}
A.~I. Titov, B.~Kämpfer, A.~Hosaka, T.~Nousch, D.~Seipt, {Determination of the
  carrier envelope phase for short, circularly polarized laser pulses}, Phys.
  Rev. D 93~(4) (2016) 045010.
\newblock \href {http://arxiv.org/abs/1512.07504} {\path{arXiv:1512.07504}},
  \href {https://doi.org/10.1103/PhysRevD.93.045010}
  {\path{doi:10.1103/PhysRevD.93.045010}}.

\bibitem{Boca:2009zz}
M.~Boca, V.~Florescu, {Nonlinear Compton scattering with a laser pulse}, Phys.
  Rev. A 80 (2009) 053403.
\newblock \href {https://doi.org/10.1103/PhysRevA.80.053403}
  {\path{doi:10.1103/PhysRevA.80.053403}}.

\bibitem{Seipt:2010ya}
D.~Seipt, B.~Kämpfer, {Non-Linear Compton Scattering of Ultrashort and
  Ultraintense Laser Pulses}, Phys. Rev. A 83 (2011) 022101.
\newblock \href {http://arxiv.org/abs/1010.3301} {\path{arXiv:1010.3301}},
  \href {https://doi.org/10.1103/PhysRevA.83.022101}
  {\path{doi:10.1103/PhysRevA.83.022101}}.

\bibitem{Mackenroth:2010jr}
F.~Mackenroth, A.~Di~Piazza, {Nonlinear Compton scattering in ultra-short laser
  pulses}, Phys. Rev. A 83 (2011) 032106.
\newblock \href {http://arxiv.org/abs/1010.6251} {\path{arXiv:1010.6251}},
  \href {https://doi.org/10.1103/PhysRevA.83.032106}
  {\path{doi:10.1103/PhysRevA.83.032106}}.

\bibitem{Ilderton:2019bop}
A.~Ilderton, B.~King, A.~J. Macleod, {Absorption cross section in an intense
  plane wave background}, Phys. Rev. D 100~(7) (2019) 076002.
\newblock \href {http://arxiv.org/abs/1907.12835} {\path{arXiv:1907.12835}},
  \href {https://doi.org/10.1103/PhysRevD.100.076002}
  {\path{doi:10.1103/PhysRevD.100.076002}}.

\bibitem{Seipt:2011zz}
D.~Seipt, B.~Kämpfer, {Scaling law for the photon spectral density in the
  nonlinear Thomson-Compton scattering}, Phys. Rev. ST Accel. Beams 14 (2011)
  040704.
\newblock \href {https://doi.org/10.1103/PhysRevSTAB.14.040704}
  {\path{doi:10.1103/PhysRevSTAB.14.040704}}.

\bibitem{Acosta:2019bvh}
U.~Hernandez~Acosta, B.~K\"ampfer, {Laser pulse-length effects in trident pair
  production}, Plasma Phys. Control. Fusion 61~(8) (2019) 084011.
\newblock \href {http://arxiv.org/abs/1901.08860} {\path{arXiv:1901.08860}},
  \href {https://doi.org/10.1088/1361-6587/ab2b1e}
  {\path{doi:10.1088/1361-6587/ab2b1e}}.

\bibitem{landauQED}
V.~B. Berestetskii, E.~M. Lifshitz, L.~P. Pitaevskii, Quantum Electrodynamics
  (second edition), Butterworth-Heinemann, Oxford, 1982.

\bibitem{abramowitzStegun}
M.~Abramowitz, I.~A. Stegun, Handbook of Mathematical Functions with Formulas,
  Graphs, and Mathematical Tables, ninth dover printing, tenth gpo printing
  Edition, Dover, New York, 1964.

\bibitem{Seipt:2016rtk}
D.~Seipt, V.~Kharin, S.~Rykovanov, A.~Surzhykov, S.~Fritzsche, {Analytical
  results for nonlinear Compton scattering in short intense laser pulses}, J.
  Plasma Phys. 82~(2) (2016) 655820203.
\newblock \href {http://arxiv.org/abs/1601.00442} {\path{arXiv:1601.00442}},
  \href {https://doi.org/10.1017/S002237781600026X}
  {\path{doi:10.1017/S002237781600026X}}.

\bibitem{Meuren:2015mra}
S.~Meuren, C.~H. Keitel, A.~Di~Piazza, {Semiclassical picture for
  electron-positron photoproduction in strong laser fields}, Phys. Rev. D
  93~(8) (2016) 085028.
\newblock \href {http://arxiv.org/abs/1503.03271} {\path{arXiv:1503.03271}},
  \href {https://doi.org/10.1103/PhysRevD.93.085028}
  {\path{doi:10.1103/PhysRevD.93.085028}}.

\bibitem{Titov:2014usa}
A.~I. Titov, B.~Kämpfer, T.~Shibata, A.~Hosaka, H.~Takabe, {Laser pulse-shape
  dependence of Compton scattering}, Eur. Phys. J. D 68~(10) (2014) 299.
\newblock \href {http://arxiv.org/abs/1408.1040} {\path{arXiv:1408.1040}},
  \href {https://doi.org/10.1140/epjd/e2014-50324-y}
  {\path{doi:10.1140/epjd/e2014-50324-y}}.

\bibitem{Seipt:2016fyu}
D.~Seipt, T.~Heinzl, M.~Marklund, S.~S. Bulanov, {Depletion of Intense Fields},
  Phys. Rev. Lett. 118~(15) (2017) 154803.
\newblock \href {http://arxiv.org/abs/1605.00633} {\path{arXiv:1605.00633}},
  \href {https://doi.org/10.1103/PhysRevLett.118.154803}
  {\path{doi:10.1103/PhysRevLett.118.154803}}.

\bibitem{1996JETP...83...14N}
N.~B. {Narozhnyi}, M.~S. {Fofanov}, {Photon emission by an electron in a
  collision with a short focused laser pulse}, Soviet Journal of Experimental
  and Theoretical Physics 83~(1) (1996) 14--23.

\bibitem{Boca:2012pz}
M.~Boca, V.~Dinu, V.~Florescu, {Electron distributions in nonlinear Compton
  scattering}, Phys. Rev. A 86 (2012) 013414.
\newblock \href {http://arxiv.org/abs/1206.6971} {\path{arXiv:1206.6971}},
  \href {https://doi.org/10.1103/PhysRevA.86.013414}
  {\path{doi:10.1103/PhysRevA.86.013414}}.

\bibitem{Krajewska:2013poa}
K.~Krajewska, J.~Z. Kaminski, {Frequency scaling law for nonlinear Compton and
  Thomson scattering: Relevance of spin and polarization effects}, Phys. Rev. A
  90~(5) (2014) 052117.
\newblock \href {http://arxiv.org/abs/1308.1663} {\path{arXiv:1308.1663}},
  \href {https://doi.org/10.1103/PhysRevA.90.052117}
  {\path{doi:10.1103/PhysRevA.90.052117}}.

\bibitem{Seipt:2014yga}
D.~Seipt, S.~G. Rykovanov, A.~Surzhykov, S.~Fritzsche, {Narrowband inverse
  Compton scattering x-ray sources at high laser intensities}, Phys. Rev. A
  91~(3) (2015) 033402.
\newblock \href {http://arxiv.org/abs/1412.2659} {\path{arXiv:1412.2659}},
  \href {https://doi.org/10.1103/PhysRevA.91.033402}
  {\path{doi:10.1103/PhysRevA.91.033402}}.

\bibitem{Krajewska:2014fwa}
K.~Krajewska, M.~Twardy, J.~Z. Kami\'nski, {Global phase and frequency comb
  structures in nonlinear Compton and Thomson scattering}, Phys. Rev. A 89~(5)
  (2014) 052123.
\newblock \href {http://arxiv.org/abs/1403.4282} {\path{arXiv:1403.4282}},
  \href {https://doi.org/10.1103/PhysRevA.89.052123}
  {\path{doi:10.1103/PhysRevA.89.052123}}.

\bibitem{Krajewska:2015hua}
K.~Krajewska, F.~Cajiao~V\'elez, J.~Z. Kami\'nski, {Generalized Klein-Nishina
  formula}, Phys. Rev. A 91~(6) (2015) 062106.
\newblock \href {http://arxiv.org/abs/1503.06531} {\path{arXiv:1503.06531}},
  \href {https://doi.org/10.1103/PhysRevA.91.062106}
  {\path{doi:10.1103/PhysRevA.91.062106}}.

\bibitem{Dinu:2015aci}
V.~Dinu, C.~Harvey, A.~Ilderton, M.~Marklund, G.~Torgrimsson, {Quantum
  radiation reaction: from interference to incoherence}, Phys. Rev. Lett.
  116~(4) (2016) 044801.
\newblock \href {http://arxiv.org/abs/1512.04096} {\path{arXiv:1512.04096}},
  \href {https://doi.org/10.1103/PhysRevLett.116.044801}
  {\path{doi:10.1103/PhysRevLett.116.044801}}.

\bibitem{2010PhRvL.105m0801H}
F.~V. {Hartemann}, F.~{Albert}, C.~W. {Siders}, C.~P.~J. {Barty},
  {Low-Intensity Nonlinear Spectral Effects in Compton Scattering}, Phys. Rev.
  Lett. 105~(13) (2010) 130801.
\newblock \href {https://doi.org/10.1103/PhysRevLett.105.130801}
  {\path{doi:10.1103/PhysRevLett.105.130801}}.

\bibitem{Harvey:2016wte}
C.~Harvey, M.~Marklund, A.~R. Holkundkar, {Focussing effects in laser-electron
  Thomson scattering}, Phys. Rev. Accel. Beams 19~(9) (2016) 094701.
\newblock \href {http://arxiv.org/abs/1606.05776} {\path{arXiv:1606.05776}},
  \href {https://doi.org/10.1103/PhysRevAccelBeams.19.094701}
  {\path{doi:10.1103/PhysRevAccelBeams.19.094701}}.

\bibitem{2013LaPhy..23g5301S}
D.~{Seipt}, B.~{Kämpfer}, {Nonlinear Compton scattering of ultrahigh-intensity
  laser pulses}, Laser Physics 23~(7) (2013) 075301.
\newblock \href {https://doi.org/10.1088/1054-660X/23/7/075301}
  {\path{doi:10.1088/1054-660X/23/7/075301}}.

\bibitem{Hartemann:1996zz}
F.~V. Hartemann, A.~L. Troha, N.~C. Luhmann, Z.~Toffano, {Spectral analysis of
  the nonlinear relativistic Doppler shift in ultrahigh intensity Compton
  scattering}, Phys. Rev. E 54 (1996) 2956--2962.
\newblock \href {https://doi.org/10.1103/PhysRevE.54.2956}
  {\path{doi:10.1103/PhysRevE.54.2956}}.

\bibitem{Krajewska:2013tla}
K.~Krajewska, M.~Twardy, J.~Z. Kami\'nski, {Supercontinuum and ultrashort-pulse
  generation from nonlinear Thomson and Compton scattering}, Phys. Rev. A
  89~(3) (2014) 032125.
\newblock \href {http://arxiv.org/abs/1311.4872} {\path{arXiv:1311.4872}},
  \href {https://doi.org/10.1103/PhysRevA.89.032125}
  {\path{doi:10.1103/PhysRevA.89.032125}}.

\bibitem{Twardy:2013jca}
M.~Twardy, K.~Krajewska, J.~Z. Kami\'nski, {Shape effects in nonlinear Thomson
  and Compton processes}, J. Phys. Conf. Ser. 497 (2014) 012019.
\newblock \href {http://arxiv.org/abs/1311.5374} {\path{arXiv:1311.5374}},
  \href {https://doi.org/10.1088/1742-6596/497/1/012019}
  {\path{doi:10.1088/1742-6596/497/1/012019}}.

\bibitem{Ilderton:2020dhs}
A.~Ilderton, B.~King, S.~Tang, {Toward the observation of interference effects
  in nonlinear Compton scattering}, Phys. Lett. B 804 (2020) 135410.
\newblock \href {http://arxiv.org/abs/2002.04629} {\path{arXiv:2002.04629}},
  \href {https://doi.org/10.1016/j.physletb.2020.135410}
  {\path{doi:10.1016/j.physletb.2020.135410}}.

\bibitem{2000JETP...90..415N}
N.~B. {Narozhny}, M.~S. {Fofanov}, {Quantum Processes in a Two-Mode Laser
  Field}, Soviet Journal of Experimental and Theoretical Physics 90~(3) (2000)
  415--427.
\newblock \href {https://doi.org/10.1134/1.559121}
  {\path{doi:10.1134/1.559121}}.

\bibitem{Wistisen:2014ysa}
T.~N. Wistisen, {Interference effect in nonlinear Compton scattering}, Phys.
  Rev. D 90~(12) (2014) 125008, [Erratum: Phys.Rev.D 91, 069903 (2015)].
\newblock \href {https://doi.org/10.1103/PhysRevD.90.125008}
  {\path{doi:10.1103/PhysRevD.90.125008}}.

\bibitem{Titov:2015pre}
A.~I. Titov, B.~Kämpfer, A.~Hosaka, H.~Takabe, {Quantum processes in short and
  intensive electromagnetic fields}, Phys. Part. Nucl. 47~(3) (2016) 456--487.
\newblock \href {http://arxiv.org/abs/1512.07987} {\path{arXiv:1512.07987}},
  \href {https://doi.org/10.1134/S1063779616030059}
  {\path{doi:10.1134/S1063779616030059}}.

\bibitem{Titov:2019kdk}
A.~I. Titov, A.~Otto, B.~Kämpfer, {Multi-photon regime of non-linear
  Breit-Wheeler and Compton processes in short linearly and circularly
  polarized laser pulses}, Eur. Phys. J. D 74~(2) (2020) 39.
\newblock \href {http://arxiv.org/abs/1907.00643} {\path{arXiv:1907.00643}},
  \href {https://doi.org/10.1140/epjd/e2020-100527-6}
  {\path{doi:10.1140/epjd/e2020-100527-6}}.

\bibitem{HernandezAcosta:2020agu}
U.~Hernandez~Acosta, A.~Otto, B.~Kämpfer, A.~I. Titov, {Nonperturbative
  signatures of nonlinear Compton scattering}, Phys. Rev. D 102~(11) (2020)
  116016.
\newblock \href {http://arxiv.org/abs/2001.03986} {\path{arXiv:2001.03986}},
  \href {https://doi.org/10.1103/PhysRevD.102.116016}
  {\path{doi:10.1103/PhysRevD.102.116016}}.

\bibitem{Kampfer:2020cbx}
B.~Kämpfer, A.~I. Titov, {Impact of laser polarization on q-exponential photon
  tails in non-linear Compton scattering}, Phys. Rev. A 103 (2021) 033101.
\newblock \href {http://arxiv.org/abs/2012.07699} {\path{arXiv:2012.07699}},
  \href {https://doi.org/10.1103/PhysRevA.103.033101}
  {\path{doi:10.1103/PhysRevA.103.033101}}.

\bibitem{Dinu:2018efz}
V.~Dinu, G.~Torgrimsson, {Single and double nonlinear Compton scattering},
  Phys. Rev. D 99~(9) (2019) 096018.
\newblock \href {http://arxiv.org/abs/1811.00451} {\path{arXiv:1811.00451}},
  \href {https://doi.org/10.1103/PhysRevD.99.096018}
  {\path{doi:10.1103/PhysRevD.99.096018}}.

\bibitem{Blackburn:2017dpn}
T.~G. Blackburn, A.~Ilderton, C.~D. Murphy, M.~Marklund, {Scaling laws for
  positron production in laser\textendash{}electron-beam collisions}, Phys.
  Rev. A 96~(2) (2017) 022128.
\newblock \href {http://arxiv.org/abs/1708.00298} {\path{arXiv:1708.00298}},
  \href {https://doi.org/10.1103/PhysRevA.96.022128}
  {\path{doi:10.1103/PhysRevA.96.022128}}.

\bibitem{Acosta:2021iyu}
U.~Hernandez~Acosta, A.~I. Titov, B.~Kämpfer, {Rise and fall of
  laser-intensity effects in spectrally resolved Compton process}, New J. Phys.
  23~(9) (2021) 095008.
\newblock \href {http://arxiv.org/abs/2105.11758} {\path{arXiv:2105.11758}},
  \href {https://doi.org/10.1088/1367-2630/ac21e0}
  {\path{doi:10.1088/1367-2630/ac21e0}}.

\bibitem{Angioi:2017ygv}
A.~Angioi, A.~Di~Piazza, {Quantum Limitation to the Coherent Emission of
  Accelerated Charges}, Phys. Rev. Lett. 121~(1) (2018) 010402.
\newblock \href {http://arxiv.org/abs/1712.01123} {\path{arXiv:1712.01123}},
  \href {https://doi.org/10.1103/PhysRevLett.121.010402}
  {\path{doi:10.1103/PhysRevLett.121.010402}}.

\bibitem{Mackenroth:2010jk}
F.~Mackenroth, A.~Di~Piazza, C.~H. Keitel, {Determining the carrier-envelope
  phase of intense few-cycle laser pulses}, Phys. Rev. Lett. 105 (2010) 063903.
\newblock \href {http://arxiv.org/abs/1001.3614} {\path{arXiv:1001.3614}},
  \href {https://doi.org/10.1103/PhysRevLett.105.063903}
  {\path{doi:10.1103/PhysRevLett.105.063903}}.

\bibitem{Corde:RevModPhys2013}
S.~Corde, K.~{Ta Phuoc}, G.~Lambert, R.~F.~V. Malka, A.~Rousse, A.~Beck,
  E.~Lefebvre, {Femtosecond x rays from laser-plasma accelerators}, Rev. Mod.
  Phys. 85 (2013) 1.

\bibitem{Rykovanov:2014ira}
S.~G. Rykovanov, C.~G.~R. Geddes, J.~L. Vay, C.~B. Schroeder, E.~Esarey, W.~P.
  Leemans, {Quasi-monoenergetic femtosecond photon sources from Thomson
  Scattering using laser plasma accelerators and plasma channels}, J. Phys. B
  47~(23) (2014) 234013.
\newblock \href {http://arxiv.org/abs/1406.1832} {\path{arXiv:1406.1832}},
  \href {https://doi.org/10.1088/0953-4075/47/23/234013}
  {\path{doi:10.1088/0953-4075/47/23/234013}}.

\bibitem{Albert:2016}
F.~Albert, A.~G.~R. Thomas,
  \href{http://stacks.iop.org/0741-3335/58/i=10/a=103001?key=crossref.2936c109ef8335882064df4972548675}{{Applications
  of laser wakefield accelerator-based light sources}}, Plasma Phys. Control.
  Fusion 58~(10) (2016) 103001.
\newblock \href {https://doi.org/10.1088/0741-3335/58/10/103001}
  {\path{doi:10.1088/0741-3335/58/10/103001}}.
\newline\urlprefix\url{http://stacks.iop.org/0741-3335/58/i=10/a=103001?key=crossref.2936c109ef8335882064df4972548675}

\bibitem{2013PhRvS..16c0705G}
I.~{Ghebregziabher}, B.~A. {Shadwick}, D.~{Umstadter}, {Spectral bandwidth
  reduction of Thomson scattered light by pulse chirping}, Physical Review
  Accelerators and Beams 16~(3) (2013) 030705.
\newblock \href {https://doi.org/10.1103/PhysRevSTAB.16.030705}
  {\path{doi:10.1103/PhysRevSTAB.16.030705}}.

\bibitem{Terzic:2013ysa}
B.~Terzi\'c, K.~Deitrick, A.~S. Hofler, G.~A. Krafft, {Narrow-Band Emission in
  Thomson Sources Operating in the High-Field Regime}, Phys. Rev. Lett. 112~(7)
  (2014) 074801.
\newblock \href {http://arxiv.org/abs/1307.0895} {\path{arXiv:1307.0895}},
  \href {https://doi.org/10.1103/PhysRevLett.112.074801}
  {\path{doi:10.1103/PhysRevLett.112.074801}}.

\bibitem{Rykovanov:2016ndn}
S.~G. Rykovanov, C.~G.~R. Geddes, C.~B. Schroeder, E.~Esarey, W.~P. Leemans,
  {Controlling the spectral shape of nonlinear Thomson scattering with proper
  laser chirping}, Phys. Rev. Accel. Beams 19~(3) (2016) 030701.
\newblock \href {https://doi.org/10.1103/PhysRevAccelBeams.19.030701}
  {\path{doi:10.1103/PhysRevAccelBeams.19.030701}}.

\bibitem{Seipt:2019yds}
D.~Seipt, V.~Y. Kharin, S.~G. Rykovanov, {Optimizing Laser Pulses for
  Narrow-Band Inverse Compton Sources in the High-Intensity Regime}, Phys. Rev.
  Lett. 122~(20) (2019) 204802.
\newblock \href {http://arxiv.org/abs/1902.10777} {\path{arXiv:1902.10777}},
  \href {https://doi.org/10.1103/PhysRevLett.122.204802}
  {\path{doi:10.1103/PhysRevLett.122.204802}}.

\bibitem{Terzic:2019eoe}
B.~Terzi\'c, A.~Brown, I.~Drebot, T.~Hagerman, E.~Johnson, G.~Krafft,
  C.~Maroli, V.~Petrillo, M.~Ruijter, {Improving performance of inverse Compton
  sources through laser chirping}, EPL 126~(1) (2019) 12003.
\newblock \href {http://arxiv.org/abs/1902.04240} {\path{arXiv:1902.04240}},
  \href {https://doi.org/10.1209/0295-5075/126/12003}
  {\path{doi:10.1209/0295-5075/126/12003}}.

\bibitem{Valialshchikov:2020dhq}
M.~A. Valialshchikov, V.~Y. Kharin, S.~G. Rykovanov, {Narrow bandwidth gamma
  comb from nonlinear Compton scattering using the polarization gating
  technique}, Phys. Rev. Lett. 126~(19) (2021) 194801.
\newblock \href {http://arxiv.org/abs/2011.12931} {\path{arXiv:2011.12931}},
  \href {https://doi.org/10.1103/PhysRevLett.126.194801}
  {\path{doi:10.1103/PhysRevLett.126.194801}}.

\bibitem{Holkundkar:2015rya}
A.~R. Holkundkar, C.~Harvey, M.~Marklund, {Thomson scattering in high-intensity
  chirped laser pulses}, Phys. Plasmas 22 (2015) 103103.
\newblock \href {http://arxiv.org/abs/1507.08265} {\path{arXiv:1507.08265}},
  \href {https://doi.org/10.1063/1.4932995} {\path{doi:10.1063/1.4932995}}.

\bibitem{Ruijter:2021scx}
M.~Ruijter, V.~Petrillo, M.~Zepf, {Decreasing the bandwidth of linear and
  nonlinear Thomson scattering radiation for electron bunches with a finite
  energy spread}, Phys. Rev. Accel. Beams 24~(2) (2021) 020702.
\newblock \href {https://doi.org/10.1103/PhysRevAccelBeams.24.020702}
  {\path{doi:10.1103/PhysRevAccelBeams.24.020702}}.

\bibitem{Kharin:2018dxa}
V.~Y. Kharin, D.~Seipt, S.~G. Rykovanov, {Higher-Dimensional Caustics in
  Nonlinear Compton Scattering}, Phys. Rev. Lett. 120~(4) (2018) 044802.
\newblock \href {https://doi.org/10.1103/PhysRevLett.120.044802}
  {\path{doi:10.1103/PhysRevLett.120.044802}}.

\bibitem{Reimann:2007}
K.~Reimann, \href{https://doi.org/10.1088/0034-4885/70/10/r02}{Table-top
  sources of ultrashort {THz} pulses} 70~(10) (2007) 1597--1632.
\newblock \href {https://doi.org/10.1088/0034-4885/70/10/r02}
  {\path{doi:10.1088/0034-4885/70/10/r02}}.
\newline\urlprefix\url{https://doi.org/10.1088/0034-4885/70/10/r02}

\bibitem{DiPiazza:2017raw}
A.~Di~Piazza, M.~Tamburini, S.~Meuren, C.~H. Keitel, {Implementing nonlinear
  Compton scattering beyond the local constant field approximation}, Phys. Rev.
  A 98~(1) (2018) 012134.
\newblock \href {http://arxiv.org/abs/1708.08276} {\path{arXiv:1708.08276}},
  \href {https://doi.org/10.1103/PhysRevA.98.012134}
  {\path{doi:10.1103/PhysRevA.98.012134}}.

\bibitem{Edwards:2020npu}
J.~P. Edwards, A.~Ilderton, {Resummation of background-collinear corrections in
  strong field QED}, Phys. Rev. D 103~(1) (2021) 016004.
\newblock \href {http://arxiv.org/abs/2010.02085} {\path{arXiv:2010.02085}},
  \href {https://doi.org/10.1103/PhysRevD.103.016004}
  {\path{doi:10.1103/PhysRevD.103.016004}}.

\bibitem{DiPiazza:2018luu}
A.~Di~Piazza, {Analytical Infrared Limit of Nonlinear Thomson Scattering
  Including Radiation Reaction}, Phys. Lett. B 782 (2018) 559--565.
\newblock \href {http://arxiv.org/abs/1804.01160} {\path{arXiv:1804.01160}},
  \href {https://doi.org/10.1016/j.physletb.2018.05.081}
  {\path{doi:10.1016/j.physletb.2018.05.081}}.

\bibitem{Baier:1975ff}
V.~N. Baier, A.~I. Milshtein, V.~M. Strakhovenko, {Interaction Between a Photon
  and a High Intensity Electromagnetic Wave}, Sov. Phys. JETP 42~(6) (1975)
  961--965.

\bibitem{King:2013zw}
B.~King, N.~Elkina, H.~Ruhl, {Photon polarisation in electron-seeded
  pair-creation cascades}, Phys. Rev. A 87 (2013) 042117.
\newblock \href {http://arxiv.org/abs/1301.7001} {\path{arXiv:1301.7001}},
  \href {https://doi.org/10.1103/PhysRevA.87.042117}
  {\path{doi:10.1103/PhysRevA.87.042117}}.

\bibitem{King:2020btz}
B.~King, S.~Tang, {Nonlinear Compton scattering of polarized photons in
  plane-wave backgrounds}, Phys. Rev. A 102~(2) (2020) 022809.
\newblock \href {http://arxiv.org/abs/2003.01749} {\path{arXiv:2003.01749}},
  \href {https://doi.org/10.1103/PhysRevA.102.022809}
  {\path{doi:10.1103/PhysRevA.102.022809}}.

\bibitem{Tang:2020xlj}
S.~Tang, B.~King, H.~Hu, {Highly polarised gamma photons from electron-laser
  collisions}, Phys. Lett. B 809 (2020) 135701.
\newblock \href {http://arxiv.org/abs/2003.03246} {\path{arXiv:2003.03246}},
  \href {https://doi.org/10.1016/j.physletb.2020.135701}
  {\path{doi:10.1016/j.physletb.2020.135701}}.

\bibitem{Wistisen:2019tgu}
T.~N. Wistisen, A.~Di~Piazza, {Numerical approach to the semiclassical method
  of radiation emission for arbitrary electron spin and photon polarization},
  Phys. Rev. D 100~(11) (2019) 116001.
\newblock \href {http://arxiv.org/abs/1909.12899} {\path{arXiv:1909.12899}},
  \href {https://doi.org/10.1103/PhysRevD.100.116001}
  {\path{doi:10.1103/PhysRevD.100.116001}}.

\bibitem{Seipt:2018adi}
D.~Seipt, D.~Del~Sorbo, C.~P. Ridgers, A.~G.~R. Thomas, {Theory of radiative
  electron polarization in strong laser fields}, Phys. Rev. A 98~(2) (2018)
  023417.
\newblock \href {http://arxiv.org/abs/1805.02027} {\path{arXiv:1805.02027}},
  \href {https://doi.org/10.1103/PhysRevA.98.023417}
  {\path{doi:10.1103/PhysRevA.98.023417}}.

\bibitem{Dinu:2019pau}
V.~Dinu, G.~Torgrimsson, {Approximating higher-order nonlinear QED processes
  with first-order building blocks}, Phys. Rev. D 102~(1) (2020) 016018.
\newblock \href {http://arxiv.org/abs/1912.11015} {\path{arXiv:1912.11015}},
  \href {https://doi.org/10.1103/PhysRevD.102.016018}
  {\path{doi:10.1103/PhysRevD.102.016018}}.

\bibitem{Torgrimsson:2020gws}
G.~Torgrimsson, {Loops and polarization in strong-field QED}, New J. Phys.
  23~(6) (2021) 065001.
\newblock \href {http://arxiv.org/abs/2012.12701} {\path{arXiv:2012.12701}},
  \href {https://doi.org/10.1088/1367-2630/abf274}
  {\path{doi:10.1088/1367-2630/abf274}}.

\bibitem{Seipt:2020diz}
D.~Seipt, B.~King, {Spin- and polarization-dependent
  locally-constant-field-approximation rates for nonlinear Compton and
  Breit-Wheeler processes}, Phys. Rev. A 102~(5) (2020) 052805.
\newblock \href {http://arxiv.org/abs/2007.11837} {\path{arXiv:2007.11837}},
  \href {https://doi.org/10.1103/PhysRevA.102.052805}
  {\path{doi:10.1103/PhysRevA.102.052805}}.

\bibitem{book:densitymatrix}
K.~Blum, {Density Matrix Theory and Applications}, 3rd Edition, Vol.~64 of
  Springer Series on Atomic, Optical, and Plasma Physics, Springer-Verlag,
  Berlin, Heidelberg, New York, 2012.

\bibitem{Seipt:2019ddd}
D.~Seipt, D.~Del~Sorbo, C.~P. Ridgers, A.~G.~R. Thomas, {Ultrafast polarization
  of an electron beam in an intense bichromatic laser field}, Phys. Rev. A
  100~(6) (2019) 061402.
\newblock \href {http://arxiv.org/abs/1904.12037} {\path{arXiv:1904.12037}},
  \href {https://doi.org/10.1103/PhysRevA.100.061402}
  {\path{doi:10.1103/PhysRevA.100.061402}}.

\bibitem{Chen:2019vly}
Y.-Y. Chen, P.-L. He, R.~Shaisultanov, K.~Z. Hatsagortsyan, C.~H. Keitel,
  {Polarized positron beams via intense two-color laser pulses}, Phys. Rev.
  Lett. 123 (2019) 174801.
\newblock \href {http://arxiv.org/abs/1904.04110} {\path{arXiv:1904.04110}},
  \href {https://doi.org/10.1103/PhysRevLett.123.174801}
  {\path{doi:10.1103/PhysRevLett.123.174801}}.

\bibitem{Baier:1967}
V.~N. Baier, V.~M. Katkov, {Radiative Polarization of Electrons in a Magnetic
  Field}, Sov. Phys. J. Exp. Theor. Phys. 25~(5) (1967) 944--947.

\bibitem{BaierKatkovFadin}
V.~N. Baier, V.~M. Katkov, V.~S. Fadin, Radiation of relativistic electrons;
  Izluchenie relyativistskikh elektronov, Atomizdat, Moscow, 1973.

\bibitem{Li:2018fcz}
Y.-F. Li, R.~Shaisultanov, K.~Z. Hatsagortsyan, F.~Wan, C.~H. Keitel, J.-X. Li,
  {Ultrarelativistic electron beam polarization in single-shot interaction with
  an ultraintense laser pulse}, Phys. Rev. Lett. 122~(15) (2019) 154801.
\newblock \href {http://arxiv.org/abs/1812.07229} {\path{arXiv:1812.07229}},
  \href {https://doi.org/10.1103/PhysRevLett.122.154801}
  {\path{doi:10.1103/PhysRevLett.122.154801}}.

\bibitem{Li:2019oxr}
Y.-F. Li, R.~Shaisultanov, Y.-Y. Chen, F.~Wan, K.~Z. Hatsagortsyan, C.~H.
  Keitel, J.-X. Li, {Polarized Ultrashort Brilliant Multi-GeV $\gamma$ Rays via
  Single-Shot Laser-Electron Interaction}, Phys. Rev. Lett. 124~(1) (2020)
  014801.
\newblock \href {http://arxiv.org/abs/1907.08877} {\path{arXiv:1907.08877}},
  \href {https://doi.org/10.1103/PhysRevLett.124.014801}
  {\path{doi:10.1103/PhysRevLett.124.014801}}.

\bibitem{Chen:2022dgo}
Y.-Y. Chen, K.~Z. Hatsagortsyan, C.~H. Keitel, R.~Shaisultanov, {Electron spin-
  and photon polarization-resolved probabilities of strong-field QED processes}
  (1 2022).
\newblock \href {http://arxiv.org/abs/2201.10863} {\path{arXiv:2201.10863}}.

\bibitem{Ivanov:2004fi}
D.~Y. Ivanov, G.~L. Kotkin, V.~G. Serbo, {Complete description of polarization
  effects in emission of a photon by an electron in the field of a strong laser
  wave}, Eur. Phys. J. C 36 (2004) 127--145.
\newblock \href {http://arxiv.org/abs/hep-ph/0402139}
  {\path{arXiv:hep-ph/0402139}}, \href
  {https://doi.org/10.1140/epjc/s2004-01861-x}
  {\path{doi:10.1140/epjc/s2004-01861-x}}.

\bibitem{Li:2019wkl}
Y.-F. Li, R.-T. Guo, R.~Shaisultanov, K.~Z. Hatsagortsyan, J.-X. Li, {Electron
  Polarimetry with Nonlinear Compton Scattering}, Phys. Rev. Applied 12~(1)
  (2019) 014047.
\newblock \href {https://doi.org/10.1103/PhysRevApplied.12.014047}
  {\path{doi:10.1103/PhysRevApplied.12.014047}}.

\bibitem{Ahrens:2021kww}
S.~Ahrens, Z.~Guan, B.~Shen, {Beam focus and longitudinal polarization
  influence on spin dynamics in the Kapitza-Dirac effect} (3 2021).
\newblock \href {http://arxiv.org/abs/2103.07121} {\path{arXiv:2103.07121}}.

\bibitem{Ekman:2020fdp}
R.~Ekman, T.~Heinzl, A.~Ilderton, {High-intensity scaling in UV-modified QED},
  Phys. Rev. D 102~(11) (2020) 116005.
\newblock \href {http://arxiv.org/abs/2010.03893} {\path{arXiv:2010.03893}},
  \href {https://doi.org/10.1103/PhysRevD.102.116005}
  {\path{doi:10.1103/PhysRevD.102.116005}}.

\bibitem{Krajewska:2012eb}
K.~Krajewska, J.~Z. Kaminski, {Breit-Wheeler Process in Intense Short Laser
  Pulses}, Phys. Rev. A 86 (2012) 052104.
\newblock \href {http://arxiv.org/abs/1209.2394} {\path{arXiv:1209.2394}},
  \href {https://doi.org/10.1103/PhysRevA.86.052104}
  {\path{doi:10.1103/PhysRevA.86.052104}}.

\bibitem{Titov:2012rd}
A.~I. Titov, H.~Takabe, B.~Kämpfer, A.~Hosaka, {Enhanced subthreshold
  electron-positron production in short laser pulses}, Phys. Rev. Lett. 108
  (2012) 240406.
\newblock \href {http://arxiv.org/abs/1205.3880} {\path{arXiv:1205.3880}},
  \href {https://doi.org/10.1103/PhysRevLett.108.240406}
  {\path{doi:10.1103/PhysRevLett.108.240406}}.

\bibitem{Breit:1934zz}
G.~Breit, J.~A. Wheeler, {Collision of two light quanta}, Phys. Rev. 46~(12)
  (1934) 1087--1091.
\newblock \href {https://doi.org/10.1103/PhysRev.46.1087}
  {\path{doi:10.1103/PhysRev.46.1087}}.

\bibitem{Karbstein:2013ufa}
F.~Karbstein, {Photon polarization tensor in a homogeneous magnetic or electric
  field}, Phys. Rev. D 88~(8) (2013) 085033.
\newblock \href {http://arxiv.org/abs/1308.6184} {\path{arXiv:1308.6184}},
  \href {https://doi.org/10.1103/PhysRevD.88.085033}
  {\path{doi:10.1103/PhysRevD.88.085033}}.

\bibitem{Blackburn:2018ghi}
T.~G. Blackburn, M.~Marklund, {Nonlinear Breit-Wheeler pair creation with
  bremsstrahlung $\gamma$ rays}, Plasma Phys. Control. Fusion 60~(5) (2018)
  054009.
\newblock \href {http://arxiv.org/abs/1802.06612} {\path{arXiv:1802.06612}},
  \href {https://doi.org/10.1088/1361-6587/aab3b4}
  {\path{doi:10.1088/1361-6587/aab3b4}}.

\bibitem{Eckey:2021rgn}
A.~Eckey, A.~B. Voitkiv, C.~M\"uller, {Strong-field Breit-Wheeler pair
  production with bremsstrahlung \ensuremath{\gamma} rays in the
  perturbative-to-nonperturbative-transition regime}, Phys. Rev. A 105~(1)
  (2022) 013105.
\newblock \href {http://arxiv.org/abs/2107.10058} {\path{arXiv:2107.10058}},
  \href {https://doi.org/10.1103/PhysRevA.105.013105}
  {\path{doi:10.1103/PhysRevA.105.013105}}.

\bibitem{Golub:2021nhj}
A.~Golub, S.~Villalba-Ch\'avez, C.~M\"uller, {Strong-field Breit-Wheeler pair
  production in QED$_{2+1}$}, Phys. Rev. D 103~(9) (2021) 096002.
\newblock \href {http://arxiv.org/abs/2103.02290} {\path{arXiv:2103.02290}},
  \href {https://doi.org/10.1103/PhysRevD.103.096002}
  {\path{doi:10.1103/PhysRevD.103.096002}}.

\bibitem{He:2012un}
L.-Y. He, B.-S. Xie, X.-H. Guo, H.-Y. Wang, {Electron-positron pair production
  in an arbitrary polarized ultrastrong laser field}, Commun. Theor. Phys. 58
  (2012) 863--871.
\newblock \href {http://arxiv.org/abs/1208.2077} {\path{arXiv:1208.2077}},
  \href {https://doi.org/10.1088/0253-6102/58/6/13}
  {\path{doi:10.1088/0253-6102/58/6/13}}.

\bibitem{Meuren:2014uia}
S.~Meuren, K.~Z. Hatsagortsyan, C.~H. Keitel, A.~Di~Piazza, {Polarization
  operator approach to pair creation in short laser pulses}, Phys. Rev. D
  91~(1) (2015) 013009.
\newblock \href {http://arxiv.org/abs/1406.7235} {\path{arXiv:1406.7235}},
  \href {https://doi.org/10.1103/PhysRevD.91.013009}
  {\path{doi:10.1103/PhysRevD.91.013009}}.

\bibitem{Mercuri-Baron:2021waq}
A.~Mercuri-Baron, M.~Grech, F.~Niel, A.~Grassi, M.~Lobet, A.~Di~Piazza,
  C.~Riconda, {Impact of the laser spatio-temporal shape on
  Breit\textendash{}Wheeler pair production}, New J. Phys. 23~(8) (2021)
  085006.
\newblock \href {http://arxiv.org/abs/2105.12458} {\path{arXiv:2105.12458}},
  \href {https://doi.org/10.1088/1367-2630/ac1975}
  {\path{doi:10.1088/1367-2630/ac1975}}.

\bibitem{Nousch:2012xe}
T.~Nousch, D.~Seipt, B.~Kämpfer, A.~I. Titov, {Pair production in short laser
  pulses near threshold}, Phys. Lett. B 715 (2012) 246--250.
\newblock \href {https://doi.org/10.1016/j.physletb.2012.07.040}
  {\path{doi:10.1016/j.physletb.2012.07.040}}.

\bibitem{Ilderton:2019vot}
A.~Ilderton, {Exact results for scattering on ultrashort plane wave
  backgrounds}, Phys. Rev. D 100~(12) (2019) 125018.
\newblock \href {http://arxiv.org/abs/1909.02484} {\path{arXiv:1909.02484}},
  \href {https://doi.org/10.1103/PhysRevD.100.125018}
  {\path{doi:10.1103/PhysRevD.100.125018}}.

\bibitem{Titov:2013kya}
A.~I. Titov, B.~K\"ampfer, H.~Takabe, A.~Hosaka, {Breit-Wheeler process in very
  short electromagnetic pulses}, Phys. Rev. A 87~(4) (2013) 042106.
\newblock \href {http://arxiv.org/abs/1303.6487} {\path{arXiv:1303.6487}},
  \href {https://doi.org/10.1103/PhysRevA.87.042106}
  {\path{doi:10.1103/PhysRevA.87.042106}}.

\bibitem{Ilderton:2019ceq}
A.~Ilderton, {Coherent quantum enhancement of pair production in the null
  domain}, Phys. Rev. D 101~(1) (2020) 016006.
\newblock \href {http://arxiv.org/abs/1910.03012} {\path{arXiv:1910.03012}},
  \href {https://doi.org/10.1103/PhysRevD.101.016006}
  {\path{doi:10.1103/PhysRevD.101.016006}}.

\bibitem{Akkermans:2011yn}
E.~Akkermans, G.~V. Dunne, {Ramsey Fringes and Time-domain Multiple-Slit
  Interference from Vacuum}, Phys. Rev. Lett. 108 (2012) 030401.
\newblock \href {http://arxiv.org/abs/1109.3489} {\path{arXiv:1109.3489}},
  \href {https://doi.org/10.1103/PhysRevLett.108.030401}
  {\path{doi:10.1103/PhysRevLett.108.030401}}.

\bibitem{Jansen:2016crq}
M.~J.~A. Jansen, C.~M\"uller, {Strong-Field Breit-Wheeler Pair Production in
  Two Consecutive Laser Pulses with Variable Time Delay}, Phys. Lett. B 766
  (2017) 71--76.
\newblock \href {http://arxiv.org/abs/1612.07137} {\path{arXiv:1612.07137}},
  \href {https://doi.org/10.1016/j.physletb.2016.12.056}
  {\path{doi:10.1016/j.physletb.2016.12.056}}.

\bibitem{Granz:2019sxb}
L.~F. Granz, O.~Mathiak, S.~Villalba-Ch\'avez, C.~M\"uller, {Electron-positron
  pair production in oscillating electric fields with double-pulse structure},
  Phys. Lett. B 793 (2019) 85--89.
\newblock \href {http://arxiv.org/abs/1903.06000} {\path{arXiv:1903.06000}},
  \href {https://doi.org/10.1016/j.physletb.2019.04.026}
  {\path{doi:10.1016/j.physletb.2019.04.026}}.

\bibitem{Titov:2018bgy}
A.~I. Titov, B.~Kämpfer, H.~Takabe, {Nonlinear Breit-Wheeler process in short
  laser double pulses}, Phys. Rev. D 98~(3) (2018) 036022.
\newblock \href {http://arxiv.org/abs/1807.04547} {\path{arXiv:1807.04547}},
  \href {https://doi.org/10.1103/PhysRevD.98.036022}
  {\path{doi:10.1103/PhysRevD.98.036022}}.

\bibitem{Krajewska:2014ssa}
K.~Krajewska, J.~Z. Kami\'nski, {Coherent combs of antimatter from nonlinear
  electron-positron-pair creation}, Phys. Rev. A 90~(5) (2014) 052108.
\newblock \href {http://arxiv.org/abs/1407.4101} {\path{arXiv:1407.4101}},
  \href {https://doi.org/10.1103/PhysRevA.90.052108}
  {\path{doi:10.1103/PhysRevA.90.052108}}.

\bibitem{Jansen:2013dea}
M.~J.~A. Jansen, C.~M\"uller, {Strongly enhanced pair production in combined
  high- and low-frequency laser fields}, Phys. Rev. A 88~(5) (2013) 052125.
\newblock \href {http://arxiv.org/abs/1309.1069} {\path{arXiv:1309.1069}},
  \href {https://doi.org/10.1103/PhysRevA.88.052125}
  {\path{doi:10.1103/PhysRevA.88.052125}}.

\bibitem{Schutzhold:2008pz}
R.~Schutzhold, H.~Gies, G.~Dunne, {Dynamically assisted Schwinger mechanism},
  Phys. Rev. Lett. 101 (2008) 130404.
\newblock \href {http://arxiv.org/abs/0807.0754} {\path{arXiv:0807.0754}},
  \href {https://doi.org/10.1103/PhysRevLett.101.130404}
  {\path{doi:10.1103/PhysRevLett.101.130404}}.

\bibitem{Dunne:2009gi}
G.~V. Dunne, H.~Gies, R.~Schutzhold, {Catalysis of Schwinger Vacuum Pair
  Production}, Phys. Rev. D 80 (2009) 111301.
\newblock \href {http://arxiv.org/abs/0908.0948} {\path{arXiv:0908.0948}},
  \href {https://doi.org/10.1103/PhysRevD.80.111301}
  {\path{doi:10.1103/PhysRevD.80.111301}}.

\bibitem{Nousch:2015pja}
T.~Nousch, D.~Seipt, B.~Kämpfer, A.~I. Titov, {Spectral caustics in laser
  assisted Breit\textendash{}Wheeler process}, Phys. Lett. B 755 (2016)
  162--167.
\newblock \href {http://arxiv.org/abs/1509.01983} {\path{arXiv:1509.01983}},
  \href {https://doi.org/10.1016/j.physletb.2016.01.062}
  {\path{doi:10.1016/j.physletb.2016.01.062}}.

\bibitem{Otto:2016fdo}
A.~Otto, T.~Nousch, D.~Seipt, B.~Kämpfer, D.~Blaschke, A.~D. Panferov, S.~A.
  Smolyansky, A.~I. Titov, {Pair production by Schwinger and
  Breit\textendash{}Wheeler processes in bi-frequent fields}, J. Plasma Phys.
  82~(3) (2016) 655820301.
\newblock \href {http://arxiv.org/abs/1604.00196} {\path{arXiv:1604.00196}},
  \href {https://doi.org/10.1017/S0022377816000428}
  {\path{doi:10.1017/S0022377816000428}}.

\bibitem{Jansen:2015loa}
M.~J.~A. Jansen, C.~M\"uller, {Pair Creation of Scalar Particles in Intense
  Bichromatic Laser Fields}, J. Phys. Conf. Ser. 594~(1) (2015) 012051.
\newblock \href {https://doi.org/10.1088/1742-6596/594/1/012051}
  {\path{doi:10.1088/1742-6596/594/1/012051}}.

\bibitem{Jansen:2015idl}
M.~J.~A. Jansen, C.~M\"uller, {Strong-Field Breit-Wheeler Pair Production in
  Short Laser Pulses: Identifying Multiphoton Interference and
  Carrier-Envelope-Phase Effects}, Phys. Rev. D 93~(5) (2016) 053011.
\newblock \href {http://arxiv.org/abs/1511.07660} {\path{arXiv:1511.07660}},
  \href {https://doi.org/10.1103/PhysRevD.93.053011}
  {\path{doi:10.1103/PhysRevD.93.053011}}.

\bibitem{Brass:2019pzr}
J.~Bra\ss{}, R.~Milbradt, S.~Villalba-Ch\'avez, G.~G. Paulus, C.~M\"uller,
  {Two-color phase-of-the-phase spectroscopy applied to nonperturbative
  electron-positron pair production in strong oscillating electric fields},
  Phys. Rev. A 101~(4) (2020) 043401.
\newblock \href {http://arxiv.org/abs/1912.04750} {\path{arXiv:1912.04750}},
  \href {https://doi.org/10.1103/PhysRevA.101.043401}
  {\path{doi:10.1103/PhysRevA.101.043401}}.

\bibitem{2015PhRvL.115d3001S}
S.~{Skruszewicz}, J.~{Tiggesb{\"a}umker}, K.~H. {Meiwes-Broer}, M.~{Arbeiter},
  T.~{Fennel}, D.~{Bauer}, {Two-Color Strong-Field Photoelectron Spectroscopy
  and the Phase of the Phase}, Phys.~Rev.~Lett 115~(4) (2015) 043001.
\newblock \href {http://arxiv.org/abs/1502.00614} {\path{arXiv:1502.00614}},
  \href {https://doi.org/10.1103/PhysRevLett.115.043001}
  {\path{doi:10.1103/PhysRevLett.115.043001}}.

\bibitem{Akal:2014eua}
I.~Akal, S.~Villalba-Ch\'avez, C.~M\"uller, {Electron-positron pair production
  in a bifrequent oscillating electric field}, Phys. Rev. D 90~(11) (2014)
  113004.
\newblock \href {http://arxiv.org/abs/1409.1806} {\path{arXiv:1409.1806}},
  \href {https://doi.org/10.1103/PhysRevD.90.113004}
  {\path{doi:10.1103/PhysRevD.90.113004}}.

\bibitem{Titov:2020taw}
A.~I. Titov, B.~Kämpfer, {Non-linear Breit\textendash{}Wheeler process with
  linearly polarized beams}, Eur. Phys. J. D 74~(11) (2020) 218.
\newblock \href {http://arxiv.org/abs/2006.04496} {\path{arXiv:2006.04496}},
  \href {https://doi.org/10.1140/epjd/e2020-10327-9}
  {\path{doi:10.1140/epjd/e2020-10327-9}}.

\bibitem{Villalba-Chavez:2012kko}
S.~Villalba-Chavez, C.~Muller, {Photo-production of scalar particles in the
  field of a circularly polarized laser beam}, Phys. Lett. B 718 (2013)
  992--997.
\newblock \href {http://arxiv.org/abs/1208.3595} {\path{arXiv:1208.3595}},
  \href {https://doi.org/10.1016/j.physletb.2012.11.035}
  {\path{doi:10.1016/j.physletb.2012.11.035}}.

\bibitem{Jansen:2016gvt}
M.~J.~A. Jansen, J.~Z. Kami\'nski, K.~Krajewska, C.~M\"uller, {Strong-field
  Breit-Wheeler pair production in short laser pulses: Relevance of spin
  effects}, Phys. Rev. D 94 (2016) 013010.
\newblock \href {http://arxiv.org/abs/1605.03476} {\path{arXiv:1605.03476}},
  \href {https://doi.org/10.1103/PhysRevD.94.013010}
  {\path{doi:10.1103/PhysRevD.94.013010}}.

\bibitem{Dai:2021vgl}
Y.-N. Dai, B.-F. Shen, J.-X. Li, R.~Shaisultanov, K.~Z. Hatsagortsyan, C.~H.
  Keitel, Y.-Y. Chen, {Photon polarization effects in polarized
  electron-positron pair production in a strong laser field} (7 2021).
\newblock \href {http://arxiv.org/abs/2107.04996} {\path{arXiv:2107.04996}}.

\bibitem{Ivanov:2004vh}
D.~Y. Ivanov, G.~L. Kotkin, V.~G. Serbo, {Complete description of polarization
  effects in $e^+e^-$ pair production by a photon in the field of a strong
  laser wave}, Eur. Phys. J. C 40 (2005) 27--40.
\newblock \href {http://arxiv.org/abs/hep-ph/0412032}
  {\path{arXiv:hep-ph/0412032}}, \href
  {https://doi.org/10.1140/epjc/s2005-02125-1}
  {\path{doi:10.1140/epjc/s2005-02125-1}}.

\bibitem{Wistisen:2020rsq}
T.~N. Wistisen, {Numerical approach to the semiclassical method of pair
  production for arbitrary spins and photon polarization}, Phys. Rev. D 101~(7)
  (2020) 076017.
\newblock \href {http://arxiv.org/abs/2002.08660} {\path{arXiv:2002.08660}},
  \href {https://doi.org/10.1103/PhysRevD.101.076017}
  {\path{doi:10.1103/PhysRevD.101.076017}}.

\bibitem{book:optics}
F.~Tr{\"{a}}ger (Ed.),
  \href{http://link.springer.com/10.1007/978-3-642-19409-2}{{Springer Handbook
  of Lasers and Optics}}, Springer Berlin Heidelberg, Berlin, Heidelberg, 2012.
\newblock \href {https://doi.org/10.1007/978-3-642-19409-2}
  {\path{doi:10.1007/978-3-642-19409-2}}.
\newline\urlprefix\url{http://link.springer.com/10.1007/978-3-642-19409-2}

\bibitem{Tang:2021vfh}
S.~Tang, {Generation of quasimonoenergetic positron beams in chirped laser
  fields}, Phys. Rev. A 104~(2) (2021) 022209.
\newblock \href {http://arxiv.org/abs/2102.01871} {\path{arXiv:2102.01871}},
  \href {https://doi.org/10.1103/PhysRevA.104.022209}
  {\path{doi:10.1103/PhysRevA.104.022209}}.

\bibitem{Blackburn:2020fqo}
T.~G. Blackburn, A.~J. MacLeod, A.~Ilderton, B.~King, S.~Tang, M.~Marklund,
  {Self-absorption of synchrotron radiation in a laser-irradiated plasma},
  Phys. Plasmas 28~(5) (2021) 053103.
\newblock \href {http://arxiv.org/abs/2005.00302} {\path{arXiv:2005.00302}},
  \href {https://doi.org/10.1063/5.0044766} {\path{doi:10.1063/5.0044766}}.

\bibitem{Tang:2019ffe}
S.~Tang, A.~Ilderton, B.~King, {One-photon pair-annihilation in pulsed
  plane-wave backgrounds}, Phys. Rev. A 100~(6) (2019) 062119.
\newblock \href {http://arxiv.org/abs/1909.01141} {\path{arXiv:1909.01141}},
  \href {https://doi.org/10.1103/PhysRevA.100.062119}
  {\path{doi:10.1103/PhysRevA.100.062119}}.

\bibitem{Lee:2021iid}
R.~N. Lee, M.~D. Schwartz, X.~Zhang, {Compton Scattering Total Cross Section at
  Next-to-Leading Order}, Phys. Rev. Lett. 126~(21) (2021) 211801.
\newblock \href {http://arxiv.org/abs/2102.06718} {\path{arXiv:2102.06718}},
  \href {https://doi.org/10.1103/PhysRevLett.126.211801}
  {\path{doi:10.1103/PhysRevLett.126.211801}}.

\bibitem{1967JETP...25..697O}
V.~P. {Ole{\v{i}}nik}, {Resonance Effects in the Field of an Intense Laser
  Beam}, Soviet Journal of Experimental and Theoretical Physics 25 (1967) 697.

\bibitem{1968JETP...26.1132O}
V.~P. {Ole{\v{i}}nik}, {Resonance Effects in the Field of an Intense Laser Ray.
  II.}, Soviet Journal of Experimental and Theoretical Physics 26 (1968) 1132.

\bibitem{Hartin:2006qnk}
A.~Hartin, {Second Order QED Processes in an Intense Electromagnetic Field},
  Ph.D. thesis, Queen Mary, U. of London (2006).
\newblock \href {http://arxiv.org/abs/1701.02906} {\path{arXiv:1701.02906}}.

\bibitem{Roshchupkin:2021yhd}
S.~P. Roshchupkin, N.~R. Larin, V.~V. Dubov, {Resonant photoproduction of
  ultrarelativistic electron-positron pairs on a nucleus in moderate and strong
  monochromatic light fields}, Phys. Rev. D 104~(11) (2021) 116011.
\newblock \href {http://arxiv.org/abs/2108.04955} {\path{arXiv:2108.04955}},
  \href {https://doi.org/10.1103/PhysRevD.104.116011}
  {\path{doi:10.1103/PhysRevD.104.116011}}.

\bibitem{Bragin:2020akq}
S.~Bragin, A.~Di~Piazza, {Electron-positron annihilation into two photons in an
  intense plane-wave field}, Phys. Rev. D 102~(11) (2020) 116012.
\newblock \href {http://arxiv.org/abs/2003.02231} {\path{arXiv:2003.02231}},
  \href {https://doi.org/10.1103/PhysRevD.102.116012}
  {\path{doi:10.1103/PhysRevD.102.116012}}.

\bibitem{Satunin:2018rdw}
P.~Satunin, {Breit-Wheeler pair production from Worldline Instantons}, EPJ Web
  Conf. 191 (2018) 02019.
\newblock \href {https://doi.org/10.1051/epjconf/201819102019}
  {\path{doi:10.1051/epjconf/201819102019}}.

\bibitem{Torgrimsson:2019sjn}
G.~Torgrimsson, {Thermally versus dynamically assisted Schwinger pair
  production}, Phys. Rev. D 99~(9) (2019) 096007.
\newblock \href {http://arxiv.org/abs/1902.07196} {\path{arXiv:1902.07196}},
  \href {https://doi.org/10.1103/PhysRevD.99.096007}
  {\path{doi:10.1103/PhysRevD.99.096007}}.

\bibitem{Brown:2015kgj}
A.~R. Brown, {Schwinger pair production at nonzero temperatures or in compact
  directions}, Phys. Rev. D 98~(3) (2018) 036008.
\newblock \href {http://arxiv.org/abs/1512.05716} {\path{arXiv:1512.05716}},
  \href {https://doi.org/10.1103/PhysRevD.98.036008}
  {\path{doi:10.1103/PhysRevD.98.036008}}.

\bibitem{Gould:2017fve}
O.~Gould, A.~Rajantie, {Thermal Schwinger pair production at arbitrary
  coupling}, Phys. Rev. D 96~(7) (2017) 076002.
\newblock \href {http://arxiv.org/abs/1704.04801} {\path{arXiv:1704.04801}},
  \href {https://doi.org/10.1103/PhysRevD.96.076002}
  {\path{doi:10.1103/PhysRevD.96.076002}}.

\bibitem{Serov:2020ots}
V.~D. Serov, S.~P. Roshchupkin, V.~V. Dubov, {Resonant Effect for
  Breit\textendash{}Wheeler Process in the Field of an X-ray Pulsar}, Universe
  6~(11) (2020) 190.
\newblock \href {https://doi.org/10.3390/universe6110190}
  {\path{doi:10.3390/universe6110190}}.

\bibitem{King:2012kd}
B.~King, H.~Gies, A.~Di~Piazza, {Pair production in a plane wave by thermal
  background photons}, Phys. Rev. D 86 (2012) 125007, [Erratum: Phys.Rev.D 87,
  069905 (2013)].
\newblock \href {http://arxiv.org/abs/1204.2442} {\path{arXiv:1204.2442}},
  \href {https://doi.org/10.1103/PhysRevD.86.125007}
  {\path{doi:10.1103/PhysRevD.86.125007}}.

\bibitem{Hu:2014ooa}
H.~Hu, J.~Huang, {Trident pair production in colliding bright x-ray laser
  beams}, Phys. Rev. A 89~(3) (2014) 033411.
\newblock \href {http://arxiv.org/abs/1308.5324} {\path{arXiv:1308.5324}},
  \href {https://doi.org/10.1103/PhysRevA.89.033411}
  {\path{doi:10.1103/PhysRevA.89.033411}}.

\bibitem{Sizykh:2021ywt}
G.~K. Sizykh, S.~P. Roshchupkin, V.~V. Dubov, {Resonant Effect of High-Energy
  Electron\textendash{}Positron Pairs Production in Collision of
  Ultrarelativistic Electrons with an X-ray Electromagnetic Wave}, Universe
  7~(7) (2021) 210.
\newblock \href {https://doi.org/10.3390/universe7070210}
  {\path{doi:10.3390/universe7070210}}.

\bibitem{Wistisen:2019pwo}
T.~N. Wistisen, {Investigation of two photon emission in strong field QED using
  channeling in a crystal}, Phys. Rev. D 100~(3) (2019) 036002.
\newblock \href {http://arxiv.org/abs/1905.05038} {\path{arXiv:1905.05038}},
  \href {https://doi.org/10.1103/PhysRevD.100.036002}
  {\path{doi:10.1103/PhysRevD.100.036002}}.

\bibitem{Titov:2021kbj}
A.~I. Titov, U.~H. Acosta, B.~Kämpfer, {Positron energy distribution in
  factorized trident process} (8 2021).
\newblock \href {http://arxiv.org/abs/2108.13043} {\path{arXiv:2108.13043}}.

\bibitem{Ritus:1972nf}
V.~I. Ritus, {Vacuum polarization correction to elastic electron and muon
  scattering in an intense field and pair electro- and muoproduction}, Nucl.
  Phys. B 44 (1972) 236--252.
\newblock \href {https://doi.org/10.1016/0550-3213(72)90282-9}
  {\path{doi:10.1016/0550-3213(72)90282-9}}.

\bibitem{Ilderton:2010wr}
A.~Ilderton, {Trident pair production in strong laser pulses}, Phys. Rev. Lett.
  106 (2011) 020404.
\newblock \href {http://arxiv.org/abs/1011.4072} {\path{arXiv:1011.4072}},
  \href {https://doi.org/10.1103/PhysRevLett.106.020404}
  {\path{doi:10.1103/PhysRevLett.106.020404}}.

\bibitem{King:2013osa}
B.~King, H.~Ruhl, {Trident pair production in a constant crossed field}, Phys.
  Rev. D 88~(1) (2013) 013005.
\newblock \href {http://arxiv.org/abs/1303.1356} {\path{arXiv:1303.1356}},
  \href {https://doi.org/10.1103/PhysRevD.88.013005}
  {\path{doi:10.1103/PhysRevD.88.013005}}.

\bibitem{King:2014wfa}
B.~King, {Double Compton scattering in a constant crossed field}, Phys. Rev. A
  91~(3) (2015) 033415.
\newblock \href {http://arxiv.org/abs/1410.5478} {\path{arXiv:1410.5478}},
  \href {https://doi.org/10.1103/PhysRevA.91.033415}
  {\path{doi:10.1103/PhysRevA.91.033415}}.

\bibitem{BaierTrident}
V.~N. Baier, V.~M. Katkov, V.~M. Strakhovenko, Higher-order effects in external
  field: pair production by a particle, Sov. J. Nucl. Phys. 14 (1972) 572.

\bibitem{Hu:2010ye}
H.~Hu, C.~Muller, C.~H. Keitel, {Complete QED theory of multiphoton trident
  pair production in strong laser fields}, Phys. Rev. Lett. 105 (2010) 080401.
\newblock \href {http://arxiv.org/abs/1002.2596} {\path{arXiv:1002.2596}},
  \href {https://doi.org/10.1103/PhysRevLett.105.080401}
  {\path{doi:10.1103/PhysRevLett.105.080401}}.

\bibitem{Mackenroth:2012rb}
F.~Mackenroth, A.~Di~Piazza, {Nonlinear Double Compton Scattering in the
  Ultrarelativistic Quantum Regime}, Phys. Rev. Lett. 110~(7) (2013) 070402.
\newblock \href {http://arxiv.org/abs/1208.3424} {\path{arXiv:1208.3424}},
  \href {https://doi.org/10.1103/PhysRevLett.110.070402}
  {\path{doi:10.1103/PhysRevLett.110.070402}}.

\bibitem{2015JPhCS.594a2024K}
K.~{Krajewska}, J.~Z. {Kami{\'n}ski}, {Circular dichroism in nonlinear
  electron-positron pair creation}, in: Journal of Physics Conference Series,
  Vol. 594 of Journal of Physics Conference Series, 2015, p. 012024.
\newblock \href {https://doi.org/10.1088/1742-6596/594/1/012024}
  {\path{doi:10.1088/1742-6596/594/1/012024}}.

\bibitem{Dinu:2019wdw}
V.~Dinu, G.~Torgrimsson, {Trident process in laser pulses}, Phys. Rev. D
  101~(5) (2020) 056017.
\newblock \href {http://arxiv.org/abs/1912.11017} {\path{arXiv:1912.11017}},
  \href {https://doi.org/10.1103/PhysRevD.101.056017}
  {\path{doi:10.1103/PhysRevD.101.056017}}.

\bibitem{Torgrimsson:2020wlz}
G.~Torgrimsson, {Nonlinear trident in the high-energy limit: Nonlocality,
  Coulomb field and resummations}, Phys. Rev. D 102~(9) (2020) 096008.
\newblock \href {http://arxiv.org/abs/2007.08492} {\path{arXiv:2007.08492}},
  \href {https://doi.org/10.1103/PhysRevD.102.096008}
  {\path{doi:10.1103/PhysRevD.102.096008}}.

\bibitem{Morozov:1975uah}
D.~A. Morozov, V.~I. Ritus, {Elastic electron scattering in an intense field
  and two-photon emission}, Nucl. Phys. B 86 (1975) 309--332.
\newblock \href {https://doi.org/10.1016/0550-3213(75)90448-4}
  {\path{doi:10.1016/0550-3213(75)90448-4}}.

\bibitem{MorozovNarozhnyiPhTr}
D.~A. Morozov, N.~B. Narozhnyi, Elastic scattering of photons in an intense
  field and the photoproduction of a pair and a photon, Sov. Phys. JETP 45
  (1977) 23.

\bibitem{Torgrimsson:2020mto}
G.~Torgrimsson, {Nonlinear photon trident versus double Compton scattering and
  resummation of one-step terms}, Phys. Rev. D 102 (2020) 116008.
\newblock \href {http://arxiv.org/abs/2010.02128} {\path{arXiv:2010.02128}},
  \href {https://doi.org/10.1103/PhysRevD.102.116008}
  {\path{doi:10.1103/PhysRevD.102.116008}}.

\bibitem{Yelatontsev:2020zfk}
V.~A. Yelatontsev, S.~P. Roshchupkin, V.~V. Dubov, {Resonant Production of an
  Ultrarelativistic Electron\textendash{}Positron Pair at the Gamma Quantum
  Scattering by a Field of the X-ray Pulsar}, Universe 6~(10) (2020) 164.
\newblock \href {https://doi.org/10.3390/universe6100164}
  {\path{doi:10.3390/universe6100164}}.

\bibitem{RevModPhys.33.8}
W.~H. McMaster, \href{https://link.aps.org/doi/10.1103/RevModPhys.33.8}{Matrix
  representation of polarization}, Rev. Mod. Phys. 33 (1961) 8--28.
\newblock \href {https://doi.org/10.1103/RevModPhys.33.8}
  {\path{doi:10.1103/RevModPhys.33.8}}.
\newline\urlprefix\url{https://link.aps.org/doi/10.1103/RevModPhys.33.8}

\bibitem{Torgrimsson:2021wcj}
G.~Torgrimsson, {Resummation of Quantum Radiation Reaction in Plane Waves},
  Phys. Rev. Lett. 127~(11) (2021) 111602.
\newblock \href {http://arxiv.org/abs/2102.11346} {\path{arXiv:2102.11346}},
  \href {https://doi.org/10.1103/PhysRevLett.127.111602}
  {\path{doi:10.1103/PhysRevLett.127.111602}}.

\bibitem{1985AmJPh..53..468B}
W.~S. {Bickel}, W.~M. {Bailey}, {Stokes vectors, Mueller matrices, and
  polarized scattered light}, American Journal of Physics 53~(5) (1985)
  468--478.
\newblock \href {https://doi.org/10.1119/1.14202} {\path{doi:10.1119/1.14202}}.

\bibitem{2000PThPh.104..769B}
E.~{Bol'Shedvorsky}, S.~{Polityko}, A.~{Misaki}, {Spin of Scattered Electrons
  in the Nonlinear Compton Effect}, Progress of Theoretical Physics 104~(4)
  (2000) 769--775.
\newblock \href {https://doi.org/10.1143/PTP.104.769}
  {\path{doi:10.1143/PTP.104.769}}.

\bibitem{1983JETP...57..935G}
Y.~T. {Grinchishin}, M.~P. {Rekalo}, {Inverse Compton effect induced by an
  intense circularly polarized wave}, Soviet Journal of Experimental and
  Theoretical Physics 57~(5) (1983) 935.

\bibitem{Li:2020bwo}
Y.-F. Li, Y.-Y. Chen, W.-M. Wang, H.-S. Hu, {Production of Highly Polarized
  Positron Beams via Helicity Transfer from Polarized Electrons in a Strong
  Laser Field}, Phys. Rev. Lett. 125~(4) (2020) 044802.
\newblock \href {http://arxiv.org/abs/2003.01547} {\path{arXiv:2003.01547}},
  \href {https://doi.org/10.1103/PhysRevLett.125.044802}
  {\path{doi:10.1103/PhysRevLett.125.044802}}.

\bibitem{Tang:2021azl}
Y.~Tang, Z.~Gong, J.~Yu, Y.~Shou, X.~Yan, {Radiative polarization dynamics of
  relativistic electrons in an intense electromagnetic field}, Phys. Rev. A
  103~(4) (2021) 042807.
\newblock \href {http://arxiv.org/abs/2107.00597} {\path{arXiv:2107.00597}},
  \href {https://doi.org/10.1103/PhysRevA.103.042807}
  {\path{doi:10.1103/PhysRevA.103.042807}}.

\bibitem{Burke:1997ew}
D.~L. Burke, et~al., {Positron production in multi - photon light by light
  scattering}, Phys. Rev. Lett. 79 (1997) 1626--1629.
\newblock \href {https://doi.org/10.1103/PhysRevLett.79.1626}
  {\path{doi:10.1103/PhysRevLett.79.1626}}.

\bibitem{Podszus:2018hnz}
T.~Podszus, A.~Di~Piazza, {High-energy behavior of strong-field QED in an
  intense plane wave}, Phys. Rev. D 99~(7) (2019) 076004.
\newblock \href {http://arxiv.org/abs/1812.08673} {\path{arXiv:1812.08673}},
  \href {https://doi.org/10.1103/PhysRevD.99.076004}
  {\path{doi:10.1103/PhysRevD.99.076004}}.

\bibitem{Ilderton:2019kqp}
A.~Ilderton, {Note on the conjectured breakdown of QED perturbation theory in
  strong fields}, Phys. Rev. D 99~(8) (2019) 085002.
\newblock \href {http://arxiv.org/abs/1901.00317} {\path{arXiv:1901.00317}},
  \href {https://doi.org/10.1103/PhysRevD.99.085002}
  {\path{doi:10.1103/PhysRevD.99.085002}}.

\bibitem{Borsellino}
A.~Borsellino, Nuovo Cimento 4 (1947) 112.

\bibitem{1959PhRv..115..672S}
K.~S. {Suh}, H.~A. {Bethe}, {Recoil Momentum Distribution in Electron Pair
  Production}, Physical Review 115~(3) (1959) 672--677.
\newblock \href {https://doi.org/10.1103/PhysRev.115.672}
  {\path{doi:10.1103/PhysRev.115.672}}.

\bibitem{1976tper.book.....J}
J.~M. {Jauch}, F.~{Rohrlich}, {The theory of photons and electrons. The
  relativistic quantum field theory of charged particles with spin one-half},
  1976.

\bibitem{Barry:1989zz}
M.~V. Berry, {Uniform Asymptotic Smoothing of Stokes's Discontinuities}, Proc.
  Roy. Soc. Lond. A 422 (1989) 7--21.
\newblock \href {https://doi.org/10.1098/rspa.1989.0018}
  {\path{doi:10.1098/rspa.1989.0018}}.

\bibitem{PhysRev.124.768}
G.~A. Baker, {Application of the Pad\'e Approximant Method to the Investigation
  of Some Magnetic Properties of the Ising Model}, Phys. Rev. 124 (1961)
  768--774.
\newblock \href {https://doi.org/10.1103/PhysRev.124.768}
  {\path{doi:10.1103/PhysRev.124.768}}.

\bibitem{CiulliFischer}
S.~{Ciulli}, J.~{Fischer}, {A convergent set of integral equations for singlet
  proton-proton scattering}, Nuclear Physics 24~(3) (1961) 465--473.
\newblock \href {https://doi.org/10.1016/0029-5582(61)90413-8}
  {\path{doi:10.1016/0029-5582(61)90413-8}}.

\bibitem{LeGuillou:1979ixc}
J.~C. Le~Guillou, J.~Zinn-Justin, {Critical Exponents from Field Theory}, Phys.
  Rev. B 21 (1980) 3976--3998.
\newblock \href {https://doi.org/10.1103/PhysRevB.21.3976}
  {\path{doi:10.1103/PhysRevB.21.3976}}.

\bibitem{BenderOrszagBook}
C.~M. Bender, S.~A. Orszag, Advanced mathematical methods for scientists and
  engineers I: asymptotic methods and perturbation theory., Springer-Verlag New
  York, 1999.

\bibitem{Kleinert:2001ax}
H.~Kleinert, V.~Schulte-Frohlinde, {Critical properties of $\phi^4$-theories},
  World Scientific, 2001.

\bibitem{ZinnJustinBook}
J.~Zinn-Justin, Quantum Field Theory and Critical Phenomena, 4th Edition,
  Clarendon press, Oxford, 2002.

\bibitem{Caliceti:2007ra}
E.~Caliceti, M.~Meyer-Hermann, P.~Ribeca, A.~Surzhykov, U.~D. Jentschura, {From
  useful algorithms for slowly convergent series to physical predictions based
  on divergent perturbative expansions}, Phys. Rept. 446 (2007) 1--96.
\newblock \href {http://arxiv.org/abs/0707.1596} {\path{arXiv:0707.1596}},
  \href {https://doi.org/10.1016/j.physrep.2007.03.003}
  {\path{doi:10.1016/j.physrep.2007.03.003}}.

\bibitem{Costin:2019xql}
O.~Costin, G.~V. Dunne, {Resurgent extrapolation: rebuilding a function from
  asymptotic data. Painlev\'e I}, J. Phys. A 52~(44) (2019) 445205.
\newblock \href {http://arxiv.org/abs/1904.11593} {\path{arXiv:1904.11593}},
  \href {https://doi.org/10.1088/1751-8121/ab477b}
  {\path{doi:10.1088/1751-8121/ab477b}}.

\bibitem{Costin:2020hwg}
O.~Costin, G.~V. Dunne, {Physical Resurgent Extrapolation}, Phys. Lett. B 808
  (2020) 135627.
\newblock \href {http://arxiv.org/abs/2003.07451} {\path{arXiv:2003.07451}},
  \href {https://doi.org/10.1016/j.physletb.2020.135627}
  {\path{doi:10.1016/j.physletb.2020.135627}}.

\bibitem{Florio:2019hzn}
A.~Florio, {Schwinger pair production from Pad\'e-Borel reconstruction}, Phys.
  Rev. D 101~(1) (2020) 013007.
\newblock \href {http://arxiv.org/abs/1911.03489} {\path{arXiv:1911.03489}},
  \href {https://doi.org/10.1103/PhysRevD.101.013007}
  {\path{doi:10.1103/PhysRevD.101.013007}}.

\bibitem{Dunne:2021acr}
G.~V. Dunne, Z.~Harris, {Higher-loop Euler-Heisenberg transseries structure},
  Phys. Rev. D 103~(6) (2021) 065015.
\newblock \href {http://arxiv.org/abs/2101.10409} {\path{arXiv:2101.10409}},
  \href {https://doi.org/10.1103/PhysRevD.103.065015}
  {\path{doi:10.1103/PhysRevD.103.065015}}.

\bibitem{Ekman:2021eqc}
R.~Ekman, T.~Heinzl, A.~Ilderton, {Reduction of order, resummation, and
  radiation reaction}, Phys. Rev. D 104~(3) (2021) 036002.
\newblock \href {http://arxiv.org/abs/2105.01640} {\path{arXiv:2105.01640}},
  \href {https://doi.org/10.1103/PhysRevD.104.036002}
  {\path{doi:10.1103/PhysRevD.104.036002}}.

\bibitem{Mera:2018qte}
H.~Mera, T.~G. Pedersen, B.~K. Nikoli\'c, {Fast summation of divergent series
  and resurgent transseries from Meijer- G approximants}, Phys. Rev. D 97~(10)
  (2018) 105027.
\newblock \href {http://arxiv.org/abs/1802.06034} {\path{arXiv:1802.06034}},
  \href {https://doi.org/10.1103/PhysRevD.97.105027}
  {\path{doi:10.1103/PhysRevD.97.105027}}.

\bibitem{Alvarez:2017sza}
G.~\'Alvarez, H.~J. Silverstone, {A new method to sum divergent power series:
  educated match}, J. Phys. Comm. 1~(2) (2017) 025005.
\newblock \href {http://arxiv.org/abs/1706.00329} {\path{arXiv:1706.00329}},
  \href {https://doi.org/10.1088/2399-6528/aa8540}
  {\path{doi:10.1088/2399-6528/aa8540}}.

\bibitem{Gavrilov:2016tuq}
S.~P. Gavrilov, D.~M. Gitman, {Vacuum instability in slowly varying electric
  fields}, Phys. Rev. D 95~(7) (2017) 076013.
\newblock \href {http://arxiv.org/abs/1612.06297} {\path{arXiv:1612.06297}},
  \href {https://doi.org/10.1103/PhysRevD.95.076013}
  {\path{doi:10.1103/PhysRevD.95.076013}}.

\bibitem{PhysRevLett.89.094801}
M.~K. Khokonov, H.~Nitta,
  \href{https://link.aps.org/doi/10.1103/PhysRevLett.89.094801}{Standard
  radiation spectrum of relativistic electrons: Beyond the synchrotron
  approximation}, Phys. Rev. Lett. 89 (2002) 094801.
\newblock \href {https://doi.org/10.1103/PhysRevLett.89.094801}
  {\path{doi:10.1103/PhysRevLett.89.094801}}.
\newline\urlprefix\url{https://link.aps.org/doi/10.1103/PhysRevLett.89.094801}

\bibitem{Chen:1988ec}
P.~Chen, K.~Yokoya, {Field Gradient Effect in Quantum Beamstrahlung}, Phys.
  Rev. Lett. 61 (1988) 1101, [Erratum: Phys.Rev.Lett. 62, 1213 (1989)].
\newblock \href {https://doi.org/10.1103/PhysRevLett.61.1101}
  {\path{doi:10.1103/PhysRevLett.61.1101}}.

\bibitem{Baier:1990pi}
V.~N. Baier, V.~M. Katkov, V.~M. Strakhovenko, {Quantum beamstrahlung and
  electroproduction of the pairs in linear colliders}, Part. Accel. 30 (1990)
  43--54.

\bibitem{Yokoya:1991qz}
K.~Yokoya, P.~Chen, {Beam-beam phenomena in linear colliders}, Lect. Notes
  Phys. 400 (1992) 415--445.
\newblock \href {https://doi.org/10.1007/3-540-55250-2_37}
  {\path{doi:10.1007/3-540-55250-2_37}}.

\bibitem{Cruz:2020vfm}
F.~Cruz, T.~Grismayer, L.~O. Silva, {Kinetic model of large-amplitude
  oscillations in neutron star pair cascades}, Astrophys. J. 908~(2) (2021)
  149.
\newblock \href {http://arxiv.org/abs/2012.05587} {\path{arXiv:2012.05587}},
  \href {https://doi.org/10.3847/1538-4357/abd2c0}
  {\path{doi:10.3847/1538-4357/abd2c0}}.

\bibitem{2018PhRvE..97d3209N}
F.~{Niel}, C.~{Riconda}, F.~{Amiranoff}, R.~{Duclous}, M.~{Grech}, {From
  quantum to classical modeling of radiation reaction: A focus on stochasticity
  effects}, Phys.~Rev.~E 97~(4) (2018) 043209.
\newblock \href {http://arxiv.org/abs/1707.02618} {\path{arXiv:1707.02618}},
  \href {https://doi.org/10.1103/PhysRevE.97.043209}
  {\path{doi:10.1103/PhysRevE.97.043209}}.

\bibitem{Burton:2014wsa}
D.~A. Burton, A.~Noble, {Aspects of electromagnetic radiation reaction in
  strong fields}, Contemp. Phys. 55~(2) (2014) 110--121.
\newblock \href {http://arxiv.org/abs/1409.7707} {\path{arXiv:1409.7707}},
  \href {https://doi.org/10.1080/00107514.2014.886840}
  {\path{doi:10.1080/00107514.2014.886840}}.

\bibitem{Blackburn:2019rfv}
T.~G. Blackburn, {Radiation reaction in electron-beam interactions with
  high-intensity lasers}, Plasma Phys. 4 (2020) 5.
\newblock \href {http://arxiv.org/abs/1910.13377} {\path{arXiv:1910.13377}},
  \href {https://doi.org/10.1007/s41614-020-0042-0}
  {\path{doi:10.1007/s41614-020-0042-0}}.

\bibitem{Baier:2003hf}
V.~N. Baier, V.~M. Katkov, {Concept of formation length in radiation theory},
  Phys. Rept. 409 (2005) 261--359.
\newblock \href {http://arxiv.org/abs/hep-ph/0309211}
  {\path{arXiv:hep-ph/0309211}}, \href
  {https://doi.org/10.1016/j.physrep.2004.11.003}
  {\path{doi:10.1016/j.physrep.2004.11.003}}.

\bibitem{Schwinger:1949ym}
J.~S. Schwinger, {On the classical radiation of accelerated electrons}, Phys.
  Rev. 75 (1949) 1912.
\newblock \href {https://doi.org/10.1103/PhysRev.75.1912}
  {\path{doi:10.1103/PhysRev.75.1912}}.

\bibitem{Baier1968}
V.~N. {Ba{\v{i}}er}, V.~M. {Katkov}, {Processes Involved in the Motion of High
  Energy Particles in a Magnetic Field}, Soviet Journal of Experimental and
  Theoretical Physics 26 (1968) 854.

\bibitem{vallee2010airy}
O.~Vall{\'e}e, M.~Soares,
  \href{https://books.google.co.uk/books?id=tNAWngEACAAJ}{Airy Functions and
  Applications to Physics}, Imperial College Press, 2010.
\newline\urlprefix\url{https://books.google.co.uk/books?id=tNAWngEACAAJ}

\bibitem{Ilderton:2018nws}
A.~Ilderton, B.~King, D.~Seipt, {Extended locally constant field approximation
  for nonlinear Compton scattering}, Phys. Rev. A 99~(4) (2019) 042121.
\newblock \href {http://arxiv.org/abs/1808.10339} {\path{arXiv:1808.10339}},
  \href {https://doi.org/10.1103/PhysRevA.99.042121}
  {\path{doi:10.1103/PhysRevA.99.042121}}.

\bibitem{Harvey:2014qla}
C.~N. Harvey, A.~Ilderton, B.~King, {Testing numerical implementations of
  strong field electrodynamics}, Phys. Rev. A 91~(1) (2015) 013822.
\newblock \href {http://arxiv.org/abs/1409.6187} {\path{arXiv:1409.6187}},
  \href {https://doi.org/10.1103/PhysRevA.91.013822}
  {\path{doi:10.1103/PhysRevA.91.013822}}.

\bibitem{DiPiazza:2018bfu}
A.~Di~Piazza, M.~Tamburini, S.~Meuren, C.~H. Keitel, {Improved
  local-constant-field approximation for strong-field QED codes}, Phys. Rev. A
  99~(2) (2019) 022125.
\newblock \href {http://arxiv.org/abs/1811.05834} {\path{arXiv:1811.05834}},
  \href {https://doi.org/10.1103/PhysRevA.99.022125}
  {\path{doi:10.1103/PhysRevA.99.022125}}.

\bibitem{Blackburn:2018sfn}
T.~G. Blackburn, D.~Seipt, S.~S. Bulanov, M.~Marklund, {Benchmarking
  semiclassical approaches to strong-field QED: nonlinear Compton scattering in
  intense laser pulses}, Phys. Plasmas 25~(8) (2018) 083108.
\newblock \href {http://arxiv.org/abs/1804.11085} {\path{arXiv:1804.11085}},
  \href {https://doi.org/10.1063/1.5037967} {\path{doi:10.1063/1.5037967}}.

\bibitem{King:2019igt}
B.~King, {Uniform locally constant field approximation for photon-seeded pair
  production}, Phys. Rev. A 101~(4) (2020) 042508.
\newblock \href {http://arxiv.org/abs/1908.06985} {\path{arXiv:1908.06985}},
  \href {https://doi.org/10.1103/PhysRevA.101.042508}
  {\path{doi:10.1103/PhysRevA.101.042508}}.

\bibitem{Dillon:2018ypt}
B.~M. Dillon, B.~King, {ALP production through non-linear Compton scattering in
  intense fields}, Eur. Phys. J. C 78~(9) (2018) 775.
\newblock \href {http://arxiv.org/abs/1802.07498} {\path{arXiv:1802.07498}},
  \href {https://doi.org/10.1140/epjc/s10052-018-6207-0}
  {\path{doi:10.1140/epjc/s10052-018-6207-0}}.

\bibitem{King:2019cpj}
B.~King, B.~M. Dillon, K.~A. Beyer, G.~Gregori, {Axion-like-particle decay in
  strong electromagnetic backgrounds}, JHEP 12 (2019) 162.
\newblock \href {http://arxiv.org/abs/1905.05201} {\path{arXiv:1905.05201}},
  \href {https://doi.org/10.1007/JHEP12(2019)162}
  {\path{doi:10.1007/JHEP12(2019)162}}.

\bibitem{DiPiazza:2016maj}
A.~Di~Piazza, {Nonlinear Breit-Wheeler pair production in a tightly focused
  laser beam}, Phys. Rev. Lett. 117~(21) (2016) 213201.
\newblock \href {http://arxiv.org/abs/1608.08120} {\path{arXiv:1608.08120}},
  \href {https://doi.org/10.1103/PhysRevLett.117.213201}
  {\path{doi:10.1103/PhysRevLett.117.213201}}.

\bibitem{Podszus:2021lms}
T.~Podszus, A.~Di~Piazza, {First-order strong-field QED processes including the
  damping of particles states} (3 2021).
\newblock \href {http://arxiv.org/abs/2103.14637} {\path{arXiv:2103.14637}}.

\bibitem{Meuren:2011hv}
S.~Meuren, A.~Di~Piazza, {Quantum electron self-interaction in a strong laser
  field}, Phys. Rev. Lett. 107 (2011) 260401.
\newblock \href {http://arxiv.org/abs/1107.4531} {\path{arXiv:1107.4531}},
  \href {https://doi.org/10.1103/PhysRevLett.107.260401}
  {\path{doi:10.1103/PhysRevLett.107.260401}}.

\bibitem{Heinzl:2021mji}
T.~Heinzl, A.~Ilderton, B.~King, {Classical Resummation and Breakdown of
  Strong-Field QED}, Phys. Rev. Lett. 127~(6) (2021) 061601.
\newblock \href {http://arxiv.org/abs/2101.12111} {\path{arXiv:2101.12111}},
  \href {https://doi.org/10.1103/PhysRevLett.127.061601}
  {\path{doi:10.1103/PhysRevLett.127.061601}}.

\bibitem{Piazza:2021vxi}
A.~D. Piazza, G.~Audagnotto, {Analytical Spectrum of Nonlinear Thomson
  Scattering Including Radiation Reaction} (2 2021).
\newblock \href {http://arxiv.org/abs/2102.11260} {\path{arXiv:2102.11260}}.

\bibitem{Torgrimsson:2021zob}
G.~Torgrimsson, {Resummation of quantum radiation reaction and induced
  polarization}, Phys. Rev. D 104~(5) (2021) 056016.
\newblock \href {http://arxiv.org/abs/2105.02220} {\path{arXiv:2105.02220}},
  \href {https://doi.org/10.1103/PhysRevD.104.056016}
  {\path{doi:10.1103/PhysRevD.104.056016}}.

\bibitem{Sevostyanov:2020dhs}
D.~G. Sevostyanov, I.~A. Aleksandrov, G.~Plunien, V.~M. Shabaev, {Total yield
  of electron-positron pairs produced from vacuum in strong electromagnetic
  fields: validity of the locally constant field approximation} (12 2020).
\newblock \href {http://arxiv.org/abs/2012.10751} {\path{arXiv:2012.10751}}.

\bibitem{Aleksandrov:2018zso}
I.~A. Aleksandrov, G.~Plunien, V.~M. Shabaev, {Locally-constant field
  approximation in studies of electron-positron pair production in strong
  external fields}, Phys. Rev. D 99~(1) (2019) 016020.
\newblock \href {http://arxiv.org/abs/1811.01419} {\path{arXiv:1811.01419}},
  \href {https://doi.org/10.1103/PhysRevD.99.016020}
  {\path{doi:10.1103/PhysRevD.99.016020}}.

\bibitem{Gavrilov:2012jk}
S.~P. Gavrilov, D.~M. Gitman, N.~Yokomizo, {Dirac fermions in strong electric
  field and quantum transport in graphene}, Phys. Rev. D 86 (2012) 125022.
\newblock \href {http://arxiv.org/abs/1207.1749} {\path{arXiv:1207.1749}},
  \href {https://doi.org/10.1103/PhysRevD.86.125022}
  {\path{doi:10.1103/PhysRevD.86.125022}}.

\bibitem{DiPiazza:2020kze}
A.~Di~Piazza, M.~A. Lopez-Lopez, {One-loop vertex correction in a plane wave},
  Phys. Rev. D 102~(7) (2020) 076018.
\newblock \href {http://arxiv.org/abs/2006.15370} {\path{arXiv:2006.15370}},
  \href {https://doi.org/10.1103/PhysRevD.102.076018}
  {\path{doi:10.1103/PhysRevD.102.076018}}.

\bibitem{Raicher:2020nkq}
E.~Raicher, Q.~Z. Lv, C.~H. Keitel, K.~Z. Hatsagortsyan, {Anomalous violation
  of the local constant field approximation in colliding laser beams}, Phys.
  Rev. Res. 3~(1) (2021) 013214.
\newblock \href {http://arxiv.org/abs/2003.06217} {\path{arXiv:2003.06217}},
  \href {https://doi.org/10.1103/PhysRevResearch.3.013214}
  {\path{doi:10.1103/PhysRevResearch.3.013214}}.

\bibitem{Lv:2021ayt}
Q.~Z. Lv, E.~Raicher, C.~H. Keitel, K.~Z. Hatsagortsyan, {Ultrarelativistic
  electrons in counterpropagating laser beams}, New J. Phys. 23~(6) (2021)
  065005.
\newblock \href {http://arxiv.org/abs/2106.01303} {\path{arXiv:2106.01303}},
  \href {https://doi.org/10.1088/1367-2630/abfa60}
  {\path{doi:10.1088/1367-2630/abfa60}}.

\bibitem{Meuren:2020nbw}
S.~Meuren, P.~H. Bucksbaum, N.~J. Fisch, F.~Fi\'uza, S.~Glenzer, M.~J. Hogan,
  K.~Qu, D.~A. Reis, G.~White, V.~Yakimenko, {On Seminal HEDP Research
  Opportunities Enabled by Colocating Multi-Petawatt Laser with High-Density
  Electron Beams} (2 2020).
\newblock \href {http://arxiv.org/abs/2002.10051} {\path{arXiv:2002.10051}}.

\bibitem{Seipt:2015rda}
D.~Seipt, A.~Surzhykov, S.~Fritzsche, B.~Kämpfer, {Caustic structures in x-ray
  Compton scattering off electrons driven by a short intense laser pulse}, New
  J. Phys. 18~(2) (2016) 023044.
\newblock \href {http://arxiv.org/abs/1507.08868} {\path{arXiv:1507.08868}},
  \href {https://doi.org/10.1088/1367-2630/18/2/023044}
  {\path{doi:10.1088/1367-2630/18/2/023044}}.

\bibitem{cain1}
K.~Yokoya, User's manual of {CAIN} version 2.35, {KEK Pub 4/96} ((2003)).

\bibitem{Blackburn:2021cuq}
T.~G. Blackburn, B.~King, {Higher fidelity simulations of nonlinear
  Breit\textendash{}Wheeler pair creation in intense laser pulses}, Eur. Phys.
  J. C 82~(1) (2022) 44.
\newblock \href {http://arxiv.org/abs/2108.10883} {\path{arXiv:2108.10883}},
  \href {https://doi.org/10.1140/epjc/s10052-021-09955-3}
  {\path{doi:10.1140/epjc/s10052-021-09955-3}}.

\bibitem{Blackburn:2021rqm}
T.~G. Blackburn, A.~J. MacLeod, B.~King, {From local to nonlocal: higher
  fidelity simulations of photon emission in intense laser pulses} (3 2021).
\newblock \href {http://arxiv.org/abs/2103.06673} {\path{arXiv:2103.06673}}.

\bibitem{bula97}
C.~Bula, {A numeric integration program to simulate nonlinear QED processes in
  Electron-laser or Photon-laser collisions},
  https://www.slac.stanford.edu/exp/e144/ps/nidoc.ps (1997).

\bibitem{etde_74290}
P.~Chen, G.~Horton-Smith, T.~Ohgaki, A.~W. Weidemann, K.~Yokoya, Cain:
  Conglomerat d\'abel et d\'interactions non-lineaires (1995).
\newblock \href {https://doi.org/10.1016/0168-9002(94)01186-9}
  {\path{doi:10.1016/0168-9002(94)01186-9}}.

\bibitem{Hartin:2018egj}
A.~Hartin, {Strong field QED in lepton colliders and electron/laser
  interactions}, Int. J. Mod. Phys. A 33~(13) (2018) 1830011.
\newblock \href {http://arxiv.org/abs/1804.02934} {\path{arXiv:1804.02934}},
  \href {https://doi.org/10.1142/S0217751X18300119}
  {\path{doi:10.1142/S0217751X18300119}}.

\bibitem{ptarmigan}
T.~G. Blackburn,
  \href{https://github.com/tgblackburn/ptarmigan}{\textsc{ptarmigan}} (2021).
\newline\urlprefix\url{https://github.com/tgblackburn/ptarmigan}

\bibitem{Wistisen:2016bry}
T.~N. Wistisen, {Leptons and photons in strong electromagnetic fields}, Ph.D.
  thesis, Aarhus U. (2016).

\bibitem{DiPiazza:2013vra}
A.~Di~Piazza, {Ultrarelativistic electron states in a general background
  electromagnetic field}, Phys. Rev. Lett. 113 (2014) 040402.
\newblock \href {http://arxiv.org/abs/1310.7856} {\path{arXiv:1310.7856}},
  \href {https://doi.org/10.1103/PhysRevLett.113.040402}
  {\path{doi:10.1103/PhysRevLett.113.040402}}.

\bibitem{DiPiazza:2015xva}
A.~Di~Piazza, {Analytical tools for investigating strong-field QED processes in
  tightly focused laser fields}, Phys. Rev. A 91~(4) (2015) 042118.
\newblock \href {http://arxiv.org/abs/1501.06475} {\path{arXiv:1501.06475}},
  \href {https://doi.org/10.1103/PhysRevA.91.042118}
  {\path{doi:10.1103/PhysRevA.91.042118}}.

\bibitem{DiPiazza:2016tdf}
A.~Di~Piazza, {First-order strong-field QED processes in a tightly focused
  laser beam}, Phys. Rev. A 95~(3) (2017) 032121.
\newblock \href {http://arxiv.org/abs/1612.04132} {\path{arXiv:1612.04132}},
  \href {https://doi.org/10.1103/PhysRevA.95.032121}
  {\path{doi:10.1103/PhysRevA.95.032121}}.

\bibitem{1991PhRvA..43.6032L}
J.~{Lindhard}, {Quantum-radiation spectra of relativistic particles derived by
  the correspondence principle}, Phys. Rev. A 43~(11) (1991) 6032--6037.
\newblock \href {https://doi.org/10.1103/PhysRevA.43.6032}
  {\path{doi:10.1103/PhysRevA.43.6032}}.

\bibitem{Wistisen:2014twa}
T.~N. Wistisen, K.~K. Andersen, S.~Yilmaz, R.~Mikkelsen, J.~L. Hansen, U.~I.
  Uggerh\o{}j, W.~Lauth, H.~Backe, {Experimental Realization of a New Type of
  Crystalline Undulator}, Phys. Rev. Lett. 112~(25) (2014) 254801.
\newblock \href {https://doi.org/10.1103/PhysRevLett.112.254801}
  {\path{doi:10.1103/PhysRevLett.112.254801}}.

\bibitem{Wistisen:2015rua}
T.~N. Wistisen, {Quantum synchrotron radiation in the case of a field with
  finite extension}, Phys. Rev. D 92~(4) (2015) 045045.
\newblock \href {https://doi.org/10.1103/PhysRevD.92.045045}
  {\path{doi:10.1103/PhysRevD.92.045045}}.

\bibitem{Raicher:2018cih}
E.~Raicher, S.~Eliezer, C.~H. Keitel, K.~Z. Hatsagortsyan, {Semiclassical
  limitations for photon emission in strong external fields}, Phys. Rev. A
  99~(5) (2019) 052513.
\newblock \href {http://arxiv.org/abs/1810.07523} {\path{arXiv:1810.07523}},
  \href {https://doi.org/10.1103/PhysRevA.99.052513}
  {\path{doi:10.1103/PhysRevA.99.052513}}.

\bibitem{Dinu:2013gaa}
V.~Dinu, T.~Heinzl, A.~Ilderton, M.~Marklund, G.~Torgrimsson, {Vacuum
  refractive indices and helicity flip in strong-field QED}, Phys. Rev. D
  89~(12) (2014) 125003.
\newblock \href {http://arxiv.org/abs/1312.6419} {\path{arXiv:1312.6419}},
  \href {https://doi.org/10.1103/PhysRevD.89.125003}
  {\path{doi:10.1103/PhysRevD.89.125003}}.

\bibitem{filon_1930}
L.~N.~G. Filon, Iii.—on a quadrature formula for trigonometric integrals,
  Proceedings of the Royal Society of Edinburgh 49 (1930) 38–47.
\newblock \href {https://doi.org/10.1017/S0370164600026262}
  {\path{doi:10.1017/S0370164600026262}}.

\bibitem{King:2012aw}
B.~King, C.~H. Keitel, {Photon-photon scattering in collisions of laser
  pulses}, New J. Phys. 14 (2012) 103002.
\newblock \href {http://arxiv.org/abs/1202.3339} {\path{arXiv:1202.3339}},
  \href {https://doi.org/10.1088/1367-2630/14/10/103002}
  {\path{doi:10.1088/1367-2630/14/10/103002}}.

\bibitem{mackenroth2014quantum}
K.~Mackenroth, \href{https://books.google.co.uk/books?id=XtYkBAAAQBAJ}{Quantum
  Radiation in Ultra-Intense Laser Pulses}, Springer Theses, Springer
  International Publishing, 2014.
\newline\urlprefix\url{https://books.google.co.uk/books?id=XtYkBAAAQBAJ}

\bibitem{2013LPB....31..503K}
K.~{Krajewska}, J.~Z. {Kami{\'n}ski}, {Spin effects in nonlinear Compton
  scattering in ultrashort linearly-polarized laser pulses}, Laser and Particle
  Beams 31~(3) (2013) 503--513.
\newblock \href {https://doi.org/10.1017/S0263034613000165}
  {\path{doi:10.1017/S0263034613000165}}.

\bibitem{Vasak:1987um}
D.~Vasak, M.~Gyulassy, H.~T. Elze, {Quantum Transport Theory for Abelian
  Plasmas}, Annals Phys. 173 (1987) 462--492.
\newblock \href {https://doi.org/10.1016/0003-4916(87)90169-2}
  {\path{doi:10.1016/0003-4916(87)90169-2}}.

\bibitem{Bialynicki-Birula:1991jwl}
I.~Bialynicki-Birula, P.~Gornicki, J.~Rafelski, {Phase space structure of the
  Dirac vacuum}, Phys. Rev. D 44 (1991) 1825--1835.
\newblock \href {https://doi.org/10.1103/PhysRevD.44.1825}
  {\path{doi:10.1103/PhysRevD.44.1825}}.

\bibitem{Zhuang:1995pd}
P.~Zhuang, U.~W. Heinz, {Relativistic quantum transport theory for
  electrodynamics}, Annals Phys. 245 (1996) 311--338.
\newblock \href {http://arxiv.org/abs/nucl-th/9502034}
  {\path{arXiv:nucl-th/9502034}}, \href
  {https://doi.org/10.1006/aphy.1996.0011} {\path{doi:10.1006/aphy.1996.0011}}.

\bibitem{Hebenstreit:2010vz}
F.~Hebenstreit, R.~Alkofer, H.~Gies, {Schwinger pair production in space and
  time-dependent electric fields: Relating the Wigner formalism to quantum
  kinetic theory}, Phys. Rev. D 82 (2010) 105026.
\newblock \href {http://arxiv.org/abs/1007.1099} {\path{arXiv:1007.1099}},
  \href {https://doi.org/10.1103/PhysRevD.82.105026}
  {\path{doi:10.1103/PhysRevD.82.105026}}.

\bibitem{Li:2014nua}
Z.~L. Li, D.~Lu, B.~F. Shen, L.~B. Fu, J.~Liu, B.~S. Xie, {Mass shift effects
  in nonperturbative multiphoton pair production for arbitrary polarized
  electric fields} (10 2014).
\newblock \href {http://arxiv.org/abs/1410.6284} {\path{arXiv:1410.6284}}.

\bibitem{Kluger:1998bm}
Y.~Kluger, E.~Mottola, J.~M. Eisenberg, {The Quantum Vlasov equation and its
  Markov limit}, Phys. Rev. D 58 (1998) 125015.
\newblock \href {http://arxiv.org/abs/hep-ph/9803372}
  {\path{arXiv:hep-ph/9803372}}, \href
  {https://doi.org/10.1103/PhysRevD.58.125015}
  {\path{doi:10.1103/PhysRevD.58.125015}}.

\bibitem{Schmidt:1998vi}
S.~M. Schmidt, D.~Blaschke, G.~Ropke, S.~A. Smolyansky, A.~V. Prozorkevich,
  V.~D. Toneev, {A Quantum kinetic equation for particle production in the
  Schwinger mechanism}, Int. J. Mod. Phys. E 7 (1998) 709--722.
\newblock \href {http://arxiv.org/abs/hep-ph/9809227}
  {\path{arXiv:hep-ph/9809227}}, \href
  {https://doi.org/10.1142/S0218301398000403}
  {\path{doi:10.1142/S0218301398000403}}.

\bibitem{Bialynicki-Birula:2011cln}
I.~Bialynicki-Birula, L.~Rudnicki, {Time evolution of the QED vacuum in a
  uniform electric Field: Complete analytic solution by spinorial
  decomposition}, Phys. Rev. D 83 (2011) 065020.
\newblock \href {http://arxiv.org/abs/1103.2666} {\path{arXiv:1103.2666}},
  \href {https://doi.org/10.1103/PhysRevD.83.065020}
  {\path{doi:10.1103/PhysRevD.83.065020}}.

\bibitem{Sheng:2018jwf}
X.-L. Sheng, R.-H. Fang, Q.~Wang, D.~H. Rischke, {Wigner function and pair
  production in parallel electric and magnetic fields}, Phys. Rev. D 99~(5)
  (2019) 056004.
\newblock \href {http://arxiv.org/abs/1812.01146} {\path{arXiv:1812.01146}},
  \href {https://doi.org/10.1103/PhysRevD.99.056004}
  {\path{doi:10.1103/PhysRevD.99.056004}}.

\bibitem{Dumlu:2009rr}
C.~K. Dumlu, {On the Quantum Kinetic Approach and the Scattering Approach to
  Vacuum Pair Production}, Phys. Rev. D 79 (2009) 065027.
\newblock \href {http://arxiv.org/abs/0901.2972} {\path{arXiv:0901.2972}},
  \href {https://doi.org/10.1103/PhysRevD.79.065027}
  {\path{doi:10.1103/PhysRevD.79.065027}}.

\bibitem{Fedotov:2010ue}
A.~M. Fedotov, E.~G. Gelfer, K.~Y. Korolev, S.~A. Smolyansky, {On the kinetic
  equation approach to pair production by time-dependent electric field}, Phys.
  Rev. D 83 (2011) 025011.
\newblock \href {http://arxiv.org/abs/1008.2098} {\path{arXiv:1008.2098}},
  \href {https://doi.org/10.1103/PhysRevD.83.025011}
  {\path{doi:10.1103/PhysRevD.83.025011}}.

\bibitem{Aleksandrov:2019ddt}
I.~A. Aleksandrov, C.~Kohlf\"urst, {Pair production in temporally and spatially
  oscillating fields}, Phys. Rev. D 101~(9) (2020) 096009.
\newblock \href {http://arxiv.org/abs/1912.09924} {\path{arXiv:1912.09924}},
  \href {https://doi.org/10.1103/PhysRevD.101.096009}
  {\path{doi:10.1103/PhysRevD.101.096009}}.

\bibitem{Blinne:2015zpa}
A.~Blinne, E.~Strobel, {Comparison of semiclassical and Wigner function methods
  in pair production in rotating fields}, Phys. Rev. D 93~(2) (2016) 025014.
\newblock \href {http://arxiv.org/abs/1510.02712} {\path{arXiv:1510.02712}},
  \href {https://doi.org/10.1103/PhysRevD.93.025014}
  {\path{doi:10.1103/PhysRevD.93.025014}}.

\bibitem{Rafelski:1993uh}
J.~Rafelski, G.~R. Shin, {Relativistic classical limit of quantum theory},
  Phys. Rev. A 48 (1993) 1869--1874.
\newblock \href {https://doi.org/10.1103/PhysRevA.48.1869}
  {\path{doi:10.1103/PhysRevA.48.1869}}.

\bibitem{Kohlfurst:2017git}
C.~Kohlf\"urst, {Phase-space analysis of the Schwinger effect in inhomogeneous
  electromagnetic fields}, Eur. Phys. J. Plus 133~(5) (2018) 191.
\newblock \href {http://arxiv.org/abs/1708.08920} {\path{arXiv:1708.08920}},
  \href {https://doi.org/10.1140/epjp/i2018-12062-6}
  {\path{doi:10.1140/epjp/i2018-12062-6}}.

\bibitem{Li:2019rex}
Z.~L. Li, B.~S. Xie, Y.~J. Li, {Boson pair production in arbitrarily polarized
  electric fields}, Phys. Rev. D 100~(7) (2019) 076018.
\newblock \href {http://arxiv.org/abs/1908.10556} {\path{arXiv:1908.10556}},
  \href {https://doi.org/10.1103/PhysRevD.100.076018}
  {\path{doi:10.1103/PhysRevD.100.076018}}.

\bibitem{Lavelle:2011yc}
M.~Lavelle, D.~McMullan, P.~Sharma, {The Factorisation of glue and mass terms
  in SU(N) gauge theories}, Phys. Rev. D 85 (2012) 045013.
\newblock \href {http://arxiv.org/abs/1110.1574} {\path{arXiv:1110.1574}},
  \href {https://doi.org/10.1103/PhysRevD.85.045013}
  {\path{doi:10.1103/PhysRevD.85.045013}}.

\bibitem{Brodsky:2014yha}
S.~J. Brodsky, G.~F. de~Teramond, H.~G. Dosch, J.~Erlich, {Light-Front
  Holographic QCD and Emerging Confinement}, Phys. Rept. 584 (2015) 1--105.
\newblock \href {http://arxiv.org/abs/1407.8131} {\path{arXiv:1407.8131}},
  \href {https://doi.org/10.1016/j.physrep.2015.05.001}
  {\path{doi:10.1016/j.physrep.2015.05.001}}.

\bibitem{Vary:2009gt}
J.~P. Vary, H.~Honkanen, J.~Li, P.~Maris, S.~J. Brodsky, A.~Harindranath, G.~F.
  de~Teramond, P.~Sternberg, E.~G. Ng, C.~Yang, {Hamiltonian light-front field
  theory in a basis function approach}, Phys. Rev. C 81 (2010) 035205.
\newblock \href {http://arxiv.org/abs/0905.1411} {\path{arXiv:0905.1411}},
  \href {https://doi.org/10.1103/PhysRevC.81.035205}
  {\path{doi:10.1103/PhysRevC.81.035205}}.

\bibitem{Zhao:2013cma}
X.~Zhao, A.~Ilderton, P.~Maris, J.~P. Vary, {Scattering in Time-Dependent Basis
  Light-Front Quantization}, Phys. Rev. D 88 (2013) 065014.
\newblock \href {http://arxiv.org/abs/1303.3273} {\path{arXiv:1303.3273}},
  \href {https://doi.org/10.1103/PhysRevD.88.065014}
  {\path{doi:10.1103/PhysRevD.88.065014}}.

\bibitem{Zhao:2013jia}
X.~Zhao, A.~Ilderton, P.~Maris, J.~P. Vary, {Non-perturbative quantum time
  evolution on the light-front}, Phys. Lett. B 726 (2013) 856--860.
\newblock \href {http://arxiv.org/abs/1309.5338} {\path{arXiv:1309.5338}},
  \href {https://doi.org/10.1016/j.physletb.2013.09.030}
  {\path{doi:10.1016/j.physletb.2013.09.030}}.

\bibitem{Li:2020uhl}
M.~Li, X.~Zhao, P.~Maris, G.~Chen, Y.~Li, K.~Tuchin, J.~P. Vary,
  {Ultrarelativistic quark-nucleus scattering in a light-front Hamiltonian
  approach}, Phys. Rev. D 101~(7) (2020) 076016.
\newblock \href {http://arxiv.org/abs/2002.09757} {\path{arXiv:2002.09757}},
  \href {https://doi.org/10.1103/PhysRevD.101.076016}
  {\path{doi:10.1103/PhysRevD.101.076016}}.

\bibitem{Li:2021zaw}
M.~Li, T.~Lappi, X.~Zhao, {Scattering and gluon emission in a color field: A
  light-front Hamiltonian approach}, Phys. Rev. D 104~(5) (2021) 056014.
\newblock \href {http://arxiv.org/abs/2107.02225} {\path{arXiv:2107.02225}},
  \href {https://doi.org/10.1103/PhysRevD.104.056014}
  {\path{doi:10.1103/PhysRevD.104.056014}}.

\bibitem{Hu:2019hjx}
B.~Hu, A.~Ilderton, X.~Zhao, {Scattering in strong electromagnetic fields:
  Transverse size effects in time-dependent basis light-front quantization},
  Phys. Rev. D 102~(1) (2020) 016017.
\newblock \href {http://arxiv.org/abs/1911.12307} {\path{arXiv:1911.12307}},
  \href {https://doi.org/10.1103/PhysRevD.102.016017}
  {\path{doi:10.1103/PhysRevD.102.016017}}.

\bibitem{Tomaras:2000ag}
T.~N. Tomaras, N.~C. Tsamis, R.~P. Woodard, {Back reaction in light cone QED},
  Phys. Rev. D 62 (2000) 125005.
\newblock \href {http://arxiv.org/abs/hep-ph/0007166}
  {\path{arXiv:hep-ph/0007166}}, \href
  {https://doi.org/10.1103/PhysRevD.62.125005}
  {\path{doi:10.1103/PhysRevD.62.125005}}.

\bibitem{Ilderton:2014mla}
A.~Ilderton, {Localisation in worldline pair production and lightfront
  zero-modes}, JHEP 09 (2014) 166.
\newblock \href {http://arxiv.org/abs/1406.1513} {\path{arXiv:1406.1513}},
  \href {https://doi.org/10.1007/JHEP09(2014)166}
  {\path{doi:10.1007/JHEP09(2014)166}}.

\bibitem{Lei:2022nsk}
Z.~Lei, B.~Hu, X.~Zhao, {Pair production in strong electric fields} (1 2022).
\newblock \href {http://arxiv.org/abs/2201.01746} {\path{arXiv:2201.01746}}.

\bibitem{2006JPhA...3914947K}
H.~J. {Korsch}, A.~{Klumpp}, D.~{Witthaut}, {On two-dimensional Bessel
  functions}, Journal of Physics A Mathematical General 39~(48) (2006)
  14947--14964.
\newblock \href {http://arxiv.org/abs/quant-ph/0608216}
  {\path{arXiv:quant-ph/0608216}}, \href
  {https://doi.org/10.1088/0305-4470/39/48/008}
  {\path{doi:10.1088/0305-4470/39/48/008}}.

\bibitem{2009PhRvE..79b6707L}
E.~{L{\"o}tstedt}, U.~D. {Jentschura}, {Recursive algorithm for arrays of
  generalized Bessel functions: Numerical access to Dirac-Volkov solutions},
  Phys.Rev.E 79~(2) (2009) 026707.
\newblock \href {http://arxiv.org/abs/0902.1099} {\path{arXiv:0902.1099}},
  \href {https://doi.org/10.1103/PhysRevE.79.026707}
  {\path{doi:10.1103/PhysRevE.79.026707}}.

\bibitem{Edwards:2021vhg}
J.~P. Edwards, C.~Schubert, {N-photon amplitudes in a plane-wave background},
  Phys. Lett. B 822 (2021) 136696.
\newblock \href {http://arxiv.org/abs/2105.08173} {\path{arXiv:2105.08173}},
  \href {https://doi.org/10.1016/j.physletb.2021.136696}
  {\path{doi:10.1016/j.physletb.2021.136696}}.

\bibitem{Affleck:1981bma}
I.~K. Affleck, O.~Alvarez, N.~S. Manton, {Pair Production at Strong Coupling in
  Weak External Fields}, Nucl. Phys. B 197 (1982) 509--519.
\newblock \href {https://doi.org/10.1016/0550-3213(82)90455-2}
  {\path{doi:10.1016/0550-3213(82)90455-2}}.

\bibitem{Lebedev:1984mei}
S.~L. Lebedev, V.~I. Ritus, {VIRIAL REPRESENTATION OF THE IMAGINARY PART OF THE
  LAGRANGE FUNCTION OF THE ELECTROMAGNETIC FIELD}, Sov. Phys. JETP 59 (1984)
  237--244.

\bibitem{Karbstein:2019wmj}
F.~Karbstein, {All-Loop Result for the Strong Magnetic Field Limit of the
  Heisenberg-Euler Effective Lagrangian}, Phys. Rev. Lett. 122~(21) (2019)
  211602.
\newblock \href {http://arxiv.org/abs/1903.06998} {\path{arXiv:1903.06998}},
  \href {https://doi.org/10.1103/PhysRevLett.122.211602}
  {\path{doi:10.1103/PhysRevLett.122.211602}}.

\bibitem{Bragin:2017yau}
S.~Bragin, S.~Meuren, C.~H. Keitel, A.~Di~Piazza, {High-Energy Vacuum
  Birefringence and Dichroism in an Ultrastrong Laser Field}, Phys. Rev. Lett.
  119~(25) (2017) 250403.
\newblock \href {http://arxiv.org/abs/1704.05234} {\path{arXiv:1704.05234}},
  \href {https://doi.org/10.1103/PhysRevLett.119.250403}
  {\path{doi:10.1103/PhysRevLett.119.250403}}.

\bibitem{King:2016jnl}
B.~King, N.~Elkina, {Vacuum birefringence in high-energy laser-electron
  collisions}, Phys. Rev. A 94~(6) (2016) 062102.
\newblock \href {http://arxiv.org/abs/1603.06946} {\path{arXiv:1603.06946}},
  \href {https://doi.org/10.1103/PhysRevA.94.062102}
  {\path{doi:10.1103/PhysRevA.94.062102}}.

\bibitem{Bohl:2015uba}
P.~B\"ohl, B.~King, H.~Ruhl, {Vacuum high harmonic generation in the shock
  regime}, Phys. Rev. A 92~(3) (2015) 032115.
\newblock \href {http://arxiv.org/abs/1503.05192} {\path{arXiv:1503.05192}},
  \href {https://doi.org/10.1103/PhysRevA.92.032115}
  {\path{doi:10.1103/PhysRevA.92.032115}}.

\bibitem{Karbstein:2017gsb}
F.~Karbstein, {Tadpole diagrams in constant electromagnetic fields}, JHEP 10
  (2017) 075.
\newblock \href {http://arxiv.org/abs/1709.03819} {\path{arXiv:1709.03819}},
  \href {https://doi.org/10.1007/JHEP10(2017)075}
  {\path{doi:10.1007/JHEP10(2017)075}}.

\bibitem{DiPiazza:2021szp}
A.~Di~Piazza, T.~P\u{a}tuleanu, {Electron mass shift in an intense plane wave},
  Phys. Rev. D 104~(7) (2021) 076003.
\newblock \href {http://arxiv.org/abs/2106.13720} {\path{arXiv:2106.13720}},
  \href {https://doi.org/10.1103/PhysRevD.104.076003}
  {\path{doi:10.1103/PhysRevD.104.076003}}.

\bibitem{1975JETP...42..400B}
V.~N. {Ba{\v{i}}er}, V.~M. {Katkov}, A.~I. {Mil'Shte{\v{i}}n}, V.~M.
  {Strakhovenko}, {The theory of quantum processes in the field of a strong
  electromagnetic wave}, Soviet Journal of Experimental and Theoretical Physics
  42 (1975) 400.

\bibitem{Ilderton:2013tb}
A.~Ilderton, G.~Torgrimsson, {Radiation reaction in strong field QED}, Phys.
  Lett. B 725 (2013) 481.
\newblock \href {http://arxiv.org/abs/1301.6499} {\path{arXiv:1301.6499}},
  \href {https://doi.org/10.1016/j.physletb.2013.07.045}
  {\path{doi:10.1016/j.physletb.2013.07.045}}.

\bibitem{Ilderton:2013dba}
A.~Ilderton, G.~Torgrimsson, {Radiation reaction from QED: lightfront
  perturbation theory in a plane wave background}, Phys. Rev. D 88~(2) (2013)
  025021.
\newblock \href {http://arxiv.org/abs/1304.6842} {\path{arXiv:1304.6842}},
  \href {https://doi.org/10.1103/PhysRevD.88.025021}
  {\path{doi:10.1103/PhysRevD.88.025021}}.

\bibitem{Krivitsky:1991vt}
V.~S. Krivitsky, V.~N. Tsytovich, {Average radiation reaction force in quantum
  electrodynamics}, Sov. Phys. Usp. 34 (1991) 250--258.
\newblock \href {https://doi.org/10.1070/PU1991v034n03ABEH002352}
  {\path{doi:10.1070/PU1991v034n03ABEH002352}}.

\bibitem{Higuchi:2004pr}
A.~Higuchi, G.~D.~R. Martin, {The Lorentz-dirac force from QED for linear
  acceleration}, Phys. Rev. D 70 (2004) 081701.
\newblock \href {http://arxiv.org/abs/quant-ph/0407162}
  {\path{arXiv:quant-ph/0407162}}, \href
  {https://doi.org/10.1103/PhysRevD.70.081701}
  {\path{doi:10.1103/PhysRevD.70.081701}}.

\bibitem{Higuchi:2005an}
A.~Higuchi, G.~D.~R. Martin, {Radiation reaction on charged particles in
  three-dimensional motion in classical and quantum electrodynamics}, Phys.
  Rev. D 73 (2006) 025019.
\newblock \href {http://arxiv.org/abs/quant-ph/0510043}
  {\path{arXiv:quant-ph/0510043}}, \href
  {https://doi.org/10.1103/PhysRevD.73.025019}
  {\path{doi:10.1103/PhysRevD.73.025019}}.

\bibitem{Holstein:2004dn}
B.~R. Holstein, J.~F. Donoghue, {Classical physics and quantum loops}, Phys.
  Rev. Lett. 93 (2004) 201602.
\newblock \href {http://arxiv.org/abs/hep-th/0405239}
  {\path{arXiv:hep-th/0405239}}, \href
  {https://doi.org/10.1103/PhysRevLett.93.201602}
  {\path{doi:10.1103/PhysRevLett.93.201602}}.

\bibitem{BaierSokolovTernov}
V.~N. {Ba{\v{i}}er}, {Radiative Polarization of Electrons in Storage Rings},
  Soviet Physics Uspekhi 14 (1971) 695.

\bibitem{Tamburini:2019tzo}
M.~Tamburini, S.~Meuren, {Efficient high-energy photon production in the
  supercritical QED regime}, Phys. Rev. D 104~(9) (2021) L091903.
\newblock \href {http://arxiv.org/abs/1912.07508} {\path{arXiv:1912.07508}},
  \href {https://doi.org/10.1103/PhysRevD.104.L091903}
  {\path{doi:10.1103/PhysRevD.104.L091903}}.

\bibitem{Blackburn:2019reaching}
T.~G. Blackburn, A.~Ilderton, M.~Marklund, C.~P. Ridgers, Reaching
  supercritical field strengths with intense lasers, New Journal of Physics
  21~(5) (2019) 053040.
\newblock \href {http://arxiv.org/abs/1807.03730} {\path{arXiv:1807.03730}},
  \href {https://doi.org/https://doi.org/10.1088/1367-2630/ab1e0d}
  {\path{doi:https://doi.org/10.1088/1367-2630/ab1e0d}}.

\bibitem{Zhang:2013ria}
S.~Zhang, {Pre-acceleration from Landau\textendash{}Lifshitz series}, PTEP
  2013~(12) (2013) 123A01.
\newblock \href {http://arxiv.org/abs/1303.7120} {\path{arXiv:1303.7120}},
  \href {https://doi.org/10.1093/ptep/ptt099} {\path{doi:10.1093/ptep/ptt099}}.

\bibitem{PhysRevE.84.056605}
S.~V. Bulanov, T.~Z. Esirkepov, M.~Kando, J.~K. Koga, S.~S. Bulanov,
  \href{https://link.aps.org/doi/10.1103/PhysRevE.84.056605}{Lorentz-abraham-dirac
  versus landau-lifshitz radiation friction force in the ultrarelativistic
  electron interaction with electromagnetic wave (exact solutions)}, Phys. Rev.
  E 84 (2011) 056605.
\newblock \href {https://doi.org/10.1103/PhysRevE.84.056605}
  {\path{doi:10.1103/PhysRevE.84.056605}}.
\newline\urlprefix\url{https://link.aps.org/doi/10.1103/PhysRevE.84.056605}

\bibitem{Moniz:1976kr}
E.~J. Moniz, D.~H. Sharp, {Radiation Reaction in Nonrelativistic Quantum
  Electrodynamics}, Phys. Rev. D 15 (1977) 2850.
\newblock \href {https://doi.org/10.1103/PhysRevD.15.2850}
  {\path{doi:10.1103/PhysRevD.15.2850}}.

\bibitem{Higuchi:2009ms}
A.~Higuchi, P.~J. Walker, {Quantum corrections to the Larmor radiation formula
  in scalar electrodynamics}, Phys. Rev. D 80 (2009) 105019.
\newblock \href {http://arxiv.org/abs/0908.2723} {\path{arXiv:0908.2723}},
  \href {https://doi.org/10.1103/PhysRevD.80.105019}
  {\path{doi:10.1103/PhysRevD.80.105019}}.

\bibitem{2008LMaPh..83..305P}
A.~D. {Piazza}, {Exact Solution of the Landau-Lifshitz Equation in a Plane
  Wave}, Letters in Mathematical Physics 83~(3) (2008) 305--313.
\newblock \href {https://doi.org/10.1007/s11005-008-0228-9}
  {\path{doi:10.1007/s11005-008-0228-9}}.

\bibitem{Kazinski:2010ce}
P.~O. Kazinski, M.~A. Shipulya, {Asymptotics of physical solutions to the
  Lorentz-Dirac equation for a planar motion in constant electromagnetic
  fields}, Phys. Rev. E 83 (2011) 066606.
\newblock \href {http://arxiv.org/abs/1012.5728} {\path{arXiv:1012.5728}},
  \href {https://doi.org/10.1103/PhysRevE.83.066606}
  {\path{doi:10.1103/PhysRevE.83.066606}}.

\bibitem{Kazinski:2013vga}
P.~O. Kazinski, {Radiation of de-excited electrons at large times in a strong
  electromagnetic plane wave}, Annals Phys. 339 (2013) 430--459.
\newblock \href {http://arxiv.org/abs/1306.1450} {\path{arXiv:1306.1450}},
  \href {https://doi.org/10.1016/j.aop.2013.09.016}
  {\path{doi:10.1016/j.aop.2013.09.016}}.

\bibitem{Elkina:2010up}
N.~V. Elkina, A.~M. Fedotov, I.~Y. Kostyukov, M.~V. Legkov, N.~B. Narozhny,
  E.~N. Nerush, H.~Ruhl, {QED cascades induced by circularly polarized laser
  fields}, Phys. Rev. ST Accel. Beams 14 (2011) 054401.
\newblock \href {http://arxiv.org/abs/1010.4528} {\path{arXiv:1010.4528}},
  \href {https://doi.org/10.1103/PhysRevSTAB.14.054401}
  {\path{doi:10.1103/PhysRevSTAB.14.054401}}.

\bibitem{Ekman:2021czy}
R.~Ekman, {Reduction of Order and Transseries Structure of Radiation Reaction}
  (12 2021).
\newblock \href {http://arxiv.org/abs/2112.10235} {\path{arXiv:2112.10235}}.

\bibitem{Landau:1975pou}
L.~D. Landau, E.~M. Lifschits, {The Classical Theory of Fields}, Vol. Volume 2
  of Course of Theoretical Physics, Pergamon Press, Oxford, 1975.

\bibitem{DelSorbo:2017fod}
D.~Del~Sorbo, D.~Seipt, T.~G. Blackburn, A.~G.~R. Thomas, C.~D. Murphy, J.~G.
  Kirk, C.~P. Ridgers, {Spin polarization of electrons by ultraintense lasers},
  Phys. Rev. A 96~(4) (2017) 043407.
\newblock \href {http://arxiv.org/abs/1702.03203} {\path{arXiv:1702.03203}},
  \href {https://doi.org/10.1103/PhysRevA.96.043407}
  {\path{doi:10.1103/PhysRevA.96.043407}}.

\bibitem{2018PPCF...60f4003D}
D.~{Del Sorbo}, D.~{Seipt}, A.~G.~R. {Thomas}, C.~P. {Ridgers}, {Electron spin
  polarization in realistic trajectories around the magnetic node of two
  counter-propagating, circularly polarized, ultra-intense lasers}, Plasma
  Physics and Controlled Fusion 60~(6) (2018) 064003.
\newblock \href {http://arxiv.org/abs/1712.08118} {\path{arXiv:1712.08118}},
  \href {https://doi.org/10.1088/1361-6587/aab979}
  {\path{doi:10.1088/1361-6587/aab979}}.

\bibitem{YokoyaCAIN}
K.~Yokoya, User's manual of cain version 2.42.

\bibitem{Wan:2019gow}
F.~Wan, R.~Shaisultanov, Y.-F. Li, K.~Z. Hatsagortsyan, C.~H. Keitel, J.-X. Li,
  {Ultrarelativistic polarized positron jets via collision of electron and
  ultraintense laser beams}, Phys. Lett. B 800 (2020) 135120.
\newblock \href {http://arxiv.org/abs/1904.04305} {\path{arXiv:1904.04305}},
  \href {https://doi.org/10.1016/j.physletb.2019.135120}
  {\path{doi:10.1016/j.physletb.2019.135120}}.

\bibitem{Song:2021wou}
H.-H. Song, W.-M. Wang, Y.-T. Li, {Generation of polarized positron beams via
  collisions of ultrarelativistic electron beams}, Phys. Rev. Res. 3~(3) (2021)
  033245.
\newblock \href {http://arxiv.org/abs/2103.10417} {\path{arXiv:2103.10417}},
  \href {https://doi.org/10.1103/PhysRevResearch.3.033245}
  {\path{doi:10.1103/PhysRevResearch.3.033245}}.

\bibitem{Wan:2020zet}
F.~Wan, Y.~Wang, R.-T. Guo, Y.-Y. Chen, R.~Shaisultanov, Z.-F. Xu, K.~Z.
  Hatsagortsyan, C.~H. Keitel, J.-X. Li, {High-energy $\gamma$ -photon
  polarization in nonlinear Breit-Wheeler pair production and $\gamma$
  polarimetry}, Phys. Rev. Res. 2~(3) (2020) 032049.
\newblock \href {http://arxiv.org/abs/2002.10346} {\path{arXiv:2002.10346}},
  \href {https://doi.org/10.1103/PhysRevResearch.2.032049}
  {\path{doi:10.1103/PhysRevResearch.2.032049}}.

\bibitem{Bell:2008zzb}
A.~R. Bell, J.~G. Kirk, {Possibility of Prolific Pair Production with
  High-Power Lasers}, Phys. Rev. Lett. 101 (2008) 200403.
\newblock \href {https://doi.org/10.1103/PhysRevLett.101.200403}
  {\path{doi:10.1103/PhysRevLett.101.200403}}.

\bibitem{Fedotov:2010ja}
A.~M. Fedotov, N.~B. Narozhny, G.~Mourou, G.~Korn, {Limitations on the
  attainable intensity of high power lasers}, Phys. Rev. Lett. 105 (2010)
  080402.
\newblock \href {http://arxiv.org/abs/1004.5398} {\path{arXiv:1004.5398}},
  \href {https://doi.org/10.1103/PhysRevLett.105.080402}
  {\path{doi:10.1103/PhysRevLett.105.080402}}.

\bibitem{Tang:2013vot}
S.~Tang, M.~A. Bake, H.-Y. Wang, B.-S. Xie, {QED cascade induced by a
  high-energy \ensuremath{\gamma} photon in a strong laser field}, Phys. Rev. A
  89~(2) (2014) 022105.
\newblock \href {http://arxiv.org/abs/1312.2317} {\path{arXiv:1312.2317}},
  \href {https://doi.org/10.1103/PhysRevA.89.022105}
  {\path{doi:10.1103/PhysRevA.89.022105}}.

\bibitem{sokolov:2010pair}
I.~V. Sokolov, N.~M. Naumova, J.~A. Nees, G.~A. Mourou, Pair creation in
  qed-strong pulsed laser fields interacting with electron beams, Phys. Rev.
  Lett. 105~(19) (2010) 195005.
\newblock \href {http://arxiv.org/abs/1009.0703v2} {\path{arXiv:1009.0703v2}},
  \href {https://doi.org/10.1103/PhysRevLett.105.195005}
  {\path{doi:10.1103/PhysRevLett.105.195005}}.

\bibitem{bulanov:2013el}
S.~S. Bulanov, C.~B. Schroeder, E.~Esarey, W.~P. Leemans, Electromagnetic
  cascade in high-energy electron, positron, and photon interactions with
  intense laser pulses, Phys. Rev. A 87~(6) (2013) 062110.
\newblock \href {http://arxiv.org/abs/1306.1260} {\path{arXiv:1306.1260}},
  \href {https://doi.org/10.1103/PhysRevA.87.062110}
  {\path{doi:10.1103/PhysRevA.87.062110}}.

\bibitem{Mironov:2014xba}
A.~A. Mironov, N.~B. Narozhny, A.~M. Fedotov, {Collapse and revival of
  electromagnetic cascades in focused intense laser pulses}, Phys. Lett. A 378
  (2014) 3254--3257.
\newblock \href {http://arxiv.org/abs/1407.6760} {\path{arXiv:1407.6760}},
  \href {https://doi.org/10.1016/j.physleta.2014.09.058}
  {\path{doi:10.1016/j.physleta.2014.09.058}}.

\bibitem{wen:2008ultrafast}
H.~Wen, M.~Wiczer, A.~M. Lindenberg, Ultrafast electron cascades in
  semiconductors driven by intense femtosecond terahertz pulses, Phys. Rev. B
  78~(12) (2008) 125203.
\newblock \href {https://doi.org/10.1103/PhysRevB.78.125203}
  {\path{doi:10.1103/PhysRevB.78.125203}}.

\bibitem{hoffmann:2009impact}
M.~C. Hoffmann, J.~Hebling, H.~Y. Hwang, K.-L. Yeh, K.~A. Nelson, Impact
  ionization in {InSb} probed by terahertz pump—terahertz probe spectroscopy,
  Phys. Rev. B 79~(16) (2009) 161201.
\newblock \href {http://arxiv.org/abs/0812.4754} {\path{arXiv:0812.4754}},
  \href {https://doi.org/10.1103/PhysRevB.79.161201}
  {\path{doi:10.1103/PhysRevB.79.161201}}.

\bibitem{hirori:2011extraordinary}
H.~Hirori, K.~Shinokita, M.~Shirai, S.~Tani, Y.~Kadoya, K.~Tanaka,
  Extraordinary carrier multiplication gated by a picosecond electric field
  pulse, Nature communications 2~(1) (2011) 1--6.
\newblock \href {https://doi.org/10.1038/ncomms1598}
  {\path{doi:10.1038/ncomms1598}}.

\bibitem{Nerush:2010fe}
E.~N. Nerush, I.~Y. Kostyukov, A.~M. Fedotov, N.~B. Narozhny, N.~V. Elkina,
  H.~Ruhl, {Laser field absorption in self-generated electron-positron pair
  plasma}, Phys. Rev. Lett. 106 (2011) 035001, [Erratum: Phys.Rev.Lett. 106,
  109902 (2011)].
\newblock \href {http://arxiv.org/abs/1011.0958} {\path{arXiv:1011.0958}},
  \href {https://doi.org/10.1103/PhysRevLett.106.035001}
  {\path{doi:10.1103/PhysRevLett.106.035001}}.

\bibitem{esirkepov:2015ac}
T.~Z. Esirkepov, S.~S. Bulanov, J.~K. Koga, M.~Kando, K.~Kondo, N.~N. Rosanov,
  G.~Korn, S.~V. Bulanov, Attractors and chaos of electron dynamics in
  electromagnetic standing waves, Phys. Lett. A 379~(36) (2015) 2044--2054.
\newblock \href {http://arxiv.org/abs/1412.6028} {\path{arXiv:1412.6028}},
  \href {https://doi.org/10.1016/j.physleta.2015.06.017}
  {\path{doi:10.1016/j.physleta.2015.06.017}}.

\bibitem{jirka:2016el}
M.~Jirka, O.~Klimo, S.~V. Bulanov, T.~Z. Esirkepov, E.~Gelfer, S.~S. Bulanov,
  S.~Weber, G.~Korn, Electron dynamics and {$\gamma$} and {$e^-e^+$} production
  by colliding laser pulses, Phys. Rev. E 93~(2) (2016) 023207.
\newblock \href {http://arxiv.org/abs/1511.04982} {\path{arXiv:1511.04982}},
  \href {https://doi.org/10.1103/PhysRevE.93.023207}
  {\path{doi:10.1103/PhysRevE.93.023207}}.

\bibitem{akhiezer:1994ki}
A.~I. Akhiezer, N.~P. Merenkov, A.~P. Rekalo, On a kinetic theory of
  electromagnetic showers in strong magnetic fields, Journal of Physics G:
  Nuclear and Particle Physics 20~(9) (1994) 1499.
\newblock \href {https://doi.org/10.1088/0954-3899/20/9/018}
  {\path{doi:10.1088/0954-3899/20/9/018}}.

\bibitem{Tamburini:2017sxg}
M.~Tamburini, A.~Di~Piazza, C.~H. Keitel, {Laser-pulse-shape control of seeded
  QED cascades}, Sci. Rep. 7~(1) (2017) 5694.
\newblock \href {https://doi.org/10.1038/s41598-017-05891-z}
  {\path{doi:10.1038/s41598-017-05891-z}}.

\bibitem{Gris2016absorption}
T.~Grismayer, M.~Vranic, J.~L. Martins, R.~A. Fonseca, L.~O. Silva, Laser
  absorption via quantum electrodynamics cascades in counter propagating laser
  pulses, Physics of Plasmas 23~(5) (2016) 056706.
\newblock \href {http://arxiv.org/abs/https://doi.org/10.1063/1.4950841}
  {\path{arXiv:https://doi.org/10.1063/1.4950841}}, \href
  {https://doi.org/10.1063/1.4950841} {\path{doi:10.1063/1.4950841}}.

\bibitem{nerush:2011an}
E.~N. Nerush, V.~F. Bashmakov, I.~Y. Kostyukov, Analytical model for
  electromagnetic cascades in rotating electric field, Physics of Plasmas
  18~(8) (2011) 083107.
\newblock \href {http://arxiv.org/abs/1105.3981v2} {\path{arXiv:1105.3981v2}},
  \href {https://doi.org/10.1063/1.3624481} {\path{doi:10.1063/1.3624481}}.

\bibitem{grismayer:2017se}
T.~Grismayer, M.~Vranic, J.~L. Martins, R.~A. Fonseca, L.~O. Silva, Seeded
  {QED} cascades in counterpropagating laser pulses, Phys. Rev. E 95~(2) (2017)
  023210.
\newblock \href {http://arxiv.org/abs/1511.07503v3}
  {\path{arXiv:1511.07503v3}}, \href
  {https://doi.org/10.1103/PhysRevE.95.023210}
  {\path{doi:10.1103/PhysRevE.95.023210}}.

\bibitem{2014PhPl...21a3105B}
V.~F. {Bashmakov}, E.~N. {Nerush}, I.~Y. {Kostyukov}, A.~M. {Fedotov}, N.~B.
  {Narozhny}, {Effect of laser polarization on quantum electrodynamical
  cascading}, Physics of Plasmas 21~(1) (2014) 013105.
\newblock \href {https://doi.org/10.1063/1.4861863}
  {\path{doi:10.1063/1.4861863}}.

\bibitem{zeldovich:1975in}
Y.~B. Zeldovich, Interaction of free electrons with electromagnetic radiation,
  Soviet Physics Uspekhi 18~(2) (1975) 79.
\newblock \href {https://doi.org/10.1070/PU1975v018n02ABEH001947}
  {\path{doi:10.1070/PU1975v018n02ABEH001947}}.

\bibitem{bulanov:2010sch}
S.~S. Bulanov, T.~Z. Esirkepov, A.~G.~R. Thomas, J.~K. Koga, S.~V. Bulanov,
  Schwinger limit attainability with extreme power lasers, Phys. Rev. Lett.
  105~(22) (2010) 220407.
\newblock \href {http://arxiv.org/abs/1007.4306} {\path{arXiv:1007.4306}},
  \href {https://doi.org/10.1103/PhysRevLett.105.220407}
  {\path{doi:10.1103/PhysRevLett.105.220407}}.

\bibitem{mironov:2021on}
A.~A. Mironov, E.~G. Gelfer, A.~M. Fedotov, Onset of electron-seeded cascades
  in generic electromagnetic fields, Phys. Rev. A 104~(1) (2021) 012221.
\newblock \href {http://arxiv.org/abs/2105.04476} {\path{arXiv:2105.04476}},
  \href {https://doi.org/10.1103/PhysRevA.104.012221}
  {\path{doi:10.1103/PhysRevA.104.012221}}.

\bibitem{Seipt:2020uxv}
D.~Seipt, C.~P. Ridgers, D.~Del~Sorbo, A.~G.~R. Thomas, {Polarized QED
  cascades}, New J. Phys. 23~(5) (2021) 053025.
\newblock \href {http://arxiv.org/abs/2010.04078} {\path{arXiv:2010.04078}},
  \href {https://doi.org/10.1088/1367-2630/abf584}
  {\path{doi:10.1088/1367-2630/abf584}}.

\bibitem{Euler:1935qgl}
H.~Euler, {\"Uber die Streuung von Licht an Licht nach der Diracschen Theorie},
  Annalen Phys. 26~(5) (1936) 398--448.
\newblock \href {https://doi.org/10.1002/andp.19364180503}
  {\path{doi:10.1002/andp.19364180503}}.

\bibitem{Furry:1937zz}
W.~H. Furry, {A Symmetry Theorem in the Positron Theory}, Phys. Rev. 51 (1937)
  125--129.
\newblock \href {https://doi.org/10.1103/PhysRev.51.125}
  {\path{doi:10.1103/PhysRev.51.125}}.

\bibitem{Fan:2017sxk}
X.~Fan, S.~Kamioka, K.~Yamashita, S.~Asai, A.~Sugamoto, {Vacuum magnetic
  birefringence experiment as a probe of the dark sector}, PTEP 2018~(6) (2018)
  063B06.
\newblock \href {http://arxiv.org/abs/1707.03609} {\path{arXiv:1707.03609}},
  \href {https://doi.org/10.1093/ptep/pty059} {\path{doi:10.1093/ptep/pty059}}.

\bibitem{Ghasemkhani:2021kzf}
M.~Ghasemkhani, V.~Rahmanpour, R.~Bufalo, A.~Soto, {Perturbative
  Euler-Heisenberg Lagrangian in a parity-violating Abelian gauge theory} (9
  2021).
\newblock \href {http://arxiv.org/abs/2109.11411} {\path{arXiv:2109.11411}}.

\bibitem{Gorghetto:2021luj}
M.~Gorghetto, G.~Perez, I.~Savoray, Y.~Soreq, {Probing CP violation in photon
  self-interactions with cavities}, JHEP 10 (2021) 056.
\newblock \href {http://arxiv.org/abs/2103.06298} {\path{arXiv:2103.06298}},
  \href {https://doi.org/10.1007/JHEP10(2021)056}
  {\path{doi:10.1007/JHEP10(2021)056}}.

\bibitem{Neves:2021tbt}
M.~J. Neves, J.~B. de~Oliveira, L.~P.~R. Ospedal, J.~A. Helay\"el-Neto,
  {Dispersion relations in nonlinear electrodynamics and the kinematics of the
  Compton effect in a magnetic background}, Phys. Rev. D 104~(1) (2021) 015006.
\newblock \href {http://arxiv.org/abs/2101.03642} {\path{arXiv:2101.03642}},
  \href {https://doi.org/10.1103/PhysRevD.104.015006}
  {\path{doi:10.1103/PhysRevD.104.015006}}.

\bibitem{Karbstein:2021obd}
F.~Karbstein, {Derivative corrections to the Heisenberg-Euler effective
  action}, JHEP 09 (2021) 070.
\newblock \href {http://arxiv.org/abs/2108.02068} {\path{arXiv:2108.02068}},
  \href {https://doi.org/10.1007/JHEP09(2021)070}
  {\path{doi:10.1007/JHEP09(2021)070}}.

\bibitem{Ritus:1975pcc}
V.~Ritus, {Lagrangian of an intense electromagnetic field and quantum
  electrodynamics at short distances}, Sov. Phys. JETP 42~(5) (1975) 774--782.

\bibitem{Gusynin:1995bc}
V.~P. Gusynin, I.~A. Shovkovy, {Derivative expansion for the one loop effective
  Lagrangian in QED}, Can. J. Phys. 74 (1996) 282--289.
\newblock \href {http://arxiv.org/abs/hep-ph/9509383}
  {\path{arXiv:hep-ph/9509383}}, \href {https://doi.org/10.1139/p96-044}
  {\path{doi:10.1139/p96-044}}.

\bibitem{Gusynin:1998bt}
V.~P. Gusynin, I.~A. Shovkovy, {Derivative expansion of the effective action
  for QED in (2+1)-dimensions and (3+1)-dimensions}, J. Math. Phys. 40 (1999)
  5406--5439.
\newblock \href {http://arxiv.org/abs/hep-th/9804143}
  {\path{arXiv:hep-th/9804143}}, \href {https://doi.org/10.1063/1.533037}
  {\path{doi:10.1063/1.533037}}.

\bibitem{Dittrich:1985yb}
W.~Dittrich, M.~Reuter, {Effective Lagrangians in quantum electrodynamics},
  Vol. 220, 1985.

\bibitem{Kim:2012vr}
S.~P. Kim, {In-out formalism for the QED effective action in a constant
  electromagnetic field}, J. Korean Phys. Soc. 61 (2012) 1206--1214.
\newblock \href {https://doi.org/10.3938/jkps.61.1206}
  {\path{doi:10.3938/jkps.61.1206}}.

\bibitem{Kim:2014iia}
S.~P. Kim, H.~K. Lee, {QED Actions in Supercritical Fields} (6 2014).
\newblock \href {http://arxiv.org/abs/1406.4292} {\path{arXiv:1406.4292}}.

\bibitem{Schubert:2001he}
C.~Schubert, {Perturbative quantum field theory in the string inspired
  formalism}, Phys. Rept. 355 (2001) 73--234.
\newblock \href {http://arxiv.org/abs/hep-th/0101036}
  {\path{arXiv:hep-th/0101036}}, \href
  {https://doi.org/10.1016/S0370-1573(01)00013-8}
  {\path{doi:10.1016/S0370-1573(01)00013-8}}.

\bibitem{Huet:2017ydx}
I.~Huet, M.~Rausch~de Traubenberg, C.~Schubert, {Asymptotic Behaviour of the
  QED Perturbation Series}, Adv. High Energy Phys. 2017 (2017) 6214341.
\newblock \href {http://arxiv.org/abs/1707.07655} {\path{arXiv:1707.07655}},
  \href {https://doi.org/10.1155/2017/6214341}
  {\path{doi:10.1155/2017/6214341}}.

\bibitem{Huet:2018ksz}
I.~Huet, M.~Rausch De~Traubenberg, C.~Schubert, {Three-loop Euler-Heisenberg
  Lagrangian in 1$+$1 QED, part 1: single fermion-loop part}, JHEP 03 (2019)
  167.
\newblock \href {http://arxiv.org/abs/1812.08380} {\path{arXiv:1812.08380}},
  \href {https://doi.org/10.1007/JHEP03(2019)167}
  {\path{doi:10.1007/JHEP03(2019)167}}.

\bibitem{Navarro-Salas:2020oew}
J.~Navarro-Salas, S.~Pla, {$(\mathcal{F},\mathcal{G})$-summed form of the QED
  effective action}, Phys. Rev. D 103~(8) (2021) L081702.
\newblock \href {http://arxiv.org/abs/2011.09743} {\path{arXiv:2011.09743}},
  \href {https://doi.org/10.1103/PhysRevD.103.L081702}
  {\path{doi:10.1103/PhysRevD.103.L081702}}.

\bibitem{Pegoraro:2021whz}
F.~Pegoraro, S.~Bulanov, {Nonlinear waves in a dispersive vacuum described with
  a high order derivative electromagnetic Lagrangian}, Phys. Rev. D 103~(9)
  (2021) 096012.
\newblock \href {http://arxiv.org/abs/2103.09744} {\path{arXiv:2103.09744}},
  \href {https://doi.org/10.1103/PhysRevD.103.096012}
  {\path{doi:10.1103/PhysRevD.103.096012}}.

\bibitem{Hattori:2020guh}
K.~Hattori, K.~Itakura, S.~Ozaki, {Note on all-order Landau-level structures of
  the Heisenberg-Euler effective actions for QED and QCD} (1 2020).
\newblock \href {http://arxiv.org/abs/2001.06131} {\path{arXiv:2001.06131}}.

\bibitem{Cho:2000ei}
Y.~M. Cho, D.~G. Pak, {Effective action: A Convergent series of QED}, Phys.
  Rev. Lett. 86 (2001) 1947--1950.
\newblock \href {http://arxiv.org/abs/hep-th/0006057}
  {\path{arXiv:hep-th/0006057}}, \href
  {https://doi.org/10.1103/PhysRevLett.86.1947}
  {\path{doi:10.1103/PhysRevLett.86.1947}}.

\bibitem{Jentschura:2001qr}
U.~D. Jentschura, H.~Gies, S.~R. Valluri, D.~R. Lamm, E.~J. Weniger, {QED
  effective action revisited}, Can. J. Phys. 80 (2002) 267--284.
\newblock \href {http://arxiv.org/abs/hep-th/0107135}
  {\path{arXiv:hep-th/0107135}}, \href {https://doi.org/10.1139/p01-139}
  {\path{doi:10.1139/p01-139}}.

\bibitem{Karbstein:2016hlj}
F.~Karbstein, {The quantum vacuum in electromagnetic fields: From the
  Heisenberg-Euler effective action to vacuum birefringence}, in: {Quantum
  Field Theory at the Limits}: {from Strong Fields to Heavy Quarks}, 2017, pp.
  44--57.
\newblock \href {http://arxiv.org/abs/1611.09883} {\path{arXiv:1611.09883}},
  \href {https://doi.org/10.3204/DESY-PROC-2016-04/Karbstein}
  {\path{doi:10.3204/DESY-PROC-2016-04/Karbstein}}.

\bibitem{Ahmad:2016vvw}
A.~Ahmad, N.~Ahmadiniaz, O.~Corradini, S.~P. Kim, C.~Schubert, {Master formulas
  for the dressed scalar propagator in a constant field}, Nucl. Phys. B 919
  (2017) 9--24.
\newblock \href {http://arxiv.org/abs/1612.02944} {\path{arXiv:1612.02944}},
  \href {https://doi.org/10.1016/j.nuclphysb.2017.03.007}
  {\path{doi:10.1016/j.nuclphysb.2017.03.007}}.

\bibitem{Edwards:2017bte}
J.~P. Edwards, C.~Schubert, {One-particle reducible contribution to the
  one-loop scalar propagator in a constant field}, Nucl. Phys. B 923 (2017)
  339--349.
\newblock \href {http://arxiv.org/abs/1704.00482} {\path{arXiv:1704.00482}},
  \href {https://doi.org/10.1016/j.nuclphysb.2017.08.002}
  {\path{doi:10.1016/j.nuclphysb.2017.08.002}}.

\bibitem{Ahmadiniaz:2017rrk}
N.~Ahmadiniaz, F.~Bastianelli, O.~Corradini, J.~P. Edwards, C.~Schubert,
  {One-particle reducible contribution to the one-loop spinor propagator in a
  constant field}, Nucl. Phys. B 924 (2017) 377--386.
\newblock \href {http://arxiv.org/abs/1704.05040} {\path{arXiv:1704.05040}},
  \href {https://doi.org/10.1016/j.nuclphysb.2017.09.012}
  {\path{doi:10.1016/j.nuclphysb.2017.09.012}}.

\bibitem{Ahmadiniaz:2021ltn}
N.~Ahmadiniaz, F.~Bastianelli, F.~Karbstein, C.~Schubert, {Tadpole contribution
  to magnetic photon-graviton conversion}, 2021.
\newblock \href {http://arxiv.org/abs/2111.01980} {\path{arXiv:2111.01980}}.

\bibitem{Edwards:2018vjd}
J.~P. Edwards, A.~Huet, C.~Schubert, {On the low-energy limit of the QED
  N-photon amplitudes: part 2}, Nucl. Phys. B 935 (2018) 198--209.
\newblock \href {http://arxiv.org/abs/1807.10697} {\path{arXiv:1807.10697}},
  \href {https://doi.org/10.1016/j.nuclphysb.2018.07.026}
  {\path{doi:10.1016/j.nuclphysb.2018.07.026}}.

\bibitem{Voskresensky:2021okp}
D.~N. Voskresensky, {Electron-positron vacuum instability in strong electric
  fields. Relativistic semiclassical approach}, Universe 7~(4) (2021) 104.
\newblock \href {http://arxiv.org/abs/2102.07182} {\path{arXiv:2102.07182}},
  \href {https://doi.org/10.3390/universe7040104}
  {\path{doi:10.3390/universe7040104}}.

\bibitem{Ahmadiniaz:2019nhk}
N.~Ahmadiniaz, J.~P. Edwards, A.~Ilderton, {Reducible contributions to quantum
  electrodynamics in external fields}, JHEP 05 (2019) 038.
\newblock \href {http://arxiv.org/abs/1901.09416} {\path{arXiv:1901.09416}},
  \href {https://doi.org/10.1007/JHEP05(2019)038}
  {\path{doi:10.1007/JHEP05(2019)038}}.

\bibitem{Karbstein:2021gdi}
F.~Karbstein, {Large $N$ external-field quantum electrodynamics}, JHEP 01
  (2022) 057.
\newblock \href {http://arxiv.org/abs/2109.04823} {\path{arXiv:2109.04823}},
  \href {https://doi.org/10.1007/JHEP01(2022)057}
  {\path{doi:10.1007/JHEP01(2022)057}}.

\bibitem{Dittrich:2000zu}
W.~Dittrich, H.~Gies, {Probing the quantum vacuum. Perturbative effective
  action approach in quantum electrodynamics and its application}, Vol. 166,
  2000.
\newblock \href {https://doi.org/10.1007/3-540-45585-X}
  {\path{doi:10.1007/3-540-45585-X}}.

\bibitem{Karbstein:2011ja}
F.~Karbstein, L.~Roessler, B.~Dobrich, H.~Gies, {Optical Probes of the Quantum
  Vacuum: The Photon Polarization Tensor in External Fields}, Int. J. Mod.
  Phys. Conf. Ser. 14 (2012) 403--415.
\newblock \href {http://arxiv.org/abs/1111.5984} {\path{arXiv:1111.5984}},
  \href {https://doi.org/10.1142/S2010194512007520}
  {\path{doi:10.1142/S2010194512007520}}.

\bibitem{Karbstein:2015cpa}
F.~Karbstein, R.~Shaisultanov, {Photon propagation in slowly varying
  inhomogeneous electromagnetic fields}, Phys. Rev. D 91~(8) (2015) 085027.
\newblock \href {http://arxiv.org/abs/1503.00532} {\path{arXiv:1503.00532}},
  \href {https://doi.org/10.1103/PhysRevD.91.085027}
  {\path{doi:10.1103/PhysRevD.91.085027}}.

\bibitem{Gies:2016czm}
H.~Gies, F.~Karbstein, N.~Seegert, {Photon merging and splitting in
  electromagnetic field inhomogeneities}, Phys. Rev. D 93~(8) (2016) 085034.
\newblock \href {http://arxiv.org/abs/1603.00314} {\path{arXiv:1603.00314}},
  \href {https://doi.org/10.1103/PhysRevD.93.085034}
  {\path{doi:10.1103/PhysRevD.93.085034}}.

\bibitem{Adler:1970gg}
S.~L. Adler, J.~N. Bahcall, C.~G. Callan, M.~N. Rosenbluth, {Photon splitting
  in a strong magnetic field}, Phys. Rev. Lett. 25 (1970) 1061--1065.
\newblock \href {https://doi.org/10.1103/PhysRevLett.25.1061}
  {\path{doi:10.1103/PhysRevLett.25.1061}}.

\bibitem{Adler:1971wn}
S.~L. Adler, {Photon splitting and photon dispersion in a strong magnetic
  field}, Annals Phys. 67 (1971) 599--647.
\newblock \href {https://doi.org/10.1016/0003-4916(71)90154-0}
  {\path{doi:10.1016/0003-4916(71)90154-0}}.

\bibitem{Papanyan:1971cv}
V.~O. Papanyan, V.~I. Ritus, {Vacuum polarization and photon splitting in an
  intense field}, Sov. Phys. JETP 34~(6) (1972) 1195--1199.

\bibitem{Papanyan:1973xa}
V.~O. Papanyan, V.~I. Ritus, {Three-photon interaction in an intense field and
  scaling invariance}, Sov. Phys. JETP 38~(5) (1974) 879--885.

\bibitem{Batalin:1971au}
I.~A. Batalin, A.~E. Shabad, {Photon green function in a stationary homogeneous
  field of the most general form}, Sov. Phys. JETP 33~(3) (1971) 483--486.

\bibitem{Becker:1974en}
W.~Becker, H.~Mitter, {Vacuum polarization in laser fields}, J. Phys. A 8~(10)
  (1975) 1638--1657.
\newblock \href {https://doi.org/10.1088/0305-4470/8/10/017}
  {\path{doi:10.1088/0305-4470/8/10/017}}.

\bibitem{Meuren:2013oya}
S.~Meuren, C.~H. Keitel, A.~Di~Piazza, {Polarization operator for plane-wave
  background fields}, Phys. Rev. D 88~(1) (2013) 013007.
\newblock \href {http://arxiv.org/abs/1304.7672} {\path{arXiv:1304.7672}},
  \href {https://doi.org/10.1103/PhysRevD.88.013007}
  {\path{doi:10.1103/PhysRevD.88.013007}}.

\bibitem{Narozhny:1968}
N.~B. Narozhny, {Propagation of plane electromagnetic waves in a constant
  field}, Sov. Phys. JETP 28~(2) (1969) 371--374.

\bibitem{Karbstein:2017jgh}
F.~Karbstein, E.~A. Mosman, {Photon polarization tensor in pulsed Hermite- and
  Laguerre-Gaussian beams}, Phys. Rev. D 96~(11) (2017) 116004.
\newblock \href {http://arxiv.org/abs/1711.06151} {\path{arXiv:1711.06151}},
  \href {https://doi.org/10.1103/PhysRevD.96.116004}
  {\path{doi:10.1103/PhysRevD.96.116004}}.

\bibitem{Karbstein:2017uzq}
F.~Karbstein, E.~A. Mosman, {Photon polarization tensor in circularly polarized
  Hermite- and Laguerre-Gaussian beams}, Mod. Phys. Lett. A 33~(07n08) (2018)
  1850044.
\newblock \href {http://arxiv.org/abs/1712.08898} {\path{arXiv:1712.08898}},
  \href {https://doi.org/10.1142/S021773231850044X}
  {\path{doi:10.1142/S021773231850044X}}.

\bibitem{Karbstein:2015xra}
F.~Karbstein, H.~Gies, M.~Reuter, M.~Zepf, {Vacuum birefringence in strong
  inhomogeneous electromagnetic fields}, Phys. Rev. D 92~(7) (2015) 071301.
\newblock \href {http://arxiv.org/abs/1507.01084} {\path{arXiv:1507.01084}},
  \href {https://doi.org/10.1103/PhysRevD.92.071301}
  {\path{doi:10.1103/PhysRevD.92.071301}}.

\bibitem{Bialynicka-Birula:1970nlh}
Z.~Bialynicka-Birula, I.~Bialynicki-Birula, {Nonlinear effects in Quantum
  Electrodynamics. Photon propagation and photon splitting in an external
  field}, Phys. Rev. D 2 (1970) 2341--2345.
\newblock \href {https://doi.org/10.1103/PhysRevD.2.2341}
  {\path{doi:10.1103/PhysRevD.2.2341}}.

\bibitem{Brezin:1971nd}
E.~Brezin, C.~Itzykson, {Polarization phenomena in vacuum nonlinear
  electrodynamics}, Phys. Rev. D 3 (1971) 618--621.
\newblock \href {https://doi.org/10.1103/PhysRevD.3.618}
  {\path{doi:10.1103/PhysRevD.3.618}}.

\bibitem{Hattori:2012ny}
K.~Hattori, K.~Itakura, {Vacuum birefringence in strong magnetic fields: (II)
  Complex refractive index from the lowest Landau level}, Annals Phys. 334
  (2013) 58--82.
\newblock \href {http://arxiv.org/abs/1212.1897} {\path{arXiv:1212.1897}},
  \href {https://doi.org/10.1016/j.aop.2013.03.016}
  {\path{doi:10.1016/j.aop.2013.03.016}}.

\bibitem{Hattori:2012je}
K.~Hattori, K.~Itakura, {Vacuum birefringence in strong magnetic fields: (I)
  Photon polarization tensor with all the Landau levels}, Annals Phys. 330
  (2013) 23--54.
\newblock \href {http://arxiv.org/abs/1209.2663} {\path{arXiv:1209.2663}},
  \href {https://doi.org/10.1016/j.aop.2012.11.010}
  {\path{doi:10.1016/j.aop.2012.11.010}}.

\bibitem{Ishikawa:2013fxa}
K.-I. Ishikawa, D.~Kimura, K.~Shigaki, A.~Tsuji, {A numerical evaluation of
  vacuum polarization tensor in constant external magnetic fields}, Int. J.
  Mod. Phys. A 28 (2013) 1350100.
\newblock \href {http://arxiv.org/abs/1304.3655} {\path{arXiv:1304.3655}},
  \href {https://doi.org/10.1142/S0217751X13501005}
  {\path{doi:10.1142/S0217751X13501005}}.

\bibitem{Denisov:2016pfu}
V.~I. Denisov, E.~E. Dolgaya, V.~A. Sokolov, {Nonperturbative QED vacuum
  birefringence}, JHEP 05 (2017) 105.
\newblock \href {http://arxiv.org/abs/1612.09086} {\path{arXiv:1612.09086}},
  \href {https://doi.org/10.1007/JHEP05(2017)105}
  {\path{doi:10.1007/JHEP05(2017)105}}.

\bibitem{Kim:2022fkt}
C.~M. Kim, S.~P. Kim, {Vacuum Birefringence in a Supercritical Magnetic Field
  and a Subcritical Electric Field} (2 2022).
\newblock \href {http://arxiv.org/abs/2202.05477} {\path{arXiv:2202.05477}}.

\bibitem{2011PhRvL.107e3604K}
G.~Y. {Kryuchkyan}, K.~Z. {Hatsagortsyan}, {Bragg Scattering of Light in Vacuum
  Structured by Strong Periodic Fields}, Phys.Rev.Lett 107~(5) (2011) 053604.
\newblock \href {http://arxiv.org/abs/1102.4013} {\path{arXiv:1102.4013}},
  \href {https://doi.org/10.1103/PhysRevLett.107.053604}
  {\path{doi:10.1103/PhysRevLett.107.053604}}.

\bibitem{Gies:2013yxa}
H.~Gies, F.~Karbstein, N.~Seegert, {Quantum Reflection as a New Signature of
  Quantum Vacuum Nonlinearity}, New J. Phys. 15 (2013) 083002.
\newblock \href {http://arxiv.org/abs/1305.2320} {\path{arXiv:1305.2320}},
  \href {https://doi.org/10.1088/1367-2630/15/8/083002}
  {\path{doi:10.1088/1367-2630/15/8/083002}}.

\bibitem{Gies:2014wsa}
H.~Gies, F.~Karbstein, N.~Seegert, {Quantum reflection of photons off
  spatio-temporal electromagnetic field inhomogeneities}, New J. Phys. 17~(4)
  (2015) 043060.
\newblock \href {http://arxiv.org/abs/1412.0951} {\path{arXiv:1412.0951}},
  \href {https://doi.org/10.1088/1367-2630/17/4/043060}
  {\path{doi:10.1088/1367-2630/17/4/043060}}.

\bibitem{Dinu:2014tsa}
V.~Dinu, T.~Heinzl, A.~Ilderton, M.~Marklund, G.~Torgrimsson, {Photon
  polarization in light-by-light scattering: Finite size effects}, Phys. Rev. D
  90~(4) (2014) 045025.
\newblock \href {http://arxiv.org/abs/1405.7291} {\path{arXiv:1405.7291}},
  \href {https://doi.org/10.1103/PhysRevD.90.045025}
  {\path{doi:10.1103/PhysRevD.90.045025}}.

\bibitem{Karbstein:2021otv}
F.~Karbstein, {Vacuum birefringence at the Gamma Factory}, in: {Physics
  Opportunities with the Gamma Factory}, 2021.
\newblock \href {http://arxiv.org/abs/2106.06359} {\path{arXiv:2106.06359}},
  \href {https://doi.org/10.1002/andp.202100137}
  {\path{doi:10.1002/andp.202100137}}.

\bibitem{DiPiazza:2006pr}
A.~Di~Piazza, K.~Z. Hatsagortsyan, C.~H. Keitel, {Light diffraction by a strong
  standing electromagnetic wave}, Phys. Rev. Lett. 97 (2006) 083603.
\newblock \href {http://arxiv.org/abs/hep-ph/0602039}
  {\path{arXiv:hep-ph/0602039}}, \href
  {https://doi.org/10.1103/PhysRevLett.97.083603}
  {\path{doi:10.1103/PhysRevLett.97.083603}}.

\bibitem{King:2010nka}
B.~King, A.~Di~Piazza, C.~H. Keitel, {A matterless double slit}, Nature Photon.
  4 (2010) 92--94.
\newblock \href {http://arxiv.org/abs/1301.7038} {\path{arXiv:1301.7038}},
  \href {https://doi.org/10.1038/nphoton.2009.261}
  {\path{doi:10.1038/nphoton.2009.261}}.

\bibitem{King:2010kvw}
B.~King, A.~Di~Piazza, C.~H. Keitel, {Double-slit vacuum polarisation effects
  in ultra-intense laser fields}, Phys. Rev. A 82 (2010) 032114.
\newblock \href {http://arxiv.org/abs/1301.7008} {\path{arXiv:1301.7008}},
  \href {https://doi.org/10.1103/PhysRevA.82.032114}
  {\path{doi:10.1103/PhysRevA.82.032114}}.

\bibitem{domenech2017implicit}
A.~P. Domenech, H.~Ruhl, An implicit ode-based numerical solver for the
  simulation of the heisenberg-euler equations in 3+1 dimensions (2017).
\newblock \href {http://arxiv.org/abs/1607.00253} {\path{arXiv:1607.00253}}.

\bibitem{grismayer2021}
T.~Grismayer, R.~Torres, P.~Carneiro, F.~Cruz, R.~A. Fonseca, L.~O. Silva,
  \href{https://doi.org/10.1088/1367-2630/ac2004}{Quantum electrodynamics
  vacuum polarization solver} 23~(9) (2021) 095005.
\newblock \href {https://doi.org/10.1088/1367-2630/ac2004}
  {\path{doi:10.1088/1367-2630/ac2004}}.
\newline\urlprefix\url{https://doi.org/10.1088/1367-2630/ac2004}

\bibitem{Lindner:2021krv}
A.~Lindner, B.~\"Olmez, H.~Ruhl, {Numerical Simulations of the Nonlinear
  Quantum Vacuum in the Heisenberg-Euler Weak-Field Expansion} (9 2021).
\newblock \href {http://arxiv.org/abs/2109.08121} {\path{arXiv:2109.08121}}.

\bibitem{Galtsov:1971xm}
D.~Galtsov, V.~Skobelev, {Photons creation by an external field}, Phys. Lett. B
  36 (1971) 238--240.
\newblock \href {https://doi.org/10.1016/0370-2693(71)90077-3}
  {\path{doi:10.1016/0370-2693(71)90077-3}}.

\bibitem{Hu:2014kda}
H.~Hu, J.~Huang, {Modified light cone condition via vacuum polarization in a
  time dependent field}, Phys. Rev. A 90~(6) (2014) 062111.
\newblock \href {http://arxiv.org/abs/1408.3258} {\path{arXiv:1408.3258}},
  \href {https://doi.org/10.1103/PhysRevA.90.062111}
  {\path{doi:10.1103/PhysRevA.90.062111}}.

\bibitem{Briscese:2017htx}
F.~Briscese, {Collective behavior of light in vacuum}, Phys. Rev. A 97~(3)
  (2018) 033803.
\newblock \href {http://arxiv.org/abs/1710.07703} {\path{arXiv:1710.07703}},
  \href {https://doi.org/10.1103/PhysRevA.97.033803}
  {\path{doi:10.1103/PhysRevA.97.033803}}.

\bibitem{Briscese:2017wuh}
F.~Briscese, {Light polarization oscillations induced by photon-photon
  scattering}, Phys. Rev. A 96~(5) (2017) 053801.
\newblock \href {http://arxiv.org/abs/1710.03338} {\path{arXiv:1710.03338}},
  \href {https://doi.org/10.1103/PhysRevA.96.053801}
  {\path{doi:10.1103/PhysRevA.96.053801}}.

\bibitem{King:2014vha}
B.~King, P.~B\"ohl, H.~Ruhl, {Interaction of photons traversing a slowly
  varying electromagnetic background}, Phys. Rev. D 90~(6) (2014) 065018.
\newblock \href {http://arxiv.org/abs/1406.4139} {\path{arXiv:1406.4139}},
  \href {https://doi.org/10.1103/PhysRevD.90.065018}
  {\path{doi:10.1103/PhysRevD.90.065018}}.

\bibitem{Karbstein:2016lby}
F.~Karbstein, C.~Sundqvist, {Probing vacuum birefringence using x-ray free
  electron and optical high-intensity lasers}, Phys. Rev. D 94~(1) (2016)
  013004.
\newblock \href {http://arxiv.org/abs/1605.09294} {\path{arXiv:1605.09294}},
  \href {https://doi.org/10.1103/PhysRevD.94.013004}
  {\path{doi:10.1103/PhysRevD.94.013004}}.

\bibitem{Karbstein:2019dxo}
F.~Karbstein, A.~Blinne, H.~Gies, M.~Zepf, {Boosting quantum vacuum signatures
  by coherent harmonic focusing}, Phys. Rev. Lett. 123~(9) (2019) 091802.
\newblock \href {http://arxiv.org/abs/1905.00858} {\path{arXiv:1905.00858}},
  \href {https://doi.org/10.1103/PhysRevLett.123.091802}
  {\path{doi:10.1103/PhysRevLett.123.091802}}.

\bibitem{Battesti:2018bgc}
R.~Battesti, et~al., {High magnetic fields for fundamental physics}, Phys.
  Rept. 765-766 (2018) 1--39.
\newblock \href {http://arxiv.org/abs/1803.07547} {\path{arXiv:1803.07547}},
  \href {https://doi.org/10.1016/j.physrep.2018.07.005}
  {\path{doi:10.1016/j.physrep.2018.07.005}}.

\bibitem{Agil:2021fiq}
J.~Agil, R.~Battesti, C.~Rizzo, {Monte Carlo study of the BMV vacuum linear
  magnetic birefringence experiment}, Eur. Phys. J. D 75~(3) (2021) 90.
\newblock \href {https://doi.org/10.1140/epjd/s10053-021-00100-z}
  {\path{doi:10.1140/epjd/s10053-021-00100-z}}.

\bibitem{Fan:2017fnd}
X.~Fan, et~al., {The OVAL experiment: A new experiment to measure vacuum
  magnetic birefringence using high repetition pulsed magnets}, Eur. Phys. J. D
  71~(11) (2017) 308.
\newblock \href {http://arxiv.org/abs/1705.00495} {\path{arXiv:1705.00495}},
  \href {https://doi.org/10.1140/epjd/e2017-80290-7}
  {\path{doi:10.1140/epjd/e2017-80290-7}}.

\bibitem{Iacopini:1979ci}
E.~Iacopini, E.~Zavattini, {Experimental Method to Detect the Vacuum
  Birefringence Induced by a Magnetic Field}, Phys. Lett. B 85 (1979) 151.
\newblock \href {https://doi.org/10.1016/0370-2693(79)90797-4}
  {\path{doi:10.1016/0370-2693(79)90797-4}}.

\bibitem{Fouche:2016qqj}
M.~Fouch\'e, R.~Battesti, C.~Rizzo, {Limits on nonlinear electrodynamics},
  Phys. Rev. D 93~(9) (2016) 093020, [Erratum: Phys.Rev.D 95, 099902 (2017)].
\newblock \href {http://arxiv.org/abs/1605.04102} {\path{arXiv:1605.04102}},
  \href {https://doi.org/10.1103/PhysRevD.93.093020}
  {\path{doi:10.1103/PhysRevD.93.093020}}.

\bibitem{Marx:2013xwa}
B.~Marx, et~al., {High-Precision X-Ray Polarimetry}, Phys. Rev. Lett. 110~(25)
  (2013) 254801.
\newblock \href {https://doi.org/10.1103/PhysRevLett.110.254801}
  {\path{doi:10.1103/PhysRevLett.110.254801}}.

\bibitem{Bernhardt:2020vxa}
H.~Bernhardt, et~al., {Ultra-high precision x-ray polarimetry with artificial
  diamond channel cuts at the beam divergence limit}, Phys. Rev. Res. 2~(2)
  (2020) 023365.
\newblock \href {https://doi.org/10.1103/PhysRevResearch.2.023365}
  {\path{doi:10.1103/PhysRevResearch.2.023365}}.

\bibitem{Ataman:2018ucl}
S.~Ataman, {Vacuum birefringence detection in all-optical scenarios}, Phys.
  Rev. A 97~(6) (2018) 063811.
\newblock \href {http://arxiv.org/abs/1807.11299} {\path{arXiv:1807.11299}},
  \href {https://doi.org/10.1103/PhysRevA.97.063811}
  {\path{doi:10.1103/PhysRevA.97.063811}}.

\bibitem{Shen:2018lbq}
B.~Shen, Z.~Bu, J.~Xu, T.~Xu, L.~Ji, R.~Li, Z.~Xu, {Exploring vacuum
  birefringence based on a 100 PW laser and an x-ray free electron laser beam},
  Plasma Phys. Control. Fusion 60~(4) (2018) 044002.
\newblock \href {https://doi.org/10.1088/1361-6587/aaa7fb}
  {\path{doi:10.1088/1361-6587/aaa7fb}}.

\bibitem{Karbstein:2018omb}
F.~Karbstein, {Vacuum birefringence in the head-on collision of x-ray
  free-electron laser and optical high-intensity laser pulses}, Phys. Rev. D
  98~(5) (2018) 056010.
\newblock \href {http://arxiv.org/abs/1807.03302} {\path{arXiv:1807.03302}},
  \href {https://doi.org/10.1103/PhysRevD.98.056010}
  {\path{doi:10.1103/PhysRevD.98.056010}}.

\bibitem{Tzenov:2019ovq}
S.~I. Tzenov, K.~M. Spohr, K.~A. Tanaka, {Dispersion Properties, Nonlinear
  Waves and Birefringence in Classical Nonlinear Electrodynamics}, J. Phys.
  Comm. 4~(2) (2020) 025006.
\newblock \href {http://arxiv.org/abs/1910.08586} {\path{arXiv:1910.08586}},
  \href {https://doi.org/10.1088/2399-6528/ab72c7}
  {\path{doi:10.1088/2399-6528/ab72c7}}.

\bibitem{Ahmadiniaz:2020lbg}
N.~Ahmadiniaz, T.~E. Cowan, R.~Sauerbrey, U.~Schramm, H.~P. Schlenvoigt,
  R.~Sch\"utzhold, {Heisenberg limit for detecting vacuum birefringence}, Phys.
  Rev. D 101~(11) (2020) 116019.
\newblock \href {http://arxiv.org/abs/2003.10519} {\path{arXiv:2003.10519}},
  \href {https://doi.org/10.1103/PhysRevD.101.116019}
  {\path{doi:10.1103/PhysRevD.101.116019}}.

\bibitem{Aleksandrov:1985}
E.~B. Aleksandrov, A.~A. Ansel'm, A.~N. Moskalev, {Vacuum birefringence in an
  intense laser radiation field}, Sov. Phys. JETP 62 (1985) 680--684.

\bibitem{Heinzl:2006xc}
T.~Heinzl, B.~Liesfeld, K.-U. Amthor, H.~Schwoerer, R.~Sauerbrey, A.~Wipf, {On
  the observation of vacuum birefringence}, Opt. Commun. 267 (2006) 318--321.
\newblock \href {http://arxiv.org/abs/hep-ph/0601076}
  {\path{arXiv:hep-ph/0601076}}, \href
  {https://doi.org/10.1016/j.optcom.2006.06.053}
  {\path{doi:10.1016/j.optcom.2006.06.053}}.

\bibitem{Inada:2017lop}
T.~Inada, T.~Yamazaki, T.~Yamaji, Y.~Seino, X.~Fan, S.~Kamioka, T.~Namba,
  S.~Asai, {Probing Physics in Vacuum Using an X-ray Free-Electron Laser, a
  High-Power Laser, and a High-Field Magnet}, Applied Science 7 (2017) 671.
\newblock \href {http://arxiv.org/abs/1707.00253} {\path{arXiv:1707.00253}},
  \href {https://doi.org/10.3390/app7070671} {\path{doi:10.3390/app7070671}}.

\bibitem{2020NIMPA.98264553X}
D.~{Xu}, B.~{Shen}, J.~{Xu}, Z.~{Liang}, {XFEL beamline design for vacuum
  birefringence experiment}, Nuclear Instruments and Methods in Physics
  Research A 982 (2020) 164553.
\newblock \href {https://doi.org/10.1016/j.nima.2020.164553}
  {\path{doi:10.1016/j.nima.2020.164553}}.

\bibitem{Mosman:2021vua}
E.~A. Mosman, F.~Karbstein, {Vacuum birefringence and diffraction at an x-ray
  free-electron laser: From analytical estimates to optimal parameters}, Phys.
  Rev. D 104~(1) (2021) 013006.
\newblock \href {http://arxiv.org/abs/2104.05103} {\path{arXiv:2104.05103}},
  \href {https://doi.org/10.1103/PhysRevD.104.013006}
  {\path{doi:10.1103/PhysRevD.104.013006}}.

\bibitem{Seino:2019wkb}
Y.~Seino, T.~Inada, T.~Yamazaki, T.~Namba, S.~Asai, {New estimation of the
  curvature effect for the X-ray vacuum diffraction induced by an intense laser
  field}, PTEP 2020~(7) (2020) 073C02.
\newblock \href {http://arxiv.org/abs/1912.01390} {\path{arXiv:1912.01390}},
  \href {https://doi.org/10.1093/ptep/ptaa084}
  {\path{doi:10.1093/ptep/ptaa084}}.

\bibitem{Karbstein:2019bhp}
F.~Karbstein, E.~A. Mosman, {X-ray photon scattering at a focused
  high-intensity laser pulse}, Phys. Rev. D 100~(3) (2019) 033002.
\newblock \href {http://arxiv.org/abs/1906.10122} {\path{arXiv:1906.10122}},
  \href {https://doi.org/10.1103/PhysRevD.100.033002}
  {\path{doi:10.1103/PhysRevD.100.033002}}.

\bibitem{Karbstein:2021hwc}
F.~Karbstein, R.~R. Q. P. T.~O. Weernink, {X-ray vacuum diffraction at finite
  spatiotemporal offset}, Phys. Rev. D 104~(7) (2021) 076015.
\newblock \href {http://arxiv.org/abs/2107.09632} {\path{arXiv:2107.09632}},
  \href {https://doi.org/10.1103/PhysRevD.104.076015}
  {\path{doi:10.1103/PhysRevD.104.076015}}.

\bibitem{Ahmadiniaz:2020kpl}
N.~Ahmadiniaz, M.~Bussmann, T.~E. Cowan, A.~Debus, T.~Kluge, R.~Sch\"utzhold,
  {Observability of Coulomb-assisted quantum vacuum birefringence}, Phys. Rev.
  D 104~(1) (2021) L011902.
\newblock \href {http://arxiv.org/abs/2012.04484} {\path{arXiv:2012.04484}},
  \href {https://doi.org/10.1103/PhysRevD.104.L011902}
  {\path{doi:10.1103/PhysRevD.104.L011902}}.

\bibitem{Ahmadiniaz:2020jgo}
N.~Ahmadiniaz, C.~Lopez-Arcos, M.~A. Lopez-Lopez, C.~Schubert, {The QED four --
  photon amplitudes off-shell: part 1} (12 2020).
\newblock \href {http://arxiv.org/abs/2012.11791} {\path{arXiv:2012.11791}}.

\bibitem{Bernard:2000ovj}
D.~Bernard, F.~Moulin, F.~Amiranoff, A.~Braun, J.~P. Chambaret, G.~Darpentigny,
  G.~Grillon, S.~Ranc, F.~Perrone, {Search for Stimulated Photon-Photon
  Scattering in Vacuum}, Eur. Phys. J. D 10 (2000) 141.
\newblock \href {http://arxiv.org/abs/1007.0104} {\path{arXiv:1007.0104}},
  \href {https://doi.org/10.1007/s100530050535}
  {\path{doi:10.1007/s100530050535}}.

\bibitem{Sangal:2021qeg}
M.~Sangal, C.~H. Keitel, M.~Tamburini, {Observing Light-by-Light Scattering in
  Vacuum with an Asymmetric Photon Collider} (1 2021).
\newblock \href {http://arxiv.org/abs/2101.02671} {\path{arXiv:2101.02671}}.

\bibitem{Kotkin:1996nf}
G.~L. Kotkin, V.~G. Serbo, {Variation in polarization of high-energy gamma
  quanta traversing a bunch of polarized laser photons}, Phys. Lett. B 413
  (1997) 122--129.
\newblock \href {http://arxiv.org/abs/hep-ph/9611345}
  {\path{arXiv:hep-ph/9611345}}, \href
  {https://doi.org/10.1016/S0370-2693(97)01086-1}
  {\path{doi:10.1016/S0370-2693(97)01086-1}}.

\bibitem{Ilderton:2016khs}
A.~Ilderton, M.~Marklund, {Prospects for studying vacuum polarisation using
  dipole and synchrotron radiation}, J. Plasma Phys. 82~(2) (2016) 655820201.
\newblock \href {http://arxiv.org/abs/1601.08045} {\path{arXiv:1601.08045}},
  \href {https://doi.org/10.1017/S0022377816000192}
  {\path{doi:10.1017/S0022377816000192}}.

\bibitem{Nakamiya:2015pde}
Y.~Nakamiya, K.~Homma, T.~Moritaka, K.~Seto, {Probing vacuum birefringence
  under a high-intensity laser field with gamma-ray polarimetry at the GeV
  scale}, Phys. Rev. D 96~(5) (2017) 053002.
\newblock \href {http://arxiv.org/abs/1512.00636} {\path{arXiv:1512.00636}},
  \href {https://doi.org/10.1103/PhysRevD.96.053002}
  {\path{doi:10.1103/PhysRevD.96.053002}}.

\bibitem{Cantatore:1991sq}
G.~Cantatore, F.~Della~Valle, E.~Milotti, L.~Dabrowski, C.~Rizzo, {Proposed
  measurement of the vacuum birefringence induced by a magnetic field on
  high-energy photons}, Phys. Lett. B 265 (1991) 418--424.
\newblock \href {https://doi.org/10.1016/0370-2693(91)90077-4}
  {\path{doi:10.1016/0370-2693(91)90077-4}}.

\bibitem{Wistisen:2013waa}
T.~N. Wistisen, U.~I. Uggerh\o{}j, {Vacuum birefringence by Compton
  backscattering through a strong field}, Phys. Rev. D 88~(5) (2013) 053009.
\newblock \href {https://doi.org/10.1103/PhysRevD.88.053009}
  {\path{doi:10.1103/PhysRevD.88.053009}}.

\bibitem{Gies:2011he}
H.~Gies, L.~Roessler, {Vacuum polarization tensor in inhomogeneous magnetic
  fields}, Phys. Rev. D 84 (2011) 065035.
\newblock \href {http://arxiv.org/abs/1107.0286} {\path{arXiv:1107.0286}},
  \href {https://doi.org/10.1103/PhysRevD.84.065035}
  {\path{doi:10.1103/PhysRevD.84.065035}}.

\bibitem{King:2018wtn}
B.~King, H.~Hu, B.~Shen, {Three-pulse photon-photon scattering}, Phys. Rev. A
  98~(2) (2018) 023817.
\newblock \href {http://arxiv.org/abs/1805.03688} {\path{arXiv:1805.03688}},
  \href {https://doi.org/10.1103/PhysRevA.98.023817}
  {\path{doi:10.1103/PhysRevA.98.023817}}.

\bibitem{Karbstein:2021ldz}
F.~Karbstein, C.~Sundqvist, K.~S. Schulze, I.~Uschmann, H.~Gies, G.~G. Paulus,
  {Vacuum birefringence at x-ray free-electron lasers}, New J. Phys. 23~(9)
  (2021) 095001.
\newblock \href {http://arxiv.org/abs/2105.13869} {\path{arXiv:2105.13869}},
  \href {https://doi.org/10.1088/1367-2630/ac1df4}
  {\path{doi:10.1088/1367-2630/ac1df4}}.

\bibitem{Blinne:2018nbd}
A.~Blinne, H.~Gies, F.~Karbstein, C.~Kohlf\"urst, M.~Zepf, {All-optical
  signatures of quantum vacuum nonlinearities in generic laser fields}, Phys.
  Rev. D 99~(1) (2019) 016006.
\newblock \href {http://arxiv.org/abs/1811.08895} {\path{arXiv:1811.08895}},
  \href {https://doi.org/10.1103/PhysRevD.99.016006}
  {\path{doi:10.1103/PhysRevD.99.016006}}.

\bibitem{Sainte-Marie:2022efx}
A.~Sainte-Marie, L.~Fedeli, N.~Za\"\i{}m, F.~Karbstein, H.~Vincenti, {Quantum
  vacuum processes in the extremely intense light of relativistic plasma mirror
  sources} (1 2022).
\newblock \href {http://arxiv.org/abs/2201.09886} {\path{arXiv:2201.09886}}.

\bibitem{2018arXiv180104812B}
A.~{Blinne}, S.~{Kuschel}, S.~{Tietze}, M.~{Zepf}, {Efficient retrieval of
  phase information from real-valued electromagnetic field data}, arXiv
  e-prints (2018) arXiv:1801.04812\href {http://arxiv.org/abs/1801.04812}
  {\path{arXiv:1801.04812}}.

\bibitem{Waters:2017tgl}
W.~J. Waters, B.~King, {On beam models and their paraxial approximation}, Laser
  Phys. 28 (2018) 015003.
\newblock \href {http://arxiv.org/abs/1705.08554} {\path{arXiv:1705.08554}},
  \href {https://doi.org/10.1088/1555-6611/aa94dc}
  {\path{doi:10.1088/1555-6611/aa94dc}}.

\bibitem{Fedotov:2006ii}
A.~M. Fedotov, N.~B. Narozhny, {Generation of harmonics by a focused laser beam
  in vacuum}, Phys. Lett. A 362 (2007) 1--5.
\newblock \href {http://arxiv.org/abs/hep-ph/0604258}
  {\path{arXiv:hep-ph/0604258}}, \href
  {https://doi.org/10.1016/j.physleta.2006.09.085}
  {\path{doi:10.1016/j.physleta.2006.09.085}}.

\bibitem{2011PhRvL.107g3602M}
Y.~{Monden}, R.~{Kodama}, {Enhancement of Laser Interaction with Vacuum for a
  Large Angular Aperture}, Phys. Rev. Lett. 107~(7) (2011) 073602.
\newblock \href {https://doi.org/10.1103/PhysRevLett.107.073602}
  {\path{doi:10.1103/PhysRevLett.107.073602}}.

\bibitem{Paredes:2014oxa}
A.~Paredes, D.~Novoa, D.~Tommasini, {Self-induced mode mixing of ultraintense
  lasers in vacuum}, Phys. Rev. A 90~(6) (2014) 063803.
\newblock \href {http://arxiv.org/abs/1412.3390} {\path{arXiv:1412.3390}},
  \href {https://doi.org/10.1103/PhysRevA.90.063803}
  {\path{doi:10.1103/PhysRevA.90.063803}}.

\bibitem{2012PhRvA..86c3810M}
Y.~{Monden}, R.~{Kodama}, {Interaction of two counterpropagating laser beams
  with vacuum}, Phys. Rev. A 86~(3) (2012) 033810.
\newblock \href {https://doi.org/10.1103/PhysRevA.86.033810}
  {\path{doi:10.1103/PhysRevA.86.033810}}.

\bibitem{Kadlecova:2018vty}
H.~Kadlecov\'a, G.~Korn, S.~V. Bulanov, {Electromagnetic shocks in the quantum
  vacuum}, Phys. Rev. D 99~(3) (2019) 036002.
\newblock \href {http://arxiv.org/abs/1807.11365} {\path{arXiv:1807.11365}},
  \href {https://doi.org/10.1103/PhysRevD.99.036002}
  {\path{doi:10.1103/PhysRevD.99.036002}}.

\bibitem{Kadlecova:2019dxv}
H.~Kadlecov\'a, S.~V. Bulanov, G.~Korn, {Properties of Finite Amplitude
  Electromagnetic Waves propagating in the Quantum Vacuum}, Plasma Phys.
  Control. Fusion 61~(8) (2019) 084002.
\newblock \href {http://arxiv.org/abs/1902.04928} {\path{arXiv:1902.04928}},
  \href {https://doi.org/10.1088/1361-6587/ab21fb}
  {\path{doi:10.1088/1361-6587/ab21fb}}.

\bibitem{Huang:2019ojh}
S.~Huang, B.~Jin, B.~Shen, {Two-beam vacuum wave mixing using high-power laser
  and x-ray free-electron laser}, Phys. Rev. D 100~(1) (2019) 013004.
\newblock \href {https://doi.org/10.1103/PhysRevD.100.013004}
  {\path{doi:10.1103/PhysRevD.100.013004}}.

\bibitem{Pegoraro:2019obj}
F.~Pegoraro, S.~V. Bulanov, {Hodograph solutions of the wave equation of
  nonlinear electrodynamics in the quantum vacuum}, Phys. Rev. D 100~(3) (2019)
  036004.
\newblock \href {http://arxiv.org/abs/1903.01733} {\path{arXiv:1903.01733}},
  \href {https://doi.org/10.1103/PhysRevD.100.036004}
  {\path{doi:10.1103/PhysRevD.100.036004}}.

\bibitem{Robertson:2020nnc}
S.~Robertson, A.~Mailliet, X.~Sarazin, F.~Couchot, E.~Baynard, J.~Demailly,
  M.~Pittman, A.~Djannati-Ata\"\i{}, S.~Kazamias, M.~Urban, {Experiment to
  observe an optically induced change of the vacuum index}, Phys. Rev. A
  103~(2) (2021) 023524.
\newblock \href {http://arxiv.org/abs/2011.13889} {\path{arXiv:2011.13889}},
  \href {https://doi.org/10.1103/PhysRevA.103.023524}
  {\path{doi:10.1103/PhysRevA.103.023524}}.

\bibitem{Roso:2021hfo}
L.~Roso, R.~Lera, S.~Ravichandran, A.~Longman, C.~Z. He, J.~A.
  P\'erez-Hern\'andez, J.~I. Api\~naniz, L.~D. Smith, R.~Fedosejevs, W.~T.
  Hill, {Towards a direct measurement of the quantum vacuum Lagrangian coupling
  coefficients using two counter propagating super-intense laser pulses} (12
  2021).
\newblock \href {http://arxiv.org/abs/2112.08302} {\path{arXiv:2112.08302}}.

\bibitem{Gies:2014jia}
H.~Gies, F.~Karbstein, R.~Shaisultanov, {Laser photon merging in an
  electromagnetic field inhomogeneity}, Phys. Rev. D 90~(3) (2014) 033007.
\newblock \href {http://arxiv.org/abs/1406.2972} {\path{arXiv:1406.2972}},
  \href {https://doi.org/10.1103/PhysRevD.90.033007}
  {\path{doi:10.1103/PhysRevD.90.033007}}.

\bibitem{PhysRevD.72.085005}
A.~Di~Piazza, K.~Z. Hatsagortsyan, C.~H. Keitel,
  \href{https://link.aps.org/doi/10.1103/PhysRevD.72.085005}{Harmonic
  generation from laser-driven vacuum}, Phys. Rev. D 72 (2005) 085005.
\newblock \href {https://doi.org/10.1103/PhysRevD.72.085005}
  {\path{doi:10.1103/PhysRevD.72.085005}}.
\newline\urlprefix\url{https://link.aps.org/doi/10.1103/PhysRevD.72.085005}

\bibitem{Bohl:2016ddu}
P.~B\"ohl, B.~King, H.~Ruhl, {Vacuum high-harmonic generation and
  electromagnetic shock}, J. Plasma Phys. 82~(2) (2016) 655820202.
\newblock \href {https://doi.org/10.1017/S0022377816000210}
  {\path{doi:10.1017/S0022377816000210}}.

\bibitem{Sasorov:2021anc}
P.~V. Sasorov, F.~Pegoraro, T.~Z.~h. Esirkepov, S.~V. Bulanov, {Generation of
  high order harmonics in Heisenberg\textendash{}Euler electrodynamics}, New J.
  Phys. 23~(10) (2021) 105003.
\newblock \href {http://arxiv.org/abs/2106.03465} {\path{arXiv:2106.03465}},
  \href {https://doi.org/10.1088/1367-2630/ac28cb}
  {\path{doi:10.1088/1367-2630/ac28cb}}.

\bibitem{Bulanov:2019wfw}
S.~V. Bulanov, P.~V. Sasorov, F.~Pegoraro, H.~Kadlecov\'a, S.~S. Bulanov, T.~Z.
  Esirkepov, N.~N. Rosanov, G.~Korn, {Electromagnetic Solitons in Quantum
  Vacuum}, Phys. Rev. D 101~(1) (2020) 016016.
\newblock \href {http://arxiv.org/abs/1910.00455} {\path{arXiv:1910.00455}},
  \href {https://doi.org/10.1103/PhysRevD.101.016016}
  {\path{doi:10.1103/PhysRevD.101.016016}}.

\bibitem{Fillion-Gourdeau:2014uua}
F.~Fillion-Gourdeau, C.~Lefebvre, S.~MacLean, {Scheme for the detection of
  mixing processes in vacuum}, Phys. Rev. A 91~(3) (2015) 031801.
\newblock \href {http://arxiv.org/abs/1407.3014} {\path{arXiv:1407.3014}},
  \href {https://doi.org/10.1103/PhysRevA.91.031801}
  {\path{doi:10.1103/PhysRevA.91.031801}}.

\bibitem{Jeong:2020vpy}
T.~M. Jeong, S.~V. Bulanov, P.~V. Sasorov, G.~Korn, J.~K. Koga, S.~S. Bulanov,
  {Photon scattering by a 4\ensuremath{\pi} -spherically-focused ultrastrong
  electromagnetic wave}, Phys. Rev. A 102~(2) (2020) 023504.
\newblock \href {https://doi.org/10.1103/PhysRevA.102.023504}
  {\path{doi:10.1103/PhysRevA.102.023504}}.

\bibitem{Karbstein:2020gzg}
F.~Karbstein, E.~A. Mosman, {Enhancing quantum vacuum signatures with tailored
  laser beams}, Phys. Rev. D 101~(11) (2020) 113002.
\newblock \href {http://arxiv.org/abs/2004.04268} {\path{arXiv:2004.04268}},
  \href {https://doi.org/10.1103/PhysRevD.101.113002}
  {\path{doi:10.1103/PhysRevD.101.113002}}.

\bibitem{Doyle:2021mdt}
L.~Doyle, P.~Khademi, P.~Hilz, A.~S\"avert, G.~Sch\"afer, J.~Schreiber,
  M.~Zepf, {Experimental estimates of the photon background in a potential
  light-by-light scattering study} (10 2021).
\newblock \href {http://arxiv.org/abs/2110.03637} {\path{arXiv:2110.03637}}.

\bibitem{Moulin:1999hwj}
F.~Moulin, D.~Bernard, {Four-wave interaction in gas and vacuum. Definition of
  a third order nonlinear effective susceptibility in vacuum:
  $\chi_{vacuum^{(3)}}$}, Opt. Commun. 164 (1999) 137--144.
\newblock \href {http://arxiv.org/abs/physics/0203069}
  {\path{arXiv:physics/0203069}}, \href
  {https://doi.org/10.1016/S0030-4018(99)00169-8}
  {\path{doi:10.1016/S0030-4018(99)00169-8}}.

\bibitem{Varfolomeev:1966}
A.~A. Varfolomeev, Induced scattering of light by light, Sov. Phys. J. Exp.
  Theor. Phys. 23~(4) (1966) 681--688.

\bibitem{Lundstrom:2005za}
E.~Lundstrom, G.~Brodin, J.~Lundin, M.~Marklund, R.~Bingham, J.~Collier, J.~T.
  Mendonca, P.~Norreys, {Using high-power lasers for detection of elastic
  photon-photon scattering}, Phys. Rev. Lett. 96 (2006) 083602.
\newblock \href {http://arxiv.org/abs/hep-ph/0510076}
  {\path{arXiv:hep-ph/0510076}}, \href
  {https://doi.org/10.1103/PhysRevLett.96.083602}
  {\path{doi:10.1103/PhysRevLett.96.083602}}.

\bibitem{Lundin:2006wu}
J.~Lundin, M.~Marklund, E.~Lundstrom, G.~Brodin, J.~Collier, R.~Bingham, J.~T.
  Mendonca, P.~Norreys, {Detection of elastic photon-photon scattering through
  four-wave mixing using high power lasers}, Phys. Rev. A 74 (2006) 043821.
\newblock \href {http://arxiv.org/abs/hep-ph/0606136}
  {\path{arXiv:hep-ph/0606136}}, \href
  {https://doi.org/10.1103/PhysRevA.74.043821}
  {\path{doi:10.1103/PhysRevA.74.043821}}.

\bibitem{Gies:2017ezf}
H.~Gies, F.~Karbstein, C.~Kohlf\"urst, N.~Seegert, {Photon-photon scattering at
  the high-intensity frontier}, Phys. Rev. D 97~(7) (2018) 076002.
\newblock \href {http://arxiv.org/abs/1712.06450} {\path{arXiv:1712.06450}},
  \href {https://doi.org/10.1103/PhysRevD.97.076002}
  {\path{doi:10.1103/PhysRevD.97.076002}}.

\bibitem{Aboushelbaya:2019ncg}
R.~Aboushelbaya, et~al., {Orbital angular momentum coupling in elastic
  photon-photon scattering}, Phys. Rev. Lett. 123~(11) (2019) 113604.
\newblock \href {http://arxiv.org/abs/1902.05928} {\path{arXiv:1902.05928}},
  \href {https://doi.org/10.1103/PhysRevLett.123.113604}
  {\path{doi:10.1103/PhysRevLett.123.113604}}.

\bibitem{Mendonca:2017tdw}
J.~T. Mendon\c{c}a, {Emission of twisted photons from quantum vacuum}, EPL
  120~(6) (2017) 61001.
\newblock \href {https://doi.org/10.1209/0295-5075/120/61001}
  {\path{doi:10.1209/0295-5075/120/61001}}.

\bibitem{2016PhRvD..93l5032T}
D.~M. {Tennant}, {Four wave mixing as a probe of the vacuum}, Phys. Rev. D
  93~(12) (2016) 125032.
\newblock \href {https://doi.org/10.1103/PhysRevD.93.125032}
  {\path{doi:10.1103/PhysRevD.93.125032}}.

\bibitem{Klar:2020ych}
L.~Klar, {Detectable Optical Signatures of QED Vacuum Nonlinearities using
  High-Intensity Laser Fields}, Particles 3~(1) (2020) 223--233.
\newblock \href {http://arxiv.org/abs/2001.05731} {\path{arXiv:2001.05731}},
  \href {https://doi.org/10.3390/particles3010018}
  {\path{doi:10.3390/particles3010018}}.

\bibitem{Gies:2021ymf}
H.~Gies, F.~Karbstein, L.~Klar, {Quantum vacuum signatures in multicolor laser
  pulse collisions}, Phys. Rev. D 103~(7) (2021) 076009.
\newblock \href {http://arxiv.org/abs/2101.04461} {\path{arXiv:2101.04461}},
  \href {https://doi.org/10.1103/PhysRevD.103.076009}
  {\path{doi:10.1103/PhysRevD.103.076009}}.

\bibitem{Erber:1966vv}
T.~Erber, {High-energy electromagnetic conversion processes in intense magnetic
  fields}, Rev. Mod. Phys. 38 (1966) 626--659.
\newblock \href {https://doi.org/10.1103/RevModPhys.38.626}
  {\path{doi:10.1103/RevModPhys.38.626}}.

\bibitem{Dremin:2002eb}
I.~M. Dremin, {Cherenkov radiation and pair production by particles traversing
  laser beams}, JETP Lett. 76 (2002) 151--154.
\newblock \href {http://arxiv.org/abs/hep-ph/0202060}
  {\path{arXiv:hep-ph/0202060}}, \href {https://doi.org/10.1134/1.1514758}
  {\path{doi:10.1134/1.1514758}}.

\bibitem{Macleod:2018zcb}
A.~J. Macleod, A.~Noble, D.~A. Jaroszynski, {Cherenkov radiation from the
  quantum vacuum}, Phys. Rev. Lett. 122~(16) (2019) 161601.
\newblock \href {http://arxiv.org/abs/1810.05027} {\path{arXiv:1810.05027}},
  \href {https://doi.org/10.1103/PhysRevLett.122.161601}
  {\path{doi:10.1103/PhysRevLett.122.161601}}.

\bibitem{Jirka:2020vih}
M.~Jirka, P.~Sasorov, S.~S. Bulanov, G.~Korn, B.~Rus, S.~V. Bulanov, {Reaching
  high laser intensity by a radiating electron}, Phys. Rev. A 103~(5) (2021)
  053114.
\newblock \href {http://arxiv.org/abs/2010.11051} {\path{arXiv:2010.11051}},
  \href {https://doi.org/10.1103/PhysRevA.103.053114}
  {\path{doi:10.1103/PhysRevA.103.053114}}.

\bibitem{Lee:2020tay}
C.-Y. Lee, {Cherenkov radiation in a strong magnetic field}, Phys. Lett. B 810
  (2020) 135794.
\newblock \href {http://arxiv.org/abs/2005.11657} {\path{arXiv:2005.11657}},
  \href {https://doi.org/10.1016/j.physletb.2020.135794}
  {\path{doi:10.1016/j.physletb.2020.135794}}.

\bibitem{Bulanov:2019gdh}
S.~V. Bulanov, P.~V. Sasorov, S.~S. Bulanov, G.~Korn, {Synergic
  Cherenkov-Compton Radiation}, Phys. Rev. D 100~(1) (2019) 016012.
\newblock \href {http://arxiv.org/abs/1904.01012} {\path{arXiv:1904.01012}},
  \href {https://doi.org/10.1103/PhysRevD.100.016012}
  {\path{doi:10.1103/PhysRevD.100.016012}}.

\bibitem{Shore:2007um}
G.~M. Shore, {Superluminality and UV completion}, Nucl. Phys. B 778 (2007)
  219--258.
\newblock \href {http://arxiv.org/abs/hep-th/0701185}
  {\path{arXiv:hep-th/0701185}}, \href
  {https://doi.org/10.1016/j.nuclphysb.2007.03.034}
  {\path{doi:10.1016/j.nuclphysb.2007.03.034}}.

\bibitem{Shabad:2021ran}
A.~E. Shabad, {Comment on Cherenkov radiation in magnetized vacuum} (11 2021).
\newblock \href {http://arxiv.org/abs/2111.05119} {\path{arXiv:2111.05119}}.

\bibitem{Artemenko:2020uxk}
I.~I. Artemenko, E.~N. Nerush, I.~Y. Kostyukov, {Quasiclassical approach to
  synergic synchrotron\textendash{}Cherenkov radiation in polarized vacuum},
  New J. Phys. 22~(9) (2020) 093072.
\newblock \href {http://arxiv.org/abs/2004.12007} {\path{arXiv:2004.12007}},
  \href {https://doi.org/10.1088/1367-2630/abb388}
  {\path{doi:10.1088/1367-2630/abb388}}.

\bibitem{Bufalo:2021myy}
R.~Bufalo, T.~Cardoso~e Bufalo, {Vacuum radiation in $z=2$ Lifshitz QED}, Phys.
  Rev. D 103~(12) (2021) 125016.
\newblock \href {http://arxiv.org/abs/2105.13848} {\path{arXiv:2105.13848}},
  \href {https://doi.org/10.1103/PhysRevD.103.125016}
  {\path{doi:10.1103/PhysRevD.103.125016}}.

\bibitem{Meuren:2014kla}
S.~Meuren, K.~Z. Hatsagortsyan, C.~H. Keitel, A.~Di~Piazza, {High-Energy
  Recollision Processes of Laser-Generated Electron-Positron Pairs}, Phys. Rev.
  Lett. 114~(14) (2015) 143201.
\newblock \href {http://arxiv.org/abs/1407.0188} {\path{arXiv:1407.0188}},
  \href {https://doi.org/10.1103/PhysRevLett.114.143201}
  {\path{doi:10.1103/PhysRevLett.114.143201}}.

\bibitem{Medina:2015qzc}
L.~Medina, M.~C. Ogilvie, {Schwinger Pair Production at Finite Temperature},
  Phys. Rev. D 95~(5) (2017) 056006.
\newblock \href {http://arxiv.org/abs/1511.09459} {\path{arXiv:1511.09459}},
  \href {https://doi.org/10.1103/PhysRevD.95.056006}
  {\path{doi:10.1103/PhysRevD.95.056006}}.

\bibitem{Gould:2018ovk}
O.~Gould, A.~Rajantie, C.~Xie, {Worldline sphaleron for thermal Schwinger pair
  production}, Phys. Rev. D 98~(5) (2018) 056022.
\newblock \href {http://arxiv.org/abs/1806.02665} {\path{arXiv:1806.02665}},
  \href {https://doi.org/10.1103/PhysRevD.98.056022}
  {\path{doi:10.1103/PhysRevD.98.056022}}.

\bibitem{Gould:2018efv}
O.~Gould, S.~Mangles, A.~Rajantie, S.~Rose, C.~Xie, {Observing Thermal
  Schwinger Pair Production}, Phys. Rev. A 99~(5) (2019) 052120.
\newblock \href {http://arxiv.org/abs/1812.04089} {\path{arXiv:1812.04089}},
  \href {https://doi.org/10.1103/PhysRevA.99.052120}
  {\path{doi:10.1103/PhysRevA.99.052120}}.

\bibitem{Korwar:2018euc}
M.~Korwar, A.~M. Thalapillil, {Finite temperature Schwinger pair production in
  coexistent electric and magnetic fields}, Phys. Rev. D 98~(7) (2018) 076016.
\newblock \href {http://arxiv.org/abs/1808.01295} {\path{arXiv:1808.01295}},
  \href {https://doi.org/10.1103/PhysRevD.98.076016}
  {\path{doi:10.1103/PhysRevD.98.076016}}.

\bibitem{Wang:2019bjp}
Y.~L. Wang, H.~B. Sang, B.~S. Xie, {Schwinger pair production correction in
  thermal system}, Phys. Rev. D 100~(11) (2019) 116016.
\newblock \href {http://arxiv.org/abs/1909.13427} {\path{arXiv:1909.13427}},
  \href {https://doi.org/10.1103/PhysRevD.100.116016}
  {\path{doi:10.1103/PhysRevD.100.116016}}.

\bibitem{2008arXiv0801.4877E}
G.~A. {Edgar}, {Transseries for beginners}, arXiv e-prints (2008)
  arXiv:0801.4877\href {http://arxiv.org/abs/0801.4877}
  {\path{arXiv:0801.4877}}.

\bibitem{Marino:2012zq}
M.~Mari\~no, {Lectures on non-perturbative effects in large $N$ gauge theories,
  matrix models and strings}, Fortsch. Phys. 62 (2014) 455--540.
\newblock \href {http://arxiv.org/abs/1206.6272} {\path{arXiv:1206.6272}},
  \href {https://doi.org/10.1002/prop.201400005}
  {\path{doi:10.1002/prop.201400005}}.

\bibitem{Aniceto:2018bis}
I.~Aniceto, G.~Basar, R.~Schiappa, {A Primer on Resurgent Transseries and Their
  Asymptotics}, Phys. Rept. 809 (2019) 1--135.
\newblock \href {http://arxiv.org/abs/1802.10441} {\path{arXiv:1802.10441}},
  \href {https://doi.org/10.1016/j.physrep.2019.02.003}
  {\path{doi:10.1016/j.physrep.2019.02.003}}.

\bibitem{Schneider_2019}
C.~Schneider,
  \href{https://duepublico2.uni-due.de/receive/duepublico_mods_00070116}{Worldline
  instantons and the sauter-schwinger effect}, Ph.D. thesis (May 2019).
\newblock \href {https://doi.org/10.17185/duepublico/70116}
  {\path{doi:10.17185/duepublico/70116}}.
\newline\urlprefix\url{https://duepublico2.uni-due.de/receive/duepublico_mods_00070116}

\bibitem{Li:2019ves}
L.~Li, T.~Nakama, C.~M. Sou, Y.~Wang, S.~Zhou, {Gravitational Production of
  Superheavy Dark Matter and Associated Cosmological Signatures}, JHEP 07
  (2019) 067.
\newblock \href {http://arxiv.org/abs/1903.08842} {\path{arXiv:1903.08842}},
  \href {https://doi.org/10.1007/JHEP07(2019)067}
  {\path{doi:10.1007/JHEP07(2019)067}}.

\bibitem{Hashiba:2020rsi}
S.~Hashiba, Y.~Yamada, J.~Yokoyama, {Particle production induced by vacuum
  decay in real time dynamics}, Phys. Rev. D 103~(4) (2021) 045006.
\newblock \href {http://arxiv.org/abs/2006.10986} {\path{arXiv:2006.10986}},
  \href {https://doi.org/10.1103/PhysRevD.103.045006}
  {\path{doi:10.1103/PhysRevD.103.045006}}.

\bibitem{Taya:2020dco}
H.~Taya, T.~Fujimori, T.~Misumi, M.~Nitta, N.~Sakai, {Exact WKB analysis of the
  vacuum pair production by time-dependent electric fields}, JHEP 03 (2021)
  082.
\newblock \href {http://arxiv.org/abs/2010.16080} {\path{arXiv:2010.16080}},
  \href {https://doi.org/10.1007/JHEP03(2021)082}
  {\path{doi:10.1007/JHEP03(2021)082}}.

\bibitem{Enomoto:2020xlf}
S.~Enomoto, T.~Matsuda, {The exact WKB for cosmological particle production},
  JHEP 03 (2021) 090.
\newblock \href {http://arxiv.org/abs/2010.14835} {\path{arXiv:2010.14835}},
  \href {https://doi.org/10.1007/JHEP03(2021)090}
  {\path{doi:10.1007/JHEP03(2021)090}}.

\bibitem{Taya:2021dcz}
H.~Taya, M.~Hongo, T.~N. Ikeda, {Analytical WKB theory for high-harmonic
  generation and its application to massive Dirac electrons}, Phys. Rev. B
  104~(14) (2021) L140305.
\newblock \href {http://arxiv.org/abs/2105.12446} {\path{arXiv:2105.12446}},
  \href {https://doi.org/10.1103/PhysRevB.104.L140305}
  {\path{doi:10.1103/PhysRevB.104.L140305}}.

\bibitem{Enomoto:2021hfv}
S.~Enomoto, T.~Matsuda, {The Exact WKB and the Landau-Zener transition for
  asymmetry in cosmological particle production} (4 2021).
\newblock \href {http://arxiv.org/abs/2104.02312} {\path{arXiv:2104.02312}}.

\bibitem{Hashiba:2021npn}
S.~Hashiba, Y.~Yamada, {Stokes phenomenon and gravitational particle production
  \textemdash{} How to evaluate it in practice}, JCAP 05 (2021) 022.
\newblock \href {http://arxiv.org/abs/2101.07634} {\path{arXiv:2101.07634}},
  \href {https://doi.org/10.1088/1475-7516/2021/05/022}
  {\path{doi:10.1088/1475-7516/2021/05/022}}.

\bibitem{Sou:2021juh}
C.~M. Sou, X.~Tong, Y.~Wang, {Chemical-potential-assisted particle production
  in FRW spacetimes}, JHEP 06 (2021) 129.
\newblock \href {http://arxiv.org/abs/2104.08772} {\path{arXiv:2104.08772}},
  \href {https://doi.org/10.1007/JHEP06(2021)129}
  {\path{doi:10.1007/JHEP06(2021)129}}.

\bibitem{Brezin:1970xf}
E.~Brezin, C.~Itzykson, {Pair production in vacuum by an alternating field},
  Phys. Rev. D 2 (1970) 1191--1199.
\newblock \href {https://doi.org/10.1103/PhysRevD.2.1191}
  {\path{doi:10.1103/PhysRevD.2.1191}}.

\bibitem{Dumlu:2011rr}
C.~K. Dumlu, G.~V. Dunne, {Interference Effects in Schwinger Vacuum Pair
  Production for Time-Dependent Laser Pulses}, Phys. Rev. D 83 (2011) 065028.
\newblock \href {http://arxiv.org/abs/1102.2899} {\path{arXiv:1102.2899}},
  \href {https://doi.org/10.1103/PhysRevD.83.065028}
  {\path{doi:10.1103/PhysRevD.83.065028}}.

\bibitem{Strobel:2013vza}
E.~Strobel, S.-S. Xue, {Semiclassical pair production rate for time-dependent
  electrical fields with more than one component: WKB-approach and world-line
  instantons}, Nucl. Phys. B 886 (2014) 1153--1176.
\newblock \href {http://arxiv.org/abs/1312.3261} {\path{arXiv:1312.3261}},
  \href {https://doi.org/10.1016/j.nuclphysb.2014.07.017}
  {\path{doi:10.1016/j.nuclphysb.2014.07.017}}.

\bibitem{Fukushima:2019iiq}
K.~Fukushima, T.~Shimazaki, {Lefschetz-thimble inspired analysis of the
  Dykhne\textendash{}Davis\textendash{}Pechukas method and an application for
  the Schwinger Mechanism}, Annals Phys. 415 (2020) 168111.
\newblock \href {http://arxiv.org/abs/1907.12224} {\path{arXiv:1907.12224}},
  \href {https://doi.org/10.1016/j.aop.2020.168111}
  {\path{doi:10.1016/j.aop.2020.168111}}.

\bibitem{Popov:2005rp}
V.~S. Popov, {Imaginary-time method in quantum mechanics and field theory},
  Phys. Atom. Nucl. 68 (2005) 686--708.
\newblock \href {https://doi.org/10.1134/1.1903097}
  {\path{doi:10.1134/1.1903097}}.

\bibitem{Kim:2019yts}
S.~P. Kim, D.~N. Page, {Equivalence between the phase-integral and
  worldline-instanton methods} (4 2019).
\newblock \href {http://arxiv.org/abs/1904.09749} {\path{arXiv:1904.09749}}.

\bibitem{Ilderton:2015lsa}
A.~Ilderton, G.~Torgrimsson, J.~W\r{a}rdh, {Pair production from residues of
  complex worldline instantons}, Phys. Rev. D 92~(2) (2015) 025009.
\newblock \href {http://arxiv.org/abs/1503.08828} {\path{arXiv:1503.08828}},
  \href {https://doi.org/10.1103/PhysRevD.92.025009}
  {\path{doi:10.1103/PhysRevD.92.025009}}.

\bibitem{Kim:2000un}
S.~P. Kim, D.~N. Page, {Schwinger pair production via instantons in a strong
  electric field}, Phys. Rev. D 65 (2002) 105002.
\newblock \href {http://arxiv.org/abs/hep-th/0005078}
  {\path{arXiv:hep-th/0005078}}, \href
  {https://doi.org/10.1103/PhysRevD.65.105002}
  {\path{doi:10.1103/PhysRevD.65.105002}}.

\bibitem{Kim:2003qp}
S.~P. Kim, D.~N. Page, {Schwinger pair production in electric and magnetic
  fields}, Phys. Rev. D 73 (2006) 065020.
\newblock \href {http://arxiv.org/abs/hep-th/0301132}
  {\path{arXiv:hep-th/0301132}}, \href
  {https://doi.org/10.1103/PhysRevD.73.065020}
  {\path{doi:10.1103/PhysRevD.73.065020}}.

\bibitem{Kim:2007pm}
S.~P. Kim, D.~N. Page, {Improved Approximations for Fermion Pair Production in
  Inhomogeneous Electric Fields}, Phys. Rev. D 75 (2007) 045013.
\newblock \href {http://arxiv.org/abs/hep-th/0701047}
  {\path{arXiv:hep-th/0701047}}, \href
  {https://doi.org/10.1103/PhysRevD.75.045013}
  {\path{doi:10.1103/PhysRevD.75.045013}}.

\bibitem{Kim:2018dsp}
S.~P. Kim, {Schwinger Pair Production via Polons and the Origin of Stokes
  Phenomena}, New Phys. Sae Mulli 68~(11) (2018) 1225--1230.
\newblock \href {https://doi.org/10.3938/NPSM.68.1225}
  {\path{doi:10.3938/NPSM.68.1225}}.

\bibitem{Dingle1973}
R.~B. Dingle, {Asymptotic Expansions: Their Derivation and Interpretation},
  Academic Press, New York and London, 1973.

\bibitem{CNP}
B.~Candelpergher, J.~C. Nosmas, F.~Pham, {Approche de la r\'{e}surgence},
  Actualit\'es math\'{e}matiques (1993).

\bibitem{AIHPA_1983__39_3_211_0}
A.~Voros, \href{www.numdam.org/item/AIHPA_1983__39_3_211_0/}{{The return of the
  quartic oscillator. The complex WKB method}}, Annales de l'I.H.P. Physique
  th\'{e}orique 39~(3) (1983) 211--338.
\newline\urlprefix\url{www.numdam.org/item/AIHPA_1983__39_3_211_0/}

\bibitem{AIF_1993__43_1_163_0}
H.~Dillinger, E.~Delabaere, F.~Pham,
  \href{www.numdam.org/item/AIF_1993__43_1_163_0/}{{R\'{e}surgence de Voros et
  p\'eriodes des courbes hyperelliptiques}}, Annales de l'Institut Fourier
  43~(1) (1993) 163--199.
\newblock \href {https://doi.org/10.5802/aif.1326}
  {\path{doi:10.5802/aif.1326}}.
\newline\urlprefix\url{www.numdam.org/item/AIF_1993__43_1_163_0/}

\bibitem{doi:10.1063/1.532206}
E.~Delabaere, H.~Dillinger, F.~Pham,
  \href{https://doi.org/10.1063/1.532206}{Exact semiclassical expansions for
  one-dimensional quantum oscillators}, Journal of Mathematical Physics 38~(12)
  (1997) 6126--6184.
\newblock \href {https://doi.org/10.1063/1.532206}
  {\path{doi:10.1063/1.532206}}.
\newline\urlprefix\url{https://doi.org/10.1063/1.532206}

\bibitem{AIHPA_1999__71_1_1_0}
E.~Delabaere, F.~Pham,
  \href{www.numdam.org/item/AIHPA_1999__71_1_1_0/}{{Resurgent methods in
  semi-classical asymptotics}}, Annales de l'I.H.P. Physique th\'{e}orique
  71~(1) (1999) 1--94.
\newline\urlprefix\url{www.numdam.org/item/AIHPA_1999__71_1_1_0/}

\bibitem{AKT1}
T.~Aoki, T.~Kawai, Y.~Takei, {The Bender-Wu Analysis and the Voros Theory},
  ICM-90 Satellite Conference Proceedings (1991) 1--29\href
  {https://doi.org/10.1007/978-4-431-68170-0_1}
  {\path{doi:10.1007/978-4-431-68170-0_1}}.

\bibitem{AKT2}
T.~Aoki, T.~Kawai, Y.~Takei, {The Bender-Wu Analysis and the Voros Theory II},
  Advanced Studies in Pure Mathematics 54 (2009) 19--94.

\bibitem{Aoki:1993ra}
T.~Aoki, T.~Kawai, Y.~Takei, {Algebraic analysis of singular perturbations: On
  exact WKB analysis} (10 1993).

\bibitem{Takei2008}
Y.~Takei, Sato’s conjecture for the weber equation and transformation theory
  for schrödinger equations with a merging pair of turning points, RIMS
  Kôkyûroku Bessatsu (01 2008).

\bibitem{Dumlu:2011cc}
C.~K. Dumlu, G.~V. Dunne, {Complex Worldline Instantons and Quantum
  Interference in Vacuum Pair Production}, Phys. Rev. D 84 (2011) 125023.
\newblock \href {http://arxiv.org/abs/1110.1657} {\path{arXiv:1110.1657}},
  \href {https://doi.org/10.1103/PhysRevD.84.125023}
  {\path{doi:10.1103/PhysRevD.84.125023}}.

\bibitem{Ilderton:2015qda}
A.~Ilderton, G.~Torgrimsson, J.~W\r{a}rdh, {Nonperturbative pair production in
  interpolating fields}, Phys. Rev. D 92~(6) (2015) 065001.
\newblock \href {http://arxiv.org/abs/1506.09186} {\path{arXiv:1506.09186}},
  \href {https://doi.org/10.1103/PhysRevD.92.065001}
  {\path{doi:10.1103/PhysRevD.92.065001}}.

\bibitem{Dunne:2006ur}
G.~V. Dunne, Q.-h. Wang, {Multidimensional Worldline Instantons}, Phys. Rev. D
  74 (2006) 065015.
\newblock \href {http://arxiv.org/abs/hep-th/0608020}
  {\path{arXiv:hep-th/0608020}}, \href
  {https://doi.org/10.1103/PhysRevD.74.065015}
  {\path{doi:10.1103/PhysRevD.74.065015}}.

\bibitem{Dumlu:2015paa}
C.~K. Dumlu, {Multidimensional quantum tunneling in the Schwinger effect},
  Phys. Rev. D 93~(6) (2016) 065045.
\newblock \href {http://arxiv.org/abs/1507.07005} {\path{arXiv:1507.07005}},
  \href {https://doi.org/10.1103/PhysRevD.93.065045}
  {\path{doi:10.1103/PhysRevD.93.065045}}.

\bibitem{Akal:2017sbs}
I.~Akal, G.~Moortgat-Pick, {Quantum tunnelling from vacuum in multidimensions},
  Phys. Rev. D 96~(9) (2017) 096027.
\newblock \href {http://arxiv.org/abs/1710.04646} {\path{arXiv:1710.04646}},
  \href {https://doi.org/10.1103/PhysRevD.96.096027}
  {\path{doi:10.1103/PhysRevD.96.096027}}.

\bibitem{Schneider:2018huk}
C.~Schneider, G.~Torgrimsson, R.~Sch\"utzhold, {Discrete worldline instantons},
  Phys. Rev. D 98~(8) (2018) 085009.
\newblock \href {http://arxiv.org/abs/1806.00943} {\path{arXiv:1806.00943}},
  \href {https://doi.org/10.1103/PhysRevD.98.085009}
  {\path{doi:10.1103/PhysRevD.98.085009}}.

\bibitem{Torgrimsson:2017cyb}
G.~Torgrimsson, C.~Schneider, R.~Sch\"utzhold, {Sauter-Schwinger pair creation
  dynamically assisted by a plane wave}, Phys. Rev. D 97~(9) (2018) 096004.
\newblock \href {http://arxiv.org/abs/1712.08613} {\path{arXiv:1712.08613}},
  \href {https://doi.org/10.1103/PhysRevD.97.096004}
  {\path{doi:10.1103/PhysRevD.97.096004}}.

\bibitem{Dunne:2006st}
G.~V. Dunne, Q.-h. Wang, H.~Gies, C.~Schubert, {Worldline instantons. II. The
  Fluctuation prefactor}, Phys. Rev. D 73 (2006) 065028.
\newblock \href {http://arxiv.org/abs/hep-th/0602176}
  {\path{arXiv:hep-th/0602176}}, \href
  {https://doi.org/10.1103/PhysRevD.73.065028}
  {\path{doi:10.1103/PhysRevD.73.065028}}.

\bibitem{Oertel:2016vsg}
J.~Oertel, R.~Sch\"utzhold, {WKB approach to pair creation in
  spacetime-dependent fields: The case of a spacetime-dependent mass}, Phys.
  Rev. D 99~(12) (2019) 125014.
\newblock \href {http://arxiv.org/abs/1603.00274} {\path{arXiv:1603.00274}},
  \href {https://doi.org/10.1103/PhysRevD.99.125014}
  {\path{doi:10.1103/PhysRevD.99.125014}}.

\bibitem{Dunne:1999uy}
G.~V. Dunne, T.~M. Hall, {Borel summation of the derivative expansion and
  effective actions}, Phys. Rev. D 60 (1999) 065002.
\newblock \href {http://arxiv.org/abs/hep-th/9902064}
  {\path{arXiv:hep-th/9902064}}, \href
  {https://doi.org/10.1103/PhysRevD.60.065002}
  {\path{doi:10.1103/PhysRevD.60.065002}}.

\bibitem{Dunne:1999vd}
G.~V. Dunne, C.~Schubert, {Two loop Euler-Heisenberg QED pair production rate},
  Nucl. Phys. B 564 (2000) 591--604.
\newblock \href {http://arxiv.org/abs/hep-th/9907190}
  {\path{arXiv:hep-th/9907190}}, \href
  {https://doi.org/10.1016/S0550-3213(99)00641-0}
  {\path{doi:10.1016/S0550-3213(99)00641-0}}.

\bibitem{Dunne:1998ni}
G.~V. Dunne, T.~Hall, {On the QED effective action in time dependent electric
  backgrounds}, Phys. Rev. D 58 (1998) 105022.
\newblock \href {http://arxiv.org/abs/hep-th/9807031}
  {\path{arXiv:hep-th/9807031}}, \href
  {https://doi.org/10.1103/PhysRevD.58.105022}
  {\path{doi:10.1103/PhysRevD.58.105022}}.

\bibitem{2007PhR...446....1C}
E.~{Caliceti}, M.~{Meyer-Hermann}, P.~{Ribeca}, A.~{Surzhykov}, U.~D.
  {Jentschura}, {From useful algorithms for slowly convergent series to
  physical predictions based on divergent perturbative expansions}, Phys. Rept.
  446~(1-3) (2007) 1--96.
\newblock \href {http://arxiv.org/abs/0707.1596} {\path{arXiv:0707.1596}},
  \href {https://doi.org/10.1016/j.physrep.2007.03.003}
  {\path{doi:10.1016/j.physrep.2007.03.003}}.

\bibitem{Torgrimsson:2017pzs}
G.~Torgrimsson, C.~Schneider, J.~Oertel, R.~Sch\"utzhold, {Dynamically assisted
  Sauter-Schwinger effect \textemdash{} non-perturbative versus perturbative
  aspects}, JHEP 06 (2017) 043.
\newblock \href {http://arxiv.org/abs/1703.09203} {\path{arXiv:1703.09203}},
  \href {https://doi.org/10.1007/JHEP06(2017)043}
  {\path{doi:10.1007/JHEP06(2017)043}}.

\bibitem{Torgrimsson:2018xdf}
G.~Torgrimsson, {Perturbative methods for assisted nonperturbative pair
  production}, Phys. Rev. D 99~(9) (2019) 096002.
\newblock \href {http://arxiv.org/abs/1812.04607} {\path{arXiv:1812.04607}},
  \href {https://doi.org/10.1103/PhysRevD.99.096002}
  {\path{doi:10.1103/PhysRevD.99.096002}}.

\bibitem{Taya:2018eng}
H.~Taya, {Franz-Keldysh effect in strong-field QED}, Phys. Rev. D 99~(5) (2019)
  056006.
\newblock \href {http://arxiv.org/abs/1812.03630} {\path{arXiv:1812.03630}},
  \href {https://doi.org/10.1103/PhysRevD.99.056006}
  {\path{doi:10.1103/PhysRevD.99.056006}}.

\bibitem{Huang:2019uhf}
X.-G. Huang, H.~Taya, {Spin-dependent dynamically assisted Schwinger
  mechanism}, Phys. Rev. D 100~(1) (2019) 016013.
\newblock \href {http://arxiv.org/abs/1904.08200} {\path{arXiv:1904.08200}},
  \href {https://doi.org/10.1103/PhysRevD.100.016013}
  {\path{doi:10.1103/PhysRevD.100.016013}}.

\bibitem{Taya:2020bcd}
H.~Taya, {Dynamically assisted Schwinger mechanism and chirality production in
  parallel electromagnetic field}, Phys. Rev. Res. 2~(2) (2020) 023257.
\newblock \href {http://arxiv.org/abs/2003.08948} {\path{arXiv:2003.08948}},
  \href {https://doi.org/10.1103/PhysRevResearch.2.023257}
  {\path{doi:10.1103/PhysRevResearch.2.023257}}.

\bibitem{Taya:2020pkm}
H.~Taya, {Mutual assistance between the Schwinger mechanism and the dynamical
  Casimir effect}, Phys. Rev. Res. 2~(2) (2020) 023346.
\newblock \href {http://arxiv.org/abs/2003.12061} {\path{arXiv:2003.12061}},
  \href {https://doi.org/10.1103/PhysRevResearch.2.023346}
  {\path{doi:10.1103/PhysRevResearch.2.023346}}.

\bibitem{Otto:2016xpn}
A.~Otto, B.~Kämpfer, {Afterglow of the dynamical Schwinger process: soft
  photons amass}, Phys. Rev. D 95~(12) (2017) 125007.
\newblock \href {http://arxiv.org/abs/1611.04438} {\path{arXiv:1611.04438}},
  \href {https://doi.org/10.1103/PhysRevD.95.125007}
  {\path{doi:10.1103/PhysRevD.95.125007}}.

\bibitem{Nikishov:1971}
A.~I. Nikishov, {Quantum Processes in a Constant Electric Field}, Soviet
  Journal of Experimental and Theoretical Physics 32 (1971) 690.

\bibitem{Keldysh:1965ojf}
L.~V. Keldysh, {Ionization in the Field of a Strong Electromagnetic Wave}, J.
  Exp. Theor. Phys. 20~(5) (1965) 1307--1314.

\bibitem{Levai:2009mn}
P.~Levai, V.~Skokov, {Nonperturbative enhancement of heavy quark-pair
  production in a strong SU(2) color field}, Phys. Rev. D 82 (2010) 074014.
\newblock \href {http://arxiv.org/abs/0909.2323} {\path{arXiv:0909.2323}},
  \href {https://doi.org/10.1103/PhysRevD.82.074014}
  {\path{doi:10.1103/PhysRevD.82.074014}}.

\bibitem{Taya:2014taa}
H.~Taya, H.~Fujii, K.~Itakura, {Finite pulse effects on $e^{+}e^{-}$ pair
  creation from strong electric fields}, Phys. Rev. D 90~(1) (2014) 014039.
\newblock \href {http://arxiv.org/abs/1405.6182} {\path{arXiv:1405.6182}},
  \href {https://doi.org/10.1103/PhysRevD.90.014039}
  {\path{doi:10.1103/PhysRevD.90.014039}}.

\bibitem{Adorno:2014bsa}
T.~C. Adorno, S.~P. Gavrilov, D.~M. Gitman, {Particle creation from the vacuum
  by an exponentially decreasing electric field}, Phys. Scripta 90 (2015)
  074005.
\newblock \href {http://arxiv.org/abs/1409.7742} {\path{arXiv:1409.7742}},
  \href {https://doi.org/10.1088/0031-8949/90/7/074005}
  {\path{doi:10.1088/0031-8949/90/7/074005}}.

\bibitem{Aleksandrov:2016lxd}
I.~A. Aleksandrov, G.~Plunien, V.~M. Shabaev, {Electron-positron pair
  production in external electric fields varying both in space and time}, Phys.
  Rev. D 94~(6) (2016) 065024.
\newblock \href {http://arxiv.org/abs/1606.06313} {\path{arXiv:1606.06313}},
  \href {https://doi.org/10.1103/PhysRevD.94.065024}
  {\path{doi:10.1103/PhysRevD.94.065024}}.

\bibitem{Taya:2016ovo}
H.~Taya, {Quark and Gluon Production from a Boost-invariantly Expanding Color
  Electric Field}, Phys. Rev. D 96~(1) (2017) 014033.
\newblock \href {http://arxiv.org/abs/1609.06189} {\path{arXiv:1609.06189}},
  \href {https://doi.org/10.1103/PhysRevD.96.014033}
  {\path{doi:10.1103/PhysRevD.96.014033}}.

\bibitem{Kohlfurst:2013ura}
C.~Kohlf\"urst, H.~Gies, R.~Alkofer, {Effective mass signatures in multiphoton
  pair production}, Phys. Rev. Lett. 112 (2014) 050402.
\newblock \href {http://arxiv.org/abs/1310.7836} {\path{arXiv:1310.7836}},
  \href {https://doi.org/10.1103/PhysRevLett.112.050402}
  {\path{doi:10.1103/PhysRevLett.112.050402}}.

\bibitem{Abdukerim:2013vsa}
N.~Abdukerim, Z.-L. Li, B.-S. Xie, {Effects of laser pulse shape and carrier
  envelope phase on pair production}, Phys. Lett. B 726 (2013) 820--826.
\newblock \href {https://doi.org/10.1016/j.physletb.2013.09.014}
  {\path{doi:10.1016/j.physletb.2013.09.014}}.

\bibitem{Sitiwaldi:2017mfh}
I.~Sitiwaldi, B.-S. Xie, {Modulation effect in multiphoton pair production},
  Phys. Lett. B 768 (2017) 174--179.
\newblock \href {http://arxiv.org/abs/1701.03586} {\path{arXiv:1701.03586}},
  \href {https://doi.org/10.1016/j.physletb.2017.02.050}
  {\path{doi:10.1016/j.physletb.2017.02.050}}.

\bibitem{Krajewska:2018lwe}
K.~Krajewska, J.~Z. Kami\'nski, {Threshold effects in electron-positron pair
  creation from the vacuum: Stabilization and longitudinal versus transverse
  momentum sharing}, Phys. Rev. A 100~(1) (2019) 012104.
\newblock \href {http://arxiv.org/abs/1811.07528} {\path{arXiv:1811.07528}},
  \href {https://doi.org/10.1103/PhysRevA.100.012104}
  {\path{doi:10.1103/PhysRevA.100.012104}}.

\bibitem{Orthaber:2011cm}
M.~Orthaber, F.~Hebenstreit, R.~Alkofer, {Momentum Spectra for Dynamically
  Assisted Schwinger Pair Production}, Phys. Lett. B 698 (2011) 80--85.
\newblock \href {http://arxiv.org/abs/1102.2182} {\path{arXiv:1102.2182}},
  \href {https://doi.org/10.1016/j.physletb.2011.02.053}
  {\path{doi:10.1016/j.physletb.2011.02.053}}.

\bibitem{Nuriman:2012hn}
A.~Nuriman, B.-S. Xie, Z.-L. Li, D.~Sayipjamal, {Enhanced electron-positron
  pair creation by dynamically assisted combinational fields}, Phys. Lett. B
  717 (2012) 465--469.
\newblock \href {https://doi.org/10.1016/j.physletb.2012.09.060}
  {\path{doi:10.1016/j.physletb.2012.09.060}}.

\bibitem{Nuriman:2013vba}
A.~Nuriman, B.-S. Xie, Z.-L. Li, D.~Sayipjamal, {Electron-positron pair
  production in a strong laser field enhanced by an assisted high frequency
  weak field}, Commun. Theor. Phys. 59 (2013) 331--334.
\newblock \href {https://doi.org/10.1088/0253-6102/59/3/15}
  {\path{doi:10.1088/0253-6102/59/3/15}}.

\bibitem{Hebenstreit:2014lra}
F.~Hebenstreit, F.~Fillion-Gourdeau, {Optimization of Schwinger pair production
  in colliding laser pulses}, Phys. Lett. B 739 (2014) 189--195.
\newblock \href {http://arxiv.org/abs/1409.7943} {\path{arXiv:1409.7943}},
  \href {https://doi.org/10.1016/j.physletb.2014.10.056}
  {\path{doi:10.1016/j.physletb.2014.10.056}}.

\bibitem{Hebenstreit:2015jaa}
F.~Hebenstreit, {The inverse problem for Schwinger pair production}, Phys.
  Lett. B 753 (2016) 336--340.
\newblock \href {http://arxiv.org/abs/1509.08693} {\path{arXiv:1509.08693}},
  \href {https://doi.org/10.1016/j.physletb.2015.12.049}
  {\path{doi:10.1016/j.physletb.2015.12.049}}.

\bibitem{Otto:2015gla}
A.~Otto, D.~Seipt, D.~Blaschke, S.~A. Smolyansky, B.~Kämpfer, {Dynamical
  Schwinger process in a bifrequent electric field of finite duration: survey
  on amplification}, Phys. Rev. D 91~(10) (2015) 105018.
\newblock \href {http://arxiv.org/abs/1503.08675} {\path{arXiv:1503.08675}},
  \href {https://doi.org/10.1103/PhysRevD.91.105018}
  {\path{doi:10.1103/PhysRevD.91.105018}}.

\bibitem{Panferov:2015yda}
A.~D. Panferov, S.~A. Smolyansky, A.~Otto, B.~Kämpfer, D.~B. Blaschke,
  L.~Juchnowski, {Assisted dynamical Schwinger effect: pair production in a
  pulsed bifrequent field}, Eur. Phys. J. D 70~(3) (2016) 56.
\newblock \href {http://arxiv.org/abs/1509.02901} {\path{arXiv:1509.02901}},
  \href {https://doi.org/10.1140/epjd/e2016-60517-y}
  {\path{doi:10.1140/epjd/e2016-60517-y}}.

\bibitem{Sitiwaldi:2018wad}
I.~Sitiwaldi, B.-S. Xie, {Pair production by three fields dynamically assisted
  Schwinger process}, Phys. Lett. B 777 (2018) 406--411.
\newblock \href {https://doi.org/10.1016/j.physletb.2017.12.060}
  {\path{doi:10.1016/j.physletb.2017.12.060}}.

\bibitem{Banerjee:2018gyt}
C.~Banerjee, M.~P. Singh, A.~M. Fedotov, {Phase control of Schwinger pair
  production by colliding laser pulses}, Phys. Rev. A 98~(3) (2018) 032121.
\newblock \href {http://arxiv.org/abs/1806.07088} {\path{arXiv:1806.07088}},
  \href {https://doi.org/10.1103/PhysRevA.98.032121}
  {\path{doi:10.1103/PhysRevA.98.032121}}.

\bibitem{Wang:2021tmo}
L.~Wang, L.-J. Li, M.~Mohamedsedik, R.~An, J.-J. Li, B.-S. Xie, F.-S. Zhang,
  {Enhancement of electron-positron pairs in combined potential wells with
  linear chirp frequency} (9 2021).
\newblock \href {http://arxiv.org/abs/2109.05399} {\path{arXiv:2109.05399}}.

\bibitem{Popov:1971iga}
V.~S. Popov, {Pair production in a variable external field (quasiclassical
  approximation)}, Zh. Eksp. Teor. Fiz. 61 (1971) 1334--1351.

\bibitem{Hebenstreit:2008ae}
F.~Hebenstreit, R.~Alkofer, H.~Gies, {Pair Production Beyond the Schwinger
  Formula in Time-Dependent Electric Fields}, Phys. Rev. D 78 (2008) 061701.
\newblock \href {http://arxiv.org/abs/0807.2785} {\path{arXiv:0807.2785}},
  \href {https://doi.org/10.1103/PhysRevD.78.061701}
  {\path{doi:10.1103/PhysRevD.78.061701}}.

\bibitem{Blaschke:2013ip}
D.~B. Blaschke, B.~K\"ampfer, S.~M. Schmidt, A.~D. Panferov, A.~V.
  Prozorkevich, S.~A. Smolyansky, {Properties of the electron-positron plasma
  created from a vacuum in a strong laser field: Quasiparticle excitations},
  Phys. Rev. D 88~(4) (2013) 045017.
\newblock \href {http://arxiv.org/abs/1301.1640} {\path{arXiv:1301.1640}},
  \href {https://doi.org/10.1103/PhysRevD.88.045017}
  {\path{doi:10.1103/PhysRevD.88.045017}}.

\bibitem{Sauter:1932gsa}
F.~Sauter, {Zum ''Kleinschen Paradoxon''}, Z. Phys. 73 (1932) 547--552.
\newblock \href {https://doi.org/10.1007/BF01349862}
  {\path{doi:10.1007/BF01349862}}.

\bibitem{Narozhnyi:1970uv}
N.~B. Narozhnyi, A.~I. Nikishov, {The Simplist processes in the pair creating
  electric field}, Sov. J. Nucl. 11 (1970) 596.

\bibitem{Nikishov:1970br}
A.~I. Nikishov, {Barrier scattering in field theory removal of klein paradox},
  Nucl. Phys. B 21 (1970) 346--358.
\newblock \href {https://doi.org/10.1016/0550-3213(70)90527-4}
  {\path{doi:10.1016/0550-3213(70)90527-4}}.

\bibitem{Karbstein:2017pbf}
F.~Karbstein, {Heisenberg-Euler effective action in slowly varying electric
  field inhomogeneities of Lorentzian shape}, Phys. Rev. D 95~(7) (2017)
  076015.
\newblock \href {http://arxiv.org/abs/1703.08017} {\path{arXiv:1703.08017}},
  \href {https://doi.org/10.1103/PhysRevD.95.076015}
  {\path{doi:10.1103/PhysRevD.95.076015}}.

\bibitem{DiPiazza:2009py}
A.~Di~Piazza, E.~Lotstedt, A.~I. Milstein, C.~H. Keitel, {Barrier control in
  tunneling e+ - e- photoproduction}, Phys. Rev. Lett. 103 (2009) 170403.
\newblock \href {http://arxiv.org/abs/0906.0726} {\path{arXiv:0906.0726}},
  \href {https://doi.org/10.1103/PhysRevLett.103.170403}
  {\path{doi:10.1103/PhysRevLett.103.170403}}.

\bibitem{Linder:2015vta}
M.~F. Linder, C.~Schneider, J.~Sicking, N.~Szpak, R.~Sch\"utzhold, {Pulse shape
  dependence in the dynamically assisted Sauter-Schwinger effect}, Phys. Rev. D
  92~(8) (2015) 085009.
\newblock \href {http://arxiv.org/abs/1505.05685} {\path{arXiv:1505.05685}},
  \href {https://doi.org/10.1103/PhysRevD.92.085009}
  {\path{doi:10.1103/PhysRevD.92.085009}}.

\bibitem{Schneider:2014mla}
C.~Schneider, R.~Sch\"utzhold, {Dynamically assisted Sauter-Schwinger effect in
  inhomogeneous electric fields}, JHEP 02 (2016) 164.
\newblock \href {http://arxiv.org/abs/1407.3584} {\path{arXiv:1407.3584}},
  \href {https://doi.org/10.1007/JHEP02(2016)164}
  {\path{doi:10.1007/JHEP02(2016)164}}.

\bibitem{Aleksandrov:2018uqb}
I.~A. Aleksandrov, G.~Plunien, V.~M. Shabaev, {Dynamically assisted Schwinger
  effect beyond the spatially-uniform-field approximation}, Phys. Rev. D
  97~(11) (2018) 116001.
\newblock \href {http://arxiv.org/abs/1805.07579} {\path{arXiv:1805.07579}},
  \href {https://doi.org/10.1103/PhysRevD.97.116001}
  {\path{doi:10.1103/PhysRevD.97.116001}}.

\bibitem{Copinger:2016llk}
P.~Copinger, K.~Fukushima, {Spatially Assisted Schwinger Mechanism and Magnetic
  Catalysis}, Phys. Rev. Lett. 117~(8) (2016) 081603, [Erratum: Phys.Rev.Lett.
  118, 099903 (2017)].
\newblock \href {http://arxiv.org/abs/1605.05957} {\path{arXiv:1605.05957}},
  \href {https://doi.org/10.1103/PhysRevLett.117.081603}
  {\path{doi:10.1103/PhysRevLett.117.081603}}.

\bibitem{Schneider:2016vrl}
C.~Schneider, R.~Sch\"utzhold, {Prefactor in the dynamically assisted
  Sauter-Schwinger effect}, Phys. Rev. D 94~(8) (2016) 085015.
\newblock \href {http://arxiv.org/abs/1603.00864} {\path{arXiv:1603.00864}},
  \href {https://doi.org/10.1103/PhysRevD.94.085015}
  {\path{doi:10.1103/PhysRevD.94.085015}}.

\bibitem{Fey:2011if}
C.~Fey, R.~Schutzhold, {Momentum dependence in the dynamically assisted
  Sauter-Schwinger effect}, Phys. Rev. D 85 (2012) 025004.
\newblock \href {http://arxiv.org/abs/1110.5499} {\path{arXiv:1110.5499}},
  \href {https://doi.org/10.1103/PhysRevD.85.025004}
  {\path{doi:10.1103/PhysRevD.85.025004}}.

\bibitem{Kim:2021dcw}
C.~M. Kim, A.~Fedotov, S.~P. Kim, {Phase-Integral Formulation of Dynamically
  Assisted Schwinger Pair Production} (9 2021).
\newblock \href {http://arxiv.org/abs/2109.10268} {\path{arXiv:2109.10268}}.

\bibitem{stuckelberg1933theorie}
E.~K.~G. St{\"u}ckelberg, Theorie der unelastischen St{\"o}sse zwischen Atomen,
  Birkh{\"a}user, 1933.

\bibitem{Shevchenko_2010}
S.~Shevchenko, S.~Ashhab, F.~Nori,
  \href{http://dx.doi.org/10.1016/j.physrep.2010.03.002}{Landau–zener–stückelberg
  interferometry}, Physics Reports 492~(1) (2010) 1–30.
\newblock \href {https://doi.org/10.1016/j.physrep.2010.03.002}
  {\path{doi:10.1016/j.physrep.2010.03.002}}.
\newline\urlprefix\url{http://dx.doi.org/10.1016/j.physrep.2010.03.002}

\bibitem{Hebenstreit:2009km}
F.~Hebenstreit, R.~Alkofer, G.~V. Dunne, H.~Gies, {Momentum signatures for
  Schwinger pair production in short laser pulses with sub-cycle structure},
  Phys. Rev. Lett. 102 (2009) 150404.
\newblock \href {http://arxiv.org/abs/0901.2631} {\path{arXiv:0901.2631}},
  \href {https://doi.org/10.1103/PhysRevLett.102.150404}
  {\path{doi:10.1103/PhysRevLett.102.150404}}.

\bibitem{Li:2014xga}
Z.-L. Li, D.~Lu, B.-S. Xie, {Multiple-slit interference effect in the time
  domain for boson pair production}, Phys. Rev. D 89~(6) (2014) 067701.
\newblock \href {https://doi.org/10.1103/PhysRevD.89.067701}
  {\path{doi:10.1103/PhysRevD.89.067701}}.

\bibitem{Li:2014psw}
Z.~L. Li, D.~Lu, B.~S. Xie, L.~B. Fu, J.~Liu, B.~F. Shen, {Enhanced pair
  production in strong fields by multiple-slit interference effect with
  dynamically assisted Schwinger mechanism}, Phys. Rev. D 89~(9) (2014) 093011.
\newblock \href {https://doi.org/10.1103/PhysRevD.89.093011}
  {\path{doi:10.1103/PhysRevD.89.093011}}.

\bibitem{Kaminski:2018ywj}
J.~Z. Kami\'nski, M.~Twardy, K.~Krajewska, {Diffraction at a time grating in
  electron-positron pair creation from the vacuum}, Phys. Rev. D 98~(5) (2018)
  056009.
\newblock \href {http://arxiv.org/abs/1807.05386} {\path{arXiv:1807.05386}},
  \href {https://doi.org/10.1103/PhysRevD.98.056009}
  {\path{doi:10.1103/PhysRevD.98.056009}}.

\bibitem{Krajewska:2019vqd}
K.~Krajewska, J.~Z. Kami\'nski, {Unitary versus pseudounitary time evolution
  and statistical effects in the dynamical Sauter-Schwinger process}, Phys.
  Rev. A 100~(6) (2019) 062116.
\newblock \href {http://arxiv.org/abs/1909.12284} {\path{arXiv:1909.12284}},
  \href {https://doi.org/10.1103/PhysRevA.100.062116}
  {\path{doi:10.1103/PhysRevA.100.062116}}.

\bibitem{Abdukerim:2015dsa}
N.~Abdukerim, Z.-L. Li, B.-S. Xie, {Electron-positron pair production in the
  low-density approximation}, Front. Phys. (Beijing) 10~(4) (2015) 101202.
\newblock \href {https://doi.org/10.1007/s11467-015-0471-3}
  {\path{doi:10.1007/s11467-015-0471-3}}.

\bibitem{Abdukerim:2017hkh}
N.~Abdukerim, Z.-L. Li, B.-S. Xie, {Enhanced electron\textendash{}positron pair
  production by frequency chirping in one- and two-color laser pulse fields},
  Chin. Phys. B 26~(2) (2017) 020301.
\newblock \href {https://doi.org/10.1088/1674-1056/26/2/020301}
  {\path{doi:10.1088/1674-1056/26/2/020301}}.

\bibitem{Olugh:2018seh}
O.~Olugh, Z.-L. Li, B.-S. Xie, R.~Alkofer, {Pair production in differently
  polarized electric fields with frequency chirps}, Phys. Rev. D 99~(3) (2019)
  036003.
\newblock \href {http://arxiv.org/abs/1811.12125} {\path{arXiv:1811.12125}},
  \href {https://doi.org/10.1103/PhysRevD.99.036003}
  {\path{doi:10.1103/PhysRevD.99.036003}}.

\bibitem{Gong:2019sbw}
C.~Gong, Z.~L. Li, B.~S. Xie, Y.~J. Li, {Electron-positron pair production in
  frequency modulated laser fields}, Phys. Rev. D 101~(1) (2020) 016008.
\newblock \href {http://arxiv.org/abs/1908.08189} {\path{arXiv:1908.08189}},
  \href {https://doi.org/10.1103/PhysRevD.101.016008}
  {\path{doi:10.1103/PhysRevD.101.016008}}.

\bibitem{Ababekri:2019qiw}
M.~Ababekri, S.~Dulat, B.~S. Xie, J.~Zhang, {Chirp effects on pair production
  in oscillating electric fields with spatial inhomogeneity}, Phys. Lett. B 810
  (2020) 135815.
\newblock \href {http://arxiv.org/abs/1912.05302} {\path{arXiv:1912.05302}},
  \href {https://doi.org/10.1016/j.physletb.2020.135815}
  {\path{doi:10.1016/j.physletb.2020.135815}}.

\bibitem{Li:2021vjf}
L.-J. Li, M.~Mohamedsedik, B.-S. Xie, {Enhanced dynamically assisted pair
  production in spatial inhomogeneous electric fields with the frequency
  chirping}, Phys. Rev. D 104~(3) (2021) 036015.
\newblock \href {http://arxiv.org/abs/2104.08828} {\path{arXiv:2104.08828}},
  \href {https://doi.org/10.1103/PhysRevD.104.036015}
  {\path{doi:10.1103/PhysRevD.104.036015}}.

\bibitem{Mohamedsedik:2021pzb}
M.~Mohamedsedik, L.-J. Li, B.~S. Xie, {Schwinger pair production in
  inhomogeneous electric fields with symmetrical frequency chirp}, Phys. Rev. D
  104~(1) (2021) 016009.
\newblock \href {http://arxiv.org/abs/2105.03018} {\path{arXiv:2105.03018}},
  \href {https://doi.org/10.1103/PhysRevD.104.016009}
  {\path{doi:10.1103/PhysRevD.104.016009}}.

\bibitem{2005PhRvL..94j0602O}
T.~{Oka}, N.~{Konno}, R.~{Arita}, H.~{Aoki}, {Breakdown of an Electric-Field
  Driven System: A Mapping to a Quantum Walk}, Phys. Rev. Lett. 94~(10) (2005)
  100602.
\newblock \href {http://arxiv.org/abs/quant-ph/0407013}
  {\path{arXiv:quant-ph/0407013}}, \href
  {https://doi.org/10.1103/PhysRevLett.94.100602}
  {\path{doi:10.1103/PhysRevLett.94.100602}}.

\bibitem{Dumlu:2010ua}
C.~K. Dumlu, G.~V. Dunne, {The Stokes Phenomenon and Schwinger Vacuum Pair
  Production in Time-Dependent Laser Pulses}, Phys. Rev. Lett. 104 (2010)
  250402.
\newblock \href {http://arxiv.org/abs/1004.2509} {\path{arXiv:1004.2509}},
  \href {https://doi.org/10.1103/PhysRevLett.104.250402}
  {\path{doi:10.1103/PhysRevLett.104.250402}}.

\bibitem{Dumlu:2010vv}
C.~K. Dumlu, {Schwinger Vacuum Pair Production in Chirped Laser Pulses}, Phys.
  Rev. D 82 (2010) 045007.
\newblock \href {http://arxiv.org/abs/1006.3882} {\path{arXiv:1006.3882}},
  \href {https://doi.org/10.1103/PhysRevD.82.045007}
  {\path{doi:10.1103/PhysRevD.82.045007}}.

\bibitem{Tanji:2010eu}
N.~Tanji, {Quark pair creation in color electric fields and effects of magnetic
  fields}, Annals Phys. 325 (2010) 2018--2040.
\newblock \href {http://arxiv.org/abs/1002.3143} {\path{arXiv:1002.3143}},
  \href {https://doi.org/10.1016/j.aop.2010.03.012}
  {\path{doi:10.1016/j.aop.2010.03.012}}.

\bibitem{Ruf:2008ahs}
M.~Ruf, G.~R. Mocken, C.~Muller, K.~Z. Hatsagortsyan, C.~H. Keitel, {Pair
  production in laser fields oscillating in space and time}, Phys. Rev. Lett.
  102 (2009) 080402.
\newblock \href {http://arxiv.org/abs/0810.4047} {\path{arXiv:0810.4047}},
  \href {https://doi.org/10.1103/PhysRevLett.102.080402}
  {\path{doi:10.1103/PhysRevLett.102.080402}}.

\bibitem{Gies:2015hia}
H.~Gies, G.~Torgrimsson, {Critical Schwinger pair production}, Phys. Rev. Lett.
  116~(9) (2016) 090406.
\newblock \href {http://arxiv.org/abs/1507.07802} {\path{arXiv:1507.07802}},
  \href {https://doi.org/10.1103/PhysRevLett.116.090406}
  {\path{doi:10.1103/PhysRevLett.116.090406}}.

\bibitem{Gies:2016coz}
H.~Gies, G.~Torgrimsson, {Critical Schwinger pair production II - universality
  in the deeply critical regime}, Phys. Rev. D 95~(1) (2017) 016001.
\newblock \href {http://arxiv.org/abs/1612.00635} {\path{arXiv:1612.00635}},
  \href {https://doi.org/10.1103/PhysRevD.95.016001}
  {\path{doi:10.1103/PhysRevD.95.016001}}.

\bibitem{Hebenstreit:2011wk}
F.~Hebenstreit, R.~Alkofer, H.~Gies, {Particle self-bunching in the Schwinger
  effect in spacetime-dependent electric fields}, Phys. Rev. Lett. 107 (2011)
  180403.
\newblock \href {http://arxiv.org/abs/1106.6175} {\path{arXiv:1106.6175}},
  \href {https://doi.org/10.1103/PhysRevLett.107.180403}
  {\path{doi:10.1103/PhysRevLett.107.180403}}.

\bibitem{Ababekri:2019dkl}
M.~Ababekri, B.-S. Xie, J.~Zhang, {Effects of finite spatial extent on
  Schwinger pair production}, Phys. Rev. D 100~(1) (2019) 016003.
\newblock \href {http://arxiv.org/abs/1905.01629} {\path{arXiv:1905.01629}},
  \href {https://doi.org/10.1103/PhysRevD.100.016003}
  {\path{doi:10.1103/PhysRevD.100.016003}}.

\bibitem{Kohlfurst:2017hbd}
C.~Kohlf\"urst, R.~Alkofer, {Ponderomotive effects in multiphoton pair
  production}, Phys. Rev. D 97~(3) (2018) 036026.
\newblock \href {http://arxiv.org/abs/1711.10766} {\path{arXiv:1711.10766}},
  \href {https://doi.org/10.1103/PhysRevD.97.036026}
  {\path{doi:10.1103/PhysRevD.97.036026}}.

\bibitem{Dong:2017vse}
S.~S. Dong, M.~Chen, Q.~Su, R.~Grobe, {Optimization of spatially localized
  electric fields for electron-positron pair creation}, Phys. Rev. A 96~(3)
  (2017) 032120.
\newblock \href {https://doi.org/10.1103/PhysRevA.96.032120}
  {\path{doi:10.1103/PhysRevA.96.032120}}.

\bibitem{Kohlfurst:2019mag}
C.~Kohlf\"urst, {Effect of time-dependent inhomogeneous magnetic fields on the
  particle momentum spectrum in electron-positron pair production}, Phys. Rev.
  D 101~(9) (2020) 096003.
\newblock \href {http://arxiv.org/abs/1912.09359} {\path{arXiv:1912.09359}},
  \href {https://doi.org/10.1103/PhysRevD.101.096003}
  {\path{doi:10.1103/PhysRevD.101.096003}}.

\bibitem{Gavrilov:2019sbt}
S.~P. Gavrilov, D.~M. Gitman, A.~A. Shishmarev, {Pair production from the
  vacuum by a weakly inhomogeneous space-dependent electric potential}, Phys.
  Rev. D 99~(11) (2019) 116014.
\newblock \href {http://arxiv.org/abs/1903.05925} {\path{arXiv:1903.05925}},
  \href {https://doi.org/10.1103/PhysRevD.99.116014}
  {\path{doi:10.1103/PhysRevD.99.116014}}.

\bibitem{Tomaras:2001vs}
T.~N. Tomaras, N.~C. Tsamis, R.~P. Woodard, {Pair creation and axial anomaly in
  light cone QED(2)}, JHEP 11 (2001) 008.
\newblock \href {http://arxiv.org/abs/hep-th/0108090}
  {\path{arXiv:hep-th/0108090}}, \href
  {https://doi.org/10.1088/1126-6708/2001/11/008}
  {\path{doi:10.1088/1126-6708/2001/11/008}}.

\bibitem{Fried:2001ur}
H.~M. Fried, R.~P. Woodard, {The One loop effective action of QED for a general
  class of electric fields}, Phys. Lett. B 524 (2002) 233--239.
\newblock \href {http://arxiv.org/abs/hep-th/0110180}
  {\path{arXiv:hep-th/0110180}}, \href
  {https://doi.org/10.1016/S0370-2693(01)01384-3}
  {\path{doi:10.1016/S0370-2693(01)01384-3}}.

\bibitem{Tanji:2011di}
N.~Tanji, K.~Itakura, {Schwinger mechanism enhanced by the
  Nielsen\textendash{}Olesen instability}, Phys. Lett. B 713 (2012) 117--121.
\newblock \href {http://arxiv.org/abs/1111.6772} {\path{arXiv:1111.6772}},
  \href {https://doi.org/10.1016/j.physletb.2012.05.043}
  {\path{doi:10.1016/j.physletb.2012.05.043}}.

\bibitem{Karabali:2019ucc}
D.~Karabali, S.~Kurkcuoglu, V.~P. Nair, {Magnetic Field and Curvature Effects
  on Pair Production II: Vectors and Implications for Chromodynamics}, Phys.
  Rev. D 100~(6) (2019) 065006.
\newblock \href {http://arxiv.org/abs/1905.12391} {\path{arXiv:1905.12391}},
  \href {https://doi.org/10.1103/PhysRevD.100.065006}
  {\path{doi:10.1103/PhysRevD.100.065006}}.

\bibitem{PhysRevLett.109.253202}
Q.~Su, W.~Su, Q.~Z. Lv, M.~Jiang, X.~Lu, Z.~M. Sheng, R.~Grobe,
  \href{https://link.aps.org/doi/10.1103/PhysRevLett.109.253202}{Magnetic
  control of the pair creation in spatially localized supercritical fields},
  Phys. Rev. Lett. 109 (2012) 253202.
\newblock \href {https://doi.org/10.1103/PhysRevLett.109.253202}
  {\path{doi:10.1103/PhysRevLett.109.253202}}.
\newline\urlprefix\url{https://link.aps.org/doi/10.1103/PhysRevLett.109.253202}

\bibitem{Adler:1969gk}
S.~L. Adler, {Axial vector vertex in spinor electrodynamics}, Phys. Rev. 177
  (1969) 2426--2438.
\newblock \href {https://doi.org/10.1103/PhysRev.177.2426}
  {\path{doi:10.1103/PhysRev.177.2426}}.

\bibitem{Bell:1969ts}
J.~S. Bell, R.~Jackiw, {A PCAC puzzle: $\pi^0 \to \gamma \gamma$ in the
  $\sigma$ model}, Nuovo Cim. A 60 (1969) 47--61.
\newblock \href {https://doi.org/10.1007/BF02823296}
  {\path{doi:10.1007/BF02823296}}.

\bibitem{Kharzeev:2007jp}
D.~E. Kharzeev, L.~D. McLerran, H.~J. Warringa, {The Effects of topological
  charge change in heavy ion collisions: 'Event by event P and CP violation'},
  Nucl. Phys. A 803 (2008) 227--253.
\newblock \href {http://arxiv.org/abs/0711.0950} {\path{arXiv:0711.0950}},
  \href {https://doi.org/10.1016/j.nuclphysa.2008.02.298}
  {\path{doi:10.1016/j.nuclphysa.2008.02.298}}.

\bibitem{Fukushima:2010vw}
K.~Fukushima, D.~E. Kharzeev, H.~J. Warringa, {Real-time dynamics of the Chiral
  Magnetic Effect}, Phys. Rev. Lett. 104 (2010) 212001.
\newblock \href {http://arxiv.org/abs/1002.2495} {\path{arXiv:1002.2495}},
  \href {https://doi.org/10.1103/PhysRevLett.104.212001}
  {\path{doi:10.1103/PhysRevLett.104.212001}}.

\bibitem{Warringa:2012bq}
H.~J. Warringa, {Dynamics of the Chiral Magnetic Effect in a weak magnetic
  field}, Phys. Rev. D 86 (2012) 085029.
\newblock \href {http://arxiv.org/abs/1205.5679} {\path{arXiv:1205.5679}},
  \href {https://doi.org/10.1103/PhysRevD.86.085029}
  {\path{doi:10.1103/PhysRevD.86.085029}}.

\bibitem{Fukushima:2015tza}
K.~Fukushima, {Simulating net particle production and chiral magnetic current
  in a $CP$-odd domain}, Phys. Rev. D 92~(5) (2015) 054009.
\newblock \href {http://arxiv.org/abs/1501.01940} {\path{arXiv:1501.01940}},
  \href {https://doi.org/10.1103/PhysRevD.92.054009}
  {\path{doi:10.1103/PhysRevD.92.054009}}.

\bibitem{Aoi:2021azo}
H.~Aoi, K.~Suzuki, {Chirality imbalance and chiral magnetic effect under a
  parallel electromagnetic field}, Phys. Rev. D 103~(3) (2021) 036002.
\newblock \href {http://arxiv.org/abs/2101.09504} {\path{arXiv:2101.09504}},
  \href {https://doi.org/10.1103/PhysRevD.103.036002}
  {\path{doi:10.1103/PhysRevD.103.036002}}.

\bibitem{Copinger:2018ftr}
P.~Copinger, K.~Fukushima, S.~Pu, {Axial Ward identity and the Schwinger
  mechanism -- Applications to the real-time chiral magnetic effect and
  condensates}, Phys. Rev. Lett. 121~(26) (2018) 261602.
\newblock \href {http://arxiv.org/abs/1807.04416} {\path{arXiv:1807.04416}},
  \href {https://doi.org/10.1103/PhysRevLett.121.261602}
  {\path{doi:10.1103/PhysRevLett.121.261602}}.

\bibitem{2012ChPhL..29b1102X}
B.-S. {Xie}, M.~{Mohamedsedik}, S.~{Dulat}, {Electron-Positron Pair Production
  in an Elliptic Polarized Time Varying Field}, Chinese Physics Letters 29~(2)
  (2012) 021102.
\newblock \href {https://doi.org/10.1088/0256-307X/29/2/021102}
  {\path{doi:10.1088/0256-307X/29/2/021102}}.

\bibitem{Blinne:2013via}
A.~Blinne, H.~Gies, {Pair Production in Rotating Electric Fields}, Phys. Rev. D
  89 (2014) 085001.
\newblock \href {http://arxiv.org/abs/1311.1678} {\path{arXiv:1311.1678}},
  \href {https://doi.org/10.1103/PhysRevD.89.085001}
  {\path{doi:10.1103/PhysRevD.89.085001}}.

\bibitem{Li:2015cea}
Z.~L. Li, D.~Lu, B.~S. Xie, {Effects of electric field polarizations on pair
  production}, Phys. Rev. D 92~(8) (2015) 085001.
\newblock \href {http://arxiv.org/abs/1508.03440} {\path{arXiv:1508.03440}},
  \href {https://doi.org/10.1103/PhysRevD.92.085001}
  {\path{doi:10.1103/PhysRevD.92.085001}}.

\bibitem{Li:2017qwd}
Z.~L. Li, Y.~J. Li, B.~S. Xie, {Momentum Vortices on Pairs Production by Two
  Counter-Rotating Fields}, Phys. Rev. D 96~(7) (2017) 076010.
\newblock \href {http://arxiv.org/abs/1707.01353} {\path{arXiv:1707.01353}},
  \href {https://doi.org/10.1103/PhysRevD.96.076010}
  {\path{doi:10.1103/PhysRevD.96.076010}}.

\bibitem{Strobel:2014tha}
E.~Strobel, S.-S. Xue, {Semiclassical pair production rate for rotating
  electric fields}, Phys. Rev. D 91 (2015) 045016.
\newblock \href {http://arxiv.org/abs/1412.2628} {\path{arXiv:1412.2628}},
  \href {https://doi.org/10.1103/PhysRevD.91.045016}
  {\path{doi:10.1103/PhysRevD.91.045016}}.

\bibitem{Wollert:2015kra}
A.~W\"ollert, H.~Bauke, C.~H. Keitel, {Spin polarized electron-positron pair
  production via elliptical polarized laser fields}, Phys. Rev. D 91~(12)
  (2015) 125026.
\newblock \href {http://arxiv.org/abs/1502.06414} {\path{arXiv:1502.06414}},
  \href {https://doi.org/10.1103/PhysRevD.91.125026}
  {\path{doi:10.1103/PhysRevD.91.125026}}.

\bibitem{Ebihara:2015aca}
S.~Ebihara, K.~Fukushima, T.~Oka, {Chiral pumping effect induced by rotating
  electric fields}, Phys. Rev. B 93~(15) (2016) 155107.
\newblock \href {http://arxiv.org/abs/1509.03673} {\path{arXiv:1509.03673}},
  \href {https://doi.org/10.1103/PhysRevB.93.155107}
  {\path{doi:10.1103/PhysRevB.93.155107}}.

\bibitem{Kim:2016yxx}
C.~M. Kim, S.~P. Kim, {Spin Resonance Effect on Pair Production in Rotating
  Electric Fields} (1 2016).
\newblock \href {http://arxiv.org/abs/1601.02468} {\path{arXiv:1601.02468}}.

\bibitem{Kohlfurst:2018kxg}
C.~Kohlf\"urst, {Spin-states in multiphoton pair production for circularly
  polarized light}, Phys. Rev. D 99~(9) (2019) 096017.
\newblock \href {http://arxiv.org/abs/1812.03130} {\path{arXiv:1812.03130}},
  \href {https://doi.org/10.1103/PhysRevD.99.096017}
  {\path{doi:10.1103/PhysRevD.99.096017}}.

\bibitem{Takayoshi:2020afs}
S.~Takayoshi, J.~Wu, T.~Oka, {Nonadiabatic Nonlinear Optics and Quantum
  Geometry -- Application to the Twisted Schwinger Effect} (5 2020).
\newblock \href {http://arxiv.org/abs/2005.01755} {\path{arXiv:2005.01755}}.

\bibitem{Huang:2019szw}
X.-G. Huang, M.~Matsuo, H.~Taya, {Spontaneous generation of spin current from
  the vacuum by strong electric fields}, PTEP 2019~(11) (2019) 113B02.
\newblock \href {http://arxiv.org/abs/1904.07593} {\path{arXiv:1904.07593}},
  \href {https://doi.org/10.1093/ptep/ptz112} {\path{doi:10.1093/ptep/ptz112}}.

\bibitem{Chu:2021eae}
C.-S. Chu, C.-H. Leung, {Induced Quantized Spin Current in Vacuum}, Phys. Rev.
  Lett. 127~(11) (2021) 111601.
\newblock \href {http://arxiv.org/abs/2105.00148} {\path{arXiv:2105.00148}},
  \href {https://doi.org/10.1103/PhysRevLett.127.111601}
  {\path{doi:10.1103/PhysRevLett.127.111601}}.

\bibitem{2020CmPhy...3...63K}
S.~{Kitamura}, N.~{Nagaosa}, T.~{Morimoto}, {Nonreciprocal Landau-Zener
  tunneling}, Communications Physics 3~(1) (2020) 63.
\newblock \href {http://arxiv.org/abs/1908.00819} {\path{arXiv:1908.00819}},
  \href {https://doi.org/10.1038/s42005-020-0328-0}
  {\path{doi:10.1038/s42005-020-0328-0}}.

\bibitem{Evans:2019zyk}
S.~Evans, J.~Rafelski, {Electron electromagnetic-mass melting in strong
  fields}, Phys. Rev. D 102~(3) (2020) 036014.
\newblock \href {http://arxiv.org/abs/1911.08714} {\path{arXiv:1911.08714}},
  \href {https://doi.org/10.1103/PhysRevD.102.036014}
  {\path{doi:10.1103/PhysRevD.102.036014}}.

\bibitem{Dunne:2006sx}
G.~V. Dunne, M.~Krasnansky, {'Background field integration-by-parts' and the
  connection between one-loop and two-loop Heisenberg-Euler effective actions},
  JHEP 04 (2006) 020.
\newblock \href {http://arxiv.org/abs/hep-th/0602216}
  {\path{arXiv:hep-th/0602216}}, \href
  {https://doi.org/10.1088/1126-6708/2006/04/020}
  {\path{doi:10.1088/1126-6708/2006/04/020}}.

\bibitem{Krasnansky:2006sx}
M.~Krasnansky, {Two-loop vacuum diagrams in background field and
  Heisenberg-Euler effective action in different dimensions}, Int. J. Mod.
  Phys. A 23 (2008) 5201--5215.
\newblock \href {http://arxiv.org/abs/hep-th/0607230}
  {\path{arXiv:hep-th/0607230}}, \href
  {https://doi.org/10.1142/S0217751X08042572}
  {\path{doi:10.1142/S0217751X08042572}}.

\bibitem{Huet:2010nt}
I.~Huet, D.~G.~C. McKeon, C.~Schubert, {Euler-Heisenberg lagrangians and
  asymptotic analysis in 1+1 QED, part 1: Two-loop}, JHEP 12 (2010) 036.
\newblock \href {http://arxiv.org/abs/1010.5315} {\path{arXiv:1010.5315}},
  \href {https://doi.org/10.1007/JHEP12(2010)036}
  {\path{doi:10.1007/JHEP12(2010)036}}.

\bibitem{Huet:2020awq}
I.~Huet, M.~Rausch~de Traubenberg, C.~Schubert, {Multiloop QED in the
  Euler-Heisenberg approach}, in: {19th Lomonosov Conference on Elementary
  Particle Physics}, 2020.
\newblock \href {http://arxiv.org/abs/2001.06667} {\path{arXiv:2001.06667}}.

\bibitem{Lan:2018xnq}
C.~Lan, Y.-F. Wang, H.~Geng, A.~Andreev, {Dynamically Assisted Schwinger Effect
  at Strong Coupling with Its Holographic Extension}, Eur. Phys. J. C 79~(11)
  (2019) 917.
\newblock \href {http://arxiv.org/abs/1811.11712} {\path{arXiv:1811.11712}},
  \href {https://doi.org/10.1140/epjc/s10052-019-7405-0}
  {\path{doi:10.1140/epjc/s10052-019-7405-0}}.

\bibitem{Gould:2019myj}
O.~Gould, D.~L.~J. Ho, A.~Rajantie, {Towards Schwinger production of magnetic
  monopoles in heavy-ion collisions}, Phys. Rev. D 100~(1) (2019) 015041.
\newblock \href {http://arxiv.org/abs/1902.04388} {\path{arXiv:1902.04388}},
  \href {https://doi.org/10.1103/PhysRevD.100.015041}
  {\path{doi:10.1103/PhysRevD.100.015041}}.

\bibitem{Gorsky:2001up}
A.~S. Gorsky, K.~A. Saraikin, K.~G. Selivanov, {Schwinger type processes via
  branes and their gravity duals}, Nucl. Phys. B 628 (2002) 270--294.
\newblock \href {http://arxiv.org/abs/hep-th/0110178}
  {\path{arXiv:hep-th/0110178}}, \href
  {https://doi.org/10.1016/S0550-3213(02)00095-0}
  {\path{doi:10.1016/S0550-3213(02)00095-0}}.

\bibitem{Semenoff:2011ng}
G.~W. Semenoff, K.~Zarembo, {Holographic Schwinger Effect}, Phys. Rev. Lett.
  107 (2011) 171601.
\newblock \href {http://arxiv.org/abs/1109.2920} {\path{arXiv:1109.2920}},
  \href {https://doi.org/10.1103/PhysRevLett.107.171601}
  {\path{doi:10.1103/PhysRevLett.107.171601}}.

\bibitem{Sato:2013iua}
Y.~Sato, K.~Yoshida, {Potential Analysis in Holographic Schwinger Effect}, JHEP
  08 (2013) 002.
\newblock \href {http://arxiv.org/abs/1304.7917} {\path{arXiv:1304.7917}},
  \href {https://doi.org/10.1007/JHEP08(2013)002}
  {\path{doi:10.1007/JHEP08(2013)002}}.

\bibitem{Sato:2013hyw}
Y.~Sato, K.~Yoshida, {Universal aspects of holographic Schwinger effect in
  general backgrounds}, JHEP 12 (2013) 051.
\newblock \href {http://arxiv.org/abs/1309.4629} {\path{arXiv:1309.4629}},
  \href {https://doi.org/10.1007/JHEP12(2013)051}
  {\path{doi:10.1007/JHEP12(2013)051}}.

\bibitem{Aleksandrov:2022rgg}
I.~A. Aleksandrov, A.~Di~Piazza, G.~Plunien, V.~M. Shabaev, {Stimulated vacuum
  emission and photon absorption in strong electromagnetic fields} (2 2022).
\newblock \href {http://arxiv.org/abs/2202.06886} {\path{arXiv:2202.06886}}.

\bibitem{Tanji:2015ata}
N.~Tanji, {Electromagnetic currents induced by color fields}, Phys. Rev. D
  92~(12) (2015) 125012.
\newblock \href {http://arxiv.org/abs/1506.08442} {\path{arXiv:1506.08442}},
  \href {https://doi.org/10.1103/PhysRevD.92.125012}
  {\path{doi:10.1103/PhysRevD.92.125012}}.

\bibitem{Blaschke:2008wf}
D.~B. Blaschke, A.~V. Prozorkevich, G.~Ropke, C.~D. Roberts, S.~M. Schmidt,
  D.~S. Shkirmanov, S.~A. Smolyansky, {Dynamical Schwinger effect and
  high-intensity lasers. realising nonperturbative QED}, Eur. Phys. J. D 55
  (2009) 341--358.
\newblock \href {http://arxiv.org/abs/0811.3570} {\path{arXiv:0811.3570}},
  \href {https://doi.org/10.1140/epjd/e2009-00156-y}
  {\path{doi:10.1140/epjd/e2009-00156-y}}.

\bibitem{Blaschke:2010vs}
D.~B. Blaschke, G.~Ropke, S.~M. Schmidt, S.~A. Smolyansky, A.~V. Tarakanov,
  {Kinetics of Photon Radiation off an e-e+ Plasma created from the Vacuum in a
  Strong Laser Field}, Contrib. Plasma Phys. 51 (2011) 451--456.
\newblock \href {http://arxiv.org/abs/1006.1098} {\path{arXiv:1006.1098}},
  \href {https://doi.org/10.1002/ctpp.201110016}
  {\path{doi:10.1002/ctpp.201110016}}.

\bibitem{Blaschke:2011is}
D.~B. Blaschke, V.~V. Dmitriev, G.~Ropke, S.~A. Smolyansky, {BBGKY kinetic
  approach for an $e^-e^+\gamma$ plasma created from the vacuum in a strong
  laser-generated electric field: The one-photon annihilation channel}, Phys.
  Rev. D 84 (2011) 085028.
\newblock \href {http://arxiv.org/abs/1105.5397} {\path{arXiv:1105.5397}},
  \href {https://doi.org/10.1103/PhysRevD.84.085028}
  {\path{doi:10.1103/PhysRevD.84.085028}}.

\bibitem{Smolyansky:2019yma}
S.~A. Smolyansky, A.~D. Panferov, S.~O. Pirogov, A.~M. Fedotov,
  {Self-consistent kinetic equations for $e^-e^+\gamma$-plasma generated from
  vacuum by strong electric field} (1 2019).
\newblock \href {http://arxiv.org/abs/1901.02305} {\path{arXiv:1901.02305}}.

\bibitem{Aleksandrov:2021ylw}
I.~A. Aleksandrov, A.~D. Panferov, S.~A. Smolyansky, {Radiation signal
  accompanying the Schwinger effect}, Phys. Rev. A 103~(5) (2021) 053107.
\newblock \href {http://arxiv.org/abs/2101.03507} {\path{arXiv:2101.03507}},
  \href {https://doi.org/10.1103/PhysRevA.103.053107}
  {\path{doi:10.1103/PhysRevA.103.053107}}.

\bibitem{Aleksandrov:2019irn}
I.~A. Aleksandrov, G.~Plunien, V.~M. Shabaev, {Photon emission in strong fields
  beyond the locally-constant field approximation}, Phys. Rev. D 100~(11)
  (2019) 116003.
\newblock \href {http://arxiv.org/abs/1909.03888} {\path{arXiv:1909.03888}},
  \href {https://doi.org/10.1103/PhysRevD.100.116003}
  {\path{doi:10.1103/PhysRevD.100.116003}}.

\bibitem{Kuchiev:2015qua}
M.~Kuchiev, J.~Ingham, {Enhanced Creation of High Energy Particles in Colliding
  Laser Beams} (11 2015).
\newblock \href {http://arxiv.org/abs/1511.06806} {\path{arXiv:1511.06806}}.

\bibitem{Villalba-Chavez:2019jqp}
S.~Villalba-Ch\'avez, C.~M\"uller, {Signatures of the Schwinger mechanism
  assisted by a fast-oscillating electric field}, Phys. Rev. D 100~(11) (2019)
  116018.
\newblock \href {http://arxiv.org/abs/1907.06438} {\path{arXiv:1907.06438}},
  \href {https://doi.org/10.1103/PhysRevD.100.116018}
  {\path{doi:10.1103/PhysRevD.100.116018}}.

\bibitem{Otto:2018jbs}
A.~Otto, H.~Oppitz, B.~Kämpfer, {Assisted Vacuum Decay by Time Dependent
  Electric Fields}, Eur. Phys. J. A 54~(2) (2018) 23.
\newblock \href {http://arxiv.org/abs/1801.09943} {\path{arXiv:1801.09943}},
  \href {https://doi.org/10.1140/epja/i2018-12473-x}
  {\path{doi:10.1140/epja/i2018-12473-x}}.

\bibitem{Corkum:1993zz}
P.~B. Corkum, {Plasma perspective on strong field multiphoton ionization},
  Phys. Rev. Lett. 71 (1993) 1994--1997.
\newblock \href {https://doi.org/10.1103/PhysRevLett.71.1994}
  {\path{doi:10.1103/PhysRevLett.71.1994}}.

\bibitem{Vampa2014}
G.~Vampa, C.~R. McDonald, G.~Orlando, D.~D. Klug, P.~B. Corkum, T.~Brabec,
  \href{https://link.aps.org/doi/10.1103/PhysRevLett.113.073901}{{Theoretical
  Analysis of High-Harmonic Generation in Solids}}, Physical Review Letters
  113~(7) (2014) 073901.
\newblock \href {https://doi.org/10.1103/PhysRevLett.113.073901}
  {\path{doi:10.1103/PhysRevLett.113.073901}}.
\newline\urlprefix\url{https://link.aps.org/doi/10.1103/PhysRevLett.113.073901}

\bibitem{Parker:1968mv}
L.~Parker, {Particle creation in expanding universes}, Phys. Rev. Lett. 21
  (1968) 562--564.
\newblock \href {https://doi.org/10.1103/PhysRevLett.21.562}
  {\path{doi:10.1103/PhysRevLett.21.562}}.

\bibitem{Pervushin:2006vh}
V.~N. Pervushin, V.~V. Skokov, {Kinetic description of fermion production in
  the oscillator representation}, Acta Phys. Polon. B 37 (2006) 2587--2600.
\newblock \href {http://arxiv.org/abs/astro-ph/0611780}
  {\path{arXiv:astro-ph/0611780}}.

\bibitem{Aleksandrov:2020mez}
I.~A. Aleksandrov, V.~V. Dmitriev, D.~G. Sevostyanov, S.~A. Smolyansky,
  {Kinetic description of vacuum $e^{+}e^{-}$ production in strong electric
  fields of arbitrary polarization}, Eur. Phys. J. ST 229~(22-23) (2020)
  3469--3485.
\newblock \href {http://arxiv.org/abs/2004.02179} {\path{arXiv:2004.02179}},
  \href {https://doi.org/10.1140/epjst/e2020-000056-1}
  {\path{doi:10.1140/epjst/e2020-000056-1}}.

\bibitem{Ilderton:2021zej}
A.~Ilderton, {Physics of adiabatic particle number in the Schwinger effect},
  Phys. Rev. D 105~(1) (2022) 016021.
\newblock \href {http://arxiv.org/abs/2108.13885} {\path{arXiv:2108.13885}},
  \href {https://doi.org/10.1103/PhysRevD.105.016021}
  {\path{doi:10.1103/PhysRevD.105.016021}}.

\bibitem{2016PhRvL.116w3603Z}
T.~{Zimmermann}, S.~{Mishra}, B.~R. {Doran}, D.~F. {Gordon}, A.~S. {Landsman},
  {Tunneling Time and Weak Measurement in Strong Field Ionization}, Phys. Rev.
  Lett. 116~(23) (2016) 233603.
\newblock \href {https://doi.org/10.1103/PhysRevLett.116.233603}
  {\path{doi:10.1103/PhysRevLett.116.233603}}.

\bibitem{Oka:2011kf}
T.~Oka, {Strong field physics in condensed matter}, 2011.
\newblock \href {http://arxiv.org/abs/1102.2482} {\path{arXiv:1102.2482}}.

\bibitem{2014JPhB...47t4030G}
S.~{Ghimire}, G.~{Ndabashimiye}, A.~D. {DiChiara}, E.~{Sistrunk}, M.~I.
  {Stockman}, P.~{Agostini}, L.~F. {DiMauro}, D.~A. {Reis}, {Strong-field and
  attosecond physics in solids}, Journal of Physics B Atomic Molecular Physics
  47~(20) (2014) 204030.
\newblock \href {https://doi.org/10.1088/0953-4075/47/20/204030}
  {\path{doi:10.1088/0953-4075/47/20/204030}}.

\bibitem{Basov2017}
D.~N. Basov, R.~D. Averitt, D.~Hsieh,
  \href{https://doi.org/10.1038/nmat5017}{Towards properties on demand in
  quantum materials}, Nature Materials 16~(11) (2017) 1077--1088.
\newblock \href {https://doi.org/10.1038/nmat5017}
  {\path{doi:10.1038/nmat5017}}.
\newline\urlprefix\url{https://doi.org/10.1038/nmat5017}

\bibitem{2018RvMP...90b1002K}
S.~Y. {Kruchinin}, F.~{Krausz}, V.~S. {Yakovlev}, {Colloquium: Strong-field
  phenomena in periodic systems}, Reviews of Modern Physics 90~(2) (2018)
  021002.
\newblock \href {http://arxiv.org/abs/1712.05685} {\path{arXiv:1712.05685}},
  \href {https://doi.org/10.1103/RevModPhys.90.021002}
  {\path{doi:10.1103/RevModPhys.90.021002}}.

\bibitem{2019ARCMP..10..387O}
T.~{Oka}, S.~{Kitamura}, {Floquet Engineering of Quantum Materials}, Annual
  Review of Condensed Matter Physics 10 (2019) 387--408.
\newblock \href {http://arxiv.org/abs/1804.03212} {\path{arXiv:1804.03212}},
  \href {https://doi.org/10.1146/annurev-conmatphys-031218-013423}
  {\path{doi:10.1146/annurev-conmatphys-031218-013423}}.

\bibitem{Allor:2007ei}
D.~Allor, T.~D. Cohen, D.~A. McGady, {The Schwinger mechanism and graphene},
  Phys. Rev. D 78 (2008) 096009.
\newblock \href {http://arxiv.org/abs/0708.1471} {\path{arXiv:0708.1471}},
  \href {https://doi.org/10.1103/PhysRevD.78.096009}
  {\path{doi:10.1103/PhysRevD.78.096009}}.

\bibitem{Zubkov:2012ht}
M.~A. Zubkov, {Schwinger pair creation in multilayer graphene}, Pisma Zh. Eksp.
  Teor. Fiz. 95 (2012) 540.
\newblock \href {http://arxiv.org/abs/1204.0138} {\path{arXiv:1204.0138}},
  \href {https://doi.org/10.1134/S0021364012090135}
  {\path{doi:10.1134/S0021364012090135}}.

\bibitem{Katsnelson:2012tp}
M.~I. Katsnelson, G.~E. Volovik, M.~A. Zubkov, {Euler - Heisenberg effective
  action and magnetoelectric effect in multilayer graphene}, Annals Phys. 331
  (2013) 160--187.
\newblock \href {http://arxiv.org/abs/1206.3973} {\path{arXiv:1206.3973}},
  \href {https://doi.org/10.1016/j.aop.2012.12.010}
  {\path{doi:10.1016/j.aop.2012.12.010}}.

\bibitem{Katsnelson:2012cz}
M.~I. Katsnelson, G.~E. Volovik, {Quantum electrodynamics with anisotropic
  scaling: Heisenberg-Euler action and Schwinger pair production in the bilayer
  graphene}, JETP Lett. 95 (2012) 411--415.
\newblock \href {http://arxiv.org/abs/1203.1578} {\path{arXiv:1203.1578}},
  \href {https://doi.org/10.1134/S0021364012080061}
  {\path{doi:10.1134/S0021364012080061}}.

\bibitem{Fillion-Gourdeau:2015dga}
F.~Fillion-Gourdeau, S.~MacLean, {Time-dependent pair creation and the
  Schwinger mechanism in graphene}, Phys. Rev. B 92~(3) (2015) 035401.
\newblock \href {https://doi.org/10.1103/PhysRevB.92.035401}
  {\path{doi:10.1103/PhysRevB.92.035401}}.

\bibitem{Akal:2016stu}
I.~Akal, R.~Egger, C.~M\"uller, S.~Villalba-Ch\'avez, {Low-dimensional approach
  to pair production in an oscillating electric field: Application to bandgap
  graphene layers}, Phys. Rev. D 93~(11) (2016) 116006.
\newblock \href {http://arxiv.org/abs/1602.08310} {\path{arXiv:1602.08310}},
  \href {https://doi.org/10.1103/PhysRevD.93.116006}
  {\path{doi:10.1103/PhysRevD.93.116006}}.

\bibitem{Akal:2018txb}
I.~Akal, R.~Egger, C.~M\"uller, S.~Villalba-Ch\'avez, {Simulating dynamically
  assisted production of Dirac pairs in gapped graphene monolayers}, Phys. Rev.
  D 99~(1) (2019) 016025.
\newblock \href {http://arxiv.org/abs/1812.03846} {\path{arXiv:1812.03846}},
  \href {https://doi.org/10.1103/PhysRevD.99.016025}
  {\path{doi:10.1103/PhysRevD.99.016025}}.

\bibitem{Szpak:2011jj}
N.~Szpak, R.~Schutzhold, {Optical lattice quantum simulator for QED in strong
  external fields: Spontaneous pair creation and the Sauter-Schwinger effect},
  New J. Phys. 14 (2012) 035001.
\newblock \href {http://arxiv.org/abs/1109.2426} {\path{arXiv:1109.2426}},
  \href {https://doi.org/10.1088/1367-2630/14/3/035001}
  {\path{doi:10.1088/1367-2630/14/3/035001}}.

\bibitem{Kasper:2015cca}
V.~Kasper, F.~Hebenstreit, M.~Oberthaler, J.~Berges, {Schwinger pair production
  with ultracold atoms}, Phys. Lett. B 760 (2016) 742--746.
\newblock \href {http://arxiv.org/abs/1506.01238} {\path{arXiv:1506.01238}},
  \href {https://doi.org/10.1016/j.physletb.2016.07.036}
  {\path{doi:10.1016/j.physletb.2016.07.036}}.

\bibitem{Linder:2015fba}
M.~F. Linder, A.~Lorke, R.~Sch\"utzhold, {Analog Sauter-Schwinger effect in
  semiconductors for spacetime-dependent fields}, Phys. Rev. B 97~(3) (2018)
  035203.
\newblock \href {http://arxiv.org/abs/1503.07108} {\path{arXiv:1503.07108}},
  \href {https://doi.org/10.1103/PhysRevB.97.035203}
  {\path{doi:10.1103/PhysRevB.97.035203}}.

\bibitem{Vajna:2015qra}
S.~Vajna, B.~D\'ora, R.~Moessner, {Nonequilibrium transport and statistics of
  Schwinger pair production in Weyl semimetals}, Phys. Rev. B 92~(8) (2015)
  085122.
\newblock \href {http://arxiv.org/abs/1505.08004} {\path{arXiv:1505.08004}},
  \href {https://doi.org/10.1103/PhysRevB.92.085122}
  {\path{doi:10.1103/PhysRevB.92.085122}}.

\bibitem{Abramchuk:2016afc}
R.~A. Abramchuk, M.~A. Zubkov, {Schwinger pair creation in Dirac semimetals in
  the presence of external magnetic and electric fields}, Phys. Rev. D 94~(11)
  (2016) 116012.
\newblock \href {http://arxiv.org/abs/1605.02379} {\path{arXiv:1605.02379}},
  \href {https://doi.org/10.1103/PhysRevD.94.116012}
  {\path{doi:10.1103/PhysRevD.94.116012}}.

\bibitem{Chernodub:2019blw}
M.~N. Chernodub, M.~A.~H. Vozmediano, {Direct measurement of a beta function
  and an indirect check of the Schwinger effect near the boundary in Dirac-Weyl
  semimetals}, Phys. Rev. Research. 1 (2019) 032002.
\newblock \href {http://arxiv.org/abs/1902.02694} {\path{arXiv:1902.02694}},
  \href {https://doi.org/10.1103/PhysRevResearch.1.032002}
  {\path{doi:10.1103/PhysRevResearch.1.032002}}.

\bibitem{2010PhRvL.105n6404E}
M.~{Eckstein}, T.~{Oka}, P.~{Werner}, {Dielectric Breakdown of Mott Insulators
  in Dynamical Mean-Field Theory}, Phys. Rev. Lett. 105~(14) (2010) 146404.
\newblock \href {http://arxiv.org/abs/1006.3516} {\path{arXiv:1006.3516}},
  \href {https://doi.org/10.1103/PhysRevLett.105.146404}
  {\path{doi:10.1103/PhysRevLett.105.146404}}.

\bibitem{Oka:2011ct}
T.~Oka, {Nonlinear doublon production in a Mott insulator: Landau-Dykhne method
  applied to an integrable model}, Phys. Rev. B 86~(7) (2012) 075148.
\newblock \href {http://arxiv.org/abs/1105.3145} {\path{arXiv:1105.3145}},
  \href {https://doi.org/10.1103/PhysRevB.86.075148}
  {\path{doi:10.1103/PhysRevB.86.075148}}.

\bibitem{PhysRevB.81.033103}
T.~Oka, H.~Aoki,
  \href{https://link.aps.org/doi/10.1103/PhysRevB.81.033103}{Dielectric
  breakdown in a mott insulator: Many-body schwinger-landau-zener mechanism
  studied with a generalized bethe ansatz}, Phys. Rev. B 81 (2010) 033103.
\newblock \href {https://doi.org/10.1103/PhysRevB.81.033103}
  {\path{doi:10.1103/PhysRevB.81.033103}}.
\newline\urlprefix\url{https://link.aps.org/doi/10.1103/PhysRevB.81.033103}

\bibitem{2012PhRvA..85c3625Q}
F.~{Queisser}, P.~{Navez}, R.~{Sch{\"u}tzhold}, {Sauter-Schwinger-like
  tunneling in tilted Bose-Hubbard lattices in the Mott phase}, Phys. Rev. A
  85~(3) (2012) 033625.
\newblock \href {http://arxiv.org/abs/1107.3730} {\path{arXiv:1107.3730}},
  \href {https://doi.org/10.1103/PhysRevA.85.033625}
  {\path{doi:10.1103/PhysRevA.85.033625}}.

\bibitem{Hongo:2020xaw}
M.~Hongo, T.~Fujimori, T.~Misumi, M.~Nitta, N.~Sakai, {Effective field theory
  of magnons: Chiral magnets and the Schwinger mechanism}, Phys. Rev. B
  104~(13) (2021) 134403.
\newblock \href {http://arxiv.org/abs/2009.06694} {\path{arXiv:2009.06694}},
  \href {https://doi.org/10.1103/PhysRevB.104.134403}
  {\path{doi:10.1103/PhysRevB.104.134403}}.

\bibitem{Solinas:2020woq}
P.~Solinas, A.~Amoretti, F.~Giazotto, {Sauter-Schwinger effect in a
  Bardeen-Cooper-Schrieffer superconductor}, Phys. Rev. Lett. 126~(11) (2021)
  117001.
\newblock \href {http://arxiv.org/abs/2007.08323} {\path{arXiv:2007.08323}},
  \href {https://doi.org/10.1103/PhysRevLett.126.117001}
  {\path{doi:10.1103/PhysRevLett.126.117001}}.

\bibitem{Pineiro:2019uzb}
A.~M. Pi\~neiro, D.~Genkina, M.~Lu, I.~B. Spielman,
  {Sauter\textendash{}Schwinger effect with a quantum gas}, New J. Phys. 21~(8)
  (2019) 083035.
\newblock \href {https://doi.org/10.1088/1367-2630/ab3840}
  {\path{doi:10.1088/1367-2630/ab3840}}.

\bibitem{Martinez:2016yna}
E.~A. Martinez, et~al., {Real-time dynamics of lattice gauge theories with a
  few-qubit quantum computer}, Nature 534 (2016) 516--519.
\newblock \href {http://arxiv.org/abs/1605.04570} {\path{arXiv:1605.04570}},
  \href {https://doi.org/10.1038/nature18318} {\path{doi:10.1038/nature18318}}.

\bibitem{Xu:2021tey}
B.~Xu, W.~Xue, {3+1 Dimension Schwinger Pair Production with Quantum Computers}
  (12 2021).
\newblock \href {http://arxiv.org/abs/2112.06863} {\path{arXiv:2112.06863}}.

\bibitem{2015PhRvB..91w5113M}
B.~{Mayer}, C.~{Schmidt}, A.~{Grupp}, J.~{B{\"u}hler}, J.~{Oelmann}, R.~E.
  {Marvel}, R.~F. {Haglund}, T.~{Oka}, D.~{Brida}, A.~{Leitenstorfer},
  A.~{Pashkin}, {Tunneling breakdown of a strongly correlated insulating state
  in V O$_{2}$ induced by intense multiterahertz excitation}, Phys. Rev. B
  91~(23) (2015) 235113.
\newblock \href {https://doi.org/10.1103/PhysRevB.91.235113}
  {\path{doi:10.1103/PhysRevB.91.235113}}.

\bibitem{2017NatMa..16.1100Y}
H.~{Yamakawa}, T.~{Miyamoto}, T.~{Morimoto}, T.~{Terashige}, H.~{Yada},
  N.~{Kida}, M.~{Suda}, H.~M. {Yamamoto}, R.~{Kato}, K.~{Miyagawa},
  K.~{Kanoda}, H.~{Okamoto}, {Mott transition by an impulsive dielectric
  breakdown}, Nature Materials 16~(11) (2017) 1100--1105.
\newblock \href {https://doi.org/10.1038/nmat4967}
  {\path{doi:10.1038/nmat4967}}.

\bibitem{2010ApPhL..97u1902S}
K.~{Shinokita}, H.~{Hirori}, M.~{Nagai}, N.~{Satoh}, Y.~{Kadoya}, K.~{Tanaka},
  {Dynamical Franz-Keldysh effect in GaAs/AlGaAs multiple quantum wells induced
  by single-cycle terahertz pulses}, Applied Physics Letters 97~(21) (2010)
  211902.
\newblock \href {https://doi.org/10.1063/1.3518483}
  {\path{doi:10.1063/1.3518483}}.

\bibitem{2013NatSR...3E1227N}
F.~{Novelli}, D.~{Fausti}, F.~{Giusti}, F.~{Parmigiani}, M.~{Hoffmann}, {Mixed
  regime of light-matter interaction revealed by phase sensitive measurements
  of the dynamical Franz-Keldysh effect}, Scientific Reports 3 (2013) 1227.
\newblock \href {http://arxiv.org/abs/1301.2083} {\path{arXiv:1301.2083}},
  \href {https://doi.org/10.1038/srep01227} {\path{doi:10.1038/srep01227}}.

\bibitem{2013Natur.493...75S}
M.~{Schultze}, E.~M. {Bothschafter}, A.~{Sommer}, S.~{Holzner},
  W.~{Schweinberger}, M.~{Fiess}, M.~{Hofstetter}, R.~{Kienberger},
  V.~{Apalkov}, V.~S. {Yakovlev}, M.~I. {Stockman}, F.~{Krausz}, {Controlling
  dielectrics with the electric field of light}, Nature 493~(7430) (2013)
  75--78.
\newblock \href {https://doi.org/10.1038/nature11720}
  {\path{doi:10.1038/nature11720}}.

\bibitem{2016Sci...353..916L}
M.~{Lucchini}, S.~A. {Sato}, A.~{Ludwig}, J.~{Herrmann}, M.~{Volkov},
  L.~{Kasmi}, Y.~{Shinohara}, K.~{Yabana}, L.~{Gallmann}, U.~{Keller},
  {Attosecond dynamical Franz-Keldysh effect in polycrystalline diamond},
  Science 353~(6302) (2016) 916--919.
\newblock \href {https://doi.org/10.1126/science.aag1268}
  {\path{doi:10.1126/science.aag1268}}.

\bibitem{2010PhRvL.105u5301K}
S.~{Kling}, T.~{Salger}, C.~{Grossert}, M.~{Weitz}, {Atomic Bloch-Zener
  Oscillations and St{\"u}ckelberg Interferometry in Optical Lattices}, Phys.
  Rev. Lett. 105~(21) (2010) 215301.
\newblock \href {https://doi.org/10.1103/PhysRevLett.105.215301}
  {\path{doi:10.1103/PhysRevLett.105.215301}}.

\bibitem{2017Natur.550..224H}
T.~{Higuchi}, C.~{Heide}, K.~{Ullmann}, H.~B. {Weber}, P.~{Hommelhoff},
  {Light-field-driven currents in graphene}, Nature 550~(7675) (2017) 224--228.
\newblock \href {http://arxiv.org/abs/1607.04198} {\path{arXiv:1607.04198}},
  \href {https://doi.org/10.1038/nature23900} {\path{doi:10.1038/nature23900}}.

\bibitem{2013Sci...340..734W}
Y.~{Wang}, D.~{Wong}, A.~V. {Shytov}, V.~W. {Brar}, S.~{Choi}, Q.~{Wu}, H.-Z.
  {Tsai}, W.~{Regan}, A.~{Zettl}, R.~K. {Kawakami}, S.~G. {Louie}, L.~S.
  {Levitov}, M.~F. {Crommie}, {Observing Atomic Collapse Resonances in
  Artificial Nuclei on Graphene}, Science 340~(6133) (2013) 734--737.
\newblock \href {http://arxiv.org/abs/1510.02890} {\path{arXiv:1510.02890}},
  \href {https://doi.org/10.1126/science.1234320}
  {\path{doi:10.1126/science.1234320}}.

\bibitem{Rafelski:2016ixr}
J.~Rafelski, J.~Kirsch, B.~M\"uller, J.~Reinhardt, W.~Greiner, {Probing QED
  Vacuum with Heavy Ions}, FIAS Interdisc. Sci. Ser. (2017) 211--251\href
  {http://arxiv.org/abs/1604.08690} {\path{arXiv:1604.08690}}, \href
  {https://doi.org/10.1007/978-3-319-44165-8_17}
  {\path{doi:10.1007/978-3-319-44165-8_17}}.

\bibitem{Mironov:2020resummation}
A.~A. Mironov, S.~Meuren, A.~M. Fedotov, Resummation of {QED} radiative
  corrections in a strong constant crossed field, Physical Review D 102~(5)
  (2020) 053005.
\newblock \href {http://arxiv.org/abs/2003.06909} {\path{arXiv:2003.06909}},
  \href {https://doi.org/https://doi.org/10.1103/PhysRevD.102.053005}
  {\path{doi:https://doi.org/10.1103/PhysRevD.102.053005}}.

\bibitem{Ritus:1972radiative}
V.~I. Ritus, Radiative corrections in quantum electrodynamics with intense
  field and their analytical properties, Ann. Phys. 69~(2) (1972) 555--582.
\newblock \href {https://doi.org/https://doi.org/10.1016/0003-4916(72)90191-1}
  {\path{doi:https://doi.org/10.1016/0003-4916(72)90191-1}}.

\bibitem{Narozhny:1980expansion}
N.~B. Narozhny, Expansion parameter of perturbation theory in intense-field
  quantum electrodynamics, Phys. Rev. D 21~(4) (1980) 1176.
\newblock \href {https://doi.org/https://doi.org/10.1103/PhysRevD.21.1176}
  {\path{doi:https://doi.org/10.1103/PhysRevD.21.1176}}.

\bibitem{Mironov:2021structure}
A.~A. Mironov, A.~M. Fedotov, Structure of radiative corrections in a strong
  constant crossed field, Phys. Rev. D 105 (2022) 033005.
\newblock \href {http://arxiv.org/abs/2109.00634} {\path{arXiv:2109.00634}},
  \href {https://doi.org/10.1103/PhysRevD.105.033005}
  {\path{doi:10.1103/PhysRevD.105.033005}}.

\bibitem{Narozhny:1979radiation}
N.~B. Narozhny, Radiation corrections to quantum processes in an intense
  electromagnetic field, Phys. Rev. D 20~(6) (1979) 1313.
\newblock \href {https://doi.org/https://doi.org/10.1103/PhysRevD.20.1313}
  {\path{doi:https://doi.org/10.1103/PhysRevD.20.1313}}.

\bibitem{Morozov:1981vertex}
D.~A. Morozov, V.~I. Ritus, N.~B. Narozhnyi,
  \href{http://www.jetp.ras.ru/cgi-bin/dn/e_053_06_1103.pdf}{Vertex function of
  electron in a constant electromagnetic field}, Sov. Phys. JETP 53~(6) (1981)
  1103.
\newline\urlprefix\url{http://www.jetp.ras.ru/cgi-bin/dn/e_053_06_1103.pdf}

\bibitem{Morozov:1981preprint}
D.~A. Morozov, V.~I. Ritus, B.~N. Narozhny,
  \href{https://cds.cern.ch/record/129715/files/CM-P00067564.pdf}{Vertex
  function of electron in a constant electromagnetic field}, preprint~84,
  Lebedev Physical Institute of AS USSR (1981).
\newline\urlprefix\url{https://cds.cern.ch/record/129715/files/CM-P00067564.pdf}

\bibitem{Yakimenko:2018kih}
V.~Yakimenko, et~al., {Prospect of Studying Nonperturbative QED with Beam-Beam
  Collisions}, Phys. Rev. Lett. 122~(19) (2019) 190404.
\newblock \href {http://arxiv.org/abs/1807.09271} {\path{arXiv:1807.09271}},
  \href {https://doi.org/10.1103/PhysRevLett.122.190404}
  {\path{doi:10.1103/PhysRevLett.122.190404}}.

\bibitem{Baumann:2019probing}
C.~Baumann, E.~N. Nerush, A.~Pukhov, I.~Y. Kostyukov, Probing non-perturbative
  {QED} with electron-laser collisions, Scientific reports 9~(1) (2019) 1--8.
\newblock \href {http://arxiv.org/abs/1811.03990} {\path{arXiv:1811.03990}},
  \href {https://doi.org/https://doi.org/10.1038/s41598-019-45582-5}
  {\path{doi:https://doi.org/10.1038/s41598-019-45582-5}}.

\bibitem{Fedeli:2021probing}
L.~Fedeli, A.~Sainte-Marie, N.~Za{\"\i}m, M.~Th{\'e}venet, J.-L. Vay, A.~Myers,
  F.~Qu{\'e}r{\'e}, H.~Vincenti, Probing strong-field {QED} with
  {D}oppler-boosted {P}etawatt-class lasers, Physical Review Letters 127~(11)
  (2021) 114801.
\newblock \href {http://arxiv.org/abs/2012.07696} {\path{arXiv:2012.07696}},
  \href {https://doi.org/https://doi.org/10.1103/PhysRevLett.127.114801}
  {\path{doi:https://doi.org/10.1103/PhysRevLett.127.114801}}.

\bibitem{Baumann:2019laser}
C.~Baumann, A.~Pukhov, Laser-solid interaction and its potential for probing
  radiative corrections in strong-field quantum electrodynamics, Plasma Physics
  and Controlled Fusion 61~(7) (2019) 074010.
\newblock \href {https://doi.org/https://doi.org/10.1088/1361-6587/ab1d2b}
  {\path{doi:https://doi.org/10.1088/1361-6587/ab1d2b}}.

\bibitem{Bradley:2021dnu}
L.~E. Bradley, M.~J.~V. Streeter, C.~D. Murphy, C.~Arran, T.~G. Blackburn,
  M.~Galletti, S.~P.~D. Mangles, C.~P. Ridgers, {Effect of laser temporal
  intensity skew on enhancing pair production in
  laser\textemdash{}electron-beam collisions}, New J. Phys. 23~(9) (2021)
  095004.
\newblock \href {https://doi.org/10.1088/1367-2630/ac1ed6}
  {\path{doi:10.1088/1367-2630/ac1ed6}}.

\bibitem{Ilderton:2019note}
A.~Ilderton, Note on the conjectured breakdown of {QED} perturbation theory in
  strong fields, Physical Review D 99~(8) (2019) 085002.
\newblock \href {http://arxiv.org/abs/1901.00317} {\path{arXiv:1901.00317}},
  \href {https://doi.org/https://doi.org/10.1103/PhysRevD.99.085002}
  {\path{doi:https://doi.org/10.1103/PhysRevD.99.085002}}.

\bibitem{Podszus:2019high}
T.~Podszus, A.~Di~Piazza, High-energy behavior of strong-field {QED} in an
  intense plane wave, Physical Review D 99~(7) (2019) 076004.
\newblock \href {http://arxiv.org/abs/1812.08673} {\path{arXiv:1812.08673}},
  \href {https://doi.org/https://doi.org/10.1103/PhysRevD.99.076004}
  {\path{doi:https://doi.org/10.1103/PhysRevD.99.076004}}.

\bibitem{Adamo:2021jxz}
T.~Adamo, A.~Ilderton, A.~J. MacLeod, {Particle-beam scattering from
  strong-field QED}, Phys. Rev. D 104~(11) (2021) 116013.
\newblock \href {http://arxiv.org/abs/2110.02567} {\path{arXiv:2110.02567}},
  \href {https://doi.org/10.1103/PhysRevD.104.116013}
  {\path{doi:10.1103/PhysRevD.104.116013}}.

\bibitem{Bassompierre:1995search}
G.~Bassompierre, G.~Bologna, D.~Boget, M.~Chevallier, S.~Costa, J.~Dufournaud,
  M.~Farizon, B.~F. Mazuy, A.~Feliciello, M.~J. Gaillard, et~al., Search for
  light neutral objects photoproduced in a crystal strong field and decaying
  into $e^+e^-$ pairs, Physics Letters B 355~(3-4) (1995) 584--594.
\newblock \href {https://doi.org/https://doi.org/10.1016/0370-2693(95)00628-X}
  {\path{doi:https://doi.org/10.1016/0370-2693(95)00628-X}}.

\bibitem{Fedotov:2019qualitative}
A.~M. Fedotov, A.~A. Mironov, Qualitative analysis of quantum-electrodynamic
  processes in a strong field, Radiophysics and Quantum Electronics 61~(12)
  (2019) 930--941.
\newblock \href {https://doi.org/https://doi.org/10.1007/s11141-019-09949-y}
  {\path{doi:https://doi.org/10.1007/s11141-019-09949-y}}.

\bibitem{Veltman:1963unitarity}
M.~Veltman, Unitarity and causality in a renormalizable field theory with
  unstable particles, Physica 29~(3) (1963) 186--207.
\newblock \href {https://doi.org/https://doi.org/10.1016/S0031-8914(63)80277-3}
  {\path{doi:https://doi.org/10.1016/S0031-8914(63)80277-3}}.

\bibitem{BagrovGitman}
V.~G. {Bagrov}, D.~M. {Gitman}, {Exact solutions of relativistic wave equations
  in external electromagnetic fields}, Tomsk State Pedagigical Institute, 1977,
  pp. 5--100.

\bibitem{Bagrov:2014rss}
V.~G. Bagrov, D.~Gitman, {The Dirac Equation and its Solutions}, De Gruyter,
  2014.

\bibitem{MendoncaPRE2011}
J.~T. {Mendon{\c{c}}a}, A.~{Serbeto}, {Volkov solutions for relativistic
  quantum plasmas}, Physical Review E 83~(2) (2011) 026406.
\newblock \href {https://doi.org/10.1103/PhysRevE.83.026406}
  {\path{doi:10.1103/PhysRevE.83.026406}}.

\bibitem{Mackenroth:2018rtp}
F.~Mackenroth, N.~Kumar, N.~Neitz, C.~H. Keitel, {Nonlinear Compton scattering
  of an ultraintense laser pulse in a plasma}, Phys. Rev. E 99~(3) (2019)
  033205.
\newblock \href {http://arxiv.org/abs/1805.01762} {\path{arXiv:1805.01762}},
  \href {https://doi.org/10.1103/PhysRevE.99.033205}
  {\path{doi:10.1103/PhysRevE.99.033205}}.

\bibitem{Heinzl:2016kzb}
T.~Heinzl, A.~Ilderton, B.~King, {Classical and quantum particle dynamics in
  univariate background fields}, Phys. Rev. D 94~(6) (2016) 065039.
\newblock \href {http://arxiv.org/abs/1607.07449} {\path{arXiv:1607.07449}},
  \href {https://doi.org/10.1103/PhysRevD.94.065039}
  {\path{doi:10.1103/PhysRevD.94.065039}}.

\bibitem{1977PhLA...60..137C}
C.~{Cronstr{\"o}m}, M.~{Noga}, {Photon induced relativistic band structure in
  dielectrics}, Physics Letters A 60~(2) (1977) 137--139.
\newblock \href {https://doi.org/10.1016/0375-9601(77)90407-8}
  {\path{doi:10.1016/0375-9601(77)90407-8}}.

\bibitem{Becker:1977}
W.~Becker, Physica 87A (1977) 601--613.

\bibitem{2014LaPhL..11a6001V}
S.~{Varr{\'o}}, {A new class of exact solutions of the Klein-Gordon equation of
  a charged particle interacting with an electromagnetic plane wave in a
  medium}, Laser Physics Letters 11~(1) (2014) 016001.
\newblock \href {http://arxiv.org/abs/1306.0097} {\path{arXiv:1306.0097}},
  \href {https://doi.org/10.1088/1612-2011/11/1/016001}
  {\path{doi:10.1088/1612-2011/11/1/016001}}.

\bibitem{Raicher:2015kaa}
E.~Raicher, S.~Eliezer, A.~Zigler, {A novel solution to the
  Klein\textendash{}Gordon equation in the presence of a strong rotating
  electric field}, Phys. Lett. B 750 (2015) 76--81.
\newblock \href {http://arxiv.org/abs/1502.03558} {\path{arXiv:1502.03558}},
  \href {https://doi.org/10.1016/j.physletb.2015.08.056}
  {\path{doi:10.1016/j.physletb.2015.08.056}}.

\bibitem{Raicher:2016bbx}
E.~Raicher, S.~Eliezer, A.~Zigler, {Nonlinear Compton scattering in a strong
  rotating electric field}, Phys. Rev. A 94~(6) (2016) 062105.
\newblock \href {http://arxiv.org/abs/1606.00476} {\path{arXiv:1606.00476}},
  \href {https://doi.org/10.1103/PhysRevA.94.062105}
  {\path{doi:10.1103/PhysRevA.94.062105}}.

\bibitem{Ekman:2021vwg}
R.~Ekman, T.~Heinzl, A.~Ilderton, {Exact solutions in radiation reaction and
  the radiation-free direction}, New J. Phys. 23~(5) (2021) 055001.
\newblock \href {http://arxiv.org/abs/2102.11843} {\path{arXiv:2102.11843}},
  \href {https://doi.org/10.1088/1367-2630/abfab2}
  {\path{doi:10.1088/1367-2630/abfab2}}.

\bibitem{King:2016oei}
B.~King, H.~Hu, {Classical and quantum dynamics of a charged scalar particle in
  a background of two counterpropagating plane waves}, Phys. Rev. D 94~(12)
  (2016) 125010.
\newblock \href {http://arxiv.org/abs/1609.08105} {\path{arXiv:1609.08105}},
  \href {https://doi.org/10.1103/PhysRevD.94.125010}
  {\path{doi:10.1103/PhysRevD.94.125010}}.

\bibitem{Hu:2015qya}
H.~Hu, J.~Huang, {Analytical solution for Klein-Gordon equation and action
  function of the solution for Dirac equation in counter-propagating laser
  waves}, Phys. Rev. A 92 (2015) 062105.
\newblock \href {http://arxiv.org/abs/1508.07597} {\path{arXiv:1508.07597}},
  \href {https://doi.org/10.1103/PhysRevA.92.062105}
  {\path{doi:10.1103/PhysRevA.92.062105}}.

\bibitem{Pauli:1932}
W.~Pauli, Helv. Phys. Acta 5 (1932) 179.

\bibitem{PhysRev.131.2789}
S.~I. Rubinow, J.~B. Keller,
  \href{https://link.aps.org/doi/10.1103/PhysRev.131.2789}{Asymptotic solution
  of the dirac equation}, Phys. Rev. 131 (1963) 2789--2796.
\newblock \href {https://doi.org/10.1103/PhysRev.131.2789}
  {\path{doi:10.1103/PhysRev.131.2789}}.
\newline\urlprefix\url{https://link.aps.org/doi/10.1103/PhysRev.131.2789}

\bibitem{Rosen1964ATW}
M.~Rosen, A three-dimensional wkb approximation for the dirac equation, Il
  Nuovo Cimento (1955-1965) 33 (1964) 1667--1679.

\bibitem{cherry1948}
T.~M. Cherry, Expansions in terms of parabolic cylinder functions, Proceedings
  of the Edinburgh Mathematical Society 8~(2) (1948) 50–65.
\newblock \href {https://doi.org/10.1017/S0013091500024792}
  {\path{doi:10.1017/S0013091500024792}}.

\bibitem{MillerGood}
S.~C. Miller, R.~H. Good,
  \href{https://link.aps.org/doi/10.1103/PhysRev.91.174}{A wkb-type
  approximation to the schr\"odinger equation}, Phys. Rev. 91 (1953) 174--179.
\newblock \href {https://doi.org/10.1103/PhysRev.91.174}
  {\path{doi:10.1103/PhysRev.91.174}}.
\newline\urlprefix\url{https://link.aps.org/doi/10.1103/PhysRev.91.174}

\bibitem{Dunne:2014bca}
G.~V. Dunne, M.~Unsal, {Uniform WKB, Multi-instantons, and Resurgent
  Trans-Series}, Phys. Rev. D 89~(10) (2014) 105009.
\newblock \href {http://arxiv.org/abs/1401.5202} {\path{arXiv:1401.5202}},
  \href {https://doi.org/10.1103/PhysRevD.89.105009}
  {\path{doi:10.1103/PhysRevD.89.105009}}.

\bibitem{Dunne:2016qix}
G.~V. Dunne, M.~Unsal, {WKB and Resurgence in the Mathieu Equation} (3 2016).
\newblock \href {http://arxiv.org/abs/1603.04924} {\path{arXiv:1603.04924}}.

\bibitem{Kohlfurst:2021skr}
C.~Kohlf\"urst, N.~Ahmadiniaz, J.~Oertel, R.~Sch\"utzhold, {Sauter-Schwinger
  effect for colliding laser pulses} (7 2021).
\newblock \href {http://arxiv.org/abs/2107.08741} {\path{arXiv:2107.08741}}.

\bibitem{Bjorken:1970ah}
J.~D. Bjorken, J.~B. Kogut, D.~E. Soper, {Quantum Electrodynamics at Infinite
  Momentum: Scattering from an External Field}, Phys. Rev. D 3 (1971) 1382.
\newblock \href {https://doi.org/10.1103/PhysRevD.3.1382}
  {\path{doi:10.1103/PhysRevD.3.1382}}.

\bibitem{Blankenbecler:1987rg}
R.~Blankenbecler, S.~D. Drell, {A Quantum Treatment of Beamstrahlung}, Phys.
  Rev. D 36 (1987) 277.
\newblock \href {https://doi.org/10.1103/PhysRevD.36.277}
  {\path{doi:10.1103/PhysRevD.36.277}}.

\bibitem{Blankenbecler:1970qd}
R.~Blankenbecler, R.~L. Sugar, {RELATIVISTIC EIKONAL EXPANSION}, Phys. Rev. D 2
  (1970) 3024.
\newblock \href {https://doi.org/10.1103/PhysRevD.2.3024}
  {\path{doi:10.1103/PhysRevD.2.3024}}.

\bibitem{Sakurai}
J.~J. Sakurai, Modern Quantum Mechanics, Cambridge University Press, 2020.

\bibitem{DiPiazza:2020wxp}
A.~Di~Piazza, {Unveiling the transverse formation length of nonlinear Compton
  scattering}, Phys. Rev. A 103~(1) (2021) 012215.
\newblock \href {http://arxiv.org/abs/2009.00526} {\path{arXiv:2009.00526}},
  \href {https://doi.org/10.1103/PhysRevA.103.012215}
  {\path{doi:10.1103/PhysRevA.103.012215}}.

\bibitem{Formanek:2021bpw}
M.~Formanek, D.~Ramsey, J.~Palastro, A.~Di~Piazza, {Radiation Reaction
  Enhancement in Flying Focus Pulses} (8 2021).
\newblock \href {http://arxiv.org/abs/2108.09651} {\path{arXiv:2108.09651}}.

\bibitem{Raicher:2019flc}
E.~Raicher, K.~Z. Hatsagortsyan, {Nonlinear QED in an ultrastrong rotating
  electric field: Signatures of the momentum-dependent effective mass}, Phys.
  Rev. Res. 2~(1) (2020) 013240.
\newblock \href {http://arxiv.org/abs/1908.05960} {\path{arXiv:1908.05960}},
  \href {https://doi.org/10.1103/PhysRevResearch.2.013240}
  {\path{doi:10.1103/PhysRevResearch.2.013240}}.

\bibitem{Wistisen:2018rlg}
T.~N. Wistisen, A.~Di~Piazza, {Impact of the quantized transverse motion on
  radiation emission in a Dirac harmonic oscillator}, Phys. Rev. A 98~(2)
  (2018) 022131.
\newblock \href {http://arxiv.org/abs/1805.05167} {\path{arXiv:1805.05167}},
  \href {https://doi.org/10.1103/PhysRevA.98.022131}
  {\path{doi:10.1103/PhysRevA.98.022131}}.

\bibitem{DiPiazza:2014uea}
A.~Di~Piazza, A.~I. Milstein, {Ultrarelativistic quasiclassical wave functions
  in strong laser and atomic fields}, Phys. Rev. A 89~(6) (2014) 062114.
\newblock \href {http://arxiv.org/abs/1404.5732} {\path{arXiv:1404.5732}},
  \href {https://doi.org/10.1103/PhysRevA.89.062114}
  {\path{doi:10.1103/PhysRevA.89.062114}}.

\bibitem{Krachkov:2019ovr}
P.~A. Krachkov, A.~Di~Piazza, A.~I. Milstein, {High-energy bremsstrahlung on
  atoms in a laser field}, Phys. Lett. B 797 (2019) 134814.
\newblock \href {http://arxiv.org/abs/1904.05094} {\path{arXiv:1904.05094}},
  \href {https://doi.org/10.1016/j.physletb.2019.134814}
  {\path{doi:10.1016/j.physletb.2019.134814}}.

\bibitem{Wistisen:2019cew}
T.~N. Wistisen, A.~Di~Piazza, {Complete treatment of single-photon emission in
  planar channeling}, Phys. Rev. D 99~(11) (2019) 116010.
\newblock \href {http://arxiv.org/abs/1904.02997} {\path{arXiv:1904.02997}},
  \href {https://doi.org/10.1103/PhysRevD.99.116010}
  {\path{doi:10.1103/PhysRevD.99.116010}}.

\bibitem{Levy1965}
J.-M. Lévy-Leblond, \href{http://eudml.org/doc/75509}{Une nouvelle limite
  non-relativiste du groupe de poincaré}, Annales de l'I.H.P. Physique
  théorique 3~(1) (1965) 1--12.
\newline\urlprefix\url{http://eudml.org/doc/75509}

\bibitem{Duval:2017els}
C.~Duval, G.~W. Gibbons, P.~A. Horvathy, P.~M. Zhang, {Carroll symmetry of
  plane gravitational waves}, Class. Quant. Grav. 34~(17) (2017) 175003.
\newblock \href {http://arxiv.org/abs/1702.08284} {\path{arXiv:1702.08284}},
  \href {https://doi.org/10.1088/1361-6382/aa7f62}
  {\path{doi:10.1088/1361-6382/aa7f62}}.

\bibitem{Zhang:2019gdm}
P.~M. Zhang, M.~Cariglia, M.~Elbistan, P.~A. Horvathy, {Scaling and conformal
  symmetries for plane gravitational waves}, J. Math. Phys. 61~(2) (2020)
  022502.
\newblock \href {http://arxiv.org/abs/1905.08661} {\path{arXiv:1905.08661}},
  \href {https://doi.org/10.1063/1.5136078} {\path{doi:10.1063/1.5136078}}.

\bibitem{Miller:2013gxa}
W.~Miller, Jr., S.~Post, P.~Winternitz, {Classical and Quantum
  Superintegrability with Applications}, J. Phys. A 46 (2013) 423001.
\newblock \href {http://arxiv.org/abs/1309.2694} {\path{arXiv:1309.2694}},
  \href {https://doi.org/10.1088/1751-8113/46/42/423001}
  {\path{doi:10.1088/1751-8113/46/42/423001}}.

\bibitem{Heinzl:2017zsr}
T.~Heinzl, A.~Ilderton, {Exact classical and quantum dynamics in background
  electromagnetic fields}, Phys. Rev. Lett. 118~(11) (2017) 113202.
\newblock \href {http://arxiv.org/abs/1701.09166} {\path{arXiv:1701.09166}},
  \href {https://doi.org/10.1103/PhysRevLett.118.113202}
  {\path{doi:10.1103/PhysRevLett.118.113202}}.

\bibitem{Heinzl:2017blq}
T.~Heinzl, A.~Ilderton, {Superintegrable relativistic systems in
  spacetime-dependent background fields}, J. Phys. A 50~(34) (2017) 345204.
\newblock \href {http://arxiv.org/abs/1701.09168} {\path{arXiv:1701.09168}},
  \href {https://doi.org/10.1088/1751-8121/aa7fa3}
  {\path{doi:10.1088/1751-8121/aa7fa3}}.

\bibitem{Nikitin:2014cfl}
A.~G. Nikitin, T.~M. Zasadko, {Superintegrable systems with position dependent
  mass}, J. Math. Phys. 56~(4) (2015) 042101.
\newblock \href {http://arxiv.org/abs/1406.2006} {\path{arXiv:1406.2006}},
  \href {https://doi.org/10.1063/1.4908107} {\path{doi:10.1063/1.4908107}}.

\bibitem{Ansell:2018dro}
L.~Ansell, T.~Heinzl, A.~Ilderton, {Superintegrable relativistic systems in
  scalar background fields}, J. Phys. A 51~(49) (2018) 495203.
\newblock \href {http://arxiv.org/abs/1805.00375} {\path{arXiv:1805.00375}},
  \href {https://doi.org/10.1088/1751-8121/aae9fb}
  {\path{doi:10.1088/1751-8121/aae9fb}}.

\bibitem{Andrzejewski:2019hub}
K.~Andrzejewski, S.~Prencel, {From polarized gravitational waves to
  analytically solvable electromagnetic beams}, Phys. Rev. D 100~(4) (2019)
  045006.
\newblock \href {http://arxiv.org/abs/1901.05255} {\path{arXiv:1901.05255}},
  \href {https://doi.org/10.1103/PhysRevD.100.045006}
  {\path{doi:10.1103/PhysRevD.100.045006}}.

\bibitem{Elbistan:2020ffe}
M.~Elbistan, N.~Dimakis, K.~Andrzejewski, P.~A. Horvathy, P.~Kos\'\i{}nski,
  P.~M. Zhang, {Conformal symmetries and integrals of the motion in pp waves
  with external electromagnetic fields}, Annals Phys. 418 (2020) 168180.
\newblock \href {http://arxiv.org/abs/2003.07649} {\path{arXiv:2003.07649}},
  \href {https://doi.org/10.1016/j.aop.2020.168180}
  {\path{doi:10.1016/j.aop.2020.168180}}.

\bibitem{Marchesiello2015}
A.~Marchesiello, L.~{\v{S}}nobl, P.~Winternitz, Three-dimensional
  superintegrable systems in a static electromagnetic field, Journal of Physics
  A: Mathematical and Theoretical 48~(39) (2015) 395206.
\newblock \href {https://doi.org/10.1088/1751-8113/48/39/395206}
  {\path{doi:10.1088/1751-8113/48/39/395206}}.

\bibitem{Marchesiello:2018kog}
A.~Marchesiello, L.~\v{S}nobl, P.~Winternitz, {Spherical type integrable
  classical systems in a magnetic field}, J. Phys. A 51~(13) (2018) 135205.
\newblock \href {https://doi.org/10.1088/1751-8121/aaae9b}
  {\path{doi:10.1088/1751-8121/aaae9b}}.

\bibitem{2012PhRvL.109f0401B}
E.~{Barnes}, S.~{Das Sarma}, {Analytically Solvable Driven Time-Dependent
  Two-Level Quantum Systems}, Phys. Rev. Lett. 109~(6) (2012) 060401.
\newblock \href {http://arxiv.org/abs/1206.0297} {\path{arXiv:1206.0297}},
  \href {https://doi.org/10.1103/PhysRevLett.109.060401}
  {\path{doi:10.1103/PhysRevLett.109.060401}}.

\bibitem{Oertel:2015yma}
J.~Oertel, R.~Sch\"utzhold, {Inverse approach to solutions of the Dirac
  equation for space-time dependent fields}, Phys. Rev. D 92~(2) (2015) 025055.
\newblock \href {http://arxiv.org/abs/1503.06140} {\path{arXiv:1503.06140}},
  \href {https://doi.org/10.1103/PhysRevD.92.025055}
  {\path{doi:10.1103/PhysRevD.92.025055}}.

\bibitem{Campos:2017waa}
A.~G. Campos, R.~Cabrera, H.~A. Rabitz, D.~I. Bondar, {Analytic Solutions to
  Coherent Control of the Dirac Equation}, Phys. Rev. Lett. 119~(17) (2017)
  173203, [Erratum: Phys.Rev.Lett. 119, 259903 (2017)].
\newblock \href {http://arxiv.org/abs/1705.02001} {\path{arXiv:1705.02001}},
  \href {https://doi.org/10.1103/PhysRevLett.119.173203}
  {\path{doi:10.1103/PhysRevLett.119.173203}}.

\bibitem{Campos:2020ked}
A.~G. Campos, R.~Cabrera, {Nondispersive analytical solutions to the Dirac
  equation}, Phys. Rev. Res. 2~(1) (2020) 013051.
\newblock \href {http://arxiv.org/abs/1911.00333} {\path{arXiv:1911.00333}},
  \href {https://doi.org/10.1103/PhysRevResearch.2.013051}
  {\path{doi:10.1103/PhysRevResearch.2.013051}}.

\bibitem{Campos:2020cza}
A.~G. Campos, K.~Z. Hatsagortsyan, C.~H. Keitel, {Construction of Dirac spinors
  for electron vortex beams in background electromagnetic fields}, Phys. Rev.
  Res. 3~(1) (2021) 013245.
\newblock \href {http://arxiv.org/abs/2011.14964} {\path{arXiv:2011.14964}},
  \href {https://doi.org/10.1103/PhysRevResearch.3.013245}
  {\path{doi:10.1103/PhysRevResearch.3.013245}}.

\bibitem{Adorno:2015ibo}
T.~C. Adorno, S.~P. Gavrilov, D.~M. Gitman, {Exactly solvable cases in QED with
  t-electric potential steps}, Int. J. Mod. Phys. A 32~(18) (2017) 1750105.
\newblock \href {http://arxiv.org/abs/1512.01288} {\path{arXiv:1512.01288}},
  \href {https://doi.org/10.1142/S0217751X17501056}
  {\path{doi:10.1142/S0217751X17501056}}.

\bibitem{Adorno:2016bjx}
T.~C. Adorno, S.~P. Gavrilov, D.~M. Gitman, {Particle creation by peak electric
  field}, Eur. Phys. J. C 76~(8) (2016) 447.
\newblock \href {http://arxiv.org/abs/1605.09072} {\path{arXiv:1605.09072}},
  \href {https://doi.org/10.1140/epjc/s10052-016-4289-0}
  {\path{doi:10.1140/epjc/s10052-016-4289-0}}.

\bibitem{Adorno:2019ffg}
T.~C. Adorno, S.~P. Gavrilov, D.~M. Gitman, {Vacuum instability in a constant
  inhomogeneous electric field. A new example of exact nonperturbative
  calculations}, Eur. Phys. J. C 80~(2) (2020) 88.
\newblock \href {http://arxiv.org/abs/1911.09809} {\path{arXiv:1911.09809}},
  \href {https://doi.org/10.1140/epjc/s10052-020-7646-y}
  {\path{doi:10.1140/epjc/s10052-020-7646-y}}.

\bibitem{Adorno:2018jbp}
T.~C. Adorno, S.~P. Gavrilov, D.~M. Gitman, {Violation of vacuum stability by
  inverse square electric fields}, Eur. Phys. J. C 78~(12) (2018) 1021.
\newblock \href {http://arxiv.org/abs/1807.01642} {\path{arXiv:1807.01642}},
  \href {https://doi.org/10.1140/epjc/s10052-018-6499-0}
  {\path{doi:10.1140/epjc/s10052-018-6499-0}}.

\bibitem{Breev:2021lpn}
A.~I. Breev, S.~P. Gavrilov, D.~M. Gitman, A.~A. Shishmarev, {Vacuum
  instability in time-dependent electric fields: New example of an exactly
  solvable case}, Phys. Rev. D 104~(7) (2021) 076008.
\newblock \href {http://arxiv.org/abs/2106.06322} {\path{arXiv:2106.06322}},
  \href {https://doi.org/10.1103/PhysRevD.104.076008}
  {\path{doi:10.1103/PhysRevD.104.076008}}.

\bibitem{Jackiw:1991ck}
R.~Jackiw, D.~N. Kabat, M.~Ortiz, {Electromagnetic fields of a massless
  particle and the eikonal}, Phys. Lett. B 277 (1992) 148--152.
\newblock \href {http://arxiv.org/abs/hep-th/9112020}
  {\path{arXiv:hep-th/9112020}}, \href
  {https://doi.org/10.1016/0370-2693(92)90971-6}
  {\path{doi:10.1016/0370-2693(92)90971-6}}.

\bibitem{Aichelburg:1970dh}
P.~C. Aichelburg, R.~U. Sexl, {On the Gravitational field of a massless
  particle}, Gen. Rel. Grav. 2 (1971) 303--312.
\newblock \href {https://doi.org/10.1007/BF00758149}
  {\path{doi:10.1007/BF00758149}}.

\bibitem{tHooft:1987vrq}
G.~'t~Hooft, {Graviton Dominance in Ultrahigh-Energy Scattering}, Phys. Lett. B
  198 (1987) 61--63.
\newblock \href {https://doi.org/10.1016/0370-2693(87)90159-6}
  {\path{doi:10.1016/0370-2693(87)90159-6}}.

\bibitem{Caron-Huot:2013fea}
S.~Caron-Huot, {When does the gluon reggeize?}, JHEP 05 (2015) 093.
\newblock \href {http://arxiv.org/abs/1309.6521} {\path{arXiv:1309.6521}},
  \href {https://doi.org/10.1007/JHEP05(2015)093}
  {\path{doi:10.1007/JHEP05(2015)093}}.

\bibitem{Lodone:2009qe}
P.~Lodone, V.~S. Rychkov, {Radiation Problem in Transplanckian Scattering},
  JHEP 12 (2009) 036.
\newblock \href {http://arxiv.org/abs/0909.3519} {\path{arXiv:0909.3519}},
  \href {https://doi.org/10.1088/1126-6708/2009/12/036}
  {\path{doi:10.1088/1126-6708/2009/12/036}}.

\bibitem{Hollowood:2015elj}
T.~J. Hollowood, G.~M. Shore, {Causality Violation, Gravitational Shockwaves
  and UV Completion}, JHEP 03 (2016) 129.
\newblock \href {http://arxiv.org/abs/1512.04952} {\path{arXiv:1512.04952}},
  \href {https://doi.org/10.1007/JHEP03(2016)129}
  {\path{doi:10.1007/JHEP03(2016)129}}.

\bibitem{Bonnor:1969gr}
W.~B. Bonnor, {The gravitational field of light}, Comm. Math. Phys. 13 (1969)
  163--174.
\newblock \href {https://doi.org/10.1007/BF01645484}
  {\path{doi:10.1007/BF01645484}}.

\bibitem{Bonnor:1969rb}
W.~B. Bonnor, {Solutions of Maxwell's equations for charge moving with the
  speed of light}, Int. J. Theor. Phys. 2 (1969) 373--379.
\newblock \href {https://doi.org/10.1007/BF00670703}
  {\path{doi:10.1007/BF00670703}}.

\bibitem{DelGaudio:2018lfm}
F.~Del~Gaudio, T.~Grismayer, R.~A. Fonseca, W.~B. Mori, L.~O. Silva, {Bright
  $\gamma$ rays source and copious nonlinear Breit-Wheeler pairs in the
  collision of high density particle beams}, Phys. Rev. Accel. Beams 22~(2)
  (2019) 023402.
\newblock \href {http://arxiv.org/abs/1807.06968} {\path{arXiv:1807.06968}},
  \href {https://doi.org/10.1103/PhysRevAccelBeams.22.023402}
  {\path{doi:10.1103/PhysRevAccelBeams.22.023402}}.

\bibitem{Fedotov:2013uja}
A.~M. Fedotov, A.~A. Mironov, {Pair creation by collision of an intense laser
  pulse with a high-frequency photon beam}, Phys. Rev. A 88~(6) (2013) 062110.
\newblock \href {http://arxiv.org/abs/1310.7258} {\path{arXiv:1310.7258}},
  \href {https://doi.org/10.1103/PhysRevA.88.062110}
  {\path{doi:10.1103/PhysRevA.88.062110}}.

\bibitem{Gavrilov:2008fv}
S.~P. Gavrilov, D.~M. Gitman, {Consistency restrictions on maximal electric
  field strength in QFT}, Phys. Rev. Lett. 101 (2008) 130403.
\newblock \href {http://arxiv.org/abs/0805.2391} {\path{arXiv:0805.2391}},
  \href {https://doi.org/10.1103/PhysRevLett.101.130403}
  {\path{doi:10.1103/PhysRevLett.101.130403}}.

\bibitem{Yang:2017xyh}
I.-S. Yang, {Secret Loss of Unitarity due to the Classical Background}, Phys.
  Rev. D 96~(2) (2017) 025005.
\newblock \href {http://arxiv.org/abs/1703.03466} {\path{arXiv:1703.03466}},
  \href {https://doi.org/10.1103/PhysRevD.96.025005}
  {\path{doi:10.1103/PhysRevD.96.025005}}.

\bibitem{Glendenning:1983qq}
N.~K. Glendenning, T.~Matsui, {CREATION OF ANTI-Q Q PAIR IN A CHROMOELECTRIC
  FLUX TUBE}, Phys. Rev. D 28 (1983) 2890--2891.
\newblock \href {https://doi.org/10.1103/PhysRevD.28.2890}
  {\path{doi:10.1103/PhysRevD.28.2890}}.

\bibitem{Kluger:1992gb}
Y.~Kluger, J.~M. Eisenberg, B.~Svetitsky, F.~Cooper, E.~Mottola, {Fermion pair
  production in a strong electric field}, Phys. Rev. D 45 (1992) 4659--4671.
\newblock \href {https://doi.org/10.1103/PhysRevD.45.4659}
  {\path{doi:10.1103/PhysRevD.45.4659}}.

\bibitem{Bloch:1999eu}
J.~C.~R. Bloch, V.~A. Mizerny, A.~V. Prozorkevich, C.~D. Roberts, S.~M.
  Schmidt, S.~A. Smolyansky, D.~V. Vinnik, {Pair creation: Back reactions and
  damping}, Phys. Rev. D 60 (1999) 116011.
\newblock \href {http://arxiv.org/abs/nucl-th/9907027}
  {\path{arXiv:nucl-th/9907027}}, \href
  {https://doi.org/10.1103/PhysRevD.60.116011}
  {\path{doi:10.1103/PhysRevD.60.116011}}.

\bibitem{Otto:2018hya}
A.~Otto, D.~Graeveling, B.~Kämpfer, {Response of the QED(2) Vacuum to a
  Quench: Long-term Oscillations of the Electric Field and the Pair Creation
  Rate}, Plasma Phys. Control. Fusion 61~(7) (2019) 074002.
\newblock \href {http://arxiv.org/abs/1812.10832} {\path{arXiv:1812.10832}},
  \href {https://doi.org/10.1088/1361-6587/ab1a21}
  {\path{doi:10.1088/1361-6587/ab1a21}}.

\bibitem{Aarts:1998td}
G.~Aarts, J.~Smit, {Real time dynamics with fermions on a lattice}, Nucl. Phys.
  B 555 (1999) 355--394.
\newblock \href {http://arxiv.org/abs/hep-ph/9812413}
  {\path{arXiv:hep-ph/9812413}}, \href
  {https://doi.org/10.1016/S0550-3213(99)00320-X}
  {\path{doi:10.1016/S0550-3213(99)00320-X}}.

\bibitem{Hebenstreit:2013qxa}
F.~Hebenstreit, J.~Berges, D.~Gelfand, {Simulating fermion production in 1+1
  dimensional QED}, Phys. Rev. D 87~(10) (2013) 105006.
\newblock \href {http://arxiv.org/abs/1302.5537} {\path{arXiv:1302.5537}},
  \href {https://doi.org/10.1103/PhysRevD.87.105006}
  {\path{doi:10.1103/PhysRevD.87.105006}}.

\bibitem{Hebenstreit:2013baa}
F.~Hebenstreit, J.~Berges, D.~Gelfand, {Real-time dynamics of string breaking},
  Phys. Rev. Lett. 111 (2013) 201601.
\newblock \href {http://arxiv.org/abs/1307.4619} {\path{arXiv:1307.4619}},
  \href {https://doi.org/10.1103/PhysRevLett.111.201601}
  {\path{doi:10.1103/PhysRevLett.111.201601}}.

\bibitem{Gelis:2013oca}
F.~Gelis, N.~Tanji, {Formulation of the Schwinger mechanism in classical
  statistical field theory}, Phys. Rev. D 87~(12) (2013) 125035.
\newblock \href {http://arxiv.org/abs/1303.4633} {\path{arXiv:1303.4633}},
  \href {https://doi.org/10.1103/PhysRevD.87.125035}
  {\path{doi:10.1103/PhysRevD.87.125035}}.

\bibitem{Kasper:2014uaa}
V.~Kasper, F.~Hebenstreit, J.~Berges, {Fermion production from real-time
  lattice gauge theory in the classical-statistical regime}, Phys. Rev. D
  90~(2) (2014) 025016.
\newblock \href {http://arxiv.org/abs/1403.4849} {\path{arXiv:1403.4849}},
  \href {https://doi.org/10.1103/PhysRevD.90.025016}
  {\path{doi:10.1103/PhysRevD.90.025016}}.

\bibitem{Gelis:2015eua}
F.~Gelis, N.~Tanji, {Quark production in heavy ion collisions: formalism and
  boost invariant fermionic light-cone mode functions}, JHEP 02 (2016) 126.
\newblock \href {http://arxiv.org/abs/1506.03327} {\path{arXiv:1506.03327}},
  \href {https://doi.org/10.1007/JHEP02(2016)126}
  {\path{doi:10.1007/JHEP02(2016)126}}.

\bibitem{Mueller:2016aao}
N.~Mueller, F.~Hebenstreit, J.~Berges, {Anomaly-induced dynamical refringence
  in strong-field QED}, Phys. Rev. Lett. 117~(6) (2016) 061601.
\newblock \href {http://arxiv.org/abs/1605.01413} {\path{arXiv:1605.01413}},
  \href {https://doi.org/10.1103/PhysRevLett.117.061601}
  {\path{doi:10.1103/PhysRevLett.117.061601}}.

\bibitem{Buyens:2016hhu}
B.~Buyens, J.~Haegeman, F.~Hebenstreit, F.~Verstraete, K.~Van~Acoleyen,
  {Real-time simulation of the Schwinger effect with Matrix Product States},
  Phys. Rev. D 96~(11) (2017) 114501.
\newblock \href {http://arxiv.org/abs/1612.00739} {\path{arXiv:1612.00739}},
  \href {https://doi.org/10.1103/PhysRevD.96.114501}
  {\path{doi:10.1103/PhysRevD.96.114501}}.

\bibitem{Gelfand:2016prm}
D.~Gelfand, F.~Hebenstreit, J.~Berges, {Early quark production and approach to
  chemical equilibrium}, Phys. Rev. D 93~(8) (2016) 085001.
\newblock \href {http://arxiv.org/abs/1601.03576} {\path{arXiv:1601.03576}},
  \href {https://doi.org/10.1103/PhysRevD.93.085001}
  {\path{doi:10.1103/PhysRevD.93.085001}}.

\bibitem{Spitz:2018eps}
D.~Spitz, J.~Berges, {Schwinger pair production and string breaking in
  non-Abelian gauge theory from real-time lattice improved Hamiltonians}, Phys.
  Rev. D 99~(3) (2019) 036020.
\newblock \href {http://arxiv.org/abs/1812.05835} {\path{arXiv:1812.05835}},
  \href {https://doi.org/10.1103/PhysRevD.99.036020}
  {\path{doi:10.1103/PhysRevD.99.036020}}.

\bibitem{Shi:2018sxm}
Y.~Shi, J.~Xiao, H.~Qin, N.~J. Fisch, {Simulations of relativistic-quantum
  plasmas using real-time lattice scalar QED}, Phys. Rev. E 97~(5) (2018)
  053206.
\newblock \href {http://arxiv.org/abs/1802.00524} {\path{arXiv:1802.00524}},
  \href {https://doi.org/10.1103/PhysRevE.97.053206}
  {\path{doi:10.1103/PhysRevE.97.053206}}.

\bibitem{Mace:2019cqo}
M.~Mace, N.~Mueller, S.~Schlichting, S.~Sharma, {Chiral Instabilities and the
  Onset of Chiral Turbulence in QED Plasmas}, Phys. Rev. Lett. 124~(19) (2020)
  191604.
\newblock \href {http://arxiv.org/abs/1910.01654} {\path{arXiv:1910.01654}},
  \href {https://doi.org/10.1103/PhysRevLett.124.191604}
  {\path{doi:10.1103/PhysRevLett.124.191604}}.

\bibitem{Berges:2020fwq}
J.~Berges, M.~P. Heller, A.~Mazeliauskas, R.~Venugopalan, {QCD thermalization:
  Ab initio approaches and interdisciplinary connections}, Rev. Mod. Phys.
  93~(3) (2021) 035003.
\newblock \href {http://arxiv.org/abs/2005.12299} {\path{arXiv:2005.12299}},
  \href {https://doi.org/10.1103/RevModPhys.93.035003}
  {\path{doi:10.1103/RevModPhys.93.035003}}.

\bibitem{Gold:2020qzr}
G.~Gold, D.~A. Mcgady, S.~P. Patil, V.~Vardanyan, {Backreaction of Schwinger
  pair creation in massive QED$_{2}$}, JHEP 10 (2021) 072.
\newblock \href {http://arxiv.org/abs/2012.15824} {\path{arXiv:2012.15824}},
  \href {https://doi.org/10.1007/JHEP10(2021)072}
  {\path{doi:10.1007/JHEP10(2021)072}}.

\bibitem{Chu:2010xc}
Y.-Z. Chu, T.~Vachaspati, {Capacitor Discharge and Vacuum Resistance in
  Massless QED$_{2}$}, Phys. Rev. D 81 (2010) 085020.
\newblock \href {http://arxiv.org/abs/1001.2559} {\path{arXiv:1001.2559}},
  \href {https://doi.org/10.1103/PhysRevD.81.085020}
  {\path{doi:10.1103/PhysRevD.81.085020}}.

\bibitem{Kluger:1991ib}
Y.~Kluger, J.~M. Eisenberg, B.~Svetitsky, F.~Cooper, E.~Mottola, {Pair
  production in a strong electric field}, Phys. Rev. Lett. 67 (1991)
  2427--2430.
\newblock \href {https://doi.org/10.1103/PhysRevLett.67.2427}
  {\path{doi:10.1103/PhysRevLett.67.2427}}.

\bibitem{Ruffini:2003cr}
R.~Ruffini, L.~Vitagliano, S.~S. Xue, {On plasma oscillations in strong
  electric fields}, Phys. Lett. B 559 (2003) 12--19.
\newblock \href {http://arxiv.org/abs/astro-ph/0302549}
  {\path{arXiv:astro-ph/0302549}}, \href
  {https://doi.org/10.1016/S0370-2693(03)00299-5}
  {\path{doi:10.1016/S0370-2693(03)00299-5}}.

\bibitem{Luttinger:1960ua}
J.~M. Luttinger, J.~C. Ward, {Ground state energy of a many fermion system.
  2.}, Phys. Rev. 118 (1960) 1417--1427.
\newblock \href {https://doi.org/10.1103/PhysRev.118.1417}
  {\path{doi:10.1103/PhysRev.118.1417}}.

\bibitem{Cornwall:1974vz}
J.~M. Cornwall, R.~Jackiw, E.~Tomboulis, {Effective Action for Composite
  Operators}, Phys. Rev. D 10 (1974) 2428--2445.
\newblock \href {https://doi.org/10.1103/PhysRevD.10.2428}
  {\path{doi:10.1103/PhysRevD.10.2428}}.

\bibitem{Berges:2004yj}
J.~Berges, {Introduction to nonequilibrium quantum field theory}, AIP Conf.
  Proc. 739~(1) (2004) 3--62.
\newblock \href {http://arxiv.org/abs/hep-ph/0409233}
  {\path{arXiv:hep-ph/0409233}}, \href {https://doi.org/10.1063/1.1843591}
  {\path{doi:10.1063/1.1843591}}.

\bibitem{Arrizabalaga:2002hn}
A.~Arrizabalaga, J.~Smit, {Gauge fixing dependence of Phi derivable
  approximations}, Phys. Rev. D 66 (2002) 065014.
\newblock \href {http://arxiv.org/abs/hep-ph/0207044}
  {\path{arXiv:hep-ph/0207044}}, \href
  {https://doi.org/10.1103/PhysRevD.66.065014}
  {\path{doi:10.1103/PhysRevD.66.065014}}.

\bibitem{Carrington:2007fp}
M.~E. Carrington, E.~Kovalchuk, {QED electrical conductivity using the 2PI
  effective action}, Phys. Rev. D 76 (2007) 045019.
\newblock \href {http://arxiv.org/abs/0705.0162} {\path{arXiv:0705.0162}},
  \href {https://doi.org/10.1103/PhysRevD.76.045019}
  {\path{doi:10.1103/PhysRevD.76.045019}}.

\bibitem{Nishiyama:2010mn}
A.~Nishiyama, A.~Ohnishi, {Entropy Production in Gluodynamics in temporal axial
  gauge in 2+1 dimensions}, Prog. Theor. Phys. 125 (2011) 775--793.
\newblock \href {http://arxiv.org/abs/1011.4750} {\path{arXiv:1011.4750}},
  \href {https://doi.org/10.1143/PTP.125.775} {\path{doi:10.1143/PTP.125.775}}.

\bibitem{Zoller:2013lfa}
T.~Z\"oller, {Nonequilibrium Formulation of Abelian Gauge Theories}, Ph.D.
  thesis, Darmstadt, Tech. Hochsch. (2013).

\bibitem{Magnusson:2019qop}
J.~Magnusson, et~al., {Multiple-colliding laser pulses as a basis for studying
  high-field high-energy physics}, Phys. Rev. A 100~(6) (2019) 063404.
\newblock \href {http://arxiv.org/abs/1906.05235} {\path{arXiv:1906.05235}},
  \href {https://doi.org/10.1103/PhysRevA.100.063404}
  {\path{doi:10.1103/PhysRevA.100.063404}}.

\bibitem{Jirka:2017cvr}
M.~Jirka, O.~Klimo, M.~Vranic, S.~Weber, G.~Korn, {QED cascade with 10 PW-class
  lasers}, Sci. Rep. 7~(1) (2017) 15302.
\newblock \href {https://doi.org/10.1038/s41598-017-15747-1}
  {\path{doi:10.1038/s41598-017-15747-1}}.

\bibitem{Slade-Lowther:2018kgv}
C.~Slade-Lowther, D.~Del~Sorbo, C.~P. Ridgers, {Identifying the
  electron\textendash{}positron cascade regimes in high-intensity laser-matter
  interactions}, New J. Phys. 21~(1) (2019) 013028.
\newblock \href {http://arxiv.org/abs/1810.04013} {\path{arXiv:1810.04013}},
  \href {https://doi.org/10.1088/1367-2630/aafa39}
  {\path{doi:10.1088/1367-2630/aafa39}}.

\bibitem{Gonoskov:2014mda}
A.~Gonoskov, S.~Bastrakov, E.~Efimenko, A.~Ilderton, M.~Marklund, I.~Meyerov,
  A.~Muraviev, A.~Sergeev, I.~Surmin, E.~Wallin, {Extended particle-in-cell
  schemes for physics in ultrastrong laser fields: Review and developments},
  Phys. Rev. E 92~(2) (2015) 023305.
\newblock \href {http://arxiv.org/abs/1412.6426} {\path{arXiv:1412.6426}},
  \href {https://doi.org/10.1103/PhysRevE.92.023305}
  {\path{doi:10.1103/PhysRevE.92.023305}}.

\bibitem{Seipt:2013hda}
D.~Seipt, B.~Kämpfer, {Laser assisted Compton scattering of X-ray photons},
  Phys. Rev. A 89~(2) (2014) 023433.
\newblock \href {http://arxiv.org/abs/1309.2092} {\path{arXiv:1309.2092}},
  \href {https://doi.org/10.1103/PhysRevA.89.023433}
  {\path{doi:10.1103/PhysRevA.89.023433}}.

\bibitem{Golub:2020kkc}
A.~Golub, S.~Villalba-Ch\'avez, H.~Ruhl, C.~M\"uller, {Linear Breit-Wheeler
  pair production by high-energy bremsstrahlung photons colliding with an
  intense X-ray laser pulse}, Phys. Rev. D 103~(1) (2021) 016009.
\newblock \href {http://arxiv.org/abs/2010.03871} {\path{arXiv:2010.03871}},
  \href {https://doi.org/10.1103/PhysRevD.103.016009}
  {\path{doi:10.1103/PhysRevD.103.016009}}.

\bibitem{HernandezAcosta:2019vok}
U.~Hernandez~Acosta, B.~K\"ampfer, {Laser pulse-length effects in trident pair
  production}, Plasma Phys. Control. Fusion 61~(8) (2019) 084011.
\newblock \href {http://arxiv.org/abs/1901.08860} {\path{arXiv:1901.08860}},
  \href {https://doi.org/10.1088/1361-6587/ab2b1e}
  {\path{doi:10.1088/1361-6587/ab2b1e}}.

\bibitem{Kosower:2018adc}
D.~A. Kosower, B.~Maybee, D.~O'Connell, {Amplitudes, Observables, and Classical
  Scattering}, JHEP 02 (2019) 137.
\newblock \href {http://arxiv.org/abs/1811.10950} {\path{arXiv:1811.10950}},
  \href {https://doi.org/10.1007/JHEP02(2019)137}
  {\path{doi:10.1007/JHEP02(2019)137}}.

\bibitem{Cristofoli:2021vyo}
A.~Cristofoli, R.~Gonzo, D.~A. Kosower, D.~O'Connell, {Waveforms from
  Amplitudes} (7 2021).
\newblock \href {http://arxiv.org/abs/2107.10193} {\path{arXiv:2107.10193}}.

\bibitem{Berson1969}
I.~Berson, Sov. Phys. JETP 29 (1969) 871.

\bibitem{Bergou:1980cp}
J.~Bergou, S.~Varro, {Nonlinear Scattering Processes in the Presence of a
  Quantized Radiation Field. 2. Relativistic Treatment}, J. Phys. A 14 (1981)
  2281--2303.
\newblock \href {https://doi.org/10.1088/0305-4470/14/9/023}
  {\path{doi:10.1088/0305-4470/14/9/023}}.

\bibitem{Heinzl:2018xnv}
T.~Heinzl, A.~Ilderton, D.~Seipt, {Mode truncations and scattering in strong
  fields}, Phys. Rev. D 98~(1) (2018) 016002.
\newblock \href {http://arxiv.org/abs/1802.05933} {\path{arXiv:1802.05933}},
  \href {https://doi.org/10.1103/PhysRevD.98.016002}
  {\path{doi:10.1103/PhysRevD.98.016002}}.

\bibitem{Lavelle:1995ty}
M.~Lavelle, D.~McMullan, {Constituent quarks from QCD}, Phys. Rept. 279 (1997)
  1--65.
\newblock \href {http://arxiv.org/abs/hep-ph/9509344}
  {\path{arXiv:hep-ph/9509344}}, \href
  {https://doi.org/10.1016/S0370-1573(96)00019-1}
  {\path{doi:10.1016/S0370-1573(96)00019-1}}.

\bibitem{Mati:2017zyc}
P.~Mati, {Quasiparticles in an interacting system of charge and monochromatic
  field}, Phys. Rev. A 95~(5) (2017) 053852.
\newblock \href {http://arxiv.org/abs/1611.02255} {\path{arXiv:1611.02255}},
  \href {https://doi.org/10.1103/PhysRevA.95.053852}
  {\path{doi:10.1103/PhysRevA.95.053852}}.

\bibitem{Endlich:2016jgc}
S.~Endlich, R.~Penco, {A Modern Approach to Superradiance}, JHEP 05 (2017) 052.
\newblock \href {http://arxiv.org/abs/1609.06723} {\path{arXiv:1609.06723}},
  \href {https://doi.org/10.1007/JHEP05(2017)052}
  {\path{doi:10.1007/JHEP05(2017)052}}.

\bibitem{Ilderton:2017xbj}
A.~Ilderton, D.~Seipt, {Backreaction on background fields: A coherent state
  approach}, Phys. Rev. D 97~(1) (2018) 016007.
\newblock \href {http://arxiv.org/abs/1709.10085} {\path{arXiv:1709.10085}},
  \href {https://doi.org/10.1103/PhysRevD.97.016007}
  {\path{doi:10.1103/PhysRevD.97.016007}}.

\bibitem{Kasper:2020akk}
V.~Kasper, G.~Juzeliunas, M.~Lewenstein, F.~Jendrzejewski, E.~Zohar, {From the
  Jaynes\textendash{}Cummings model to non-abelian gauge theories: a guided
  tour for the quantum engineer}, New J. Phys. 22~(10) (2020) 103027.
\newblock \href {http://arxiv.org/abs/2006.01258} {\path{arXiv:2006.01258}},
  \href {https://doi.org/10.1088/1367-2630/abb961}
  {\path{doi:10.1088/1367-2630/abb961}}.

\bibitem{Ekman:2020vsc}
R.~Ekman, A.~Ilderton, {Backreaction in strong field QED: A toy model}, Phys.
  Rev. D 101~(5) (2020) 056022.
\newblock \href {http://arxiv.org/abs/2002.03759} {\path{arXiv:2002.03759}},
  \href {https://doi.org/10.1103/PhysRevD.101.056022}
  {\path{doi:10.1103/PhysRevD.101.056022}}.

\bibitem{ritus-jetp69}
V.~I. Ritus, \href{http://jetp.ras.ru/cgi-bin/dn/e_029_03_0532.pdf}{Effect of
  an electromagnetic field on decays of elementary particles}, Sov. Phys. JETP
  29~(3) (1969) 532.
\newline\urlprefix\url{http://jetp.ras.ru/cgi-bin/dn/e_029_03_0532.pdf}

\bibitem{Becker83}
W.~Becker, G.~Moore, R.~Schlicher, M.~Scully,
  \href{https://www.sciencedirect.com/science/article/pii/0375960183903663}{A
  note on total cross sections and decay rates in the presence of a laser
  field}, Physics Letters A 94~(3) (1983) 131--134.
\newblock \href {https://doi.org/https://doi.org/10.1016/0375-9601(83)90366-3}
  {\path{doi:https://doi.org/10.1016/0375-9601(83)90366-3}}.
\newline\urlprefix\url{https://www.sciencedirect.com/science/article/pii/0375960183903663}

\bibitem{Nikishov83}
A.~Nikishov, V.~Ritus, {Effect of laser field on beta decays of nuclei}, Sov.
  Phys. JETP 58 (1983) 14.

\bibitem{Akhmedov:1983gts}
E.~K. Akhmedov, {BETA DECAY IN THE FIELD OF AN INTENSE ELECTROMAGNETIC WAVE},
  Sov. Phys. JETP 58 (1983) 883--889.

\bibitem{Akhmedov:2010ee}
E.~K. Akhmedov, {Beta decay and other processes in strong electromagnetic
  fields}, Phys. Atom. Nucl. 74 (2011) 1299--1315.
\newblock \href {http://arxiv.org/abs/1011.3776} {\path{arXiv:1011.3776}},
  \href {https://doi.org/10.1134/S1063778811080035}
  {\path{doi:10.1134/S1063778811080035}}.

\bibitem{Wistisen:2020czu}
T.~N. Wistisen, C.~H. Keitel, A.~Di~Piazza, {Transmutation of protons in a
  strong electromagnetic field}, New J. Phys. 23~(6) (2021) 065007.
\newblock \href {http://arxiv.org/abs/2011.08031} {\path{arXiv:2011.08031}},
  \href {https://doi.org/10.1088/1367-2630/abf705}
  {\path{doi:10.1088/1367-2630/abf705}}.

\bibitem{Cortes:2012tu}
H.~M.~C. Cortes, C.~Muller, C.~H. Keitel, A.~Palffy, {Nuclear recollisions in
  laser-assisted $\alpha$ decay}, Phys. Lett. B 723 (2013) 401--405.
\newblock \href {http://arxiv.org/abs/1207.2395} {\path{arXiv:1207.2395}},
  \href {https://doi.org/10.1016/j.physletb.2013.05.025}
  {\path{doi:10.1016/j.physletb.2013.05.025}}.

\bibitem{Qi:2018fwx}
J.~Qi, T.~Li, R.~Xu, L.~Fu, X.~Wang, {\ensuremath{\alpha} decay in intense
  laser fields: Calculations using realistic nuclear potentials}, Phys. Rev. C
  99~(4) (2019) 044610.
\newblock \href {http://arxiv.org/abs/1810.07331} {\path{arXiv:1810.07331}},
  \href {https://doi.org/10.1103/PhysRevC.99.044610}
  {\path{doi:10.1103/PhysRevC.99.044610}}.

\bibitem{Qi:2020yhj}
J.~Qi, L.~Fu, X.~Wang, {Nuclear fission in intense laser fields}, Phys. Rev. C
  102~(6) (2020) 064629.
\newblock \href {http://arxiv.org/abs/2008.03498} {\path{arXiv:2008.03498}},
  \href {https://doi.org/10.1103/PhysRevC.102.064629}
  {\path{doi:10.1103/PhysRevC.102.064629}}.

\bibitem{Palffy:2019scn}
A.~P\'alffy, S.~V. Popruzhenko, {Can Extreme Electromagnetic Fields Accelerate
  the $\alpha$ Decay of Nuclei?}, Phys. Rev. Lett. 124~(21) (2020) 212505.
\newblock \href {http://arxiv.org/abs/1909.07826} {\path{arXiv:1909.07826}},
  \href {https://doi.org/10.1103/PhysRevLett.124.212505}
  {\path{doi:10.1103/PhysRevLett.124.212505}}.

\bibitem{Queisser:2019nuh}
F.~Queisser, R.~Sch\"utzhold, {Dynamically assisted nuclear fusion}, Phys. Rev.
  C 100~(4) (2019) 041601.
\newblock \href {http://arxiv.org/abs/1902.04905} {\path{arXiv:1902.04905}},
  \href {https://doi.org/10.1103/PhysRevC.100.041601}
  {\path{doi:10.1103/PhysRevC.100.041601}}.

\bibitem{Queisser:2020mns}
F.~Queisser, R.~Sch\"utzhold, {Comment on ''Enhanced deuterium-tritium fusion
  cross sections in the presence of strong electromagnetic fields''} (3 2020).
\newblock \href {http://arxiv.org/abs/2003.02661} {\path{arXiv:2003.02661}}.

\bibitem{Wang:2020woc}
X.~Wang, {Substantially enhanced deuteron-triton fusion probabilities in
  intense low-frequency laser fields}, Phys. Rev. C 102~(1) (2020) 011601.
\newblock \href {http://arxiv.org/abs/2006.09634} {\path{arXiv:2006.09634}},
  \href {https://doi.org/10.1103/PhysRevC.102.011601}
  {\path{doi:10.1103/PhysRevC.102.011601}}.

\bibitem{Kohlfurst:2021dfk}
C.~Kohlf\"urst, F.~Queisser, R.~Sch\"utzhold, {Dynamically assisted tunneling
  in the impulse regime}, Phys. Rev. Res. 3~(3) (2021) 033153.
\newblock \href {http://arxiv.org/abs/2102.07474} {\path{arXiv:2102.07474}},
  \href {https://doi.org/10.1103/PhysRevResearch.3.033153}
  {\path{doi:10.1103/PhysRevResearch.3.033153}}.

\bibitem{Muller:2013xaa}
S.~J. M\"uller, C.~H. Keitel, C.~M\"uller, {Higgs Boson Creation in
  Laser-Boosted Lepton Collisions}, Phys. Lett. B 730 (2014) 161--165.
\newblock \href {http://arxiv.org/abs/1307.6751} {\path{arXiv:1307.6751}},
  \href {https://doi.org/10.1016/j.physletb.2014.01.047}
  {\path{doi:10.1016/j.physletb.2014.01.047}}.

\bibitem{Muller:2014iza}
S.~J. M\"uller, C.~H. Keitel, C.~M\"uller, {Particle Production Reactions in
  Laser-Boosted Lepton Collisions}, Phys. Rev. D 90~(9) (2014) 094008.
\newblock \href {http://arxiv.org/abs/1408.2991} {\path{arXiv:1408.2991}},
  \href {https://doi.org/10.1103/PhysRevD.90.094008}
  {\path{doi:10.1103/PhysRevD.90.094008}}.

\bibitem{Gies:2000tc}
H.~Gies, R.~Shaisultanov, {On the axial current in an electromagnetic field and
  low-energy neutrino - photon interactions}, Phys. Rev. D 62 (2000) 073003.
\newblock \href {http://arxiv.org/abs/hep-ph/0003144}
  {\path{arXiv:hep-ph/0003144}}, \href
  {https://doi.org/10.1103/PhysRevD.62.073003}
  {\path{doi:10.1103/PhysRevD.62.073003}}.

\bibitem{Gies:2000wc}
H.~Gies, R.~Shaisultanov, {Neutrino interactions with a weak slowly varying
  electromagnetic field}, Phys. Lett. B 480 (2000) 129--134.
\newblock \href {http://arxiv.org/abs/hep-ph/0009342}
  {\path{arXiv:hep-ph/0009342}}, \href
  {https://doi.org/10.1016/S0370-2693(00)00424-X}
  {\path{doi:10.1016/S0370-2693(00)00424-X}}.

\bibitem{Mohammadi:2013ksa}
R.~Mohammadi, S.-S. Xue, {Laser photons acquire circular polarization by
  interacting with a Dirac or Majorana neutrino beam}, Phys. Lett. B 731 (2014)
  272--278.
\newblock \href {http://arxiv.org/abs/1312.3862} {\path{arXiv:1312.3862}},
  \href {https://doi.org/10.1016/j.physletb.2014.02.031}
  {\path{doi:10.1016/j.physletb.2014.02.031}}.

\bibitem{Meuren:2015iha}
S.~Meuren, C.~H. Keitel, A.~Di~Piazza, {Nonlinear neutrino-photon interactions
  inside strong laser pulses}, JHEP 06 (2015) 127.
\newblock \href {http://arxiv.org/abs/1504.02722} {\path{arXiv:1504.02722}},
  \href {https://doi.org/10.1007/JHEP06(2015)127}
  {\path{doi:10.1007/JHEP06(2015)127}}.

\bibitem{Formanek:2017mbv}
M.~Formanek, S.~Evans, J.~Rafelski, A.~Steinmetz, C.-T. Yang, {Strong fields
  and neutral particle magnetic moment dynamics}, Plasma Phys. Control. Fusion
  60 (2018) 074006.
\newblock \href {http://arxiv.org/abs/1712.07698} {\path{arXiv:1712.07698}},
  \href {https://doi.org/10.1088/1361-6587/aac06a}
  {\path{doi:10.1088/1361-6587/aac06a}}.

\bibitem{Dvornikov:2018tmm}
M.~Dvornikov, {Spin-flavor oscillations of Dirac neutrinos in a plane
  electromagnetic wave}, Phys. Rev. D 98~(7) (2018) 075025.
\newblock \href {http://arxiv.org/abs/1806.08719} {\path{arXiv:1806.08719}},
  \href {https://doi.org/10.1103/PhysRevD.98.075025}
  {\path{doi:10.1103/PhysRevD.98.075025}}.

\bibitem{Dvornikov:2019pxd}
M.~Dvornikov, {Spin-flavor oscillations of Dirac neutrinos in matter under the
  influence of a plane electromagnetic wave}, Phys. Rev. D 99~(3) (2019)
  035027.
\newblock \href {http://arxiv.org/abs/1901.01022} {\path{arXiv:1901.01022}},
  \href {https://doi.org/10.1103/PhysRevD.99.035027}
  {\path{doi:10.1103/PhysRevD.99.035027}}.

\bibitem{Dvornikov:2019sfo}
M.~Dvornikov, {Neutrino spin oscillations in external fields in curved
  spacetime}, Phys. Rev. D 99~(11) (2019) 116021.
\newblock \href {http://arxiv.org/abs/1902.11285} {\path{arXiv:1902.11285}},
  \href {https://doi.org/10.1103/PhysRevD.99.116021}
  {\path{doi:10.1103/PhysRevD.99.116021}}.

\bibitem{Huang:2015oca}
X.-G. Huang, {Electromagnetic fields and anomalous transports in heavy-ion
  collisions --- A pedagogical review}, Rept. Prog. Phys. 79~(7) (2016) 076302.
\newblock \href {http://arxiv.org/abs/1509.04073} {\path{arXiv:1509.04073}},
  \href {https://doi.org/10.1088/0034-4885/79/7/076302}
  {\path{doi:10.1088/0034-4885/79/7/076302}}.

\bibitem{Fukushima:2016xgg}
K.~Fukushima, {Evolution to the quark\textendash{}gluon plasma}, Rept. Prog.
  Phys. 80~(2) (2017) 022301.
\newblock \href {http://arxiv.org/abs/1603.02340} {\path{arXiv:1603.02340}},
  \href {https://doi.org/10.1088/1361-6633/80/2/022301}
  {\path{doi:10.1088/1361-6633/80/2/022301}}.

\bibitem{Fukushima:2018grm}
K.~Fukushima, {Extreme matter in electromagnetic fields and rotation}, Prog.
  Part. Nucl. Phys. 107 (2019) 167--199.
\newblock \href {http://arxiv.org/abs/1812.08886} {\path{arXiv:1812.08886}},
  \href {https://doi.org/10.1016/j.ppnp.2019.04.001}
  {\path{doi:10.1016/j.ppnp.2019.04.001}}.

\bibitem{Gelis:2021zmx}
F.~Gelis, {Some Aspects of the Theory of Heavy Ion Collisions}, Rept. Prog.
  Phys. 84~(5) (2021) 056301.
\newblock \href {http://arxiv.org/abs/2102.07604} {\path{arXiv:2102.07604}},
  \href {https://doi.org/10.1088/1361-6633/abec2e}
  {\path{doi:10.1088/1361-6633/abec2e}}.

\bibitem{Kovchegov:2012mbw}
Y.~V. Kovchegov, E.~Levin, {Quantum chromodynamics at high energy}, Vol.~33,
  Cambridge University Press, 2012.
\newblock \href {https://doi.org/10.1017/CBO9781139022187}
  {\path{doi:10.1017/CBO9781139022187}}.

\bibitem{Gelis:2012ri}
F.~Gelis, {Color Glass Condensate and Glasma}, Int. J. Mod. Phys. A 28 (2013)
  1330001.
\newblock \href {http://arxiv.org/abs/1211.3327} {\path{arXiv:1211.3327}},
  \href {https://doi.org/10.1142/S0217751X13300019}
  {\path{doi:10.1142/S0217751X13300019}}.

\bibitem{Albacete:2014fwa}
J.~L. Albacete, C.~Marquet, {Gluon saturation and initial conditions for
  relativistic heavy ion collisions}, Prog. Part. Nucl. Phys. 76 (2014) 1--42.
\newblock \href {http://arxiv.org/abs/1401.4866} {\path{arXiv:1401.4866}},
  \href {https://doi.org/10.1016/j.ppnp.2014.01.004}
  {\path{doi:10.1016/j.ppnp.2014.01.004}}.

\bibitem{Lappi:2006fp}
T.~Lappi, L.~McLerran, {Some features of the glasma}, Nucl. Phys. A 772 (2006)
  200--212.
\newblock \href {http://arxiv.org/abs/hep-ph/0602189}
  {\path{arXiv:hep-ph/0602189}}, \href
  {https://doi.org/10.1016/j.nuclphysa.2006.04.001}
  {\path{doi:10.1016/j.nuclphysa.2006.04.001}}.

\bibitem{Bzdak:2011yy}
A.~Bzdak, V.~Skokov, {Event-by-event fluctuations of magnetic and electric
  fields in heavy ion collisions}, Phys. Lett. B 710 (2012) 171--174.
\newblock \href {http://arxiv.org/abs/1111.1949} {\path{arXiv:1111.1949}},
  \href {https://doi.org/10.1016/j.physletb.2012.02.065}
  {\path{doi:10.1016/j.physletb.2012.02.065}}.

\bibitem{Deng:2012pc}
W.-T. Deng, X.-G. Huang, {Event-by-event generation of electromagnetic fields
  in heavy-ion collisions}, Phys. Rev. C 85 (2012) 044907.
\newblock \href {http://arxiv.org/abs/1201.5108} {\path{arXiv:1201.5108}},
  \href {https://doi.org/10.1103/PhysRevC.85.044907}
  {\path{doi:10.1103/PhysRevC.85.044907}}.

\bibitem{Hirono:2012rt}
Y.~Hirono, M.~Hongo, T.~Hirano, {Estimation of electric conductivity of the
  quark gluon plasma via asymmetric heavy-ion collisions}, Phys. Rev. C 90~(2)
  (2014) 021903.
\newblock \href {http://arxiv.org/abs/1211.1114} {\path{arXiv:1211.1114}},
  \href {https://doi.org/10.1103/PhysRevC.90.021903}
  {\path{doi:10.1103/PhysRevC.90.021903}}.

\bibitem{Voronyuk:2014rna}
V.~Voronyuk, V.~D. Toneev, S.~A. Voloshin, W.~Cassing, {Charge-dependent
  directed flow in asymmetric nuclear collisions}, Phys. Rev. C 90~(6) (2014)
  064903.
\newblock \href {http://arxiv.org/abs/1410.1402} {\path{arXiv:1410.1402}},
  \href {https://doi.org/10.1103/PhysRevC.90.064903}
  {\path{doi:10.1103/PhysRevC.90.064903}}.

\bibitem{Baur:2008hn}
G.~Baur, {Coherent photon-photon interactions in very peripheral relativistic
  heavy ion collisions}, Eur. Phys. J. D 55 (2009) 265--269.
\newblock \href {http://arxiv.org/abs/0810.1400} {\path{arXiv:0810.1400}},
  \href {https://doi.org/10.1140/epjd/e2009-00019-7}
  {\path{doi:10.1140/epjd/e2009-00019-7}}.

\bibitem{Peroutka:2017esw}
B.~Peroutka, K.~Tuchin, {Quantum diffusion of electromagnetic fields of
  ultrarelativistic spin-half particles}, Nucl. Phys. A 966 (2017) 64--72.
\newblock \href {http://arxiv.org/abs/1703.02606} {\path{arXiv:1703.02606}},
  \href {https://doi.org/10.1016/j.nuclphysa.2017.05.104}
  {\path{doi:10.1016/j.nuclphysa.2017.05.104}}.

\bibitem{BAUR20071}
G.~Baur, K.~Hencken, D.~Trautmann,
  \href{https://www.sciencedirect.com/science/article/pii/S037015730700347X}{Electron–positron
  pair production in ultrarelativistic heavy ion collisions}, Physics Reports
  453~(1) (2007) 1--27.
\newblock \href {https://doi.org/https://doi.org/10.1016/j.physrep.2007.09.002}
  {\path{doi:https://doi.org/10.1016/j.physrep.2007.09.002}}.
\newline\urlprefix\url{https://www.sciencedirect.com/science/article/pii/S037015730700347X}

\bibitem{Obraztsov:2021tip}
I.~V. Obraztsov, A.~I. Milstein, {Quadrupole radiation and e+e\ensuremath{-}
  pair production in the collision of nonrelativistic nuclei}, Phys. Lett. B
  820 (2021) 136514.
\newblock \href {http://arxiv.org/abs/2103.00439} {\path{arXiv:2103.00439}},
  \href {https://doi.org/10.1016/j.physletb.2021.136514}
  {\path{doi:10.1016/j.physletb.2021.136514}}.

\bibitem{Francener:2021wzx}
R.~Francener, B.~D. Moreira, V.~P. Goncalves, {Photoproduction of relativistic
  QED bound states in hadronic collisions} (10 2021).
\newblock \href {http://arxiv.org/abs/2110.03466} {\path{arXiv:2110.03466}}.

\bibitem{Voitkiv:2009ef}
A.~B. Voitkiv, B.~Najjari, A.~Di~Piazza, {Bound-bound pair production in
  relativistic collisions}, New J. Phys. 12 (2010) 063011.
\newblock \href {http://arxiv.org/abs/0906.1540} {\path{arXiv:0906.1540}},
  \href {https://doi.org/10.1088/1367-2630/12/6/063011}
  {\path{doi:10.1088/1367-2630/12/6/063011}}.

\bibitem{Brandenburg:2021lnj}
J.~D. Brandenburg, W.~Zha, Z.~Xu, {Mapping the electromagnetic fields of
  heavy-ion collisions with the Breit-Wheeler process}, Eur. Phys. J. A 57~(10)
  (2021) 299.
\newblock \href {http://arxiv.org/abs/2103.16623} {\path{arXiv:2103.16623}},
  \href {https://doi.org/10.1140/epja/s10050-021-00595-5}
  {\path{doi:10.1140/epja/s10050-021-00595-5}}.

\bibitem{Kawai:1985xq}
H.~Kawai, D.~C. Lewellen, S.~H.~H. Tye, {A Relation Between Tree Amplitudes of
  Closed and Open Strings}, Nucl. Phys. B 269 (1986) 1--23.
\newblock \href {https://doi.org/10.1016/0550-3213(86)90362-7}
  {\path{doi:10.1016/0550-3213(86)90362-7}}.

\bibitem{Bern:2008qj}
Z.~Bern, J.~J.~M. Carrasco, H.~Johansson, {New Relations for Gauge-Theory
  Amplitudes}, Phys. Rev. D 78 (2008) 085011.
\newblock \href {http://arxiv.org/abs/0805.3993} {\path{arXiv:0805.3993}},
  \href {https://doi.org/10.1103/PhysRevD.78.085011}
  {\path{doi:10.1103/PhysRevD.78.085011}}.

\bibitem{Bern:2010ue}
Z.~Bern, J.~J.~M. Carrasco, H.~Johansson, {Perturbative Quantum Gravity as a
  Double Copy of Gauge Theory}, Phys. Rev. Lett. 105 (2010) 061602.
\newblock \href {http://arxiv.org/abs/1004.0476} {\path{arXiv:1004.0476}},
  \href {https://doi.org/10.1103/PhysRevLett.105.061602}
  {\path{doi:10.1103/PhysRevLett.105.061602}}.

\bibitem{Bern:2010yg}
Z.~Bern, T.~Dennen, Y.-t. Huang, M.~Kiermaier, {Gravity as the Square of Gauge
  Theory}, Phys. Rev. D 82 (2010) 065003.
\newblock \href {http://arxiv.org/abs/1004.0693} {\path{arXiv:1004.0693}},
  \href {https://doi.org/10.1103/PhysRevD.82.065003}
  {\path{doi:10.1103/PhysRevD.82.065003}}.

\bibitem{Bern:2019prr}
Z.~Bern, J.~J. Carrasco, M.~Chiodaroli, H.~Johansson, R.~Roiban, {The Duality
  Between Color and Kinematics and its Applications} (9 2019).
\newblock \href {http://arxiv.org/abs/1909.01358} {\path{arXiv:1909.01358}}.

\bibitem{White:2021gvv}
C.~D. White, {The double copy: from optics to quantum gravity} (5 2021).
\newblock \href {http://arxiv.org/abs/2105.06809} {\path{arXiv:2105.06809}}.

\bibitem{Coleman:1977ps}
S.~R. Coleman, {Nonabelian Plane Waves}, Phys. Lett. B 70 (1977) 59--60.
\newblock \href {https://doi.org/10.1016/0370-2693(77)90344-6}
  {\path{doi:10.1016/0370-2693(77)90344-6}}.

\bibitem{Adamo:2017nia}
T.~Adamo, E.~Casali, L.~Mason, S.~Nekovar, {Scattering on plane waves and the
  double copy}, Class. Quant. Grav. 35~(1) (2018) 015004.
\newblock \href {http://arxiv.org/abs/1706.08925} {\path{arXiv:1706.08925}},
  \href {https://doi.org/10.1088/1361-6382/aa9961}
  {\path{doi:10.1088/1361-6382/aa9961}}.

\bibitem{Adamo:2017sze}
T.~Adamo, E.~Casali, L.~Mason, S.~Nekovar, {Amplitudes on plane waves from
  ambitwistor strings}, JHEP 11 (2017) 160.
\newblock \href {http://arxiv.org/abs/1708.09249} {\path{arXiv:1708.09249}},
  \href {https://doi.org/10.1007/JHEP11(2017)160}
  {\path{doi:10.1007/JHEP11(2017)160}}.

\bibitem{Adamo:2018mpq}
T.~Adamo, E.~Casali, L.~Mason, S.~Nekovar, {Plane wave backgrounds and
  colour-kinematics duality}, JHEP 02 (2019) 198.
\newblock \href {http://arxiv.org/abs/1810.05115} {\path{arXiv:1810.05115}},
  \href {https://doi.org/10.1007/JHEP02(2019)198}
  {\path{doi:10.1007/JHEP02(2019)198}}.

\bibitem{Adamo:2019zmk}
T.~Adamo, A.~Ilderton, {Gluon helicity flip in a plane wave background}, JHEP
  06 (2019) 015.
\newblock \href {http://arxiv.org/abs/1903.01491} {\path{arXiv:1903.01491}},
  \href {https://doi.org/10.1007/JHEP06(2019)015}
  {\path{doi:10.1007/JHEP06(2019)015}}.

\bibitem{Einstein:1937qu}
A.~Einstein, N.~Rosen, {On Gravitational waves}, J. Franklin Inst. 223 (1937)
  43--54.
\newblock \href {https://doi.org/10.1016/S0016-0032(37)90583-0}
  {\path{doi:10.1016/S0016-0032(37)90583-0}}.

\bibitem{Brinkmann:1925fr}
H.~W. Brinkmann, {Einstein spapces which are mapped conformally on each other},
  Math. Ann. 94 (1925) 119--145.
\newblock \href {https://doi.org/10.1007/BF01208647}
  {\path{doi:10.1007/BF01208647}}.

\bibitem{Ward1987}
R.~S. Ward, Progressing waves in flat spacetime and in plane-wave spacetimes
  4~(3) (1987) 775--778.
\newblock \href {https://doi.org/10.1088/0264-9381/4/3/034}
  {\path{doi:10.1088/0264-9381/4/3/034}}.

\bibitem{Adamo:2020qru}
T.~Adamo, A.~Ilderton, {Classical and quantum double copy of back-reaction},
  JHEP 09 (2020) 200.
\newblock \href {http://arxiv.org/abs/2005.05807} {\path{arXiv:2005.05807}},
  \href {https://doi.org/10.1007/JHEP09(2020)200}
  {\path{doi:10.1007/JHEP09(2020)200}}.

\bibitem{Monteiro:2014cda}
R.~Monteiro, D.~O'Connell, C.~D. White, {Black holes and the double copy}, JHEP
  12 (2014) 056.
\newblock \href {http://arxiv.org/abs/1410.0239} {\path{arXiv:1410.0239}},
  \href {https://doi.org/10.1007/JHEP12(2014)056}
  {\path{doi:10.1007/JHEP12(2014)056}}.

\bibitem{Goldberger:2017frp}
W.~D. Goldberger, S.~G. Prabhu, J.~O. Thompson, {Classical gluon and graviton
  radiation from the bi-adjoint scalar double copy}, Phys. Rev. D 96~(6) (2017)
  065009.
\newblock \href {http://arxiv.org/abs/1705.09263} {\path{arXiv:1705.09263}},
  \href {https://doi.org/10.1103/PhysRevD.96.065009}
  {\path{doi:10.1103/PhysRevD.96.065009}}.

\bibitem{Goldberger:2016iau}
W.~D. Goldberger, A.~K. Ridgway, {Radiation and the classical double copy for
  color charges}, Phys. Rev. D 95~(12) (2017) 125010.
\newblock \href {http://arxiv.org/abs/1611.03493} {\path{arXiv:1611.03493}},
  \href {https://doi.org/10.1103/PhysRevD.95.125010}
  {\path{doi:10.1103/PhysRevD.95.125010}}.

\bibitem{Andrzejewski:2018zby}
K.~Andrzejewski, S.~Prencel,
  \href{https://doi.org/10.1088/1361-6382/ab2394}{Niederer's transformation,
  time-dependent oscillators and polarized gravitational waves}, Classical and
  Quantum Gravity 36~(15) (2019) 155008.
\newblock \href {http://arxiv.org/abs/1810.06541} {\path{arXiv:1810.06541}}.
\newline\urlprefix\url{https://doi.org/10.1088/1361-6382/ab2394}

\bibitem{Bialynicki-Birula:2004bvr}
I.~Bialynicki-Birula, {Particle beams guided by electromagnetic vortices: New
  solutions of the Lorentz, Schrodinger, Klein-Gordon and Dirac equations},
  Phys. Rev. Lett. 93 (2004) 020402.
\newblock \href {http://arxiv.org/abs/physics/0403078}
  {\path{arXiv:physics/0403078}}, \href
  {https://doi.org/10.1103/PhysRevLett.93.020402}
  {\path{doi:10.1103/PhysRevLett.93.020402}}.

\bibitem{Ilderton:2018lsf}
A.~Ilderton, {Screw-symmetric gravitational waves: a double copy of the
  vortex}, Phys. Lett. B 782 (2018) 22--27.
\newblock \href {http://arxiv.org/abs/1804.07290} {\path{arXiv:1804.07290}},
  \href {https://doi.org/10.1016/j.physletb.2018.04.069}
  {\path{doi:10.1016/j.physletb.2018.04.069}}.

\bibitem{Bialynicki-Birula:2015mvf}
I.~Bialynicki-Birula, Z.~Bialynicka-Birula, {Gravitational waves carrying
  orbital angular momentum}, New J. Phys. 18~(2) (2016) 023022.
\newblock \href {http://arxiv.org/abs/1511.08909} {\path{arXiv:1511.08909}},
  \href {https://doi.org/10.1088/1367-2630/18/2/023022}
  {\path{doi:10.1088/1367-2630/18/2/023022}}.

\bibitem{Bialynicki-Birula:2018nnk}
I.~Bialynicki-Birula, S.~Charzy\'nski, {Trapping and guiding bodies by
  gravitational waves endowed with angular momentum}, Phys. Rev. Lett. 121~(17)
  (2018) 171101.
\newblock \href {http://arxiv.org/abs/1810.02219} {\path{arXiv:1810.02219}},
  \href {https://doi.org/10.1103/PhysRevLett.121.171101}
  {\path{doi:10.1103/PhysRevLett.121.171101}}.

\bibitem{Zhang:2018srn}
P.~M. Zhang, C.~Duval, G.~W. Gibbons, P.~A. Horvathy, {Velocity Memory Effect
  for Polarized Gravitational Waves}, JCAP 05 (2018) 030.
\newblock \href {http://arxiv.org/abs/1802.09061} {\path{arXiv:1802.09061}},
  \href {https://doi.org/10.1088/1475-7516/2018/05/030}
  {\path{doi:10.1088/1475-7516/2018/05/030}}.

\bibitem{Zhang:2018msv}
P.~M. Zhang, M.~Cariglia, C.~Duval, M.~Elbistan, G.~W. Gibbons, P.~A. Horvathy,
  {Ion Traps and the Memory Effect for Periodic Gravitational Waves}, Phys.
  Rev. D 98~(4) (2018) 044037.
\newblock \href {http://arxiv.org/abs/1807.00765} {\path{arXiv:1807.00765}},
  \href {https://doi.org/10.1103/PhysRevD.98.044037}
  {\path{doi:10.1103/PhysRevD.98.044037}}.

\bibitem{Zhang:2021lrw}
P.-M. Zhang, M.~Elbistan, P.~A. Horvathy, {Particle motion in circularly
  polarized vacuum pp waves}, Class. Quant. Grav. 39~(3) (2022) 035008.
\newblock \href {http://arxiv.org/abs/2108.00838} {\path{arXiv:2108.00838}},
  \href {https://doi.org/10.1088/1361-6382/ac43d2}
  {\path{doi:10.1088/1361-6382/ac43d2}}.

\bibitem{LIGOScientific:2016aoc}
B.~P. Abbott, et~al., {Observation of Gravitational Waves from a Binary Black
  Hole Merger}, Phys. Rev. Lett. 116~(6) (2016) 061102.
\newblock \href {http://arxiv.org/abs/1602.03837} {\path{arXiv:1602.03837}},
  \href {https://doi.org/10.1103/PhysRevLett.116.061102}
  {\path{doi:10.1103/PhysRevLett.116.061102}}.

\bibitem{LIGOScientific:2016sjg}
B.~P. Abbott, et~al., {GW151226: Observation of Gravitational Waves from a
  22-Solar-Mass Binary Black Hole Coalescence}, Phys. Rev. Lett. 116~(24)
  (2016) 241103.
\newblock \href {http://arxiv.org/abs/1606.04855} {\path{arXiv:1606.04855}},
  \href {https://doi.org/10.1103/PhysRevLett.116.241103}
  {\path{doi:10.1103/PhysRevLett.116.241103}}.

\bibitem{Ribeyre1}
X.~Ribeyre, V.~Tikhonchuk, High frequency graviational waves generation in
  laser plasma interaction, The Twelfth Marcel Grossmann Meeting
  1640--1642\href {https://doi.org/10.1142/9789814374552_0292}
  {\path{doi:10.1142/9789814374552_0292}}.

\bibitem{Pustovoit:2020ksx}
V.~Pustovoit, V.~Gladyshev, V.~Kauts, A.~Morozov, V.~Gorelik, I.~Fomin,
  D.~Portnov, E.~Sharandin, A.~Kayutenko, {High frequency gravitational waves
  generation by optical methods}, J. Phys. Conf. Ser. 1557 (2020) 012034.
\newblock \href {https://doi.org/10.1088/1742-6596/1557/1/012034}
  {\path{doi:10.1088/1742-6596/1557/1/012034}}.

\bibitem{Gelfer:2015fbj}
E.~Gelfer, H.~Kadlecov\'a, O.~Klimo, S.~Weber, G.~Korn, {Gravitational waves
  generated by laser accelerated relativistic ions}, Phys. Plasmas 23 (2016)
  093107.
\newblock \href {http://arxiv.org/abs/1512.07188} {\path{arXiv:1512.07188}},
  \href {https://doi.org/10.1063/1.4962520} {\path{doi:10.1063/1.4962520}}.

\bibitem{MorozovGravStanding}
A.~N. {Morozov}, V.~I. {Pustovoit}, I.~V. {Fomin}, {Generation of Gravitational
  Waves by a Standing Electromagnetic Wave}, Gravitation and Cosmology 27~(1)
  (2021) 24--29.
\newblock \href {https://doi.org/10.1134/S020228932101014X}
  {\path{doi:10.1134/S020228932101014X}}.

\bibitem{Lageyre:2021cxe}
P.~Lageyre, X.~Ribeyre, E.~D'Humieres, {Gravitational influence of high power
  laser pulses} (7 2021).
\newblock \href {http://arxiv.org/abs/2108.00896} {\path{arXiv:2108.00896}}.

\bibitem{Kadlecova:2016fbx}
H.~Kadlecov\'a, O.~Klimo, S.~Weber, G.~Korn, {Gravitational wave generation by
  interaction of high power lasers with matter using shock waves}, Eur. Phys.
  J. D 71~(4) (2017) 89.
\newblock \href {http://arxiv.org/abs/1602.08904} {\path{arXiv:1602.08904}},
  \href {https://doi.org/10.1140/epjd/e2017-70586-y}
  {\path{doi:10.1140/epjd/e2017-70586-y}}.

\bibitem{Affleck:1981ag}
I.~K. Affleck, N.~S. Manton, {Monopole Pair Production in a Magnetic Field},
  Nucl. Phys. B 194 (1982) 38--64.
\newblock \href {https://doi.org/10.1016/0550-3213(82)90511-9}
  {\path{doi:10.1016/0550-3213(82)90511-9}}.

\bibitem{Gould:2017zwi}
O.~Gould, A.~Rajantie, {Magnetic monopole mass bounds from heavy ion collisions
  and neutron stars}, Phys. Rev. Lett. 119~(24) (2017) 241601.
\newblock \href {http://arxiv.org/abs/1705.07052} {\path{arXiv:1705.07052}},
  \href {https://doi.org/10.1103/PhysRevLett.119.241601}
  {\path{doi:10.1103/PhysRevLett.119.241601}}.

\bibitem{Gould:2021bre}
O.~Gould, D.~L.~J. Ho, A.~Rajantie, {Schwinger pair production of magnetic
  monopoles: Momentum distribution for heavy-ion collisions}, Phys. Rev. D
  104~(1) (2021) 015033.
\newblock \href {http://arxiv.org/abs/2103.14454} {\path{arXiv:2103.14454}},
  \href {https://doi.org/10.1103/PhysRevD.104.015033}
  {\path{doi:10.1103/PhysRevD.104.015033}}.

\bibitem{Ho:2019ads}
D.~L.~J. Ho, A.~Rajantie, {Classical production of \textquoteright{}t
  Hooft\textendash{}Polyakov monopoles from magnetic fields}, Phys. Rev. D
  101~(5) (2020) 055003.
\newblock \href {http://arxiv.org/abs/1911.06088} {\path{arXiv:1911.06088}},
  \href {https://doi.org/10.1103/PhysRevD.101.055003}
  {\path{doi:10.1103/PhysRevD.101.055003}}.

\bibitem{Ho:2021uem}
D.~L.~J. Ho, A.~Rajantie, {Instanton solution for Schwinger production of
  \textquoteright{}t Hooft-Polyakov monopoles}, Phys. Rev. D 103~(11) (2021)
  115033.
\newblock \href {http://arxiv.org/abs/2103.12799} {\path{arXiv:2103.12799}},
  \href {https://doi.org/10.1103/PhysRevD.103.115033}
  {\path{doi:10.1103/PhysRevD.103.115033}}.

\bibitem{Acharya:2021ckc}
B.~Acharya, et~al., {Search for magnetic monopoles produced via the Schwinger
  mechanism}, Nature 602~(7895) (2022) 63--67.
\newblock \href {http://arxiv.org/abs/2106.11933} {\path{arXiv:2106.11933}},
  \href {https://doi.org/10.1038/s41586-021-04298-1}
  {\path{doi:10.1038/s41586-021-04298-1}}.

\bibitem{Kobayashi:2021des}
T.~Kobayashi, {Monopole-antimonopole pair production in primordial magnetic
  fields}, Phys. Rev. D 104~(4) (2021) 043501.
\newblock \href {http://arxiv.org/abs/2105.12776} {\path{arXiv:2105.12776}},
  \href {https://doi.org/10.1103/PhysRevD.104.043501}
  {\path{doi:10.1103/PhysRevD.104.043501}}.

\bibitem{Jaeckel:2010ni}
J.~Jaeckel, A.~Ringwald, {The Low-Energy Frontier of Particle Physics}, Ann.
  Rev. Nucl. Part. Sci. 60 (2010) 405--437.
\newblock \href {http://arxiv.org/abs/1002.0329} {\path{arXiv:1002.0329}},
  \href {https://doi.org/10.1146/annurev.nucl.012809.104433}
  {\path{doi:10.1146/annurev.nucl.012809.104433}}.

\bibitem{Essig:2013lka}
R.~Essig, et~al., {Working Group Report: New Light Weakly Coupled Particles},
  in: {Community Summer Study 2013}: {Snowmass on the Mississippi}, 2013.
\newblock \href {http://arxiv.org/abs/1311.0029} {\path{arXiv:1311.0029}}.

\bibitem{Graham:2015ouw}
P.~W. Graham, I.~G. Irastorza, S.~K. Lamoreaux, A.~Lindner, K.~A. van Bibber,
  {Experimental Searches for the Axion and Axion-Like Particles}, Ann. Rev.
  Nucl. Part. Sci. 65 (2015) 485--514.
\newblock \href {http://arxiv.org/abs/1602.00039} {\path{arXiv:1602.00039}},
  \href {https://doi.org/10.1146/annurev-nucl-102014-022120}
  {\path{doi:10.1146/annurev-nucl-102014-022120}}.

\bibitem{Irastorza:2018dyq}
I.~G. Irastorza, J.~Redondo, {New experimental approaches in the search for
  axion-like particles}, Prog. Part. Nucl. Phys. 102 (2018) 89--159.
\newblock \href {http://arxiv.org/abs/1801.08127} {\path{arXiv:1801.08127}},
  \href {https://doi.org/10.1016/j.ppnp.2018.05.003}
  {\path{doi:10.1016/j.ppnp.2018.05.003}}.

\bibitem{Tajima:2012mx}
T.~Tajima, K.~Homma, {Fundamental Physics Explored with High Intensity Laser},
  Int. J. Mod. Phys. A 27 (2012) 1230027.
\newblock \href {http://arxiv.org/abs/1209.2822} {\path{arXiv:1209.2822}},
  \href {https://doi.org/10.1142/S0217751X1230027X}
  {\path{doi:10.1142/S0217751X1230027X}}.

\bibitem{Chen:2006cd}
S.-J. Chen, H.-H. Mei, W.-T. Ni, {Q \& A experiment to search for vacuum
  dichroism, pseudoscalar-photon interaction and millicharged fermions}, Mod.
  Phys. Lett. A 22 (2007) 2815--2831.
\newblock \href {http://arxiv.org/abs/hep-ex/0611050}
  {\path{arXiv:hep-ex/0611050}}, \href
  {https://doi.org/10.1142/S0217732307025844}
  {\path{doi:10.1142/S0217732307025844}}.

\bibitem{CAST:2017uph}
V.~Anastassopoulos, et~al., {New CAST Limit on the Axion-Photon Interaction},
  Nature Phys. 13 (2017) 584--590.
\newblock \href {http://arxiv.org/abs/1705.02290} {\path{arXiv:1705.02290}},
  \href {https://doi.org/10.1038/nphys4109} {\path{doi:10.1038/nphys4109}}.

\bibitem{Armengaud:2014gea}
E.~Armengaud, et~al., {Conceptual Design of the International Axion Observatory
  (IAXO)}, JINST 9 (2014) T05002.
\newblock \href {http://arxiv.org/abs/1401.3233} {\path{arXiv:1401.3233}},
  \href {https://doi.org/10.1088/1748-0221/9/05/T05002}
  {\path{doi:10.1088/1748-0221/9/05/T05002}}.

\bibitem{Redondo:2010dp}
J.~Redondo, A.~Ringwald, {Light shining through walls}, Contemp. Phys. 52
  (2011) 211--236.
\newblock \href {http://arxiv.org/abs/1011.3741} {\path{arXiv:1011.3741}},
  \href {https://doi.org/10.1080/00107514.2011.563516}
  {\path{doi:10.1080/00107514.2011.563516}}.

\bibitem{Bahre:2013ywa}
R.~B\"ahre, et~al., {Any light particle search II \textemdash{}Technical Design
  Report}, JINST 8 (2013) T09001.
\newblock \href {http://arxiv.org/abs/1302.5647} {\path{arXiv:1302.5647}},
  \href {https://doi.org/10.1088/1748-0221/8/09/T09001}
  {\path{doi:10.1088/1748-0221/8/09/T09001}}.

\bibitem{GammeV:2008cqp}
A.~S. Chou, et~al., {A Search for chameleon particles using a photon
  regeneration technique}, Phys. Rev. Lett. 102 (2009) 030402.
\newblock \href {http://arxiv.org/abs/0806.2438} {\path{arXiv:0806.2438}},
  \href {https://doi.org/10.1103/PhysRevLett.102.030402}
  {\path{doi:10.1103/PhysRevLett.102.030402}}.

\bibitem{Afanasev:2008jt}
A.~Afanasev, O.~K. Baker, K.~B. Beard, G.~Biallas, J.~Boyce, M.~Minarni,
  R.~Ramdon, M.~Shinn, P.~Slocum, {New Experimental limit on Optical Photon
  Coupling to Neutral, Scalar Bosons}, Phys. Rev. Lett. 101 (2008) 120401.
\newblock \href {http://arxiv.org/abs/0806.2631} {\path{arXiv:0806.2631}},
  \href {https://doi.org/10.1103/PhysRevLett.101.120401}
  {\path{doi:10.1103/PhysRevLett.101.120401}}.

\bibitem{OSQAR:2015qdv}
R.~Ballou, et~al., {New exclusion limits on scalar and pseudoscalar axionlike
  particles from light shining through a wall}, Phys. Rev. D 92~(9) (2015)
  092002.
\newblock \href {http://arxiv.org/abs/1506.08082} {\path{arXiv:1506.08082}},
  \href {https://doi.org/10.1103/PhysRevD.92.092002}
  {\path{doi:10.1103/PhysRevD.92.092002}}.

\bibitem{SAPPHIRES:2021vkz}
K.~Homma, et~al., {Search for sub-eV axion-like resonance states via stimulated
  quasi-parallel laser collisions with the parameterization including fully
  asymmetric collisional geometry} (5 2021).
\newblock \href {http://arxiv.org/abs/2105.01224} {\path{arXiv:2105.01224}}.

\bibitem{Gies:2008wv}
H.~Gies, {Strong laser fields as a probe for fundamental physics}, Eur. Phys.
  J. D 55 (2009) 311--317.
\newblock \href {http://arxiv.org/abs/0812.0668} {\path{arXiv:0812.0668}},
  \href {https://doi.org/10.1140/epjd/e2009-00006-0}
  {\path{doi:10.1140/epjd/e2009-00006-0}}.

\bibitem{Villalba-Chavez:2013bda}
S.~Villalba-Chavez, A.~Piazza, {Axion-induced birefringence effects in laser
  driven nonlinear vacuum interaction}, JHEP 11 (2013) 136.
\newblock \href {http://arxiv.org/abs/1307.7935} {\path{arXiv:1307.7935}},
  \href {https://doi.org/10.1007/JHEP11(2013)136}
  {\path{doi:10.1007/JHEP11(2013)136}}.

\bibitem{Villalba-Chavez:2013goa}
S.~Villalba-Ch\'avez, {Laser-driven search of axion-like particles including
  vacuum polarization effects}, Nucl. Phys. B 881 (2014) 391--413.
\newblock \href {http://arxiv.org/abs/1308.4033} {\path{arXiv:1308.4033}},
  \href {https://doi.org/10.1016/j.nuclphysb.2014.01.021}
  {\path{doi:10.1016/j.nuclphysb.2014.01.021}}.

\bibitem{Shakeri:2020sin}
S.~Shakeri, D.~J.~E. Marsh, S.-S. Xue, {Light by Light Scattering as a New
  Probe for Axions} (2 2020).
\newblock \href {http://arxiv.org/abs/2002.06123} {\path{arXiv:2002.06123}}.

\bibitem{Villalba-Chavez:2016hxw}
S.~Villalba-Ch\'avez, T.~Podszus, C.~M\"uller, {Polarization-operator approach
  to optical signatures of axion-like particles in strong laser pulses}, Phys.
  Lett. B 769 (2017) 233--241.
\newblock \href {http://arxiv.org/abs/1612.07952} {\path{arXiv:1612.07952}},
  \href {https://doi.org/10.1016/j.physletb.2017.03.043}
  {\path{doi:10.1016/j.physletb.2017.03.043}}.

\bibitem{Dobrich:2010hi}
B.~Dobrich, H.~Gies, {Axion-like-particle search with high-intensity lasers},
  JHEP 10 (2010) 022.
\newblock \href {http://arxiv.org/abs/1006.5579} {\path{arXiv:1006.5579}},
  \href {https://doi.org/10.1007/JHEP10(2010)022}
  {\path{doi:10.1007/JHEP10(2010)022}}.

\bibitem{Beyer:2021xql}
K.~A. Beyer, G.~Marocco, C.~Danson, R.~Bingham, G.~Gregori, {Parametric
  co-linear axion photon instability} (8 2021).
\newblock \href {http://arxiv.org/abs/2108.01489} {\path{arXiv:2108.01489}}.

\bibitem{Burton:2017bxi}
D.~A. Burton, A.~Noble, {Plasma-based wakefield accelerators as sources of
  axion-like particles}, New J. Phys. 20~(3) (2018) 033022.
\newblock \href {http://arxiv.org/abs/1710.01906} {\path{arXiv:1710.01906}},
  \href {https://doi.org/10.1088/1367-2630/aab475}
  {\path{doi:10.1088/1367-2630/aab475}}.

\bibitem{Huang:2020lxo}
S.~Huang, B.~Shen, Z.~Bu, X.~Zhang, L.~Ji, S.~Zhai, {Axionlike-particle
  generation by laser-plasma interaction} (5 2020).
\newblock \href {http://arxiv.org/abs/2005.02910} {\path{arXiv:2005.02910}}.

\bibitem{Evans:2018qwy}
S.~Evans, J.~Rafelski, {Virtual axion-like particle complement to
  Euler-Heisenberg-Schwinger action}, Phys. Lett. B 791 (2019) 331--334.
\newblock \href {http://arxiv.org/abs/1810.06717} {\path{arXiv:1810.06717}},
  \href {https://doi.org/10.1016/j.physletb.2019.03.008}
  {\path{doi:10.1016/j.physletb.2019.03.008}}.

\bibitem{King:2018qbq}
B.~King, {Electron-seeded ALP production and ALP decay in an oscillating
  electromagnetic field}, Phys. Lett. B 782 (2018) 737--743.
\newblock \href {http://arxiv.org/abs/1802.07507} {\path{arXiv:1802.07507}},
  \href {https://doi.org/10.1016/j.physletb.2018.06.016}
  {\path{doi:10.1016/j.physletb.2018.06.016}}.

\bibitem{Dillon:2018ouq}
B.~M. Dillon, B.~King, {Light scalars: coherent nonlinear Thomson scattering
  and detection}, Phys. Rev. D 99~(3) (2019) 035048.
\newblock \href {http://arxiv.org/abs/1809.01356} {\path{arXiv:1809.01356}},
  \href {https://doi.org/10.1103/PhysRevD.99.035048}
  {\path{doi:10.1103/PhysRevD.99.035048}}.

\bibitem{Vogel:2013raa}
H.~Vogel, J.~Redondo, {Dark Radiation constraints on minicharged particles in
  models with a hidden photon}, JCAP 02 (2014) 029.
\newblock \href {http://arxiv.org/abs/1311.2600} {\path{arXiv:1311.2600}},
  \href {https://doi.org/10.1088/1475-7516/2014/02/029}
  {\path{doi:10.1088/1475-7516/2014/02/029}}.

\bibitem{Gies:2006ca}
H.~Gies, J.~Jaeckel, A.~Ringwald, {Polarized Light Propagating in a Magnetic
  Field as a Probe of Millicharged Fermions}, Phys. Rev. Lett. 97 (2006)
  140402.
\newblock \href {http://arxiv.org/abs/hep-ph/0607118}
  {\path{arXiv:hep-ph/0607118}}, \href
  {https://doi.org/10.1103/PhysRevLett.97.140402}
  {\path{doi:10.1103/PhysRevLett.97.140402}}.

\bibitem{Villalba-Chavez:2016hht}
S.~Villalba-Ch\'avez, S.~Meuren, C.~M\"uller, {Minicharged particles search by
  strong laser pulse-induced vacuum polarization effects}, Phys. Lett. B 763
  (2016) 445--453.
\newblock \href {http://arxiv.org/abs/1608.08879} {\path{arXiv:1608.08879}},
  \href {https://doi.org/10.1016/j.physletb.2016.10.068}
  {\path{doi:10.1016/j.physletb.2016.10.068}}.

\bibitem{Villalba-Chavez:2013txu}
S.~Villalba-Ch\'avez, C.~M\"uller, {Searching for minicharged particles via
  birefringence, dichroism and Raman spectroscopy of the vacuum polarized by a
  high-intensity laser wave}, Annals Phys. 339 (2013) 460--483.
\newblock \href {http://arxiv.org/abs/1306.6456} {\path{arXiv:1306.6456}},
  \href {https://doi.org/10.1016/j.aop.2013.08.013}
  {\path{doi:10.1016/j.aop.2013.08.013}}.

\bibitem{Villalba-Chavez:2014nya}
S.~Villalba-Ch\'avez, C.~M\"uller, {Light dark matter candidates in intense
  laser pulses I: paraphotons and fermionic minicharged particles}, JHEP 06
  (2015) 177.
\newblock \href {http://arxiv.org/abs/1412.4678} {\path{arXiv:1412.4678}},
  \href {https://doi.org/10.1007/JHEP06(2015)177}
  {\path{doi:10.1007/JHEP06(2015)177}}.

\bibitem{Villalba-Chavez:2015xna}
S.~Villalba-Ch\'avez, C.~M\"uller, {Light dark matter candidates in intense
  laser pulses II: the relevance of the spin degrees of freedom}, JHEP 02
  (2016) 027.
\newblock \href {http://arxiv.org/abs/1510.00222} {\path{arXiv:1510.00222}},
  \href {https://doi.org/10.1007/JHEP02(2016)027}
  {\path{doi:10.1007/JHEP02(2016)027}}.

\bibitem{Dobrich:2012sw}
B.~Dobrich, H.~Gies, N.~Neitz, F.~Karbstein, {Magnetically amplified tunneling
  of the 3rd kind as a probe of minicharged particles}, Phys. Rev. Lett. 109
  (2012) 131802.
\newblock \href {http://arxiv.org/abs/1203.2533} {\path{arXiv:1203.2533}},
  \href {https://doi.org/10.1103/PhysRevLett.109.131802}
  {\path{doi:10.1103/PhysRevLett.109.131802}}.

\bibitem{Dobrich:2012jd}
B.~D\"obrich, H.~Gies, N.~Neitz, F.~Karbstein, {Magnetically amplified
  light-shining-through-walls via virtual minicharged particles}, Phys. Rev. D
  87~(2) (2013) 025022.
\newblock \href {http://arxiv.org/abs/1203.4986} {\path{arXiv:1203.4986}},
  \href {https://doi.org/10.1103/PhysRevD.87.025022}
  {\path{doi:10.1103/PhysRevD.87.025022}}.

\bibitem{Fortin:2019npr}
J.-F. Fortin, K.~Sinha, {Photon-dark photon conversions in extreme background
  electromagnetic fields}, JCAP 11 (2019) 020.
\newblock \href {http://arxiv.org/abs/1904.08968} {\path{arXiv:1904.08968}},
  \href {https://doi.org/10.1088/1475-7516/2019/11/020}
  {\path{doi:10.1088/1475-7516/2019/11/020}}.

\bibitem{Born:1933qff}
M.~Born, {Modified field equations with a finite radius of the electron},
  Nature 132~(3329) (1933) 282.1.
\newblock \href {https://doi.org/10.1038/132282a0}
  {\path{doi:10.1038/132282a0}}.

\bibitem{Born:1934gh}
M.~Born, L.~Infeld, {Foundations of the new field theory}, Proc. Roy. Soc.
  Lond. A 144~(852) (1934) 425--451.
\newblock \href {https://doi.org/10.1098/rspa.1934.0059}
  {\path{doi:10.1098/rspa.1934.0059}}.

\bibitem{Ellis17}
J.~Ellis, N.~E. Mavromatos, T.~You,
  \href{https://link.aps.org/doi/10.1103/PhysRevLett.118.261802}{Light-by-light
  scattering constraint on born-infeld theory}, Phys. Rev. Lett. 118 (2017)
  261802.
\newblock \href {https://doi.org/10.1103/PhysRevLett.118.261802}
  {\path{doi:10.1103/PhysRevLett.118.261802}}.
\newline\urlprefix\url{https://link.aps.org/doi/10.1103/PhysRevLett.118.261802}

\bibitem{Davila:2013wba}
J.~M. Davila, C.~Schubert, M.~A. Trejo, {Photonic processes in Born-Infeld
  theory}, Int. J. Mod. Phys. A 29 (2014) 1450174.
\newblock \href {http://arxiv.org/abs/1310.8410} {\path{arXiv:1310.8410}},
  \href {https://doi.org/10.1142/S0217751X14501747}
  {\path{doi:10.1142/S0217751X14501747}}.

\bibitem{Kadlecova:2021bmp}
H.~Kadlecov\'a, {Electromagnetic waves in Born Electrodynamics} (3 2021).
\newblock \href {http://arxiv.org/abs/2103.03575} {\path{arXiv:2103.03575}}.

\bibitem{Kadlecova:2021jqq}
H.~Kadlecov\'a, {Electromagnetic waves in Born--Infeld electrodynamics} (7
  2021).
\newblock \href {http://arxiv.org/abs/2107.12249} {\path{arXiv:2107.12249}}.

\bibitem{Rebhan:2017zdx}
A.~Rebhan, G.~Turk, {Polarization effects in light-by-light scattering:
  Euler\textendash{}Heisenberg versus Born\textendash{}Infeld}, Int. J. Mod.
  Phys. A 32~(10) (2017) 1750053.
\newblock \href {http://arxiv.org/abs/1701.07375} {\path{arXiv:1701.07375}},
  \href {https://doi.org/10.1142/S0217751X17500531}
  {\path{doi:10.1142/S0217751X17500531}}.

\bibitem{Bai:2021dgm}
Z.~Bai, et~al., {LUXE-NPOD: new physics searches with an optical dump at LUXE}
  (7 2021).
\newblock \href {http://arxiv.org/abs/2107.13554} {\path{arXiv:2107.13554}}.

\bibitem{Tizchang:2018mzr}
S.~Tizchang, R.~Mohammadi, S.-S. Xue, {Probing Lorentz violation effects via a
  laser beam interacting with a high-energy charged lepton beam}, Eur. Phys. J.
  C 79~(3) (2019) 224.
\newblock \href {http://arxiv.org/abs/1811.00486} {\path{arXiv:1811.00486}},
  \href {https://doi.org/10.1140/epjc/s10052-019-6716-5}
  {\path{doi:10.1140/epjc/s10052-019-6716-5}}.

\bibitem{Tizchang:2016hml}
S.~Tizchang, S.~Batebi, M.~Haghighat, R.~Mohammadi, {Using an intense laser
  beam in interaction with muon/electron beam to probe the Noncommutative QED},
  JHEP 02 (2017) 003.
\newblock \href {http://arxiv.org/abs/1608.01231} {\path{arXiv:1608.01231}},
  \href {https://doi.org/10.1007/JHEP02(2017)003}
  {\path{doi:10.1007/JHEP02(2017)003}}.

\bibitem{Fresneda:2015zya}
R.~Fresneda, D.~M. Gitman, A.~E. Shabad, {Photon propagation in noncommutative
  QED with constant external field}, Phys. Rev. D 91~(8) (2015) 085005.
\newblock \href {http://arxiv.org/abs/1501.04987} {\path{arXiv:1501.04987}},
  \href {https://doi.org/10.1103/PhysRevD.91.085005}
  {\path{doi:10.1103/PhysRevD.91.085005}}.

\bibitem{Ilderton:2010rx}
A.~Ilderton, J.~Lundin, M.~Marklund, {Strong Field, Noncommutative QED}, SIGMA
  6 (2010) 041.
\newblock \href {http://arxiv.org/abs/1003.4184} {\path{arXiv:1003.4184}},
  \href {https://doi.org/10.3842/SIGMA.2010.041}
  {\path{doi:10.3842/SIGMA.2010.041}}.

\bibitem{Cohen:2006ir}
A.~G. Cohen, S.~L. Glashow, {A Lorentz-Violating Origin of Neutrino Mass?} (5
  2006).
\newblock \href {http://arxiv.org/abs/hep-ph/0605036}
  {\path{arXiv:hep-ph/0605036}}.

\bibitem{Cohen:2006ky}
A.~G. Cohen, S.~L. Glashow, {Very special relativity}, Phys. Rev. Lett. 97
  (2006) 021601.
\newblock \href {http://arxiv.org/abs/hep-ph/0601236}
  {\path{arXiv:hep-ph/0601236}}, \href
  {https://doi.org/10.1103/PhysRevLett.97.021601}
  {\path{doi:10.1103/PhysRevLett.97.021601}}.

\bibitem{Ilderton:2016rqk}
A.~Ilderton, {Very Special Relativity as a background field theory}, Phys. Rev.
  D 94~(4) (2016) 045019.
\newblock \href {http://arxiv.org/abs/1605.04967} {\path{arXiv:1605.04967}},
  \href {https://doi.org/10.1103/PhysRevD.94.045019}
  {\path{doi:10.1103/PhysRevD.94.045019}}.

\bibitem{Ari:2015tca}
V.~Ar\i{}, A.~A. Billur, S.~C. \.Inan, M.~K\"oksal, {Anomalous
  WW\ensuremath{\gamma} couplings with beam polarization at the Compact Linear
  Collider}, Nucl. Phys. B 906 (2016) 211--230.
\newblock \href {http://arxiv.org/abs/1506.08998} {\path{arXiv:1506.08998}},
  \href {https://doi.org/10.1016/j.nuclphysb.2016.02.029}
  {\path{doi:10.1016/j.nuclphysb.2016.02.029}}.

\bibitem{Fujii:2018mli}
K.~Fujii, et~al., {The role of positron polarization for the inital $250$ GeV
  stage of the International Linear Collider} (1 2018).
\newblock \href {http://arxiv.org/abs/1801.02840} {\path{arXiv:1801.02840}}.

\bibitem{2016SurSR..71..547H}
C.~{Hugenschmidt}, {Positrons in surface physics}, Surface Science Reports
  71~(4) (2016) 547--594.
\newblock \href {http://arxiv.org/abs/1611.04430} {\path{arXiv:1611.04430}},
  \href {https://doi.org/10.1016/j.surfrep.2016.09.002}
  {\path{doi:10.1016/j.surfrep.2016.09.002}}.

\bibitem{Torgrimsson:2016ant}
G.~Torgrimsson, J.~Oertel, R.~Sch\"utzhold, {Doubly assisted Sauter-Schwinger
  effect}, Phys. Rev. D 94~(6) (2016) 065035.
\newblock \href {http://arxiv.org/abs/1607.02448} {\path{arXiv:1607.02448}},
  \href {https://doi.org/10.1103/PhysRevD.94.065035}
  {\path{doi:10.1103/PhysRevD.94.065035}}.

\bibitem{Gusynin:1999dynamical}
V.~P. Gusynin, V.~A. Miransky, I.~A. Shovkovy, Dynamical chiral symmetry
  breaking in {QED} in a magnetic field: Toward exact results, Physical review
  letters 83~(7) (1999) 1291.
\newblock \href {http://arxiv.org/abs/9811079} {\path{arXiv:9811079}}, \href
  {https://doi.org/https://doi.org/10.1103/PhysRevLett.83.1291}
  {\path{doi:https://doi.org/10.1103/PhysRevLett.83.1291}}.

\bibitem{Coquereaux:1981fermionic}
R.~Coquereaux, Fermionic expansion in quantum electrodynamics, Physical Review
  D 23~(10) (1981) 2276.
\newblock \href {https://doi.org/https://doi.org/10.1103/PhysRevD.23.2276}
  {\path{doi:https://doi.org/10.1103/PhysRevD.23.2276}}.

\bibitem{Palanques:1984The}
A.~Palanques-Mestre, P.~Pascual, The $1/n_f$ expansion of the $\gamma$ and
  $\beta$ functions in {QED}, Communications in mathematical physics 95~(3)
  (1984) 277--287.
\newblock \href {https://doi.org/https://doi.org/10.1007/BF01212398}
  {\path{doi:https://doi.org/10.1007/BF01212398}}.

\bibitem{Bialynicki-Birula:2021yvi}
I.~Bialynicki-Birula, {New solutions of the Dirac, Maxwell, and Weyl equations
  from the fractional Fourier transform}, Phys. Rev. D 103~(8) (2021) 085001.
\newblock \href {http://arxiv.org/abs/2101.03325} {\path{arXiv:2101.03325}},
  \href {https://doi.org/10.1103/PhysRevD.103.085001}
  {\path{doi:10.1103/PhysRevD.103.085001}}.

\bibitem{Ahmadiniaz:2021fey}
N.~Ahmadiniaz, F.~M. Balli, C.~Lopez-Arcos, A.~Q. Velez, C.~Schubert,
  {Color-kinematics duality from the Bern-Kosower formalism}, Phys. Rev. D
  104~(4) (2021) L041702.
\newblock \href {http://arxiv.org/abs/2105.06745} {\path{arXiv:2105.06745}},
  \href {https://doi.org/10.1103/PhysRevD.104.L041702}
  {\path{doi:10.1103/PhysRevD.104.L041702}}.

\bibitem{Edwards:2022qiw}
J.~P. Edwards, {Graviton scattering amplitudes in first quantisation}, in:
  {19th Mexican School on Particles and Fields}, 2022.
\newblock \href {http://arxiv.org/abs/2201.01697} {\path{arXiv:2201.01697}}.

\bibitem{Behnke:2013xla}
{The International Linear Collider Technical Design Report - Volume 1:
  Executive Summary} (6 2013).
\newblock \href {http://arxiv.org/abs/1306.6327} {\path{arXiv:1306.6327}}.

\bibitem{Baer:2013cma}
{The International Linear Collider Technical Design Report - Volume 2: Physics}
  (6 2013).
\newblock \href {http://arxiv.org/abs/1306.6352} {\path{arXiv:1306.6352}}.

\bibitem{Aicheler:2012bya}
{A Multi-TeV Linear Collider Based on CLIC Technology}: {CLIC Conceptual Design
  Report} (10 2012).
\newblock \href {https://doi.org/10.5170/CERN-2012-007}
  {\path{doi:10.5170/CERN-2012-007}}.

\bibitem{Roloff:2018dqu}
{The Compact Linear e$^+$e$^-$ Collider (CLIC): Physics Potential} (12 2018).
\newblock \href {http://arxiv.org/abs/1812.07986} {\path{arXiv:1812.07986}}.

\bibitem{Cros:2019tns}
B.~Cros, P.~Muggli, {ALEGRO input for the 2020 update of the European Strategy}
  (1 2019).
\newblock \href {http://arxiv.org/abs/1901.08436} {\path{arXiv:1901.08436}}.

\end{thebibliography}

\end{document}